\documentclass[12pt,openany]{book}
\usepackage[margin=1in]{geometry} 
\usepackage{fontawesome}
\usepackage{import}
\usepackage[parfill]{parskip}  
\usepackage{graphicx}
\usepackage{amssymb}
\usepackage{braket}
\usepackage{epstopdf}
\usepackage{mathtools}
\usepackage{subfig}
\usepackage{crimson}
\usepackage[toc,page]{appendix}
\usepackage{cases}
\usepackage{tikz} 
\tikzset{
    ->-/.style={decoration={
  markings,
  mark=at position .5 with {\arrow{>}}},postaction={decorate}},
    -<-/.style={decoration={
  markings,
  mark=at position .5 with {\arrow{<}}},postaction={decorate}},
    ->/.style={decoration={
  markings,
  mark=at position .4 with {\arrow{>}}},postaction={decorate}},
}
\usetikzlibrary{shapes,arrows,positioning,automata,backgrounds,calc,er,patterns}
\usetikzlibrary{calc,positioning,shadows.blur,decorations.pathreplacing}
\usepackage{etoolbox}

\tikzset{%
        brace/.style = { decorate, decoration={brace, amplitude=5pt} },
       mbrace/.style = { decorate, decoration={brace, amplitude=5pt, mirror} },
        label/.style = { black, midway, scale=0.7, align=center },
     toplabel/.style = { label, above=.5em, anchor=south },
    leftlabel/.style = { label,rotate=-90,left=.5em,anchor=north },   
  bottomlabel/.style = { label, below=.5em, anchor=north },
        force/.style = { rotate=-90,scale=0.6 },
        round/.style = { rounded corners=2mm },
       legend/.style = { right,scale=0.6 },
        nosep/.style = { inner sep=0pt },
   generation/.style = { anchor=base }
}

\newcommand\particle[7][white]{%
  \begin{tikzpicture}[scale=1.4,x=1cm, y=1cm]
    \path[fill=#1,blur shadow={shadow blur steps=5}] (0.1,0) -- (0.9,0)
        arc (90:0:1mm) -- (1.0,-0.9) arc (0:-90:1mm) -- (0.1,-1.0)
        arc (-90:-180:1mm) -- (0,-0.1) arc(180:90:1mm) -- cycle;
    \ifstrempty{#7}{}{\path[fill=purple!50!white]
        (0.6,0) --(0.7,0) -- (1.0,-0.3) -- (1.0,-0.4);}
    \ifstrempty{#6}{}{\path[fill=green!50!black!50] (0.7,0) -- (0.9,0)
        arc (90:0:1mm) -- (1.0,-0.3);}
    \ifstrempty{#5}{}{\path[fill=orange!50!white] (1.0,-0.7) -- (1.0,-0.9)
        arc (0:-90:1mm) -- (0.7,-1.0);}
    \draw[\ifstrempty{#2}{dashed}{black}] (0.1,0) -- (0.9,0)
        arc (90:0:1mm) -- (1.0,-0.9) arc (0:-90:1mm) -- (0.1,-1.0)
        arc (-90:-180:1mm) -- (0,-0.1) arc(180:90:1mm) -- cycle;
    \ifstrempty{#7}{}{\node at(0.825,-0.175) [rotate=-45,scale=0.3] {#7};}
    \ifstrempty{#6}{}{\node at(0.9,-0.1)  [nosep,scale=0.23] {#6};}
    \ifstrempty{#5}{}{\node at(0.9,-0.9)  [nosep,scale=0.25] {#5};}
    \ifstrempty{#4}{}{\node at(0.1,-0.1)  [nosep,anchor=west,scale=.4]{#4};}
    \ifstrempty{#3}{}{\node at(0.1,-0.85) [nosep,anchor=west,scale=0.4] {#3};}
    \ifstrempty{#2}{}{\node at(0.1,-0.5)  [nosep,anchor=west,scale=1.5] {#2};}
  \end{tikzpicture}
}

\newcommand{\gE}{E}

\newcommand{\bb}{\boldsymbol}
\newcommand{\gVI}{V^{\textrm{int}}}
\newcommand{\gVE}{V^{\textrm{ext}}}

\newcommand{\EI}{\mathcal{E}}
\newcommand{\EE}{\mathcal{R}}
\renewcommand{\AA}{a}
\newcommand{\sunny}{\text{\faSunO}}
\newcommand{\shady}{\text{\faMoonO}}

\newcommand{\dd}{\textrm{d}}
\newcommand{\pp}{\textrm{p}}

\newcommand{\rr}{\textrm{r}}
\DeclareMathOperator{\Sym}{Sym}
\usepackage{xcolor}
\definecolor{col1}{RGB}{100,143,255}
\definecolor{col2}{RGB}{120, 94, 240}
\definecolor{col3}{RGB}{254,97,0}
\definecolor{col4}{RGB}{220, 38, 127}
\definecolor{col5}{RGB}{255, 176, 0}

\newcommand{\op}{\mathcal{O}}

\newcommand{\mF}{\mathcal{F}}

\newcommand{\mO}{\mathcal{O}}

\usepackage[compat=1.1.0]{tikz-feynman} 

\usetikzlibrary{positioning}
\usetikzlibrary{automata,positioning,arrows,matrix,backgrounds,calc}
\usetikzlibrary{decorations.text}
\usetikzlibrary{decorations.pathmorphing}
\usetikzlibrary{shapes.geometric}
\usetikzlibrary{arrows.meta}

\usepackage{pgf}
\usepackage{relsize}

\usetikzlibrary{decorations,patterns,arrows.meta}

\usetikzlibrary{decorations.pathreplacing,decorations.markings}
\usepackage{empheq}
\usepackage{slashed}
\usepackage{changepage}
\usepackage{stackrel}
\newcommand{\ba}{\begin{align}}
\newcommand{\ea}{\end{align}}

\usepackage{color} 
\usepackage{braket}
\usepackage[sorting=none]{biblatex}
\newcommand\ddfrac[2]{\frac{\displaystyle #1}{\displaystyle #2}}

\renewcommand{\arraystretch}{1.5} 

\graphicspath{{D:/Users/alexandresalas/Documents/Imagenes Latex/}}
\usepackage[colorlinks=true, pdfstartview=FitV, linkcolor=blue, citecolor=blue, urlcolor=blue]{hyperref}

\newcommand{\be}{\begin{equation}}
\newcommand{\ee}{\end{equation}}

\newcommand{\ep}{\epsilon}

\usepackage{amssymb}
\usepackage[normalem]{ulem}
\newcommand{\ff}{\mathfrak{f}}

\begin{document}
\begin{titlepage}
	\centering
{\LARGE Analytical Quantum Field methods\\
in\\
\vspace{.5cm}
Particle Physics}\\
\vspace{1cm}
\vspace{1cm}

\includegraphics[width=50mm]{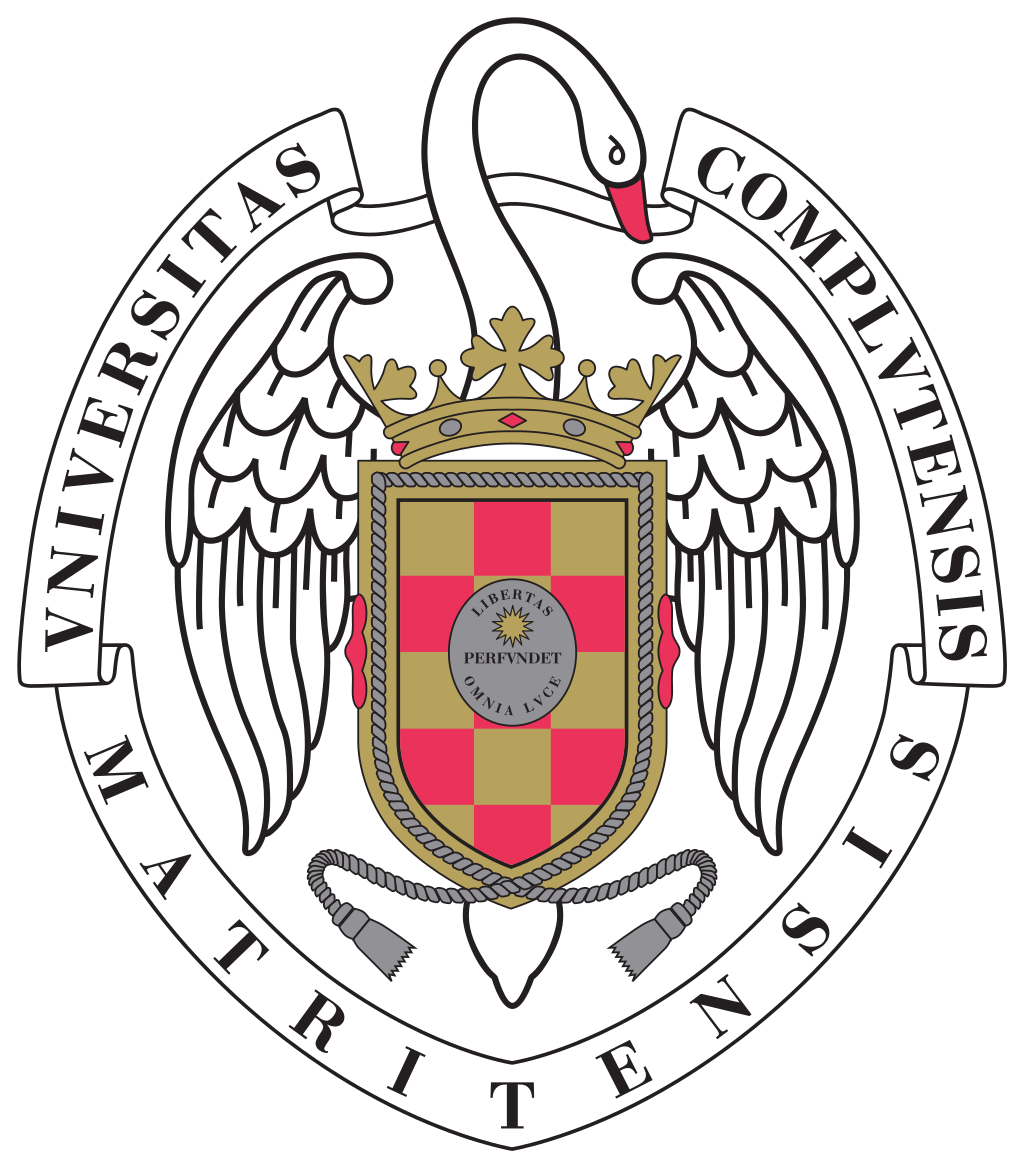}

	{\scshape\Large Universidad Complutense de Madrid\par}
 {\scshape\large Facultad de Ciencias Físicas\par}
	\vspace{1cm}
	{\scshape\Large \par}
	\vspace{1.5cm}
	{\huge\itshape Alexandre Salas Bern\'ardez\par}
	\vfill
	\vfill
	\vfill
	\vfill
	 Supervised by Felipe Llanes-Estrada

	\vfill

	{\large Mayo 2023\par}
\end{titlepage}
\pagenumbering{gobble}
\frontmatter
\frontmatter
\newpage $\;$\pagenumbering{gobble}\newpage
\newpage $\;$\pagenumbering{gobble}\chapter*{Agradecimientos/Acknowledgments}
Quiero agradecer profundamente a Felipe José Llanes Estrada todos estos años de trabajo conjunto y su magnífica dirección. Ha sido una experiencia muy agradable a la par que intelectualmente estimulante. Quiero también agradecer a mis colaboradores Juan José Sanz Cillero y Raquel Gómez Ambrosio el haber hecho de un trabajo de investigación una vivencia amena y divertida. Doy también las gracias a todos y todas mis compañeras de estudios doctorales por haber creado un ambiente cómodo y acogedor en el departamento.

I also want to thank Eric Laenen for his sustained support over the years and his excellent guidance. Additionally, I am extremely thankful for Michael Borinsky's passion and goodwill during our shared research project. I also want to acknowledge the kind hospitality extended to me by Reinhard Alkofer, Aleksi Kurkela and their respective organizations.

Quiero expresar el más profundo agradecimiento a mi madre, cuyo trabajo de sostén y cuidado y sus consejos han sido esenciales para mi éxito académico. Agradezco también a mi padre su apoyo moral y amor incondicional. Asimismo, quiero agradecer a Juan Ferrera su gran paciencia y dedicación al enseñarme el noble arte de las matemáticas. A mis amigos y amigas  les agradezco de corazón su presencia, su gran amor, su apoyo y todos los grandes momentos compartidos.

\newpage
\pagenumbering{gobble}
\tableofcontents

\chapter{List of publications}
\setcounter{page}{1}
\pagenumbering{roman}

The research activity conducted during the duration of this thesis has produced the following peer-reviewed publications:
\begin{enumerate}
    \item \textit{Distinguishing electroweak EFTs with $W_L W_L\to n\times h$}, R.~G\'omez-Ambrosio, F.~J.~Llanes-Estrada, A.~Salas-Bern\'ardez and J.~J.~Sanz-Cillero,
\textbf{Phys. Rev. D 106, no.5, 5 (2022)}.
\item \textit{Systematizing and addressing theory uncertainties of unitarization with the Inverse Amplitude Method}, A.~Salas-Bern\'ardez, F.~J.~Llanes-Estrada, J.~Escudero-Pedrosa and J.~A.~Oller, \textbf{SciPost Phys. \textbf{11}, no.2, 020 (2021)}.
\item \textit{Chiral symmetry breaking for fermions charged under large Lie groups}, F.~J.~Llanes-Estrada and A.~Salas-Bern\'ardez,
\textbf{Commun. Theor. Phys. \textbf{71}, no.4, 410-416 (2019)}.
\item \textit{Explicit computation of jet functions in coordinate space}, A.~Salas-Bern\'ardez,
\textbf{Nucl. Phys. B \textbf{985}, 116024 (2022)}.
\item 
\textit{Flow-oriented perturbation theory}, M.~Borinsky, Z.~Capatti, E.~Laenen and A.~Salas-Bern\'ardez, \textbf{JHEP \textbf{01}, 172 (2023)}.
\end{enumerate}
The contents included in this thesis have been presented in various conferences, producing the following Proceedings:
\begin{enumerate}
    \item \textit{SMEFT as a slice of HEFT's parameter space}, A.~Salas-Bernardez, J.~J.~Sanz-Cillero, F.~J.~Llanes-Estrada and R.~Gomez-Ambrosio, \textbf{EPJ Web Conf. \textbf{274}, 08013 (2022)}.
    \item \textit{How to predict a new physics scale and its uncertainty with the Inverse Amplitude Method}, A.~Salas-Bern\'ardez, \textbf{PoS \textbf{EPS-HEP2021}, 346 (2022)}.
    \item \textit{Assessment of systematic theory uncertainties in IAM unitarization}, J.~Escudero-Pedrosa, F.~J.~Llanes-Estrada, J.~A.~Oller and A.~Salas-Bern\'ardez, \textbf{Nucl. Part. Phys. Proc. \textbf{312-317}, 82-86 (2021)}.
    \item \textit{Resonances in unitarized HEFT at the LHC}, A.~Dobado, F.~J.~Llanes-Estrada, A.~Salas-Bernardez and R.~L.~Delgado, 
\textbf{PoS \textbf{EPS-HEP2019}, 362 (2020)}.
    \end{enumerate}

    Other unpublished documents are:
    \begin{itemize}
        \item \textit{SMEFT is falsifiable through multi-Higgs measurements (even in the absence of new light particles)}, R.~G\'omez-Ambrosio, F.~J.~Llanes-Estrada, A.~Salas-Bern\'ardez and J.~J.~Sanz-Cillero, 2207.09848.
    \end{itemize}
\newpage

\listoffigures

\listoftables

\chapter{Preface}
This thesis is based mainly on the candidate's publications mentioned above. After a general introduction, the original contributions are included in chapters 3, 4 and 5. With some further original contributions in subsection \ref{subsc:CTP}. 

\section*{Notations and conventions}
Throughout this thesis the units used set
\begin{equation*}
    \hbar=c=1\;,
\end{equation*}
where $c$ is the speed of light in the vacuum and $\hbar$ is the reduced Planck constant. 

With the exception of section \ref{sec:explicitcomputationjetCS}, the convention used for the metric of spacetime is the ``mostly minus'' metric $g_{\mu\nu}=\text{diag } (1,-1,-1,-1)$; four vectors are denoted by light italic type; three-vectors by boldface type (except for that section, where they expressed with an arrow on top); unit (four)-vectors have a hat on top. As an example $x^\mu=(x^0,\boldsymbol{x})$ and $x_\mu=(x^0,-\boldsymbol{x})$, with $\partial_\mu=(\partial_0,\boldsymbol{\nabla})$. 

The Pauli and Dirac matrices (in the Dirac representation), Dirac spinors, as well as the conventions on Fourier transforms are taken from those of \cite{Peskin:1990zt}.

\chapter{Introduction}
  \setcounter{chapter}{1}
\setcounter{page}{1}
\pagenumbering{arabic} 
Around the beginning of the 20th century, the limitations of the classical approach to describe microscopic physical processes were becoming more and more apparent, and the first quantum descriptions of phenomena started appearing: Planck's hypothesis on the quantization of energy levels to model black body radiation \cite{Planck}, Einstein's explanation of the photoelectric effect \cite{Einstein}, Bohr's description of the hydrogen's atom spectral lines \cite{Bohr}, and many more. These findings opened a new exciting highway of research in physics that culminated in the condensation of Quantum Mechanics into a few postulates
by Von Neumann \cite{Neumann}. Quantum Mechanics was a brilliant and intriguing way of describing microscopic phenomena at low energies. However, the unification of the Quantum theory with Einstein's theory of special relativity \cite{Einstein:1905ve} was not an easy task.  Given the success of the models presented independently by Pauli \cite{Pauli:1927qhd} and Darwin \cite{Darwin:1927du} describing the electron as a spin-$\frac{1}{2}$ particle to explain hydrogen-line fine-splitting spectra, Paul Dirac managed to describe the nature of the electron resorting only to its quantum behaviour and the basic principle of special relativity in his paper \textit{``The Quantum Theory of the Electron''} \cite{Dirac:1928hu}. At the same time, many efforts were made in the direction of describing the radiation produced by electrons and the photon was already being described as a field. This perspective led to the description of arbitrary particles as quantum fields by Heisenberg and Pauli \cite{Heisenberg:1929xj,Heisenberg:1930xk}. It is also worth mentioning that Heisenberg and Pauli introduced the Lagrangian formulation of Quantum Field Theory (QFT), which is widely used in modern particle physics and throughout this thesis. This formulation helped deriving the equations of motion for the fields via an action principle, while they quantized the fields by promoting them to operators on a certain (later called) Fock space.

From here on, addressing multiple new problems (interpretation of probabilities, infinities in the theory and many more), the use of quantum field theories to describe the interactions of elementary particles started to become  standard. 

At the same time, the ``gauge'' properties of the fundamental forces of Nature became, with time, evident, to the point that, nowadays, all forces known (the electromagnetic, the weak and strong nuclear forces) are expressed in terms of a gauge quantum field theory. Also, the gauge essence of the theory of gravity (which is not a force of Nature in the sense of the others) was also used and noticed by Hilbert, who derived in 1915 Einstein's field equations for General Relativity by assuming the gauge invariance of the action under a general coordinate transformation \cite{Hilbert:1915tx}.

The unprecedented experimental accuracy of quantum field theories has proven it to be a very powerful method for understanding the microscopic world. In this thesis, the application of the Quantum Field Theory framework to a wide variety of topics in Particle Physics proves again the potentiality of this language in the research of cutting edge physics topics.

\section{The Standard Model and beyond}

Throughout the 20th century, many developments and discoveries led to the vast understanding of the theories describing the nature of particles and interactions in the observable universe that we have at the present day. The model describing all known forces and fundamental particles is called the Standard Model (SM), term introduced in \cite{Pais:1975gn}. This model characterizes in detail the electroweak and strong interactions and is compatible with the vast majority of experimental observations and data \cite{ParticleDataGroup:2022pth}. We will briefly discuss the main features of the SM and also mention the experimental hints that point to the possibility of having physics beyond the Standard Model (BSM).

The SM is a renormalizable QFT (\textit{i.e.} infinities in the theory can be cured and correctly interpreted, see Sect. \ref{sec:PTandrenormalization}) where both the strong and electroweak interactions are described by gauge theories. It is based on the Lie group (for an introduction to Lie groups, see \cite{Zee:2016fuk}) symmetries
\begin{equation}
SU(3)_c\otimes SU(2)_L\otimes U(1)_Y\;,\label{eq:SMLieGroups}
\end{equation}
where $SU(3)_c$ denotes the color interaction responsible
for the strong force, $SU(2)_L$ the isospin coupling of lefthanded fermions and $U(1)_Y$ the hypercharge group. 

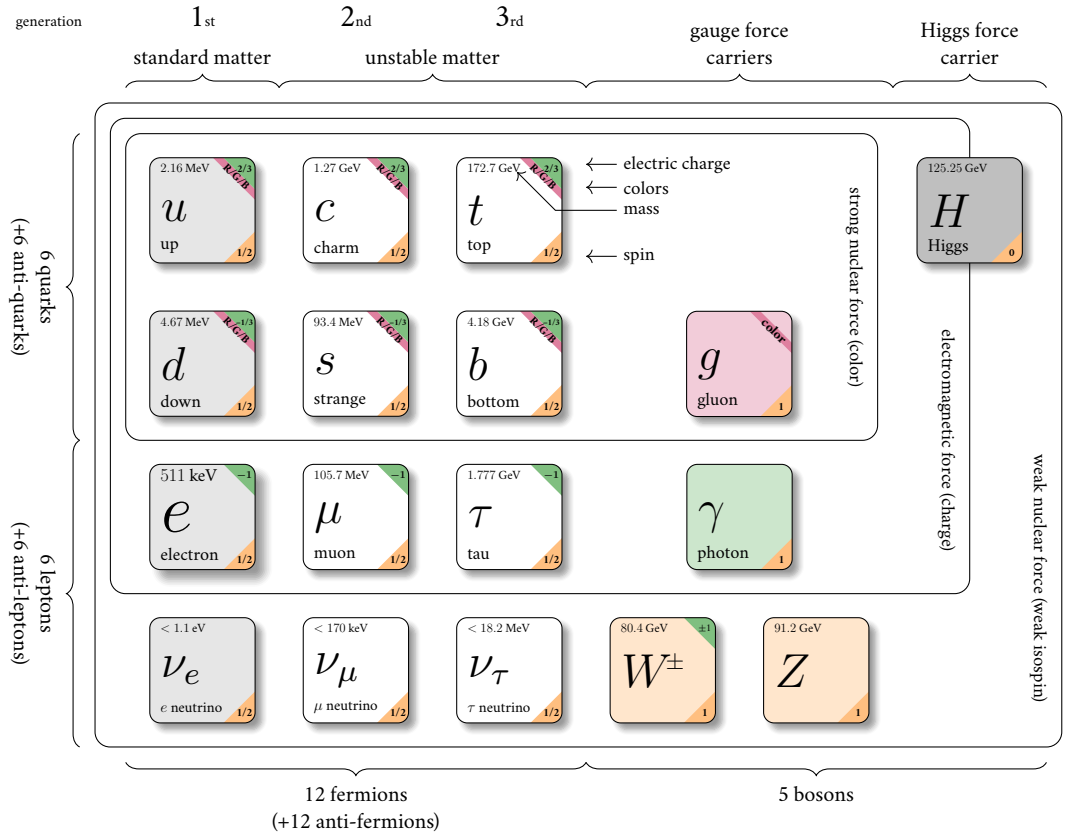
\begin{figure}[ht!]
\centering
\begin{tikzpicture}[scale=1.7,x=1.2cm, y=1.2cm]
  \draw[round] (-0.5,0.5) rectangle (4.4,-1.5);
  \draw[round] (-0.6,0.6) rectangle (5.0,-2.5);
  \draw[round] (-0.7,0.7) rectangle (5.6,-3.5);

  \node at(0, 0)   {\particle[gray!20!white]
                   {$u$}        {\Large up}       {$2.16$ MeV}{\Large \textbf{1/2} }{\Large\textbf{2/3}}{\large\textbf{R/G/B}}};
  \node at(0,-1)   {\particle[gray!20!white]
                   {$d$}        {\Large down}    {$4.67$ MeV}{\Large \textbf{1/2}}{\large\textbf{--1/3}}{\large\textbf{R/G/B}}};
  \node at(0,-2)   {\Large \particle[gray!20!white]
                   {$e$}        {\Large electron}       {$511$ keV}{\Large\textbf{1/2}}{\Large$\boldsymbol{-1}$}{}};
  \node at(0,-3)   {\particle[gray!20!white]
                   {$\nu_e$}    {\large $e$ neutrino}         {$<1.1$ eV}{\Large\textbf{1/2}}{}{}};
  \node at(1, 0)   {\particle
                   {$c$}        {\Large charm}   {$1.27$ GeV}{\Large\textbf{1/2}}{\Large\textbf{2/3}}{\large\textbf{R/G/B}}};
  \node at(1,-1)   {\particle 
                   {$s$}        {\Large strange}  {$93.4$ MeV}{\Large\textbf{1/2}}{{\large\textbf{--1/3}}}{\large\textbf{R/G/B}}};
  \node at(1,-2)   {\particle
                   {$\mu$}      {\Large muon}         {$105.7$ MeV}{\Large\textbf{1/2}}{\Large$\boldsymbol{-1}$}{}};
  \node at(1,-3)   {\particle
                   {$\nu_\mu$}  {\large$\mu$ neutrino}    {$<170$ keV}{\Large\textbf{1/2}}{}{}};
  \node at(2, 0)   {\particle
                   {$t$}        {\Large top}    {$172.7$ GeV}{\Large\textbf{1/2}}{\Large\textbf{2/3}}{\large\textbf{R/G/B}}};
  \node at(2,-1)   {\particle
                   {$b$}        {\Large bottom}  {$4.18$ GeV}{\Large\textbf{1/2}}{\large\textbf{--1/3}}{\large\textbf{R/G/B}}};
  \node at(2,-2)   {\particle
                   {$\tau$}     {\Large tau}          {$1.777$ GeV}{\Large\textbf{1/2}}{\Large$\boldsymbol{-1}$}{}};
  \node at(2,-3)   {\particle
                   {$\nu_\tau$} {\large$\tau$ neutrino}  {$<18.2$ MeV}{\Large\textbf{1/2}}{}{}};
  \node at(3,-3)   {\particle[orange!20!white]
                   {$W^{\hspace{-.3ex}\scalebox{.5}{$\pm$}}$}
                                {}              {$80.4$ GeV}{\Large\textbf{1}}{\large$\boldsymbol{\pm}$\Large1}{}};
  \node at(4,-3)   {\particle[orange!20!white]
                   {$Z$}        {}                    {$91.2$ GeV}{\Large\textbf{1}}{}{}};
  \node at(3.5,-2) {\particle[green!50!black!20]
                   {$\gamma$}   {\Large photon}                        {}{\Large\textbf{1}}{}{}};
  \node at(3.5,-1) {\particle[purple!20!white]
                   {$g$}        {\Large gluon}                    {}{\Large\textbf{1}}{}{\Large\textbf{ color}}};
  \node at(5,0)    {\particle[gray!50!white]
                   {$H$}        {\Large Higgs}              {$125.25$ GeV}{\Large\textbf{0}}{}{}};

  \node at(4.25,-0.5) [force]      {strong nuclear force (color)};
  \node at(4.85,-1.5) [force]    {electromagnetic force (charge)};
  \node at(5.45,-2.4) [force] {weak nuclear force (weak isospin)};

  \draw [<-] (2.5,0.3)   -- (2.7,0.3)          node [legend] {electric charge};
  \draw [<-] (2.5,0.15)  -- (2.7,0.15)         node [legend] {colors};
  \draw [<-] (2.05,0.25) -- (2.3,0) -- (2.7,0) node [legend]   {mass};
  \draw [<-] (2.5,-0.3)  -- (2.7,-0.3)         node [legend]   {spin};

  \draw [mbrace] (-0.8,0.5)  -- (-0.8,-1.5)
                 node[leftlabel] {6 quarks\\(+6 anti-quarks)};
  \draw [mbrace] (-0.8,-1.5) -- (-0.8,-3.5)
                 node[leftlabel] {6 leptons\\(+6 anti-leptons)};
  \draw [mbrace] (-0.5,-3.6) -- (2.5,-3.6)
                 node[bottomlabel]
                 {12 fermions\\(+12 anti-fermions)};
  \draw [mbrace] (2.5,-3.6) -- (5.5,-3.6)
                 node[bottomlabel] {5 bosons};

  \draw [brace] (-0.5,.8) -- (0.5,.8) node[toplabel]         {standard matter};
  \draw [brace] (0.5,.8)  -- (2.5,.8) node[toplabel]         {unstable matter};
  \draw [brace] (2.5,.8)  -- (4.5,.8) node[toplabel]          {gauge force\\
  carriers};
  \draw [brace] (4.5,.8)  -- (5.5,.8) node[toplabel]       {Higgs force\\carrier};

  \node at (0,1.2)   [generation] {1\tiny st};
  \node at (1,1.2)   [generation] {2\tiny nd};
  \node at (2,1.2)   [generation] {3\tiny rd};
  \node at (-1,1.2) [generation] {\tiny generation};
\end{tikzpicture}
\caption[Particle content in the SM]{ Particle content in the SM. The quoted particle's masses are taken from \cite{ParticleDataGroup:2022pth} and we omit their uncertainties. In this thesis we will treat with all these particles except for the leptons, included here for completeness. (Modified from \href{https://texample.net/tikz/examples/model-physics/}{\nolinkurl{texample.net}}).}\label{fig:SMcontent}
\end{figure}

The SM contains three parts (see Figure \ref{fig:SMcontent}): a matter sector described by fermionic fields, the force sector described by vector boson gauge fields, and the so called electroweak symmetry breaking sector (EWSBS). The matter sector contains three generations, each one made of two quark flavors (up and down-like quarks) and two leptons (left-handed neutrino and electron-like), with all their antiparticles. The gauge fields  describing the strong interactions are called gluons $g$, whereas the electroweak gauge fields are, after symmetry breaking, the $W^\pm$ and $Z$ massive weak force bosons and the familiar $\gamma$ electromagnetic photon. The Electroweak Symmetry Breaking Sector (EWSBS) is responsible for the massive nature of the weak force bosons and has the famous Higgs boson as a physical particle, theorized in 1964 \cite{Englert:1964et,Higgs:1964pj,Guralnik:1964eu} and finally discovered in 2012 \cite{ATLAS:2012yve}. The couplings of all matter fields (except neutrinos) with the Higgs boson are encoded in the so called Yukawa interactions, which generate the ``current'' masses of these matter fields \cite{ParticleDataGroup:2022pth}. 

Up to date, almost all experiments are consistent with the SM. Some exceptions are:
\begin{itemize}
\item The \textbf{measured value of the muon's gyromagnetic ratio}, first at Brookhaven \cite{Muong-2:2006rrc} and confirmed recently at Fermilab \cite{Muong-2:2021ojo}, is in some tension ($4.2\sigma$) with the SM theoretical prediction from \cite{Aoyama:2020ynm}. Lattice BMW predictions are in better agreement with the SM, theorists do not seem to agree with each other on the SM theoretical value. See \cite{kesha} for a detailed review.
\item  The \textbf{mass measurement of the $W^\pm $} mass by the CDF collaboration \cite{CDF:2022hxs} with an uncertainty band that leaves well outside the SM prediction. Nonetheless, recent ATLAS results are in agreement with the SM value for the $W^\pm$ mass \cite{ATLAS:2023fsi}.
\end{itemize}

Apart from these experimental tensions, the SM seems not to be the fundamental theory for Nature's particle interactions at all energies for several other reasons:
\begin{itemize}
    \item \textbf{Neutrino's masses:} from the discovery of the electron neutrino in 1956 \cite{Cowan:1956rrn} it was widely believed that neutrinos were massless. This was in agreement with the fact that neutrinos were observed to be only left-handed and not right-handed (regarding chirality). However, when the Super-Kamiokande experiment \cite{Super-Kamiokande:1998kpq} confirmed the oscillation of neutrino flavours, the need for neutrino's masses was reinstated (and hence the existence of right-handed neutrinos).   This poses the question, how can the SM accommodate the needed right-handed neutrinos?

    \item \textbf{The naturalness problem:} if one views the SM as an effective field theory at low energies, valid below a scale where new degrees of freedom might appear, naturalness is the statement that the nature of this effective field theory should not be too much dependent on the corresponding high energy complete theory. However, having a scalar particle as a fundamental field in the SM poses a problem regarding the effect of quantum radiative corrections to the Higgs boson's mass. These corrections to its mass become very large at high energies (if a cut-off scale in the theory is assumed, f.e. the Planck scale) and hence the ``small'' value of the Higgs boson mass is viewed as a \textit{fine tuning problem}.  See \cite{Dine:2015xga} for a detailed modern review.
    
    \item \textbf{Cosmic inflation and matter-antimatter asymmetry:} the theory of inflation is very appealing since it helps solving several problems in fundamental physics such as the flatness problem and the horizon problem \cite{Guth:1980zm,Linde:1984ir}, but it might require new physics. Also, measurements (COBE \cite{Boggess:1992xla} or more recently WMAP \cite{WMAP:2012nax}) addressing the cosmological evolution of our universe support the occurrence of an inflationary period at its very first moments. This and the fact that there is more matter than antimatter in the Universe cannot be explained by the SM alone \cite{Riotto:1999yt}.
    \item \textbf{Dark matter evidence:}  Furthermore, throughout the last century, there has been mounting experimental evidence that the majority of matter in the universe is not part of the SM (recent data from Planck collaboration suggest there is roughly 5 times more dark matter than ordinary SM matter \cite{Planck:2018vyg}). Signaling the possibility that there are new stable dark particles.
\end{itemize}
To address these problems, many researchers have pursued specific models that can accommodate these tensions with the SM. Some examples are: the Two Higgs Doublet Model \cite{Bhattacharyya:2015nca}, modelling flavor changing neutral currents; technicolor \cite{Weinberg:1975gm,}, the composite Higgs models \cite{Kaplan:1983fs} or supersymmetric models, which could alleviate the naturalness problem (see \cite{ParticleDataGroup:2022pth} for a review); the dilaton model for inflation and baryon asymmetry \cite{Bruggisser:2022ofg}; and many more). Many of these models are inspired by the faint hint of a possible grand unification theory of the strong and electroweak forces, since the running couplings of the gauge groups in eq. (\ref{eq:SMLieGroups}) seem to become comparable around a scale of $10^{15}$ GeV. We will use this hint to address why the SM has those specific symmetries and no other in section \ref{subsc:CTP}.

In this thesis and specifically in Chapters 3 and 4, our point of view is more humble, we do not seek for specific completions of the SM. Instead, we will assume that there is New Physics at a very high energy compared to the energies hitherto reached by accelerators and treat with an effective field theory (EFT) built up with the observed particle content. Within these EFTs, the hope is to observe new forces among the known particles that may be mediated by very heavy dark or new unknown particles (since direct detection of these may be impossible at present). We hence rely on the improvement of precision on current or near-future experiments (such as High-Luminosity LHC \cite{Bruning:2015dfu}) in order to be able to extrapolate low energy data to assess the nature of physics beyond the Standard Model. To this end, we concentrate on the EWSBS since it is deeply related to most of the issues with the SM raised above. In Chapter 3, we focus on how to experimentally distinguish the different EFTs for this sector and, in Chapter 4, on how to put to use dispersive methods to predict the energies that must be reached in order to directly access the New Physics. 

The search for new physics also requires accurate control of the SM pieces. Whereas Electromagnetism and the Weak nuclear force can almost always be perturbatively treated (reaching hence high precision), the Strong nuclear force does not. At low energies this interaction is so intense that its particle degrees of freedom, the quarks and gluons, appear confined inside hadrons (such as the proton, the neutron or the pions) through the process called Confinement. This process is still not completely understood and poses one of the biggest problems in modern theoretical physics, we will study this problem in section \ref{sc:Confinement}. 
In this low-energy strongly-interacting regime of the strong nuclear force, the dispersive methods studied in Chapter 4 and the assessment of their uncertainties is crucial for achieving precision. On the other hand, at high energies this force can be perturbatively treated and in Chapter 5 we study this regime from the coordinate space viewpoint of QFTs, introducing also a novel and promising formalism called Flow Oriented Perturbation Theory.

In Chapter 2 we will introduce the main tools that are used in this thesis to study QFTs both in the perturbative and non-perturbative regimes.

\chapter{Quantum Field Theories: perturbative and non-perturbative regimes}
\setcounter{chapter}{2}
\setcounter{section}{0}
\setcounter{equation}{0}
\setcounter{table}{0}
\setcounter{figure}{0}

In this chapter we will review, as a theoretical framework, the main definitions and concepts of quantum field theories used in this thesis. When relevant, we will introduce some novel treatments that have been researched during this thesis' duration (specifically section \ref{subsc:CTP}).

\section{The gauge principle}\label{sec:electrodynamics}

As early as 1865, Maxwell introduced his equations to describe electromagnetism \cite{Maxwell:1865zz} and noticed that there was some freedom in the choice of the electromagnetic potentials (the electric potential $\phi$ and the vector potential $\boldsymbol{A}$, usually cast as $A^\mu=(\phi,\boldsymbol{A})$) which define the electric $\boldsymbol{E}=-\boldsymbol{\nabla}\phi-\partial_t \boldsymbol{A}$ and magnetic $\boldsymbol{B}=\boldsymbol{\nabla}\times \boldsymbol{A}$ fields: one can shift $\phi\to \phi+\partial_t f$ and $\boldsymbol{A}\to \boldsymbol{A}-\boldsymbol{\nabla}f$ without changing the electric and magnetic fields. This freedom is called the \textit{gauge freedom} in the description of the electromagnetic interactions (this term was introduced by Weyl in 1918 in an attempt to unify gravity and electromagnetism \cite{Weyl:1918ib}).

One can exploit this freedom for choosing a condition over $A^\mu$ that simplifies the problem at hand. This choice is usually regarded as \emph{fixing a gauge} and effectively reduces the available degrees of freedom (four) to match the fact that electromagnetic waves have only two polarizations, which are transversal to the direction of propagation (one degree of freedom is eliminated through the gauge fixing and the other through the condition that waves propagate at the speed of light). Maxwell's equations can be written in covariant form as
\begin{equation}
\partial^\mu F_{\mu\nu}=Qe j_\nu  \;\;\;\text{ with } \;\;\; F_{\mu\nu}=\partial_\mu A_\nu-\partial_\nu A_\mu\;,\label{eq:Maxwelleq}
\end{equation}
where $j^\mu=(\rho,\boldsymbol{j})$ is the electromagnetic four-current, $Q$ is the amount of the proton's elementary charge, $e$, and $\rho$ and $\boldsymbol{j}$ are the charge distribution and electric current respectively. On the other hand the charged particles (electrons or protons) are fermions, described by the Dirac field $\psi$ obeying the Dirac equation
\begin{equation}
 (i\gamma^\mu \partial_\mu-m )\psi=0\;.
 \end{equation}
In order to treat with full Electrodynamics (with both photons and charged particles), one needs to couple the Dirac field to the electromagnetic field. To do so, working in analogy with classical electrodynamics, where, in order to reproduce the Lorentz force $Qe(\boldsymbol{E}+\boldsymbol{p}\times \boldsymbol{B}/m)$ one needs to replace the momentum $\boldsymbol{p}$ by $\boldsymbol{p}-Qe\boldsymbol{A}$ and the Hamiltonian $H$ by $H-Qe\phi$, one must modify the corresponding quantum operators as $\partial_\mu \to \partial_\mu+iQe A_\mu$ to obtain the interacting Dirac equation
\begin{equation}\label{eq:interactingDirac}
\big[i\gamma^\mu \left(\partial_\mu+iQe A_\mu \right)-m\big]\psi=0\;. 
\end{equation}
Both equations (\ref{eq:Maxwelleq}) and (\ref{eq:interactingDirac}) are the basic equations of electrodynamics (ED). To obtain these equations from an action principle, one can define the Lagrangian density
\begin{equation}
\mathcal{L}_{\text{ED}}=-\frac{1}{4}F_{\mu\nu} F^{\mu\nu}+\bar{\psi} \left[i\gamma^\mu\left(\frac{1}{2}\overset{\text{\tiny$\leftrightarrow$}}{\partial}_\mu+iQeA_\mu\right)-m\right]\psi\label{eq:Lagrangianelectrodyn}\;,
\end{equation}
where $\bar{\psi}=\psi^\dagger \gamma^0$ and $a\overset{\text{\tiny$\leftrightarrow$}}{\partial}_\mu b=a\partial_\mu b-(\partial_\mu a)b$. This Lagrangian is Hermitean, Lorentz invariant and Maxwell's and the interacting Dirac equations can be derived from the Euler-Lagrange equations
\begin{equation}
    \partial_\mu \frac{\partial \mathcal{L}}{\partial (\partial_\mu A_\nu)}-\frac{\partial \mathcal{L}}{\partial A_\nu}=0\;\;,\;\;\;\partial_\mu \frac{\partial \mathcal{L}}{\partial (\partial_\mu \bar{\psi}_\alpha )}-\frac{\partial \mathcal{L}}{\partial \bar{\psi}_\alpha}=0\;,
\end{equation}
where $\alpha=1,..., 4$ is the index selecting the component of the Dirac field.

Also, the Lagrangian in (\ref{eq:Lagrangianelectrodyn}) is invariant under the global $U(1)$ (the unitary group in one dimension) transformation $\psi\to e^{-iQe\theta}\psi$ for real $\theta$. According to Noether's theorem this invariance gives rise to the conserved current $j^\mu=\bar{\psi}\gamma^\mu \psi$, $\partial_\mu j^\mu=0$ ensuring the conservation of the electric charge.\

Furthermore, the $\theta$ spinor phase can be made coordinate dependent, $\theta=\theta(x)$, and the Lagrangian in (\ref{eq:Lagrangianelectrodyn}) is still invariant under the local transformation $\psi\to e^{-iQ\theta(x)}\psi$ provided $A^\mu$ transforms as $A^\mu\to A^\mu+\partial^\mu \theta /e$ (which is nothing but the gauge transformation we discussed at the beginning of this section). The combination $D_\mu\equiv(\partial_\mu+iQeA_\mu)$ is called the \textit{covariant derivative} since $D_\mu\psi\to e^{-iQ\theta(x)}D_\mu \psi$. Summing up, classical electrodynamics can be cast as a $U(1)$ \textbf{local gauge field theory} with fermions as matter fields as in (\ref{eq:Lagrangianelectrodyn}). 

We now turn to the discussion on how to quantize scalar field theories and will leave the quantization of gauge (Abelian or non-Abelian) field theories for section \ref{sec:YangMills}.

\section{Quantization of scalar field theories}\label{sec:Quantization}

Following quantum mechanics, Heisenberg and Pauli \cite{Heisenberg:1929xj} studied quantum field theories by promoting the fields to be operators with certain canonical commutation relations. This procedure is called the canonical operator formalism, and all the Green's functions of the theory (\textit{i.e.} the functions that help solving the inhomogeneous equations for the fields) are expressed in terms of a vacuum expectation value of field operators. For example, take the free scalar field with Lagrangian 
\begin{equation}
\mathcal{L}_{\text{KG}}=\frac{1}{2}(\partial_\mu \varphi\partial^\mu \varphi-m^2\varphi^2)\;,
\end{equation}
with the Klein-Gordon (KG) equation as equation of motion (with $\Box\equiv \partial_\mu \partial^\mu$)
\begin{equation}
    \Box \varphi +m^2\varphi=0\;.
\end{equation}
Similarly to the harmonic oscillator in QM, this equation is solved in terms of Fourier modes and creation and annihilation operators as (defining $E_{\boldsymbol{p}}\equiv\sqrt{\boldsymbol{p}^2+m^2}$)
\begin{align}
    {\varphi}(\boldsymbol{x},t)=\int \frac{d^3\boldsymbol{p}}{(2\pi)^3\sqrt{2E_{\boldsymbol{p}}}} (a({\boldsymbol{p}})e^{-i(E_{\boldsymbol{p}}t-\boldsymbol{p}\cdot \boldsymbol{x})}+a^\dagger({\boldsymbol{p}})e^{i(E_{\boldsymbol{p}}t-\boldsymbol{p}\cdot \boldsymbol{x})})\;.\label{eq:ScalarFourier}
\end{align}

This theory is quantized by imposing the canonical commutation relations
\begin{equation}
[\hat{\varphi}(x),\hat{\pi}(y)]=i\delta^{(4)}(x-y)\;,
\end{equation}
where the conjugate momentum operator is obtained from promoting the conjugate momentum $\pi(x)=\frac{\delta \mathcal{L}_0}{{\delta (\partial_0\varphi)}}=\partial_t \varphi(x)$ and the field $\varphi$ to operators. Among the Fourier modes, the creation and annihilation operators obey the only non-trivial commutation relation 
\begin{equation}
    [\hat{a}({\boldsymbol{p}}),\hat{a}^\dagger({\boldsymbol{q}})]=(2\pi)^3\delta^{(3)}(\boldsymbol{p}-\boldsymbol{q})\;.\label{eq:commutation}
\end{equation}
The vacuum of the free theory, $\ket{0}$, is such that $ \hat{a}({\boldsymbol{p}})\ket{0}=0$ for all $\boldsymbol{p}$. The one-particle state with momentum $\boldsymbol{p}$ is defined as $\ket{\boldsymbol{p}}=\sqrt{2E_{\boldsymbol{p}}}\hat{a}^\dagger({\boldsymbol{p}})\ket{0}$, so that amplitudes like the ortogonality relation $\braket{\boldsymbol{p}|\boldsymbol{q}}=2E_{\boldsymbol{p}}(2\pi)^3\delta^{(3)}(\boldsymbol{p}-\boldsymbol{q})$ are Lorentz invariant. With these definitions one can interpret the state $\varphi(x)\ket{0}$ as creating a particle at the spacetime position $x$ (or destroying an antiparticle there) \cite{Peskin:1995ev}.

Hence, for calculating the probability amplitude of creating a particle at $x$ and finding it at $y$ one must compute $\braket{0|\varphi(y)\varphi(x)|0}$, which is one of the so called Green's functions of the theory. This is so because it can be shown that $(\partial^2+m^2)\braket{0|\varphi(y)\varphi(x)|0}=-i\delta^{(4)}(x-y)$ \cite{Peskin:1995ev}, which is by definition the Green's function for the Klein-Gordon equation.

\subsection{Path Integral Quantization}\label{subsec:PathIntegral}

The quantization procedure of a field theory is not unique. A modern standard method is to quantize the field theory using the Feynman path integral method (introduced by Feynman for quantum mechanics \cite{Feynman:1948ur}). This method of quantization is very useful because it does not need the operator formalism, it quantizes the theory (\textit{i.e.} it is able to compute Green's functions) through functional integrals of classical fields. 

Let us describe the path integral formulation of QFTs with the simple example of an interacting massive scalar field theory with the Lagrangian density
\begin{equation}
\mathcal{L}=\frac{1}{2}(\partial_\mu \phi\partial^\mu \phi-m^2\phi^2)-V(\phi)\label{eq:KGScalarLagrangian}\;,
\end{equation}
where $V(\phi)$ is the potential energy density that couples the field to itself. We will treat the field operator $\hat{\phi}(x)$ in the Heisenberg representation, so that its time evolution is set by the usual equation
\begin{equation}
    -i\frac{\partial \hat{\phi}(x)}{\partial t}=[\hat{H},\hat{\phi}(x)]\;.
\end{equation}
So that 
\begin{equation}
    \hat{\phi}(x)= e^{i\hat{H}t} \hat{\phi}(0,\boldsymbol{x})e^{-i\hat{H}t}\;,
\end{equation}
where $\hat{H}$ is the Hamiltonian operator of the system in the canonical operator formalism. Take $\ket{{\phi}(0,\boldsymbol{x})}$ to be an eigenstate of $\hat{\phi}(0,\boldsymbol{x})$. We can prove that the state $\ket{{\phi}(0,\boldsymbol{x}),t}\equiv e^{i\hat{H}t} \ket{{\phi}(0,\boldsymbol{x})}$ is an eigenstate of $\hat{\phi}(x)$ with eigenvalue ${\phi}(0,\boldsymbol{x})$.

Let us compute the probability amplitude of one initial field configuration, $\phi^i$ ($\phi(0,\boldsymbol{x})$ at each spatial point $\boldsymbol{x}$), at an initial time, $t_i$, evolving into a final field configuration $\phi^f$ at a final time, $t_f$,
\begin{equation}
    \braket{\phi_f,t_f|{\phi_i,t_i}}=\braket{\phi_f|e^{-i\hat{H} (t_f-t_i)}|{\phi^i}}\;.\label{eq:amplitudefields}
\end{equation}
Now we need to discretize spacetime, since otherwise $\hat{\phi}(x)$ will have uncountably infinite degrees of freedom. We will hence take a finite volume of space $V$ and divide it into $N$ small cells of volume $v$ such that $V=Nv$. At the end of the process we will take the limits $N\to\infty$ and $v\to 0$ with $V$ fixed and then take $V\to \infty$ to finally recover the continuous Euclidean three-dimensional space. In this way we will have a finite number of fields $\phi^j\equiv \phi(0,\boldsymbol{x}_j)$ where $\boldsymbol{x}_j$ ($j=1,...,N)$ is the point at the center of the $j$th cell. We will slice the time interval $t_f-t_i$ into $M+1$ equal intervals of duration $\tau$ as well. \\

Using the completeness of the state $\ket{\phi^j_l}\equiv\ket{\phi^j_l,t_l} $
\begin{equation}
    \int d\phi^j_l \ket{\phi^j_l} \bra{\phi_j^l}=1\;,
\end{equation}
on eq. (\ref{eq:amplitudefields}) (expressing $\braket{\phi_{f}^1|\phi_{i}^1}\otimes \braket{\phi_{f}^2|\phi_{i}^2}\equiv\braket{\phi_{f}^1,\phi_{f}^2|\phi_{i}^1,\phi_{i}^2}$), we obtain
\begin{align}
   &\braket{\phi_f|\phi_i}=  \lim_{N\to\infty}  \braket{\phi_{f}^1,\phi_{f}^2,..., \phi_{f}^N,t_f|{\phi_{i}^1,\phi_{i}^2,..., \phi_{i}^N,t_i}}=\nonumber\\
   =&\lim \int \prod_{j=1}^N{d\phi_M^j...d\phi^j_i}\braket{\phi_{f}^j|\phi_{M}^j}\braket{\phi_{M}^j|\phi_{M-1}^j}...\braket{\phi_{1}^j|\phi_{i}^j}\;.\label{eq:PathIntegral1}
\end{align}
Where the limit in the last line takes first $N\to \infty$ with $V$ finite, followed by the limit $V\to\infty $ and $M\to\infty$ with $M\tau$ finite. Each of these factors, when expanded in powers of $\tau$, equals
\begin{align}
\braket{\phi_{m+1}^j|\phi_{m}^j}=&\braket{\phi_{m+1}^j|e^{-i\hat{H}\tau}|\phi_{m}^j}=\braket{\phi_{m+1}^j|\phi_{m}^j}-i\tau \braket{\phi_{m+1}^j|\hat{H}|\phi_{m}^j}+\mathcal{O}(\tau^2)\;.\label{eq:PathIntegral4}
\end{align}
At this point, we need to discretize the Hamiltonian as $\hat{H}=\int d^3\boldsymbol{x}\,\hat{\mathcal{H}}(\boldsymbol{x})=v\sum_j \hat{\mathcal{H}}(\boldsymbol{x}_j)$ so that, using the completeness relation for the conjugate momentum eigenstates $\ket{\pi^j}$ of the operator $\hat{\pi}^j$ with eigenvalue $\pi^j$, we obtain from eq. (\ref{eq:PathIntegral4})
\begin{align}
   \braket{\phi_{m+1}^j|\phi_{m}^j}=&\int d\pi^j_m \braket{\phi_{m+1}^j|\pi^j_m}\braket{\pi^j_m|\phi_{m}^j}\nonumber \\
   &-i\tau \int d\pi^j_m d{\pi'}_m^j \braket{\phi_{m+1}^j|\pi^j_m}\braket{\pi^j_m|\hat{H}|{\pi'}^j_m}\braket{{\pi'}^j_m|\phi_{m}^j}+\mathcal{O}(\tau^2)\;.
\end{align}
Notice that $\braket{\pi^j|\hat{H}|{\pi'}^j}$ is the Hamiltonian in momentum representation, which is proportional to a $\delta(\pi^j-{\pi'}^j)$ due to conservation of momentum (translational invariance of the theory). Remembering that $\braket{\pi_m^j|\phi_{m}^j}=e^{-i\pi_m^j \phi_m^j}/(2\pi)$ from the usual conjugate representation $\ket{\phi^j_m}=\int d\pi e^{-i\pi_m^j \phi_m^j}\ket{\pi_m^j}/(2\pi) $, we find
\begin{align}
    \braket{\phi_{m+1}^j|\phi_{m}^j}=&\int d\pi_m^j  e^{i\pi_m^j(\phi_{m+1}^j-\phi_{m}^j)}-iv\tau \int d\pi_m^j e^{i\pi_m^j(\phi_{m+1}^j-\phi_{m}^j)} \braket{\pi_m^j|\hat{\mathcal{H}}(\boldsymbol{x}_{j})|\pi_m^j}+\mathcal{O}(\tau^2)\nonumber\\
    =&\int \frac{d\pi_m^j}{2\pi} e^{i\pi_m^j(\phi_{m+1}^j-\phi_{m}^j)-i\tau v\mathcal{H}^j_m}+\mathcal{O}(\tau^2)\;,\label{eq:PathIntegral2}
\end{align}
where $\mathcal{H}^j_m\equiv \braket{\pi_m^j|\hat{\mathcal{H}}(\boldsymbol{x}_{j})|\pi_m^j}$. Now we insert eq. (\ref{eq:PathIntegral2}) into eq. (\ref{eq:PathIntegral1}) to obtain at $\mathcal{O}(\tau^2)$,
\begin{align}
   &\braket{\phi_f,t_f|{\phi_i,t_i}}=\nonumber \\
   =&\lim \int \prod_{j=1}^N \Big( \prod_{m=1}^{M} d\phi_m^j \prod_{n=0}^{M} \frac{vd\pi_m^j}{2\pi} \Big)\exp\left \{i\sum_{m=0}^M \tau \sum_{j=1}^N v\left(\pi_m^j\frac{\phi_{m+1}-\phi_m}{\tau}-\mathcal{H}_m^j\right)\right\}\label{eq:PathIntegral3}\;.
\end{align}
We can now identify the discretized time derivative of $\phi$ to write eq. (\ref{eq:PathIntegral3}) in the compact notation ($[d\phi]$ denotes the measure as in eq. (\ref{eq:PathIntegral3}) ) 
\begin{align}
   &\braket{\phi_f,t_f|{\phi_i,t_i}}=\int [d\phi] \left[\frac{vdp}{2\pi}\right] \exp{\left\{i\int_{t_i}^{t_f} dt \int d^3\boldsymbol{x} (\pi(x)\partial_t \phi(x)-\mathcal{H}(x))\right\}}\;,
 \end{align}
and, transforming the Hamiltonian density $\mathcal{H}(x)$ to the Lagrangian density in eq. (\ref{eq:KGScalarLagrangian}) while performing a Gaussian integration in $\pi$, we finally find 
\begin{align}
   &\braket{\phi_f,t_f|{\phi_i,t_i}}=C\int [d\phi]  \exp{\left\{i\int_{t_i}^{t_f} dt \int d^3\boldsymbol{x} \mathcal{L}(x)\right\}}\label{eq:PathIntegral10}\;,
 \end{align}
where $C$ is a numerical constant.

We could also have followed the same derivation for quantities such as $\braket{\phi_f,t_f|\hat{\phi}(x)|{\phi_i,t_i}}$ or the more interesting $\braket{\phi_f,t_f|\mathcal{T}[\hat{\phi}(x)\hat{\phi}(y)]|{\phi_i,t_i}}$, where $\mathcal{T}$ is the time ordering operator, to obtain 
\begin{align}
   &\braket{\phi_f,t_f|\mathcal{T}[\hat{\phi}(x_1)...\hat{\phi}(x_n)]|{\phi_i,t_i}}=C\int [d\phi]\phi(x_1)...\phi(x_n)  \exp{\left\{i\int_{t_i}^{t_f} dt \int d^3\boldsymbol{x} \mathcal{L}(x)\right\}}\;.\label{eq:PathIntegral5}
 \end{align}
 As seen in the introduction of this section,  objects like $\braket{0|\mathcal{T}[\hat{\phi}(x)\hat{\phi}(y)]|0}$ (Time-ordered Green's functions) tell us the probability amplitude of creating particles at some initial spacetime point and measuring them at some other final point. Hence, We would like to extract the ground state or vacuum contribution of eq. (\ref{eq:PathIntegral5}). For doing so we will use the completeness of the (energy) eigenstates of the Hamiltonian $\ket{E_m}$ fulfilling $\hat{H}\ket{E_m}=E_m\ket{E_m}$. We take the ground state energy to be $E_0=0$ (assuming this vacuum is unique), finding that 
\begin{align}
   &\braket{\phi_f,t_f|\mathcal{T}[\hat{\phi}(x_1)...\hat{\phi}(x_n)]|{\phi_i,t_i}}=\nonumber\\
   =&\sum_{m,m'=0}^{\infty} e^{iE_mt_i-iE_{m'}t_f}\braket{\phi_f,0|E_{m'}}\braket{E_{m}|\phi_i,0} \braket{E_{m'}|\mathcal{T}[\hat{\phi}(x_1),...,\hat{\phi}(x_n)]|E_m}\;.\label{eq:PathIntegral6}
 \end{align}
 Now we will perform a Wick rotation and set $t_i=-iT$ and $t_f=iT$, also taking $T\to \infty$. When taking this limit, all exponentials in eq. (\ref{eq:PathIntegral6}) vanish except for $E_m=E_{m'}=0$. Hence,
 \begin{align}
   &\lim_{T\to\infty}\braket{\phi_f,iT|\mathcal{T}[\hat{\phi}(x_1),...,\hat{\phi}(x_n)]|{\phi_i,-iT}}=\braket{\phi_f,0|0}\braket{0|\phi_i,0} \braket{0|\mathcal{T}[\hat{\phi}(x_1),...,\hat{\phi}(x_n)]|0}\;.\label{eq:PathIntegral7}
 \end{align}
 We can also notice that 
 \begin{align}
   &\lim_{T\to\infty}\braket{\phi_f,iT|{\phi_i,-iT}}=\braket{\phi_f,0|0}\braket{0|\phi_i,0}\;.\label{eq:PathIntegral8}
 \end{align}
Combining eqns. (\ref{eq:PathIntegral10}), (\ref{eq:PathIntegral5}), (\ref{eq:PathIntegral7}) and (\ref{eq:PathIntegral8}) while transforming back to real time, we find the path integral expression for Green's functions for scalar QFTs,
 \begin{align}
   &\braket{0|\mathcal{T}[\hat{\phi}(x_1),...,\hat{\phi}(x_n)]|0}=\ddfrac{\int [d\phi]\phi(x_1)...\phi(x_n)  \exp{\left\{i \int d^4{x} \mathcal{L}(x)\right\}}}{\int [d\phi]\exp{\left\{i \int d^4{x} \mathcal{L}(x)\right\}}}\;.\label{eq:PathIntegral9}
 \end{align}
 This expression greatly resembles the computation of correlation functions in statistical physics. In fact, just as in statistical physics,  one can define a partition function (functional)
 \begin{equation}
     Z[J]=\int [d\phi] \exp{\left\{i \int d^4{x} (\mathcal{L}(x)+\phi(x)J(x))\right\}}\label{eq:partitionfunctionalscalar}\;,
 \end{equation}
and take functional derivatives thereof to obtain the Green's functions ($n$-point correlation functions) for the field as 
 \begin{equation}
    \braket{0|\mathcal{T}[\hat{\phi}(x_1),...,\hat{\phi}(x_n)]|0} =\frac{(-i)^n}{Z[0]}\frac{\delta^n Z[J]}{\delta J(x_1)...\delta J(x_n)}\Big|_{J=0}\;.\label{eq:greensfunctionsandZ}
 \end{equation}
 
 In this way, we have quantized scalar QFT, and expressed Green's functions of the theory as functional integrals of the classical action and fields. We will use this language again when quantizing non-Abelian QFTs in Sec. \ref{sec:YangMills}. We will now turn to make a connection between experimentally measurable quantities and computable quantities in quantum field theories (Green's functions/$n$-point correlation functions).

\section{Cross sections and the Scattering Matrix}
 The most successful method used to the present day for  unraveling the structure of matter in the universe are collider or scattering experiments (from Rutherford's experiment to the LHC). In these experiments one accelerates a bunch of known particles (such as electrons, protons or heavy nuclei) in a collimated beam and collides them with another bunch of particles. This collision (scattering in the language of waves) will create other particles that are eventually detected in the experiment. These processes elucidate the nature of microscopic interactions between different particles and the more energetic the collision is, the smaller the structure of particles that is resolved.
 
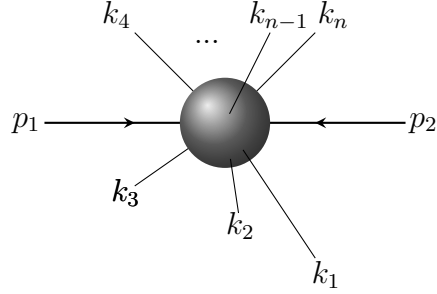
\begin{figure}[ht!]
     \centering
 \begin{tikzpicture}[>=stealth,scale=1.2]
  \draw[->-,thick] (-2,0)--(0,0);
    \draw[->-,thick] (2,0)--(0,0);
    
    \draw (0,0)--(1,1);
    \draw (0,0)--(-1,1);
    \draw (0,0)--(-1,-.7);
    \shade[ball color=gray]  (0,0) circle (.5cm);
    \draw (0.05,.1)--(10*0.05,10*.1);
    \draw (2*0.03,2*-.2)--(5*0.03,5*-.2);
    \draw (0.2,-.3)--(5*0.2,5*-.3);
    \draw (5.6*0.2,5.6*-.3) node {$k_1$};
    \draw (5.7*0.03,5.7*-.2) node {$k_2$};
    \draw (11.9*0.05,11.9*.1) node {$k_{n-1}$};
    \draw (1.1*-1,1.1*-.7) node {$k_{3}$};
    \draw (1.1*-1,1.1*-.7) node {$k_{3}$};
    \draw (1.2*-1,1.2*1) node {$k_{4}$};
    \draw (1.2*-1+1,1.2*1-.3) node {$...$};
    \draw (2.2,0) node {$p_2$};
        \draw (-2.2,0) node {$p_1$};
    \draw (1.2*1,1.2*1) node {$k_{n}$};
\end{tikzpicture}
     \caption[Representation of a $2\to2$ scattering process]{Representation of the scattering process of eq. (\ref{eq:scatteringprocess}) where two incoming particles with momentum $p_1$ and $p_2$ collide producing $n$ outgoing particles travelling freely with momentum $k_i$ each.}
     \label{fig:collider}
 \end{figure}

 Consider as an example a scattering process between two particles in the prepared beams producing $n$ particles in the final state as depicted in Fig. \ref{fig:collider}, 
 \begin{equation}
     (p_1,J_1)+(p_2,J_2)\longrightarrow (k_1,j_1)+...+(k_n,j_n)\;,\label{eq:scatteringprocess}
 \end{equation}
 where the two colliding particles in the initial state carry momentum $p_1$ and $p_2$ and angular momentum $J_1$ and $J_2$ respectively. Likewise, the final-state particles are produced with momenta $k_i$ and angular momenta $j_i$.
 
 The aim is to construct a measurable quantity that will be independent of the size of the beam and its intensity (number of particles per unit volume), so that different collider experiments are able to access it. In our particular case we want to study how likely the process of eq. (\ref{eq:scatteringprocess}) is, normalized by the number of incoming particles in the incident beam, the density of target particles and the time and volume where the scattering is taking place. This quantity will have units of area, and is called the \textbf{\textit{cross section}} because it represents the effective area that a target particle is ``showing'' to the incident particles.
 
 With these considerations, the total cross section for the process in eq. (\ref{eq:scatteringprocess}) amounts to \cite{Muta:2010xua}
 \begin{equation}
\sigma_{\text{tot}}=\sum_{\text{pol}}\frac{W}{F D}\;,
 \end{equation}
 where the sum represents the sum over the polarizations of the incoming and outgoing particles (averaged over the product of the total number of polarizations of each incoming particle), $W$ is the transition probability per unit time and unit volume for the given process (encoding the probability of the initial state evolving into the final state) which will be computed in the next paragraph, $F$ is the incident particle flux and $D$ the density of target particles. In the frame where one of them (the target) sits at rest, these quantities equal 
 \begin{align}
     &F(s)=2{p_1}_0|\vec{v}_1-\vec{v}_2|=K(s)/m_2\\
     &D=2{p_2}_0=2m_2\;,
 \end{align}
 where the center-of-mass squared energy equals
 \begin{equation}
  s=(p_1+p_2)^2\;,    
 \end{equation}
 and the kinematical factor $K(s)=\sqrt{(s-(m_1+m_2)^2)(s-(m_1-m_2)^2)}$.\\

 \paragraph{The S-matrix:}
 Now we turn to computing the probability of the initial state evolving with the time-evolution operator $e^{-i\hat{H}t}$ to the final state. Both the final and initial state particles are supposed to be well described by wavepackets that are narrow in momenta (\textit{i.e.} they have a well defined momentum). The incoming (initial) $\ket{\boldsymbol{p}_1,\boldsymbol{p}_2 \text{ in}}$ and outgoing (final) $\ket{\boldsymbol{k}_1,...,\boldsymbol{k}_n\text{ out}}$ states are supposed to have the incoming particles produced at $T\to-\infty$ and the outgoing ones to be observed at $T\to+\infty$, so that (omitting polarizations) their overlap is
 \begin{equation}
\braket{\boldsymbol{k}_1,...,\boldsymbol{k}_n \text{ out}|\boldsymbol{p}_1,\boldsymbol{p}_2\text{ in}}=\lim_{t\to \infty } \braket{\boldsymbol{k}_1,...,\boldsymbol{k}_n,+ t|\boldsymbol{p}_1,\boldsymbol{p}_2, -t}=\lim_{t\to\infty}\braket{\boldsymbol{k}_1,...,\boldsymbol{k}_n|e^{-2i\hat{H}t}|\boldsymbol{p}_1,\boldsymbol{p}_2}\;,
 \end{equation}
 where in the last expression all states are defined at a common reference time ($t=0$). This limiting procedure of time-evolution  operators is called the S-matrix, and its matrix elements are defined as 
 \begin{equation}\label{eq:Smatrixelements}
\braket{\boldsymbol{k}_1,...,\boldsymbol{k}_n \text{ out}|\boldsymbol{p}_1,\boldsymbol{p}_2\text{ in}}\equiv \braket{\boldsymbol{k}_1,...,\boldsymbol{k}_n|S|\boldsymbol{p}_1,\boldsymbol{p}_2}.
\end{equation}
This operator is unitary by definition, meaning that it conserves total probability (there is no loss of information).

 Hence, the transition probability per unit time and volume for the process is related to the interacting part of the S-matrix, $(\text{S}-\mathbb{I})$, as
 \begin{align}
     W=\frac{1}{V_4}\int \prod_{i=1}^n \frac{d^3\boldsymbol{k}_i}{(2\pi)^32E_{\boldsymbol{k}_i}}\sum_{\mu_1,...,\mu_n}\left|\langle \boldsymbol{k}_1\mu_1,..., \boldsymbol{k}_n\mu_n|(\text{S}-\mathbb{I})|\boldsymbol{p}_1\lambda_1,\boldsymbol{p}_2\lambda_2\rangle\right|^2\;,
 \end{align}
 where $V_4=\int d^4 x=(2\pi)^4 \delta (0)$ is the spacetime volume where the scattering takes place. The interacting part of the S-matrix is given its own name: the \emph{transition matrix}, $T$,  and (extracting the total momentum conservation law) the scattering amplitude, 
 $\mathcal{M}$, as
 \begin{align}
    \label{eq:transitionmatrix} &\langle \boldsymbol{k}_1\mu_1,..., \boldsymbol{k}_n\mu_n|(\text{S}-\mathbb{I})|\boldsymbol{p}_1\lambda_1,\boldsymbol{p}_2\lambda_2\rangle\equiv \langle \boldsymbol{k}_1\mu_1,..., \boldsymbol{k}_n\mu_n|iT|\boldsymbol{p}_1\lambda_1,\boldsymbol{p}_2\lambda_2\rangle\equiv\\
    &\equiv i(2\pi)^4 \delta^4\left(\sum_{i=1}^n k_i-p_1-p_2\right) \mathcal{M}(p_1\lambda_1,p_2\lambda_2\to k_1\mu_1,..., k_n\mu_n) \;.
 \end{align} 
 Summing up, the total cross section for the process amounts to 
 \begin{align}
    \sigma= \frac{1}{2K(s)}\frac{1}{(2J_1+1)(2J_2+1)}
&\sum_{\lambda_1,\lambda_2,\mu_1,...,\mu_n} \int \prod_{i=1}^n \frac{d^3\boldsymbol{k}_i}{(2\pi)^32E_{\boldsymbol{k}_i}}(2\pi)^4 \delta^{(4)}\left(\sum_{i=1}^n k_i-p_1-p_2\right)\times\nonumber \\
    &\times \left| \mathcal{M}(p_1\lambda_1,p_2\lambda_2\to k_1\mu_1,..., k_n\mu_n)\right |^2\;.
 \end{align}
 
 It finally remains to connect the transition matrix element above to computable quantities in the quantum field theory, \textit{i.e.} the correlation functions in the field theory. To do so we will employ the Lehmann-Symmanzik-Zimmerman (LSZ) reduction formula \cite{Lehmann:1954rq}. This procedure will help us linking the transition matrix element to Green's functions for the quantum fields. 

 To expose the LSZ formalism, for simplicity, we will restrict ourselves to the case of elastic scattering (\textit{i.e.} a $2\to 2$ scattering) of neutral particles with mass $m$ (described by the Lagrangian in eq. (\ref{eq:KGScalarLagrangian})). The S-matrix element of eq. (\ref{eq:Smatrixelements}) for this process equals
 \begin{equation}
\braket{\boldsymbol{k}_1,\boldsymbol{k}_2|S|\boldsymbol{p}_1,\boldsymbol{p}_2}=\braket{\boldsymbol{k}_1,\boldsymbol{k}_2 \text{ out}|\boldsymbol{p}_1,\boldsymbol{p}_2\text{ in}}\;.\label{eq:smatrix2to2}
 \end{equation}
We define the incoming and outgoing fields as  asymptotic limits of the Heisenberg interacting field $\hat{\phi}(x)$ as
\begin{equation}\label{eq:asymptoticfields}
    \hat{\varphi}_{ \genfrac{}{}{0pt}{}{\text{out}}{\text{in}}}(x)= \lim_{t\to \pm \infty} \hat{\phi} (x)\;,
\end{equation}
where we are assuming that interactions are gradually turned off as $t\to \pm \infty$ (as particles get far apart their interaction decreases, \textit{i.e.} we assume short range interactions) to justify the use of asymptotic free fields. 

 The ``in'' states are constructed by applying creation operators to the vacuum state as
 \begin{equation}
\ket{\boldsymbol{p}_1,\boldsymbol{p}_2\text{ in}}=\sqrt{2E_{\boldsymbol{p}_1}}\sqrt{2E_{\boldsymbol{p}_2}}\hat{a}_{\text{in}}^\dagger(\boldsymbol{p}_1)\hat{a}_{\text{in}}^\dagger(\boldsymbol{p}_2)\ket{0}\;,
 \end{equation}
where the operators $\hat{a}^\dagger_{\text{in}}(\boldsymbol{p})$ are the Fourier coefficients of the incoming free scalar field $\hat{\varphi}_{\text{in}}(x)$ of eq. (\ref{eq:asymptoticfields}) as in eq. (\ref{eq:ScalarFourier}) with the commutation relations of eq. (\ref{eq:commutation}). Similar relations hold for outgoing operators $\hat{a}^\dagger_{\text{out}}(\boldsymbol{k})$. It is easy to check that
\begin{equation}
    \sqrt{2E_{\boldsymbol{p}}}\hat{a}_{\text{in}}(\boldsymbol{p})= i\int d^3\boldsymbol{x} e^{i(tE_{\boldsymbol{p}}-\boldsymbol{x}\cdot \boldsymbol{p})} \overset{\text{\tiny$\leftrightarrow$}}{\partial}_0\hat{\varphi}_{\text{in}}(x)\;.
\end{equation}
With an analogous relation for the ``out'' operators. In this way we manipulate the $S$-matrix element of eq. (\ref{eq:smatrix2to2}) as
\begin{align}
&\braket{\boldsymbol{k}_1,\boldsymbol{k}_2 \text{ out}|\boldsymbol{p}_1,\boldsymbol{p}_2\text{ in}}=\braket{\boldsymbol{k}_1| \sqrt{2E_{\boldsymbol{k}_2}}\hat{a}_{\text{out}}(\boldsymbol{k}_2)|\boldsymbol{p}_1,\boldsymbol{p}_2\text{ in}}=\nonumber\\
=&i \lim_{x_0\to+\infty}\int d^3\boldsymbol{x} e^{i(tE_{\boldsymbol{p}}-\boldsymbol{x}\cdot \boldsymbol{p})} \overset{\text{\tiny$\leftrightarrow$}}{\partial}_0\braket{\boldsymbol{k}_1|\hat{\phi}(x) |\boldsymbol{p}_1,\boldsymbol{p}_2\text{ in}}\;=\nonumber\\
=& \braket{\boldsymbol{k}_1| \sqrt{2E_{\boldsymbol{k}_2}}\hat{a}_{\text{in}}(\boldsymbol{k}_2)|\boldsymbol{p}_1,\boldsymbol{p}_2\text{ in}}\nonumber+i \int d^4{x} {\partial}_0\left( e^{i(tE_{\boldsymbol{p}}-\boldsymbol{x}\cdot \boldsymbol{p})} \overset{\text{\tiny$\leftrightarrow$}}{\partial}_0\braket{\boldsymbol{k}_1|\hat{\phi}(x) |\boldsymbol{p}_1,\boldsymbol{p}_2\text{ in}}\right)\;,
\end{align}
where we use that the single-particle incoming Hilbert space is equal to the single-particle outgoing Hilbert space, and from the second to the third line we have discarded surface terms (one can justify this by using wave-packets instead of plane waves \cite{Muta:2010xua,Peskin:1995ev}). Using now that $\partial_0(f\overset{\text{\tiny$\leftrightarrow$}}{\partial}_0g)=f(\overset{\text{\tiny$\leftrightarrow$}}{\partial}_0)^2g$ and that the plane waves fulfil the Klein-Gordon equation $\partial_0^2e^{ik_2\cdot x}=(\boldsymbol{\nabla}^2-m^2)e^{ik_2\cdot x}$ we find,
\begin{align}
    \braket{\boldsymbol{k}_1,\boldsymbol{k}_2 \text{ out}|\boldsymbol{p}_1,\boldsymbol{p}_2\text{ in}}=&\braket{\boldsymbol{k}_1,\boldsymbol{k}_2 \text{in}|\boldsymbol{p}_1,\boldsymbol{p}_2\text{ in}}+\nonumber\\
    &+i\int d^4{x} e^{i(E_{\boldsymbol{k}_2}t- \boldsymbol{k}_2\cdot \boldsymbol{x})} (\Box+m^2)\braket{\boldsymbol{k}_1|\hat{\phi}(x) |\boldsymbol{p}_1,\boldsymbol{p}_2\text{ in}}\;.
\end{align}
Using this procedure repeatedly we find the so called \textit{reduction formula} \cite{Muta:2010xua}
\begin{align}
    &\braket{\boldsymbol{k}_1,\boldsymbol{k}_2 \text{ out}|\boldsymbol{p}_1,\boldsymbol{p}_2\text{ in}}=\braket{\boldsymbol{k}_1,\boldsymbol{k}_2 \text{in}|\boldsymbol{p}_1,\boldsymbol{p}_2\text{ in}}+i^4\int d^4{x_1}d^4{x_2}d^4{y_1}d^4{y_2} e^{i(k_1\cdot x_1+k_2\cdot x_2-p_1\cdot y_1-p_2\cdot y_2)}\times\nonumber\\
    &\times (\Box_{x_1}+m^2)(\Box_{x_2}+m^2)(\Box_{y_1}+m^2)(\Box_{y_2}+m^2)\braket{0|\mathcal{T}[\hat{\phi}(x_1)\hat{\phi}(x_2)\hat{\phi}(y_1)\hat{\phi}(y_2) ]|0}\;.
\end{align}
Noticing now that $\braket{\boldsymbol{k}_1,\boldsymbol{k}_2 \text{ out}|\boldsymbol{p}_1,\boldsymbol{p}_2\text{ in}}-\braket{\boldsymbol{k}_1,\boldsymbol{k}_2 \text{in}|\boldsymbol{p}_1,\boldsymbol{p}_2\text{ in}}=\braket{\boldsymbol{k}_1,\boldsymbol{k}_2|(S-\mathbb{I})|\boldsymbol{p}_1,\boldsymbol{p}_2}$, we finally connect the transition matrix of eq. (\ref{eq:transitionmatrix}) for the $2\to2$ scattering to a fields' Green's function 
\begin{align}
&\braket{\boldsymbol{k}_1,\boldsymbol{k}_2|iT|\boldsymbol{p}_1,\boldsymbol{p}_2} =i^4\int d^4{x_1}d^4{x_2}d^4{y_1}d^4{y_2} e^{i(k_1\cdot x_1+k_2\cdot x_2-p_1\cdot y_1-p_2\cdot y_2)}\times\nonumber\\
    &\times (\Box_{x_1}+m^2)(\Box_{x_2}+m^2)(\Box_{y_1}+m^2)(\Box_{y_2}+m^2)\braket{0|\mathcal{T}[\hat{\phi}(x_1)\hat{\phi}(x_2)\hat{\phi}(y_1)\hat{\phi}(y_2) ]|0}\;.\label{eq:LSZreduction1}
\end{align}

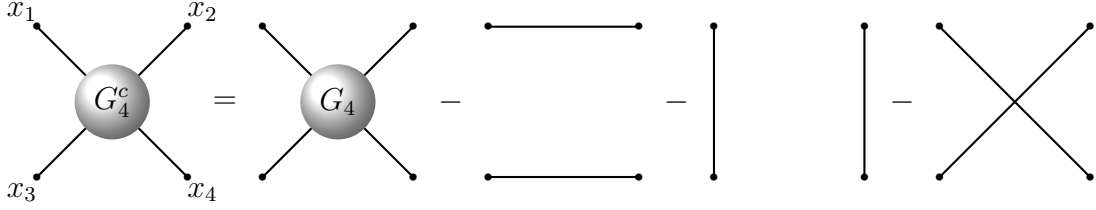
\begin{figure}[ht!]
    \centering
   \begin{tikzpicture}
\draw (-1,-1) node {\tiny$\bullet$};
\draw (1,1) node {\tiny$\bullet$};
\draw (1,-1) node {\tiny$\bullet$};
\draw (-1,1) node {\tiny$\bullet$};

\draw (-1+3,-1) node {\tiny$\bullet$};
\draw (1+3,1) node {\tiny$\bullet$};
\draw (1+3,-1) node {\tiny$\bullet$};
\draw (-1+3,1) node {\tiny$\bullet$};

\draw (-1+6,-1) node {\tiny$\bullet$};
\draw (1+6,1) node {\tiny$\bullet$};
\draw (1+6,-1) node {\tiny$\bullet$};
\draw (-1+6,1) node {\tiny$\bullet$};

\draw (-1+9,-1) node {\tiny$\bullet$};
\draw (1+9,1) node {\tiny$\bullet$};
\draw (1+9,-1) node {\tiny$\bullet$};
\draw (-1+9,1) node {\tiny$\bullet$};

\draw (-1+12,-1) node {\tiny$\bullet$};
\draw (1+12,1) node {\tiny$\bullet$};
\draw (1+12,-1) node {\tiny$\bullet$};
\draw (-1+12,1) node {\tiny$\bullet$};

   \draw[thick] (-1,-1)--(0,0);
   \draw[thick] (1,1)--(0,0);
   \draw[thick] (1,-1)--(0,0);
   \draw[thick] (-1,1)--(0,0);
       \shade[ball color=white]  (0,0) circle (.5cm);
\draw (1.5,0) node {$=$};       
   \draw[thick] (-1+3,-1)--(0+3,0);
   \draw[thick] (1+3,1)--(0+3,0);
   \draw[thick] (1+3,-1)--(0+3,0);
   \draw[thick] (-1+3,1)--(0+3,0);
       \shade[ball color=white]  (0+3,0) circle (.5cm);
\draw (1.5+3,0) node {$-$};    

   \draw[thick] (-1+6,1)--(1+6,1);
   \draw[thick] (1+6,-1)--(-1+6,-1);
   \draw (1.5+6,0) node {$-$};  
      \draw[thick] (1+9,-1)--(1+9,1);
   \draw[thick] (-1+9,-1)--(-1+9,1);
   \draw (1.5+9,0) node {$-$};  
      \draw[thick] (-1+12,-1)--(1+12,1);
   \draw[thick] (1+12,-1)--(-1+12,1); 

   \draw (0,0) node {$G_4^c$};
   \draw (3,0) node {$G_4$};
   
   \draw (-1.2,-1.2) node {$x_3$};
\draw (1.2,1.2) node {$x_2$};
\draw (1.2,-1.2) node {$x_4$};
\draw (-1.2,1.2) node {$x_1$};
   \end{tikzpicture}
    \caption[Representation of the four point connected Green's function]{Representation of eq. (\ref{eq:exampleconnected4}) for the coordinate-space connected four-point Green's function, $G_4^c(x_1,x_2,x_3,x_4)$. In all five diagrams, the black lines represent two point Green's functions $G_2=G_2^c$.}
    \label{fig:4pointGF}
\end{figure}

At this point, it is useful to define the momentum-space Green's functions, $\Tilde{G}_n(p_1,...,p_{n})$, as Fourier transforms of coordinate space Green's functions, $G_n(x_1,...,x_n)=\braket{0|\mathcal{T}[\hat{\phi}(x_1)...\hat{\phi}(x_n) ]|0}$, 
\begin{equation}
(2\pi)^4\delta^{(4)}(p_1+...+p_n) \Tilde{G}(p_1,...,p_{n-1})= \int d^4 x_1...d^4 x_n e^{-i(p_1\cdot x_1+...+p_n\cdot x_n)}   G_n(x_1,...,x_n)\;.
\end{equation}
It is also useful to define the \textit{truncated} Green's functions, $\Tilde{G}_n^t$, from which incoming/outgoing one-particle poles are removed, by the relation
\begin{equation}
  \Tilde{G}_n(p_1,...,p_{n-1})=  \Tilde{G}_2(p_1)...\Tilde{G}_2(p_n) \Tilde{G}_n^t(p_1,...,p_{n-1})\;.\label{eq:truncatedGF}
\end{equation}
With these definitions at hand we can integrate by parts in eq. (\ref{eq:LSZ2}) to obtain the LSZ reduction formula 
\begin{align}
\braket{\boldsymbol{k}_1,\boldsymbol{k}_2|iT|\boldsymbol{p}_1,\boldsymbol{p}_2} =&i^4 (2\pi)^4\delta^{(4)}(k_1+k_2-p_1-p_2)\times\nonumber\\
\times\Big[(m^2-k_1^2)&(m^2-k_2^2)(m^2-p_1^2)(m^2-p_2^2)\Tilde{G}_4(k_1,k_2,p_1) \Big]\Big|_{k_1^2=k_2^2=p_1^2=p_2^2=m^2}\label{eq:LSZ2}\;.
\end{align}
As we will see shortly, the propagator $\Tilde{G}_2(p)$, near the mass shell behaves as $\Tilde{G}_2(p)\sim iZ_\phi/(p^2-m^2)$ where $Z_\phi$ is the so called field renormalization constant (a factor of $\sqrt{Z_\phi}$ comes from each field inside the correlation function). This means that
\begin{align}
\braket{\boldsymbol{k}_1,\boldsymbol{k}_2|iT|\boldsymbol{p}_1,\boldsymbol{p}_2} =&Z_\phi^4 \Tilde{G}_4^t(k_1,k_2,p_1)\;,
\end{align}
so that computing transition matrix elements amounts to calculating truncated Green's functions in momentum-space. Actually, it can be shown \cite{Peskin:1995ev,Muta:2010xua} that the only contributing Green's functions to the $T$ matrix elements are not only truncated but also connected (this comes from normalizing the correlation functions with the vacuum contribution, $\braket{0|0}$). The connected Green's functions, $G_n^c$, can be extracted from the functional $Z[J]=\exp{\{{i W[J]}\}}$ with 
\begin{equation}
    G_n^c(x_1,...x_n)=(-i)^{n-1} \frac{\delta^n W[J]}{\delta J(x_1) ... \delta J(x_n)}\Big|_{J=0}\;.
\end{equation}
As an example one can show that 
\begin{equation}
    \frac{\delta^2 W[J]}{\delta J(x_1) \delta J(x_2)}\Big|_{J=0}= \frac{-i}{Z[0]}\frac{\delta^2 Z[J]}{\delta J(x_1) \delta J(x_2)}\Big|_{J=0}=i\braket{0|\mathcal{T}[\hat{\phi}(x_1)\hat{\phi}(x_1)]|0}=iG_2(x_1,x_2)\label{eq:twopointisconnected}\;,
\end{equation}
and, for the connected four-point Green's function,
\begin{align}
    G_4^c(x_1,...,x_4)=&G_4(x_1,...,x_4)-\nonumber\\
   &-G_2(x_1,x_2)G_2(x_3,x_4)-G_2(x_1,x_3)G_2(x_2,x_4)-G_2(x_1,x_4)G_2(x_2,x_3)\;.\label{eq:exampleconnected4}
    \end{align}
  One can see from eq. (\ref{eq:exampleconnected4}) how the connected 4-point Green's function is the full four-point function minus the disconnected pieces connecting the external points (see Fig. \ref{fig:4pointGF}). This fact generalizes for any $n$-point connected Green's function.

 The truncated and connected Green's functions, $G_n^{tc}$, can be easily calculated in perturbation theory with the language of Feynman diagrams (anticipated by the representation in Fig. \ref{fig:4pointGF}). We now turn to describe this method.

\subsection{Perturbation theory and Feynman graphs}\label{sec:PTphi4}
Let us now describe the perturbative method for computing Green's functions in scalar field theories. This method assumes that the field theory under treatment is perturbed from the free scalar theory of eq. (\ref{eq:KGScalarLagrangian}) through small corrections sourced by the potential $V(\phi)$, for example $V(\phi)=\lambda \phi^4/(4!)$. Strictly speaking, perturbation theory assumes that the Green's functions of the theory are suitable to be approximated by an asymptotic series in the couplings of the theory ($\lambda$ for the example). This means that, computations of increasing order in the coupling are expected to subsequently improve the precision on the prediction of observable quantities (such as the cross section).

To start, let us conveniently cast the partition functional of scalar QFT in eq. (\ref{eq:partitionfunctionalscalar}) as
\begin{equation}
    Z[J]=\exp{\left\{-i\int d^4 y V(\phi)\right \}} Z_0[J]\;,\label{eq:PTforscalars}
\end{equation}
where $Z_0[J]$ is the partition functional of the free scalar field theory (after integration by parts)
\begin{equation}
    Z_0[J]=\int [d\phi] \exp{\left\{-\frac{i}{2} \int d^4x d^4y \phi(x) K(x,y)\phi(y)+i\int d^4x\phi(x)J(x) \right\}}\;,\label{eq:freescalarPT}
\end{equation}
and the \textit{kernel} $K(x,y)$ 
\begin{equation}
    K(x,y)=\delta^{(4)}(x-y) (\Box_y +m^2)\;.
\end{equation}
eq. (\ref{eq:freescalarPT}) is a Gaussian-like  integral which can be computed (analogously to completing squares in a usual Gaussian integral) to get, by neglecting an overall numerical constant \cite{Muta:2010xua}, 
\begin{equation}
    Z_0[J]= \exp{\left\{-\frac{1}{2} \int d^4 x d^4y J(x) D(x,y) J(y)\right\}}\;,\label{eq:freescalarsolvedfunctional}
\end{equation}
with the free scalar-field Feynman-propagator in coordinate space $D(x,y)$ being the inverse operator of the kernel $K(x,y)$, \textit{i.e.}
\begin{equation}
    \int d^4 z K(x,z) D(z,y)=-i \delta^{(4)}(x-y)\;. \label{eq:inverseKG}
\end{equation}
As is section \ref{sec:Quantization}, $D(x,y)$, is nothing more than the two-point Green's function of the free scalar field $\hat{\varphi}$. In fact, from eq. (\ref{eq:freescalarsolvedfunctional}) one easily finds
\begin{equation}
    G_2^{\text{free}}(x,y)=\braket{0|\mathcal{T}[\hat{\varphi}(x)\hat{\varphi}(y)]|0}=\frac{(-i)^2}{Z_0[0]} \frac{\delta ^2 Z_0[J]}{\delta J(x) \delta J(y)}\Big|_{J=0}= D(x,y)\;.
\end{equation}
Using the Fourier representation $D(x,y)=\int d^4 p e^{-ip\cdot(x-y)}\tilde{D}(p) /(2\pi)^4$ in eq. (\ref{eq:inverseKG}) one finds
\begin{equation}
    D(x,y)=\int \frac{d^4 p}{(2\pi)^4} \frac{i e^{-ip\cdot(x-y)}}{p^2-m^2+i\epsilon}\;\;\Rightarrow \;\; \tilde{D}(p)=\frac{i}{p^2-m^2+i\epsilon}\;.
\end{equation}

To continue, let use take $V(\phi)=\lambda \phi^4/(4!)$ to write eq. (\ref{eq:PTforscalars}) as a perturbative series in $\lambda$,
\begin{equation}
    Z[J]=\left(1-\frac{i\lambda}{4!}\int d^4 y \left( \frac{\delta}{\delta J(y)}\right)^4 -\frac{1}{2} \left(\frac{\lambda}{4!}\right )^2 \left(\int d^4 y \left( \frac{\delta}{\delta J(y)}\right)^4\right)^2+...\right) Z_0[J]\;,
\end{equation}
where we used that $\phi(y)=\delta/(i\delta J(y))$ (as operators acting inside the path integral of $Z_0[J]$). By computing the functional derivatives we find
\begin{equation}
    \frac{\delta^4 Z_0[J]}{(\delta J(y))^4}=\left(3 (D(y,y))^2-6 D(y,y) \left (\int d^4 x D(x,y)J(x)\right)^2+ \left (\int d^4 x D(x,y)J(x)\right)^4\right) Z_0[J]\,,
\end{equation}
Which, for example, helps us computing the first order correction to the two-point Green's function
\begin{align}
    G_2(x_1,x_2)=\frac{(-i)^2}{Z[0]}& \frac{\delta ^2 Z[J]}{\delta J(x_1)\delta J(x_2)}\Big|_{J=0}=\frac{1}{Z[0]}\Big [D(x_1,x_2)\left( 1-\frac{i \lambda}{8}\int d^4 y (D(y,y))^2\right)+\nonumber\\
    &- \frac{i\lambda}{2}   \int d^4 y D(x_1,y)D(y,y)D(y,x_2) +\mathcal{O}(\lambda^2)\Big]\;.\label{eq:twopointfunction1}
\end{align}
Using that
\begin{equation}
    Z[0]=1-\frac{i \lambda}{8}\int d^4 y (D(y,y))^2+\mathcal{O}(\lambda^2)\label{eq:vacuumfirstorder}\;,
\end{equation}
we finally find
\begin{align}
    G_2(x_1,x_2)=D(x_1,x_2)- \frac{i\lambda}{2}   \int d^4 y D(x_1,y)D(y,y)D(y,x_2) +\mathcal{O}(\lambda^2)\;.\label{eq:twopointlambdafour}
\end{align}
These equations can be pictorially represented using Feynman Graphs as in Fig. \ref{fig:FeynmanGraphs1}. 

\begin{figure} [hb!]
    \centering
    \begin{tikzpicture}
    \draw[] (-2.05,0) node {$ G_2(x_1,x_2)^{\mathcal{O}(\lambda)}=$};
        \draw[] (0,0)-- (2,0);
        \draw[] (3,0)-- (5,0);
        \draw[] (6,0)-- (8,0);
        \draw[]  (3.75,.5) ellipse (7pt and 7pt);
        \draw[]  (4.25,.5) ellipse (7pt and 7pt);
        \draw[] (2.5,0) node {$+$};
        \draw[] (5.5,0) node {$+$};
        \draw[]  (7,.25) ellipse (7pt and 7pt);
        \draw (0,0) node {\tiny$\bullet$};
        \draw (2,0) node {\tiny$\bullet$};
        \draw (3,0) node {\tiny$\bullet$};
        \draw (5,0) node {\tiny$\bullet$};
        \draw (6,0) node {\tiny$\bullet$};
        \draw (8,0) node {\tiny$\bullet$};
        \draw (7,0) node {\tiny$\bullet$};
        \draw (4,.5) node {\tiny$\bullet$};
        \draw (8.25,0) node {\Big)};
        \draw (8.5,0) node {\Big/};
        \draw (-.25,0) node {\Big(};
        \draw (8.75,0) node {\Big(};
        \draw (9,0.0) node {$1$};
        \draw (9.5,0) node {$+$};
        \draw (11.25,0) node {\Big)};
        \draw[]  (10.25,0) ellipse (7pt and 7pt);
        \draw[]  (10.75,0) ellipse (7pt and 7pt);
        \draw (10.5,0) node {\tiny$\bullet$};

\draw[] (0,-1)-- (2,-1);
       \draw (0,-1) node {\tiny$\bullet$};
        \draw (2,-1) node {\tiny$\bullet$}; 
        \draw[]  (4,-0.75) ellipse (7pt and 7pt);
        \draw[] (3,-1)-- (5,-1);
                \draw (3,-1) node {\tiny$\bullet$};
        \draw (5,-1) node {\tiny$\bullet$};
        \draw (4,-1) node {\tiny$\bullet$};
        \draw (-.72,-1) node {$=$};
    \end{tikzpicture}
    \caption[Feynman graph representation]{First line: Representation using Feynman graphs of the RHS of eq. (\ref{eq:twopointfunction1}), using the value of $Z[0]$ from (\ref{eq:vacuumfirstorder}). Second line: representation of the result in eq. (\ref{eq:twopointlambdafour}). All at first order in $\lambda$.}
    \label{fig:FeynmanGraphs1}
\end{figure}
Since the two point Green's function is connected by definition (as shown in eq. (\ref{eq:twopointisconnected})), the contributing diagrams to it are also connected (\textit{i.e.} there are no disjoint parts in each diagram). This statement is true for any $n$-point connected Green's function.

The relation between Fig. \ref{fig:FeynmanGraphs1} and eq. (\ref{eq:twopointlambdafour}) is made explicit by the use of the following Feynman rules in coordinate space (see further details in section \ref{sec:fopt}):
\begin{enumerate}
    \item To compute the $\lambda^m$ order contribution to an $n$-point connected Green's function in the coupling, sum all possible connected graphs having $n$ external points and $m$ interaction vertices (or internal points).
    \item Each edge on a graph connecting points $x$ and $y$ (external or internal) corresponds to a propagator $D(x,y)$.
    \item For each interaction vertex at a given graph's point, $y_i$, integrate by $-i\lambda \int d^4 y_i$. 
    \item Divide each graph by its corresponding symmetry factor (see \cite{Peskin:1995ev,Muta:2010xua} for details). 
\end{enumerate}

Given these rules, we can compute Green's functions perturbatively in a very intuitive and easy way using Feynman graphs or diagrams. 

As seen trough the use of the LSZ reduction formula in eq. (\ref{eq:LSZreduction1}), it is common to compute connected and truncated Green's functions in momentum space. Hence, we will translate the Feynman rules of scalar field theories to the momentum space picture:
\begin{enumerate}
    \item To compute the order $\lambda^m$ contribution to an $n$-point connected Green's function, sum all possible connected graphs having $n$ external edges and $m$ interaction vertices.
    \item For each edge in the graph with assigned momentum $k$, multiply by $\tilde{D}(k)$ and, for each interaction vertex, multiply by $-i\lambda$. 
    \item Apply conservation of four-momentum at each vertex.
    \item For each loop in a diagram, integrate with the measure $\int \frac{d^4k}{(2\pi)^4}$.
    \item Divide each contribution by its symmetry factor.
    \item Truncate the connected Green's function according to eq. (\ref{eq:truncatedGF}).
\end{enumerate}
For example, eq. (\ref{eq:twopointlambdafour}) becomes in momentum space
\begin{equation}
    G^{c}_2(p)= \tilde{D}(p)-\frac{i\lambda }{2}(\tilde{D}(p))^2\int \frac{d^4k}{(2\pi)^4} \tilde{D}(k)+\mathcal{O}(\lambda^2)\;,
\end{equation}
so that the truncated two point function is
\begin{equation}
    G^{tc}_2(p)=\frac{1}{G^{c}_2(p)} = (\tilde{D}(p))^{-1}+\frac{i\lambda }{2}\int \frac{d^4k}{(2\pi)^4} \tilde{D}(k)+\mathcal{O}(\lambda^2)\;.\label{eq:correctiontoGF}
\end{equation}

\subsection{Renormalization of scalar field theories}\label{sec:PTandrenormalization}
When computing perturbative corrections to Green's functions for the fields, one finds that there are divergent loop integrals, such as the integral in the RHS of eq. (\ref{eq:correctiontoGF}). These integrals usually have the so called UV divergences, \textit{i.e.} divergences occurring for large loop momenta. To make sense of these divergences one needs to extract them somehow, and, to do so, one can employ the method of Dimensional Regularization \cite{Muta:2010xua}. This method extends the field theory at hand to $D=4-2\epsilon$ dimensions in order to regulate the divergences of loop integrals and to extract them as a Laurent series in $\epsilon$. Once extracted, the strategy is to absorb the divergent terms by redefining the fields as well as the couplings and masses. For the scalar theory in eq. (\ref{eq:KGScalarLagrangian}) with $V(\phi)=-\lambda \phi^4/4!$ one can define
\begin{equation}
    \phi=\sqrt{Z_\phi} \phi_R\;\;\;\;\lambda=Z_\lambda \lambda_R\;\;\;\;m=\sqrt{Z_m} m_R\;,
\end{equation}
where the subscript $R$ denotes renormalized quantities. In this way one can express the Lagrangian as
\begin{align}
    \mathcal{L}=& \frac{1}{2}(\partial_\mu \phi_R\partial^\mu \phi_R-m_R^2\phi_R^2)-\frac{\lambda_R}{4!}\phi_R^4+\nonumber\\
    &+\frac{1}{2}[(Z_\phi-1)\partial_\mu \phi_R\partial^\mu \phi_R-(Z_\phi Z_m-1)m_R^2\phi_R^2]-(Z_\lambda-1)\frac{\lambda_R}{4!}\phi_R^4\;.\label{eq:renormalizedscalar}
\end{align}
Now one can define the renormalization constants, $Z_\phi$, $Z_\lambda$ and $Z_\phi$, in such a way that the second line in eq. (\ref{eq:renormalizedscalar}) acts as a \textit{counterterm} Lagrangian. This means that one chooses these renormalization constants in order to exactly cancel the divergences obtained in perturbative calculations with the part of the Lagrangian in the first line of (\ref{eq:renormalizedscalar}) (which are the divergences one would have obtained from the original unrenormalized Lagrangian). 

When using dimensional regularization, extending the field theory to $D=4-2\epsilon$ spacetime dimensions, one needs to properly adjust the mass dimension of the Lagrangian, since it must equal $D$. A scalar field in $D$ dimensions has mass dimension of $[\phi]=1-\epsilon$, so that $[m]=1$ and $[\lambda]=2\epsilon$. We can hence extract the mass-dimension dependence on $\epsilon$ as
\begin{align}
\lambda=\lambda_0\mu_0^{2\epsilon}\;,\;\;\;\;&\;\;\;\;\lambda_R=\lambda_r \mu^{2\epsilon}\;\;\;\;\;\Rightarrow\;\;\;\;\; \lambda_r=\left(\frac{\mu_0}{\mu}\right)^{2\epsilon} Z_\lambda^{-1} \lambda_0\label{eq:renormalizationphi4}\;,
\end{align}
where $\lambda_0$ and $\lambda_r$ are dimensionless couplings, $\mu_0$ is a fixed energy scale, and $\mu$ is the (variable) renormalization energy scale. Using eq. (\ref{eq:renormalizationphi4}) and $m=\sqrt{Z_m} m_R=\sqrt{Z_m} m_r$ we obtain the renormalization group equations 
\begin{align}
    &\mu \frac{d \lambda_r}{d\mu}=\beta (\mu) &\text{with}\;\;\;\;\;\;\;\;&\beta(\mu)=-2\epsilon \lambda_r-\frac{\mu}{Z_\lambda} \frac{d Z_\lambda}{d \mu} \lambda_r\;,\label{eq:betaphi4}\\
  &\mu \frac{d m_r}{d\mu}=-m_r\gamma_m(\mu)   &\text{with}\;\;\;\;\;\;\;\;&\gamma_m(\mu)=\frac{\mu}{\sqrt{Z_m}} \frac{d \sqrt{Z_m}}{d\mu}\;.\label{eq:anomalousphi4}
\end{align}
Both the $\beta(\mu)$ function and the anomalous mass dimension $\gamma_m(\mu)$ are finite functions, since the divergences in the renormalization constants exactly cancel in both equations (\ref{eq:betaphi4}) and (\ref{eq:anomalousphi4}). These equations predict that the masses and the couplings of particles in QFT depend on the renormalization energy scale $\mu$. The interpretation of this fact is due to Wilson \cite{Wilson:1971bg}, which explains that once that a mass and the coupling of a particle are measured at one particular energy, at a different energy regime the quantum corrections behave differently, hence shifting the value of the coupling and masses. The equations (\ref{eq:betaphi4}) and (\ref{eq:anomalousphi4}) set the so called ``running'' of both the coupling and mass over different energy scales.

The treatment of scalar theories that we have described above is easily translatable to more complex theories such as Quantum Chromodynamics or the Electroweak force, but helps understanding most of the tools used in those realistic QFTs. We move on to describe them since are the building blocks of the Standard Model.
\section{Yang-Mills theories and QCD}\label{sec:YangMills}
In 1954, Yang and Mills extended Abelian field theories, such as Electrodynamics, to more general non-Abelian field theories \cite{Yang:1954ek}. This idea turned out to be very successful to describe fundamental properties of the strong interaction. First, asymptotic freedom (needed to explain Bjorken scaling \cite{Bjorken:1968dy}, which suggested that there were almost-free point-like particles inside the proton, \textit{i.e.} quarks). Second, the need for an extra quantum number for quarks in order to preserve Pauli's exclusion principle from the discovery in 1952 of hadronic resonances like the $\Delta^{++}$ \cite{Anderson:1952nw,Ashkin:1956rah} (the solution to this problem was proposed by Han and Nambu in \cite{Han:1965pf}, this quantum number for quarks was later called color). After the proof of its asymptotic freedom by Gross, Wilczek \cite{Gross:1973id} and Politzer \cite{Politzer:1973fx} in 1973 and of its renormalizability by 't Hooft in 1972 \cite{tHooft:1972tcz}, the Yang-Mills theory for strong interactions coupled to quarks, Quantum-Chromodynamics, was finally confirmed to be the correct description for the dynamics of the strong nuclear force (after extensive testing in the 1980s at collider experiments).  Nowadays, Yang-Mills theories are well established and used to describe all known interactions of matter (strong and electroweak forces). In this section we will briefly introduce these theories and their quantization.

\subsubsection{Classical Yang-Mills theories}
Let us now extend the classical approach to electrodynamics of section \ref{sec:electrodynamics} to generic non-Abelian groups. Take a fermionic field $\psi(x)$ with $N$ (``color'' or ``weak isospin'') components, $\psi_i(x)$, which transforms in the so called fundamental representation of a Lie group $G$ as
\begin{equation}
\psi'_i(x)=U_{ij}\psi_j(x) \;\;\text{ with }\;\; U=\exp{\left(-iT^a \theta^a(x)\right)}\;,\label{eq:gaugetransformationPsi}
\end{equation}
where all repeated indices are summed over and the $\theta^a(x)$ are arbitrary real functions.
Here, the matrices $T^a$, with $a=1,...,n$, are the generators of the Lie algebra corresponding to $G$ which obey the commutation relations
\begin{equation}
    [T^a,T^b]=if^{abc} T^c\;,
\end{equation}
where the $f^{abc}$ are the structure constants that characterize the Lie algebra of $G$. Analogously to the electrodynamics case, defining the covariant derivative in the fundamental representation of $G$ as $D_\mu=\partial_\mu-ig T^a A_\mu^a (x)$ and imposing that it fulfils $(D_\mu\psi)'=UD_\mu \psi$ we have that the gauge fields must transform as
\begin{equation}
T^a(A_\mu^a)'=U\left(T^a A_\mu ^a-\frac{i}{g}U^{-1}\partial_\mu U\right)U^{-1}\;.\label{eq:gaugetransformationA}
\end{equation}
From computing the curvature tensor, $F_{\mu\nu}$, as the commutator of two covariant derivatives $[D_\mu,D_\nu]=-igT^a F^a_{\mu\nu}$ one finds the gauge covariant field strength tensor for non-Abelian field theories
\begin{equation}
    F_{\mu\nu}^a=\partial_\mu A_\nu^a-\partial_\nu A_\mu^a+gf^{abc} A_\mu^b A^c_\nu\;.
\end{equation}
In this way, we can construct the Lagrangian for Yang-Mills theories 
\begin{equation}
    \mathcal{L}_{\text{YM}}=-\frac{1}{4}F_{\mu\nu}^a F^{a\,\mu\nu}+\bar{\psi}(i\gamma^\mu D_\mu -m)\psi\label{eq:classicalYM}\;,
\end{equation}
which is invariant under the transformations of eqns. (\ref{eq:gaugetransformationPsi}) and (\ref{eq:gaugetransformationA}).

Relevant for the SM of particle physics is the special case of Chromodynamics (CD) where $G$ is the special unitary group in three dimensions $SU(3)$ ($3\times 3$ complex matrices with unit determinant) and there are six quark flavors, $f=u,d,c,s,b,t$ (corresponding to the ones in Fig. \ref{fig:SMcontent}),  
\begin{equation}
    \mathcal{L}_{\text{CD}}=-\frac{1}{4}F_{\mu\nu}^a F^{a\,\mu\nu}+\sum_{f}\bar{\psi}_f(i\gamma^\mu D_\mu -m_f)\psi_f\label{eq:classicalCD}\;,
\end{equation}
which describes the classical strong interaction of ``colored'' quarks.

\subsubsection{Quantization of gauge theories}
Let us focus in quantizing pure non-Abelian gauge theories with Lagrangian
\begin{equation}
    \mathcal{L}_{\text{gauge}}=-\frac{1}{4}F_{\mu\nu}^a F^{a\,\mu\nu}\;.\label{eq:pureYM}
\end{equation}
If one tries to quantize this theory in the operator formalism with $A_\mu^a$ as the quantization variable, one will find the problem that one component of the canonical momentum, $\Pi_0 ^a(x)\equiv \frac{\delta \mathcal{L}}{\delta (\partial_0 A^{a\,0})}$, vanishes. This means that no canonical commutation relations can be imposed on this operator. This is due to the fact that there is still the gauge freedom on the field $A_\mu^a$ and one needs eliminate this freedom by choosing a gauge. In electrodynamics, it is possible to directly introduce a gauge-fixing term in the Lagrangian (such as $(\partial_\mu A^\mu)^2$). However, gauge invariance on amplitudes for non-Abelian gauge fields is spoiled by the appearance of three-body and four-body interactions in the Lagrangian of eq. (\ref{eq:pureYM}) even after choosing the gauge in this way \cite{Muta:2010xua}. 

At this point, the path integral quantization procedure of section \ref{subsec:PathIntegral} becomes very useful for correctly fixing the gauge. Translating the result of eq. (\ref{eq:partitionfunctionalscalar}), we can build the generating functional for non-Abelian gauge fields as
\begin{equation}
    Z[J]=\int [dA] \exp{\left\{i \int d^4{x} (\mathcal{L}_{\text{gauge}}(x)+A_\mu^a(x)J^{a\,\mu}(x))\right\}}\label{eq:partitionfunctionalpureYM}\;.
\end{equation}
One can show that the measure $[dA]$ is gauge invariant, so that $Z[0]$ is too. However the presence of $J$ spoils the gauge invariance of $Z[J]$. Consequently, let us first deal with $Z[0]$. One of the first shortcomings of dealing with $Z[0]$ comes from the fact that the integrand is gauge invariant and hence constant over each \textit{gauge orbit} (which is the set of field configurations, $A^{a\, }_\mu(\theta)$, that one can reach by applying all possible gauge transformations $U(\theta)$ on a given $A_\mu^a$). The region of integration corresponding to every gauge orbit is infinite, making hence $Z[0]$ divergent and meaningless. In consequence, we would like to factor out this infinite measure, choosing a class representative $A_\mu ^a$ for each gauge orbit $A^{a}_\mu{ (\theta)}$. For doing so we require that the equation,
\begin{equation}
    G^\mu A_\mu^a(\theta)=B^a \label{eq:gaugeorbitcond}\;,
\end{equation}
yields a unique solution $\theta^a$\footnote{This requirement is not guaranteed in the non-perturbative regime of Yang-Mills theories, where the gauge condition in eq. (\ref{eq:gaugeorbitcond}) may have several solutions, called Gribov copies \cite{Gribov:1977wm}. This is closely related to Confinement and will be discussed in section \ref{sc:Confinement}.}, where $G^\mu$ is a gauge condition and $B^a$ an arbitrary function. To introduce this gauge fixing condition in the functional integral one can define the Fadeev-Popov determinant \cite{FADDEEV196729}, $\Delta_G[A]$, through the equation
\begin{equation}
   \Delta_G[A]\int [dg] \delta^{(n)}(G^\mu A_\mu^a(\theta)-B^a) =1\;,\label{eq:Fadeevtrick}
\end{equation}
where $[dg]=\prod_{a} [d\theta^a]$ is the gauge-invariant group measure (see \cite{Muta:2010xua} for the proof). We can solve for the operator $\Delta_G[A]$ with the usual rule for $\delta$ functions that define a system of equations to get
\begin{equation}
 \Delta_G[A]= \det M_G\;\; \text{ with }\;\; (M_G(x,y))^{ab}=\frac{\delta [G^\mu A_\mu^a(\theta)] (x)}{\delta \theta^b(y)}\;. \label{eq:FPoperator}
\end{equation}

We can introduce eq. (\ref{eq:Fadeevtrick}) into $Z[0]$ obtaining
\begin{equation}
    Z[0]=\int [dA] [dg ] \delta^{(n)}(G^\mu A_\mu^a(\theta)-B^a) \Delta_G[A] e^{i\int d^4 x \mathcal{L}_{\text{gauge} }(x)}\;,\label{eq:fadeevalmost}
\end{equation}
and, using the fact that $\Delta_G[A]$ is gauge invariant and also $Z[0]$ and hence so is the argument inside the delta function in eq. (\ref{eq:fadeevalmost}), we can factor out the integral volume over the group measure, $[dg]$, and redefine $Z[0]$ as
\begin{equation}
    Z[0]=\int [dA] \delta^{(n)}(G^\mu A_\mu^a-B^a) \Delta_G[A] e^{i\int d^4 x \mathcal{L}_{\text{gauge} }(x)}\label{eq:Zfunctionalalmost}\;.
\end{equation}
To fix the gauge we can integrate $Z[0]$ in eq. (\ref{eq:Zfunctionalalmost}) with respect to $B^a$ with the Gaussian weight 
\begin{equation}
    \exp{\left\{-i\frac{1}{2\alpha}\int d^4 x (B^a(x))^2\right\}}\;,
\end{equation}
to finally obtain meaningful gauge-fixed expression for the partition functional $Z[J]$,
\begin{equation}
    Z[J]=\int [dA] \det M_G \exp{\left\{ i\int d^4 x(\mathcal{L}_{\text{gauge}}-\frac{1}{2\alpha}(G^\mu A_\mu ^a)^2+A_\mu ^a J^{a\,\mu}) \right\}}\;.\label{eq:gaugepartitionfunctional}
\end{equation}
In this way we have fixed the gauge in a consistent manner, obtaining the non-trivial factor $\det M_G$ that would have been lost by naive gauge fixing. The fact that the naive gauge fixing works for electrodynamics comes from the fact that $\det{M_G}$ is a constant for Abelian gauge fields, making the canonical quantization work. Particular useful gauges are the \textit{Lorentz gauge} where $G^\mu=\partial^\mu$ (when $\alpha=0$ this gauge choice is regarded to as \textit{Landau gauge}) and the \textit{Coulomb gauge} where $G^\mu=(0,\boldsymbol{\nabla})$.

\subsubsection{Quantum Chromodynamics}
At this point, to quantize Yang-Mills theories coupled to fermions, we wish to include fermions in the functional integral for gauge fields. In order to do so one needs the tools so called Grassmann algebra: integration of anti-commuting functions (this is so because fermion fields anticommute). In this algebra, the so called Berezin integrals \cite{Berezin:1966nc} are useful for exponentiating $\det{M_G}$ in eq. (\ref{eq:gaugepartitionfunctional}), 
\begin{equation}
    \det{M_G}=\int [d\xi] [d\xi^*] \exp{-i \left\{\int d^4 x d^4 y \xi^{\ast\, a} (x) (M_G(x,y))^{ab} \xi^b(y) \right\}}\;,
\end{equation}
as well as defining a functional integral for fermion fields. The auxiliary fields $\xi$ and $\xi^*$ are called \textit{ghost fields} or Fadeev-Popov ghosts. They account for the restriction imposed by the gauge choice and are unphysical fields (hence the name), since they are scalar and they anticommute, violating the spin-statistics theorem as a consequence (we cannot build a meaningful Hamiltonian from ghost fields). This means that in practical calculations ghost fields must be taken into account in intermediate steps and never as final states. 

In the Lorentz gauge $(M_G(x,y))^{ab}=(\delta^{ab}\Box -gf^{abc}\partial^\mu A^c_\mu) \delta^{(4)}(x-y)$, and therefore 
\begin{equation}
    \int d^4 x d^4 y \xi^{\ast\, a} (x) (M_G(x,y))^{ab} \xi^b(y) = -\int d^4 x(\partial^\mu \xi^*(x))D_\mu^{ab} \xi^b(x),
\end{equation}
where we integrated by parts in $x$ and $D^{ab}_\mu = \delta^{ab}\partial_\mu -g f^{abc}A_\mu^c$ is the covariant derivative in the adjoint representation of the Lie group $G$. 

Gathering up all the pieces we finally find the partition functional of Yang-Mills theories and QCD
\begin{align}
    Z[J,\chi^*,\chi,\eta,\Bar{\eta}]=\int [dA][d\xi^*][d\xi][d\psi][d\Bar{\psi}]\exp{\{i\int d^4 x (\mathcal{L}_{\text{QCD}}+ AJ+\chi^*\xi+\xi^*\chi+\bar{\psi}\eta+\bar{\eta} \psi) \}}\label{eq:partitionfunctionalQCD},
\end{align}
where $\chi^a$ and $\chi^{\ast\,a}$ are anticommuting source functions, and $\eta$ and $\Bar{\eta}$ are fermionic (we omit color indices). The Lagrangian for QCD (for now we treat with only one quark flavor and a generic semi-simple Lie group $G$) equals
\begin{equation}
    \mathcal{L}_{\text{QCD}}=\mathcal{L}_{\text{YM}}+\mathcal{L}_{\text{GF}}+\mathcal{L}_{\text{FP}}\;,
\end{equation}
with $\mathcal{L}_{\text{YM}}$ defined in eq. (\ref{eq:classicalYM}), the gauge fixing Lagrangian being $\mathcal{L}_{\text{GF}}=-(\partial^\mu A^{a}_\mu)^2/2\alpha$ and the Fadeev-Popov Lagrangian being $\mathcal{L}_{\text{FP}}=(\partial^\mu \xi^{\ast\,a}) D_\mu^{ab}\xi^b$. 

\subsection{Perturbative treatment of Yang-Mills theories}
As we will see shortly, QCD in the SM is a force of opposite nature to QED. The electromagnetic force becomes very intense at short distances (high energies), whereas the strong force grows with distance. As a consequence, quarks and gluons of QCD cannot be directly observed as final state particles, this is due to the confining properties of the strong force (see section \ref{sc:Confinement}). Only hadrons (f.e. the proton or the pion), bound states of quarks and gluons, can be observed as final states. The ``hadronization'' of quarks is a highly non-perturbative phenomenon. However, at very high energy processes (such as Deep Inelastic Scattering \cite{Blumlein:2012bf}), sensitivity to QCD perturbative interactions is reached, since for these processes the hadronic final states are averaged over (totally inclusive processes).

As in the scalar case of section \ref{sec:PTphi4}, to compute perturbative calculations in QCD, we separate the free part in the partition functional of Yang-Mills in eq. (\ref{eq:partitionfunctionalQCD}) from the interacting part as in eq. (\ref{eq:PTforscalars}). All the details for these computations can be found in \cite{Muta:2010xua,Itzykson:1980rh}. Here we limit ourselves to present the Feynman rules for Yang-Mills theories in momentum space.

The Feynman rules for QCD in momentum space are very similar to the ones presented for the scalar case with the following rules for propagators and vertices:
\begin{itemize}
    \item Each gluon propagator with momentum $k$,
    
    \begin{tikzpicture}
    \centering
\draw[decoration={aspect=0.4, segment length=1mm, amplitude=1mm,coil},decorate] (0,0)--(3,0); 
\draw (-6,0) node {$\,$};
\draw (3.4,0) node {$b\, \nu\;,$};
\draw (-.4,0) node {$a\, \mu$};
\draw (1.5,.5) node {$k$};
    \end{tikzpicture}
    
    corresponds to $-i\delta_{ab} d_{\mu\nu}(k)/k^2$ with $d_{\mu\nu}(k)=g_{\mu\nu} -(1-\alpha) k_\mu k_\nu $.

    \item Each ghost propagator with momentum $k$,
    
    \begin{tikzpicture}[>=stealth]
    \centering
\draw[-<-,dashed,line width=0.4mm] (0,0)--(3,0); 
\draw (-6,0) node {$\,$};
\draw (3.4,0) node {$b\;,$};
\draw (-.4,0) node {$a$};
\draw (1.5,.5) node {$k$};
    \end{tikzpicture}
    
corresponds to $i\delta_{ab}/k^2$.
\item Each quark propagator with momentum $k$,
    
    \begin{tikzpicture}[>=stealth]
    \centering
\draw[-<-,line width=0.4mm] (0,0)--(3,0); 
\draw (-6,0) node {$\,$};
\draw (3.4,0) node {$j\;,$};
\draw (-.4,0) node {$i$};
\draw (1.5,.5) node {$k$};
    \end{tikzpicture}
    
corresponds to $i\delta_{ij}/(\slashed{k}-m)$.
\item There are three-gluon and four-gluon vertices, and a gluon-ghost-ghost vertex whose Feynman rules can be found in Appendix D of \cite{Muta:2010xua}.
\item Since it is of interest for this thesis, the tree-level quark-quark-gluon ($q\bar{q}g$) vertex,

    \begin{tikzpicture}[>=stealth]
    \centering
\draw[-<-, line width=0.4mm] (0,0)--(1.5,0); 
\draw[-<-,line width=0.4mm] (1.5,0)--(3,0); 
\draw (-6,0) node {$\,$};
\draw (3.4,0) node {$j\;,$};
\draw (-.4,0) node {$i$};
\draw (1.9,1) node {$a\;\mu$};
\draw[decoration={aspect=0.4, segment length=1mm, amplitude=1mm,coil},decorate] (1.5,0)--(1.5,1); 
    \end{tikzpicture}

corresponds to $ig\gamma_\mu T^a_{ij}$.
\item Each outgoing (incoming) fermion with momentum $p$ corresponds to a spinor $u(p)$ ($\bar{u}(p)$). Each outgoing (incoming) antifermion with momentum $p$ corresponds to a spinor $\bar{v}(p)$ (${v}(p)$). Also, outgoing and incoming gluon or vector lines with momentum $k$ correspond to a polarization vector $\varepsilon_\mu(k)$.
\end{itemize}

With these rules one can now perform any perturbative calculation in QCD, we will for example use the calculations at order $g^5$ in section \ref{subsc:CTP} to obtain the renormalized running masses and couplings for different non-Abelian groups. In section \ref{sec:explicitcomputationjetCS} we will use the corresponding Feynman rules in coordinate space for QCD to perform some explicit computations for factorized QCD amplitudes. 

\subsection{Renormalization of Yang-Mills theories}
The strategy to renormalize QCD order by order is the same as for the scalar case in section \ref{sec:PTandrenormalization}. One redefines the fields as
\begin{equation}
    A_\mu^a=\sqrt{Z_A} A_{R\,\mu}^a\;\;\;,\;\xi=\sqrt{Z_\xi} \xi_R\;\;,\;\;\xi^{\ast}=\sqrt{Z_\xi} \xi^{\ast}_R\;\;,\;\;\psi=\sqrt{Z_\psi} \psi_R\;,
\end{equation}
and the parameters as
\begin{equation}
    g=Z_g g_R\;\;,\;\;\alpha=Z_A \alpha_R\;\;,\;\; m=Z_m m_R\;.\label{eq:renormalizedqcdparameters}
\end{equation}
Notice we choose the renormalization of the gauge parameter $\alpha$ in order to keep the gauge condition unaltered; this condition is crucial for obtaining the so called Slavnov-Taylor identities. These ensure the universality of the renormalized coupling $g_R$, \textit{i.e.} in all parts of the Lagrangian where $g$ appears, this coupling renormalizes in a unique way. The Slavnov-Taylor identities can be obtained from the BRS \cite{Becchi:1975nq} symmetry of the QCD Lagrangian, a symmetry of the quantum Lagrangian inherited by the gauge symmetry of the classical one. BRS symmetry is crucial for the proof that QCD is renormalizable at all orders in perturbation theory. 

Here we will only concentrate on the running of the gauge coupling $g_r(\mu)=g_R \mu^{-\epsilon}$ and the quark mass $m_r=m_R$. As in the scalar case, the renormalization equations for these quantities are
\begin{align}
    &\mu \frac{d g_r}{d\mu}=\beta (\mu) &\text{with}\;\;\;\;\;\;\;\;&\beta(\mu)=-\epsilon g_r-\frac{\mu}{Z_g} \frac{d Z_g}{d \mu} g_r\;,\label{eq:betaQCD}\\
  &\mu \frac{d m_r}{d\mu}=-m_r\gamma_m(\mu)   &\text{with}\;\;\;\;\;\;\;\;&\gamma_m(\mu)=\frac{\mu}{\sqrt{Z_m}} \frac{d \sqrt{Z_m}}{d\mu}\;.\label{eq:anomalousQCD}
\end{align}
One can compute the running functions $\beta(\mu)$ and $\gamma(\mu)$ perturbatively as
\begin{align}
    \beta(\mu)=&-\beta_1 g_r^3-\beta_2 g_r^5+\mathcal{O}(g^7)\label{eq:beta2loops}\;,\\
\gamma_m(\mu)=&\gamma_{m\,1}g^2_r+\gamma_{m\,2}g^4_r+\mathcal{O}(g_r^6)\;.
\end{align}
For example, the first and second correction to the running of the coupling $g_r$, $\beta_1$, can be computed from the ghost self energy, the ghost-gluon vertex and the gluon self energy (owing to the Slavnov-Taylor identities) at two loops to give, in the MS renormalization scheme \cite{Muta:2010xua},
 \begin{align}
\beta_1=&\frac{1}{(4\pi)^2}\Big(\frac{11C_G-N_f}{3}\Big)\;,\label{eq:couplingrunning1}\\
\beta_2=&\frac{1}{(4\pi)^4}\Big(\frac{34}{3}C_G^2-2\Big(\frac{5}{3}C_G+C_F\Big)N_f\Big)\label{eq:couplingrunning2}\;.
\end{align}
Where $\sum_{a}(T^a T^a)_{ij}\equiv \delta_{ij} C_F$ and $C_A=\sum_{cd} f^{acd} f^{bcd}=\delta^{ab}C_G$ are the fundamental and adjoint Casimir for the group $G$. $N_f$ is the number of quark flavors. 

Likewise, the first and second corrections to the running of the current mass $m_r$ are
\begin{align}
    \gamma_{m\,1}=&\frac{6C_F}{(4\pi)^2}\label{eq:massrunning1}\;,\\
    \gamma_{m\,2}=&\frac{C_F}{(4\pi)^2}\left(3 C_F+\frac{97}{3}C_G-\frac{10}{3} N_f\right)\label{eq:massrunning2}\;.
\end{align}

These results, especially eq. (\ref{eq:couplingrunning1}), are important to prove asymptotic freedom in QCD (where $C_G=3$ and $N_f=6$), \textit{i.e.} the fact that the coupling for the strong force, $g_s(\mu)$, in the SM vanishes as $\mu\to\infty$, making quarks behave freely at high energies. We will use these equations in section \ref{subsc:CTP} to study the running of couplings and masses for different Lie groups.

\subsection{Dyson-Schwinger equations}
The functional treatment of QFTs will help us describing and dealing with non-Abelian QFTs in their non-perturbative regime through the Dyson-Schwinger equations (derived for Electrodynamics by those two authors \cite{Dyson:1949ha,Schwinger:1951hq} and extended to non-Abelian theories in \cite{Baker:1976vz}). In this section we will, for simplicity, first derive Dyson-Schwinger equations (DSE) in QED and then move forward to present them in QCD. These equations are the wave equations of a QFT, formulated as identities among its exact Green's functions. They are most easily obtained by noticing that a total derivative inside a path integral vanishes, \textit{i.e} 
\begin{equation}
\int [d\phi] \frac{\delta}{\delta \phi} (F[\phi])=0\label{eq:totalderivativefunctional}\;,
\end{equation}   
for any arbitrary bounded functional, $F$, of the field $\phi$ (this assumes that the functional integral brings no boundary terms for arbitrary large $|\phi|$ field configurations). Choosing $F$ to be the usual exponential of the action, $S[\phi]=\int d^4 x \mathcal{L}[\phi]$, plus the source terms we obtain
\begin{equation}
\int [d\phi] \frac{\delta}{\delta \phi} \exp{\left\{i S[\phi]+\int d^4 x \phi J \right\}}=\int [d\phi] i\left[\frac{\delta S}{\delta \phi} [\phi]+J\right] \exp{\left\{i S(\phi)+\int d^4 x\, \phi J \right\}}=0\;.
\end{equation}
Which, in terms of the partition functional $Z[J]$, can be expressed as (changing $\phi\to\frac{\delta}{i\delta J}$)
\begin{equation}
   \left[ \frac{\delta S}{\delta \phi} \left[ \frac{\delta}{i\delta J}\right] +J  \right] Z[J]=0\;.\label{eq:DSE1}
\end{equation}

For the case of QED we use in eq. (\ref{eq:DSE1}) the corresponding partition functional of eq. (\ref{eq:partitionfunctionalQCD}) with the gauge group being $U(1)$ (so that there are no ghost fields or color indices) and $g=-e$, finding (omitting spinor indices)
\begin{equation}
    \left[ \eta(x)+\left( i\slashed \partial -m-e\gamma^\mu \frac{\delta}{i \delta J^\mu (x)}\right)\frac{\delta }{i\delta \bar{\eta}(x)}\right]Z[J,\eta,\bar{\eta}]=0\;.\label{eq:QEDSDE1}
\end{equation}
If, at this point, we take another functional derivative as $\frac{\delta }{-i\delta \eta (y)}$ in eq. (\ref{eq:QEDSDE1}) and set $\eta=\bar{\eta}=0$, we get 
\begin{equation}
    \delta^{(4)}(x-y) Z[J,0,0]-\left( i\slashed \partial -m-e\gamma^\mu \frac{\delta}{i \delta J^\mu (x)}\right)Z[J,0,0]S_F(x-y;J)=0\;,
\end{equation}
where 
\begin{equation}
   S_F(x-y;J)\equiv \frac{1}{Z[J,0,0]}\left(\frac{\delta^2 Z[J,\eta,\bar{\eta}]}{i\delta  \bar{\eta}(y) (-i)\delta \eta (x)} \right)\Big|_{\eta=\bar{\eta}=0}\;, 
\end{equation}
 is the fermion propagator in the presence of the source $J$. 
 
 Now we use the definition of the generating functional for connected Green's functions $Z[J,\eta,\bar{\eta}]=\exp{iW[J,\eta,\bar{\eta}]}$ to obtain 
\begin{equation}
    \exp{iW[J,0,0]}\left[\delta^{(4)}(x-y)-\left( i\slashed \partial -m-e\gamma^\mu \frac{\delta W[J,0,0]}{ \delta J^\mu (x)}-e\gamma^\mu \frac{\delta}{i \delta J^\mu (x)}\right)S_F(x-y;J)\right]=0\;.\label{eq:almostDSE}
\end{equation}

In order to simplify eq. (\ref{eq:almostDSE}), let us express the classical fields as vacuum expectation values of field operators (similarly to eq. (\ref{eq:greensfunctionsandZ})) in the presence of sources:
\begin{equation}
    A^\mu_{\text{cl}}(x;J,{\eta},\bar{\eta})\equiv \frac{1}{Z[J,\eta,\bar{\eta}]}\frac{\delta }{i\delta J_\mu(x) } Z[J,\eta,\bar{\eta}]=\frac{\delta W[J,\eta,\bar{\eta}]}{\delta J_\mu(x)}\;,\label{eq:classicalAsources}
\end{equation}
\begin{align}
    \psi_{\text{cl}}(x;J,{\eta},\bar{\eta})&\equiv \frac{1}{Z[J,\eta,\bar{\eta}]}\frac{\delta }{i\delta \bar{\eta}(x) } Z[J,\eta,\bar{\eta}]=\frac{\delta W[J,\eta,\bar{\eta}]}{\delta \bar{\eta}(x)}\;,\label{eq:classicalpsources}\\
    \bar{\psi}_{\text{cl}}(x;J,{\eta},\bar{\eta})&\equiv \frac{1}{Z[J,{\eta},\bar{\eta}]}\frac{\delta }{-i\delta {\eta}(x) } Z[J,{\eta},\bar{\eta}]=\frac{\delta W[J,{\eta},\bar{\eta}]}{-\delta{\eta}(x)}\;.\label{eq:classicalpbarsources}
\end{align}

Now we need to define the \textit{effective action} as a Legendre transform of $W$,
\begin{equation}
    \Gamma[A_{\text{cl}},\psi_{\text{cl}},\bar{\psi}_{\text{cl}}]\equiv W[J,\eta,\bar{\eta}]-\int d^4y (\bar{\eta}\psi_{\text{cl}}+\bar{\psi}_{\text{cl}}\eta+J_\mu A^\mu_{\text{cl}} )\label{eq:effectiveactionQED}\;,
\end{equation}
which is a functional of the classical fields \cite{Itzykson:1980rh}. From eq. (\ref{eq:effectiveactionQED}) we have that (dropping the $\text{cl}$ subindices and the dependence on the source fields)
\begin{equation}
    \frac{\delta \Gamma}{\delta A_\mu(x)}=-J^\mu (x)\;,\;\;\;\frac{\delta \Gamma}{\delta \psi(x)}=\bar{\eta} (x)\;,\;\;\;\frac{\delta \Gamma}{\delta \bar{\psi}(x)}=-\eta (x)\;. \label{eq:derivativeseffectiveaction}
\end{equation}
At this point we need to use the following identity
\begin{align}
    \delta^{4}(x-y)=&\frac{\delta \psi(x)}{\delta \psi(y)}=\int d^4 z \frac{\delta {\eta}(z)}{\delta \psi(y)} \frac{\delta \psi (x)}{\delta {\eta}(z)}=\nonumber\\
    =&\int d^4 z\left(\frac{\delta^2 \Gamma}{\delta \psi(y) \delta \bar{\psi}(z)}\right)\Big|_{\psi=\bar{\psi}=0}\left(\frac{\delta^2 W}{\delta \eta(z) \delta \bar{\eta}(x)}\right)\Big|_{\eta=\bar{\eta}=0}\;,\label{eq:inversepropagatoridentity}
\end{align}
obtained by the use of the chain rule, the eqns. (\ref{eq:classicalpsources}), (\ref{eq:classicalpbarsources}) and (\ref{eq:derivativeseffectiveaction}), and setting afterwards $\eta$ and $\bar{\eta}$ to zero for convenience (which also means setting $\psi$ and $\bar{\psi}$ to zero). Eq. (\ref{eq:inversepropagatoridentity}) means that the inverse fermion propagator equals
\begin{equation}
    S^{-1}_F(x-y;J)=\left(\frac{-i\delta^2 W}{\delta \eta(z) \delta \bar{\eta}(x)}\right)^{-1}\Big|_{\eta=\bar{\eta}=0}=\left(\frac{-i\delta^2 \Gamma}{\delta \psi(x) \delta \bar{\psi}(y)}\right)\Big|_{\psi=\bar{\psi}=0}\;.
\end{equation}
This relation will help us compute the last term in eq. (\ref{eq:almostDSE}),
\begin{align}
    \frac{\delta}{i \delta J^\mu (x)}S_F(x-y;J)=\int d^4 z \frac{\delta A_\nu(z)}{\delta J^\mu (x)} \frac{\delta }{\delta A_\nu (z)} \left(\frac{\delta^2 \Gamma}{\delta \psi(x) \delta \bar{\psi}(y)}\Big|_{\psi=\bar{\psi}=0}\right)^{-1}\label{eq:intermediateDSE}\;,
\end{align}
since we have the following relations \cite{Itzykson:1980rh},
\begin{equation}
    \left(\frac{\delta }{\delta A_\mu (x)} \frac{\delta^2 \Gamma}{\delta \bar{\psi}(y) \delta {\psi}(z)}\right)\Big|_{A=\bar{\psi}={\psi}=0}=-ie \Gamma^\mu (x;y,z)\label{eq:irrvertex}\;,
\end{equation}
with $\Gamma^\mu$ being the irreducible vertex function (which is the sum of all proper/1PI diagrams contributing to the photon-electron-positron ($\gamma e\bar{e}$/$g q\bar{q}$) vertex), and the usual definition,
\begin{equation}
  \left(  \frac{\delta A_\nu(z)}{\delta J^\mu (x)}\right)\Big|_{J=\bar{\eta}={\eta}=0}=\left(\frac{\delta^2 W[J,\eta,\bar{\eta}]}{\delta J^\mu (x)\delta J^\nu (z)}\right)\Big|_{J=\bar{\eta}={\eta}=0}=i D_{\mu\nu} (x-z)\;,\label{eq:photonpropagator}
\end{equation}
where $D_{\mu\nu} (x)$ is the full photon propagator. 

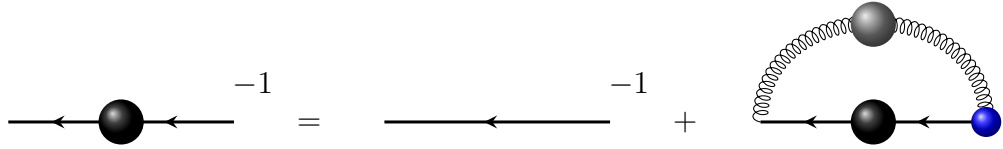
\begin{figure}[ht!]
    \centering
    \begin{tikzpicture}[>=stealth]
        \draw[-<-,line width=0.4mm] (0,0)--(1.5,0); 
        \draw[-<-,line width=0.4mm] (1.5,0)--(3,0); 
        \shade[ball color=black]  (1.5,0) circle (.3cm);
        \draw (4,-0.025) node {$=$};
        \draw (3.25,0.5) node {$-1$};
        \draw[-<-,line width=0.4mm] (5,0)--(8,0); 
        \draw (8.25,0.5) node {$-1$};
        \draw (9,-0.) node {$+$};
        \draw[-<-,line width=0.4mm] (10,0)--(11.5,0); 
        \draw[-<-,line width=0.4mm] (11.5,0)--(13,0); 
        \draw[decoration={aspect=0.5, segment length=1mm, amplitude=1mm,coil},decorate] (10,0) .. controls (10,1.75) and (13,1.75) .. (13,0);
        \shade[ball color=black]  (11.5,0) circle (.3cm);
        \shade[ball color=gray]  (11.5,1.3) circle (.3cm);
        \shade[ball color=blue]  (13,0) circle (.2cm);
    \end{tikzpicture}
    \caption[Representation of the quark propagator DSE]{Representation of the DSE in eq. (\ref{eq:DSEcoordinatespace}) for the electron (quark) propagator in QED (QCD). The black blob in between fermion lines represents the full fermion propagator, the gray blob in between gluon lines represents the full photon/gluon propagator, and the blue blob represents the $\gamma e\bar{e}$ ($g q\bar{q}$) dressed vertex.}
    \label{fig:DSE}
\end{figure}

Using eq. (\ref{eq:irrvertex}) and (\ref{eq:photonpropagator}) into eq. (\ref{eq:intermediateDSE}), we have finally from eq. (\ref{eq:almostDSE}), after setting $J=0$ (which means $A_{\text{cl}}(x)=0$) and multiplying by a $S_F^{-1}(y,y')$ and integrating over $y$ and relabeling $y'\to y$, the DSE for the electron propagator in coordinate-space
\begin{align}
    S_F^{-1}(x,y)=(i\slashed{\partial}-m) \delta^{(4)}(x-y)-e^2\int d^4z\,d^4v\,\gamma^\mu D_{\mu\nu}(x-z) S_F(z-v) \Gamma^\mu(z;v,y) \;.\label{eq:DSEcoordinatespace}
\end{align}
This DSE for the electron propagator in QED is depicted using Feynman diagrams in Fig. \ref{fig:DSE}.

The DSE for the quark propagator in eq. (\ref{eq:DSEcoordinatespace}) is nothing but the electron's field equation in full QED since it is equivalent to
\begin{equation}
    [(i\slashed{\partial}-m-\Sigma) S_F](x-y)=\delta^{(4)}(x-y)\;,
\end{equation}
where $\Sigma$ is the self energy operator equal to the last term in the RHS of eq. (\ref{eq:DSEcoordinatespace}).

In QCD, the boundary terms that could arise from the total derivative in eq. (\ref{eq:totalderivativefunctional}) also vanish since $\det M_G$ vanishes at the Gribov Horizon \cite{Greensite:2011zz} for Coulomb or Landau gauges (see the next section \ref{sc:Confinement} for details). So that the same derivation of the DSE can be performed for QCD to obtain, similarly to eq. (\ref{eq:DSEcoordinatespace}) transformed to momentum space, \begin{equation}
    \tilde{S}^{-1}(p)=Z_\psi(\tilde{S}_{0}(p))^{-1}-Z_{1F} {g^2 C_F}\int \frac{d^4 k}{(2\pi)^4}\gamma^\mu \tilde{D}_{\mu\nu}(p-k) \tilde{S}(k) \Gamma^\mu(p,k)\;,\label{momentumfullDSE}
\end{equation} 
where now $ \tilde{S}_0(p^2)=\frac{i}{\slashed p-m_c}\;,$ is the free fermion propagator with current mass $m_c$ and $\tilde{D}_{\mu\nu}$ is the gluon propagator both in momentum space, $\Gamma^\mu$ is the renormalized quark-antiquark-gluon ($q\bar{q}g$) vertex (without its color structure), and $Z_{1F}=Z_g Z_\psi\sqrt{ Z_A}$ is the vertex's renormalization constant.

In order to have a complete set of coupled equations for solving QCD in its non-perturbative regime, one needs to obtain the DSE for the ghost and gluon propagators and also for the $q\bar{q}g$ and the ghost-antighost-gluon vertex. These equations can be found in the literature (\textit{e.g.} \cite{Greensite:2011zz} or \cite{Swanson:2010pw}) and we will not present them here.  

Now we turn to use QCD in both the perturbative and non-perturbative regimes for studying the effect on dynamical mass generation on fermions charged under different Lie groups.

\subsection{Why is the SM formulated to have such specific symmetries?}\label{subsc:CTP}
A peculiar feature of the SM's symmetry groups in eq. (\ref{eq:SMLieGroups}) is the small size of the numbers 1-2-3. Why do we observe such symmetry groups? Why not larger groups like $SU(7)$ or $Sp(8)$? 

To address these questions, based on our original publication \cite{Llanes-Estrada:2018azk}, we study how hypothetical quarks colored under different Lie groups acquire masses from a Grand Unified Theory (GUT) scale where all groups under consideration are chosen to have the same couplings and quark masses, down to lower energies where the interaction becomes strong. To do so, we will use the Renormalization Group Equations (RGE) at two loops in eqns. (\ref{eq:beta2loops}) to (\ref{eq:massrunning2}).  Then, when interactions become strong, we treat the DSE in the so called ``rainbow ladder'' approximation and omitting the wave-function renormalization of the quarks. This is enough for qualitatively studying mass generation at the lowest scales.

\begin{figure}[!ht]
\includegraphics[width=.5\columnwidth]{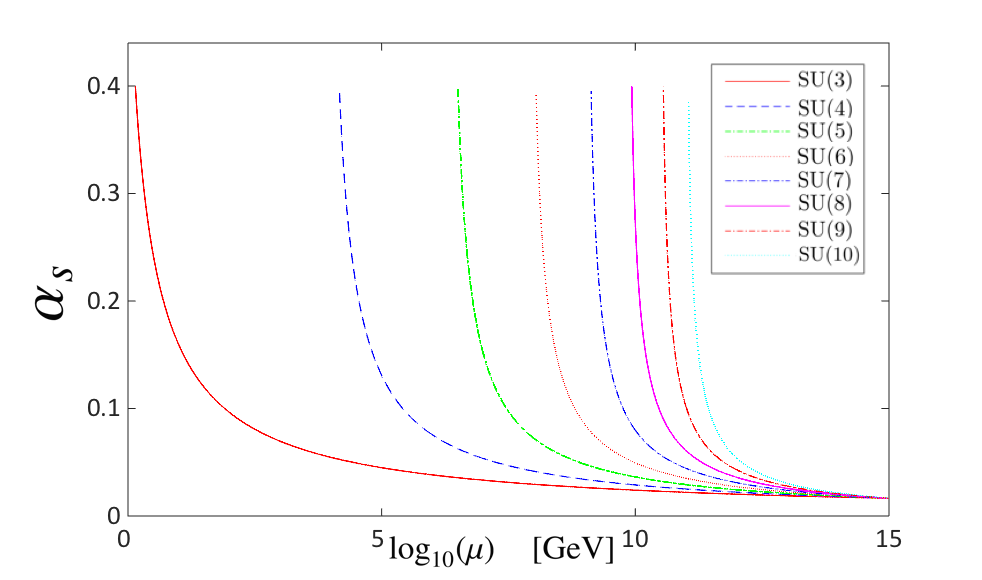}
\includegraphics[width=.5\columnwidth]{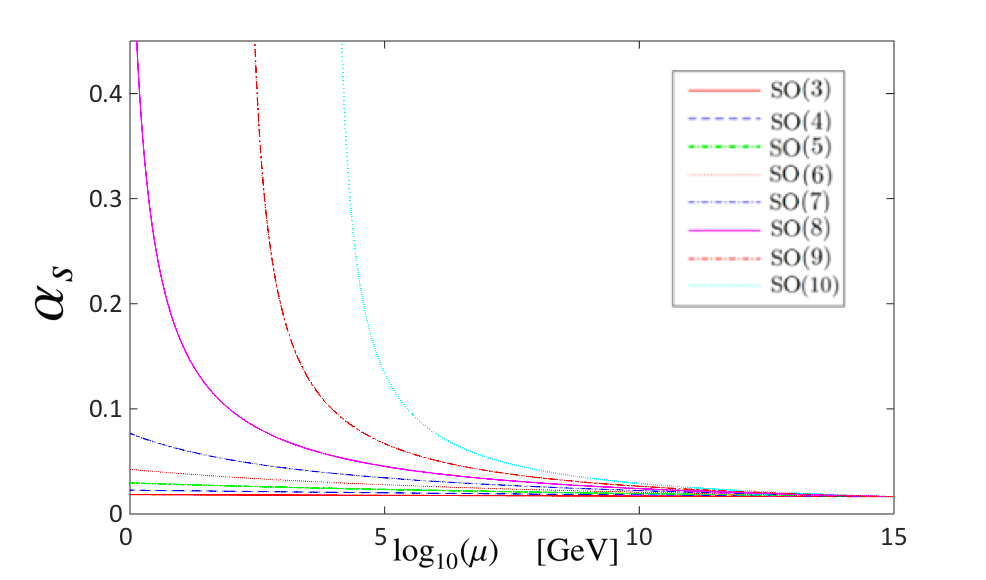}
\includegraphics[width=.5\columnwidth]{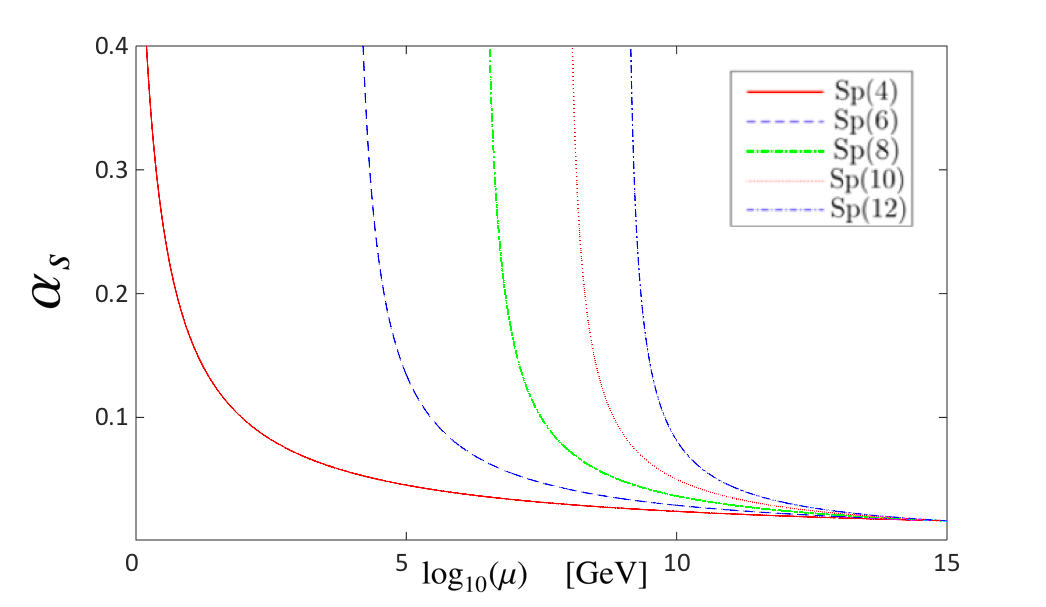}
\includegraphics[width=.5\columnwidth]{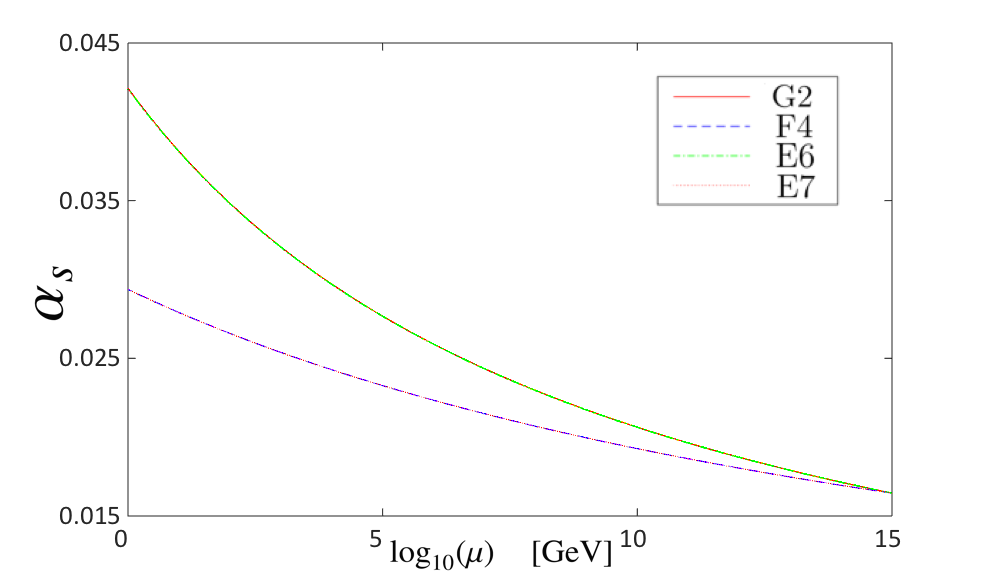}
  \caption[Running couplings for different Lie groups]{Running couplings, $\alpha_s=g^2/4\pi$, for the families $SU(N)$, $SO(N)$, $Sp(N)$ and most of the Exceptional Lie groups. \label{fig:runningcoupling}}
\end{figure}

The evolution for the couplings for the different color groups from the Grand Unification scale of $\mu_{\text{GUT}}=10^{15}$ GeV to the point where interactions become strong (at a scale $\sigma$) for each group (we choose $C_F\,\alpha_s(\sigma)=0.4$), is presented in Fig. \ref{fig:runningcoupling}.

\begin{figure}[!ht]
\includegraphics[width=.5\columnwidth]{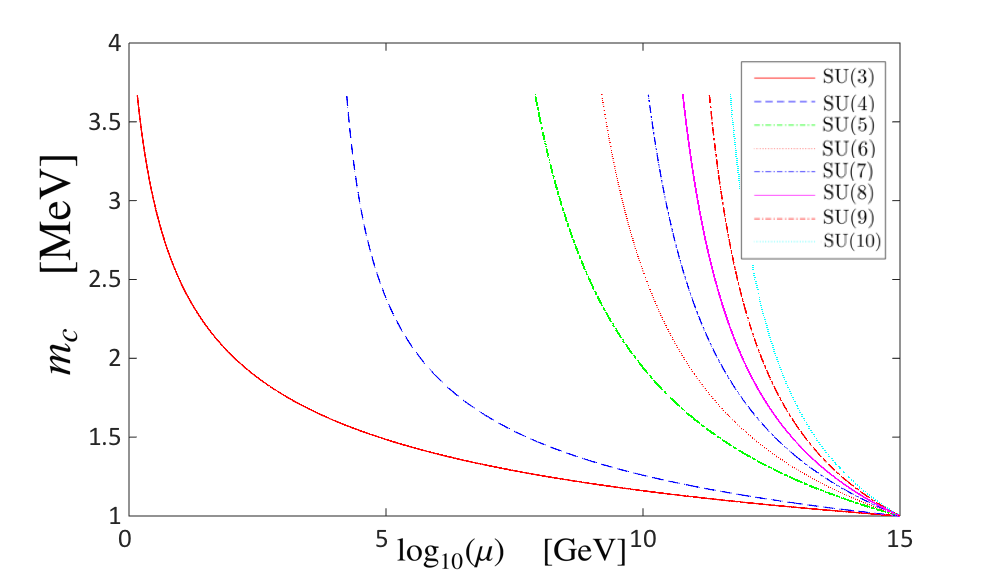}
\includegraphics[width=.5\columnwidth]{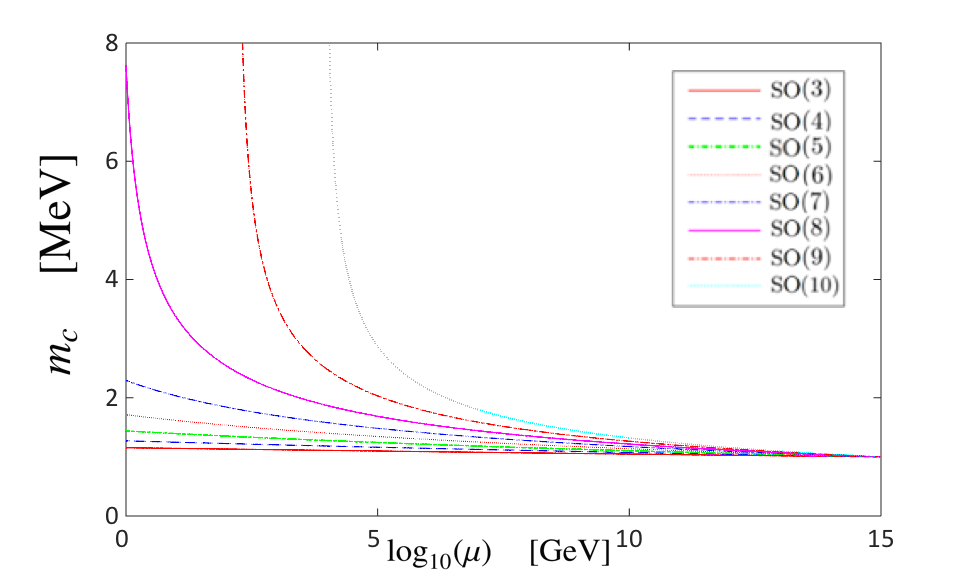}
\includegraphics[width=.5\columnwidth]{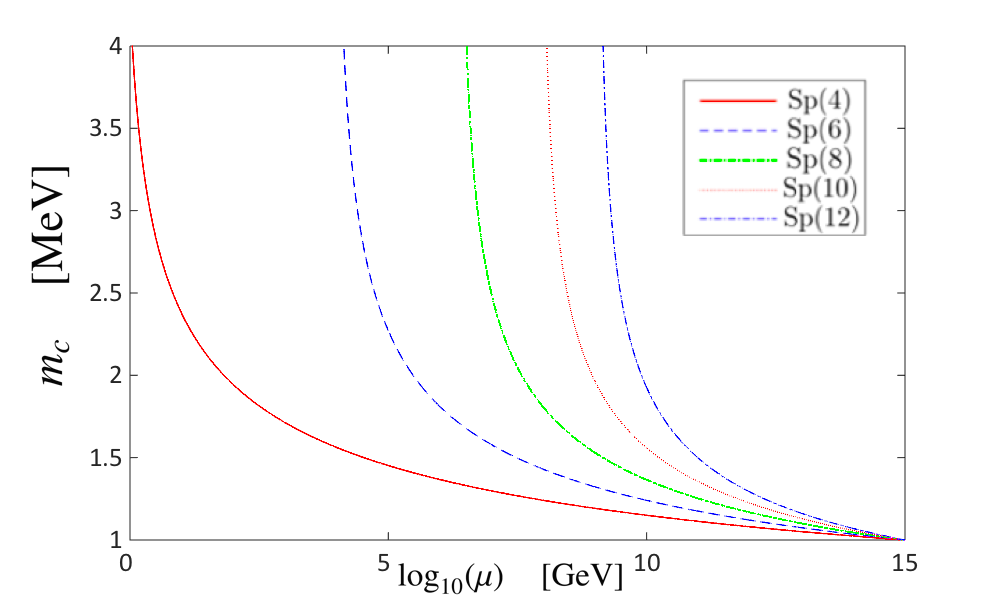}
\includegraphics[width=.5\columnwidth]{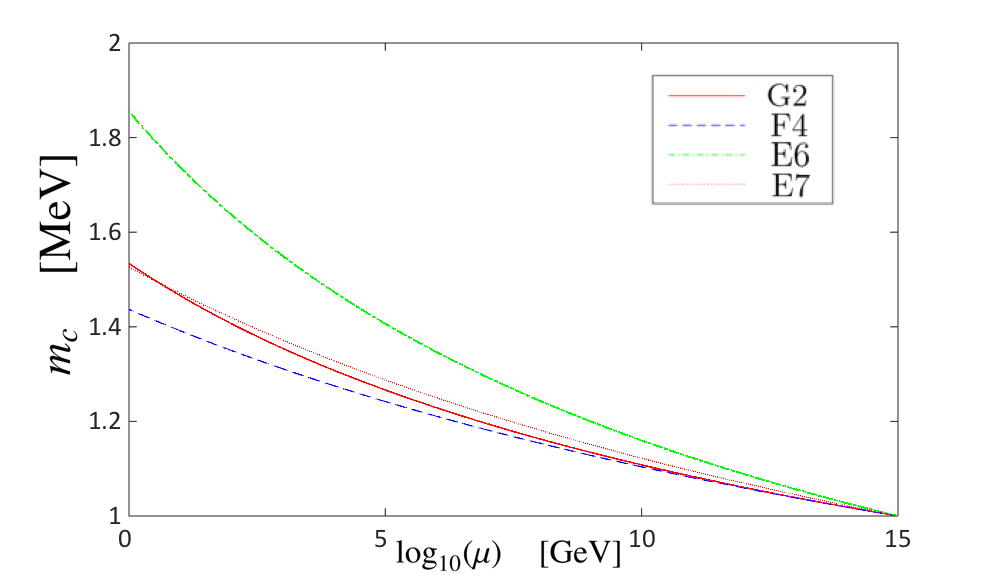}
  \caption[Running of fermion masses for different Lie groups]{Running masses for the families $SU(N)$, $SO(N)$, $Sp(N)$ and four out of five Exceptional Lie groups.\label{fig:runningmass}}
\end{figure}

At the GUT  starting scale of the RGEs we choose a fermion mass $m_{c}(\mu_{\text{GUT}})=1$ MeV and fix the coupling $\alpha_{s}(\mu_{\text{GUT}})\equiv g_s({\mu_{\text{GUT}}})^2/4\pi=0.0165$ to broadly reproduce the isospin average mass for the $SU(3)_C$ quarks of the first generation at the scale $\mu=2\;\text{GeV}$,
\begin{equation}
\overline{m}(2\,\text{GeV})=\frac{m_u(2\;\text{GeV})+m_d(2\;\text{GeV})}{2}\simeq3.5\;\;\text{MeV}\;.
\end{equation}
These initial conditions are taken to be the same for all Lie groups, as suggested by the concept of GUT. Then, the mass running for the various Lie groups, with color factors taken from \cite{Llanes-Estrada:2018azk} is plotted in Figure~\ref{fig:runningmass}.

\begin{figure}[!ht]\centering
\begin{tikzpicture}
     \draw[-<-,line width=0.4mm] (10,0)--(11.5,0); 
        \draw[-<-,line width=0.4mm] (11.5,0)--(13,0); 
        \draw[decoration={aspect=0.5, segment length=1mm, amplitude=1mm,coil},decorate] (10,0) .. controls (10,1.75) and (13,1.75) .. (13,0);
        \shade[ball color=black]  (11.5,0) circle (.3cm);
        \draw (15,.5) node {${=\Sigma_{\text{rainbow}}(p)}$};
\end{tikzpicture}
  \caption[Rainbow truncation of the DSE]{The rainbow truncation 
  of the complete DSE in Fig. (\ref{fig:DSE}) approximates $\Sigma(p)$ (last term in the RHS of Fig. \ref{fig:DSE}) with $\Sigma_{\text{rainbow}}(p)$.\label{fig:rainbow}}
\end{figure}
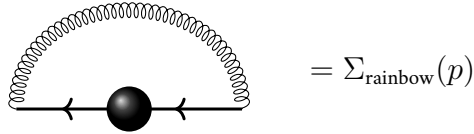

Once the interactions become strong, perturbation theory breaks down and we need to use the DSE. We will adopt the simplest possible DSE for the quark self energy where the dressed $gq\bar{q}$ vertex and the gluon propagator (see Fig. \ref{fig:rainbow}). This is the rainbow ladder approximation and owes its name to the fact that it only sums all rainbow-shaped diagrams (see \cite{Llanes-Estrada:2018azk}), ignoring the rest of contributions. 

In the non-perturbative regime, the free propagator of a fermion with current mass $m_c$, $ \tilde{S}_0(p^2)$,
 becomes a fully dressed one $ \tilde{S}(p^2)=\frac{i}{A(p^2)\slashed p-B(p^2)}\;$. 
Being only interested in qualitative features of spontaneous mass generation, we can approximate $A(p^2)=1$ which leaves the physical mass (\textit{i.e.} the pole of the propagator) as $M(p^2)\equiv B(p^2)$.
From the DSE of eq. (\ref{momentumfullDSE}), 
\begin{equation}
\tilde{S}(p^2)=\tilde{S}_0(p^2)\,(1+\Sigma(p) \tilde{S}_0(p^2))^{-1}\;,
\end{equation}
 we see that
\begin{equation} \tilde{S}^{-1}(p^2)=\tilde{S}_0(p^2)^{-1}+\Sigma(p) \Rightarrow M(p^2)=m_c+\Sigma(p)\;. 
\end{equation}

 Passing to Euclidean space with $k^0_E=ik^0$, 
$p^0_E=ip^0$, and omitting the renormalization constant (we will employ a renormalization procedure where $Z_{1F}=Z_\psi=1$, see below), the rainbow self energy, $\Sigma_{\text{rainbow}}(p)$, is given by
\begin{align}
\Sigma_{\rm rainbow}(p)=&-C_Fg^2_s\int\frac{d^4k}{(2\pi)^4}\gamma^\mu\frac{i}{\slashed k-M(k^2)}\gamma^\nu \frac{\eta_{\mu\nu}}{(k-p)^2}\nonumber\\
= C_F&g^2_s\int_0^\infty\frac{dk_E \,k_E^3}{\pi^3}\frac{M(k^2)}{M^2(k^2)+k_E^2}\int_{-1}^{+1}\sqrt{1-x^2}\frac{dx}{(k_E^2-2|k_E||p_E|\,x+p^2_E)}\;,\label{eq:euclideanrainbow}
\end{align}
in Feynman gauge ($\alpha=1$ for the gluon propagator). We define the last integral in $x$ of eq. (\ref{eq:euclideanrainbow}) as the averaged gluon propagator $D^0_{k-p}$ (in the Feynman Gauge) over the four dimensional polar angle. Hence, we conclude that the Dyson-Schwinger equation in the rainbow approximation for the quark propagator is
\begin{equation}\label{eq:DSEeuclid}
M(p^2)=m_c+C_Fg^2_s\int_0^\infty\frac{dq \,q^3}{\pi^3}\frac{M(q^2)}{M^2(q^2)+q^2}D^0_{q-p}\;.
\end{equation}

The integral in (\ref{eq:DSEeuclid}) is divergent and must be regularized. We could employ a simple cutoff regularization cutting this integral at a scale $\Lambda$; instead we would like to preserve Lorentz invariance and exhibit renormalizability. Following \cite{Llanes-Estrada:2018azk,GarciaFernandez:2015jmn}, we introduce renormalization constants $Z(\Lambda^2,\mu^2)$ to absorb infinities and any dependence on the cutoff $\Lambda$ as
\begin{equation}
\tilde{S}^{-1}(p^2,\mu^2)\equiv Z_\psi \tilde{S}_0^{-1}(p^2)+\Sigma(p^2,\mu^2)\;,
\end{equation}
where the dependence of $\Sigma$ on $\mu$ is given by the fermion and gluon propagators. Apart from the wave-function renormalization $Z_\psi$ we introduce $Z_m$ for the bare quark mass. The relation between the (cutoff dependent) unrenormalized mass $m_c(\Lambda^2)$ and the renormalized mass at the renormalization scale $\mu$, $m_R(\mu^2)$, is that of eq. (\ref{eq:renormalizedqcdparameters}),
\begin{equation}
m_c(\Lambda^2)=Z_m(\Lambda^2,\mu^2)m_R(\mu^2).
\end{equation}
Since we will maintain the restriction $A(p^2)=1$, renormalization of the quark wave-function is not necessary, therefore $Z_\psi=1$. The only renormalization condition is to fix the  mass function at $p^2=\mu^2$. The DSE is then
\begin{equation}
M(p^2)=Z_m m_R(\mu^2)+\Sigma(p^2,\mu^2)\;.\label{EQ:RMM}
\end{equation}
Evaluating (\ref{EQ:RMM}) at $p^2=\mu^2$ and subtracting it again to (\ref{EQ:RMM}), taking $\mu$ parallel to $p$, we obtain,
\begin{align}\label{eq:DSEMOM}
M(p^2)=&M(\mu^2)+C_Fg^2_s\int_0^\infty\frac{dq \,q^3}{\pi^3}\frac{M(q^2)}{M^2(q^2)+q^2}\Big(D^0_{q-p}-D^0_{q-\mu}\Big)\;.
\end{align}
It is easy to show that asymptotically~\cite{Llanes-Estrada:2018azk,GarciaFernandez:2015jmn},
\begin{equation}
\frac{\partial M(p^2)}{\partial \Lambda}\propto\frac{M(\Lambda^2)(p-\mu)}{\Lambda^2}\;.
\end{equation}
Therefore, for large $\Lambda$, $M(p^2)$ stops depending on the cutoff (which can be taken {\it e.g.} to $\Lambda=10^{10}\; \text{GeV}$), and renormalization is achieved.

\begin{figure}[!ht]
\centering
\includegraphics[width=.7\columnwidth]{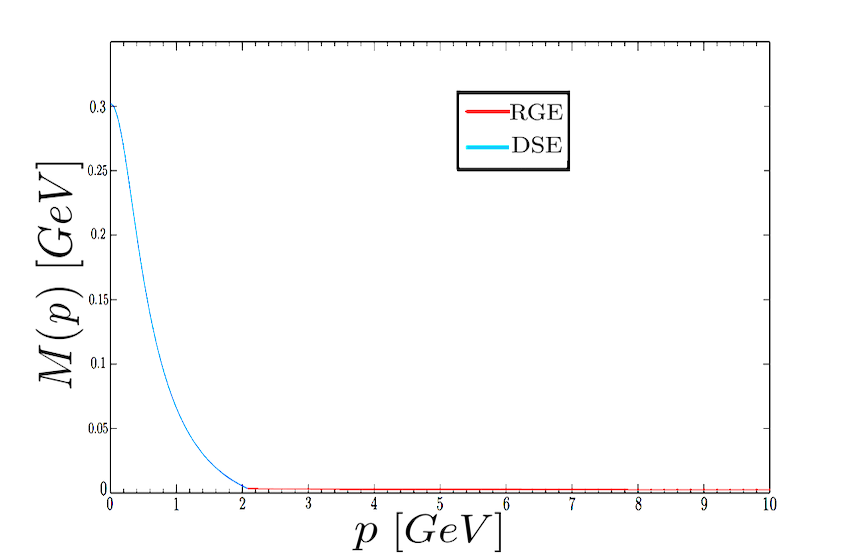}
\caption[Matching of RGE and DSE solutions for the fermion masses]{Matching of RGE and DSE solutions of the Mass Running for $SU(3)$.}\label{fg:DSERGE}
\end{figure}

Now we are ready to obtain the quark constituent masses for all the groups studied. We match the RGE solution (high scales) to the DSE solution (low scales) at the matching energy $\sigma$ where interactions become strong,  $C_F\alpha_s(\sigma)=0.4$ for each group, as advertised. 
For $SU(3)$ ($C_F=\frac{4}{3}$), the scale where $\alpha_s(\sigma)=0.3$ is $\sigma=2.09\; \text{GeV}$. From this point down in scale we freeze $\alpha_s$.  A constant vertex factor of order 7 is applied to the DSE to guarantee sufficient chiral symmetry breaking at low scales, determined by requiring the constituent quark mass $M(0)$ to be close to $300\;\text{MeV}$ (roughly one third of the proton's mass) using the subtracted DSE of eq. (\ref{eq:DSEMOM}). This is supposed to mock up the effect of vertex-corrections not included, and is known to scale with $N$~\cite{Kizilersu:2006et} for large $N$, the group's fundamental dimension.
Finally, the $M(p)$ obtained is plotted in Figure \ref{fg:DSERGE}.

\begin{figure}[!ht]
\centering
\includegraphics[width=0.9\columnwidth]{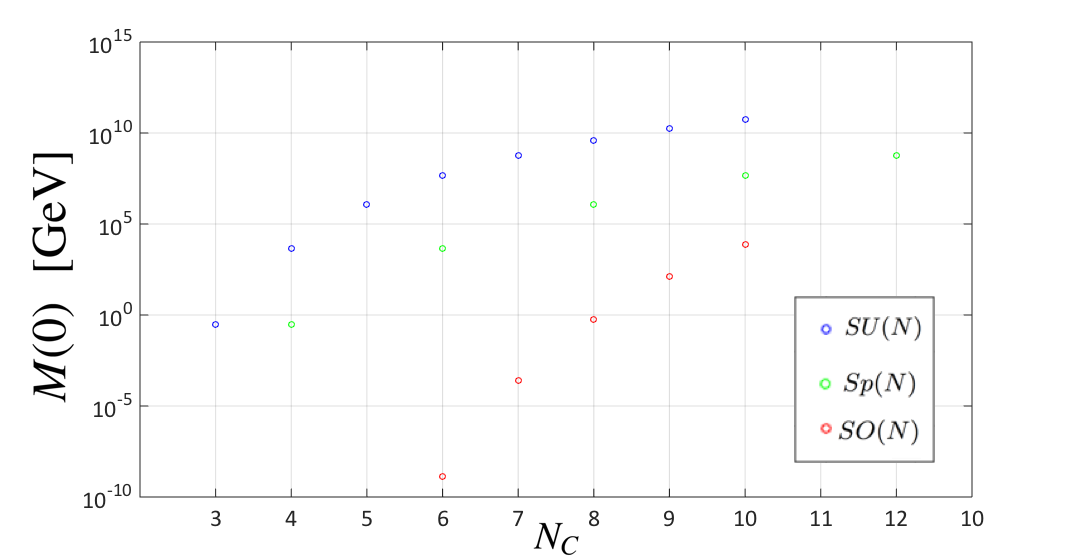}
\caption[Constituent quark masses for different Lie groups]{Constituent Masses for the groups which break chiral symmetry in RGE before $10^{-5}\;\text{GeV}$.}\label{constituent}
\end{figure}

To obtain the constituent fermion masses for the different Lie Groups we use a trick presented in \cite{Llanes-Estrada:2018azk,GarciaFernandez:2015jmn}: to perform a scale transformation 
\begin{equation}
p^2\to \lambda^2p^2\;, \;\; \sigma^2\to \lambda^2\sigma^2\;,
\end{equation}
on the DSE of eq. (\ref{eq:DSEMOM}). Changing the integration variable $q^2\to \lambda^2q^2$, giving $d^4q\to \lambda^4d^4q$, 
the modified DSE equation is satisfied by a modified $\tilde{M}$  and the relation between the constituent masses is simply $ M(0)=\frac{\tilde{M}(0)}{\lambda}\;.$
Now, taking $\lambda$ as the ratio of the scales where interactions become strong for $SU(3)$ and another group,
\begin{equation}
\frac{\sigma_{group}}{\sigma_{SU(3)}}=\lambda\;,
\end{equation}
the mass function scales in the same way,
\begin{equation}
\frac{M_{group}(0)}{M_{SU(3)}(0)}=\lambda\;.
\end{equation}
Hence, eliminating the auxiliary $\lambda$, we find
\begin{equation}
\frac{M_{group}(0)}{M_{SU(3)}(0)}=\frac{\sigma_{group}}{\sigma_{SU(3)}}\,.
\end{equation}
Using these results we compute the constituent masses for the quarks charged under the different groups (Fig. \ref{constituent}). 

Summing up, combining the methods of RGE and DSE and requiring that the constituent masses of $SU(3)_C$ colored quarks to be 300 MeV has allowed us to obtain the constituent masses of hypothetical fermions charged under different groups from a Grand Unification Scale of $10^{15}\;\text{GeV}$. 
From this treatment we can conclude that groups belonging to the $SU(N)$ and $Sp(N)$ families, with $N>4$, generate masses of order or above the few TeV. Notwithstanding the crude approximations we have employed, our computation gives about 5 TeV to $SU(4)$-charged fermions, which would not be far out of reach of mid-future experiments provided the GUT conditions apply. It appears from our simple work that larger groups (except the Exceptional Groups and $SO(N)$ with $N<10$) might endow fermions with a mass too high to make them detectable in the foreseeable future. 

Hence,  the question ``\textit{Why the symmetry group of the Standard Model, $S U (3)_C \otimes SU(2)_L \otimes U(1)_Y$ , contains only small-dimensional subgroups?}'' can be partially answered: It happens that, upon equal conditions at a large Grand Unification scale, large-dimension groups in the classical $SO(N)$, $SU(N)$ and $Sp(N)$ families force dynamical mass generation at higher scales because their coupling runs faster. Should fermions charged under these groups exist, they would appear in the spectrum at much higher energies than hitherto explored.

\section{Confinement in QCD}\label{sc:Confinement}

At low energies and low temperatures, quarks and gluons are confined inside hadrons (such as the proton or the neutron). All known hadrons are color singlets under the $SU(3)_C$ color group, and many physicists refer to confinement as the absence of asymptotic colored particle-states (more specifically \textit{color confinement}). Nonetheless, in gauge theories with a Higgs mechanism (see section \ref{subsec:EWEFTs} below), where forces are of Yukawa-type and the dynamics resemble those of the weak interactions, asymptotic states are also color singlets. This is known as the FSOS Theorem \cite{Fradkin:1978dv,Osterwalder:1977pc}, which states that observables in a gauge-Higgs theory are continuously connected between a Higgs-like phase and a confinement phase, so that in both phases one has colorless asymptotic states. 

\begin{figure}
    \centering
\includegraphics[width=0.6\columnwidth]{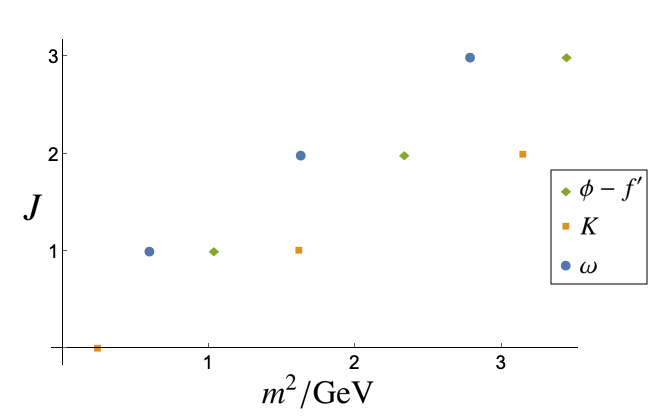}
    \caption[Regge trajectories]{Empirical Regge trajectories for some of the meson families (data from \cite{Bali:2000gf}). Each family of mesons lies approximately on a straight line (the Regge trajectory).}
    \label{fig:reggetrajectories}
\end{figure}

Hence, what is the main difference between the dynamics of QCD and the electroweak theory? One peculiarity of the QCD spectrum is that if one plots the total angular momentum of mesons and baryons against their squared mass one finds that they fall into the so called Regge trajectories (straight lines) as shown in Fig. \ref{fig:reggetrajectories}.

There is a very simple model that explains Regge trajectories, it describes mesons as ``spinning sticks'' with mass $m$, mass per unit length $\sigma$ (which is called \textit{string tension}) and length $L=2R$,
\begin{equation*}
   \begin{tikzpicture}[ scale=2]
   \draw [<->] (-1,.3)--(1,.3);
   \draw (0,.4) node {$2R$};
   \draw[inner color=white,outer color=cyan,draw=black] (0,0) ellipse (26pt and 5pt);
   \draw[inner color=white,outer color=magenta,draw=black] (1,0) ellipse (5pt and 5pt);
   \draw[inner color=white,outer color=yellow,draw=black] (-1,0) ellipse (5pt and 5pt);
   \end{tikzpicture}\;,
\end{equation*}
which rotates with the endpoints moving at the speed of light (this means assuming that the quarks sitting at the endpoints of the stick are massless), so that the linear velocity for each point on the stick is $v(r)=r/R$ in natural units.

The relativistic energy for the spinning stick is
\begin{equation}
m=E=2\int_0^R\frac{\sigma dr}{\sqrt{1-v^2}}=2\int_0^R\frac{\sigma dr}{\sqrt{1-\frac{r^2}{R^2}}}=\pi \sigma R\;,
\end{equation}
and the relativistic angular momentum is
\begin{equation}
J=2\int_0^R\frac{\sigma r v(r) dr}{\sqrt{1-\frac{r^2}{R^2}}}=\frac{1}{2}\pi \sigma R^2\;.
\end{equation}
Hence, with this simple model we reproduce to some extent the linear Regge trajectories for light mesons shown in Fig. \ref{fig:reggetrajectories}:
\begin{equation}
J=\frac{1}{2\pi\sigma}m^2\;.
\end{equation}

\begin{figure}[ht!]
    \centering
    \includegraphics[width=0.5\columnwidth]{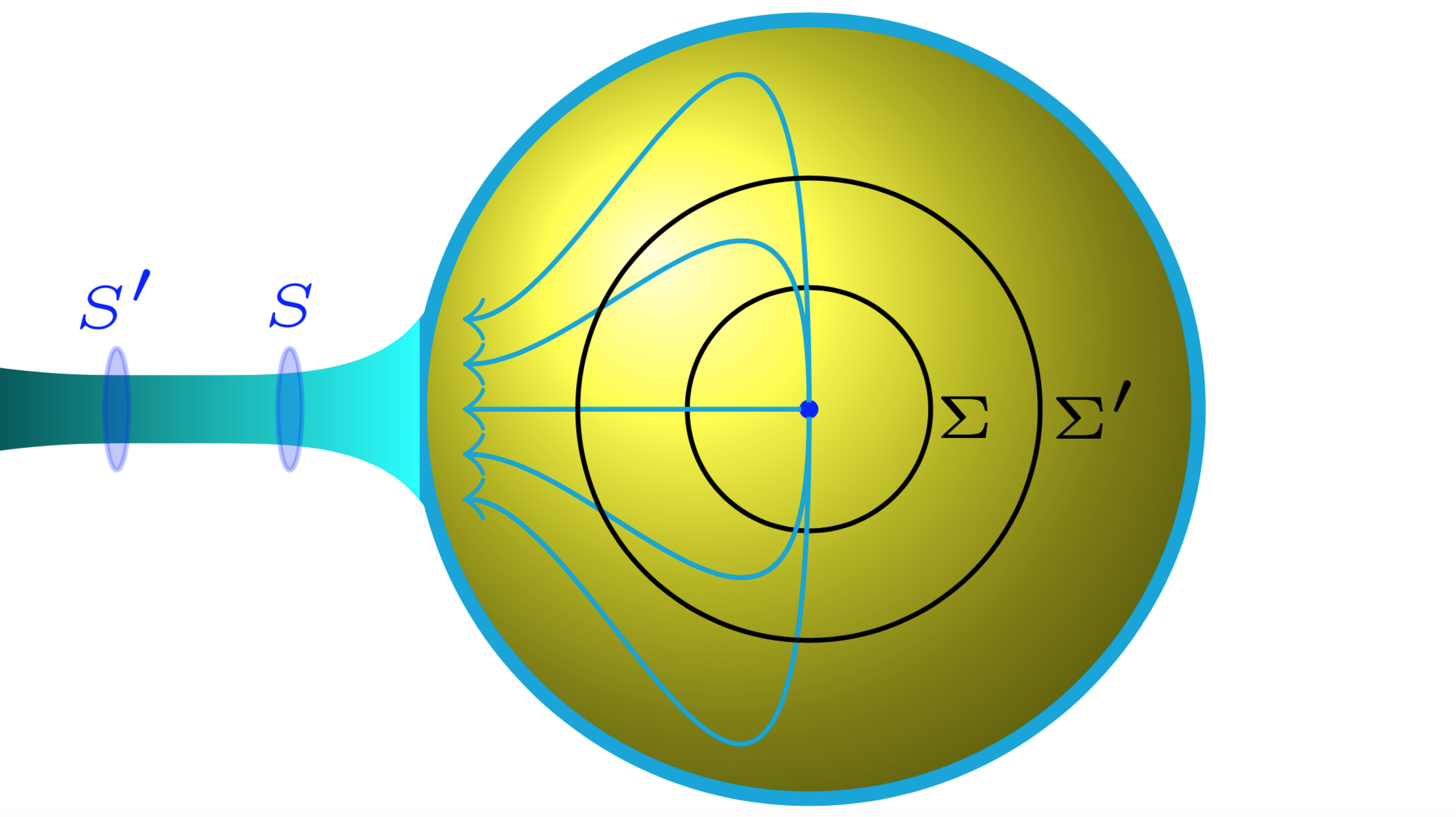}
    \caption[Flux tube field]{The creation of a flux tube, \textit{i.e.} the color electric field (blue lines) being squeezed into a cylinder, interpolates between a Coulomb-like potential at short distances and a linear rising potential at greater distances. This is so because the chromo-electric field between $\Sigma$ and $\Sigma'$ decreases as $1/r^2$ and it is constant between $S$ and $S'$, since the total electric flux remains constant.}
    \label{fig:fluxtubepotential}
\end{figure}

The spinning stick model is an oversimplification of mesons which describes qualitatively Regge trajectories, but the important question here is how could a stick-like (or string-like) object emerge from QCD. A possible answer is that the color electric field between the constituent quarks of a meson is compressed into a cylindrical region, with a cross-sectional area that remains nearly constant as the quark separation increases. In that case, the energy stored in the color electric field grows linearly with the quark separation \cite{Greensite:2011zz}. Also, as expected by asymptotic freedom at short distances, the potential must behave like a Coulomb potential (see Fig. \ref{fig:fluxtubepotential}, where the formation of a chromoelectric flux tube explains the linear rising potential at large separations and Coulomb-like potential at short distances). 

Lattice simulations in $SU(3)$ gauge theory with static quarks do support the flux tube picture \cite{Bali:2000gf}. In real QCD with dynamical quarks, the linear potential must eventually become flat (independent of the inter-quark distance) when the energy stored in the flux tube is sufficient to extract a $q\bar{q}$ out of the vacuum, breaking the flux tube (this is the mechanism for meson decay, that we will discuss in the next section). There also some other features of the confining force (supported by lattice simulations) that a full picture of confinement should explain. These are: the Casimir scaling at intermediate distances \cite{Bali:2000un} and the $N$-ality (the number of boxes in a Young Tableau \cite{Zee:2016fuk}) dependence of the potential at large distances for quarks in different representations of the gauge group (see \cite{Greensite:2011zz} for details), among others.

There are several partial explanations for confinement in QCD, all depending on different gauge choices, the first one that appeared historically was suggested by 't Hooft \cite{tHooft:1995isn} and Madelstam \cite{Mandelstam:1974pi}, inspired by solid state physics. In a Type II superconductor, an external magnetic field cannot enter the superconductor (Meissner effect). However, above a critical value of the external magnetic field, the material lets some magnetic flux tubes go through (generating a linear rising potential among hypothetical magnetic monopoles). This picture suggested that the vacuum of QCD could consist of magnetic Cooper pairs (being hence a dual Type II superconductor) and Polyakov was able to prove confinement in compact QED in 2+1 dimensions using this mechanism in \cite{Polyakov:1987ez}. In order to be in compliance with the $N$-ality dependence of the inter-quark potential, the monopole picture needs that the magnetic monopoles' worldlines lie on the so called \textit{vortex sheets} with their magnetic fields collimated along \textit{center vortices} \cite{Greensite:2011zz}. 

These last two concepts are related to one of the most successful, yet intriguing, explanations of the properties of the confining force: the Center Vortex mechanism. The center vortex mechanism for confinement is based on a unbroken realization of a global center symmetry \cite{Greensite:2011zz} (the center of a group consists of the group elements that commute with all the elements of the group). The so called Polyakov loops \cite{Polyakov:1975rs} are order parameters  for the spontaneous breaking of the center symmetry as well as being related to the free energy of a quark (\textit{i.e.} the energy that costs to introduce an isolated quark in the QCD vacuum): if the expectation value of the Polyakov loop is zero it turns out that center symmetry is unbroken and the free energy of an isolated quark infinite (see \cite{Greensite:2011zz} for details). In the lattice, the study of time-like Wilson loop operators elucidates the nature of the potential between static quarks (they represent the creation, time evolution and destruction of a quark and an antiquark, both static): a linear rising potential means that Wilson loops must fall off with an area law \cite{Greensite:2011zz}. The Wilson loop operator, which creates a unit of electric flux, is dual to the 't Hooft operator, which creates a a thin line of magnetic flux (also called a thin center vortex). 't Hooft showed in \cite{tHooft:1977nqb} that a perimeter falloff for the 't Hooft loops implies an area falloff of Wilson loops, meaning that totally uncorrelated magnetic fluxes or center vortices means heavily correlated electric fluxes (\textit{i.e.} magnetic disorder is related to the creation of flux tubes). The center vortex mechanism has passed several tests in the lattice, it is possible to locate center vortices and if one removes them by hand, the confining properties of the strong force disappear (see \cite{Greensite:2011zz} for details). This mechanism can also accommodate the $N$-ality dependence and the Casimir scaling of the potential, and the fact that at high energies there is a deconfinement phase transition for QCD. On the other hand this mechanism cannot account for theories with broken center symmetry, which is the case for quarks in the fundamental representation of $SU(N)$ groups, although center vortices seem to be responsible of intermediate scale string tensions and allow flux tube breaking \cite{Greensite:2011zz}. 

Other attempts to prove confinement are: the Gribov-Zwanziger \cite{Gribov:1977wm,Zwanziger:1991gz} mechanism in Coulomb gauge and the DSE approach in Landau gauges. 

\begin{figure}[ht!]
    \centering
   \begin{tikzpicture}
      \draw (0,0)--(3*.8660,3/2);
      \draw (0,0)-- (8,0);
      \draw (3*.8660,3/2)--(3*.8660+8,3/2);
      \draw (3*.8660+8,3/2)-- (8,0);
      \draw[inner color=white,outer color=white,draw=orange] (5,3/4) ellipse (70pt and 15pt);
      \draw[color=blue,thick] (3.2,.8) .. controls (3.4,2) and (5,2) .. (5,.8);
      \draw[dotted,color=blue,thick] (3.2+3,.8) .. controls (3+3,.2) and (5.2,.2) .. (5,.8);
      \draw[dotted,color=blue,thick] (2.5,0) .. controls (2.7,.4) and (3.2,.4) .. (3.2,.8);
      \draw[dotted,color=blue,thick] (2.5,0) .. controls (2.7,.3) and (3.2,.4) .. (3.2,.8);
      \draw[color=blue,thick] (2.5,0) .. controls (2,-.4) and (0.5,-.8) .. (0,-.8);
      \draw[color=blue] (3.2,.8) node {\tiny$\times$};
      \draw[color=blue] (3.2+3,.8) node {\tiny$\times$};
      \draw[color=blue] (5,.8) node {\tiny$\times$};
      \draw[color=blue,thick] (8,2) .. controls (7.5,1.5) and (3.2+3.2,1.4) .. (3.2+3,.8);
      \draw [thick] (6-.2,.25) .. controls (6.6-.2,1) and (6.1-.2,1) .. (6-.2,1.25);
      \draw [thick] (4-.2+.5,.25) .. controls (4.6-.2+.5,1) and (4.1-.2+.5,1) .. (4-.2+.5,1.25);
      \draw (7,-.3) node {Gauge-fixing hypersurface};
      \draw[color=orange] (8.7,1) node {Gribov Horizon};
      \draw[color=blue] (9,2) node {Gauge orbit};
      \draw (5.45,1) node {FMR};
      \draw (6.7,.75) node {GR};
      \draw (3.7,.65) node {GR};
   \end{tikzpicture}
    \caption[Gribov and FM Regions]{The rectangular region represents the subspace of the space of all gauge configurations that satisfy the (Coulomb or Landau) gauge condition. The region inside the ellipse (the Gribov Horizon) in orange is known as the Gribov Region (GR). A generic gauge orbit it blue intersects the GR several times: the intersections (marked with $\times$ in blue) are the Gribov copies. The Fundamental Modular Region (FMR) is the region where each gauge orbit intersects the gauge-fixing hypersurface only once \cite{Greensite:2011zz}.}
    \label{fig:gribovregion}
\end{figure}
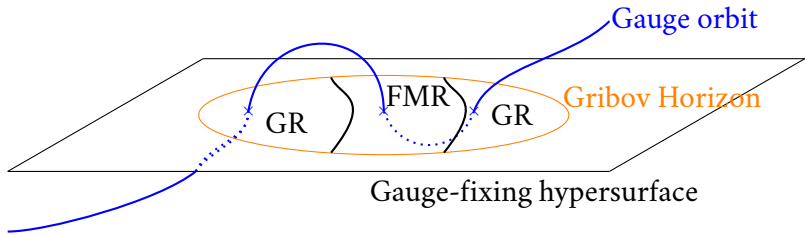

The first mechanism argues that the inter-quark Coulomb potential for QCD is so enhanced in the infrared that it becomes linear at large distances. This is closely related to the fact that in Coulomb (or Landau) gauge each gauge field is a member of a gauge orbit and this gauge orbit may intersect the subspace generated by the gauge restriction more than once, these are called \textit{Gribov copies} of a gauge configuration. At each Gribov copy, the Fadeev-Popov (FP) determinant of eq. (\ref{eq:FPoperator}) may have different signs with the same absolute value, producing the possibility that the functional integral vanishes. This is not desirable and Gribov argued that the domain of the functional integration should be restricted to the \textit{Gribov region}, a region in $A_\mu^a$ field-space where the FP determinant is positive, so that different Gribov copies cannot cancel each other's contribution. The boundary of the Gribov region is called the \textit{Gribov horizon} (defined by the region where the first eigenvalue of the FP operator vanishes) and it concentrates most of the volume of the Gribov region (see Fig \ref{fig:gribovregion}).  All this discussion is relevant because the Coulomb potential is related to the inverse of the FP operator, $M$ of eq. (\ref{eq:FPoperator}), and since close to the Gribov horizon the eigenvalues of $M$ are very close to zero, they will greatly enhance the Coulomb potential. Coulomb confinement is strongly supported by lattice simulations \cite{Greensite:2011zz}, although it is only a necessary but not sufficient condition for confinement (since the Coulomb potential is an upper bound of the static quark potential).

The DSE approach tries to obtain solutions to the fully coupled equations that support a confinement-like behaviour of QCD in the infrared region (see \cite{vonSmekal:1997ohs,Fischer:2002hna,Fischer:2006ub, Alkofer:2008tt}, among many others), and the discussion is still ongoing. Recent works also relate this framework to the Center Vortex mechanism for confinement \cite{Leinweber:2022ukj}.  

Summing up, the confinement problem in non-Abelian gauge theories is very rich and fundamental, with very diverse explanations (each for a different gauge) that must converge to the same basic physics phenomenology. In a future publication we will obtain a gauge field solution in Landau gauge that resembles an infinitely long flux tube and try to address the longstanding $^3P_0$ problem of meson decay, coupling the flux tube to a $q\bar{q}$ pair through the parametrization of the $gq\bar{q}$ vertex obtained from solving the fully coupled DSEs.

\section{Effective Field Theories for the Standard Model and beyond}
From Rutherford's experiment to the ongoing upgrade of the LHC (HL-LHC), the goal in accelerator physics has been accessing higher and higher energies with increasing precision, aiming at resolving the structure of matter and its interactions at smaller scales (actually Rutherford only had a natural source of $\alpha^{++}$ particles). At present, we are reaching a point where direct discovery of new particles outside the SM seems more and more unlikely, either because these particles interact too feebly with ordinary particles or because they are too heavy to be produced at current accelerators. In this thesis we will take the second assumption: new physics sits at an energy scale far above the energies hitherto reached at experiment and our hope is to observe, not new particles, but new forces among the known particles that are not described by the SM. This is the natural setting for effective field theories (EFTs), where one parameterizes the effect of new physics with operators built up with the known particles only, hence describing new forces among these. In section \ref{subsec:CHPT} we will introduce the EFT used in low energy QCD, where QCD is strongly coupled and the useful degrees of freedom are hadrons: chiral perturbation theory (ChPT). The study of this theory becomes very useful for new physics searches because, as we will see in section \ref{subsec:EWEFTs} and more in detail in Chapter 3, chiral perturbation theory is the perfect language to describe deviations from the SM in the Electroweak sector. It is also important to mention that usually, the amplitudes calculated through effective Lagrangians are at odds with unitarity at high energies. 

Usually, in order to make sense of the effective theory and restore its predictive power, one needs to ``unitarize'' the amplitudes by some procedure. In Chapter 4, we will discuss one of the most successful unitarization methods: the Inverse Amplitude Method. The aim is to assess the systematic uncertainties of this to put its predictive power at use with LHC data.

\subsection{Strong force EFT}\label{subsec:CHPT}

Since QCD becomes non-perturbative at low energies, the relevant degrees of freedom are not quarks and gluons but their bound states: the hadrons. The (approximate) symmetries of QCD constrain the properties of hadrons, especially the chiral symmetry of QCD and its spontaneous breaking. The fermionic part of the Lagrangian in eq. (\ref{eq:classicalCD}) is, taking the quark masses to zero, invariant under $SU(N_f)$ transformations in flavor space for both left handed and right handed chiral projections of the quark fields (as vectors in flavor space),
\begin{align}
    &\psi_L\equiv \frac{1}{2}(1-\gamma_5)\psi \;\text{ with }\; \psi_L\to L \psi_L\;,\\
    &\psi_R\equiv \frac{1}{2}(1+\gamma_5)\psi \;\text{ with }\; \psi_R\to R \psi_R\;,
\end{align}
where $(L,R)\in SU(N_f)_L\otimes SU(N_f)_R$. This symmetry produces the conservation of the vector, $V^{\mu\,a}=\bar{\psi}\gamma^\mu T^a\psi$, and axial, $V_5^{\mu\,a}=\bar{\psi}\gamma^5\gamma^\mu T^a\psi$, currents, so that for each hadron multiplet there should be a degenerated multiplet with opposite parity and approximately the same mass.

Since the up and down quarks are very light, $SU(2)_L\otimes SU(2)_R$ should be a good approximate symmetry of the corresponding hadron spectrum. However, experiments only observe $SU(2)_V$ triplets like the rho $\rho=(\rho^-,\rho^0,\rho^+)$ or the pion $\pi=(\pi^-,\pi^0,\pi^+)$ and not the corresponding triplets with opposite parity. This means that that the axial generators do not annihilate the vacuum and hence chiral symmetry is spontaneously broken to its $SU(2)_V$ subgroup as 
\begin{equation}
    SU(2)_L\otimes SU(2)_R\longrightarrow SU(2)_V\;,\label{eq:SSBpattern1}
\end{equation}
producing, due to Goldstone's theorem \cite{Nambu:1960xd,Goldstone:1962es}, three pseudoscalar massless bosons. These Nambu-Goldstone bosons are the pions, which are not exactly massless due to the explicit breaking of the chiral symmetry produced by the current quark masses. 

To describe the relevant physics for hadron interactions at low energies, one can build up an effective theory with the NGB (the pions for the $SU(2)$ case). One can cast the NGB into a unitary matrix, $U$,
\begin{equation}
    U=\exp \left( i\frac{\phi(x)}{f_0}\right)\;,\;\;\text{ with }\;\;{\phi}= \boldsymbol{\phi}\cdot \boldsymbol{\sigma}/2=\begin{pmatrix}
           \pi^0&\sqrt{2}\pi^+ \\
           \sqrt{2}\pi^-& -\pi^0
         \end{pmatrix}\;, \label{eq:UmatrixChPT}
\end{equation}
for the $SU(2)$ case, where $f_0$ is a constant with mass dimension (this constant is approximately equal to the pion decay constant, $f_\pi$). In this way one can build up the most general Lagrangian compatible with the symmetries of QCD and its chiral symmetry breaking pattern in terms of the matrix $U$. 

This Lagrangian has an infinite number of compatible terms organized as a perturbative power expansion in the number of derivatives (a term with $k$ derivatives will produce a contribution proportional to $p^k/\Lambda_\chi$, where $p$ is the typical momentum of the process and $\Lambda_\chi=4\pi f_0$ is the symmetry breaking scale). Only even derivatives are compatible with Lorentz invariance of the Lagrangian, so that the ChPT Lagrangian has the form
\begin{equation}
\mathcal{L}_{\text{ChPT}}=\mathcal{L}_0+\mathcal{L}_2+\mathcal{L}_4+...\;,\label{eq:LChpt}
\end{equation}
with $\mathcal{L}_0$ containing no derivatives, $\mathcal{L}_2$ two derivatives, and so forth. The term with two derivatives is
\begin{equation}
    \mathcal{L}_2=\frac{f_0^2}{4} \langle \partial_\mu U^\dagger \partial^\mu U\rangle\;,\label{eq:L2ChPT}
\end{equation}
with the brackets representing a color trace, and produces the leading order (LO) terms in the chiral expansion. At higher orders one needs to take into account $\mathcal{L}_4$, $\mathcal{L}_6$, ..., and loop corrections; with each Feynman diagram contributing to an order given by Weinberg's power counting rule \cite{Weinberg:1978kz}. Each term in the Lagrangian will be multiplied by some coefficients called low energy constants (LEC) in hadron physics or Wilson coefficients in the EFT generic language. 

Since the focus of this thesis in on the EW SBS sector and its EFT extensions, we will move on to introduce them in the next section. It turns out that ChPT is the perfect language for this task and the QCD symmetry breaking pattern of eq. (\ref{eq:SSBpattern1}) is very similar to that of the EW sector. In Chapter 4 we will also use hadron physics ChPT to assess the systematic uncertainties of one particular unitarization method, in order to put it to use in the EW sector for predicting possible new physics resonances.

\subsection{Electroweak Symmetry Breaking Sector}\label{subsec:EWEFTs}
Historically, when obtaining the fundamental properties of the forces in the SM, gauge invariance of the theories became a crucial tool to unravel the structure of these forces. When the $W^\pm$ and the $Z$ bosons of the nuclear weak force where understood to be massive, there was a big theoretical effort to reconcile this fact with gauge invariance (since mass terms for vector bosons break it). This effort culminated in the well known Higgs mechanism, explaining the masses of the gauge bosons via a spontaneous symmetry breaking mechanism. In this section we will explain this mechanism and present the effective field theories that are relevant for new physics in this symmetry breaking sector (SBS).

In the SM the electroweak interactions are based in the gauge group $SU(2)_L\otimes U(1)_Y$, where the subscript $L$ denotes the fact that all chirally left handed fermions in the SM are organized into $SU(2)_L$ (weak)-isospin doublets as
\begin{align}
    \mathcal{Q} =& \begin{pmatrix}
           \mathcal{U}\\
           \mathcal{D}
         \end{pmatrix}\;, &\mathcal{L}= \begin{pmatrix}
           \mathcal{N}\\
           \mathcal{E}
         \end{pmatrix}\;, 
  \end{align}
  where $\mathcal{Q}$ is the quark doublet with up-like quarks, $\mathcal{U}$, and down-like quarks, $\mathcal{D}$, for each family. The same happens for the lepton doublet, $\mathcal{L}$, with neutrinos, $\mathcal{N}$, and electron-like fermions, $\mathcal{E}$. The left-handed part of these doublets, $\mathcal{Q}_L$ and $\mathcal{L}_L$, transform under the usual $SU(2)$ gauge transformation (with generators $T^a=\sigma^a/2$, where $\sigma^a$ are Pauli matrices and with the upper and lower members of the doublets having weak isospin $T^3$ components $1/2$ and $-1/2$ respectively). All right handed components $\mathcal{Q}_R$ and $\mathcal{L}_R$ are singlets under the $SU(2)_L$ transformation. Finally, for a given fermion $\psi=e,\nu_e,u,d,...$ the $U(1)_Y$ hypercharge acts with charge $y^\psi_L$ ($y^\psi_R$) on its left (right) handed chiral components. All hypercharge, weak isospin and electric charge (defined as $Q=Y+T^3$) assignments for all fermions in the SM are listed in Table \ref{tab:SMhypercharges}.
  \begin{table}[ht!]
      \centering
      \begin{tabular}{|c|c|c|c|c|c|c|c|}
          \hline&  \;\;$\mathcal{U}_R$\;\; &\;\;$\mathcal{U}_L$ \;\; &\;\;$\mathcal{D}_R$ \;\;&\;\;$\mathcal{D}_L$ \;\;&\;\;$\mathcal{N}_L$ \;\;&\;\;$\mathcal{E}_R$ \;\;&\;\;$\mathcal{E}_L$  \;\;   \\\hline
         $y$ & $2/3$ &$1/6$ & $-1/3$ & $1/6$ & $-1/2$ & $-1$ & $-1/2 $ \\\hline
         $T^3$ & $0$ &$1/2$ & $0$ & $-1/2$ & $1/2$ & $0$ & $-1/2 $ \\\hline
         $Q$ & $2/3$ &$2/3$ & $-1/3$ & $-1/3$ & $0$ & $-1$ & $-1 $ \\
         \hline
      \end{tabular}
      \caption[Hypercharge and weak isospin assignments in the SM]{Hypercharge and weak isospin projection assignments for all matter fields in the SM. }
      \label{tab:SMhypercharges}
  \end{table}

  The electrically neutral component of the gauge fields for the weak $SU(2)_L$, $W^a_\mu$ with coupling $g$, and for the hypercharge, $B_\mu$ with coupling $g'$, mix to produce the physical photon, $A$, and the electroweak boson, $Z$, as
  \begin{equation}
      A_\mu=\sin \theta_W W^3_\mu+\cos \theta_W B_\mu\;,\;\;\;\;Z_\mu=\cos \theta_W W^3_\mu-\sin \theta_W B_\mu\;.
  \end{equation}
  where $\theta_W=\arctan (g'/g)$ is the ``Weinberg angle''.
These bosons have zero electric charge since the electric charge, commutes with $T^3$ and the hypercharge $Y$. There are two other combinations with well defined electric charge which are the electroweak $W^\pm$ bosons
\begin{equation}
    W_\mu^{\pm}=\frac{1}{\sqrt{2}}(W^1_\mu\mp iW_\mu^2)\;.
\end{equation}

\subsubsection{The SM Electroweak SBS and the Higgs mechanism}
In order to give masses to the electroweak bosons $W^\pm$ and $Z$ while preserving gauge invariance one can couple the SM to the SBS. The simplest way to do this is to define a complex scalar $SU(2)_L$ doublet $H^T=(\phi^+,\phi^0)$ with hypercharge $Y_\phi=1/2$, so that $Q\phi^+=\phi^+$ and $Q\phi^0=0$ and Lagrangian
\begin{equation}
    \mathcal{L}_{\text{SBS}}=(D_\mu H)^\dagger (D^\mu H)-V(H)+\mathcal{L}_{\text{YK}}\;.\label{eq:SBSlagrangian}
\end{equation}
Where $D_\mu H=\left(\partial_\mu +i\frac{g'}{2} B_\mu-i\frac{g}{2}\sigma^a W^a_\mu\right)H$ and $\mathcal{L}_{\text{YK}}$ are the Yukawa terms coupling the SBS to matter fields. Now we choose \textit{ad hoc} the potential $V(H)$ to spontaneously break the global $SU(2)_L\times U(1)_Y$ down to $U(1)_\text{EM}$ as
\begin{equation}
    V(H)=-\mu^2(H^\dagger H)+\lambda (H^\dagger H)^2\;,\label{eq:SMpotential11}
\end{equation}
with $\mu^2,\lambda>0$. This potential has a minimum at $(H^\dagger H)=\mu^2/2\lambda\equiv v^2/2$ and produces a family of degenerate vacua related by an $SU(2)$ transformation. The so called Higgs vacuum expectation value $v$, is according to \cite{ParticleDataGroup:2022pth} $v\simeq 246 $ GeV. As a result we need to choose a vacuum and a useful choice is $H_{\text{vac}}^T=(0,v/\sqrt{2})$. This choice spontaneously breaks the global Electroweak symmetry. Replacing $H'=H-H_{\text{vac}}$ in eq. (\ref{eq:SBSlagrangian}), the covariant derivatives of $H$ produce mass terms for the $W$ and $Z$ bosons. The mass for these bosons at tree level equals
\begin{align}
M_W=\frac{g v}{2}\;,\;\;\;\;\;\;\;M_Z=\frac{M_W}{\cos \theta_W}\;.
\end{align}
The experimental values for these masses are listed in Fig. \ref{fig:SMcontent}.

\subsubsection{General considerations for the EW EFTs}
To finish this section let us note that if one sets the gauge charges $g'$ and $g$ to zero, one can show that the purely scalar part in the Lagrangian of eq. (\ref{eq:SBSlagrangian}) has a larger symmetry, namely $SU(2)_L\otimes SU(2)_R$. This means that, without gauge couplings, the symmetry breaking pattern is $SU(2)_L\otimes SU(2)_R\to SU(2)_{L+R}$ producing three massless Nambu-Goldstone bosons (NBG) \cite{Peskin:1995ev,Dobado:1997jx}. These three NGBs become, through the Higgs mechanism, the longitudinal polarizations of the weak gauge bosons (polarizations that would not appear if they were massless). The $SU(2)_{L+R}$ symmetry is regarded to as the ``custodial symmetry'' since it keeps the $\rho$ parameter, measuring the relative strength of  of the charged and neutral weak currents, equal to one. Apart from the CDF measurement mentioned in the introduction, the custodial symmetry seems a well established approximate symmetry of the SM. Nonetheless, the $U(1)_Y$ coupling does break explicitly the custodial symmetry, inducing corrections to $\rho$ as $\rho=1+\mathcal{O}(g^2)$ \cite{Dobado:1997jx}. 

All the above considerations leave us with only one possibility for building up an effective Lagrangian for the EW SBS \cite{Dobado:1997jx}, \textit{i.e.} there has to be a new physics sector coupled to the SM that contains a symmetry breaking pattern from a global group $G=SU(2)_L\otimes SU(2)_R$ to a subgroup $H=SU(2)_{L+R}$ that leaves three NBG (giving masses to the three electroweak gauge bosons) that are embedded in the manifold/coset $K=SU(2)_{L-R}$, which is an axial group, and hence are pseudoscalar particles (just as the pions for QCD). In the next chapter we will study in much more detail the EFT setting for the EW SBS.


\setcounter{chapter}{3}\label{ch:EW}
\setcounter{section}{0}
\setcounter{equation}{0}
\setcounter{equation}{0}
\setcounter{table}{0}
\setcounter{figure}{0}
\chapter{Electroweak Symmetry Breaking Sector Effective Field Theories}

There are many aspects of the Standard Model and beyond that can be studied at accelerators (the many-parameter flavor structure in both lepton and quark sectors, the Higgs couplings to the fermions, the CP violating phases, QCD processes, ...) but, at the energy frontier, the most important aspect of physics that is being clarified right now is the nature of the Electroweak Symmetry Breaking mechanism: whether it happens as the well-known discussion in the Standard Model paradigm presented in subsection \ref{subsec:EWEFTs}, or whether new particles or interactions influence the global $SU(2)\times SU(2)\to SU(2)$ breaking pattern that is crucial for the Electroweak interactions. When extending the EW SBS and assuming there is new physics sitting at a large scale $\Lambda$, two effective field theories have been proposed: Standard Model EFT (SMEFT) and the Higgs EFT (HEFT). 

Throughout most of this chapter we will discuss these theories in a regime where the energies, $E$, of the scattered particles are much higher than their masses, $m_h$, and much smaller than the new physics scale,
 so that $m_h\ll E\ll \Lambda $. In such regime and in the presence of new physics that would yield derivative couplings with $\partial \sim E\gg m_h$, the much discussed Higgs potential $V(h)$ is actually a correction that does not play the pivotal role
it enjoys in the SM. As we will see shortly, the crucial function to distinguish between SMEFT and HEFT will be the Flare function $\mF(h)$ (that name originated in our first publication on the topic \cite{Gomez-Ambrosio:2022qsi}).

 For a while, it was often stated that the SMEFT~\cite{Jenkins:2013zja,Alonso:2013hga,Brivio:2017vri,Ghezzi:2015vva,Alioli:2022fng,Marzocca:2020jze} and the HEFT~\cite{Brivio:2013pma,Eboli:2021unw,Kozow:2019txg,Dobado:2019fxe,Carena:2021onl} must encode similar physics, just in different coordinates (related by a simple change between Cartesian-like to spherical-like coordinates), with HEFT being perhaps a bit more general because it does not incorporate the Higgs boson $h$ into an $SU(2)$ multiplet.

 However, the work of the San Diego and Oregon groups~\cite{Alonso:2016oah,Alonso:2016btr,Alonso:2015fsp,Cohen:2020xca} has sharpened the differences between both formulations. Every Lagrangian in SMEFT form, such as that in eq.~(\ref{SMEFTEWlagrangianV}) below, can be recast in the shape of HEFT. An example in the relevant energy region is given in  eq.~(\ref{FbosonLagrangianLO}). Nevertheless, the converse statement is not true, not every HEFT can be cast as a SMEFT. From the work at San Diego it has become apparent that this can be achieved only if the Flare function $\mathcal{F}(h)$ presented shortly in eq.~(\ref{expandF}) has a zero for some real value of the classical field $h$, $\exists\ h_\ast\in\mathbb{R} \arrowvert \mathcal{F}(h_\ast)=0$. (The precise and complete conditions as presently understood for this SMEFT $\leftrightarrow$ HEFT equivalence are presented in section \ref{sec:convert}). This function controls high-energy processes with multiple Higgs bosons in the final state: achieving a good control over it requires measurements with an increasingly large number of them, that would look in a detector like a flare of Higgs bosons, whence the name of $\mF$.

 A summary of our present understanding is given by the following scheme:
 \begin{center}\boxed{
 \begin{tabular}{ccccc}
       &          &                    &          & Specific correlations\\
 Valid &          & Double zero        &          & between $a_i$ \\ 
 SMEFT &$\implies$& of $\mathcal{F}(h)$&$\implies$& coefficients \\
       &          & at some $h_\ast$   &          & from expanding $\mathcal{F}(h)$\\
       &          &                    &          & around $h=0$
 \end{tabular}}
 \end{center}

Both aspects, the double zero of $\mathcal{F}(h)$ (see subsection~\ref{subsec:noSMEFT} below, for example) and the possibility of using correlations between its coefficients, $a_i$, to distinguish SMEFT from HEFT from experimental data (see subsection~\ref{subsec:correlations} and section \ref{sc:correlationsexplicit}) have been separately discussed in the last years. We here provide an integrated discussion with full detail, putting less weight on geometric aspects and more in field-theory and particle physics ones and make several new contributions.  
Finally, it is also worth mentioning that similar relations exist between the coefficients of the nonderivative $V_{\rm HEFT}$
potential, since a valid SMEFT description of $V$ must obey a series of conditions analogous to those for the flare function  $\mF(h)$~\cite{Cohen:2020xca}. Assuming SMEFT's validity would then also impose important correlations between the coefficients of the potential --trilinear, quartic, etc.-- (see e.g.~\cite{Gonzalez-Lopez:2020lpd} for a HEFT phenomenological analysis). This and correlations that SMEFT produces on Yukawa couplings are presented in subsection~\ref{coeffspotential}.

The rest of this chapter is organized as follows: we provide a basic discussion in section \ref{sc:cosmetic}, where some superficial differences between the electroweak SMEFT and HEFT theories are studied at the level of the Lagrangian.  Section~\ref{sec:convert} is devoted to identifying the effective operators  in SMEFT (of dimension 6 and 8) with leading TeV-scale contributions to $W_LW_L\to n \times h$ scattering and to expressing them in HEFT coordinates (here $W_L$ stands for the longitudinal polarizations of electroweak bosons). In doing so, we expose the correlations on the HEFT parameters induced by assuming SMEFT’s validity. 
Then, in section~\ref{Geometricsection} we review the field-redefinition invariant criteria to distinguish SMEFT from HEFT. We also present a novel and simple analytic way of deriving the conditions on the flare function, $\mathcal{F}(h)$, that allow for the deployment of SMEFT around the symmetric point. Continuing, section \ref{sec:generic-properties} deals with the generic properties of the flare function $\mathcal{F}(h)$ such as its positivity, the general correlations (and their experimental bounds in section \ref{sc:correlationsexplicit}) induced on its coefficients by assuming the existence of a symmetric point (where $\mathcal{F}(h^*)=0$) and analyticity  around, it allowing a continuation to our physical vacuum. Also, some example functions are provided as illustration. Section~\ref{nhproduction} cursorily discusses how to experimentally extract the value of the coefficients of the flare function, presenting the amplitudes of the scattering of two Goldstone bosons to $n$ Higgs bosons (with $n=1,2,3$, although an automated amplitude generator for $n\in \mathbb{N}$ can be provided on demand). 
We also observe that to first nontrivial order, the structure of SMEFT makes amplitudes with more than two Higgses in the final state vanish. With the first non-zero contributions appearing at the next order, which is not the case for a generic HEFT. 
Section~\ref{Zeroessection} explores the status of the symmetric-point knowledge based on current bounds on HEFT parameters by ATLAS and CMS, and also under what conditions Schwarz’s Lemma can guarantee the existence of such a point. Finally, section~\ref{sec:farfuture} takes on a more speculative road by studying the far-future possibility of producing a large number of Higgs bosons at non-zero temperature, hence giving access to an arbitrary expansion order of the flare function. 
 Our final conclusions are provided in section~\ref{sec:conclusions} and some technical details are relegated to Appendix~\ref{app:Lemma}. 

\section{Cosmetic differences between EW EFTs}\label{sc:cosmetic}
In this section we will review the differences between the SMEFT framework and the HEFT setting at the Lagrangian level, to afterwards move on to show explicitly how a SMEFT can be cast as a HEFT through field redefinitions.
\subsection{Key aspects of the SMEFT Lagrangian} \label{SMEFTLL}

The first of the theories extending the SM EW SBS is the Standard Model EFT (SMEFT) electroweak Lagrangian. As in subsection \ref{subsec:EWEFTs}, its symmetry breaking sector is expressed in terms of the $SU(2)$ doublet $H^T=\frac{1}{\sqrt{2}} (\phi_1+i\phi_2, \phi_4+i\phi_3) $ (that can also be collected as an $O(4)$ quartet  $\boldsymbol\phi=(\phi_1,\phi_2,\phi_3, \phi_4)$) and takes the general form
 \begin{equation}
     \mathcal{L}_{\rm SMEFT}= A(|H|^2)|\partial H|^2+\frac{1}{2}B(|H|^2)(\partial (|H|^2))^2-V(|H|^2)+\mathcal{O}(\partial ^4) 
     \label{SMEFTEWlagrangianV} \ .
 \end{equation}
Where  both functions $A(|H|^2)$ and $B(|H|^2)$ are real, and analytic around $|H|^2=0$. The SM is retrieved by choosing $A(|H|^2)=1$ and $B(|H|^2)=0$. 
 Thus, the organization of the theory is carried around an electroweak symmetric vacuum, instead of the energy minimum at $\langle \phi_4\rangle = v$.
  
SMEFT arranges the order of usage of operators in terms of their canonical-dimension counting, so that the leading corrections to the SM are composed of dimension six operators (each multiplied by a Wilson coefficient and divided by the new physics scale squared, $\Lambda^2$).

In the TeV region, the derivative terms multiplying $B$ become much larger than $V$, and we can neglect this potential. It will be shown that the piece of most importance for this chapter is that function $B$, that contains the electroweak symmetry breaking physics in the TeV region in the presence of new physics, particularly the dimension-6 operator 
 $\mathcal{O}_{H\Box} =   |H|^2 \Box |H|^2 $.

\subsection{HEFT Lagrangian (for the EW SBS in the TeV region)} \label{HEFTLL}
What has come to be called the Higgs EFT (HEFT) Lagrangian (the second to bear that name) is an evolution of the Electroweak chiral Lagrangian (inspired in eqns. (\ref{eq:LChpt})-(\ref{eq:L2ChPT})). Its degrees of freedom are built from a Cartesian to spherical-like change of coordinates %
\begin{equation}\label{eq:EWNGB-spherical-coord}
\boldsymbol\phi=(1+h/v)\boldsymbol n  \, ,
\end{equation}
where $\boldsymbol n=(\omega_1,\omega_2,\omega_3,\sqrt{v^2-\omega_1^2-\omega_2^2-\omega_3^2})$, so that $\boldsymbol n\cdot\boldsymbol n=v^2$. It couples the $\omega_i$ Goldstone bosons to an additional low-energy Higgs field singlet, $h$, that is not assumed to be part of the $SU(2)$ 
Goldstone triplet. 
At leading order in the chiral counting, the scalar sector of the HEFT Lagrangian  (in EW Goldstone spherical coodinates $\omega^i$ in~eq.(\ref{eq:EWNGB-spherical-coord})) is given by  
 \begin{eqnarray} \label{FbosonLagrangianLO}
{\cal L}_{\text{LO HEFT}} =&& \frac{1}{2}\mathcal{F}(h)
\partial_\mu\omega^i\partial^\mu\omega^j\left(\delta_{ij}+\frac{\omega^i\omega^j}{v^2-\boldsymbol{\omega}^2}\right)
+\frac{1}{2}\partial_\mu h\partial^\mu h  \, ,
\end{eqnarray}
where the function $\mathcal{F}$ scales the scattering amplitudes involving two, four, and generically an even number of Goldstone bosons.  
Thus, the flare function $\mathcal{F}(h)$ relates the EW Goldstone processes to amplitudes with an arbitrary number of Higgs bosons (since these are of the same order in the chiral counting):
\begin{equation} \label{expandF}
    {\mathcal F}(h)=1+\sum_{n=1}^{\infty}{a_n}\Big(\frac{h}{v}\Big)^n\;.
\end{equation}
Usually, since only the first terms of the $\mathcal{F}$ function are known, the Lagrangian is expressed~\cite{Delgado:2015kxa,Delgado:2013hxa} in terms of $\mathcal{F}(h)\simeq \left[1+2a\frac{h}{v}+b\left(\frac{h}{v}\right)^2\right]$, with 
$a_1=2a$ and $ a_2=b\;.$
In the TeV regime, the leading corrections to this Lagrangian are not of order $m_W$ or $m_h$, both in the $100$ GeV range, but rather derivative couplings. This means that $V(h)$ is irrelevant and electroweak symmetry breaking in the TeV region is more naturally discussed in terms of the coefficients of the Higgs-flare function $\mathcal{F}(h)$.

At NLO, the Lagrangian relevant to study unitarity and resonances in the TeV regime 
acquires two further derivatives (so that amplitudes receive terms of order $s^2$) and
becomes
\begin{eqnarray} \label{bosonLagrangian}
{\cal L}_{\text{NLO HEFT}} =&& \frac{1}{2}\left[1+2a\frac{h}{v}+b\left(\frac{h}{v}\right)^2\right]
\partial_\mu\omega^i\partial^\mu\omega^j\left(\delta_{ij}+\frac{\omega^i\omega^j}{v^2-\boldsymbol{\omega}^2}\right)
+\frac{1}{2}\partial_\mu h\partial^\mu h %
\nonumber\\
 &+& \frac{4\alpha_4}{v^4}\partial_\mu \omega^i\partial_\nu \omega^i\partial^\mu\omega^j\partial^\nu\omega^j
+\frac{4\alpha_5}{v^4}\partial_\mu\omega^i\partial^\mu\omega^i\partial_\nu\omega^j\partial^\nu\omega^j
+\frac{g}{v^4}(\partial_\mu h\partial^\mu h)^2
\nonumber\\
 &+& \frac{2d}{v^4}\partial_\mu h\partial^\mu h\partial_\nu\omega^i\partial^\nu\omega^i
+\frac{2e}{v^4}\partial_\mu h\partial^\nu h\partial^\mu\omega^i\partial_\nu\omega^i
\ ,
\end{eqnarray}
which will be used in the next chapter. Here we will concentrate on the LO Lagrangian (tree-level amplitudes $\propto s$) in  eq.~(\ref{FbosonLagrangianLO}) with the Taylor series of ${\mathcal F}$ around the physical vacuum $h=0$ (with zero number of physical Higgs particles, that is, with $\phi_i = \langle \phi_4 \rangle\delta_{i4} = v\delta_{i4}$ in terms of the SMEFT coordinates) given by eq.~(\ref{expandF}). The NLO coefficients of the second and third lines in eq.~(\ref{bosonLagrangian}) should eventually encode similar physics to the neglected $\mO({\partial^4})$ terms of dimension 8 in eq.~(\ref{SMEFTEWlagrangianV}), but we here concentrate on the comparison between $A(|H|^2)$, $B(|H|^2)$ and ${\mathcal F(h)}$.

\section{TeV-scale relevant EW SMEFT in HEFT form}\label{sec:convert}
We now show the explicit transformation to polar coordinates, and then to HEFT, of the
SMEFT electroweak Lagrangian in eq.~(\ref{SMEFTEWlagrangianV}) in terms of the $SU(2)$ doublet $H$, neglecting the gauge couplings and $V(H)$ as is appropriate for TeV-scale physics where $E\gg m_{h,W,Z}$ , following the discussion in~\cite{Cohen:2020xca}.
This is achieved by decomposing the doublet $H$ of the SMEFT framework in the spherical polar coordinates of eq.~(\ref{eq:EWNGB-spherical-coord}),
\begin{eqnarray}
H =&& \left(1+\frac{h}{v}\right)\,  U(\omega) \,     H_{\text{vac}}   \, ,
\end{eqnarray}
 where $h$ denotes the radial Higgs-boson field in the SMEFT framework. The $SU(2)$ matrix $U(\omega)$ contains the EW Goldstone bosons analogously to eq. (\ref{eq:UmatrixChPT}).
The VEV modification due to higher-order corrections can always be later incorporated to the analysis by considering a  shift in the Higgs field~\footnote{This removes terms linear in $h$ and recenters the Higgs field expansion around the potential minimum.} $h\to h+\Delta$. 

Substituting $H^\dagger H= (v+h)^2/2$ in $A$ and $B$ of eq. (\ref{SMEFTEWlagrangianV}) and neglecting the potential $V$, we get
\begin{eqnarray}
\mathcal{L}_{\rm polar-SMEFT} =&& \frac{1}{2}(v+h)^2{A(h)}(\partial_\mu  \boldsymbol{n}\cdot \partial^\mu  \boldsymbol{n}) 
+ \frac{1}{2}\Big (A(h)+(v+h)^2 B(h)\Big )(\partial h)^2
\nonumber \\ 
=&&
\frac{1}{4} (v+h)^2 \, A(h) \, \langle \partial_\mu U^\dagger \partial^\mu U\rangle 
+ \frac{1}{2} (A(h)  + (v+h)^2 B(h)) (\partial h)^2  \, ,
\label{SMEFTtoHEFT1}
\end{eqnarray}
with $A((h+v)^2/2)\equiv \tilde{A}(h)\to A(h)$ now a function of $h$ to avoid cumbersome notation.

The SM, with $A=1$ and $B=0$, is the first and simplest of the family of SMEFT Lagrangians in eq.~(\ref{SMEFTtoHEFT1}), and in this form it reads
\begin{eqnarray}
\mathcal{L}_{\rm SM} =   |\partial H|^2  
\, =\, 
\frac{1}{4} (v+h)^2 \,   \langle \partial_\mu U^\dagger \partial^\mu U\rangle  + \frac{1}{2}  (\partial h)^2  \, . 
\label{eq:SM-Lagr}
\end{eqnarray}

In the general case, even though
the coordinates of eq.~(\ref{SMEFTtoHEFT1}) are now those of HEFT, the Lagrangian is not yet in its canonical form because, by convention, the HEFT Higgs field's $h$ kinetic term needs to be fixed to its free-wave standard expression 
\begin{eqnarray}
\mathcal{L}_{\rm HEFT} = 
\frac{v^2}{4} \mathcal{F}(h_1) \,   \langle \partial_\mu U^\dagger \partial^\mu U\rangle 
+
\frac{1}{2} (\partial h_1)^2  \, , 
\label{eq:HEFT-Lagr}
\end{eqnarray}
which requires a further change of the $h$ variable. Finding an $h_1$ field that absorbs the multiplicative factor in eq.~(\ref{SMEFTtoHEFT1}) and that becomes the Higgs field in the HEFT framework, implies solving the differential condition~\cite{Giudice:2007fh} 
 \begin{equation}\label{htoh1}
     dh_1\, =\, \sqrt{A(h)+(v+h)^2 B(h)}\,\, dh \, ,
 \end{equation} 
that will collect all factors of $h_1$ to multiply only the Goldstone term to the right of eq.~(\ref{SMEFTtoHEFT1}), from which the $\mathcal{F}$ of eq.~(\ref{FbosonLagrangianLO}) can be read off,
\begin{equation}
     v^2\mathcal{F}(h_1)\, =\, (v+h(h_1))^2\, A(h(h_1))\ .
     \label{FfromSMEFT}
 \end{equation}
in terms of $h_1=h_1[A,B](h)$. In the next subsection~\ref{subsec:noA} we will show that non-trivial $A$ terms are unnecessary, so we can set $A=1$ and employ $B$ alone, which will determine the relation $h_1=h_1(h)$.

Once $h$ has been expressed in terms of $h_1$, the Lagrangian will have reached its HEFT form and the subindex in $h_1$ may be dropped~\footnote{Note we are dropping here the possible shift $h_1\to h_1 + \Delta$, required if there are modification to the SM Higgs potential. We are interested in high-energy effects and ignore non-derivative operators in $\mathcal{L}_{\rm HEFT}$. This shift can easily be incorporated if needed.}. With this method, 
the coefficients of eq.~(\ref{expandF}) expanding the generic HEFT Lagrangian radial function can be retrieved from the initial SMEFT.

To complete this discussion we will quickly digress, in the next subsection, to show that $A=1$ can be consistently taken, afterwards proceeding to carry out the transformation $h\to h_1$ for the relevant SMEFT pieces for the electroweak sector in the TeV energy regime where $m_h\ll E\sim \partial \ll \Lambda$.

\subsection{\texorpdfstring{$A(H)$}{A(H)} is not really necessary for \texorpdfstring{$\sqrt{s}\gg m_h$}{s>>mh}} \label{subsec:noA}

We here quickly show that it is possible, and can be more convenient, to eliminate the $n^{\rm th}$-power operators (for $n\geq 1$) obtained in an expansion of $A$, by means of a partial integration. For this, note that, up to a total divergence, 
\begin{eqnarray}
\underbrace{(H^\dagger H)^n |\partial H|^2}_{A\text{-type operator}} =
-\frac{n}{2}\underbrace{(H^\dagger H)^{n-1}   (\partial | H|^2)^2}_{B\text{-type operator}}   -\frac{1}{2} (H^\dagger H)^n \left(  (\partial^2 H^\dagger)H+ H^\dagger (\partial^2 H)\right) \,,
\label{eq:A-simpl}
\end{eqnarray}
obtained by using the relation $\partial^2 |H|^2 = 2 |\partial H|^2 + (\partial^2 H^\dagger)H + H^\dagger (\partial^2 H)$.  
This teaches us that we can always convert (by partial integration) any $n^{\rm}$-power operator of $A$-type into an $(n-1)$-power operator of $B$-type. 
The price to pay includes an irrelevant total derivative and a couple of terms proportional to $\partial^2 H$ and $\partial^2 H^\dagger$. However, the classical equations of motion of $H$ can be used to trade the derivative operators for $\partial^2 H $ (and its conjugate) by operators without derivatives, up to correction of higher dimension in $1/\Lambda^2$. In this way, the $A$-type of operators can be removed from the theory and transformed into $B$-operators at fixed dimension 6, 8, etc. 
Employing this freedom, we will set $\Delta A_{\rm BSM}=0$ and just keep the leading operator, $A=A_{\rm SM}=1$. Hence SMEFT can  be formulated in polar coordinates $(h,\omega^a)$ as     
\begin{eqnarray} \label{polarSMEFT1}
\mathcal{L}_{\rm polar-SMEFT} =&& 
\frac{v^2}{4} \bigg(1+\frac{h}{v}\bigg)^2 \,   \langle \partial_\mu U^\dagger \partial^\mu U\rangle 
+
\frac{1}{2} \bigg(1  + (v+h)^2 B(h) \bigg) (\partial h)^2  \, ,
\end{eqnarray}
instead of eq.~(\ref{SMEFTtoHEFT1}).
The change of variables in the Higgs field, $h=h(h_1)$ of eq.~(\ref{htoh1}) then becomes 
\begin{eqnarray}
dh_1 =&& \bigg(1  + (v+h)^2 B(h) \bigg) ^{1/2}  dh \, .
\end{eqnarray}
This change determines $\mathcal{F}$ in the form 
\begin{eqnarray}\label{cerodoble}
\mathcal{F}(h_1)   =&& \bigg(1+\frac{h(h_1)}{v}\bigg)^2 \, ,
\end{eqnarray}
with $h$ implicitly given~\cite{Giudice:2007fh} by the relation
\begin{eqnarray}
h_1 =&& \int_0^{h} \bigg(1  + (v+h)^2 B(h) \bigg) ^{1/2}  dh \, .
\end{eqnarray}

\subsection{Explicit computation with SMEFT's power expansion of \texorpdfstring{$B(|H|^2)$}{B(|H|2)}}

\subsubsection{Order 6 in the SMEFT counting}

The SMEFT Lagrangian is an alternative parametrization of SM deviations, that assumes the SM symmetries and fields, and particularly assumes the traditional doublet structure for the Higgs field. The Higgs sector of this Lagrangian was introduced in eq.~\eqref{SMEFTEWlagrangianV}, and can more generally be written as 
\begin{equation}
    \mathcal{L}_{\text{SMEFT}} =
    \mathcal{L}_{\text{SM}} + 
    \sum_{n=5}^{\infty}
    \sum_i
    \frac{c_i^{(n)}}{\Lambda^{n-4}} \op_i^{(n)}  \, .
\end{equation}

 At dimension 6, there are three operators of the SMEFT Warsaw basis~\cite{Grzadkowski:2010es} that directly distort the Standard Model's Electroweak Symmetry Breaking Lagrangian, which written in terms of the Higgs field doublet $H$ appropriate for SMEFT are ($\partial^2\equiv\Box$)
\begin{align} \label{Warsawbasis}
&  \op_H = (H^\dagger H)^3  \, ,
 & \op_{HD} = (H^\dagger D_{\mu} H)^*  (H^\dagger D^{\mu} H) \, ,  \nonumber \\
 & \op_{H \Box} = (H^\dagger H) \Box (H^\dagger H)\ .
\end{align}
They can of course be reexpressed in terms of the singlet field for the Higgs boson via $(H^\dagger H) = (h + v)^2/2$ (in polar coordinates this is manifestly gauge-independent).  
Those three operators are actually all that is needed for  Higgs-Goldstone boson scattering up to dimension 6 in the SMEFT counting. Moreover, $\op_{HD}$ breaks custodial symmetry so that it can be counted as higher order due to the small size of the corrections to Peskin-Takeuchi \cite{Peskin:1990zt} observables in the SM at LEP \cite{Kennedy:1990ib}

 We would like to remark that not only at dimension-6 but also at dimension-8 there is an additional operator with two derivatives acting only on a product of Higgs doublets. However, these terms violate custodial symmetry and they actually contribute to an independent type of HEFT operator, 
 Longhitano's $a_0$ Lagrangian term~\cite{Longhitano:1980iz,Longhitano:1980tm}. Consistently, this $a_0$ operator is related to the experimentally suppressed oblique $T$--parameter. Thus, we will no longer consider this type of custodial breaking operators in this chapter, although a similar study can be worked out if this kind of corrections needed to be included.
 
In turn, $\op_{H}$ is not a derivative operator, so that it does not contribute to the flare function that we are pursuing (though it does affect the Higgs self-coupling, namely the Higgs SM potential, and the vacuum expectation value, important near threshold, its impact in the TeV region is much smaller than that of the derivative operator).

In summary, only the $\mO_{H\Box}$ operator contributes to $\mF(h)$ at  order $\mathcal{O}(\Lambda^{-2})$.
Moreover, it has been shown in~\cite{Alonso:2021rac}, by geometric arguments, that only one operator is needed at this order, which is consistent with our discussion.
The rest of the electroweak operators of the Warsaw basis that the reader may be wondering about,
\begin{align}
&  \op_W = \epsilon_{ijk} W_{\mu }^{\nu i} W_{\nu}^{\rho j }
W^{\mu k}_{\rho}  \, ,
& \op_{HW} = (H^\dagger H) W_{\mu \nu}^i W^{\mu \nu i} \, , \nonumber \\
 & \op_{H B} = (H^\dagger H) B_{\mu \nu} B^{\mu \nu}    \, ,
 & \op_{H W B} = (H^\dagger \tau^i H) W_{\mu \nu}^i B^{\mu \nu}   \, ,
\end{align}
are necessary only if one intends to couple the transverse electroweak gauge bosons~\cite{Gonzalez-Lopez:2020lpd,Maas:2020kda}, but they are of no concern for our purposes of studying the TeV-region electroweak-symmetry breaking Lagrangian that requires only, by the Equivalence Theorem~\cite{Veltman:1989ud,Dobado:1993dg} in the TeV region, the Goldstone bosons, which encode the relevant physics of the longitudinal $W_L$, $Z_L$.
Further, a generic basis could also contain
an operator of the form
$\partial_{\mu}  (H^\dagger H) \partial^{\mu}  (H^\dagger H)$, but this is eliminated in the standard Warsaw treatment because it is equivalent to $\op_{H \Box}$ in eq.~(\ref{Warsawbasis}) up to a total divergence, in analogy to eq.~(\ref{eq:A-simpl}),
\begin{align}
 (X^2) \Box (X^2) = - \partial_{\mu} X^2 \partial^{\mu} X^2 + \underbrace{ \partial_{\mu} ( X^2  \partial^{\mu} X^2 )}_{\rm{surface \, \, term}} \, ,
\end{align}
or in terms of the $h$ singlet,
\begin{align}
  \op_{H \Box} & = (H^\dagger H) \Box (H^\dagger H) = - \partial_{\mu}  (H^\dagger H) \partial^{\mu}  (H^\dagger H) + \partial_\mu(...) = - (h +  v)^2 \partial_{\mu} h \partial^{\mu} h + \partial_\mu(...)\ .
\end{align}

Therefore, the only contributing dimension-six operator of the Warsaw basis that preserves custodial symmetry is
 \begin{align}
 \label{convierteops}
  \op_{H \Box} & = (H^\dagger H) \Box (H^\dagger H) = - \partial_{\mu}  (H^\dagger H) \partial^{\mu}  (H^\dagger H)\, , 
\end{align}
that in the Lagrangian appears multiplied by the Wilson coefficient $c_{H\Box}$ and is suppressed by two powers of the high-energy scale $\Lambda$ respect to the dimension-4 Lagrangian.
Comparing with eq.~(\ref{SMEFTEWlagrangianV}) we read the (constant) values $A(|H|^2)=1$ and $B(|H|^2)=-2\frac{c_{H\Box}}{\Lambda^2}$, that contain no fields.

\subsubsection{The role of $c_{H\Box}$ in SMEFT and bounds on its size from experimental data}

Because SMEFT has been used for a few years now to analyze  LHC data, there already exist bounds on the coefficient $c_{H\Box}$ from Run 2 of the machine (even if the associated operator $\mathcal{O}_{H\Box}$ is quite elusive in LHC fits);  we now recall those bounds.

The best overall constraints on the dimension-6 basis arise from 
Higgs-sector observables (production and decay)~\cite{Ellis:2020unq,Ethier:2021bye}, but it is only when combined with other electroweak channels that this  $c_{H\Box}$ coefficient can be well constrained. 
The reason for this is the way that $\mO_{H \Box}$ enters in the $B$ piece of the Lagrangian in eq.~(\ref{SMEFTEWlagrangianV}). Its effect is to change the
Higgs wave-function normalization 
\begin{equation} \label{eq:chboxshift}
  \mathcal{L}_{\text{SMEFT}} = 
 \frac{1}{2}  \left( 1  -  \frac{2 c_{H \Box} v^2}{\Lambda^2}  
  \right) \partial_{\mu} h \partial^{\mu} h \, +\, ...
\end{equation}
instead of the Higgs couplings to other particles, that are not directly affected. Hence, the contribution of this operator to any on-shell production or decay process of a single Higgs boson appears as a kinematics-independent shift, as evident from eq.~\eqref{eq:chboxshift}.  In particular, for the reference value of $\Lambda = 1$ TeV used in most analysis, this overall shift for the several processes considered, whether decay width or production cross-section, becomes
 \begin{equation} \label{ecudel012}
\frac{\sigma_{H,\ \text{SMEFT}}}{\sigma_{H,\ \text{SM}}}   \propto    \frac{\Gamma_{H,\ \text{SMEFT}}}{\Gamma_{H,\ \text{SM}}}  \propto  
  1 + 2 \frac{ c_{H \Box} v^2}{\Lambda^2}
  = 1 + 0.12 c_{H \Box}\, ,
\end{equation}
which was already numerically observed by the ATLAS collaboration and reported, for example in Table 1 of~\cite{ATLAS:2019dhi}. 
It is obvious that the numbers there, between 0.115 and 0.125, just reflect the exact 0.12 factor of eq.~(\ref{ecudel012}). This is true in any process involving only one on-shell Higgs boson (as will also be the case in our eq.~(\ref{amp:h}) below); but events with two or multiple $h$ particles such as eq.~(\ref{amp:2h}) and following have different dependences on $c_{H\Box}$ and will allow a cleaner separation within SMEFT.
Also, the cross sections above in $\frac{\sigma_{H,\ \text{SMEFT}}}{\sigma_{H,\ \text{SM}}} $ are implicitly understood as their narrow Higgs-width approximation, where one Higgs is produced on-shell and then cascades into the final products. For off-shell Higgs studies the dependency would be different.

In consequence, this kinematically not very exciting $\mathcal{O}_{H\Box}$ operator is often overlooked and few works actually constrain it. 
Still, the works of Ellis {\it et al.}~\cite{Ellis:2020unq} and Ethier {\it et al.}~\cite{Ethier:2021bye} offer quite interesting bounds on $c_{H\Box}$, that at 95\% confidence level, and rounding off to the precision of the leading digit of the uncertainty, read as follows,
\begin{eqnarray}
c_{H\Box} \simeq -0.3\pm 0.7 \ \ {\rm (individual)} \\
c_{H\Box} \simeq -1 \pm 2 \ \ {\rm (marginalized)} \ .
\end{eqnarray}
Both are compatible with the Standard Model value $c_{H\Box}=0$.

\subsubsection{Operator of order 8 in the SMEFT counting}
\label{subsec:dim8op}

Going one order further in the $1/\Lambda^2$ power counting makes the SMEFT parametrization more interesting~\cite{Passarino:2019yjx,Dawson:2021xei,Hays:2018zze,Henning:2015alf}. In particular, the full dimension-8 basis in SMEFT was recently published in \cite{Murphy:2020rsh,Li:2020gnx}. To find the dimension-8 operator that contributes $\mathcal{O}(\Lambda^{-4})$ corrections to the flare function $\mathcal{F}(h)$ we can return to 
the $B(H)$ term in eq.~(\ref{SMEFTEWlagrangianV})  remembering that $A(H)$ is irrelevant as per section~\ref{subsec:noA},  $ \frac{1}{2}B(|H|^2)(\partial (|H|^2))^2$, and set $B(|H|^2)\propto |H|^2$ instead of 1.
Therefore, at dimension-8 we only find the following operator:
\begin{equation}
    \op_{H \Box}^{(8)}  = |H|^4 \Box  |H|^2 =   - 2 |H|^2 (\partial (|H|^2))^2 = - (h + v)^4 \partial_{\mu} h \partial^{\mu} h  \ .
\end{equation}
 We have chosen this operator's normalization for convenience and resemblance to $\mO_{H\Box}^{(6)}$ in eq.~(\ref{convierteops}) when expressed in terms of the Higgs doublet modulus $(v+h)/\sqrt{2}$. Through partial integration, it can be easily rewritten in other forms considered in the bibliography (up to a total derivative): 
\begin{equation}
    \op_{H \Box}^{(8)}\,=\,  - 2 |H|^2 (\partial (|H|^2))^2   
    \,=\,  |H|^4 \Box  |H|^2 
    \,=\,  2 |H|^4 |\partial H|^2  +   |H|^4 ( (\Box H^\dagger) H+H^\dagger (\Box H)) \,,    
\end{equation}
with the last contribution $|H|^4 ( (\Box H^\dagger) H+H^\dagger (\Box H))$ being proportional to the Higgs equations of motion, so they can be removed from the effective action and transformed into operators including EW Goldstone bosons and also fermions, as well as the $Q_{H^6}^{(1)}=|H|^4 |DH|^2 $ operator 
chosen for the dimension-8 basis in Ref.~\cite{Murphy:2020rsh}. The other possible two-derivative dimension-8 operator, $Q_{H^6}^{(2)}$, breaks  custodial symmetry and will not be discussed.

In the next section, we will first work out the precise change of Higgs variable $h\to h_1$ 
from SMEFT to HEFT up to dimension-6. 
Afterwards, in subsection~\ref{subsec:dim8}, we will proceed up to the next order and provide the NNLO modification induced by  
this dimension--8 operator. 

\subsection{Change of coordinates \texorpdfstring{$h^{\rm SMEFT}\to h_1^{\rm HEFT}$}{hSMEFT->h1HEFT}}
Once we have identified the SMEFT relevant operators at the TeV scale, we will expose the correlations on HEFT parameters produced by SMEFT's structure.

\subsubsection{Derivation and  result at dimension-6}

The first step to put the dimension-6 relevant SMEFT Lagrangian into HEFT form is then to change the variable as in eq.~(\ref{SMEFTtoHEFT1}) yielding
  \begin{equation}
     \mathcal{L}_{\rm SMEFT}= \frac{1}{2}\Big (1-2(v+h)^2\frac{c_{H\Box}}{\Lambda^2}\Big )(\partial_\mu h)^2+\frac{1}{2}(v+h)^2(\partial_\mu  \boldsymbol{n}\cdot \partial^\mu  \boldsymbol{n})\;.
 \end{equation}

We next have to take the Higgs kinetic energy to canonical form.  
This requires integrating~(\ref{htoh1}) 
($t$ being the integration variable taking the place of $h$):
\begin{equation}
  h_1=\int_0^h  \sqrt{1-(v+t)^2\frac{2 c_{H\Box}}{\Lambda^2}}dt=\int_v^{v+h} \sqrt{1-t^2\frac{2 c_{H\Box}}{\Lambda^2}}dt=\sqrt{\frac{\Lambda^2}{2c_{H\Box}}}\int_{\theta_0}^{\theta_1} |\cos{\theta}|\cos{\theta}d\theta\;,
  \label{integral1}
\end{equation}
where $\theta_0=\arcsin{\sqrt{\frac{2c_{H\Box}}{\Lambda^2}}v}$ and $\theta_1=\arcsin{\sqrt{\frac{2c_{H\Box}}{\Lambda^2}}(v+h)}$.
Now, since $0\leq\theta_0\leq\theta_1\ll\pi/2$ (because the EFT coefficient is very small by current bounds \cite{Buchalla:2015qju}) we can assume that the cosine in eq. (\ref{integral1}) is positive and hence find

\begin{eqnarray}    h_1=&&\frac{1}{2}\sqrt{\frac{\Lambda^2}{2c_{H\Box}}}\Big( \theta +\frac{\sin{2\theta}}{2}\Big)\Big|_{\theta_0}^{\theta_1}=\frac{1}{2}\sqrt{\frac{\Lambda^2}{2 c_{H\Box}}}\Big( \theta + \sin{\theta}\sqrt{1-\sin^2\theta}\Big)\Big|_{\theta_0}^{\theta_1}
\nonumber=\\
&& =\frac{1}{2}\left((v+h)\sqrt{1-\frac{2 c_{H\Box}(h+v)^2}{\Lambda^2}} -v\sqrt{1-\frac{2 c_{H\Box}v^2}{\Lambda^2}}  \right)+\nonumber\\
&&+\frac{1}{2}\sqrt{\frac{\Lambda^2}{2c_{H\Box}}}\Big(\arcsin{\sqrt{\frac{2c_{H\Box}}{\Lambda^2}}(v+h)}-\arcsin{\sqrt{\frac{2c_{H\Box}}{\Lambda^2}}v}\Big)\, .\label{result1}
\end{eqnarray}
Such expression is not particularly useful, especially taking into account that we need to invert it to obtain $h(h_1)$, so we explore it by 
expanding eq. (\ref{result1}) to leading order in $c_{H\Box}/\Lambda^2$, finding
\begin{equation}
    h_1=h - \frac{1}{3}\Big(\frac{c_{H\Box}}{\Lambda^2}\Big)(h^3+3h^2v+3hv^2)+\mathcal{O}\Big(\frac{c_{H\Box}^2}{\Lambda^4}\Big)\, ,
\label{eq:h1-from-h-NLO}
\end{equation}
which can be iteratively inverted, yielding
\begin{equation}
    h=h_1+\frac{1}{3}\Big(\frac{c_{H\Box}}{\Lambda^2}\Big)(h_1^3+3h
    _1^2v+3h_1v^2)+\mathcal{O}\Big(\frac{c_{H\Box}^2}{\Lambda^4}\Big)
    \;.
    \label{eq:h-from-h1-NLO}
\end{equation}
Finally, we can use eq. (\ref{FfromSMEFT}) and (\ref{eq:h-from-h1-NLO}) to obtain ${\mathcal F}(h_1)$,
\begin{eqnarray} \label{F_SMEFT6}
    \mathcal{F}(h_1)  
    \,=&&\, \left(1+ \frac{h(h_1)}{v}\right)^2=\nonumber\\
    =&&\left(1+\frac{h_1}{v}\right)^2 +  \frac{2v^2  c_{H\Box}}{\Lambda^2}\left(1+\frac{h_1}{v}\right)  \left(\frac{h_1^3}{3v^3}+\frac{h
    _1^2}{v^2}+\frac{h_1}{v}\right )  +\mathcal{O}\Big(\frac{c_{H\Box}^2}{\Lambda^4}\Big)= \nonumber \\ 
   =&&  1 +   \left( \frac{h_1}{v} \right)  
  \left( 2 + 2 \frac{ c_{H \Box} v^2}{\Lambda^2} \right)  \ 
  +  \left( \frac{h_1}{v} \right) ^2 \left( 1 + 4 \frac{ c_{H \Box} v^2}{\Lambda^2}  \right)+  \nonumber \\ 
  & &+   \left( \frac{h_1}{v} \right) ^3  \left(  8 \frac{ c_{H \Box} v^2}{3 \Lambda^2} \right)  +  \left( \frac{h_1}{v} \right) ^4 \left( 2 \frac{c_{H \Box} v^2}{3 \Lambda^2}  \right)  
  +\mathcal{O}\Big(\frac{c_{H\Box}^2}{\Lambda^4}\Big)  
 \, ,
\label{eq:SMEFT-F-Lambda2}
\end{eqnarray}
which expresses the expansion coefficients of HEFT's $\mathcal{F}$ in terms of the SMEFT Wilson coefficient (in the philosophy of the appendix of~\cite{Buchalla:2018yce}),
\begin{eqnarray}\label{coefsSMEFTHEFT}
a_1 = 2a=2\left(1 + v^2\frac{c_{H\Box}}{\Lambda^2}\right)\; , \;\;\;
a_2 = b=1+{4v^2}\frac{c_{H\Box}}{\Lambda^2}\;, \;\;\;
a_3 = \frac{8v^2}{3}\frac{c_{H\Box}}{\Lambda^2}\;, \;\;\;
a_4 = \frac{2v^2}{3}\frac{c_{H\Box}}{\Lambda^2}\, .\label{eq:firstBoxcorrections}
\end{eqnarray}
These relations expose the inclusion of SMEFT into HEFT: the $a_i$ coefficients, independent parameters in the latter, are correlated in SMEFT up to a given order, as all of the first four $a_i$ are given in terms of only one Wilson coefficient $c_{H\Box}$.
This feature has been suggested as a handle to discern, from upcoming experimental data, whether SMEFT will be applicable later on (in the presence of any separation from the SM values $a=b=1$).  
Measurements of the $\omega\omega\to n h$ scattering process (see section~\ref{nhproduction})
would allow the determination of the $a_i$ and can probe the SMEFT-predicted correlations~\cite{Agrawal:2019bpm} (or the SM ones~\cite{Arganda:2018ftn,Gonzalez-Lopez:2020lpd}). 
In the absence of such correlations, it is plausible that a HEFT formulation would be needed. 

However, this difference can be put into question in the presence of unnatural Wilson coefficients. 
If the dimension-8 operators contribute at an order similar to that of the dimension-6 operator, because the coefficients are not of order 1 or because $\Lambda$ is not large enough to significantly suppress them, additional SMEFT parameters would appear in the expressions of eq.~(\ref{coefsSMEFTHEFT}), 
decorrelating the $a_i$ coefficients and complicating the analysis.
Therefore, though perhaps illustrative, given that naturalness may have already failed as a safe guiding principle in view of the light Higgs boson mass, we have provided more robust criteria that helps systematize the correlations to distinguish SMEFT from HEFT, and to them we turn in the next section.

To finish, let us comment on the shift on the value of the symmetric point. The position of the symmetric point $h_\ast$ that satisfies $\mF(h_\ast)=0$ is always 
given by $h(h_1)=-v$ (with $|H|=(h+v)/\sqrt{2}$), as seen directly from eq.~(\ref{cerodoble}). 
In turn, the position of the symmetric point in HEFT coordinates becomes displaced from its SM value $h_\ast^{\rm SM}=-v$  by an   $\mO(\Lambda^{-2})$ SMEFT correction,
 \begin{equation} 
\label{sympoint2}
\frac{h_\ast}{v}\, =\, \frac{h_1(h)}{v}\bigg|_{h=-v}
\,=\, -1 + \frac{c_{H \Box}  v^2 }{3\Lambda^2}   \,,
\end{equation}
where we use the relation $h_1=h_1(h)$ in eq.~(\ref{eq:h1-from-h-NLO}). 
  This procedure provides the only real root of $F(h_1)$ up to $\mO(1/\Lambda^2)$, with $\mF(h_1)=F(h_1)^2$ (in addition, there are two more complex-conjugate roots). However, this information gets blurred if we instead consider $\mF$ up to $\mO(1/\Lambda^2)$, which has the real roots $h_1/v=-1$ and $h_1/v=-1 + 2 c_{H\Box} v^2/\Lambda^2$ (in addition to a pair of complex-conjugate roots which escape to infinity for $\Lambda\to \infty$). 
Notice that, due to the $1/\Lambda^2$ truncation in eq.~(\ref{F_SMEFT6}), none of the four zeroes of eq.~(\ref{F_SMEFT6}), a fourth-order polynomial, seems to be double. 
This means that inside a small interval of values of $h_1$ of width suppressed by $1/\Lambda^2$, eq.~(\ref{F_SMEFT6}) can violate positivity (see subsection~\ref{subsec:positivity} below). Furthermore, we find that $\mF(h_1)=\mO(1/\Lambda^4)$ for $h_1/v =-1 + x \, v^2/\Lambda^2$ $\forall x\in \mathbb{R}$ \cite{Gomez-Ambrosio:2022qsi}. Hence, it looks that the perturbative analysis of $\mF$ alone does not allow us to extract any information beyond $h_\ast=-v +\mO(1/\Lambda^2)$. 

We note, however, that eq.~(\ref{sympoint2}) is indeed the correct solution of $h(h_1)=-v$ up to order $c_H/\Lambda^2$, even though the $\mO(1/\Lambda^2)$ Flare-function in eq.~(\ref{F_SMEFT6}) fails to recover that precise expression at that perturbative order. 
This ambiguity will be solved in the $\mO(1/\Lambda^4)$ analysis in the next subsection.   

\subsubsection{Change of variables and symmetric point at dimension-8} \label{subsec:dim8}

We now quote the result of adding the 
operator of dimension-8 in eq.~(\ref{subsec:dim8op}); the calculation follows along the same lines of the previous paragraph so we only quote the combined result for the flare function $\mathcal{F}$, which reads 
\begin{align} \label{F_SMEFT8}
   \mF(h_1)
    \,=\, \left(1+ \frac{h(h_1)}{v}\right)^2 \, 
     =  1 &+ \left( \frac{h_1}{v} \right)  
  \left( 2 + 2 \frac{ c_{H \Box}^{(6)} v^2}{\Lambda^2} 
  + 3 \frac{(c_{H \Box}^{(6)})^2 v^4}{\Lambda^4}  + 
      2 \frac{c_{H \Box}^{(8)} v^4}{\Lambda^4}  \right) + \nonumber \\
  &+
   \left( \frac{h_1}{v} \right)^2 \left( 1 + 
   4 \frac{ c_{H \Box}^{(6)} v^2}{\Lambda^2} + 
   12 \frac{(c_{H \Box}^{(6)})^2 v^4}{\Lambda^4} + 
    6 \frac{c_{H \Box}^{(8)} v^4}{\Lambda^4}  \right) + 
 \nonumber \\ &+
    \left( \frac{h_1}{v} \right)^3 
    \left(  8 \frac{ c_{H \Box}^{(6)} v^2}{3 \Lambda^2} + 
    56 \frac{(c_{H \Box}^{(6)})^2 v^4}{3 \Lambda^4}  +
    8 \frac{ c_{H \Box}^{(8)} v^4}{\Lambda^4}   \right) +
\nonumber \\ &+
 \left( \frac{h_1}{v} \right)^4 \left( 
 2 \frac{c_{H \Box}^{(6)} v^2}{3 \Lambda^2} +
 44 \frac{(c_{H \Box}^{(6)})^2 v^4}{3 \Lambda^4} + 
 6 \frac{c_{H \Box}^{(8)} v^4}{\Lambda^4} \right) + 
  \nonumber \\ 
  &+
   \left( \frac{h_1}{v} \right)^5
   \left( 88 \frac{(c_{H \Box}^{(6)})^2 v^4}{15 \Lambda^4} +
    12 \frac{ c_{H \Box}^{(8)} v^4}{5 \Lambda^4}  \right) + 
  \nonumber \\
  &+ \left( \frac{h_1}{v} \right)^6 
  \left( 44 \frac{(c_{H \Box}^{(6)})^2 v^4}{45 \Lambda^4} + 2 \frac{c_{H \Box}^{(8)} v^4}{ 5 \Lambda^4}  \right) + \op{(\Lambda^{-6})} \, .
\end{align}
Note that the bracket in each line provides the corresponding $a_j$ up to and including $\mO(\Lambda^{-4})$.  
Also, to make the order manifest and avoid confusion, we have denoted $c_{H\Box}$ by $c^{(6)}_{H\Box}$ in this paragraph 
and below whenever it may be needed. 

As for the symmetric point around which SMEFT is built, eq.~(\ref{sympoint2}) takes a further correction of $\mO(\Lambda^{-4})$ that may take it away from the Standard Model value. 
 Again, eq.~(\ref{cerodoble}) shows that the $SU(2)\times SU(2)$ fixed-point point condition $\mF(h_\ast)=0$ has always the solution $h=-v$, which in the in the HEFT coordinates up to and including $\mO(\Lambda^{-4})$ SMEFT corrections is given by  
\begin{equation}
\frac{h_\ast}{v}\, =\, \frac{h_1(h)}{v}\bigg|_{h=-v} 
\,=\, -1 + \frac{c_{H \Box}^{(6)} v^2 }{3\Lambda^2}   +    \left( (c_{H \Box}^{(6)})^2  + 2 c_{H \Box}^{(8)} \right)\frac{v^4}{10\Lambda^4} \, .
\label{eq:SMEFTd8-sym-point}
\end{equation} 
where we used the relation between SMEFT and HEFT coordinates at this order:  
\begin{equation}
h_1 \,=\, h 
\, +\, \frac{c_{H\Box}^{(6)}}{3\Lambda^2}\left(v^3-(v+h)^3\right) 
\, +\, \frac{((c_{H\Box}^{(6)})^2+2c_{H\Box}^{(8)})}{10\Lambda^4}\left(v^5-(v+h)^5\right)\,+\,\mO\left(\frac{1}{\Lambda^6}\right)\, ,  
\end{equation}
 which can be iteratively inverted, yielding 
\begin{eqnarray}
h =&& h_1 
+\frac{c_{H\Box}^{(6)}}{3\Lambda^2}\left((v+h)^3-v^3\right) 
+\frac{(c_{H\Box}^{(6)})^2}{30\Lambda^4}\left(13 (v+h_1)^5- 10 v^3(v+h_1)^2 -3 v^5\right) \nonumber\\
&&+\frac{c_{H\Box}^{(8)}}{5\Lambda^4}\left((v+h)^5-v^5\right)\,+\,\mO\left(\frac{1}{\Lambda^6}\right)\; .
\end{eqnarray}

The study of the $\mO(1/\Lambda^4)$ Flare-function in eq.~(\ref{F_SMEFT8}) shows that $\mF(h_1)=\mO(\Lambda^{-4})$ for $h_1/v=  -1 +x\,  v^2 /\Lambda^2 $ for all $x\in \mathbb{R}$ except for $h_1/v=  -1 +c_{H \Box}^{(6)} v^2 /3\Lambda^2 $, where we find $\mF(h_1)=\mO(\Lambda^{-6})$.   
Hence, requiring $\mF(h_1)=0$ up to $\mO(1/\Lambda^4)$ leads to the real solution $h_1/v=  -1 +c_{H \Box}^{(6)} v^2 /3\Lambda^2 + \mO(1/\Lambda^4)$, solving the $\mO(1/\Lambda^2)$ ambiguity of the previous subsection. However, the perturbative expression in eq.~(\ref{F_SMEFT8}) is not able to recover the $\mO(1/\Lambda^4)$ expansion of the symmetric point~(\ref{eq:SMEFTd8-sym-point}). In order to fix the $\mO(1/\Lambda^4)$ term of $h_\ast$ there are two ways to proceed: analyze the Flare-function constraint $\mF(h_1)=0$ up to $1/\Lambda^6$; or solve the equation $F=1+ h(h_1)/v=0$ up to $\mO(1/\Lambda^4)$. 

\section{Geometric and analytic distinction between HEFT and SMEFT}
\label{Geometricsection}
\label{sec:geometry}

This section exposes the precise theoretical conditions allowing to discern between SMEFT and HEFT, summarizing the main results of several articles~\cite{Dobado:2019fxe,Alonso:2016oah,Alonso:2016btr,Alonso:2015fsp,Cohen:2020xca} in their geometrical aspects and adding an extended analytical discussion about the function $\mathcal{F}$ of our own. 
Much of the past confusion between the two EFT formulations arose from the fact that there are two coordinate systems to describe the same set of particles. For this reason, the San Diego and Oregon groups employed a geometric perspective to be able to make coordinate-invariant statements. 

\subsection{Flat SM geometry}
The $O(4)$ components in the scalar field of eq.~(\ref{eq:EWNGB-spherical-coord}) used for the SM Higgs sector, $\boldsymbol{\phi}=(\phi^1,\phi^2,\phi^3,\phi^4)$ are taken to represent coordinates in a (momentarily flat, later in the next subsection curved) geometric manifold $\mathcal{M}$. 
$\boldsymbol{\phi}$ contains the Higgs field and the three ``eaten'' Goldstone bosons and has a Lagrangian (turning off gauge fields)
\begin{equation}
    \mathcal{L}_{\rm SM}=\frac{1}{2}\partial_\mu\boldsymbol{\phi}\cdot\partial^\mu \boldsymbol{\phi}-\frac{\lambda}{4}(\boldsymbol{\phi}\cdot\boldsymbol{\phi}-v^2)^2\;.
\end{equation}
In these Cartesian coordinates the global $O(4)$ transformations should act \textit{linearly}
\begin{equation}
    \boldsymbol{\phi}\to O\boldsymbol{\phi}\;,\;\;\;\;\;\;O^TO=\boldsymbol{1}\;.\label{linearsym}
\end{equation}
The field breaks the global electroweak symmetry $O(4)$ down to $O(3)$ by acquiring a vacuum expectation value
$$\langle\boldsymbol{\phi\cdot\phi}\rangle=v^2\ , $$
where $v\simeq 246$ GeV. Usually, the vacuum expectation value is chosen to be $\langle\phi_4\rangle=v$ while $\langle\phi_1\rangle=\langle\phi_2\rangle=\langle\phi_3\rangle=0$ and the Higgs field $h$ in these Cartesian coordinates is defined through the relation $\phi_4=v+h$. \par

The alternative coordinate system in which HEFT is based expresses the SM Higgs sector Lagrangian in polar form 
\begin{equation}
    \boldsymbol{\phi}=\Big(1+\frac{h}{v}\Big)\boldsymbol{n}(\boldsymbol\omega), \;\;\;\;\;\;\;\; \boldsymbol n\cdot\boldsymbol n=v^2\;.
\end{equation}
Clearly, the constraint $\boldsymbol n\cdot\boldsymbol n=v^2$ makes the $O(4)$ symmetry to be realized in a \textit{non-linear} way.
This comes about because the four components of $\boldsymbol n=(\omega^1,\omega^2,\omega^3,\pm \sqrt{v^2-{\boldsymbol \omega}^2})$
rotate linearly with $O$ in eq. (\ref{linearsym}),
imposing a non-linear transformation law on the polar-coordinate Goldstone bosons, $w^a$ ($a=1,2,3$). In these polar coordinates, the Higgs sector SM Lagrangian of eq. (\ref{eq:SBSlagrangian}) is~\footnote{Note the difference with eq.(2.12) in~\cite{Alonso:2016oah} where the $\omega_i\omega_j$ piece is absent. It is unnecessary unless amplitudes with more than two Goldstone bosons are analyzed, which we leave for future investigation.}
\begin{equation}
    \mathcal{L}_{\text{SM}}=\frac{1}{2}\Big(1+\frac{h}{v}\Big)^2(\partial_\mu \omega^i\partial_\mu  \omega^j )\left(\delta_{ij}+\frac{\omega_i\omega_j}{v^2-\boldsymbol{\omega}^2}\right)+\frac{1}{2}(\partial_\mu h)^2-\frac{\lambda}{4} (h^2+2vh)^2\;.
\end{equation}
In the SM, we thus see that
\begin{equation}
    \mathcal{F}(h=h_1) = 
   \left(1+\frac{h}{v}\right)^2
    \label{SMF}
\end{equation}

This exercise enlightens the fact that that the SM Higgs sector $O(4)$ symmetry can be realized both in a linear or a non-linear manner. That is why, when studying EFTs that extend the SM, one should  concentrate on objects which are invariant under field redefinitions~\cite{Passarino:2016saj}.

\subsection{Beyond SM: curved \texorpdfstring{$\phi$}{f} geometry}

The kinetic term of the Lagrangian (giving the classical equations of motion as the geodesics of that manifold) in eq.~(\ref{FbosonLagrangianLO}) has the form
$$\frac{1}{2}g_{ij}(\boldsymbol{\phi})\partial_\mu \phi^i\partial^\mu \phi^j\;,$$
and can be interpreted in terms of a metric tensor $g_{ij}$ that provides lengths in the geometrical space of the $\boldsymbol{\phi}$ fields,  $g_{ij}d\phi_id\phi_j$ in any of the different coordinate choices. 
This is the point of contact between physics experiments at colliders, \textit{i.e.}  production and scattering amplitudes derived from this Lagrangian, and the field geometry, where the amplitudes are represented in terms of curvature invariants. We now briefly recall the results presented in~\cite{Alonso:2016oah,Cohen:2020xca}.

In the SM, the metric  is just a Kronecker delta $g_{ij}(\boldsymbol{\phi})_{SM}=\delta_{ij}$, but the SMEFT of eq.~(\ref{SMEFTEWlagrangianV}) extends it to a more general form
\begin{equation}
    g_{ij}\arrowvert_{\tiny SMEFT}=A\Big(\frac{\boldsymbol{\phi}\cdot\boldsymbol{\phi}}{\Lambda^2}\Big)\delta_{ij}+B\Big(\frac{\boldsymbol{\phi}\cdot\boldsymbol{\phi}}{\Lambda^2}\Big)\frac{\phi_i\phi_j}{\Lambda^2}
\end{equation}
where $\Lambda$ is the new physics scale.
 It is always possible to express the  SMEFT metric (and in particular its SM limit) in HEFT form by changing to polar coordinates followed by a field redefinition to make the kinetic term of $h$ canonical  as in section~\ref{sec:convert}. This makes the HEFT metric take the generic form
 \begin{equation}
     g_{ij}(h,\boldsymbol{\omega})_{HEFT}=\begin{bmatrix}
\mathcal{F}(h)g_{ab}(\boldsymbol \omega) & 0 \\
0 & 1
\end{bmatrix}\label{HEFTmetric}
 \end{equation}
 where $g_{ab}(\boldsymbol \omega )$ is the $O(3)$ invariant metric on the scalar submanifold $O(4)/O(3)=S^3$ described by the Goldstone bosons in angular coordinates (corresponding to the manifold $SU(2)_{L-R}$ introduced in subsection \ref{subsec:EWEFTs}). 
The SM is the special case with a flat scalar manifold $\mathcal{M}$  (since its metric is just $\delta_{ij}$ for all values of the field: there exist global Riemann coordinates). 
For both SMEFT and HEFT, $\mathcal{M}$ has curvature. But what makes SMEFT different from HEFT?

eq.~(\ref{HEFTmetric}) allows to interpret the function $\mathcal{F}(h)$ in the manifold $\mathcal{M}\ni \boldsymbol{\phi}$ as 
a scale factor akin to the $a(t)$ one in the Friedmann-Robertson-Walker metric. For each value of $h$  (in units of $v$) there is an $S^3$ submanifold (parametrized by the $\omega_i$), away from the origin of $\mathcal{M}$ by an amount $\mathcal{F}(h)$, that acts as a radial distance.\par
One can in that way always write SMEFT in HEFT form, but as we will see, the converse is not always true. This means that
 $$\text{SM}\subset\text{SMEFT}\subset\text{ HEFT}\;.$$
 This is so because, in order to write a HEFT as a SMEFT, there must exist a point in $\mathcal{M}$ which is invariant under the $O(4)$ symmetry. This translates into the condition that the function $\mathcal{F}(h)$ must vanish for some $h_\ast$, $\mathcal{F}(h_\ast)=0$. Hence this $h_\ast$, an invariant point under the $O(4)$ symmetry,  plays the role of an origin for the Cartesian-like SMEFT coordinates on $\mathcal{M}$. 
 
This invariant point is a necessity to deploy a linear representation of the $O(4)$ group around it, just as in eq.~(\ref{linearsym}), and hence to write down HEFT as a SMEFT.
But it happens that the existence of such zero of $\mathcal{F}(h)$ is not a sufficient condition for SMEFT to be deployed. Non-analyticities might arise at the fixed point $h_\ast$ or between that $h_\ast$ and $h=0$ (the physical vacuum), spoiling the possibility of constructing a viable SMEFT around $h_\ast$ that is applicable in particle physics. This is why we must require that the metric and thus $\mathcal{F}(h)$ be analytic in a sufficient domain (see \cite{Cohen:2020xca} for further detail). 
In a lower energy regime $m_h\sim E$ where the potential $V$ makes a relevant contribution, the same considerations also apply to $V$.  

Once the basics of the situation have been understood, instead of obtaining these results by further following the powerful yet intricate geometric methods just mentioned, we will continue with coordinate-dependent field theory in the spirit of staying close to the phenomenological formulation that can be brought to bear at accelerator experiments.

\subsection{Zero and analyticity of \texorpdfstring{$\mathcal{F}$}{F} upon passing from HEFT to SMEFT}
\label{subsec:noSMEFT}

We are going to show in this subsection how to proceed from HEFT to SMEFT, and under what condition this is possible in an analytical manner in terms of the field coordinates. 
For this we need to combine the Higgs $h_1$ and the EW Goldstone bosons $\omega^a$ appropriate for HEFT into the complex doublet $H$ used in SMEFT. Problems can arise about whether the resulting Lagrangian will obey minimal physical requirements from the theoretical (\textit{e.g.}, analyticity) and the phenomenological (\textit{e.g.}, perturbation theory convergence) points of view.  

The first step to reconstruct the SMEFT form 
from the HEFT Lagrangian in eq.~(\ref{eq:HEFT-Lagr})
is to define a new Higgs variable $h$ from the HEFT one, $h_1$ in this subsection,  by the condition 
\begin{eqnarray}
\mathcal{F}(h_1(h)) = F^2(h_1(h)) \,=\, \bigg(1+\frac{h}{v}\bigg)^2\, , 
\qquad \qquad 
F(h_1(h)) \,=\,  1+\frac{h}{v} \, , 
\end{eqnarray}
implying the inverse relations
\begin{eqnarray} \label{Finverses}
h_1   \,=\, \mathcal{F}^{-1}\left((1+h/v)^2\right)\, , 
\qquad \qquad 
h_1 \, =\,  F^{-1}( 1+ h/v) \, . 
\end{eqnarray}

This change of variable unravels the standard HEFT normalization of the Higgs kinetic term in eq.~(\ref{bosonLagrangian}), turning the Lagrangian in the ``polar-SMEFT'' form, 
\begin{eqnarray} \label{polarSMEFT2}
\mathcal{L}_{\text{polar-SMEFT}} =&& 
\frac{v^2}{4} \bigg(1+\frac{h}{v}\bigg)^2 \,   \langle \partial_\mu U^\dagger \partial^\mu U\rangle 
+
\frac{1}{2} \bigg(   \frac{1}{v} (F^{-1})'(1+h/v)  \bigg)^2 (\partial h)^2  \, .
\end{eqnarray}  
The second, $h$-kinetic term can also be expressed in terms of the square of $F$, that is, $\mathcal{F}$.
To achieve it, we can simply replace $(F^{-1})'(1+h/v)$  by  $  2 (1+h/v)\, (\mathcal{F}^{-1})'((1+h/v)^2)  $.  
We note that $F^{-1}$ and $\mathcal{F}^{-1}$ are the inverse functions of $F$ and $\mathcal{F}$, respectively, and $F(h_1)=1+h(h_1)/v$.

Up to this point there is no concerning issue; this polar-coordinate form half way between HEFT and SMEFT, that we also find when calculating in the opposite direction in eq.~(\ref{polarSMEFT1}) is 
still a valid Lagrangian (if we postpone for a later moment the discussion on the convergence of the perturbative series) totally equivalent to HEFT.  

The possible obstacle to this conversion can however arise when trying to reconstruct the Higgs-doublet field $H$ from the EW Goldstone bosons $\omega^a$ in $U$ and the Higgs scalar field $h$,  making use~\cite{Cohen:2020xca} of
\begin{eqnarray}
\label{transformaSMEFT}
|H|^2 =&& \frac{(v+h)^2}{2}\, ,
\nonumber\\
|\partial H|^2 =&& \frac{(v+h)^2}{4}   \langle \partial_\mu U^\dagger \partial^\mu U\rangle +  \frac{1}{2}   (\partial h)^2 \, ,
\nonumber\\
(\partial|H|^2)^2 =&& (v+h)^2 \,   (\partial h)^2\,=\, 2 |H|^2 \,   (\partial h)^2 \, .
\end{eqnarray} 
The inversion of these equations to express eq.~(\ref{polarSMEFT2}) in terms of $H$
brings about a possible $|H|^{-2}$ singularity in the SMEFT Lagrangian,
\begin{eqnarray}
\mathcal{L}_{\rm SMEFT} =&&  \underbrace{ |\partial H|^2}_{= \mathcal{L}_{\rm SM}} \quad +\quad 
\underbrace{ \frac{1}{2} \bigg[ \bigg(   \frac{1}{v} (F^{-1})'\left(\sqrt{2| H |^2/v^2}\right)  \bigg)^2 \,\,\,-\,\,\, 1\bigg] \, \frac{(\partial| H |^2)^2}{2 | H |^2}  }_{=\Delta \mathcal{L}_{\text{BSM}}}\, .
\label{eq:SMEFT-from-HEFT}
\end{eqnarray}

That divergence from the $|H|^2$ denominator is incompatible with a power-expansion in powers of $|H|$ as needed to deploy the SMEFT counting.
The only way that eq.~(\ref{eq:SMEFT-from-HEFT}) can provide an analytical Lagrangian around $|H|=0$ to allow a valid SMEFT expansion in powers of $H$ is by restricting the $\mathcal{F}=F^2$ function of eq.~(\ref{FbosonLagrangianLO}) to fulfill the condition 
\begin{eqnarray}   \bigg[
\frac{1}{v} (F^{-1})'\left(\sqrt{2| H |^2/v^2}\right)  \,\,\,-\,\,\, 1 \bigg] 
&\stackrel{| H |\to 0}{=}&  0 \,\,\, + \,\,\,   \mathcal{O}(| H |^2)  
\label{eq:analyticity-condition0}\;,
\end{eqnarray}
so that this zero cancels the $|H|^{-2}$ denominator in eq.~(\ref{eq:SMEFT-from-HEFT}).
Furthermore, even if the zero is cancelled, the analyticity of the SMEFT Lagrangian at any order implies that the $\mathcal{O}(|H|^2)$ remnant must also have an analytic expansion in integer powers of $|H|^2$ (from the square-root argument) . Otherwise an expansion-breaking nonanalyticity is present and SMEFT becomes a theory that is not systematically improvable by furthering the expansion.  This remarkable fact can be traced to eq.~(\ref{Finverses}), where the change of variables $h\to h_1$ happens at the level of individual singlet particles, whereas the doublet $H$ employed in SMEFT (and in the SM) needs to be squared to $|H|^2$ to produce an electroweak singlet, forcing the square root upon us.

The relation~(\ref{eq:analyticity-condition0}) is a differential equation for $(F^{-1})$ in the variable $z:=\sqrt{2|H|^2/v^2}$, whose integration leads to 
\begin{eqnarray}
 F^{-1}(z) 
&\stackrel{|H|\to 0}{=}&  F^{-1}(0) \,\, +\,\, v z \,\,  \,+\,\mathcal{O}(z^3)\,.  
\label{eq:Fm1}
\end{eqnarray}
 The analyticity of the SMEFT Lagrangian at all orders implies that the $\mathcal{O}(z^3)$ remainder has an analytic expansion that only contains odd powers of $z$.
We solve for the $z$ variable around $z=0$ in terms of $F^{-1}(z)$, 
and invert to recover the original function
 $F=(F^{-1})^{-1}$ around the point $h_1^*\equiv F^{-1}(0)$, remembering from eq.~(\ref{Finverses}) that $F^{-1}$ is a HEFT-Higgs $h_1$ value: 
\begin{eqnarray}
z = F(h_1) 
&\stackrel{| H |\to 0}{=}&   \frac{1}{v}(h_1\, -\,h_1^*) \,\, \,\,+\,\,\, \,\mathcal{O}((h_1-h_1^*)^3)\,,   
\label{eq:zeroF-cond}
\end{eqnarray}
where the $\mathcal{O}(z^2)$ remnant can be put in $\mathcal{O}((h_1-h_1^*)^2)$ form up to higher orders.  In terms of $\mathcal{F}$ the relation would be given by $(1+h/v)^2= 2|H|^2/v^2=z^2=F(h_1)^2=\mathcal{F}(h_1)$.   
 Moreover, the analyticity of the SMEFT Lagrangian at all orders implies that the solution of eq.~(\ref{eq:Fm1}) for $z$ --shown in~(\ref{eq:zeroF-cond})-- has an analytical expansion around $h_1=h_1^*$ that only contains odd powers of $(h_1-h_1^*)$.   

The existence of that zero $h_1^*\equiv F^{-1}(0)$ of $F$ --and thus of its square $\mathcal{F}$--, and the analyticity required for a power series expansion (both of $\mathcal{F}$ and the Higgs potential $V$), broadly constitute the necessary and sufficient requirements for a given HEFT Lagrangian density  characterized by  $\mathcal{F}$ to be expressible as a SMEFT.
Let us summarize and make these findings, that agree with the ones presented in~\cite{Cohen:2020xca}, somewhat more precise:
\begin{enumerate}
    \item{} $F(h_1)$ must have at least a simple zero at some $h_1^*$, {\it i.e.},   $F(h_1^*)=0$. This implies that the function in the HEFT Lagrangian density of eq.~(\ref{FbosonLagrangianLO}) $\boxed{\mathcal{F}(h_1^*)=F(h_1^*)^2=0}$ must have a double zero.  
    
    \item{} At that point $h_1^*$, $F$ must have the slope  $F'(h_1^*)=\frac{1}{v}$.  This translates into two conditions over $\mathcal{F}$, namely 
    $$\boxed{\mathcal{F}'(h_1^*)=0\ , \ \mathcal{F}''(h_1^*)=\frac{2}{v^2}}\;.$$ 
    
    \item{} 
 Finally, it is possible to exploit the analyticity of the SMEFT Lagrangian at higher orders, if the expansion is to be continued and be systematically improvable. In general, analyticity as an all-order requirement forces all even derivatives to vanish at the symmetric point: $F^{(\ell)}(h_1^\ast)=0$ for even $\ell$. From the point of view of $\mathcal{F}$ this implies the vanishing of all odd derivatives,
    $\boxed{\mathcal{F}^{(2\ell+1)}(h_1^*)=0}$ .  
\end{enumerate}

The first two conditions mean that the HEFT  $\mathcal{F}$ flare function must be an upward-bending parabola if an equivalent SMEFT is to exist. In the next subsection,  Figure~\ref{fig:Fstatus} shall expose that current knowledge is compatible with it, and allows to estimate how intensely one of the HEFT coefficients needs to separate from the SM or SMEFT for the latter not to be applicable.

Should eq.~(\ref{eq:zeroF-cond}) fail, the SMEFT Lagrangian would not have an analytical expansion in powers of the doublet field $H$. 
Moreover, it is important to remark that in order to avoid a singularity, at least at dimension-6, the remnant in~(\ref{eq:analyticity-condition0}) must be at least  $\mathcal{O}(|H|^2)$, or equivalently, the remnant in~(\ref{eq:zeroF-cond}) must be at least $\mathcal{O}((h_1-h_1^*)^3)$. 

Various examples will be provided in subsection~\ref{subsec:examples} below.
 We will deal with the possibility of experimentally finding such zeroes $h_1^*$ in section~\ref{Zeroessection}.

\section{Generic properties of the flare function \texorpdfstring{$\mathcal{F}(h)$}{F(h)}}
\label{sec:generic-properties}

\subsection{Current knowledge of \texorpdfstring{$\mathcal{F}(h)$}{F(h)}}

In particle physics language, the appearance of the $\mathcal{F}(h)$ function in eq.~(\ref{FbosonLagrangianLO}) controls 
the (derivative) coupling of a pair of $\omega\simeq W_L$ longitudinal gauge bosons to any number of Higgs bosons.

\begin{figure}[!ht] 
\centering
\includegraphics[width=0.48\columnwidth]{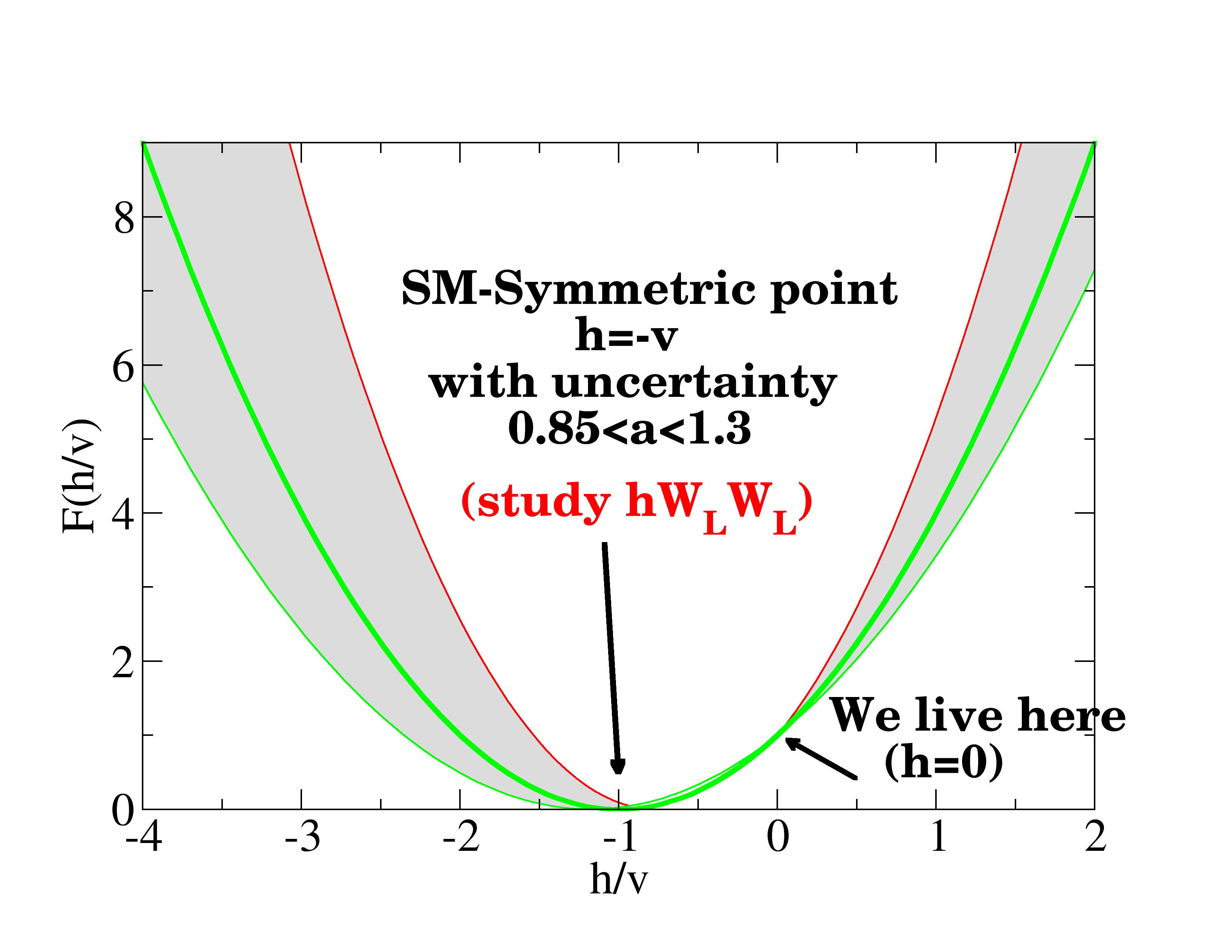}
\includegraphics[width=0.48\columnwidth]{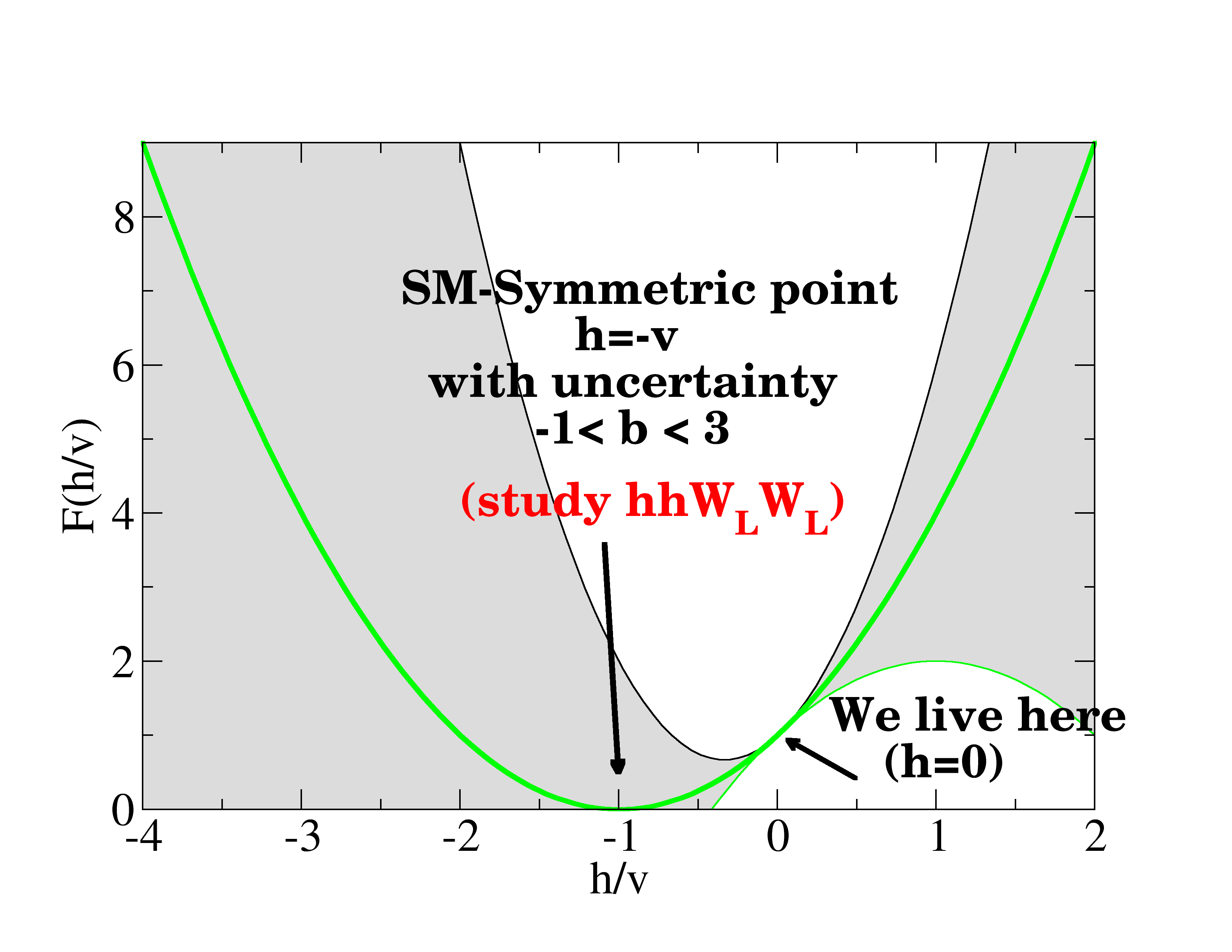}
\includegraphics[width=0.55\columnwidth]{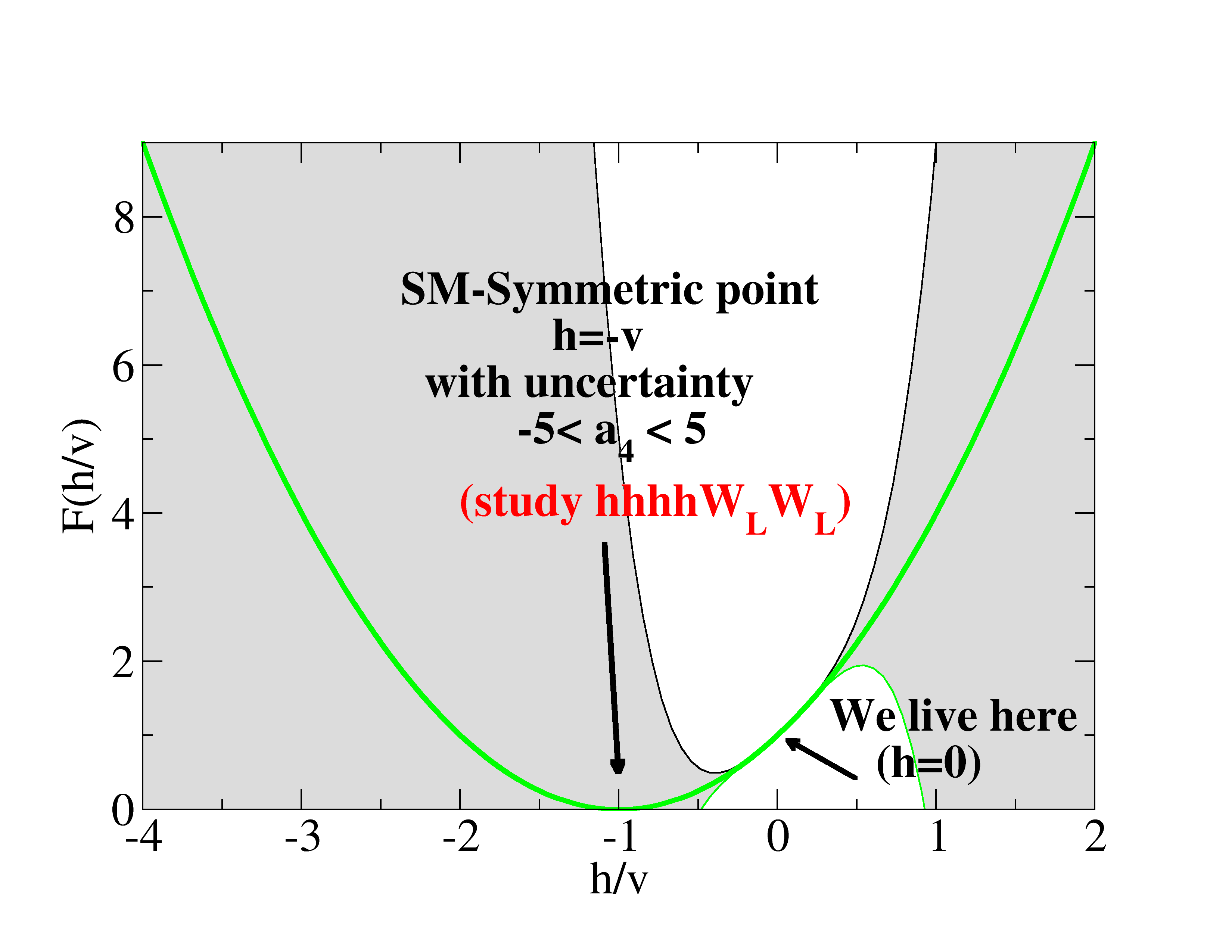}
\caption[Sensitivity of $\mathcal{F}(h)$ to typical  parameter ranges]{\label{fig:Fstatus} {\small Sensitivity of $\mathcal{F}(h)$ to typical  parameter ranges.
The solid parabola (green) is the SM prediction $1+2\frac{h}{v}+\left(\frac{h}{v}\right)^2$. The grey bands, in the order given,
show the uncertainty due to our present knowledge of the $a_i$ coefficients  that we vary one at a time around the SM value.
Respectively, the coefficients $a_1=2a$, $a_2=b$ and $a_4$, couple $\omega\omega$ to $h$ (the best constrained one, $a$), $hh$ (controlled by $b$), and in view that a third order polynomial always has a zero (and we ignore $a_3$ by itself), $hhhh$ with an even number of powers as the next most interesting one.} 
}
\end{figure}

While this coefficient is the dominant Higgs production in the TeV region, multiboson processes at the LHC in the hundred GeV energy regime already constrain, although not tightly, the coefficients of the $\mathcal{F}(h_1)$ expansion. 
Since, for the rest of the main body of the chapter, we will be concentrating on HEFT and there can be no confusion with the SMEFT $h$ field, \textbf{we will drop the subindex $h_1\to h$} and make it explicit whenever a change of coordinates between SMEFT and HEFT is used.
We give a graphical representation of the present status of $\mathcal{F}$ in Figure~\ref{fig:Fstatus}.

At the time of writing there is no significant deviation from the Standard Model, which is a particular case of SMEFT, meaning that there is no particular reason to doubt the applicability of SMEFT: the uncertainty bands in Fig.~\ref{fig:Fstatus} by no means exclude a zero of $\mathcal{F}$, possibly where the Standard Model requires it, at $h=-v$. 
The SM line, as a particular SMEFT case, is a parabola with vertex at $\mathcal{F}=0$, as discussed in subsection~\ref{subsec:noSMEFT}.
The reason why we have cut off values $\mathcal{F}<0$ in Fig.~\ref{fig:Fstatus} will become clear in the next subsection.

\subsubsection{Positivity from boundedness of the Hamiltonian} \label{subsec:positivity}

From the LO HEFT Lagrangian in eq.~(\ref{FbosonLagrangianLO}) we can construct the Hamiltonian of the theory, finding 
\begin{equation}\label{HamiltonianHEFT}
    H_{\text{HEFT}}=\int d^3\boldsymbol x \; \frac{1}{2}\left[\partial_0 h\partial_0 h+\boldsymbol\nabla h  \cdot \boldsymbol \nabla h \, +\,      \mathcal{F}(h) \, \left( \partial_0 \omega^i\partial_0 \omega^j+\boldsymbol\nabla \omega^i \cdot  \boldsymbol \nabla \omega^j \right)\left(\delta_{ij}+\frac{\omega_i\omega_j}{v^2-\boldsymbol{\omega}^2}\right) \right]\;.
\end{equation}
 From the condition that the Hamiltonian $H_{\text{HEFT}}$ must be bounded from below for vacuum stability one obtains that $\mathcal{F}(h)\geq 0$. This justifies the common usage of the form $F^2(h)$ instead of simply $\mathcal{F}$. 
 While a matter of taste, it is not clear what in particle physics is the quantity $F$ being squared (a radial distance in the $\omega^a$ field space), so we prefer $\mathcal{F}$ for most of the discussion.

One can also often see that,
after expanding the function $\mathcal{F}(h)$ around our physical low-energy vacuum $h=0$ as in eq.~(\ref{expandF}),  
the positivity condition on $\mathcal{F}(h)$ is forgotten or not explicitly mentioned, although in those approaches employing $F(h)$ with $\mathcal{F}=F^2(h)$ it is automatically incorporated. 

Therefore we are going to distinguish three cases. First let us mention that if 
$\mathcal{F}$ requires an infinite expansion, the information about positiveness is intricately hidden in the coefficients $a_i$. 

The second case that we next address corresponds to the treatment of experimental data within order by order EFT; $\mathcal{F}$ is truncated to a few terms and the customary assumption that $\mathcal{F}=F^2$ is accepted. However, because the most general polynomial of degree $n$ cannot be written as a square \cite{Polya}, we briefly discuss, as a third case, the possibility that $\mathcal{F}$ is well approximated by a polynomial, but this needs to be decomposed as  $\mathcal{F}=F_1^2+F_2^2$ that holds in all generality (because it corresponds to $|F|^2$, the modulus square of a complex function). 
For the sake of simplicity, we will express $h$ in units of $v$ in the discussion of this section~\ref{sec:generic-properties}. 

\begin{figure}[!ht]
\raisebox{0.3\height}{\includegraphics[width=90mm]{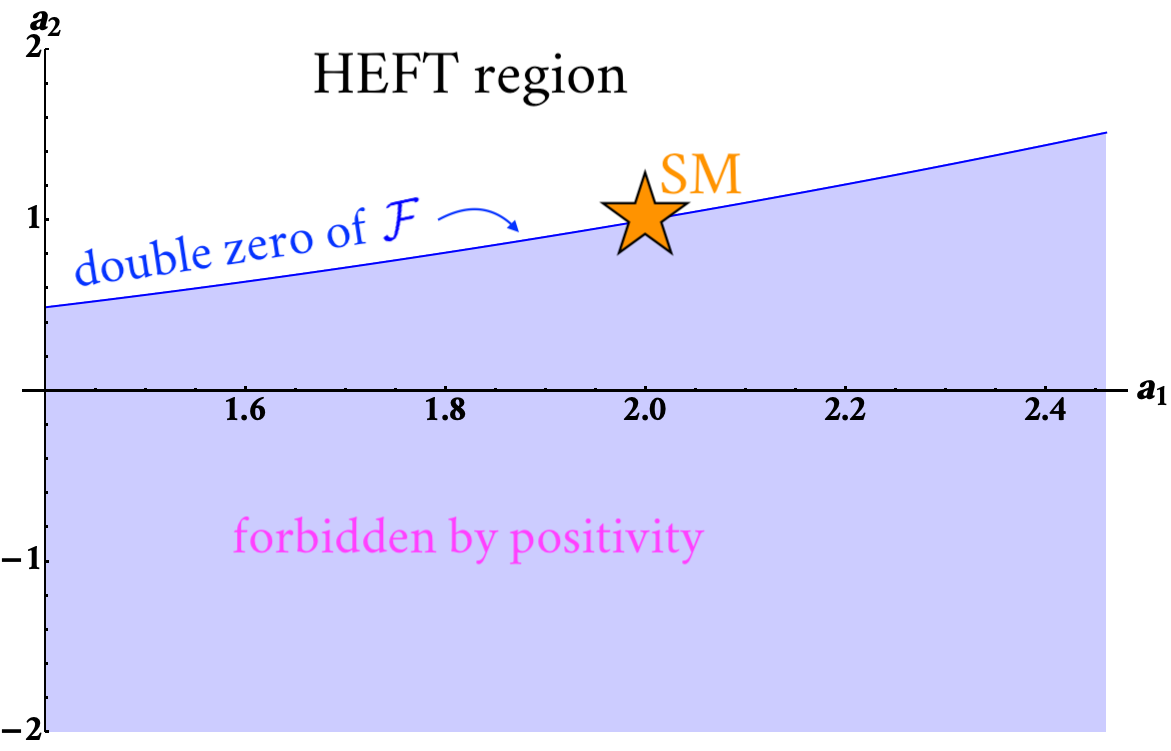}}\hfill
\includegraphics[width=75mm]{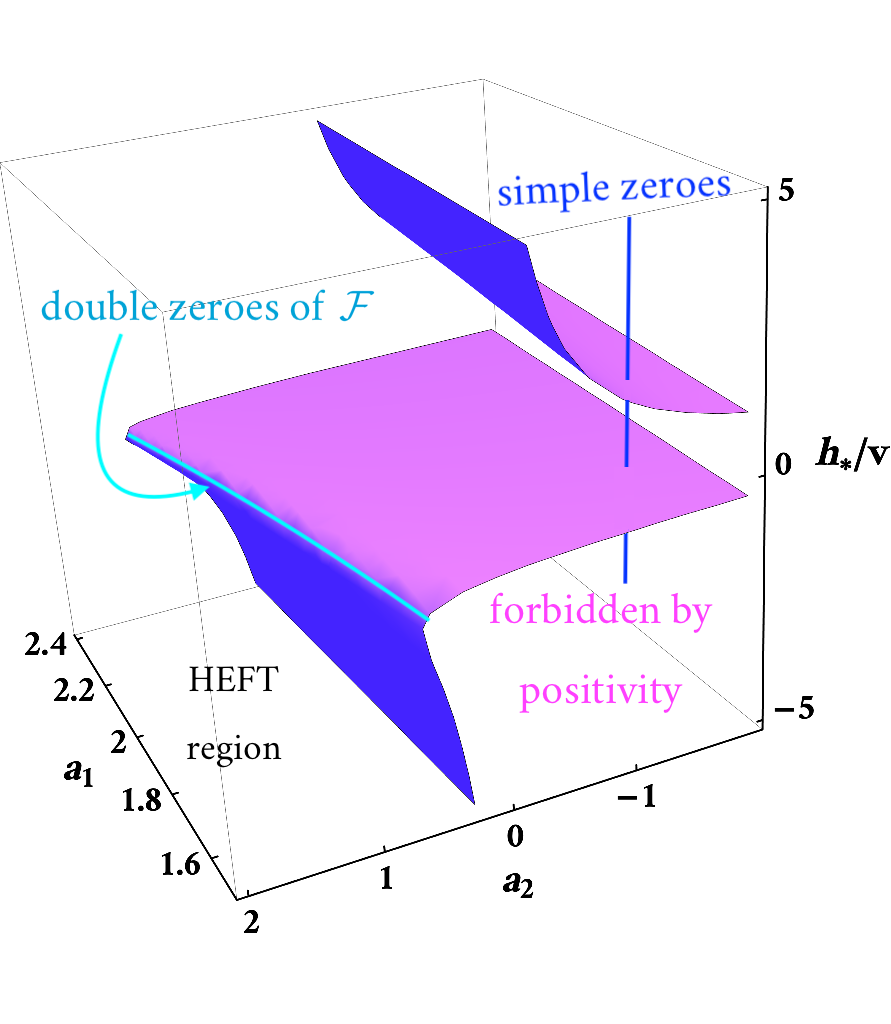}
    \caption[$(a_1,a_2)$ plane]{$(a_1,a_2)$ plane (\textbf{left plot}) and $(a_1,a_2,h_\ast)$ three-dimensional representation (\textbf{right plot)}, showing the regions where HEFT is valid; those where there is a double zero (that follow {\it e.g. } from the dilaton model discussed in subsection~\ref{subsec:exampleFnoSMEFT}); and the region where there are two simple zeroes and a quadratic polynomial $\mathcal{F}$ corresponds to no valid theory.}
    \label{fig:positivity}
\end{figure}

Figure~\ref{fig:positivity} displays, in its left plot, the $(a_1,a_2)$ plane corresponding to a quadratic $\mathcal{F}$. The positivity-forbidden region is the lower, shaded area (blue). The white area at the top respects positivity and supports a HEFT formulation. The dividing line is that where the Higgs flare function  presents a double zero, so it is of the form
$\mathcal{F}=(1+\frac{a_1}{2}h)^2$. 

On the right panel of the Figure \ref{fig:positivity} we show a three-dimensional plot where, additionally to $(a_1,a_2)$, the value of $h_\ast$ is represented. There are regions where there is no zero of $\mathcal{F}$ (so only HEFT is applicable), a line (along the lower surface's fold) where the zero of $\mathcal{F}$ is double. However, SMEFT is not an acceptable description along this line (at dimension 6) since non-analyticities arise when reconstructing the Higgs doublet $H$ for $a\neq1$ because, in that case, the curvature of the Flare function at the symmetric point is not equal to $2/v^2$ as should be for SMEFT (see the previous section; an example of this is the dilaton model, explained below around eq. (\ref{eq:dilatonexample})).
There are also regions where there are two simple real roots of $\mathcal{F}=0$ that entail a sign change, and thus a violation of positivity. An example of this is shown by a vertical line piercing both sheets.

\paragraph{$\mathcal{F}=F^2$ is assumed.}

In this situation, there are restrictions among the coefficients $a_i$ that guarantee positivity of $\mathcal{F}$. We obtain them up to fourth order, by squaring the expansion of $F$.
If the expansion ended at first order, that is, normalizing $h$ by $v$,
\begin{equation}
    F(h)=1+\alpha h \implies \mathcal{F}(h)=F^2(h)=1+2 \alpha  h+h^2 \alpha ^2\;,
\end{equation}
we obtain the relation 
\begin{equation}
a_2\, =\,  \frac{a_1^2}{4}\,,
\label{eq:posit-F-rel-h1}
\end{equation} 
or $b=a^2$ which is exactly the correct one for $a$ and $b$ in the Standard Model.

If that relation is experimentally found to be broken, then at least one more order is necessary in the expansion of $F$. This then implies that the degree of the flare function $\mF$ must be at least two orders higher,
\begin{equation}\label{poscorrels1}
F(h)=1+\alpha_1 h +\alpha_2 h^2
\implies 
\mathcal{F}(h)=1+2 \alpha_1  h+h^2 \left(\alpha_1 ^2+2 \alpha_2 \right)+ 2 \alpha_1  \alpha_2 h^3+
\alpha_2 ^2 h^4\, ,
\end{equation}
implying two correlations: 
\begin{equation}
    2a_3=a_1\left(a_2-\frac{a_1^2}{4} \right)\, ,  \ \ \ \ \ \ \  4a_4=  \left(a_2-\frac{a_1^2}{4} \right)^2\, ,
\label{eq:posit-F-rel-h2}
\end{equation}
These new correlations~(\ref{eq:posit-F-rel-h2}) would substitute~(\ref{eq:posit-F-rel-h1}), being a smoking gun of the presence of further BSM 3-Higgs and 4-Higgs vertices in the effective Lagrangian, which could then be measured in further collider experiments.

Iterating the analysis procedure, any eventual experimental deviations from  the relations in eq.~(\ref{eq:posit-F-rel-h2}), 
immediately imply the presence of higher order coefficients in the function $F$, and therefore also in the flare function $\mF$ that provides the $WW\to n h$ effective vertices.

To  have enough freedom to accommodate the SMEFT values of the $a_i$ coefficients in eq.~(\ref{coefsSMEFTHEFT}) to $\mathcal{O}(\Lambda^{-2})$, the expansion of $F$ needs to be run up to fourth order in $h$,
\begin{equation}
F(h)=1+\alpha_1 h +\alpha_2 h^2 +\alpha_3 h^3 +\alpha_4 h^4 
\dots
\end{equation} 
where the Greek names of the coefficients mimic those of the expansion of $\mathcal{F}$.
The positivity conditions on the coefficients $a_i$ of $\mathcal{F}$ can be obtained from
\begin{equation}
\mathcal{F}(h)=F^2(h)=1+2 \alpha_1  h+h^2 \left(\alpha_1^2+2 \alpha_2 \right)+h^3 (2 \alpha_1  \alpha_2 +2 \alpha_3 )+
h^4 \left(2 \alpha_1  \alpha_3 +\alpha_2^2+2 \alpha_4 \right)\dots
\label{eq:Fc=F2}
\end{equation}   
and essentially leave the first four $a_i$ undetermined, while those with $i=5\dots 8$ become dependent of those earlier four. Once more, an experiment that does not respect the corresponding correlations points to a higher term in the expansion of $F$ and so on. There is a tower of positivity correlations that should be experimentally tested as multiHiggs data in the correct kinematic region becomes available.

\vspace{1cm}
\paragraph{Most general non-negative polynomial
satisfying $\mathcal{F}=F_1^2+F_2^2$.} 

In this case, the flare function $\mathcal{F}(h)$ is the most general nonnegative polynomial $\mathcal{F}(h)\geq 0$  $\forall h\in\mathbb{R}$, and therefore, even of degree $2d$,
$\mF(h)=\sum^{2d}_{n=0} a_n h^n$.

We invoke the theorem \cite{Polya} that states that such a nonnegative polynomial can be decomposed as $\mF(h)=F_1^2(h)+F_2^2(h)$ 
in terms of two polynomials\footnote{
To demonstrate it, we will note that the most general form of a polynomial of order $2d$, that is $\mF(h)=a_n\prod_{i=1}^d(h-h^\ast_i)(h-\bar{h}^\ast_i)$, can be restricted because
positivity and real analyticity demand that both the real and complex roots must be double, 
 $$\mF(h):=| F(h)|^2\;\;\text{ with } F(h)=\sqrt{a_n}\prod_{i=1}^n(h-h^\ast_i)\ ;$$
the theorem then follows from taking the real and imaginary parts, $F_1(h)=\text{Re } [F(h)]$ and $F_2(h)=\text{Im }[F(h)]$. (That the real roots are double thwarts any sign change near them and guarantees positivity).}
$F_1(h)$ and $F_2(h)$ of degree $d$.
The equivalent of~eq.(\ref{poscorrels1}) then becomes
\begin{align}
\mF(h)=&1 + 2 (\alpha_1 \cos\theta+\beta_1 \sin\theta) h + (\alpha_1^2+\beta_1^2+2 \alpha_2 \cos\theta+2 \beta_2 \sin\theta) h^2 +\nonumber\\
&+2( \alpha_1 \alpha_2+\beta_1 \beta_2 ) h^3+ (\alpha_2^2+\beta_2^2) h^4  \,,
\end{align}
having expanded $F_1$ up to second order  with coefficients $\alpha_i$, $F_2$ with coefficients $\beta_i$, and having noted that because of the normalization of the kinetic term, $\mathcal{F}(0)=F_1(0)^2+F_2(0)^2=1$, a practical parametrization is $F_1(0)=\cos\theta$, $F_2(0)=\sin\theta$ for some angle $\theta$.
The number of free parameters is now high enough
(five, $\theta,\, \alpha_{1,2},\, \beta_{1,2}$, for four orders that depend on them) so that, in an order by order expansion of the polynomial, the correlations are weaker than in the simplified case $\mathcal{F}=F^2$ with $F$ real.

Nevertheless, one should note that if, given an experimental situation, the highest given order $h^{2d}$ has a negative coefficient $a_{2d}$, then higher orders are necessary. This can be used in experiment to detect (further) new physics. Currently, the sign of $b=a_2$ is not known; discerning whether it is positive or negative is therefore an interesting experimental analysis exercise that, if it turned out to be negative, would immediately and of necessity point out to new coefficients $a_3$ and $a_4$ (and of course, indicate new physics, since in the Standard Model $b=1$).

\subsection{Restrictions on \texorpdfstring{$\mathcal{F}(h)$}{F(h)}'s coefficients from assuming SMEFT}
\label{subsec:correlations}
The restrictions over $\mathcal{F}$ at the symmetric point $h_\ast$ that guarantee the existence of a SMEFT at the end of subsection~\ref{subsec:noSMEFT} translate into conditions over the $a_i$ at the physical vacuum $h=0$, that are constrained by experiment. Here we will write down the known ones. 
Let us express the series expansion around $h=0$ setting $a_0:=1$, and take $h$ normalized to $v$, so that $v=1$,
\begin{equation}
    \mathcal{F}(h) = \sum_{i=0}^n a_i h^i\ .
\end{equation}
 
Since the conditions over $\mathcal{F}$ are taken at $h_\ast$, we reexpand around that point, and in terms of $a^\ast_j=\mF^{(j)}(h_\ast)/j!$, find
\begin{equation}
    \mathcal{F}(h) = \sum_{j=0}^n a^\ast_j (h-h_\ast)^j\ .
\end{equation}
By matching the two expansions around the two different points it is easy to read off the coefficients $a^\ast_j$ (on which the conditions over $\mathcal{F}$ are expressed) in terms of the $a_i$ (more directly accessible to experiment).  The relation reads
\begin{equation}
    a^\ast_j = \sum_{k=0}^\infty a_{k+j} h_\ast^k \cdot b_{jk}\ .\label{eq:recursion1}
\end{equation}
The coefficients of this expansion can be recursively 
obtained,
\begin{eqnarray}
    b_{0k} = 1\ \forall k \,, \nonumber \\
    b_{jk} = \sum_{l=0}^k b_{j-1\ l} \ .
\end{eqnarray}
The closed formula that solves eq. (\ref{eq:recursion1}) as
\begin{equation}
        a_\ell^\ast = A_{\ell j} a_j\,,
\end{equation}
requires the following simple auxiliary matrix
\begin{eqnarray}
A_{\ell j}= \left\{
\begin{tabular}{ccc}
    $0$ & if & $j<\ell$\,, \\ 
    $\left(\begin{array}{ccc}   j \\ \ell  \end{array}\right)  \, h_*^{j-\ell}$ &  if &  $j\geq \ell$\, . 
\end{tabular}    
\right.   
\end{eqnarray}

We can now deploy the four straightforward conditions 
$\mathcal{F}(h_\ast)=\mathcal{F}'(h_\ast)=\mathcal{F}^{'''}(h_\ast)=0$, 
$\mathcal{F}^{''}(h_\ast)=2/v^2 \to 2$ as four linear constraints on the coefficients around the physical vacuum, namely
\begin{align} \label{matchcoeffs}
&\sum_{k=0}^\infty h_\ast^k a_k \cdot b_{0k} = a_0^\ast = 0\,, \nonumber & &\sum_{k=0}^\infty h_\ast^k a_{k+1} \cdot b_{1k} = a_1^\ast = 0\,, \nonumber \\
&\sum_{k=0}^\infty h_\ast^k a_{k+2} \cdot b_{2k} = a_2^\ast = 1\,, \qquad & &\sum_{k=0}^\infty h_\ast^k a_{k+3} \cdot b_{3k} = a_3^\ast = 0 \,.
\end{align}
\paragraph{Square-matrix four-coefficient truncation}
A first possible truncation of the series is to keep the terms with the first four  $(a_1,a_2,a_3,a_4)$  coefficients (the zeroth coefficient is identically $a_0=\mF(0)=1$ by construction of the HEFT formalism so we include it in the inhomogeneous term together with the $(a_1^\ast,a_2^\ast,a_3^\ast,a_4^\ast)$ values from the conditions on $\mathcal{F}$). These are the coefficients that collect dimension-6 SMEFT corrections to the SM as shown in eq.~(\ref{F_SMEFT6}), and the linear system becomes
\begin{equation}\label{matrixforas}
\begin{pmatrix}
h_\ast & h_\ast^2 &h_\ast^3 &h_\ast^4 \\
1 & 2h_\ast & 3h_\ast^2 & 4h_\ast^3 \\
0 & 1 & 3 h_\ast & 6 h_\ast^2 \\
0 & 0 & 1 & 4h_\ast \\
\end{pmatrix} \begin{pmatrix} a_1\\ a_2 \\ a_3\\ a_4 \end{pmatrix}=
\begin{pmatrix} -1 \\ 0 \\ 1 \\ 0  \end{pmatrix} \ .
\end{equation}
The matrix has determinant $h_\ast^4$, so that barring the zero at
$h_\ast=0$ (the physical vacuum, where the coefficients $a_i$ and $a^\ast_i$ coincide), the system has a unique solution for each $h_\ast$.  
Such solution for the $a_i$ is that of eq.~(\ref{F_SMEFT6}), with the symmetric point of SMEFT 
$h^*=-v+v \frac{c_{H\Box}^{(6)}v^2}{3\Lambda^2} $ that of eq.~(\ref{sympoint2}), as can be easily checked by substitution.

\paragraph{Systematic order by order truncation}
Instead of that truncation, one could be more systematic and match the two Taylor expansions in powers of $h$ at an arbitrary (but equal) order, say $N$. 
In that case the system  $A\, \vec{a}_\ast =\vec{a}$ has $N$ unknowns but $(N+1)$ equations and compatibility becomes an issue. The criterion of Rouche-Frobenius guaranteeing an algebraic solution then links possible values of the  $h_\ast$ with the unknown $a_{2n}^\ast$ that can appear on the right hand side of the equivalent system.

Taking $\mF(h)$ as a polynomial of order $h^4$, this compatibility condition is, 
\begin{equation}
1\,=\,\mF(0)\,=\, \frac{h_\ast^2}{v^2} \,  + \, a_4^\ast  \, \frac{h_\ast^4}{v^4} \,.
\label{eq:rel-polh4}
\end{equation}
Without extra work, the vanishing of $a_n^\ast$ for odd $n=1,3,5...$ yields the same relation even if $\mF(h)$ is a polynomial of order $h^5$. 
For a flare function $\mF(h)$, still polynomial but now of order $h^6$ (or even $h^7$) the constraint takes one more term,  
\begin{equation}
1\,=\,\mF(0)\,= \frac{h_\ast^2}{v^2} \,  + \, a_4^\ast  \, \frac{h_\ast^4}{v^4}\,  + \, a_6^\ast  \, \frac{h_\ast^6}{v^6} \,.
\label{eq:rel-polh6}
\end{equation} 
Let us recall that a non-singular SMEFT Lagrangian requires $a_0^\ast=0$, $a_2^\ast=1$  
and $a_n^{\ast}=0$ for odd $n=1,3,5...$, as shown earlier in subsection~\ref{subsec:noSMEFT}.
It is not difficult to check that SMEFT fulfills these relations~(\ref{eq:rel-polh4}) and (\ref{eq:rel-polh6}) at $\mO(\Lambda^{-2})$ and $\mO(\Lambda^{-4})$, respectively, as that effective theory predicts (see Eqs.~(\ref{F_SMEFT8}) and~(\ref{eq:SMEFTd8-sym-point})):
\begin{eqnarray}
 \frac{ h_\ast }{v} =&&  \frac{ \mF^{-1}(0) }{v} \,=\,   -1 + \frac{c_{H \Box}^{(6)} v^2 }{3\Lambda^2}   +    \left( (c_{H \Box}^{(6)})^2  + 2 c_{H \Box}^{(8)} \right)\frac{v^4}{10\Lambda^4} +\mO(\Lambda^{-6})\,,
 \nonumber\\
a_4^\ast =&& \frac{1}{4!}\mF^{(4)}(h_\ast)\,=\, \frac{2 c_{H \Box}^{(6)} v^2 }{3\Lambda^2} +\mO(\Lambda^{-6})\, ,
\nonumber\\
a_6^\ast =&& \frac{1}{6!}\mF^{(6)}(h_\ast)\,=\, \left(\frac{44 (c_{H \Box}^{(6)})^2  }{45} +\frac{2 c_{H \Box}^{(8)}   }{5}  \right)  \frac{ v^4 }{ \Lambda^4} +\mO(\Lambda^{-6})\, , 
\end{eqnarray}
with all these properties determined by the precise form of the flare function $\mF. $

If we extended the analysis to include $a_5$ and $a_6$, which is easily done and omitted for briefness, we would have two more Lagrangian parameters but only one further constraint over $\mathcal{F}$, namely the vanishing of its fifth derivative. This means that SMEFT would have a second parametric degree of freedom that could take any value. And in fact, this is precisely the case in eq.~(\ref{F_SMEFT8}), that depends on the additional parameter $c^{(8)}_{H\Box}$ from the dimension-8 relevant Lagrangian. \\

\paragraph{Resulting testable correlations}
Table~\ref{tab:correlations}  collects the correlations between the $a_i$ coefficients of HEFT that we have worked out at order $\Lambda^{-2}$ and $\Lambda^{-4}$ (further correlations are possible from the higher odd derivatives of $\mathcal{F}$ vanishing, and all become a bit weaker numerically if yet higher orders in $1/\Lambda$ are studied, by the need of introducing further $a_i$ coefficients).

The correlation in the first row, second column of Table~\ref{tab:correlations} originates in a  quadratic one
 $2(\Delta a_2-2 \Delta a_1) -\frac{3}{4}\left(a_3-\frac{4}{3}\Delta a_1\right)=\left(-3\Delta a_1+\frac{5}{2}\Delta a_2-\frac{9}{8}a_3\right)^2$
 with two solutions for $a_3$, a small and a large one. In keeping near the SM value $a_3=0$ we take this second one and reexpand to linearize in $a_3$ so that it can be related to $a_1$ and $a_2$ in a straightforward manner; the difference is more suppressed than $\mathcal{O}(\Lambda^{-4})$ in the SMEFT 
 expansion.

 The remarkable property of these equations is that they are independent of the SMEFT parameters $c^{(n)}_{H\Box}$, that is, they are tests of the SMEFT theory framework itself, up to a given order in $1/\Lambda$, that cannot be rewritten away in terms of its parameters.

\begin{table}
    \caption[Correlations between the $a_i$ HEFT coefficients necessary for SMEFT to exist]{\small Correlations between the $a_i$ HEFT coefficients necessary for SMEFT to exist, at order $\Lambda^{-2}$ and $\Lambda^{-4}$. They are given in terms of $\Delta a_1:=a_1-2=2a-2$ and $\Delta a_2:=a_2-1=b-1$. This way, all the objects in the table vanish in the Standard Model, with all the equalities becoming $0=0$. 
    Notice that the r.h.s. of each identity in the second column shows the $\mO(\Lambda^{-4})$ corrections to the relations of the first column. The third one assumes the perturbativity of the SMEFT expansion.
    \label{tab:correlations}}
    \centering
    \begin{tabular}{|c|c|c|}\hline
        \textbf{Correlations}  & \textbf{Correlations}  & 
       ${\Lambda^{-4}}$ \textbf{Assuming}  
        \\ 
        \textbf{accurate at order} $\Lambda^{-2}$ & \textbf{accurate at order} $\Lambda^{-4}$ & \textbf{SMEFT perturbativity}
        \\ \hline
        $\Delta a_2=2\Delta a_1$ &  &   $|\Delta a_2| \leq 5 |\Delta a_1|$
        \\
        $a_3=\frac{4}{3} \Delta a_1$ & 
         $\left(a_3-\frac{4}{3}\Delta a_1\right) =\frac{8}{3}(\Delta a_2-2 \Delta a_1)   -\frac{1}{3}\left(\Delta a_1\right)^2$ 
        & 
        \\  
        $a_4=\frac{1}{3} \Delta a_1$ & 
        $\left(a_4-\frac{1}{3}\Delta a_1\right) = \frac{5}{3}\Delta a_1 - 2\Delta a_2 +\frac{7}{4} a_3=  $
        & 
        those for $a_3$, $a_4$, $a_5$, $a_6$
        \\
         & $ \phantom{\left(a_4-\frac{1}{3}\Delta a_1\right)}=\frac{8}{3}(\Delta a_2-2 \Delta a_1)-\frac{7}{12}\left(\Delta a_1\right)^2  $ &
        \\ 
        $a_5=0$ &  
        $a_5 = \frac{8}{5}\Delta a_1 -\frac{22}{ 15} \Delta a_2 +a_3=$ 
        & 
        all the same 
        \\ 
         &  $\phantom{a_5}= \frac{6}{5}  (\Delta a_2-2 \Delta a_1) -\frac{1}{3}\left(\Delta a_1\right)^2  $   & 
        \\ 
       $a_6=0$ & 
        $a_6=\frac{1}{6}a_5$
        &
         \\ 
        \hline
    \end{tabular}
\end{table}
 
These equations can be experimentally tested looking for the consistency of SMEFT.
Given tight experimental bounds on $a_1$, these relations (and those from $\mathcal{F}\geq 0 $) can already predict how the next HEFT coefficients will look like if SMEFT is valid.
This we will delay until subsection~\ref{subsec:numeritos} below.

The $1/\Lambda^2$ relations in the first column of Table~\ref{tab:correlations}, all hanging from $\Delta a_1$, are rather constraining given that one-Higgs production is well known. Those in the second column, as they depend also on $\Delta a_2$, which is much less well measured, are not very useful; but they can be further tightened by imposing perturbativity of the SMEFT expansion. \\

\paragraph{Perturbativity constraints} Perturbativity can be deployed by recalling that, at $\mathcal{O}(\Lambda^{-4})$,
\begin{eqnarray}
a_1 =&& 
  \left( 2 + 2 \frac{ c_{H \Box}^{(6)} v^2}{\Lambda^2} 
  + 3 \frac{(c_{H \Box}^{(6)})^2 v^4}{\Lambda^4}  + 
      2 \frac{c_{H \Box}^{(8)} v^4}{\Lambda^4}  \right)  \nonumber \\
a_2 =&&  \left( 1 + 
   4 \frac{ c_{H \Box}^{(6)} v^2}{\Lambda^2} + 
   12 \frac{(c_{H \Box}^{(6)})^2 v^4}{\Lambda^4} + 
    6 \frac{c_{H \Box}^{(8)} v^4}{\Lambda^4}  \right) \ .
\end{eqnarray}
For clarity, let us shorten notation for the rest of the paragraph, writing
\begin{eqnarray} \label{shorthanda1a2}
 \Delta a_1 =&& 2x +3x^2 +2y \nonumber \\
 \frac{\Delta a_2}{2} =&& 2x + 6x^2 +3y = \Delta a_1 +3x^2 + y\ . 
\end{eqnarray}
In general, there are two free parameters, $x$ and $y$. What perturbativity suggests is that each of the terms of the $\mathcal{O}(\Lambda^{-4})$ should not be larger than the $\mathcal{O}(\Lambda^{-2})$ term (this is akin to the Cauchy criterion for convergence of a sequence, but of course there is no guarantee that it will be satisfied at a fixed order; again, it is only a perturbativity argument, similar to the one in~\cite{Hays:2018zze}). 
Taking this at face value, it must be that $3x^2\leq 2|x|$  (by the way, this means that $|x|\leq 2/3$, that however is of little value as experimental constraints are much tighter) and that $|y|<|x|$. 

Returning to the first of eq.~(\ref{shorthanda1a2}) and separately analyzing the positive and negative $x$ cases, we find
\begin{equation}
    |x|< {\rm max}\left(|\Delta a_1^-|,\frac{1}{2}|\Delta a_1^+|\right)
\end{equation}
and noting that half the upper 95\% uncertainty $\Delta a_1^+/2$ is larger than the lower one $\Delta a_1^-$ as discussed around Table~\ref{tab:correl-exp-bounds} below, leads us to 
\begin{equation} \label{corra2_c}
    |\Delta a_2| \leq 2|\Delta a_1| + 2\cdot \frac{3}{2} |\Delta a_1^+|\implies
    |\Delta a_2| \leq 5 |\Delta a_1^+|\ ,
\end{equation}
relation which we elevate to the third column of Table~\ref{tab:correlations}, in the understanding that the uncertainty there is the maximum ($+$) of the two asymmetric uncertainties, and where the absolute value bars have been at last dropped.

In the order in which experimental data can be used,

\begin{itemize}
\item  A nonzero measurement of $\Delta a_2$ signals new physics. SMEFT or HEFT are needed.
\item If additionally the stronger correlation $\Delta a_2 = 2 \Delta a_1$ is violated,  
severe corrections to $1/\Lambda^2$ SMEFT are suggested.
\item If the weaker correlation in eq.~(\ref{corra2_c}), $\Delta a_2 \leq 5 \Delta a_1$  is violated,
those correlations make SMEFT unnatural and put its perturbative use into question but they do not necessarily rule it out as discussed in the next paragraph.
\item If the weakest correlation in the second column of Table~\ref{tab:correlations} is broken, the first two orders of SMEFT do not make much sense and the theory is falsified for all practical purposes.
\end{itemize}

To close this subsection, we note that the presence of the zero (and minimum) of $\mathcal{F}$ at $h^\ast$ is a distinguishing property in the TeV region, for near-threshold physics the Higgs potential $V(h)$ is also important. The SMEFT potential needs to be analytic too so that a power-expansion makes sense. The relevant theory regarding $V$ is  briefly discussed in subsection ~\ref{coeffspotential}.

\subsection{Example functions to illustrate HEFT vs SMEFT differences}
\label{subsec:examples}

Let us illustrate the whole discussion with a few simple example functions 
(as opposed to the more ambitious construction of entire UV completions shown in \cite{Cohen:2020xca,Grinstein:2007mp}).

\subsubsection{Example flare functions $\mathcal{F}$ where SMEFT is applicable}

A couple of examples of HEFT flare functions that lead to regular SMEFT Lagrangians are:

\begin{itemize}
    \item The SM has $\mathcal{F}(h)=(1+h/v)^2$ that of course is analytic, possesses a zero at $h_\ast=-v$ and trivially fulfills all correlations in Table~\ref{tab:correlations} since $\Delta a_1=0=\Delta a_2$, $a_i=0\ \forall i>2$.
    \item The Minimally Composite Higgs Model with symmetry breaking pattern $SO(5)/SO(4)$ ~\cite{Contino:2011np}, with 
    $\mathcal{F}(h)=\frac{f^2}{v^2}\sin^2\left(\frac{h}{f} +\arcsin{\frac{v}{f}}\right)$, which expanded to fourth order in $h/v$ and second in $v/f$ yields~\footnote{
    Up to one more order, $\mO(v^4/f^4)$, the flare function in the MCHM can be given by the polynomial 
        $
    \mF(h)=1 + \left(2-\frac{v^2}{f^2}- \frac{v^4}{f^4}\right)\frac{h}{v} 
    +  \left(1-\frac{2 v^2}{f^2}\right)\frac{h^2}{v^2} 
    +  \left(-\frac{4 v^2}{3 f^2}+ \frac{2v^4}{3f^4}\right)  \frac{h^3}{v^3}  
    +  \left(-\frac{ v^2}{3 f^2}+ \frac{2v^4}{3f^4}\right) \frac{h^4}{v^4}  
    +  \left(\frac{4 v^4}{15 f^4}\right)  \frac{h^5}{v^5}  
    +  \left(\frac{ 2 v^4}{45 f^4}\right) \frac{h^6}{v^6}\, .$  
    This result is fully  consistent with  the $\mO(\Lambda^{-4})$ SMEFT flare function  in eq.~(\ref{F_SMEFT8}) for the relations $c_{H\Box}^{(6)}= -\Lambda^2/(2f^2)$, 
    $c_{H\Box}^{(8)} = - \Lambda^4/(2f^4)$. 
    }:
 \begin{eqnarray}
    \mF(h) =&& 1 + \left(2-\frac{v^2}{f^2}\right)\frac{h}{v} 
    +  \left(1-\frac{2 v^2}{f^2}\right)\frac{h^2}{v^2} 
    -  \frac{4 v^2}{3 f^2} \frac{h^3}{v^3} 
    -  \frac{v^2}{3 f^2} \frac{h^4}{v^4} \, .
    \end{eqnarray}    
    It is easy to observe that this is a particular case of the SMEFT flare function at $\mO(\Lambda^{-2})$ in eq.~(\ref{eq:SMEFT-F-Lambda2}) after the identification 
    $c_{H\Box}  =  -\Lambda^2/(2f^2)$. 
\end{itemize}

\subsubsection{Example flare functions $\mathcal{F}$ where SMEFT is not applicable} \label{subsec:exampleFnoSMEFT}

 Examples of HEFT Lagrangians that transform to non-regular SMEFT Lagrangians are given by the models with $\mathcal{F}=e^{2h/v}$ or $\mathcal{F}= 1+\frac{1}{2}\sin(4h/v)$. Such models fail to have a zero of $\mathcal{F}$.

However, as seen in section \ref{Geometricsection}, this condition is not sufficient to have an appropriate SMEFT Lagrangian in terms of $H$: we illustrate this with the dilatonic model~\cite{Halyo:1991pc,Goldberger:2007zk,Hernandez-Leon:2017kea}, that has a HEFT function $\mathcal{F}=(1+ a h/v)^2 $ that does present a zero at $h^*= - v/a$. Nonetheless, the corresponding SMEFT Lagrangian~(\ref{eq:SMEFT-from-HEFT}) happens to be singular for $a\neq 1$, with a pole at $H^{\dagger} H=0$:
\begin{align}
\mathcal{L}_{\text{SMEFT}}=& \mathcal{L}_{\text{SM}} +  
\frac{1}{2} \bigg[ \bigg(   \frac{1}{v} (F^{-1})'(1+h/v)  \bigg)^2 -1\bigg] \, (\partial h)^2 
\nonumber=\\
=&\, \mathcal{L}_{\text{SM}} + \frac{1}{2} \frac{(1-a)}{a} (\partial h)^2 
= \, \mathcal{L}_{\text{SM}} + \frac{1}{2} \frac{(1-a)}{a}  \, \frac{(\partial|H|^2)^2}{2 |H|^2} \, .\label{eq:dilatonexample}
\end{align}

It might be tempting to consider that the divergence of the second term in the second line in eq. (\ref{eq:dilatonexample}) could be removed by an appropriate rescaling of $h$, but this would disarray the operators in $\mathcal{L}_{\rm SM}$, which would not come together anymore to conform $\mathcal{L}_{\rm SM}$. In this case, it happens that there is a zero in $F(h)=1 + \frac{a h}{v}$ at $h^*=-\frac{v}{a}$ but the slope of $F$ is not $\frac{1}{v}$ but rather $F'(h^*)=\frac{a}{v} \neq \frac{1}{v}$ for $a\neq 1$.  

From a completely different approach, based on the phenomenology of the effective couplings, we could observe that the dilaton is not compatible with the SMEFT expansion, since SMEFT --in eq.~(\ref{coefsSMEFTHEFT})-- predicts $\Delta b= 4 \Delta a$ up to $1/\Lambda^4$ NNLO corrections \cite{Agrawal:2019bpm,Sanz-Cillero:2017jhb}, while the dilatonic model predicts that we should be observing $\Delta b= 2 \Delta a$~\cite{Halyo:1991pc,Goldberger:2007zk,Hernandez-Leon:2017kea}, 
with $\Delta a\equiv a-1,\, \Delta b\equiv b-1$.  
The only way SMEFT could be able to reproduce the ``dilatonic-line data'' is through a 100\% correction from operators of dimension-8 and greater, indicating a breakdown of the $1/\Lambda$ expansion.

\subsubsection{Example of potentials $V$ where SMEFT is applicable}
Next, we propose two Higgs-Higgs self-interaction potentials that lead to regular SMEFT Lagrangians (corresponding to the correlations explained in subsection \ref{coeffspotential} below) , for example

\begin{itemize}
    \item The SM potential (with $\lambda, -\mu^2$ both positive)
    is given by  
    \begin{equation}
    V_{\text{SM}}(H)=\mu^2H^\dagger H+\lambda (H^\dagger H)^2\, ,
    \end{equation}
    
    which, in HEFT coordinates, becomes\footnote{In this case, the correlations of table~\ref{tab:corV} in subsection~\ref{coeffspotential} below are trivially satisfied, because the variables there defined $\Delta v_3=\Delta v_4=\dots 0$ all vanish.} 
    \begin{equation} 
    V_{\text{SM}}(h)=\frac{m_h^2}{2}\left({h^2} +\frac{h^3}{v}+\frac{h^4}{4v^2}\right) \,, 
    \end{equation}
    with $v^2=-\mu^2/\lambda$ and $m_h^2=2 |\mu|^2$.

    \item The SMEFT potential with the correlations obtained in section~\ref{coeffspotential} will need to have an expansion which, up to $\mathcal{O}(\Lambda^{-2})$ has the form
\begin{align}
V(h) = \frac{m_h^2}{2}&\bigg[h^2+\frac{h^3}{v} \left({1}+\epsilon\right)+\frac{h^4}{v^2} \left(\frac{1}{4}+\frac{3}{2}\epsilon \right)+\frac{3\epsilon}{4}\frac{h^5}{v^3}+\frac{\epsilon }{8}\frac{ h^6}{v^4}\bigg]\, .
\label{eq:SMEFT-V-correlations}
\end{align}
Including the custodial-invariant SMEFT operator without derivatives, $\mathcal{O}_H$, of eq. (\ref{Warsawbasis}) leaves the potential as 
\begin{equation} 
V_{\rm SMEFT}(H)=\mu^2H^\dagger H+\lambda (H^\dagger H)^2 -  \frac{c_H}{\Lambda^2} (H^\dagger H)^3\, .
\label{eq:V_SMEFT}
\end{equation}
By expanding $H$ around its minimum, and expressing the SMEFT potential in HEFT coordinates, one reproduces the structure in eq.~(\ref{eq:SMEFT-V-correlations})
with $m_h^2=-2\mu^2 \left(1+ 3\epsilon/4\right)$ and $2\langle |H|^2\rangle =v^2= v_0^2\left(1-3\epsilon/4\right) $, where we made use of the lowest order vev $v_0^2=\, -\mu^2/\lambda$ and the $\mO(\Lambda^{-2})$ correction $\epsilon= - 2c_H v^4/m_h^2\Lambda^2= \, \mu^2 c_H    /(\lambda^2  \Lambda^2)$. Notice that, for sake of clarity in the illustration, here we have taken $c_{H\Box}=0$, so there is no Higgs field renormalization: treating only terms in the potential, \textit{i.e.} non-derivative couplings implies, up to a constant shift, $h=h_1$. 
\end{itemize}

\subsubsection{Example of potentials $V$ where SMEFT is not applicable}

An example of a potential which can not be written as a SMEFT is
\begin{equation}
V(H) =V_{\text{SM}}(H)+ \frac{\varepsilon}{H^\dagger H}\;,
\end{equation}
with $\varepsilon$ a constant small enough so as to avoid unsettling the potential away from $h=0$ by a finite fraction of $v$ 
now there is no symmetric $O(4)$ point where the function is analytic, there is a divergence at the origin. 
Consistently with the symmetric-point criterion, SMEFT cannot be used: this model does not reproduce eq.~(\ref{eq:SMEFT-V-correlations}).

\subsection{Unitarity imposes no constraint on the coefficients, causality may}
It has recently been proposed that unitarity violations in the effective theory could be used to describe the space of theories that can be characterized as HEFT but that, due to non-analyticities, can not be brought up to SMEFT form~\cite{Cohen:2021ucp}. 
While this may deserve further study, we are not very sure about that program. 

The reason is that the HEFT Lagrangian yields a properly Hermitian Hamiltonian, and therefore a unitary scattering matrix. Truncating an expansion of a partial wave amplitude in perturbation theory is indeed a procedure that violates unitarity, but this has nothing to do with the theory itself, but with the truncation. For example, as we will see in detail in the next chapter, in the well-known case of two-body elastic scattering one can, instead of the partial wave amplitude, expand first the inverse partial wave amplitude to one loop in the EFT
\begin{equation}
    \frac{1}{t^{IJ}} = \frac{1}{t^{IJ}_0+t^{IJ}_1}
\end{equation}
and then invert back to obtain 
\begin{equation}
    t^{IJ} \simeq \frac{(t_0^{IJ})^2}{t^{IJ}_0-t^{IJ}_1}\ .
\end{equation}
This expansion of the inverse amplitude, that can be carried out order by order, 
can also be derived from a dispersion relation, so it satisfies all analyticity properties expected from an elastic scattering amplitude. 
Additionally, elastic unitarity over the physical cut of the amplitude is exact,
no matter how strong the interaction, as long as the low-energy theory has the structure of HEFT (or Chiral Perturbation Theory or other similar theories with derivative couplings). 
This has been documented at length in the literature\cite{Dobado:1989gr,Delgado:2013loa,Espriu:2014jya,Delgado:2015kxa,Corbett:2015lfa,Salas-Bernardez:2022xk} so we will not delve any longer on the issue here. The point is that the failure (or not) of unitarity is not really about the theory, whether SMEFT, HEFT or another, but about the way to treat it to obtain observables. This is an ancient observation dating at least to the Effective Range Expansion~\cite{Bethe:1949yr} that needs to be discussed more often in  the context of high-energy physics.

On the other hand, causality does impose limits on the parameters of an effective Lagrangian, though they have not been very thoroughly studied and perhaps we will attempt this in future work. These come about because a scattered wave packet in the forward direction cannot precede the incoming wave packet (though this is possible at wide angles~\cite{Llanes-Estrada:2019ktp}). Perturbatively, Wigner's bound for the derivative of the phase shift of any partial wave $\delta_J$ respect to the centre-of-mass three-momentum $k$, in terms of the scatterer's radius $R$ is a well-known low-energy result~\cite{Pelaez:1996wk}, 
\begin{equation}
\frac{d\delta_{J}}{dk} \geq -R \,.
\end{equation}
However, what should be used
for $R$ in a relativistic scattering theory is less well understood.
Such a set of bounds on the scattering matrix (one for each of its partial wave projections) yields one-sided bounds on the $a_i$ coefficients.
Employing unitarized methods one can immediately set constraints by demanding that no poles of the amplitude lay on the first Riemann sheet~\cite{Espriu:2014jya,Delgado:2015kxa} of $s$, which also violate causality. But these poles typically fall in regions where the uncertainties of the unitarized amplitude~\cite{Salas-Bernardez:2020hua} are large. In all, we think that this deserves a separate investigation, as it is not clear that it affects our main thrust of distinguishing SMEFT from HEFT.

\section{SMEFT-induced bounds on HEFT parameters}\label{sc:correlationsexplicit}
In this section we will use the data of ATLAS and CMS experiments for setting bounds on HEFT parameters through the correlations induced by assuming SMEFT's structure.

\subsection{SMEFT bounds on the \texorpdfstring{$\mF(h)$}{F(h)} coefficients}

\begin{figure}[!ht]
\centering
\makebox[\textwidth][c]{\includegraphics[width=120mm]{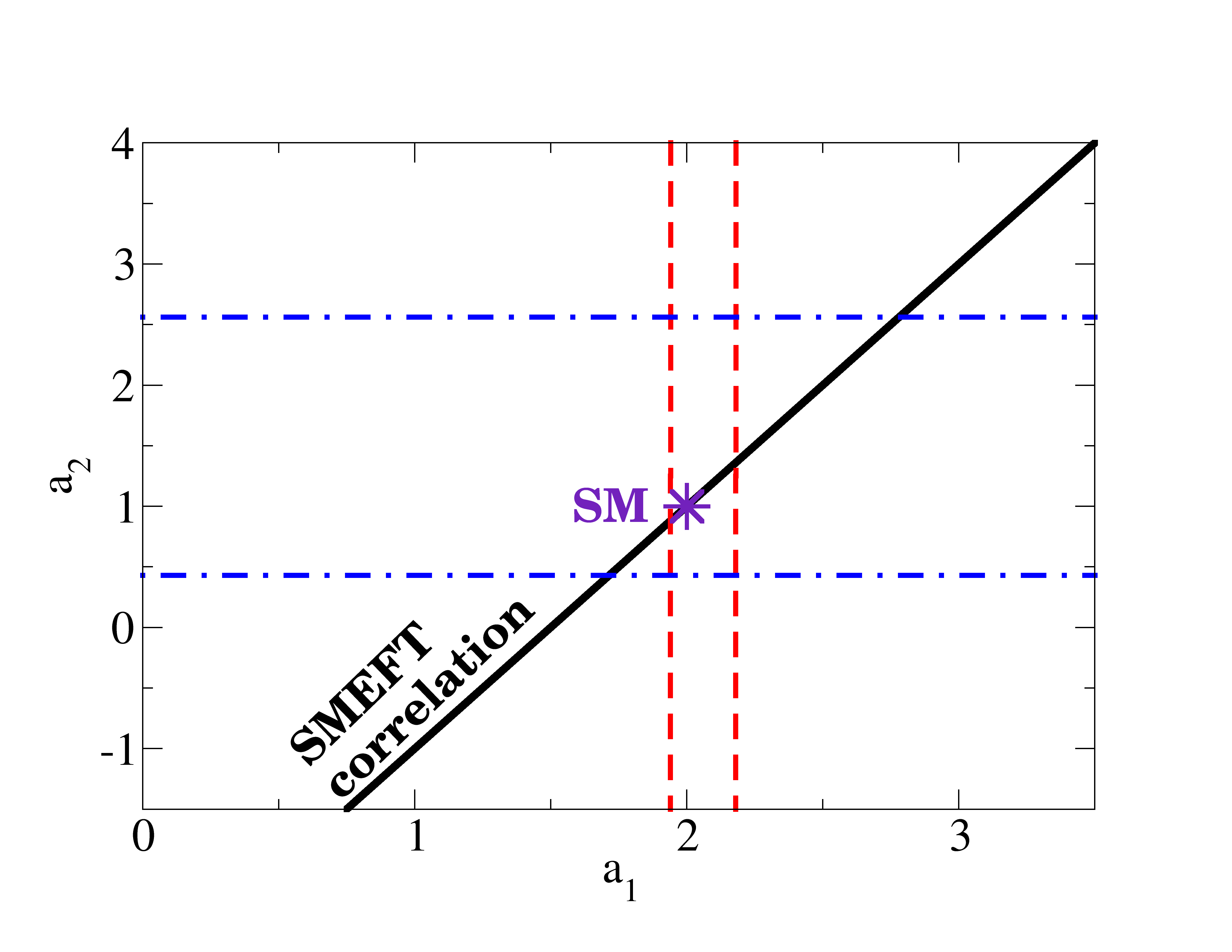}}
\caption[SMEFT correlation on HEFT parameters at order $1/\Lambda^2$]{SMEFT at order $1/\Lambda^2$ predicts the correlation $a_2=2a_1-3$ from the first column in Table~\ref{tab:correlations}, which is plotted against the current 95\% confidence intervals for these two HEFT parameters~\cite{ATLAS:2020qdt,CMS:2022cpr}.}\label{SOFP}
\end{figure}
\subsubsection{From one and two Higgs production}

Knowledge of the $a_i$ coefficients is rapidly evolving, as they directly correspond to the $\kappa_i$ scaling cross sections respect to the Standard Model ones.
A data-driven constraint for $a_1$  based on LHC run-I data can be found in~\cite{Brivio:2016fzo}; at 2$\sigma$, those authors conclude that $a\in [0.7,1.23]$.
A bound on $b$ was originally  obtained by examining the absence of a resonance in $W_L W_L$ scattering below 700 GeV \cite{Delgado:2014dxa} (according to \cite{Salas-Bernardez:2020hua}, presented in the next chapter, the dispersive methods used for obtaining these bounds have a 10-20\% uncertainty on the position of the resonance).
Direct ATLAS and CMS work has improved those earlier limits, and
the latest bounds on the first two $a_i$ coefficients are discussed next in subsection~\ref{subsec:numeritos}; those coefficients $a_1=2a$ and $a_2=b$ remain the only ones  with current experimental constraints.

In Figure \ref{SOFP}, a straight line shows the SMEFT correlation obtained in the first column of Table \ref{tab:correlations}. The rest of the plane corresponds at most to HEFT theory. The SM is the point in the center of the figure. Finally the 95$\%$ confidence bands for the $a_1$ and $a_2$ parameters are presented as dashed lines with the numbers taken from the caption of Table \ref{tab:correl-exp-bounds}.

\subsubsection{Multiple Higgs production} \label{subsec:numeritos}


We employ the correlations found earlier in Table~\ref{tab:correlations}, in conjunction with current direct experimental bounds on deviations of $a_1$ and $a_2$ from their Standard Model values, to propagate the information to other coefficients of $\mathcal{F}$ that are presently unconstrained {\it  provided SMEFT holds}.

These are then quoted in Table~\ref{tab:correl-exp-bounds}, an interesting new contribution of this chapter to the phenomenology of HEFT. If, for example, $a_3$ is measured to be different from zero, this would immediately establish new physics (which is known); but additionally, if it exceeds the bounds given in the table, it would mean that SMEFT correlations are being violated and the EFT has to be extended into HEFT.

\begin{table}[!t]
    \caption[ATLAS and CMS bounds on HEFT coefficients produced by SMEFT's assumption]{\small
    We input the 95\% confidence-level experimental bounds $a_1/2=a\in [0.97,1.09]$~\cite{ATLAS:2020qdt}  and, for the middle column,  $a_2=b=\kappa_{2V}\in[-0.43,2.56]$~\cite{ATLAS:2020jgy} (see the second erratum), by the ATLAS collaboration (top row) or the CMS collaboration (bottom row) interval of 
    $a_2=b=\kappa_{2V}\in[-0.1,2.2]$~\cite{CMS:2022cpr}.
    With them we have calculated and show here the expected corresponding 95\% CL intervals for several $W_LW_L\sim \omega\omega \to n h$ coupling, $a_n$, employing the relations of Table~\ref{tab:correlations}. Violations of the intervals in the first column would sow doubt on the SMEFT adequacy at $\mathcal{O}(\Lambda^{-2})$; surpassing any in the third column, on its perturbativity; and those of the middle column would void SMEFT of much significance as an EFT. They can be further tightened with improved experimental data for $\kappa_{2V}$.} 
    \label{tab:correl-exp-bounds}  
    \centering   
    \begin{tabular}{|c|c|c|}\hline
       \textbf{ Consistent SMEFT} &  \textbf{Consistent SMEFT} &  \textbf{Perturbativity of}\\
        \textbf{range at order} $\Lambda^{-2}$ & \textbf{range at order}  $\Lambda^{-4}$ & $\Lambda^{-4}$  \textbf{SMEFT} \\ \hline
        $ \Delta a_2\in [-0.12,0.36] $       
        & ATLAS  &  ATLAS 
        \\
        $a_3\in [-0.08,0.24] $  &  $a_3\in [-4.1,4.0]$  &  $a_3\in [-3.1,1.7]$
        \\  
        $a_4\in[-0.02,0.06]    $       & 
        $a_4\in [-4.2,3.9]$     & $a_4\in [-3.3,1.5]$ 
        \\
        $a_5=0$ &  $a_5 \in[-1.9,1.8]$ & $a_5 \in[-1.5,0.6]$ 
        \\
        $a_6=0$ &  $a_6=a_5 $ & $ a_6=a_5$
        \\ 
        \hline
& CMS & CMS  \\ 
& $a_3\in [-3.2,3.0]$
&$a_3\in [-3.1,1.7]$  \\
&$a_4\in [-3.3,3.0]$
&$a_4\in [-3.3,1.5]$   \\
&$a_5\in [-1.5,1.3]$
&$a_5\in [-1.5,0.6]$   \\
&$a_6=a_5$
&$a_6=a_5$   \\ \hline
\end{tabular}
\end{table}

\begin{figure}[ht!]
\centering
    \includegraphics[width=0.5\textwidth]{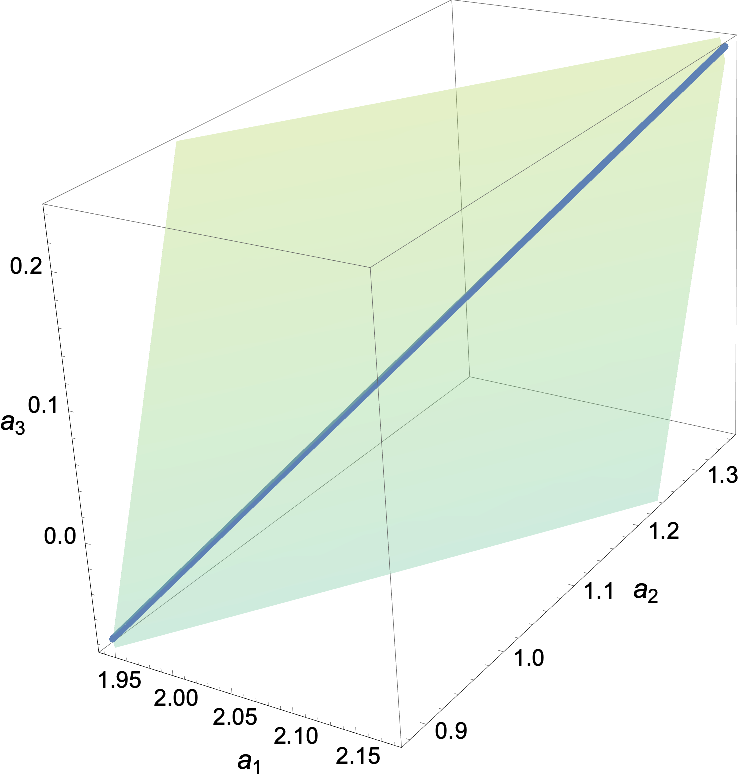}
    \caption[SMEFT correlation on HEFT parameters at order $\Lambda^{-4}$.]{ Correlations for the HEFT coefficients from Table~\ref{tab:correlations} that need to be satisfied for SMEFT to be a valid EFT of the electroweak symmetry breaking sector. The solid diagonal is the correlation  of order $\Lambda^{-2}$, that becomes broadened as the indicated band at order $\Lambda^{-4}$.}   \label{fig:aicorrelations}
\end{figure}

The constrains in the first column assume the validity of SMEFT up to order $1/\Lambda^2$, $\mO(\Lambda^{-2})$; because of  the tight experimental bounds on the $WW\to h$ coupling $a_1=2a$, the remaining $a_n$ couplings are strongly limited.
If SMEFT is considered up to $1/\Lambda^4$,  $\mO(\Lambda^{-4})$ (as we do in the second column),
the $WW\to hh$ coupling $a_2=b$ becomes independent, as seen in Table~\ref{tab:correlations} and Fig. \ref{fig:aicorrelations}; its experimental bounds are then also an input. Being poorly measured so far, it introduces a large uncertainty in the higher correlations. Thus, the bounds on the second column of Table~\ref{tab:correl-exp-bounds} are much looser.  
Those large uncertainties can be much reduced by improving the experimental knowledge of $a_2$: a decrease of its uncertainty by an order of magnitude scales almost linearly  and makes these errors roughly a factor $10$ smaller.

Notice that the values in third column in Table~\ref{tab:correl-exp-bounds} are identical within the precision quoted for ATLAS and CMS. The reason is that when the experimental uncertainty of $\Delta a_2$ is very large, at the practical level, its only limitation comes from the constraint $|\Delta a_2|\leq 5 |\Delta a_1|$, this is, $\mbox{min}(5a_{1}^{\rm-},-5a_1^{\rm +}) \leq \Delta a_2\leq \mbox{max}(-5a_{1}^{\rm -},5a_1^{\rm +})$. 
Since effectively the bounds just depend on the allowed values for $a_1$ we are obtaining the same outcomes for ATLAS and CMS in the third column.

\newpage
\subsection{Correlations and bounds in the Potential and Yukawa couplings}\label{coeffspotential}

As stated above, the SMEFT-induced correlations arise from the need for
consistency of the SMEFT formulation when a change of variable $h_{\rm HEFT}\to h_{\rm SMEFT}$ is performed. 
This change affects any other pieces of the Lagrangian involving the Higgs bosons, such as the Yukawa couplings to fermions, saliently the top quark, or the interactions among Higgs bosons themselves (both of which we examine here), as well as couplings to transversal gauge bosons (that we leave for future works). 

\begin{figure}[!ht]
    \centering
    \includegraphics[width=0.6\columnwidth]{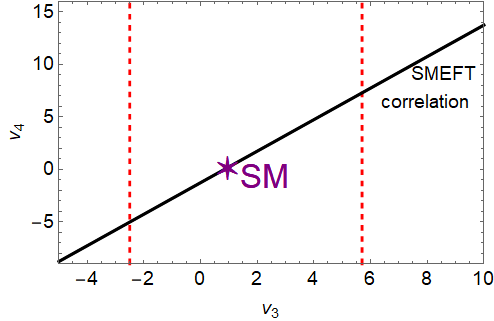}
    \caption[SMEFT correlations on the coefficients of the Higgs potential]{\small The correlation $v_4=\frac{3}{2} v_3 -\frac{5}{4} -\frac{1}{6}\Delta a_1$ that SMEFT predicts at $\mathcal{O}(1/\Lambda^2)$ is plotted making use of current 95\% confidence interval for $v_3\in[-2.5,5.7]$~\cite{ATLAS:2021jki}. The experimental $a_1$ uncertainty~\cite{ATLAS:2020qdt,CMS:2022cpr},  $a_1/2\in[0.97,1.09]$, is numerically negligible and allows to  predict a SMEFT band given by the solid black line. An experimental measurement for $v_4$ is still missing. }
    \label{fig:a1a22}
\end{figure}

The much discussed $V(H)$ Higgs-potential of eq. (\ref{eq:SMpotential11}), experimentally accessible at ``low'' $\sqrt{s}$ because it contains no derivative couplings, acquires in HEFT additional non-renormalizable couplings  organized in a power-series expansion
\begin{align}
    V_{\rm HEFT}= \frac{m_h^2 v^2}{2}  \Bigg[&   \left(\frac{h_{\rm HEFT}}{v}\right)^2 +  v_3   \left(\frac{h_{\rm HEFT}}{v}\right)^3+ v_4    \left(\frac{h_{\rm HEFT}}{v}\right)^4 + \dots \Bigg]\,,\label{eq:expandV}
\end{align}
with $v_3=1$, $v_4=1/4$ and $v_{n\geq 5}=0$ in the SM. 
Its coefficients also need to satisfy constraints that are exposed in Table~\ref{tab:corV} and Figure~\ref{fig:a1a22} if and when SMEFT applies. These correlations arise from the fact that the potential must only have even powers of $H$ (see \cite{Gomez-Ambrosio:2022why}).

\begin{figure}[!ht]
    \centering
    \includegraphics[width=0.6\columnwidth]{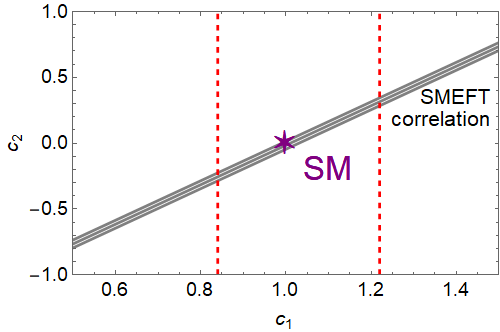}
    \caption[SMEFT correlations on the Yukawa couplings]{\small 
     The correlation , $c_2=\frac{3}{2} (c_1 -1) -\frac{1}{4}\Delta a_1$, that SMEFT predicts at $\mathcal{O}(1/\Lambda^2)$ is plotted making use of current 95\% confidence interval for the top Yukawa coupling $c_1\in[0.84,1.22]$\cite{deBlas:2018tjm} (dashed red lines). The experimental $a_1$ uncertainty~\cite{ATLAS:2020qdt,CMS:2022cpr},  $a_1/2\in[0.97,1.09]$ (at 95\% CL), is reflected in the width of the gray band with the SMEFT correlation. An experimental determination of the $ t\bar{t}\to hh$ coupling $c_2$ is still missing.   
   However, should it be measured, a test of SMEFT is possible by comparing to $c_2\in[-0.27,0.35]$  (from linearly adding uncertainties with the existing data and analyses).}
    \label{fig:c1c2}
\end{figure}

Similarly, the SM piece coupling the top quark, $\psi_t$, to the Higgs boson is extended in HEFT~\cite{Castillo:2016erh} by a multiplicative function $\mathcal{G}(h)$ 
\begin{equation}
\mathcal{L}_Y= -\mathcal{G}(h) m_t \bar{\psi}_t \psi_t 
\sqrt{1-\frac{\boldsymbol{\omega}^2}{v^2}}\; ,
\end{equation}
with a Taylor expansion around the physical $h=0$ vacuum given by
\begin{equation} \label{Yukawa}
    \mathcal{G}(h_{\rm HEFT})= 1 + c_1 \frac{h_{\rm HEFT}}{v} + c_2 \left( \frac{h_{\rm HEFT}}{v} \right)^2+\dots \end{equation}  
    (with $c_1=1$, $c_{i\geq 2}= 0$ 
in the Standard Model).
The correlations among these coefficients induced by SMEFT at order $1/\Lambda^2$  are then again given in Table~\ref{tab:corV} and Figure~\ref{fig:c1c2}. These come again from the fact that in SMEFT $\mathcal{G}$ must only contain odd powers of $H$.

\begin{table}[!ht]
\setlength{\arrayrulewidth}{0.3mm} 
\setlength{\tabcolsep}{0.2cm}  
\renewcommand{\arraystretch}{1.4} 
\caption[SMEFT correlations on other HEFT functions]{\small Correlations among the coefficients $\Delta v_3:=v_3-1$, $\Delta v_4:=v_4-1/4$, $v_5$ and $v_6$ of the HEFT Higgs potential expansion in eq.~(\ref{eq:expandV}) that need to hold, at $\mO(1/\Lambda^2)$, if SMEFT is a valid description of the electroweak sector.
Based  on the current bound $\Delta v_3\in[-2.5,5.7]$ in~\cite{ATLAS:2021jki}, $\mO(1/\Lambda^2)$ SMEFT predicts the coefficient intervals in the last column, testable in few-Higgs final states. A coupling 
$c_{H\Box}\neq 0$ induces the correction $\Delta a_1\propto c_{H\Box}$, nevertheless numerically negligible since experimental uncertainties from $v_3$ much exceed those of $a_1$. Likewise, we include the leading correlations for the Yukawa $\mathcal{G}(h)$ function of eq.~(\ref{Yukawa}), constraining $c_2$ and $c_3$ by $c_1$ and $a_1$ (from the correction to the value of the symmetric point $h_*$). We make use of current 95\% confidence interval for the top Yukawa coupling $c_1\in[0.84,1.22]$~\cite{deBlas:2018tjm}.}   
\label{tab:corV}
\begin{center}
 \begin{tabular}{|c|c|} \hline 
        $\Delta v_4=\frac{3}{2}\Delta v_3  -\frac{1}{6}\Delta a_1$ & $\Delta v_4\in[-3.8,8.6]$\\[2ex]
        $ v_5=6v_6=\frac{3}{4}\Delta v_3 -\frac{1}{8}\Delta a_1 $ 
        & $v_5=6 v_6 \in[-1.9,4.3]$ \\ \hline 
        $c_2
       = 3 c_3  
        =\frac{3}{2} (c_1 -1) -\frac{1}{4}\Delta a_1$ & $c_2
        = 3 c_3  
        \in[-0.27,0.35]$\\
         \hline    
    \end{tabular}
\end{center}  
\end{table}

In terms of specific operators in the SMEFT Warsaw basis, the potential $V(h_{\rm HEFT})$ is affected by both $\mO_{H\Box}$ and $\mathcal{O}_H$ of eq. (\ref{Warsawbasis}), so that in terms of their Wilson coefficients,
\begin{align}
v_3 =&1 +  \frac{3v^2 c_{H\Box}}{\Lambda^2} +\epsilon_{c_H}\;,
 &v_4= \frac{1}{4} +  \frac{25v^2c_{H\Box} }{6\Lambda^2} +\frac{3}{2}\epsilon_{c_H}&
\nonumber\\ 
v_5 =&    \frac{2v^2c_{H\Box} }{\Lambda^2} +\frac{3}{4}\epsilon_{c_H}\;,
 &v_6=  \frac{v^2c_{H\Box} }{3\Lambda^2} +\frac{1}{8}\epsilon_{c_H}\;,\;\;\;\;\;\;\;\;\; &v_{n\geq 7} = 0 \;,
\end{align}
with  $m_h^2=\, -\mu^2 \left(2+\frac{4 c_{H\Box}v^2}{\Lambda^2}+ \frac{6}{4}\epsilon_{c_H}\right)$, $\langle |H|^2\rangle =\frac{v^2}{2}= -\frac{\mu^2}{\lambda}\left(1-\frac{3}{4}\epsilon_{c_H}\right) $ and $\epsilon_{c_H}= \frac{\mu^2 c_H}{\lambda^2  \Lambda^2}$. 

Also, the $c_i$ in $\mathcal{G}(h_{\rm HEFT})$ modifying the Yukawa coupling receive analogous contributions from both SMEFT coefficients $c_{H\square}$ and $c_{uH}$ in standard notation, the second alternatively named $c_{tH^+}$ in~\cite{Brod:2022bww}. The correction
$$c_1 = 1 -\frac{v^3}{\sqrt{2}m_t} \frac{c_{tH^+}}{\Lambda^2} 
  +\frac{c_{H\square}v^2}{\Lambda^2}+\mO(1/\Lambda^4)  $$
can be carried on to the higher coefficients using the relations in Table~\ref{tab:corV} 
(with $\Delta a_1=2 c_{H\square} v^2/\Lambda^2+\mO(1/\Lambda^4)$.)

\section{\texorpdfstring{$ww\to n\times h$}{ww->nxh} for all \texorpdfstring{$n$}{n} in HEFT as the telltale process}
\label{nhproduction}

In this section we will indicate how to extract the coefficients of the flare function $\mathcal{F}$ in a process where $n$ Higgses are produced in the final state.

\begin{figure}[ht!]
\centering
     \begin{tikzpicture}[scale=1]
     \draw[decoration={aspect=0, segment length=1.8mm, amplitude=0.7mm,coil},decorate] (-1,1) -- (0,0)-- (-1,-1);
     \draw[] (0,0)-- (1,1);
     \draw[] (1.25,1) node {$h_1$};
          \draw[] (1.25,-1) node {$h_n$};
     \draw[] (0,0)-- (.98,0.6);
          \draw[] (1.25,0.6) node {$h_2$};
     \draw[] (.75,-.15) node {.};
     \draw[] (.75,0.) node {.};
     \draw[] (.75,-0.3) node {.};
     \draw[] (0,0)-- (1,-1);
       \draw[] (3,0) node {$\displaystyle{  \, =\,  -\frac{n!a_n}{2v^n} \,  s }$};  
\end{tikzpicture}\;\;\;\;\;\;\;\;\;\;\;\;
 \begin{tikzpicture}[scale=1]
     \draw[decoration={aspect=0, segment length=1.8mm, amplitude=0.7mm,coil},decorate] (-1,1) -- (0,.6) -- (0,-.6)-- (-1,-1);
     \draw[] (-0.0,0.6)-- (1,1);
     \draw[] (1.25,1) node {$h_1$};
          \draw[] (1.25,-1) node {$h_n$};
     \draw[] (-0.0,0.6)-- (.98,0.6);
          \draw[] (1.25,0.6) node {$h_2$};
     \draw[] (.75,-.15) node {.};
     \draw[] (.75,0.) node {.};
     \draw[] (.75,-0.3) node {.};
     \draw[] (-0.0,-0.66)-- (1,-1);
       \draw[] (3.25,0) node {$t/u$-channel type};  
\end{tikzpicture}
\caption[The flare function controls $\omega\omega\to n\times h$ processes]{\textbf{Left}: The $a_n$ coefficients of the flare function $\mathcal{F}$ control the contact piece of $\omega\omega\to n\times h$ processes. A large number $n$ of Higgs bosons in the final state would appear as a flare of them in the detector read out, whence the nickname of the function. \textbf{Right}: $t/u$-channel type diagrams also contribute to the  $\omega\omega\to n\times h$ process, they produce terms proportional to a product of $a_i$ coefficients with $i<n$ as explained in the main text.}
\end{figure}
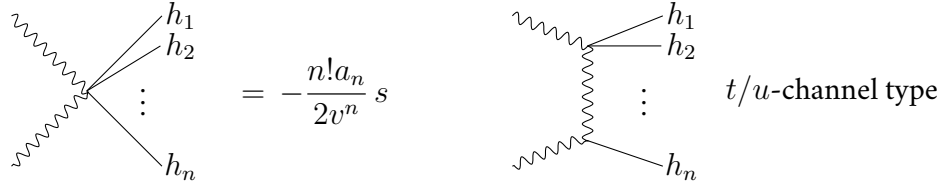

First we start by noticing that the measurement of the $\omega^+\omega^-\to h$ total cross section  gives us information the value of the first nontrivial coefficient of $\mF(h)$, $a_1=2a$. The value of $a$ is well constrained and hence we move on to identify the processes where the subsequent coefficients of the flare function can be measured.

Generalizing to $n>1$ Higgs bosons in the final state, the contributions to the amplitude will come from the contact diagram and the $t$-channel and $u$-channel diagrams. The contact diagram will give a contribution of $n! s a_n/(2v^n)$ whereas the $t/u$-channel will produce a string proportional to all the coefficients of $\mathcal{F}(h)$, $a_m$, for $1\leq m\leq n-1$. Therefore, for generic $n$, the amplitude will take the form
\begin{equation}
    T_{\omega\omega\to n \times h}=\frac{s}{v^n}\sum _{i=1}^{p(n)}\left(\psi_i(q_1,q_2,\{p_k\})\prod_{j=1}^{|\text{IP}[n]_{i}|}a_{\text{IP}[n]_{i}^j}\right)\;,
\end{equation}
where $\psi_i(q_1,q_2,\{p_k\})$ are functions depending on all four-momenta involved in the process (the two Goldstone bosons having momenta $q_1$ and $q_2$ and the $k$-th Higgs boson with momentum $p_k$) which will be made explicit below. These functions contribute to the angular integration used to obtain the total cross section of the process.  The symbol $\text{IP}[n]$ represents the integer partitions of $n$ and it is a collection of $p(n)$ vectors with length $|\text{IP}[n]_i|$ each, and components $\text{IP}[n]_i^j$. For example, for $n=4$
(see \cite{Gomez-Ambrosio:2022qsi}), $\text{IP}[4]=\{\{4\},\{3,1\},\{2,2\},\{2,1,1\},\{1,1,1,1\}\}$ and hence $p(4)=5$, and the lengths  $|\text{IP}[4]_i|=\{1,2,2,3,4\}$ and $\text{IP}[4]_2^1=3$. In that case the amplitude takes the form
\begin{align}
    T_{\omega\omega\to 4 \times h}=&\frac{s}{v^4}\Big(4!a_4+a_3a_1\psi_2(q_1,q_2,\{p_k\})+a_2^2\psi_3(q_1,q_2,\{p_k\})+\nonumber\\
    &+a_2 a_1^2 \psi_4(q_1,q_2,\{p_k\})+a_1^4\psi_5(q_1,q_2,\{p_k\})\Big)\;.
    \end{align}

The strategy is to fit to data each $a_n$ with increasing $n$ starting form the one-Higgs boson production, then fit two-Higgs boson production, etc. We have developed a small program for the computation of the amplitudes $T_{\omega\omega\to n\times h}$ that can be provided by the author on request. We present in the next subsection~\ref{subsec:amplitudes} the amplitudes for the production of one, two, three and four Higgs bosons.

\subsection{Amplitudes of \texorpdfstring{$\omega\omega \to n\times h$}{ww->nxh} with \texorpdfstring{$n=1,2,3,4$}{n=1,2,3,4}} 
\label{subsec:amplitudes}
Formally, the amplitude $\omega\omega \to h$ with the LO HEFT Lagrangian in eq.~(\ref{FbosonLagrangianLO}) is 
given by
\begin{equation} \label{amp:h}
T_{\omega\omega\to h}=-\frac{a_1 s}{2 v}\;.
\end{equation}
There is no on-shell cross section associated to this amplitude (because of the impossibility to satisfy four-momentum conservation with three on-shell massless particles). 
Therefore we move on and quote the amplitude with two Higgs bosons in the final state, that is simply~\cite{Delgado:2015kxa}
\begin{equation} \label{amp:2h}
 T_{\omega\omega\to hh} \,=\, \frac{s}{v^2}(a^2-b) \, = \frac{s}{v^2}\left(\frac{a_1^2}{4}-a_2\right).
\end{equation}

The tree-level amplitude with a larger number of Higgs bosons
can be obtained (by automated means); the one with
three Higgs bosons in the final state is,
\begin{align} \label{amp:3h}
    T&_{\omega\omega\to hhh}= \frac{s}{v^3}\left(-\frac{a_1^3}{2}+2a_1a_2-3a_3\right)\;.
\end{align}

The amplitude with four Higgs bosons in the final state, $T_{\omega\omega\to hhhh}$, is complicated enough that it is worth to look it up in our publication \cite{Gomez-Ambrosio:2022qsi}. We have not yet found a a compact form of it.

A check of these  amplitudes is to take the limit to the Standard Model
by setting the $a_i$  coefficients to the values 
$a_1=2$, $a_2=1$, $a_3=0$ and $a_4=0$.
Because the SM is renormalizable and unitary, these derivative terms must vanish, as indeed our computation reproduces,
having eq.~(\ref{amp:2h}) and eq.~(\ref{amp:3h}) above as well as the amplitude of four Higgs bosons satisfy
\begin{equation}
 T^{SM}_{\omega\omega\to hh} 
 =0\; ;\ \ \
 T^{SM}_{\omega\omega\to hhh}    
 =0\; ;\ \ \ T^{SM}_{\omega\omega\to hhhh} 
 =0\; ,
\end{equation}
where conservation of momentum has been used. 
 For a more detailed comparison, we refer to Ref.~\cite{Arganda:2018ftn}, which provides the physical $WW\to hh$ cross section in the SM, with $\sigma\sim 10^2$~pb (this cross section gets reduced to $\sigma\sim \mO(\mbox{fb})$ for the actual LHC process $pp\to hh$+2~jets containing this vector boson scattering).

\subsection{Cross-sections}

Equations~(\ref{amp:h})-(\ref{amp:3h}) and successive for an increasing number of Higgs bosons are what is needed for a phenomenological extraction of the $a_i$ coefficients in the TeV region.
From single Higgs production, through eq.~(\ref{amp:h}),
$a_1$ is already constrained (see subsection~\ref{subsec:numeritos}), so current work focuses on two-Higgs processes which allows to address $a_2=b$ in eq.~(\ref{amp:2h}). The $a_1$ appears squared (and is known to 10\% precision) and $a_2$ appears linearly, interference in this latter amplitude is possible and the sign of the deviations of $a_2$ from the SM value is at hand.

With $a_1$ and $a_2$ already constrained, it would become feasible to in turn constrain $a_3$ (null in the Standard Model) with eq.~(\ref{amp:3h}) and so forth for higher coefficients with higher-point processes with more bosons in the final state. Since each successive amplitude is linear in the highest appearing coefficient, their signs can be determined if a separation from the SM value is found.

An important correlation that allows to ascertain whether SMEFT is at play comes from the observation that at order $1/\Lambda^{2}$ all the deviations in $\mF$ from the SM in $a_1$ through $a_4$ stem from the same operator (see eq.~(\ref{coefsSMEFTHEFT})). Note also that the amplitudes in subsection~\ref{subsec:amplitudes} are the net deviations from the Standard Model in HEFT, since their SM prediction is zero. In fact, since the $\mathcal{O}_{H\Box}$ operator only has four boson legs, the amplitudes with more than two boson legs in the final state vanish at leading order in the SMEFT counting,
\begin{equation}
    T_{\omega\omega\to n\times h}=\mathcal{O}(\Lambda^{-4})\;\;\text{ for }\;\; n>2\;. 
\end{equation}
Hence, cross sections for these processes will be strongly suppressed by the new physics scale, and the non-zero contributions appear at dimension 8. On the other hand, in a general HEFT case the cross sections are non-zero: for three Higgses in the final state, for example, we have
\begin{equation}
\sigma_{\omega\omega\to hhh}=\frac{1}{s}{\left(\frac{a_1^3}{2}-2a_1a_2+3a_3\right)^2  \, 
\frac{4\pi^3}{3}\left(\frac{s}{16\pi^2 v^2}\right)^2
}   
\, .
\end{equation}
 This can be a way of distinguishing whether SMEFT is  applicable or not, from ``low'' energy data, without access to the underlying UV completion of any new physics. One can compare the predictions with experimental data, and see whether indeed there is no room for a discrepancy at order $1/\Lambda^2$. 

\begin{figure}[ht!]
    \centering
\includegraphics[width=0.7\textwidth]{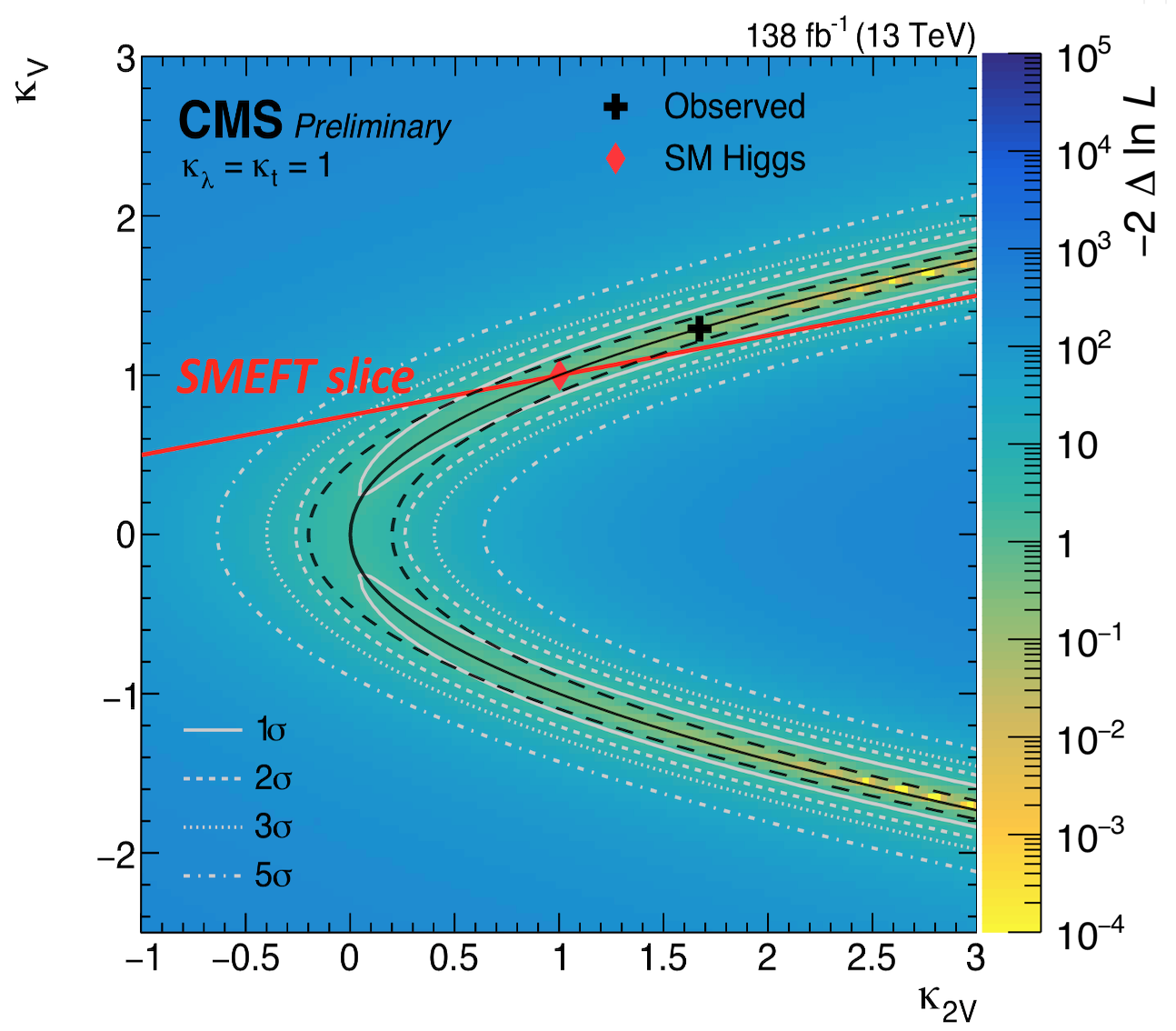}
    \caption[CMS sensitivity to the HEFT amplitudes]{{  CMS experimental confidence regions for the $hWW$ coupling $\kappa_V=a=a_1/2$ and for the $hhWW$ coupling $\kappa_{2V}=b=a_2$~\cite{CMS:2022nmn} (white lines and colour map).  
    We have superimposed the correlation between $a_1$ and $a_2$ in SMEFT at $\mathcal{O}(1/\Lambda^2)$ (red line). In addition, we have also plotted the parabolas $a_2-a_1^2/4 =0$~(solid black) and $a_2-a_1^2/4=\pm 0.2$~(dashed black), which determine the $WW\to hh$ scattering in the naive equivalence theorem. 
    }
}
    \label{fig:CMS1}  
\end{figure}

Finally, a comment on the use of GB amplitudes is due at this point: Scattering amplitudes based on Goldstone bosons can be related to scattering amplitudes of longitudinal gauge bosons at the expense of $\mO(m_W/\sqrt{s})$ corrections. At $WW$ center-of-mass energy of 800~GeV (challenging but not in the far future), this becomes a 1\% expansion parameter, of the size of $\alpha_{EM}$. It is therefore a very reasonable starting point for energies well enough over the threshold. However, the Goldstone-Higgs couplings discussed in the chapter ($a_1, a_2, a_3,...$) parameterize these processes in terms of the physical $W$’s as well. For this example, this happens with $\omega^+\omega^-\to hh$: $W^+W^- \to hh$ is also fully determined by $a_1=2a$ and $a_2=b$ at LO in HEFT~\cite{Arganda:2018ftn,Quezada-Calonge:2022lop}, even beyond the naive Equivalence Theorem~\cite{Veltman:1989ud,Dobado:1993dg}. Recently, the full process has been calculated at one-loop in \cite{Herrero:2021iqt,Herrero:2022krh}, and at NLO in \cite{Domenech:2022uud}.  Notice in Fig. \ref{fig:CMS1} how the CMS experiment is sensitive to the amplitude computed in the naive equivalence theorem when exclusively studying two-Higgs production.

Furthermore, a complete comparison with data needs to take the transverse gauge bosons into account. This is no problem, nevertheless, as HEFT’s leading order Lagrangian already incorporates all possible $W$ interactions, determined at that order by the same $a_1, a_2,...$.  
Still, the possibility of separating longitudinal and transverse $W$ polarizations in data analysis would improve the efficiency of theory-experiment comparisons (see f.e. \cite{Maina:2021xpe}) 
Thus, precise enough experimental data should be able to recover the Flare-function coefficients that set the amplitudes.

\section{Finding out whether the \texorpdfstring{$\mathcal{F}(h)$}{F(h)} function has a zero}
\label{Zeroessection}
\label{sec:zeroes}

Among the precise conditions that allow to express a HEFT as a SMEFT,  thoroughly studied in \cite{Cohen:2020xca}, the first necessary requirement among those spelled out in subsection~\ref{subsec:noSMEFT}
is the existence of an $O(4)$ symmetric point $h=h_\ast$. 
This requires a zero, that recalling the Taylor expansion in eq.~(\ref{expandF}) yields the relation
$$
\mathcal{F}(h_\ast)=1+\sum_{n=1}^{\infty}{a_n}\frac{h_\ast^n}{v^n}=0\;.
$$

In this section we will try to address what can be done, empirically and assuming that any UV physics is not known or understood (bottom-up approach) to improve the knowledge of whether such zero $h_\ast$ could be present.

\subsection{When Schwarz's Lemma guarantees a function's zero }
\label{subsec:Schwarz}
In this subsection we examine and adapt a known result from complex-variable analysis that guarantees the existence of a zero of a complex function: in the case of $\mathcal{F}(h)$ this would be an $O(4)$ fixed point candidate around which SMEFT could be built. 

The information that we would eventually need to have at hand to exploit the theorem would be a number of coefficients of the Taylor series,
depending on any future accelerators energy reach (subsection~\ref{nhproduction}). 
To avoid too large a mathematical digression, Schwarz's Lemma and two of its corollaries are detailed in Appendix \ref{app:Lemma}. What can guarantee a zero of $\mathcal{F}$ is the second corollary.
The needed hypotheses are as follows:
\begin{itemize}
    \item{} First, the function $\mathcal F(h)$ (extended to be a complex function of a complex $h$ argument, \underline{in units of $v$} throughout this whole section) needs to be analytic inside a disc of radius $|h|=R$ around the vacuum $h=0$. This disk has to be large enough to reach the possible symmetric point  (\textit{i.e.}, $h=h_\ast$ or, in SMEFT, $|H|=0$) from the observed vacuum ({\it i.e.}, $\langle |H|\rangle=1/\sqrt{2}$, or in HEFT  $\langle h\rangle=0$), where one constructs the flare function $\mF(h)$.    
    
    \item{} Second, the image of that disc (the set of  possible values of $\mathcal{F}$) has to be contained inside another disc of radius $M$ (the maximum value of $|\mathcal{F}|$) centered at $\mathcal{F}(0)=1$. Finally, the derivative of the function is assumed to have been measured, so that  $\mathcal{F}'(0)=a_1$ is known.

\end{itemize}

The second corollary then guarantees  that a disc of radius $R_{\rm min }:= R^2a_1^2/
\gamma M$ centered at $\mathcal{F}=1$ is completely contained in the image of $\mathcal{F}$. Here $\gamma$ is 
\begin{equation}
    \gamma=\frac{(\sqrt{2}+2)(\sqrt2+1)}{\sqrt{2}}\simeq 5.83\;.
\end{equation}

Therefore, a zero of $\mathcal{F}(h)$ is ensured if that radius $R_{\rm min}$ is greater than one (so that $\mathcal{F}=0$ can be reached from $\mathcal{F}=1$),
\begin{equation} \label{conditionfor0}
R^2a_1^2/\gamma M > 1 \implies \exists h^\ast\ | \
\mathcal{F}(h^\ast)=0\ .
\end{equation}

Depending on how large the $a_i$ coefficients end up being, this lemma could provide a tool to extract a scale at which one is sure that there exists an $O(4)$ fixed point candidate. 

To use that second corollary in Appendix \ref{app:Lemma}, notice that by construction we have that ${\mathcal F}(0)=1$ and hence we can employ the auxiliary $g(h)\equiv \mathcal{F}(h)-1$ satisfying $g(0)=0$ and $g'(0)=\mathcal{F}'(0)=a_1$, which is the $g$ to which the corollary applies. This means that, if the function $\mathcal{F}(h)$ is analytic in the open disc of radius $R$, denoted as {$D(0,R)$}, then we will have that the condition for the existence of at least one (complex)
value of the Higgs field $h^\ast \in D(0,R)$ such that $\mathcal{F}(h^\ast)=0$ is (see eq.~(\ref{2nddisk}) below)
\begin{equation} R^2 > \frac{\gamma M}{a_1^2}
\end{equation}
where $M$ is the maximum value that $|\mathcal{F}(h)|$ takes for $h\in D(0,R)$. For clarification see Figure \ref{fig:possiblezero}. Regrettably, the application of the lemma will give a definite positive answer to the existence of a zero if $M(R)$ is at most $R^2$, which means that we can only profit from the lemma for polynomials of order up to two (due to analyticity). This still leaves room for some cases that we explore below, saliently including the variations around the SM that are conceivable in the near future, with $\mathcal{F}$ up to order 4.

\begin{figure}[!t]
    \centering
    \begin{tikzpicture}[scale=1]
 \draw[dotted,fill=orange,opacity=0.5] (2,0) arc (0:360:2);
       \draw (1,2.2) node {$h\in \mathbb{C}$};
       
       \draw[blue] (.8,.9) node {$R$};
    \draw [very thin] (-2.3,0) -- (2.3,0);
    \draw [very thin] (0,2.3) -- (0,-2.3);
     \draw [color=blue,<->] (0,0) -- (1,1.7);

 \draw [color=blue,<->] (0,0) -- (1,1.7);
 
 \draw [thick, ->] (2.1,1) .. controls (2.5,1.2) and (3.6,1.2) .. (4,1);
 \draw (3,1.5) node {$\mathcal{F}(h)$};
\end{tikzpicture}
    \begin{tikzpicture}[scale=1]
 \draw[dotted,fill=cyan,opacity=0.1] (2.7,0) arc (0:360:1.7);
       \draw (-1,2.2) node {$\mathcal{F}(h)\in \mathbb{C}$};
       \draw (1,-.2) node {1};
       \draw[thick,red] (0,0) node {$\times$};
       \draw[thick,red] (0,0.2) node {\tiny possible fixed point};
       \draw[blue] (1.95,.8) node {$\frac{R^2c_1^2}{\gamma M}$};
    \draw [very thin] (-1.8,0) -- (2.3,0);
    \draw [very thin] (0,2.3) -- (0,-2.3);
     \draw [color=blue,<->] (1,0) -- (1.75,1.5);
      \draw[fill=blue,opacity=0.1] plot [smooth cycle] coordinates {(4,-1) (0.2,-2) (-.6,-1) (-1,0) (-.5,2) (2,2)};
      
      \draw [color=black,<->,opacity=0.6] (1.05,-0.05) -- (4,-1);
       \draw[black,opacity=0.6] (2.2,-.8) node {$M$};
      \draw[blue,opacity=0.4] (1,1.95) node  {$F(D(0,R))$};
\end{tikzpicture}
    \caption[Depiction of Scharz's lemma]{{\small \textbf{Left-hand side:} the  disc of radius $R$, $D(0,R)$ (orange), is the region where the Taylor approximation of the $\mathcal{F}(h)$ function is supposedly trusted, which can only be experimentally assessed. \textbf{Right-hand side:} the grayish outer region is the image of $D(0,R)$, namely $\mathcal{F}(D(0,R))$, and $M$ is the maximum distance of $\mathcal{F}(D(0,R))$ to 1 (thus, the maximum value of $|\mathcal{F}-1|$ for $|h|\leq R$). 
    Under the conditions of applicability for Schwarz's lemma, we can assure that the disc on the right (bluish), $D(1,\frac{R^2c_1^2}{\gamma M})$, is contained in $\mathcal{F}(D(0,R))$, {\it i.e.} $D(1,\frac{R^2a_1^2}{\gamma M})\subset  \mathcal{F}(D(0,R))$.\\
    When it happens that $\frac{R^2c_1^2}{\gamma M}>1$, the radius of the disc in the image $\mathcal{F}$ plane around $\mathcal{F}(h=0)=1$, we are assured that $\mathcal{F}(h)$ has a zero for some $h^\ast\in D(0,R)$.}}
    \label{fig:possiblezero}
\end{figure}
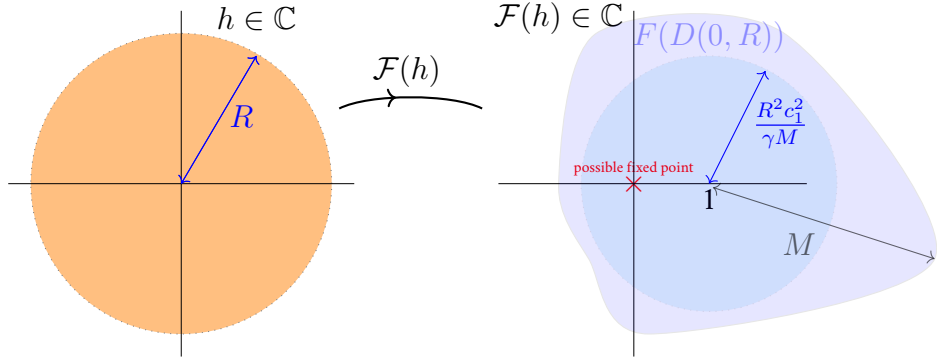$\,$
Experimentally, the full $\mathcal{F}(h)$ can not be measured. It is only its Taylor expansion that can be accessed in practice (unless the SM UV-completion is directly observed, of course). Hence, we must follow the logic:
\begin{enumerate}
\item First we must assume that $\mathcal{F}(h)$ is analytic in a neighborhood of the $h=0$ physical vacuum (and hence its Taylor expansion, and thus HEFT, makes sense). This region can be taken as the open disk $D(0,R)$.
\item Suppose we measure $k$ coefficients 
of the Taylor expansion of the function $\mathcal{F}(h)$ such that, for $h$ in units of $v$,
$$\mathcal{F}(h)=1+\sum_{i=1}^{k}a_i h^i+R_k(h)$$
where of course we trust the expansion up to an energy scale such that we know that, for $h\in D(0,R)$, $\mathcal{F}(h)$ is analytic~\footnote{The difficulty here is that, unlike in analyticity in Mandelstam  $s$ that ultimately follows from causality via Titchmarsh's theorem~\cite{Llanes-Estrada:2019ktp}, it is hard to find a guiding principle in $h$-- space that justifies assuming analyticity. At least we are exposing the necessary hypothesis, that is often taken for granted when writing down a SMEFT.}.
Here $R_k(h)$ is the remainder of the Taylor Expansion. We must neglect this Taylor remainder since, although it is bounded by $\{\max_{|h|=R} |\mathcal{F}(h)|\}\beta^{k+1}/(1-\beta)$ for $\beta\in[|h|/R,1)$, it cannot be experimentally accessed.
\item  Assign $$M=\max_{|h|=R} |\sum_{i=1}^{k}a_i h^i|.$$
The zeroth order coefficient is omitted because we must use the maximum of $g(h)\equiv \mathcal{F}(h)-1$, as described in the appendix.
Notice that the maximum modulus of $\sum_{i=1}^{k}a_i h^i_1$ is reached at the boundary of its domain thanks to the \textit{Maximum modulus principle}.
\item Using the second corollary we will have that we can assure the presence of an $O(4)$ fixed point if we reach a field intensity such that
\begin{equation}\label{criterionSchwarz}
{|h|^2}=R^2>  \frac{\gamma M}{a_1^2}
\end{equation}
\end{enumerate}

\paragraph{Standard Model case}
This discussion has been quite abstract, so let us try to apply eq.~(\ref{conditionfor0}) in practice. The first obvious example is the Standard Model. 

We can  apply Schwarz's lemma  to either the $F(h)$ function, in the SM $F(h)=1+{h}$, or its square $\mathcal{F}$, the flare function. In the first case we see that $F(h)$ is analytic for all $h\in\mathbb{C}$, and hence we can take $R$ as big as we want. It is immediate to see that $M=R$ so that we can assure the presence of a zero of $F(h)$ whenever 
\begin{equation}
R^2 >  {\gamma}R\;,
\end{equation}
which can be met for $R=|h|>\gamma$ (in units of $v$).

If instead we apply Schwarz's lemma directly to the flare function $\mathcal{F}(h)=F^2(h)$, we find no useful information, as can be understood from the result in the next example. \\

\paragraph{Generic second-order polynomial}
Taking $\mathcal{F}(h)=1+a_1 {h}+a_2 {h^2}$, the condition to assure the presence of a fixed-point candidate becomes
\begin{equation}
    R^2>\frac{\gamma}{a_1^2}(|a_1|R+|a_2|R^2)\geq\frac{\gamma}{a_1^2}{{\max_{|h|=R}}\Big(\Big|a_1 {h}+a_2{h^2}\Big|\Big)}
\end{equation}
So that for $R$ sufficiently large, the condition will be met if $1>\frac{\gamma|a_2|}{a_1^2}$, i.e. $a_1^2/a_2>\gamma$ assures the presence of a fixed-point candidate.  For the known $a_1=2$ central value we obtain that
\begin{equation}
    \frac{4}{\gamma}>|a_2|\;\;\Rightarrow \;\;\text{if }a_2\in (-0.68,0.68) \;\Rightarrow\;\text{  zero of } \mathcal{F} \text{  assured}
\end{equation}
This result is in agreement with the condition of positivity on the discriminant of the polynomial which gives $\frac{a_1^2}{4}>a_2$ (SMEFT region in Fig. \ref{SOFP}) and hence guarantees a zero of $\mathcal{F}(h^\ast)=0$ for $h^\ast\in\mathbb{R}$.

Comparing to the interval for $a_2$ given by experiment and quoted in Table~\ref{tab:correl-exp-bounds}, we see that for negative $a_2$, the experimental bound is already inside the Schwarz's lemma limit; if the upper experimental limit also drops into the 0.68 boundary (which is not unthinkable, only a factor 3 better than the current LHC extraction), then Schwarz's lemma will tell us that a zero of $\mathcal{F}$ is at hand unless new discoveries of higher $a_i$ coefficients require further scrutiny.
Because a third-order polynomial always has a real zero, this takes us to a fourth order one, discussed in the next paragraph.

\paragraph{Perfect-square, fourth-order polynomial} 
Taking now a quadratic 
$F(h)=1+\alpha{h}+\beta {h^2}$ 
entails a quartic flare function 
\begin{equation}
    \mathcal{F}(h)=1+2\alpha  {h}+(\alpha^2+2\beta) {h^2}+2\alpha\beta {h^3}+\beta^2{h^4}\ .
\end{equation} 
In this case, the condition of eq.~(\ref{criterionSchwarz}) that guarantees the presence of a symmetric point candidate $h_\ast$ becomes
$$\frac{\alpha^2}{\gamma}>\beta\,.$$
Squaring the above relation, for the central value $a_1=2$, we get the bound on the fourth order coefficient
$$\frac{a_1^4}{4 \gamma^2}>a_4\;\;\Rightarrow \;\;\text{if }a_4\in (-0.118,0.118) \;\Rightarrow\;\text{  zero of } \mathcal{F} \text{  assured}.$$
Notice of course that if $a_4$ is measured to be negative, higher order terms will be needed in the expansion of $\mathcal{F}$ (see Subsection \ref{subsec:positivity}) to guarantee its positivity.

\section{Far future: multiple Higgs production in extreme-\texorpdfstring{$T$}{T} collisions}
\label{sec:farfuture}

The pion was first discovered in 1947~\cite{Perkins:1947mf,Lattes:1947mw} when precious few events from cosmic rays were obtained in photographic emulsions taken at high altitudes; nowadays, they are routinely produced by the thousands per event in central heavy-ion collisions at the LHC~\cite{ALICE:2021hkc}. Whereas currently multiple Higgs-boson events (or for that matter, multiple longitudinal gauge-boson ones) are not possible, one day they might come within reach. 
At that point, the entire $\mathcal{F}(h)$ function (or at least, to a very large order in the Taylor expansion) may become part of potential observables.
We wish to illustrate the possibility of accessing it with such future work in this subsection.

The idea of a large number of Higgs bosons has been put forward before~\cite{Khoze:2017tjt,Khoze:2018mey,Khoze:2018qhz} although in a different context, in a proposal to solve the hierarchy problem. Here, we notice that the appearance of $\mathcal{F}(h)$ in eq.~(\ref{FbosonLagrangianLO}) makes it that, in a thermal medium with temperatures of order the hundreds of GeV (over two orders of magnitude beyond what is possible today, but not an arbitrarily large scale, and within the validity of the EFT), the process $X\to n\times h + m\times W_L/Z_L$ with a thermal distribution becomes possible. In the next lines we propose a very schematic analysis chain that proceeds according to the following flow diagram:

 \begin{center}
\setlength{\arrayrulewidth}{0.3mm} 
\setlength{\tabcolsep}{0.1cm}  
\renewcommand{\arraystretch}{1.} 
 \boxed{
 \begin{tabular}{cccccccccccc}
         &          &                     &           &                 &         &Measure  & & 
 Use                & & Compare & \\
 Measure &          & Fit to it           &            &Obtain &         &volume $V$           & &  
 $V$, $T$,          &  & to \\ 
 $p_T$   &$\implies$& $T$ and             & $\implies$& $\mathcal{F}(h_1)$ &$\implies$ & using HBT     & $\implies$         & $E_{\boldsymbol{k}}$& $\implies$ & measured \\
         &          & $E_{\boldsymbol{k}}$&           &    from $E_{\boldsymbol{k}}$           &        &  interf. &  &
 to predict         & & $N$ \\
         &          &                      &            &              &         & & &
 $N$                & & as check 
 \end{tabular}}
 \end{center}
 \vspace{0.3cm}

\paragraph{Transverse-momentum distribution}
As pioneered by Hagedorn~\cite{Hagedorn:1966mqw,Hagedorn:1965st}, an observable revealing the statistical distribution is the {$p_T$-distribution of} the bosons produced. 

In the case of a free-boson gas with Lagrangian $\frac{1}{2}\left( (\partial h)^2 + (\partial \omega_i)^2)\right)$ this is given~\cite{Gupta:2020naz} by 
\begin{equation} \label{free_dist}
\left. \frac{d^2 N}{2\pi p_T dp_T d\eta} \right|_0 =
m_T \frac{gV}{(2\pi)^3} e^{\frac{\mu-E}{T}}\ .
\end{equation}
While $\mu$ is the chemical potential associated to a conserved particle number (which can be left out if events with different number of bosons are considered), $V$ is the volume of the source and $g=1$, 3 or 4 depending on what is measured ($h$, $V_L$ or both), we want to call attention to the transverse mass-like quantity $m_T= \sqrt{m^2+p_T^2}\simeq p_T$. 

Integrating eq.~(\ref{free_dist}) over the longitudinal momentum (or rapidity) then yields a typical $p_T$ distribution
\begin{equation} \label{freeptspectrum}
    f_{\rm Bose}(p_T)dp_T=  {\rm constant}\ \times \ p_T dp_T
    \int_0^\infty dp_x \frac{1}{e^{\frac{1}{T} \sqrt{p_x^2+m_T^2}}-1}
\end{equation}

In the simplest free gas described by eq.~(\ref{free_dist}), $f_{\rm Bose}(p_T)$ falls off as a simple exponential. This Boltzmann-like dependence is obtained from mean occupation numbers
\begin{equation}
    \bar{n}_{\alpha}(k)= X_{\alpha}(k) \frac{\partial \log Z}{\partial X_{\alpha}(k)}
\end{equation}
with $X_{\alpha}(k)= exp(-E_\alpha(k)/T)$ and the partition function expressed~\cite{Hagedorn:1966mqw} as
\begin{equation}
    Z = \sum_n \prod_{\alpha k} X_{\alpha}(k)^{\bar{n}_\alpha(k)}\ .
\end{equation}
Here $k=0,1,2,\dots \infty$ as corresponds to a boson occupation number. The momentum distribution can then be obtained from the density of states 
$\frac{V}{2\pi^2}d^3p$.

The $p_T$ dependence of eq.~(\ref{freeptspectrum}) is modified in the interacting theory: this is what gives access to the function $\mathcal{F}(h)$. 

In case there is an interacting Hamiltonian, containing the  $\mF(h)$ function, 
it is possible, through the statistical distribution of bosons
to access it almost completely or at least to a very high degree in the $a_i$ expansion.  Such statistical distribution~\cite{Fetter} (see eq. (26.6) in page 251 there, though in a nonrelativistic treatment) will amount to
\begin{equation} \label{ptspectrum}
    \frac{dN}{d^3 \boldsymbol{k}}=\frac{-gV}{(2\pi)^3} T\sum_{n\in\mathbb{Z}}\frac{1}{i\omega_n-\sqrt{E_{\boldsymbol{k}}^2+\Sigma(\boldsymbol k,i\omega_n)}}
\end{equation}
where $E_{\boldsymbol{k}}=\sqrt{\boldsymbol{k}^2+m^2}$, the summation is carried over Matsubara frequencies $\omega_n=2\pi n T$,  and
$\Sigma(\boldsymbol k,k_0)$ is the self energy of the (Higgs or Goldstone) bosons defined through, just as in the DSE of chapter 2,
\begin{equation} \label{selfE}
    [G(\boldsymbol{k},k_0)]^{-1}=\left[ [G^0(\boldsymbol k,k_0)]^{-1} +\Sigma(\boldsymbol k,k_0)\right]^{-1}\ .
\end{equation}

The propagator $G(\boldsymbol{k},i\omega_n)$ (see \cite{Pawlowski:2015mia,Pawlowski:2017gxj} for a detailed discussion on analytic continuation of propagators) can be computed from the euclidean partition function, that has a path-integral representation,
\begin{align}
    Z=&\int \mathcal{D} h\mathcal{D}\boldsymbol{\omega}\exp{\Bigg\{-\int_0^\beta d\tau   \int d^3{x} \left[ \frac{1}{2}\mathcal{F}(h)
\partial_\mu\omega^i\partial_\mu\omega^j\left(\delta_{ij}+\frac{\omega^i\omega^j}{v^2-\boldsymbol{\omega}^2}\right)
+\frac{1}{2}\partial_\mu h\partial_\mu h  \right] \Bigg\}} \label{HEFTZ}
\end{align}
where summation in the Euclidean $\mu$ indices is assumed. We then  directly see how the $\mathcal{F}(h)$ function affects the statistical distribution of bosons through their self energy. The coordinate-space representation of this propagator for Higgs bosons will simply amount to
\begin{align}
    G(\boldsymbol{x},\tau)=&\int \mathcal{D} h\mathcal{D}\boldsymbol{\omega}\;h(\boldsymbol{x},\tau)h(\boldsymbol{0},0)\nonumber\times\\
    &\times\exp{\Bigg\{-\int_0^\beta d\tau   \int d^3{x}\left[ \frac{1}{2}\mathcal{F}(h)
\partial_\mu\omega^i\partial_\mu\omega^j\left(\delta_{ij}+\frac{\omega^i\omega^j}{v^2-\boldsymbol{\omega}^2}\right)
+\frac{1}{2}\partial_\mu h\partial_\mu h  \right] \Bigg\}}\;.
\end{align}
Once this path integration has been estimated on the lattice or by other means, the self energy from eq.~(\ref{selfE}) can be extracted, and substituting it into eq.~(\ref{ptspectrum}), a $p_T$ spectrum directly comparable with experiment can be obtained as a functional of $\mF$.

\paragraph{Number of Higgs bosons}

Additionally, we can try to get an idea of what is the number of Higgs bosons that should be produced in an experiment in order to access the SM $O(4)$ symmetric point $h_\ast=-v$. The SM Higgs potential is
\begin{equation}
    V_{SM}(\phi)=\frac{1}{2}\mu^2\boldsymbol\phi\cdot \boldsymbol{\phi}+\frac{1}{4}\lambda(\boldsymbol\phi\cdot \boldsymbol\phi)^2
\end{equation}
where $-\mu^2,\lambda>0$ and $\boldsymbol\phi\cdot \boldsymbol\phi=\phi_1^2+\phi_2^2+\phi_3^2+\phi_4^2$. Choosing the unitary gauge, the SM vacuum sits at $\phi_1=\phi_2=\phi_3=0$ and $\phi_4=v$ and $\phi=\phi_4=\sqrt{-\mu^2/\lambda}$. After redefining $\phi_4=v+h$ the physical Higgs mass amounts to $m_h=\sqrt{2\lambda v^2}$, which using $m_h=125.3\;\text{GeV}$ and $v=246\;\text{GeV}$ gives $\lambda\simeq 0.13$.

The invariant point under $O(4)$ in field space is the origin $\phi=0$. The difference of potential energy density between the SM vacuum and the SM $O(4)$ invariant point is
$$\Delta V=V(0)-V(v)=\frac{1}{4}\lambda v^4\simeq1.19\cdot 10^8\; \text{GeV}^4=1.49\times 10^{10}\;\text{GeV}/\text{fm}^3. $$\par
Now we wish to translate this energy density into a temperature, for doing so we look for $T$ such that
\begin{equation}
    \varepsilon(T)\equiv \frac{1}{(2\pi)^3}\int d^3\mathbf{k}\frac{E_\mathbf{k}}{e^{E_\mathbf{k}/k_B T}-1}=\Delta V\;.
\end{equation}
where $E_\mathbf{k}=\sqrt{c^2\hbar^2\mathbf{k}^2+m^2c^4}$ is the relativistic energy of a boson with three-momentum $\mathbf{k}$.
This gives a temperature of 
$k_B T=140\;\text{GeV}$ (which matches the EWSB second order phase transition critical temperature~\cite{Ramsey-Musolf:2019lsf}). Using this temperature we are ready to compute the number density of Higgs bosons at a temperature where the SM symmetric point is reached
\begin{equation}
    n(T)=\frac{N}{V}= \frac{1}{(2\pi)^3}\int d^3\mathbf{k}\frac{1}{e^{E_\mathbf{k}/k_B T}-1}=258\;\text{Higgs\ bosons}/(0.1\text{fm})^3.
\end{equation}
This is certainly a daunting concentration of energy and particle number that is not expected in a foreseeable future. But when/if it is achieved, the absolute number can serve as cross-check of the $p_T$ spectrum line shape to extract a temperature (hopefully the same) if the volume of the hot source, addressed next, is known.

\paragraph{Obtaining the volume of a multi-Higgs source}
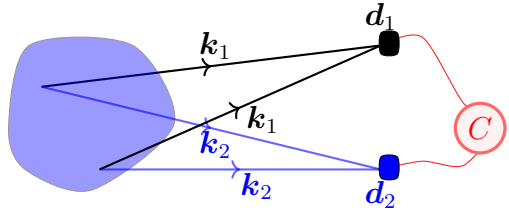
\begin{figure}[ht!]
\centering
   \begin{tikzpicture}[scale=1.1]
\draw[] (4,1.35) node  {$\boldsymbol{d}_1$};
\draw[] (2,1) node  {$\boldsymbol{k}_1$};
\draw[color=blue] (2,-.2) node  {$\boldsymbol{k}_2$};
\draw[] (2.57,0.1) node  {$\boldsymbol{k}_1$};
\draw[color=blue] (2.5,0.1-.8) node  {$\boldsymbol{k}_2$};
\draw[color=blue] (4,-.8) node  {$\boldsymbol{d}_2$};
\draw[color=red,opacity=0.9] plot [smooth] coordinates {(4,1) (4.5,1.1) (5,.3) (5.4,0)};
\draw[color=red,opacity=0.9] plot [smooth] coordinates {(4,-0.5) (4.5,1.1-1.5) (5,.3-.75) (5.4,0)};
     \draw [color=black,thick,->-] (-.1,0.5) -- (4,1);
      \draw[scale=0.5,fill=blue,opacity=0.4] plot [smooth cycle] coordinates {(3,.1) (1,-1.5) (-.6,-1) (-1,0) (-.5,2) (2,2)};
      
      \draw [color=blue,opacity=0.6,thick,->-] (0.6,-0.5) -- (4,-0.5);
      \draw [color=blue,opacity=0.6,thick,->-](-.1,0.5) -- (4,-0.5);
      
      \draw [color=black,thick,->-] (0.6,-0.5) -- (4,1);
 \draw[scale=0.5,fill=blue] plot [smooth cycle] coordinates {(8,-1.2) (8,.5-1.2) (8.4,.5-1.2) (8.4,-1.2)};
  \draw[scale=0.5,fill=black] plot [smooth cycle] coordinates {(8,+1.8) (8,.5+1.8) (8.4,.5+1.8) (8.4,+1.8)};
  \filldraw[color=red!60, fill=red!5, very thick] (5.2,0) circle (.3);
  \draw[color=red] (5.2,0) node  {$C$};
\end{tikzpicture}
    \caption[HBT/GGLP Interferometry]{HBT/GGLP Interferometry: Detecting two particles with momenta $\boldsymbol{k}_1$ and $\boldsymbol{k}_2$ at the respective  detection points $\boldsymbol{d}_1$ and $\boldsymbol{d}_2$ and studying their correlation gives information about the dimensions of the homogeneity region of source in bluish.}
    \label{fig:HBT}
\end{figure}
It remains to guess what would be the hot source volume that a
future multi-Higgs factory, ({\it i.e.} a collider capable of producing statistically significant numbers of Higgs bosons) could achieve. 
This type of machine would allow us to explore the existence of such a symmetric point by directly heating the electroweak sector to populate it, and to explore the properties of the EW phase transition. 

With the data in hand, the volume could
be obtained by using Hanbury-Brown-Twiss (HBT) or its particle physics analogue Goldhaber-Goldhaber-Lee-Pais (GGLP) interferometry: the Higgs bosons exiting the collision would retain memory (by interference) of the radius of the source that emitted them. The technique is routinely used in astrophysics to establish the size of astrophysical objects from the emitted photons, and in nuclear collisions by analyzing pions. That future electroweak collider could likewise obtain the radius of a hot electroweak ball from the Goldstone and Higgs bosons emitted.

Let us state schematically how this interferometry works (a comprehensive review can be found in \cite{Weiner:1999th}). 
Suppose that a source has emission points continuously distributed in a  space-time volume $V_4$ with an emission probability amplitude, $\Pi({r};\boldsymbol{k})$, of emitting a particle with momentum $\boldsymbol{k}$ (on the mass-shell with plane-wave wave-function $\psi_{\boldsymbol{k}}(\boldsymbol{r})\propto e^{i\boldsymbol{k}\cdot \boldsymbol{r}}$) at the space-time point $r$. Hence, the total probability of observing the emission of one particle with momentum $\boldsymbol k$ from the source is $P(\boldsymbol{k})=\int_{V_4} d^4{r} | \Pi({r};\boldsymbol{k})|^2$.\\
Likewise, the total probability of measuring two particles with momenta $\boldsymbol{k}_1$ and $\boldsymbol{k}_2$, assuming the two emissions are uncorrelated, \textit{i.e.} $\Pi({r}_1,{r}_2;\boldsymbol{k}_1,\boldsymbol{k}_2)=\Pi({r}_1;\boldsymbol{k}_1)\Pi({r}_2;\boldsymbol{k}_2)$, amounts to
\begin{align}
P(\boldsymbol{k}_1,\boldsymbol{k}_2)=&\int_{V_4} d^4{r}_1d^4{r}_2 \left|
\frac{\psi_{\boldsymbol{k}_1}({ \boldsymbol{r}}_1)\psi_{{\boldsymbol{k}}_2}({\boldsymbol{r}}_2)+\psi_{{\boldsymbol{k}}_1}({\boldsymbol{r}}_2)\psi_{{\boldsymbol{k}}_2}({\boldsymbol{r}}_1)}{\sqrt{2}}
\right|^2 |\Pi({r_1};\boldsymbol{k}_1)|^2 |\Pi({r_2};\boldsymbol{k}_2)|^2\nonumber\\
=&P(\boldsymbol{k}_1)P(\boldsymbol{k}_2)+\int_{V_4} d^4{r}_1d^4{r}_2\cos{\big[(\boldsymbol{r}_1-\boldsymbol{r}_2)\cdot(\boldsymbol{k}_1-\boldsymbol{k}_2)\big]}|\Pi({r_1};\boldsymbol{k}_1)|^2 |\Pi({r_2};\boldsymbol{k}_2)|^2\;.
\end{align}

The GGLP experiment could be adapted to measuring, at two detection points $\boldsymbol{d}_1$ and $\boldsymbol{d}_2$, two Higgs bosons with precise momentum $\boldsymbol{k}_1$ and $\boldsymbol{k}_2$ respectively (see Fig. \ref{fig:HBT}). The correlation function among the two momenta is
\begin{equation}
C(\boldsymbol{k}_1,\boldsymbol{k}_2):=\frac{P(\boldsymbol{k}_1,\boldsymbol{k}_2)}{P(\boldsymbol{k}_1)P(\boldsymbol{k}_2)}=1+\frac{\int_{V_4} d^4{r}_1d^4{r}_2\cos{\big[(\boldsymbol{r}_1-\boldsymbol{r}_2)\cdot(\boldsymbol{k}_1-\boldsymbol{k}_2)\big]}|\Pi({r_1};\boldsymbol{k}_1)|^2 |\Pi({r_2};\boldsymbol{k}_2)|^2}{P(\boldsymbol{k}_1)P(\boldsymbol{k}_2)}.\label{eq:correlationHBT}
\end{equation}
Under the assumptions explained thoroughly in \cite{Lisa:2005dd} (neglection of higher order symmetrization, smoothness and equal time approximations, useful for large
 (RHIC-like) sources) the correlation function in eq. (\ref{eq:correlationHBT}) simplifies to 
\begin{align}
    C(\boldsymbol{k}_1,\boldsymbol{k}_2)-1=&\int d^3\boldsymbol{r}'\mathcal{S}_{\boldsymbol{K}}(\boldsymbol{r}')\cos{\big[(\boldsymbol{r}_1-\boldsymbol{r}_2)\cdot(\boldsymbol{k}_1-\boldsymbol{k}_2)\big]}\nonumber\\
    \mathcal{S}_{\boldsymbol{K}}(\boldsymbol{r}'):=&\frac{\int_{V_4} d^4{r}_1d^4{r}_2|\Pi({r_1};\boldsymbol{k}_1)|^2 |\Pi({r_2};\boldsymbol{k}_2)|^2\delta(\boldsymbol{r}'-\boldsymbol{r}_1+\boldsymbol{r}_2)}{P(\boldsymbol{k}_1)P(\boldsymbol{k}_2)},
\end{align}
where $\boldsymbol{K}=\boldsymbol{k}_2+\boldsymbol{k}_1$ is the total momentum of the pair of outgoing particles. The function $\mathcal{S}_{\boldsymbol{K}}(\boldsymbol{r}')$ encodes \textit{``the distribution of relative positions of particles with identical velocities and total momentum} $\boldsymbol{K}$'' \cite{Lisa:2005dd} and it gives information about the size of the region of homogeneity of a source (\textit{i.e.} the region where the equilibrium assumptions can be taken). The curvature of $C(\boldsymbol{k}_1,\boldsymbol{k}_2)$ at $\boldsymbol q:=\boldsymbol{k}_1-\boldsymbol{k}_2=0$ is related to the mean-square separation of the three-dimensional quadrupolar moments \cite{Lisa:2005dd}
\begin{equation}
    -\frac{C(\boldsymbol{k}_1,\boldsymbol{k}_2)}{dq_i dq_j}\Bigg|_{\boldsymbol q=0}=\int d^3\boldsymbol{r}S_{\boldsymbol{K}}(\boldsymbol{r})r_i r_j\;.
\end{equation}
In this way we can obtain the volume of the region where the source can be considered homogeneous and the equilibrium conditions apply.

\section{Summary}\label{sec:conclusions}
In this chapter we have bridged between the SMEFT and HEFT formalisms inspired by the work 
of other groups \cite{Alonso:2015fsp,Alonso:2016btr,Alonso:2016oah,Cohen:2020xca,Cohen:2021ucp}. 
We have focused on the Higgs-flare function $\mathcal{F}$ that controls the derivative couplings of two Goldstone bosons $\omega_i$ to any number of Higgs bosons. 
We have exhaustively studied this flare function $\mathcal{F}$ and particularly addressed the existence of its key zero at a symmetric point in the $(\omega_i,h)$ field space. 
In what follows we summarise the novel results presented here.

Saliently, we have provided a simple derivation of the previously known conditions on the flare function $\mathcal{F}(h)$ and its derivatives around the EW symmetric point necessary to deploy SMEFT. By comparing the Taylor expansions between the symmetric point and our vacuum, the correlations induced by SMEFT’s validity on HEFT coefficients are exposed at fixed order in the EFT expansion. We believe that this will extend the clarity of  the criteria for SMEFT to exist, so far mostly discussed in geometric terms.

We extend previous results concerning the expression of a few coefficients of this function in terms of the $c_{H\Box}$ Wilson coefficient of SMEFT 
 in section~\ref{sec:convert};  
in this chapter we have addressed a larger number of such coefficients, we have employed the Warsaw basis, in contrast with earlier analyses, and we have proceeded to the next  order ($1/\Lambda^4$) in the SMEFT expansion. 
We have identified the relevant TeV-scale SMEFT operators at dimensions 6 and 8; we have then employed the correlations that we here report, together with ATLAS and CMS constraints in order to identify the HEFT coefficient space where SMEFT at either dimension 6 or dimension 8 can be deployed.

With the latest ATLAS and CMS bounds on the $a_1$ (also known as $2\kappa_{V}$) and $a_2$ (known as $\kappa_{2V}$) coefficients we have explicitly given 95\% confidence intervals for a few $a_i$, $i>2$ ones, that if exceeded would automatically rule out SMEFT, at least to the orders here considered, and point out to the need of extending the SMEFT framework.

Further, we have completely eliminated the Wilson coefficients and obtained correlations that are solely expressed in terms of the HEFT parameters 
and can be used to falsify SMEFT itself from within the wrapping theory 
(sections~\ref{sec:geometry} and~\ref{sec:generic-properties}).  
These correlations provide simple tests that analysts following upcoming and future experimental data  can employ to test the framework of SMEFT. It may be useful for those analysis to have explicit expressions of the $\omega\omega\to n h$ amplitudes in HEFT and therefore we explicitly provide 
in section~\ref{subsec:amplitudes} 
the leading order in perturbation theory of those with lowest Higgs number  $n=1,2,3$. An automated program applicable to generic $n$ can be provided on demand. We have also noticed that, to leading order in the SMEFT counting and in the TeV-scale, amplitudes with $n>2$ Higgses in the final state vanish.

Finally, a few additional original contributions of this work are listed here:  
 section~\ref{sec:generic-properties} provides a thorough study of the flare function $\mathcal{F}(h)$ and its properties such as positivity from the boundedness-from-below of the HEFT Hamiltonian; 
 section~\ref{sec:zeroes} studies what can be said about a possible zero of this function combining Schwarz’s lemmas of complex analysis and the current knowledge of the first two coefficients; 
 and section~\ref{sec:farfuture} discusses how far-future colliders could access the full function by producing Higgs bosons at finite temperature. While $\omega\omega\to n h$ processes allows one to access $\mathcal{F}$ order by order, a future collider that could substantially increase the temperature of the collision environment would open the entire function via the $p_T$-spectrum of the emitted Higgs bosons.

In section~\ref{coeffspotential}  we have given similar correlations that we have extracted among the coefficients of the $V(h)$ nonderivative Higgs potential and the Yukawa function $\mathcal{G}(h)$. This is attractive because it does not require the Equivalence Theorem (the process $\omega\omega\to nh$ needs to be extracted from $W_LW_L\to nh$ data and corrections are needed at low-energy, while $V(h)$ does not involve the Goldstone bosons) and is already accessible at LHC energies. Interestingly, it is affected by the properties of $\mathcal{F}$ since it is this function which controls the change of variable between HEFT and SMEFT, $h_1\to h$.

\setcounter{chapter}{4}
\setcounter{section}{0}
\setcounter{equation}{0}
\setcounter{equation}{0}
\setcounter{table}{0}
\setcounter{figure}{0}
\chapter{Unitarization Methods for QCD and the Electroweak Sector}\label{ch:IAM}

Experimental data \cite{ParticleDataGroup:2022pth} support the fact that any new physics scale may be beyond the reach of present colliders and no new particles are directly produced. Assuming there is a new physics scale not too much higher up in energy, the effects of the new physics can be felt through the coefficients of an Effective Field Theory based on the Standard Model particles. As introduced in the previous chapter, these coefficients, typically multiplied by powers of the momentum that grow with the reaction energy, eventually entail unitarity violations as is well known from hadron physics. Chiral perturbation (ChPT) theory offers there a model-independent characterization of $\pi$, $K$ and $\eta$ interactions in the shape of an expansion on meson masses and momenta due to derivative couplings in the effective Lagrangian (introduced in subsection \ref{subsec:CHPT}). The use of ChPT is restricted to energy ranges of about 200 MeV above the first production threshold. Perturbation theory based on that Effective Theory then fails, again due to fast unitarity violations (see Fig. \ref{fig:amps} below).
These are well known to appear in  the resonant $J=0$,1 $\pi\pi$ phase shifts~\cite{Ananthanarayan:2000ht,Garcia-Martin:2011iqs}, with the low-energy  scalar resonance $f_0(500)$ around 500~MeV (the threshold being nearby at 280 MeV); 
in $\eta\to3\pi$ decays~\cite{Gan:2020aco}; in $\gamma\gamma\to \pi^0\pi^0$~\cite{Oller:2007sh}, etc.

Unitarization techniques such as the Inverse Amplitude Method (IAM) reviewed in this chapter allow the reliable computation of amplitudes at higher energies, at least up to and including the first resonance in each channel, (although often more resonances can also be generated like the $f_0(500)$ and $f_0(980)$ in the isoscalar scalar mesonic sector \cite{Pelaez:2004xp,Oller:1997ng,Oller:1997ti}.) The same method has been deployed for NLO Higgs Effective Field Theory (HEFT) of eq. (\ref{FbosonLagrangianLO}) in much recent work \cite{Delgado:2014dxa}.

More generically, a widely used strategy in hadron physics~\cite{Yao:2020bxx} is to construct ``unitarized" amplitudes that extrapolate to those higher energy regimes satisfying unitarity exactly (for elastic processes). Popular ones are the K-matrix method, that is already being incorporated into the Monte Carlo simulations of high-energy processes at the LHC; the $N/D$ method \cite{Chew:1960iv,Oller:1998zr}; or variations of the Bethe-Salpeter equation \cite{Oller:1997ti,Nieves:1999bx}. The approach has not yet been widely adopted by experimental collaborations, but the lack of a yet higher energy collider means extrapolation of electroweak results by means of unitarity remains on the table. Among the problems to circumvent~\cite{Delgado:2019ucx}
 is the fact that unitarity is best expressed in terms of partial waves, while the simulation chain of high-energy experiments is based on Feynman diagrams. \newpage

The most powerful unitarization methods are based on dispersion relations, incorporating known analyticity properties of scattering amplitudes. These methods solidly extrapolate the low-energy theory to the resonance region;
a noteworthy approach among them, which has been broadly used, is the Inverse Amplitude Method~\footnote{Truong has provided an interesting historic perspective of early developments~\cite{Truong:2010wa}. For a recent review devoted to the unitarization techniques in general the reader can consult \cite{Oller:2020guq}.} (IAM) \cite{Truong:1988zp,Dobado:1992ha,Dobado:1989qm,Dobado:1996ps}.

In these dispersive approaches, shortly described in section~\ref{sec:disp}, the Effective Theory is used to fix the subtraction constants of the dispersion relation, and because this relation can incorporate in principle all the model-independent information that first principles impose on the amplitude, the information contained in those 
low-energy coefficients is maximally exploited, generating an energy dependence valid at much higher values of the energy than originally expected. 

For example, one can predict the mass, width, and couplings of the first resonance of the $W_L W_L$ scattering amplitude \cite{Dobado:2015goa} (a tell-tale of what Higgs mechanism is at work, as the equivalence theorem guarantees that the Goldstone boson scattering amplitude coincides with that for $W_L W_L$ at high energy), for each angular momentum, $J$, and weak isospin, $I$, channel.

A check on the reliability of the IAM has been carried out in \cite{Corbett:2015lfa} by eliminating a heavy scalar particle from the theory and trying to reconstruct its mass (successfully) from its imprint in the low-energy parameters (see also \cite{Dobado:1989gv}).
To date however, the uncertainties of the dispersive approaches have not systematically been laid out, so that for some colleagues there remain question marks about the reliability of the unitarization methods.  
A classic way of analyzing different unitarization methods is to use several of them for the same problem and with the same perturbative amplitude, to map out a reasonable spread of possible results~\cite{Delgado:2015kxa,Garcia-Garcia:2019oig}. Instead, we try to follow the strategy of trying to {\it a priori} constrain the uncertainty in the unitarization method.

It is the purpose of this work (see the original publication \cite{Salas-Bernardez:2020hua}) to start analyzing the systematic uncertainties of the Inverse Amplitude Method to put its eventual predictions (if ever  any separations from the SM couplings at the HL-LHC are discovered) on a firmer footing~\footnote{The reader may wonder why not adopt the more precise Roy equations: the reason is that they require abundant data over the resonance region and beyond, which does not currently look like a reasonable expectation at any new electroweak physics sector at the LHC.}. 

What we here undertake is to carefully review the derivation of the IAM method (section~\ref{sec:IAMder}) and to discuss the uncertainty which may be assigned to each of the approximations therein (section~\ref{sec:uncertainties}). 
 Section~\ref{sec:outlook} presents a minimal outlook and distills the main conclusions of the analysis in Table~\ref{tab:wrapup}.
A few more details and derivations are left for the appendices of this thesis.

\section{Unitarity and Dispersion Relations} \label{sec:disp}
\subsection{Reminder of unitarity for partial waves}
Conservation of wavefunction-probability in two-particle to two-particle scattering processes is expressed as a nonlinear integral relation for the scattering matrix  $T_I(s,t,u)$ (where $s$, $t$ and $u$ are the well known Mandelstam variables and $I$ the isospin index) of definite isospin (in hadron physics) or electroweak isospin (in HEFT).

If we decompose $T_I$ in terms of partial waves of angular momentum $J$, the expression of unitarity is much simpler, see eq.~(\ref{OpticalTh}). This decomposition reads
\be
T_I(s,t,u)=16\eta \pi\sum_{J=0}^\infty(2J+1)t_{IJ}(s)P_{J}(\cos\theta_s)
\ee
and converges for physical $s$ and scattering angle $x\equiv \cos \theta_s \in [-1,1]$; additionally, convergence succeeds over the Lehmann ellipse for unphysical $\cos \theta$ where the behavior of the Legendre polynomials $P_J$ is controllable \cite{Lehmann:1958ita}. 
If the scattering particles are identical, $\eta=2$, otherwise  $\eta=1$.  The explicit expression for the partial waves $t_{IJ}(s)$ is, inverting,
\be
t_{IJ}(s)=\frac{1}{32\pi\eta}\int_{-1}^{+1}dx\, P_{J}(x)T_I(s,t(x),u(x))\ .
\ee

\par
The analytic structure of the amplitude in terms of the complex variable $s$ in the physical or first Riemann sheet is so that $T_I(s,t,u)$, as well as $t_{IJ}(s)$, are real in a segment of the real axis and they develop cuts, (and eventually bound state poles, though not in the physical systems here considered). For identical particles of mass $m_\pi$, such as in $\pi\pi$ scattering or $W_LW_L$ scattering, one such cut develops from the production threshold $s_{\rm th}=4m_\pi^2$,   up to $+\infty$. (Sometimes the approximation $s_{\rm th}\simeq 0$ is adopted.)  This discontinuity is referred to as the right cut (RC). In the case that two pions appear also in the final state we will have that the partial waves inherit the same analytic structure as the amplitude, which, by crossing symmetry, develops a branch cut in the whole negative real axis which we call left cut (LC). The LC comes from the regions where physical particles are exchanged in the $t$- and $u$-channels and there is no possibility of avoiding the singularities of the amplitude (i.e. at the endpoints $x=-1,+1$). 
Inelastic channels, due to $2n$-pion states, $K\bar{K}$, $\eta\eta$, etc. in hadron physics, or $2n$-$w_i$, $hh$, etc. in the electroweak sector, would accrue additional cuts extending from
$(2n)^2m_\pi^2$, $4m_K^2,\,4m^2_\eta,\,...$ to $+\infty$, respectively. 

The unitarity of the $S$ matrix makes the partial waves, for physical $Re(s)>0$, obey the relation, akin to the optical theorem,
\be
\text{Im} \,t_{IJ}(s)=\sigma(s)|t_{IJ}(s)|^2\label{OpticalTh}\;.
\ee
which has the merit of being purely algebraic, though nonlinear in the partial waves.
Here $\sigma(s)=\sqrt{1-4m^2/s} $ is the phase-space factor (which equals one for massless incoming particles). 
Since eq.~(\ref{OpticalTh}) is nonlinear, it restricts the modulus of the amplitude to satisfy 
\be
|t_{IJ}(s)|\leq1/\sigma(s) \ ,  \label{unitarity}
\ee
for physical $s$ above threshold.

The key observation for the Inverse Amplitude Method is that the unitarity condition for purely elastic processes in (\ref{OpticalTh}) also fixes the inverse of the partial wave for physical values of $s$ above the first threshold at $s_{th}$ as
\begin{equation}
\text{Im}\frac{1}{t_{IJ}(s)}=-\sigma(s) \text{   for   } s>s_{\text{th}}\ .\label{unitarityinverse}
  \end{equation}
  
This is a remarkable exact statement about a non-perturbative amplitude inasmuch as inelastic channels such as $W_L  W_L\to W_L W_L W_L W_L$ for HEFT or $\pi\pi\to \pi\pi\pi\pi$ in hadron physics can be neglected and we will dedicate subsection~\ref{subsec:fourpi} to assess the uncertainty due to this hypothesis, which is exact below the four-particle threshold and deteriorates as energy sufficiently increases beyond that.
If the inelasticity stems from an additional two-body channel such as $K\bar{K}$ in hadron physics or $hh$ in the HEFT, the IAM requires a matrix extension. A brief recount is provided in subsection~\ref{subsec:KK} below.

The immediate question that comes to mind, given that the imaginary part of the inverse amplitude is known from kinematics alone, is how can one bring in the dynamical information to also obtain its real part. To this end we dedicate the next subsection~\ref{InverseAmp}.
\subsection{Exploiting the analytical properties of the inverse amplitude}
\label{InverseAmp}
For over a century, dispersion relations have been known to link the imaginary and real parts of ``causal'' functions satisfying Cauchy's theorem in appropriate complex-$E$ (here, $s$) plane regions.  
To exhibit and exploit the analytic structure of the inverse amplitude $1/t_{IJ}(s)$ we will deploy the appropriate dispersion relation (see {\it e.g.}~\cite{Hoferichter:2015hva} for a recent detailed introduction and the book \cite{Oller:2019rej} for a pedagogical account). 
The Cauchy theorem can be applied to any function $f(s)$ which is analytic in a complex-plane domain. 
  \begin{figure}[ht!]
  \centering
\includegraphics[width=0.6\columnwidth]{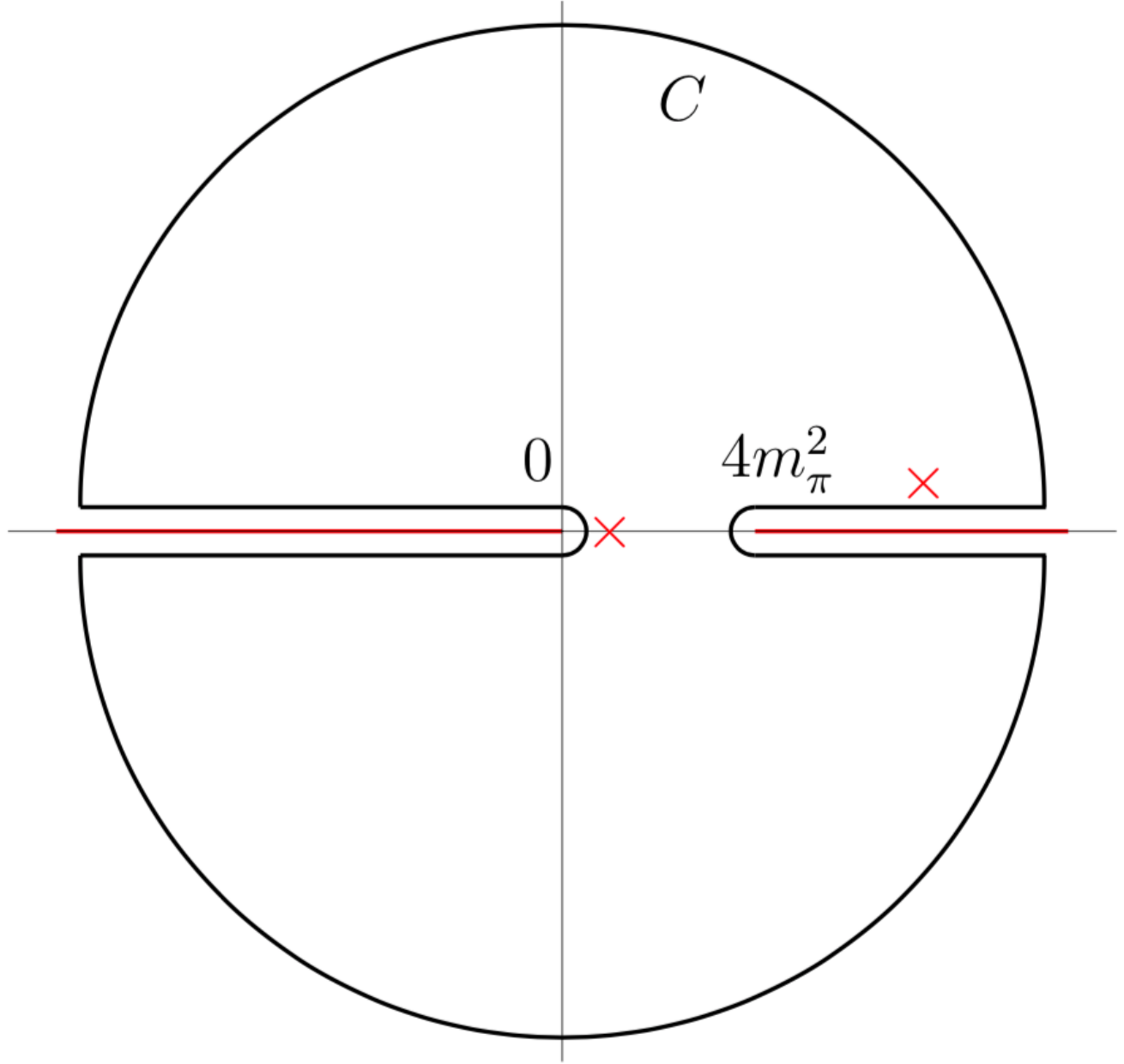}
\caption[Analytic structure of elastic scattering partial waves for pions]{Analytic structure of elastic scattering partial waves for pions and the contour $C$ in the complex-$s$ plane that will be used to write a dispersion relation for the inverse amplitude. The red lines represent the discontinuity cuts in the partial wave amplitude. The red crosses additionally represent the $n$-th order pole at $z=\epsilon$ and the simple pole at $z=s+i\epsilon$ (with $s>4m^2$) coming from the denominators in eq.~(\ref{I}).}\label{contour}
\end{figure}

Application of the theorem is convenient for the following integral,
\begin{equation}
I(s)\equiv \frac{1}{2\pi i}\int_C dz \frac{f(z)}{(z-\epsilon)^n (z-s)}\label{I}\;,
\end{equation}
where $f(z)$ is taken to have two branch cuts extending from $4m^2$ to $+\infty$ and from $0$ to $-\infty$ (just as the partial wave amplitude $t(s)$), and the contour of integration $C$ is taken as depicted in Fig. \ref{contour}.
In eq. (\ref{I}), $s$ is above the RC (i.e. $s$ stands for $s+i\epsilon$ with $s\geq4m^2_\pi$).\par
If the function $f$ is polynomially bounded as $|z|\to \infty$ (which, though not the case for partial wave amplitudes that can diverge exponentially for composite particles~\cite{Llanes-Estrada:2019ktp}, is satisfied for their inverse amplitude in eq.~(\ref{def:inverse}) below, that does fall as $e^{-az}$ with $a$ fixed)
we are able to neglect the contribution to eq. (\ref{I}) coming from the two large semi-circumferences in Fig. \ref{contour}. Due to Schwartz's reflection principle $f(s+i\epsilon)=f^\ast(s-i\epsilon)$, we are left in (\ref{I}) with the integrals of the imaginary part of $f$ over the LC and RC,
\begin{eqnarray}
I(s)=&&\frac{1}{\pi}\int_{-\infty}^0 ds'
\frac{\text{Im}f(s')}{(s'-\ep)^n (s'-s)}+\frac{1}{\pi}\int_{4m^2_\pi}^\infty dz \frac{\text{Im}f(s')}{(s'-\ep)^n (s'-s)}~,
\end{eqnarray}
where $\text{Im}f(s)$ is the imaginary part of $\lim_{\varepsilon\to 0^+} f(s+i\varepsilon)$ with $s\in$ RC or LC. 
 On the other hand, $I(s)$ equals the sum of its residues coming from the simple pole at $z=s$ and the $n$-th order pole at $z=\epsilon$, with $\ep\in(0,4m_\pi^2)$.  In this way we find the $n$-times subtracted dispersion relation for $f(s)$,
\begin{align}
f(s)=&\sum_{k=0}^{n-1}\frac{f^{(k)}(\ep)}{k!}(s-\ep)^k+\nonumber\\
&+\frac{(s-\ep)^n}{\pi}\int_{-\infty}^0 ds' \frac{\text{Im}f(s')}{(s'-\ep)^n (s'-s)}+\frac{(s-\ep)^n}{\pi}\int_{4m^2_\pi}^\infty ds' \frac{\text{Im}f(s')}{(s'-\ep)^n (s'-s)}
\end{align}
(which is valid safe at branch points where the multiple derivatives $f^{(k)}$ could fail to exist; this is generally of no concern).

\section{The Inverse Amplitude Method: derivation} \label{sec:IAMder}

In ChPT the partial wave amplitude $t_{IJ}$ accepts a Taylor-like expansion in powers of $s$ (modified by logarithms) for small real $s$ as (dropping the $IJ$ subindices) $t\simeq t_0+t_1+\mathcal{O}(s^3)$, where $t_0=a+bs$, and the leading behavior of each term in the series is $t_i\sim s^{i+1}$.
Work in the seventies revealed the appeal of writing down a dispersion relation for the inverse amplitude~\footnote{With equal right one could, instead of expanding $t\simeq t_0+t_1$, expand $t^{-1}\simeq \frac{1}{t_0+t_1}\simeq \frac{1}{t_0} - \frac{t_1}{t_0^2}$ whose inverse leads directly to eq.~(\ref{usualIAM}) below; but this expansion, while a direct mnemonic rule, is less conducive to an analysis of the uncertainties incurred, and does not expose the validity of the IAM in the complex $s$-plane as the dispersive derivation does, so we adopt the dispersive framework\cite{Pelaez:2015qba}.}   for pion-pion or electroweak Goldstone boson scattering, see~\cite{Dobado:1992ha,Truong:2010wa,Dobado:1989gr,Oller:2020guq,Dobado:1996ps}. 
It is customary since the last of these references to define the function, probably introduced by Lehmann~\cite{Lehmann:1972kv},
  \begin{equation}\label{def:inverse}
  G(s)\equiv \frac{t_0(s)^2}{t(s)}\ . 
  \end{equation}
  This function has the same analytic structure as $t_{IJ}$ except some additional poles coming from zeros of $t$. At low energies these zeros are known as Adler zeroes \cite{Adler:1964um} and will indeed appear in scalar waves. For this function we make a third order subtraction
  (the order being the minimum compatible with the order of the EFT to which we work, given that the perturbative amplitudes' leading growth is polynomial) so that the dispersion relation for $G$ reads~\footnote{{\it Strictu senso} we choose a subtraction point slightly separated from $s=0$ so that the factors become $(s-\epsilon)$ and $\frac{1}{z-\epsilon}$ to avoid the divergence at $z=0$ which is included in the interval of integration of the left cut. This plays little role in the derivation that follows.},
  \begin{align}
  G(s)=& G(\ep)+ G'(\ep) (s-\ep)+\frac{1}{2}G''(\ep)(s-\ep)^2+PC(G)+
   \nonumber\\ &+\frac{(s-\ep)^3}{\pi}\int_{LC}ds'\frac{\text{Im}\,G(s')}{(s'-\ep)^3(s'-s)}+\frac{(s-\ep)^3}{\pi}\int_{RC}ds'\frac{\text{Im}\,G(s')}{(s'-\ep)^3(s'-s)}
   \label{disprel}\; .
  \end{align}
  Proceeding backwards in this formula, we first encounter $PC(G)$, the contribution due to those Adler zeroes of $t$. 
   The standard IAM method at Next to Leading Order (NLO) neglects their contribution at this order in ChPT since their size on the physical-$s$ half axis counts as NNLO~\cite{GomezNicola:2007qj}. However, there is no special difficulty in including them: the uncertainty introduced by neglecting the Adler zeroes is ``\textit{much less than the uncertainties (mostly of systematic origin) of the existing data on meson-meson scattering}''~\cite{GomezNicola:2007qj} and that modified methods taking these into account differ in $\mathcal{O}(10^{-3})$ from the standard IAM in the physical region (see subsection~\ref{subsec:mIAM} where we delve on the uncertainty in neglecting these zeroes and its remedy). Putting those Adler zeroes (``Pole Contributions'' or $PC$) aside for now, we continue examining the remaining contributions to (\ref{disprel}).  
   
   Second, the right cut  (RC) of the integral  is treated exactly by the IAM as long as only the elastic two-body cut contributes, and this is the one that the basic IAM includes. Inelasticities can be due to both two and four-body additional channels. The two-body ones can be treated with a coupled-channel IAM \cite{Guerrero:1998ei,GomezNicola:2001as},    that has a less crisp theoretical basis: here we adopt the philosophy of staying within the one-channel IAM and use its coupled channel extension to estimate the uncertainty from omitting that channel, as long as this is sensible (see subsection~\ref{subsec:KK}). The four-body channel, as far as we know, is not tractable, so the IAM just leaves it out; but its unknown contribution can be controlled as it is short of phase space until quite high energies (see subsection~\ref{subsec:fourpi}). 
   
   Third, proceeding to the subtraction constants, we adopt three of them $G(\ep)$, $G'(\ep)$ and $G''(\ep)$ which suffices in hadron physics thanks to the saturation of the known Froissart-Martin bound~\cite{Froissart:1961ux,Martin:1962rt} controlling the growth of the physical cross section and the polynomial behavior of $t_0^2\propto s^2$.  
   In the NLO IAM, their values are taken from NLO ChPT,  a valid approximation because they are taken with $s$ around zero, where the EFT is valid (the uncertainty therein is discussed in subsection~\ref{subsec:pert}). In this NLO approximation we can safely set $\epsilon=0$ in the argument of the subtraction constants since $t_0$ and $t_1$ are essentially polynomials at such low $s$.
   This results in, 
  \begin{align} 
  \label{intermediateG}
  G(s)&\equiv\frac{t_0(s)^2}{t(s)}=t_0(s)-t_1(s)+\frac{s^3}{\pi}\int_{LC}ds'\frac{\text{Im}\,G(s')+\text{Im}\,t_1(s')}{s'^3(s'-s)}\;.
  \end{align}
  
In the fourth place and finally, this dispersion relation can be further simplified if we approximate the left cut contribution by taking the NLO chiral approximation of the discontinuity of $G$, $\text{Im}\,G\simeq-\text{Im}\,t_1$. Then the integral vanishes and a remarkable formula,
\begin{equation}
 t_{IAM}\equiv \frac{t_0^2}{(t_0-t_1)} \label{usualIAM}     \,,
 \end{equation}
is obtained: the usual IAM amplitude at NLO.
In this step of the derivation, an uncertainty is introduced upon approximating the left cut, which is further examined in subsection~\ref{subsec:LCbound} below. This is the most difficult one, as the left cut extends to $s=-\infty$ where the EFT is not valid: replacing $\text{Im}\,G$ by $-\text{Im}\,t_1$ is only sensible if the amplitude is wanted on the right-hand complex plane with $\text{Im}\,s>0$ where the influence of the left cut is smaller. It must be this larger distance between the left cut and the resonance region over right cut that binds the introduced uncertainty.  
This smaller contribution of the left-hand cut is readily observed when numerically comparing it to that of the right cut for the IAM amplitude itself at the $\rho$-resonance mass $m_\rho=770$ MeV,
\begin{equation}
\int_{-\Lambda^2}^{-\lambda^2}dz\frac{\text{Im } t_{IAM}(z)}{z^3(z-m_\rho^2)}\Big/\Big|\int_{4m_\pi^2}^{\Lambda^2}dz\frac{\text{Im } t_{IAM}(z)}{z^3(z-m_\rho^2)}\Big|\simeq0.5\%
\end{equation}
where we choose $\lambda=470$ MeV, i.e. the scale where ChPT is known to be reasonably accurate (see Fig. \ref{fig:amps}), and the cutoff as $\Lambda=20$ GeV (where the value of the integrals becomes independent of this cutoff). 
This very small value of the uncertainty is however obtained by assuming the IAM also along the left cut, which is an unwarranted use thereof: we will later strive to obtain {\it a priori} bounds that do not assume the IAM's validity there.\\

The central quantity for the discussion in this chapter is the relative separation between the approximate IAM amplitude $t_{IAM}$ and the exact one, $t$, which $t_{ IAM}$ approximates,
\be
\Delta(s) = \Big(\frac{t_{IAM}(s)-t(s)}{t_{IAM}(s)}\Big)\ .
\ee
Therein, the contribution due to approximating the LC follows from eq.~(\ref{intermediateG}) to be
\be
 \Delta(s) G(s)\equiv\Big(\frac{t_{IAM}-t}{t_{IAM}}\Big)\frac{t_0^2}{t}=\frac{s^3}{\pi}\int_{LC}ds'\frac{\text{Im}\,G+\text{Im}\,t_1}{{s'}^3(s'-s)}\label{IAMerr}\ .
\ee
Trying to set bounds on this integral is the goal of subsection \ref{subsec:LCbound}.
\par

When $t(s)$ is of slow variation, there is a chance that the relative uncertainty $\Delta(s)$ is numerically small. However, near a narrow resonance such as the $\rho$ or an equivalent $Z'$-like one in the electroweak sector, the amplitude is very sensitive to small changes of the pole position. There, $\Delta(m_\rho^2)$ can be of order 1, which is not very relevant: what is interesting then is to constrain the uncertainty incurred in computing that pole position, that is, the mass and width of the resonance.

We can discuss the position of a resonance in two ways. If it is isolated and narrow, a first approximation to $s_R$
is to use the saturation of unitarity  $|t(s_R)|=1/\sigma(s_R)$ over the real, physical $s$-axis.
In the IAM, this reduces to solving for $s_R$ the simple algebraic equation~\cite{Dobado:2019fxe}
\be \label{poleposition}
t_0(s_R)-t_1(s_R)+i\sigma(s_R) t_0^2(s_R)=0 \ ,
  \ee
which is equivalent to 
  \be \label{polepositionb}
t_0(s_R)-\text{Re}t_1(s_R)=0 \ ,
  \ee

The uncertainty introduced by this IAM approximation instead of employing the exact amplitude is equivalent to a nonvanishing quantity on the right hand side (RHS),
  \be
t_0(s_R)-t_1(s_R)+i\sigma(s_R) t_0^2(s_R)=-\Delta(s_R) G(s_R)\;.\label{corrpoleposition}
  \ee
  Where all the functions are evaluated over the real $s$ RC.\par
  We see therein that constraining the uncertainty of the amplitude ($\Delta$) also helps in constraining the uncertainty in the position of any resonance.

If one is willing to discuss $s_R$ with an imaginary part, an alternative starting point  could be the relation between an amplitude in the first and in the second Riemann sheets $t^{II}(s) = \frac{-t^I(s)}{1+2i{\sigma}t^I(s)}$ so that the complex position of the pole can be obtained from the amplitude in the first sheet by $t(s_R)
=\frac{i}{2{\sigma}}$. 
This yields a variant of eq.~(\ref{poleposition})
\be \label{poleposition_b}
t_0(s_R)-t_1(s_R)+2i {\sigma}(s_R) t_0^2(s_R)=0 \ ,
\ee
but now for $s_R$ complex (in this case we have to choose the determination $\sigma(s^\ast)=-\sigma(s)^\ast$), with the equivalent of  eq.~(\ref{corrpoleposition}) being
\be
t_0(s_R)-t_1(s_R)+2i{\sigma}(s_R) t_0^2(s_R)=-\Delta(s_R) G(s_R)\;\label{corrpoleposition_b}\ .
\ee
Either can be used to discuss new-physics resonances and, once one chosen, the relative uncertainties of one or the other method are very similar.

We will try to quantify the uncertainty in the position of a resonance following eq.~(\ref{corrpoleposition}); and we will exemplify with the vector-isovector resonance in hadron physics (where all quantities are known) except in subsections~\ref{subsec:KK} and~\ref {subsec:fourpi} where the mass versus masslessness of the pions/Goldstone bosons in ChPT/HEFT respectively, do make a difference due to phase space.

\section{Sources and estimates of uncertainty}\label{sec:uncertainties}
As we have shown in section~\ref{sec:IAMder},
the basic NLO-IAM treatment  disregards contributions from Adler zeroes, from higher orders in perturbation theory, and from inelastic channels; and it approximates the contribution of the left cut. We wish to compute how the mass (and, only briefly around figure~\ref{fig:movpolo} below, the width) of a higher-energy resonance depends on these contributions and eventually put bounds on them to have systematic uncertainties under control. If we recover the full expression for $G(s)$ taking into account inelasticities, Adler zeroes and the next order in PT for the subtraction constants, we have to add terms to eq.~(\ref{IAMerr}) to read
\begin{align}
    \Delta(s) G(s)=\frac{s^3}{\pi}\big(LC(G+t_1)\big)+3^{rd}{PT}+PC(G)+{\mathcal I_z},\label{alluncertainties}
\end{align}
where ${\mathcal I_z}$ takes into account the correction coming from inelasticities on the discontinuity of $t$ over the right cut (see the subsections \ref{subsec:KK} and \ref{subsec:fourpi} and specifically equation (\ref{inelasticim})), $LC$ represents the three-times-subtracted left-cut integral of the discontinuity in its argument, $3^{rd}PT$ takes into account the next order in PT correction to the displacement of the pole (see subsection \ref{subsec:pert})).
\subsection{Uncertainty when neglecting the poles of the inverse amplitude} \label{subsec:mIAM}

In deriving eq.~(\ref{usualIAM}) through the inverse amplitude $1/t$, the possible zeroes of the amplitude in eq.~(\ref{disprel}) were neglected. 
In chiral perturbation theory, such zeroes appear when one of the pions is taken with (near) zero mass and (near) zero energy and are referred to as ``Adler zeroes''~\cite{Adler:1964um}. 
Beyond these, one could also have extra zeroes which we distinctively refer to as  ``Castillejo-Dyson-Dalitz'' poles~\cite{Castillejo:1955ed}, and we treat both in the next two paragraphs.  Such a CDD pole can be seen as a zero of $t$ not predicted by a subtracted dispersion relation, given both discontinuities along the LHC and RHC and fixing the subtraction constants by some known values of $T$. That is, dispersion relations do not have unique solutions and the CDD poles reflect it~\cite{Castillejo:1955ed,Dyson:1957rgq}.  In this sense, we consider that indeed  an Adler zero is also a CDD pole, because it is LO in ChPT, while the discontinuities are at least NLO, unable to drive the correct energy dependence of the Adler zero. 


\subsubsection{Adler zeroes}

This first shortcoming of the Adler zeroes was addressed in~\cite{GomezNicola:2007qj} and demonstrated 
to be quantitatively small; moreover, it is a systematic uncertainty that can be disposed of if extremely high precision was needed. In view of the larger ones that follow, we believe that this is unnecessary and we will limit ourselves 
to discussing it in this subsection, ignoring it thereafter.

Near threshold, the amplitude accepts the chiral expansion in Effective Theory, 
$t\simeq t_0+t_1+t_2\dots \simeq a + b s + \varepsilon s^2 +\dots$ up to logarithms. 
$1/t$ does not exist when $t$ vanishes; at LO, this happens at the Adler zero of $t$ that lies at $s=-a/b$. 
Keeping higher orders, its position slightly shifts. The inverse amplitude develops a pole there, but its effect is tiny once $s$ becomes even marginally larger than this value. 
 The reason is that the dispersion relation is intelligently formulated in terms of $G=t_0^2/t$.

First, let us examine the chiral limit ($a=0$)  around which the effective theory is built. $G$ becomes a function of $s$ alone, independent of $m_\pi$ or, schematically,
$G\simeq b^2 s^2 /(s(b+\varepsilon s)) = b^2 s /(b+\varepsilon s)$ and the numerator's $s^2$ power has eliminated the zero of the denominator at threshold and there is no such pole. 

Outside the chiral limit ($a\propto m_\pi^2$) 
\begin{equation}
    G\simeq \frac{(a+bs)^2}{a+bs+\varepsilon s^2} \ .
\end{equation}
Once more, at LO, $G\simeq t_0^2/(t_0+t_1) \simeq t_0$ presents no pole. 

But the LO is insufficient near the zero of the denominator when $t_0$ and $t_1$ are nearly cancelling out.
Then, the double zero of the numerator is slightly displaced with respect to the zero of the denominator, and the pole at low-$s$ is not exactly cancelled. 
To see it, let us factorize the denominator of $G$,
\begin{equation}
  G\simeq b^2 \frac{(s+a/b)^2}{\varepsilon (s-s_+)(s-s_-)}
\end{equation}
with $s_\pm = -\frac{b}{2\varepsilon}(1\pm \sqrt{1-4\varepsilon a/b^2})$. 
Taking into account the chiral counting, 
$a\sim m_\pi^2$, $\varepsilon \sim 1/\Lambda^2$, we see that
there is a pole near the original one, $s_-\simeq -\frac{a}{b} + O\left(\frac{m_\pi^4}{\Lambda^2}\right)$ and a pole that comes from infinity, at
$s_+\simeq -\frac{b}{\epsilon}\sim O(\Lambda^2)$. This last one is outside the range of validity of the theory and can be safely discarded.
The first one is unavoidable, but remains below the threshold, and when $s$ is in the physical zone, its effect on $G$ is of order $\frac{m_\pi^4}{s\Lambda^2}$:
the memory of the Adler zero is very small outside its very proximity. Since the dispersion relation for the right cut is weighted for $s$ far from this zero of $t$ (pole of $G$), its presence becomes a numerically small correction.

Summarizing so far: the Adler zero causes no encumbrance in the chiral limit, which is often the approximation used for HEFT~\cite{Delgado:2015kxa}; and if masses are kept, it introduces a very small uncertainty.

In any case, this one uncertainty of the basic method can be disposed of, if need be, by use of the ``Modified Inverse Amplitude Method'' of~\cite{GomezNicola:2007qj}. Here, the amplitude is represented by 
\begin{equation} \label{mIAM0}
  t_{\rm mIAM}\equiv \frac{t_0^2}{t_0-t_1+A_{\rm mIAM}}
\end{equation}
with a modification term in the denominator that uses the position of the Adler zero computed in chiral perturbation theory (appropriate since $s$ is very low),
$s_A\simeq s_0+s_1+\dots$
\begin{equation}\label{mIAM}
  A_{\rm mIAM}  = t_1(s_0) -\frac{(s_0-s_A)(s-s_0)}{s-s_A} (t'_0(s_0)-t'_1(s_0))\ .   
\end{equation}

At the position of a resonance, $s=s_R\gg s_0,s_A$, the Mandelstam variable $s$ drops out and $A_{\rm mIAM}\sim \mathcal{O}(s_0^2, s_1)$ produces a small constant shift of the pole position upon substituting it in the denominator of eq.~(\ref{mIAM0}). 
The relative uncertainty in that pole position is therefore $\mathcal{O}(s_0^2/s_R^2)$.

The largest such uncertainty will affect the channel with the lightest resonance, the $IJ=00$ scalar, isoscalar partial wave that has its Adler zero at $s_0=m_\pi^2/2$ in ChPT (see for example eq.~(9.3.21) of~\cite{Yndurain:2002ud} for the value of $s_1$). 
Thus, the uncertainty at the $f_0(500)$ pole in this scalar channel is of $\mathcal{O}(m_\pi^4/M_{f_0}^4)\simeq 0.6\%$, at the few per mille level in hadron physics.
For the Electroweak Chiral Lagrangian or HEFT, taking $(m_Z/1{\rm TeV})^4\simeq 7\times 10^{-5}$, we  see that the uncertainty is totally negligible.

\subsubsection{Castillejo-Dalitz-Dyson poles of the inverse amplitude}

It remains to discuss what happens if $t_0+t_1$, now a polynomial of second order (up to a logarithm) develops an additional zero above the threshold. 
These zeroes give rise to so-called Castillejo-Dyson-Dalitz (CDD) poles~\cite{Castillejo:1955ed} of the inverse amplitude. A dispersion relation for $G\propto 1/t$ would need an additional pole contribution with treatment parallel to that of the Adler zero just discussed.

There is little that one can do to avoid these CDD poles from contributing if/when they are present,  
as they represent new physics which is neither obvious nor generated from the boson-boson scattering dynamics itself \cite{Dyson:1957rgq,Oller:1998zr};
nevertheless, we do not consider them a source of uncertainty since if 
they are separately treated: given the measurement of the low--energy coefficients, one can try to identify the presence of the CDD pole and subtract it as shown shortly.

Such CDD pole often appears in the $I=J=0$ $\pi\pi$ partial wave: in many parametrizations, the phase shift is larger than $\pi$ below the $K\bar{K}$ threshold \cite{Oller:2007xd}, where the amplitude is considered to be elastic. Therefore, both real and imaginary parts of $t$ vanish, $t(s)=\sigma e^{i\delta}\sin\delta$ is zero at the point $\delta(s_C)=\pi$ with $\sqrt{s_C}<2m_K$. 

The basic IAM runs into trouble when the zero of $t$ at the CDD pole happens near 
a resonance, because then two contradictory equations need to be satisfied: the resonance condition  (vanishing of the denominator in $t_{IAM}(s)$ at $s=s_R$) $t_0(s_R)-\text{Re}t_1(s_R)= 0$  and the CDD-pole condition from the ChPT expansion at $s=s_C$, $t_0(s_C)+\text{Re}t_1(s_C)=0$. 
Their simultaneous fulfillment would imply that $\text{Re}t_1$ and $t_0$ vanish at close values of $s$, which is not possible anywhere near the resonance region, since $t_0(s)$ is zero only at the Adler zero which is below threshold. Not taking care of this CDD pole was found to make the prediction of a new resonance to be off by as much as 25\% in Ref.~\cite{Oller:1999me}, so care should be exercised. This is shown with an explicit example in Appendix~\ref{CDDfailure}.

\subsubsection{How to handle a CDD pole if present} \label{subsec:modifyIAMCDD}

The difficulty can be overcome by studying the behavior of $t_0+t_1$ in each case and, if suspicion of a CDD pole of $G\propto t^{-1}$ arises, by appropriately modifying the IAM as we next sketch.

Hence, the first thing to do is to detect the possibility of a CDD pole from the low energy chiral expansion.
This expansion is such that  the imaginary part of $t_1$ does not vanish above $4m_\pi^2$, the two-pion threshold; thus, it is necessary to study its real part. 

The question to be examined upon deploying the IAM for any partial wave is whether there is a value $s=s_C$ above threshold  
such that its real part does vanish
\begin{align}
\label{200725.9}
t_0(s_C)+\text{Re}t_1(s_C)=0~.
\end{align}

This is a practical test to try to detect the presence of a CDD pole. When focused on the toy-model-amplitude  $t(s)$ of eq.~(\ref{200725.1}) in Appendix~\ref{CDDfailure},
it actually yields the exact result, detecting the CDD pole of the inverse amplitude at $s_C=M_0^2$, as evident in eq.~(\ref{200725.5}).~\footnote{This example also illustrates that different results can be obtained upon looking for poles in the second Riemann sheet versus
establishing the position of a resonance by requiring the vanishing of the real part of the partial wave.}

Of course, the presence of a zero in $t(s)$ requires both real and imaginary parts to vanish whereas eq.~\eqref{200725.9} is a condition on the real part alone. 
One can ask to what extent it is a good criterion for detecting a zero in the partial wave from the knowledge of the NLO amplitude alone. The reason it is useful is that 
elastic partial waves $t(s)=e^{i\delta}\sin\delta$ only have one independent real function, 
the phase shift $\delta(s)$.  ${\rm Re}(t(s))= \cos \delta \sin\delta$ vanishes for $\delta=0$ or $\pi/2~\text{mod}~\pi$, but the second case corresponds to maximum (nonzero) imaginary part and a resonant amplitude, and nothing should be said about it from the purely perturbative $t$. 
Thus, $\delta=0 \,\text{mod}\,\pi$, implying  $\sin\delta=0$ and $|t(s)|=0$ is the only relevant case, and because the amplitude is small near the zero, eq.~\eqref{200725.9} is generally adequate.

Once such CDD pole at $s_C$ has been identified, a second step  is to introduce an auxiliary function $t^C(s)$ without the related zero at $s_C$, that can be minimally defined as 
\begin{align}
\label{200725.10}
t^C(s)=&\frac{t(s)}{s-s_C}~.
\end{align}

It is the real part of the inverse of this auxiliary function $\text{Re}(1/t^C(s))$
that is chirally expanded  up to ${\cal O}(s^2)$ or NLO. 
To do it, we note that $s_C={\cal O}(p^0)$ because $s_C$ is not an Adler zero, it is a large scale. Such expansion gives
\begin{align}
\label{201018.1}
\text{Re}\frac{1}{t^C(s)}=\text{Re}\frac{s-s_C}{t_0+t_1}=\frac{s-s_C}{t_0}+s_C\frac{\text{Re}t_1}{t_0^2}+{\cal O}(s)~.
\end{align}
In turn, the imaginary part of $1/t^C(s)$ is fixed to $-i(s-s_C)\sigma(s)$ by elastic two-body unitarity and, added to the real part, yields eq.~\eqref{modIAM2} below.

This discussion can be wrapped up in two simple routine steps: check for the presence of a CDD pole with eq.~(\ref{200725.9}), and if one is present, then make a substitution in the IAM,
\begin{equation}\label{modIAM2}
t_{\rm IAM} = \frac{t_0^2}{t_0-t_1} \to
\frac{t_0^2}{t_0-t_1+\frac{s}{s-s_c}{\rm Re}(t_1)} \ .
\end{equation}

The additional piece in the denominator of this equation, from applying the IAM to $t^C$ instead of $t$, guarantees the CDD pole at the correct position, and does not affect the good unitarity behavior that the IAM enjoys; in the chiral counting, the difference between both formulae starts at order $s^3$.

Applying this procedure to the toy amplitude of eq.~(\ref{200725.1}) gives (the next equality implying an approximation in chiral perturbation theory)
\begin{align}
\label{200725.11}
G^C:=\frac{1}{t^C(s)}&\ = \frac{f^4}{s}-i\sigma(s-s_C)~,
\end{align}
from which one can immediately recover the exact $t(s) = (s-s_C)t^C(s)$. 

Once the expansion of $t^C(s)$ is at hand,
the IAM is applied thereto,  and upon completion, the zero that was factored out is multiplied back to reconstruct the approximation to the amplitude. In the example in Appendix~\ref{CDDfailure}  this is
\begin{align}
\label{200725.12}
t_{IAM}(s)=&\frac{1}{\frac{f^4}{s(s-M_0^2)}-i\sigma}\,.
\end{align}
that indeed reproduces the exact $t(s)$ of  eq.~(\ref{200725.1}). The conclusion, therefore, is that the unhandled presence of a CDD pole of the inverse amplitude leads to uncertainties of order $M_R^2/M_0^2 \sim \mathcal{O}(1)$, but these can be dealt with, analogously to the Adler zeroes, to eliminate the problem. This observation is elevated to table ~\ref{tab:wrapup}. Another example is also worked out in the Appendix~\ref{CDDfailure} for the $I=J=1$  channel with HEFT \cite{Delgado:2015kxa}, by choosing the low-energy constants to generate a zero in the partial-wave amplitude.

\subsection{From additional two-body channels} \label{subsec:KK}

Strictly speaking, the unitarity condition in eq.~(\ref{OpticalTh}) is valid below any inelastic threshold. Generically, in the presence of several elastic and inelastic channels, that unitarity relation should be modified as 
\begin{equation}
    \text{Im}\;t=\sum_i\sigma_i(s)|t_{\pi\pi\to i}|^2\theta(s-(\sum_j m_j)^2) \label{unitaritymod}
\end{equation}
where $\sum_j m_j$ represents the sum of the masses of the intermediate state particles and $\sigma_i(s)$ the phase space factor, all in the $i$th channel (also, it is intended that the symbol $\sum_i$ sums or integrates over remaining quantum numbers in channel $i$). 

The first (strongly interacting) channel to open, as allowed by $G$-parity, appears when four pions can go on-shell in the intermediate state (around 550 MeV). We discuss this in the next subsection~\ref{subsec:fourpi}.  
\par
Here we start by discussing the uncertainty introduced by the first \emph{two-body} inelastic channel, with two kaons in the intermediate state, $K\bar{K}$, and with threshold around 985 MeV. Above this energy, the unitarity relation for the inverse amplitude in eq.~(\ref{unitarityinverse}) is modified to read
\begin{equation}
    \text{Im}\;\frac{1}{t_{\pi\pi}}=-\sigma_{\pi\pi}\Big(1+\frac{\sigma_{K\bar{K}}}{\sigma_{\pi\pi}}\frac{|t_{\pi\pi \to K\bar{K}}|^2}{|t_{\pi\pi\to\pi\pi}|^2}\Big)\;.\label{inelasticim}
\end{equation}
There are two regimes that allow to disregard the inelastic channel in this modified unitarity equation: when the ratio of the squared amplitudes is small, or when the ratio of the phase-space factors is the one suppressing the amplitude.

The second term inside the parenthesis of eq.~(\ref{inelasticim}) is the correction to the right cut discontinuity of $t$ related to the inelastic ``${\mathcal I_z}$'' contribution to eq.~(\ref{alluncertainties}),
\begin{equation}
   {\mathcal I_z}(s)=\frac{s^3}{\pi}\int_{RC} dz \frac{-t_0^2\sigma_{\pi\pi}}{z^3(z-s)}\frac{\sigma_{K\bar{K}}}{\sigma_{\pi\pi}}\frac{|t_{\pi\pi \to K\bar{K}}|^2}{|t_{\pi\pi\to\pi\pi}|^2}\;.\label{Inel}
\end{equation}\par

In this form we additionally see that the three subtractions bias the integral in eq.~(\ref{Inel}) towards the low energy region so that the effect of the coupled channels is even less prominent if the elastic and inelastic amplitudes have the same scaling with $z$ so that its powers cancel out in the ratio leaving only logarithms (which is quite the case in ChPT). Finally, the $z-s$ factor in the denominator enhances the region of $z$ around the external value of $s$ which is the argument of  ${\mathcal I_z}(s)$. (This we often take as the $\rho$-resonance mass, \emph{i.e.} $m_\rho=770$ MeV). \\

\paragraph{\textbf{In hadron physics,}} all these mechanisms are suppressing the inelastic coupling of 
$\pi\pi$ and $K\bar{K}$ below 1.2 GeV~\cite{GomezNicola:2001as}, except near
the $f_0(980)$ 
resonance~\cite{Oller:1997ng}.
Because those happen close to the $K\bar{K}$ threshold, and the first resonances in $\pi\pi$ scattering ($\sigma$ and $\rho$) lie below it, the uncertainty in the latter's masses is well controlled.

In practice, the dominant contribution to the integrated uncertainty ${\mathcal I_z}(s)$ of eq.~(\ref{Inel}) in meson scattering comes from the energy region 
up to $1.2$ GeV where we can constrain it with
the available data for $|t_{\pi\pi\to K\bar{K}}|^2/|t_{\pi\pi\to\pi\pi}|^2$ in the $J=I=1$ channel as the integrand will be heavily suppressed above (and much below) $m_\rho$.
The factor
\begin{equation}
 \frac{\sigma_{K\bar{K}}}{\sigma_{\pi\pi}} \frac{{|t^{IJ=11}_{\pi\pi\to K\bar{K}}|^2}}{{|t^{IJ=11}_{\pi\pi\to\pi\pi}|^2}}=\frac{\sigma_{\pi\pi}}{\sigma_{K\bar{K}}}\frac{1-\eta_{11}^2}{\eta_{11}^2-2\cos{2\delta_{\pi\pi}}+1}
\end{equation}
is less than $0.08$ below 1.2 GeV
 (because
the vector-isovector elasticity, $\eta_{11}$, is relatively close to one in the 1-1.2 GeV energy region, $\eta_{11}\simeq0.99$, \cite{GomezNicola:2001as,Pelaez:2019eqa} and the pion-pion channel phase shift, $\delta_{\pi\pi}$, varies slowly in this energy region).

We will therefore concentrate on the one-channel IAM that entirely neglects the contribution of the second channel, and use the coupled-channel IAM only to estimate the uncertainty therein.

Introducing the value above allows us to set a bound to the displacement of the pole from the $K\bar{K}$ intermediate state up to $s=(1.2\text{ GeV})^2$. Since we know that, below 1.2 GeV,
\begin{equation}
   \frac{\sigma_{KK}|t^{IJ=11}_{\pi\pi\to K\bar{K}}|^2}{\sigma_{\pi\pi}|t^{IJ=11}_{\pi\pi\to\pi\pi}|^2}\leq 0.08\;,\label{patapaf}
\end{equation}

we can use eq.~(\ref{corrpoleposition}), accounting for the uncertainty due to the two-body inelastic contribution to the displacement of the pole, ${\mathcal I_z}$ in (\ref{alluncertainties}), (\ref{Inel}) and (\ref{patapaf}) above to find

\begin{equation}
    |\Delta(s)G(s)|\leq 0.08\frac{s^3}{\pi}|RC(t_1)|(s)\simeq 1 \cdot 10^{-4}\;.\label{2body}
\end{equation}
with the expression evaluated at $s=m_\rho^2$. 
(Though this is below the two-kaon threshold, so that at the resonance mass $\sigma_{KK}=0$, the uncertainty does not vanish because the integral in $RC$ extends through $\infty$.)
In eq.~\eqref{2body} $RC(f)(s)=\int_{RC}dz\, \text{Im}f(z)/[z^3(z-s)]$. 
We take such potential displacement of the $\rho$-pole due to this uncertainty to Table \ref{tab:wrapup}. \\

\paragraph{ \textbf{Within the Electroweak Standard Model},} the coupled-channel IAM has also been deployed in a series of articles~\cite{Delgado:2013loa}. Because in the $s\sim 1$ TeV$^2$ region (where the HEFT would be used) the $W$, $Z$ and $h$ boson squared masses of order $0.01$ TeV$^2$ are all negligible, they have equal phase space (the ratio $\sigma_{hh}/\sigma_{\omega\omega}$ is close to 1), providing no suppression to eq.~(\ref{inelasticim});
only the equivalent ratio of squared amplitudes $|t_{ww\to hh}|^2/|t_{ww{\to ww}}|^2$ suppresses the inter-channel coupling. This is, in the notation of~\cite{Delgado:2015kxa}, proportional to the parameter combination $(a^2-b)^2$, that vanishes in the Standard Model (but its value in strongly interacting theories of BSM physics is of course unknown). 
What we can state is that, if low-energy measurements reveal this combination of low-energy constants to be numerically small, the coupled-channel uncertainty introduced into the single-channel problem is immediately under control.

We lean on the  IAM method for coupled channels to generate the uncertainty of the one-channel IAM. The coupled-channel case~\cite{Oller:1997ng,Oller:1998hw,GomezNicola:2001as} takes
the same form as the standard IAM but in terms of matrix-valued amplitudes. For $n$ coupled channels, the amplitude will be an $n$-by-$n$ matrix, $\mathbf{t}$, so that the IAM expression becomes 
\be
\mathbf{t}_{IAM}=\mathbf{t}_0(\mathbf{t}_0-\mathbf{t}_1)^{-1}\mathbf{t}_0\;.
\ee

Concentrating on the HEFT with two channels, upon disregard of the contributions to the IAM partial wave amplitude of $ww\to ww$ coming from the coupled channels $ww\to hh$ and $hh\to hh$, an error $\delta$ is incurred,
\begin{equation}
    \delta=\frac{(t_1^{11} t_0^{12} - t_0^{11} t_1^{12}) (t_1^{11} t_0^{21} - t_0^{11} t_1^{21})(t_0^{11} - t_1^{11})^{-1}}{ (t_1^{12} (t_0^{21} - t_1^{21}) + t_0^{12} (t_1^{21}-t_0^{21}) + (t_0^{11} - t_1^{11}) (t_0^{22} - t_1^{22}))}\;,
    \label{uncCoupledChannel}
\end{equation}

where the superindices denote matrix elements of $\mathbf{t}$. We are parametrizing $t_{IAM}^{11}=t_{IAM}+\delta$ so that $\delta$ takes into account all contributions from coupled channels to the one-channel IAM partial wave amplitude, $t_{IAM}$. Note that, neglecting logarithms,  eq.~(\ref{uncCoupledChannel}) is a fraction of second-order polynomials in $s$ and the zeroes of the denominator correspond to zeroes of  ${\rm det}\left(\mathbf{t}_{IAM}(s)^{-1}\right)$ and, therefore, to poles in the amplitudes (resonances).

Reading the coefficients from \cite{Delgado:2015kxa} for the $IJ=00$ channel (neglecting logarithmic contributions), the expression for $\delta$ at LO in $(a^2-b)^2$  and $(3d+e)$ is

\begin{equation}
    \delta(s)\simeq s^2 \left(a^2-b\right)^2\left[\frac{-9 \left(29(a^2-1)^2+5376 \pi ^2 a_4+8448 \pi ^2 a_5\right)^2}{2949440 \pi  g \big( 576\pi^2v^2(a^2-1)+5376 \pi ^2 a_4 s+8448 \pi ^2 a_5 s+101 \left(a^2-1\right)^2 s\big)^2}\right]\label{Helostodos}
\end{equation}

We see from eq.~(\ref{Helostodos}) that, as the coupling between channels $({a^2-b})$ approaches zero, the uncertainty from neglecting the coupled channel vanishes as expected. The denominator of  eq.~(\ref{Helostodos}) carries a dependence on the $hhhh$ interaction parameter $g$ coming from inverting the $G^{ij}$ matrix~\footnote{This cannot be set to zero for invertibility, but as the channels decouple its value becomes irrelevant; thus, we take this parameter $g$ to be of order one with little loss of precision on the $\omega\omega$ channel.}. 

A particularity of this HEFT theory is that a small numerical factor suppresses $\delta$. Evaluating it for $s=1 \text{ TeV}$, with $g=1$, and the example values in table~\ref{tab:parametros} for the remaining 
coefficients of the HEFT Effective Lagrangian (consistent with current LHC bounds), 
we obtain a tiny
\begin{equation}
    |\delta(s=1 \text{ TeV}^2)|\simeq 5\cdot 10^{-3}\;.
\end{equation}

\begin{table}[ht!]
    \begin{center}
    \begin{tabular}{c|c|c}
      HEFT Parameter   & Bounds known to us  & Example value taken \\ \hline
       $|a-1|$    & $<0.15$~\cite{Delgado:2017cls,Dobado:2019fxe} & $a=0.9$ \\
       $b-1$      & $\in (-1,3)$~\cite{Delgado:2014dxa}   & $b=1.5$ \\
       $a_4$  & $<6\cdot 10^{-4}$ \cite{CMS:2019qfk}& $5\cdot 10^{-4}$  \\
       $a_5$  & $<8\cdot 10^{-4}$ \cite{CMS:2019qfk} & $5\cdot 10^{-4}$  \\
    \end{tabular}
    \end{center}
    \caption[Other bounds on the parameters of the HEFT Lagrangian]{Bounds on the parameters of the HEFT Lagrangian and the allowed example values that we have used to estimate the uncertainty in eq.~(\ref{uncCoupledChannel}). We write down a sensible value from considering the 95\% confidence bounds, with no attempt at combining different experiments or theoretical analysis. Additionally, $v=246$ GeV as is well known from the electroweak theory.}
    \label{tab:parametros}
\end{table}\par

This two-body uncertainty propagates to the pole position of the $IJ=00$ channel resonance, whose real part (new physics mass) becomes fuzzed by
\begin{align}
    |\Delta(s)G(s)|=&\big|\frac{-\delta}{t_{IAM}}\frac{t_0^2}{t_{IAM}+\delta}\Big|\simeq 2\cdot10^{-3}\label{HEFTdelta}
\end{align}
at 1 TeV; this is an order of magnitude larger than the hadron physics counterpart in eq.~(\ref{2body}). It  is not unexpected because of the heavy phase space suppression in the earlier case. Even so, with present knowledge, if a scale of new physics strongly affecting vector-boson scattering is discovered, the corresponding inelasticity does not appear to be a worry.


\subsection{From inelastic channels with additional identical particles}
\label{subsec:fourpi}
Let us start discussing the case of hadron physics. 
As mentioned at the beginning of Section \ref{subsec:KK}, a four-pion channel opens around 550 MeV. \emph{A posteriori}, the accuracy of the IAM would be suggestive of a small contribution by the four-pion channel (a four-pion intermediate state is a three-loop
and higher-order effect in ChPT), just as the $K\bar{K}$-one discussed in subsection~\ref{subsec:KK}. Phenomenologically, the effect of that $4\pi$ channel for $\sqrt{s}<1$~GeV seems to be very small as indicated by a small experimental inelasticity $\eta\approx 1$, as well as from explicit calculations~\cite{Albaladejo:2008qa}.
We will address the uncertainty introduced by neglecting this channel \emph{a priori}, basically controlling it due to reduced phase space. 

\par
The massive $n$-particle phase space differential is
\begin{equation}
d\phi_n=\delta^{(4)}\left(p-\sum_{i=1}^n q_i\right)\prod_{i=1}^n\frac{d^3q_i}{(2\pi)^32E_i}\;.
\end{equation}
Integrating this four particle phase space, having used the three-dimensional delta function, we arrive at
\begin{align}
    \phi_4=&\int_0^{+\infty}\Big(\prod_{i=1}^3\frac{d|p_i||p_i|^2}{2E_i}\Big)\int_{-1}^{+1}dx_1dx_2\int_0^{2\pi}d\varphi\frac{\delta(\sqrt{s}-E_1-E_2-E_3-E_4)}{(2\pi)^6 E_4}\;.\label{4PS}
\end{align}

Here 
\begin{align}
    E_4=&\Big(|p_1|^2+|p_2|^2+|p_3|^2+2|p_1||p_2|x_2+2|p_2||p_3|x_3+\nonumber\\
    &+2|p_2||p_3|\left(x_2x_3+sin\varphi\sqrt{1-x_2^2}\sqrt{1-x_3^2}\right)+m^2\Big)^{\frac{1}{2}}
\end{align}

\begin{figure}[ht!]
\centering
\includegraphics[width=100mm]{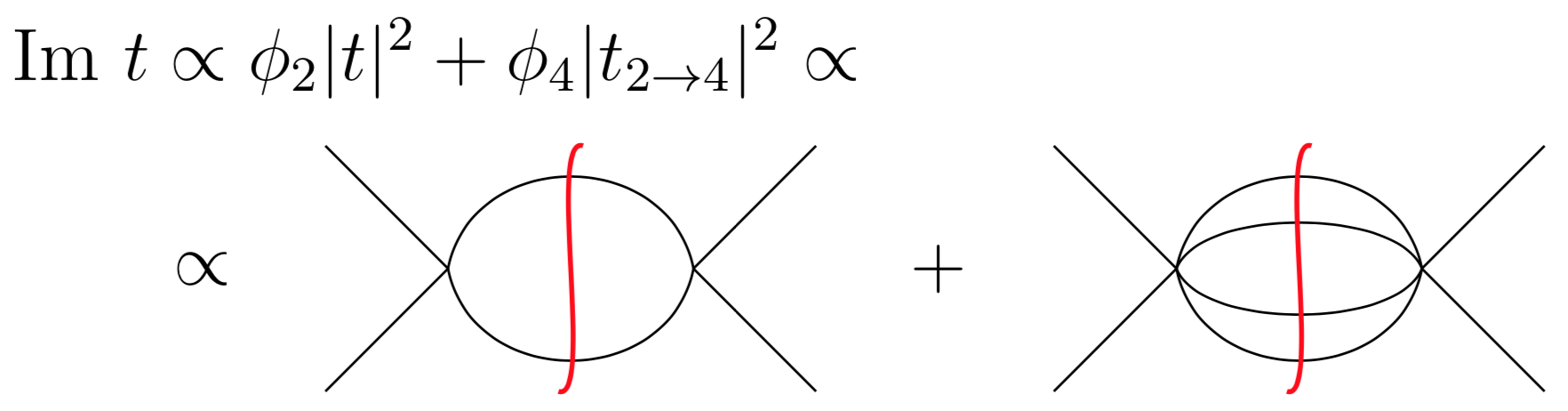}
\caption[Factors when comparing Phase spaces]{When considering each of the 2- or 4-body phase spaces, we have assigned $2\pi$ factors according to the normalization of one-pion states. For each pion in an intermediate state there is a factor $(2\pi)^{-4}$. However, when calculating the imaginary part of the amplitude, each cut line will produce an extra $2\pi$ (together with a $f_\pi^{-1}$). This leaves the typical $(2\pi)^{-3}$ normalization factor for each pion. Hence, we must only use $f_\pi^4$ to compare the two-body and four-body phase spaces.}\label{IMPart}
    \end{figure}

After numerically integrating the expression in eq.~(\ref{4PS}), we can compare the four-particle phase phase, $\phi_4$, with the two-body one, $\phi_2$, which is related to $\sigma(s)$ defined in eq.~(\ref{OpticalTh}) as $\phi_2(s)=\sigma(s)/8\pi$.  For a fair comparison, the two-body phase space needs to be multiplied by a power of $f_\pi$, $\phi_2\times f_\pi^4$, to match the dimensions of the four-body one (here the pion decay constant is taken to be $f_\pi=93$ MeV). This $f_\pi^4$ dimensionful factor is extracted from the two-body amplitude (that correspondingly has  different dimensions from the inelastic two to four body one). The reasoning behind this choice is explained in Fig. \ref{IMPart}.

\begin{figure}[ht!]
\centering
   \includegraphics[width=110mm]{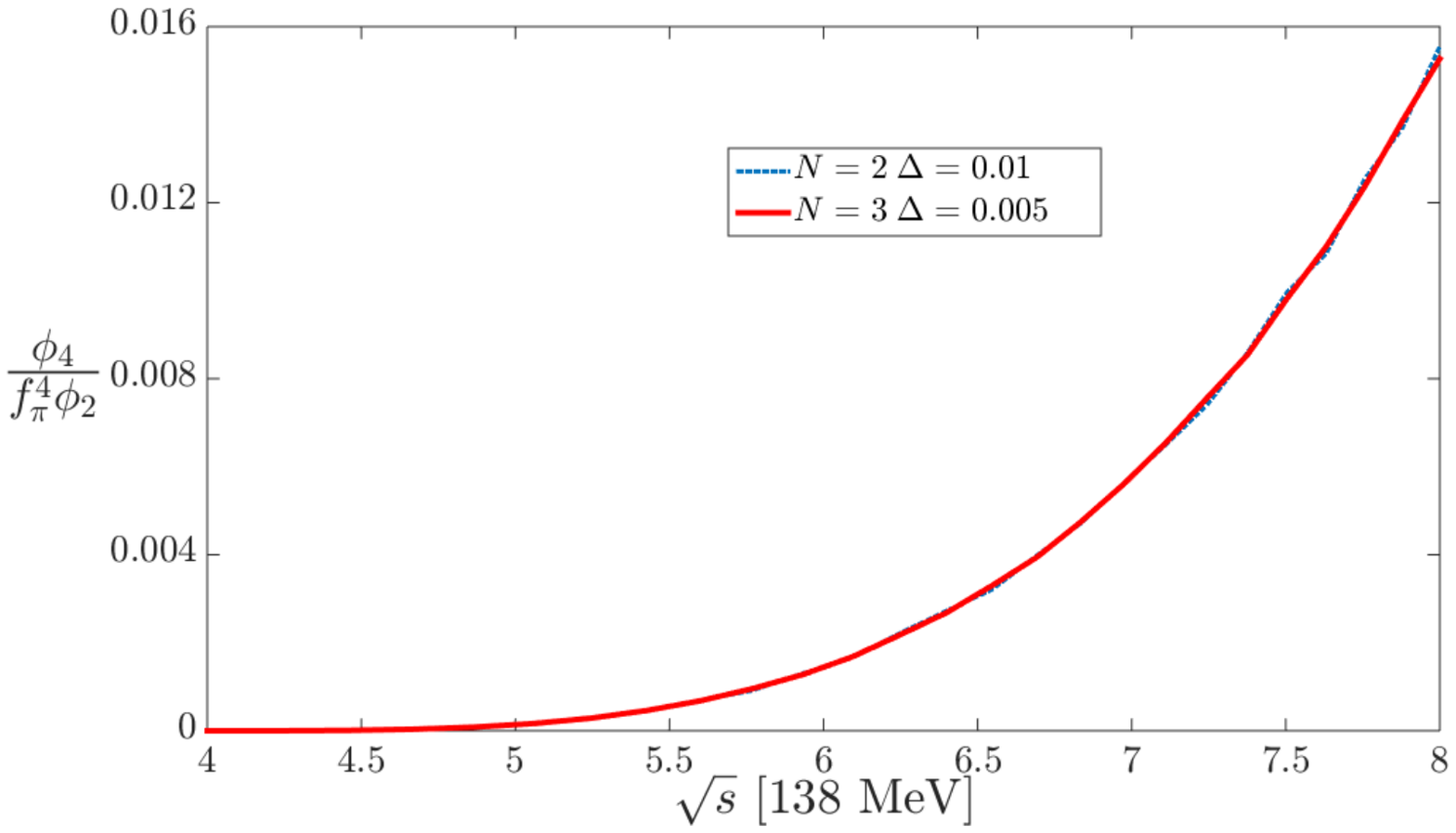}
    \caption[Numerical computation of the four- to two-particle phase-space ratio]{Numerical computation of the four- to two-particle phase-space ratio from  threshold $\sqrt{s}=4m_\pi$ up to $\sqrt{s}=1.1\text{ GeV}$. 
    (The 4-body integral is calculated in successively better approximation by replacing the energy delta function in Eq~(\ref{4PS}) by a Gaussian of decreasing width $\Delta$. $N$ is the number of 20-Gaussian point partitions of each variable's integration interval. 
    Convergence is excellent and both curves are barely distinguishable.)}
    \label{fig:4ParticlePS}
\end{figure}

The resulting ratio, which is the meaningful measure of the relative weight of four- and two-body states, $\phi_4/(\phi_2 f_\pi^4)$, is  plotted in Fig.~\ref{fig:4ParticlePS}. We see  that, up to $s=(1.1 \text{ GeV})^2$, the four-particle phase space is heavily suppressed compared to the two particle one~\footnote{This is easily understood by the analogous simple relation $ \int_0^\infty dx_1dx_2dx_3dx_4 \delta(\sum_i x_i -1) < \int_0^\infty dy_1 dy_2 \delta(\sum_i y_i -1)$. }.

\begin{figure}[ht!]
\centering
\includegraphics[width=115mm]{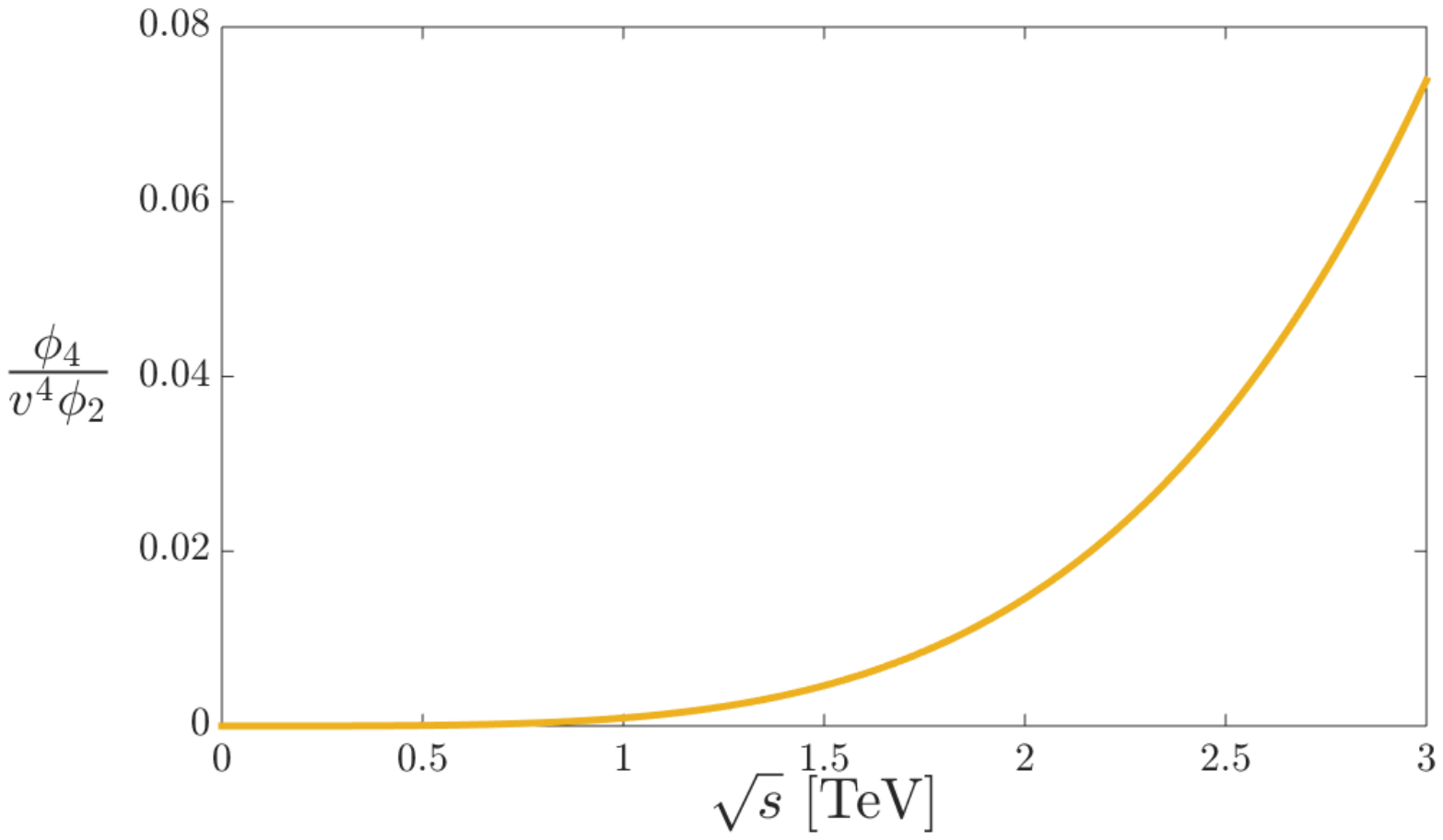}
    \caption[Ratio between the massless four-particle and two-particle phase]{Ratio between the massless four-particle and two-particle phase spaces up to 3 TeV, that can be analytically evaluated from eq.~\eqref{psanalytic}. Here $v=246$ GeV.}
    \label{fig:4pHEFT}
\end{figure}

\begin{figure}[ht!]
    \centering
    \includegraphics[width=85mm]{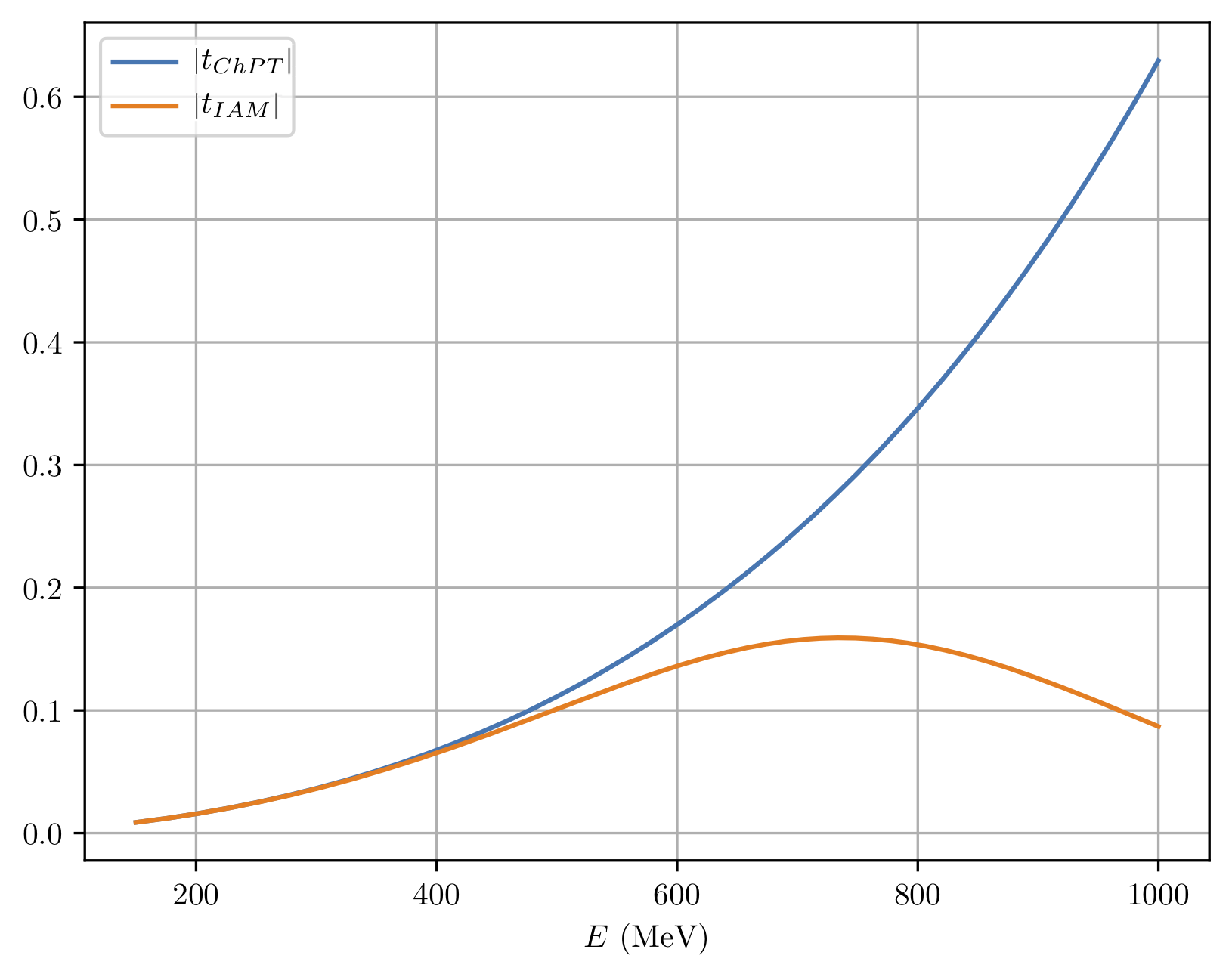}
    \caption[Comparison between the moduli of $t_{\text{ChPT}}$ and $t_{\text{IAM}}$]{Comparison between the moduli of $t_{\text{ChPT}}$ and $t_{\text{IAM}}$ at $\mathcal{O}(p^4)$ for the $IJ=11$ channel. The two amplitudes separate around a scale 200 MeV above the two pion threshold.
    \label{fig:amps}}
\end{figure}

For the HEFT case in the TeV region it is an excellent approximation to adopt massless particles, which does away with the need for numerical computation since the $n$-particle phase space has a simple analytic expression
\begin{equation} \label{psanalytic}
    \phi_n=\frac{1}{2(4\pi)^{2n-3}}\frac{{s}^{n-2}}{{\Gamma(n})\Gamma(n-1)}\;.
\end{equation}
We plot in Fig.~\ref{fig:4pHEFT} the ratio analogous to figure~\ref{fig:4ParticlePS}, substituting $f$ by $v=0.246$ TeV as appropriate for the electroweak sector.  Therein we see that, up to 3 TeV, four particle phase space is again much smaller than the two particle phase space.

Now we can compute the displacement of the pole in both hadron physics and the electroweak HEFT, taking the ratio $|t_{\pi\pi\to\pi\pi\pi\pi}|^2/|t_{\pi\pi\to\pi\pi}|^2$ and $|t_{ww\to wwww}|^2/|t_{ww\to ww}|^2$ to be of order one, with  the phase space ratios from Fig. \ref{fig:4ParticlePS} and Fig. \ref{fig:4pHEFT}. Taking this information to eq.~(\ref{alluncertainties}) and in analogy with eq.~(\ref{inelasticim}) for the four body intermediate states, we find
\[
    |\Delta(s)G(s)|\leq \left\{   \begin{array}{l}
2\cdot 10^{-2} \frac{s^3}{\pi}|RC(t_1)|(s)\;\;\text{for hadron physics}\\
8\cdot 10^{-2}\frac{s^3}{\pi}|RC(t_1)|(s)\;\;\text{for the HEFT}\;.
              \end{array}
    \right.
  \]
  This means that for hadron physics, after computing the right cut integral, the channel $IJ=11$ receives an uncertainty of order 
  \begin{equation}
      |\Delta(s)G(s)|\leq 4\cdot10^{-5}\;\;\;\text{for $s=m_\rho^2$}\;.\label{4body}
  \end{equation}
  Where $m_\rho= 770$ MeV. This is much smaller than the other terms of eq.~(\ref{corrpoleposition}). For example, the first term $t_0 = \frac{s}{192\pi f_\pi^2}$ is about 0.12 at the $\rho$ meson pole; this amounts to a relative uncertainty from the correction at the level of four in ten thousand! Hence, the influence of the multi-pion cut in the $\rho$ pole position is very small.

This smallness is also reflected in the scaling of this source of uncertainty with the mass of the resonance. Taking the massless limit for simplicity for $\phi_4$ in eq.~\eqref{psanalytic} we see that $\phi_4/\phi_2\sim s^2$, so that if $s^3/\pi \,|RC(t_1)|(s)$ scales as $s^2$ (a typical NLO ChPT contribution), then $\Delta(s)G(s)$ for the four-body intermediate contribution scales as $s^4$ (a three-loop effect). Therefore, this  uncertainty to the mass of the resonances scales as $(m_{\rm resonance}/\Lambda_U)^8$, with $\Lambda_{\rm U}$ the unitarity cutoff the theory,  $4\pi f_\pi$ for hadron physics and $4\pi v$ for HEFT. Then, the influence of this uncertainty is very much diminished for $m_{{\rm resonance}}<\Lambda_{\rm U}$, though it raises strongly for higher energies.

\subsection{From truncating perturbation theory for the elastic amplitude} \label{subsec:pert}

Perturbative expansions, and among them Chiral Perturbation Theory, are organized as a geometric series homogeneous in an invariant energy squared $E^2$, that acts as a counting parameter. The naive uncertainty estimate upon truncating the series at order $n$ is of order $(E^2)^{n+1}$, which assumes that the typical coefficient multiplying this power is not excessively large, in order not to alter the relative geometric size of the term. This is suspected to fail, for example, when the number of allowed Feynman diagrams grows factorially, but it is widely accepted as the organizing principle and so we also adopt it
(although a recent study \cite{Bonvini:2020xeo} lays out an interesting probabilistic method to account for the uncertainty in the next order of such an expansion).  

Increasing the order in the expansion of the EFT to improve the behavior of the unitarized method is a strategy employed, for example, in the $N/D$ method for $NN$ scattering~\cite{Oller:2014uxa}. 
Here we attempt to quickly estimate effects that appear at NNLO and have not been discussed in the other subsections of this section~\ref{sec:uncertainties}.

In the elastic Inverse Amplitude Method at NLO, two orders of the expansion have been kept. Can we make a statement about the third, neglected order, even if the method is not organized as a geometric expansion?
If the next $t_2\sim O(p^6)$ order of chiral-like perturbation theory for a given process and channel becomes known (which is the case in hadron physics~\cite{Bijnens:1998fm,Niehus:2020gmf} , but not yet for the electroweak HEFT), one can expand the inverse amplitude to that higher order, using 
\begin{equation}
G(s)=\frac{t^{2}_0}{t}\simeq \frac{t^{2}_0}{t_{0}+t_1+t_2}
\end{equation}
and expanding to dispose of yet higher order contributions 
\begin{equation}\label{Gtoorder6}
G(s)=\frac{t^{2}_0}{t}\simeq t_{0}-t_{1}-t_{2}+\frac{t^{2}_1}{t_0}\ .
\end{equation}
(Some authors like thinking of the IAM as a Pad\'e approximation, and at times criticize the inherent ambiguity in how to choose the right sequence of Pad\'e approximants. It is the dispersive derivation that selects what is the appropriate sequence of approximations, by applying the EFT expansion to $G$.)

However, we wish to quantify how much can the pole get displaced by including the NNLO correction, not actually calculate to this order in every instance.  This displacement can be captured by the low energy constants of the third order perturbation theory reflecting the imprint of a resonance on them. The relation between those constants and the resonance is known in Resonance Effective Theory~\cite{Guo:2009hi,Ecker:1988te,Rosell:2015bra}(by integration of heavy degrees of freedom) and enters in the equation for the displacement of the pole in eq.~(\ref{alluncertainties}) as
\begin{align}
    3^{rd}PT=& G(0)+ G'(0) s+\frac{1}{2}G''(0)s^2- (t_0-t_1)(0)+ (t_0-t_1)'(0) s+\frac{1}{2}(t_0-t_1)''(0)s^2\nonumber\\
    =&\big(\frac{t^{2}_1}{t_0}-t_{2}\big)(0)+ \big(\frac{t^{2}_1}{t_0}-t_{2}\big)'(0) s+\frac{1}{2}\big(\frac{t^{2}_1}{t_0}-t_{2}\big)''(0)s^2\;,\label{substrNNLO}
\end{align}
for the NNLO correction of $G$.
(The derivatives at $s=0$ exist even in the chiral limit as we are focusing on the polynomial contributions from the higher-order counterterms only.)
This means that, for $IJ=11$, 
\begin{align}
       |\Delta(s)G(s)|=&\frac{m_\pi^2}{480 f^6_ \pi} \Big|\Big(4 m^4_\pi (5  r_2 + 20  r_3 - 20  r_4 + 72  r_5- 8  r_6 - 
     10  r_F)+ \nonumber\\
       & + m^2_\pi (-5  r_2 - 40  r_3 + 80  r_4 - 216  r_5 + 24  r_6 + 10  r_F) s+\nonumber\\
     & + (-80  l_1^2 + 80  l_1  l_2 - 20  l_2^2 + 5  r_3 -15  r_4 + 54  r_5 + 14  r_6) s^2\Big)\Big|\label{3PT}
\end{align}
Example values of the constants in eq.~(\ref{3PT}) are given in Table \ref{tab:constants}. 

\begin{table}[]
    \centering
\begin{tabular}{c|c}      
Constants & Values consistent with \cite{Guo:2009hi,Nebreda:2012ve}\\ \hline\hline
    $(l_1,l_2,l_3)$ &  $(2,4,10)\cdot 10^{-3}$\\
   $(r_1,r_2,r_3,r_4,r_5,r_6,r_F)$  & $(-17,17,-4,0.0,0.9,0.25,-1)\cdot 10^{-4}$\\
   \hline
    \end{tabular}
    \caption[NNLO renormalized scattering constants in hadron physics]{Renormalized scattering constants up to NNLO which we interpret as evaluated at the renormalization scale $\mu=\rho=770 \text{ MeV}$, this being the mass of the most prominent high-energy resonance  eliminated to obtain them.}
    \label{tab:constants}
\end{table}
Evaluating eq.~(\ref{3PT}) at the $\rho$ mass, $s=m_\rho^2$, we find
\begin{equation}
    |\Delta(s)G(s)|\simeq 6\cdot 10^{-3}\;.
\end{equation}
Because the IAM matches the EFT at low $s$, $m_\pi$ to NLO order,
we expect the uncertainty to scale with the typical NNLO $s^3$ behavior for (quasi)massless Goldstone bosons.

However, the explicit power of $m_\pi^2$  in eq.~(\ref{substrNNLO}) entails a
$m_\pi^2 s^2$ contribution to the uncertainty (that we take to table~\ref{tab:wrapup} shown later on).

\subsection{From approximate left cut}\label{subsec:LCbound}

The integral in the RHS of eq.~(\ref{IAMerr}) is the most difficult piece of the IAM derivation to be bound or constrained, and it also contributes the largest uncertainty. 
We will divide this integral into different energy regimes distinguished by how the amplitude scales with energy. It will transpire that the largest contribution is due to the low-energy part of the integration region, and since ChPT is expected to reasonably approximate the amplitude there, we will be able to bind the induced uncertainty. 

In non-relativistic~\cite{Oller:2018zts} theory, the left cut contains the information about the interaction potentials whereas the right cut contains the physical particles, so there is an intrinsic difference of treatment.
In the relativistic theory on the other hand, the left cut of a given partial wave is related to the right cuts of other channels and partial waves due to crossing symmetry, as encoded for example in Roy-Steiner equations~\cite{Ananthanarayan:2000ht,Garcia-Martin:2011iqs}.

If the integrand of eq.~(\ref{IAMerr}) vanished, that is, ${\rm Im}\, G= -{\rm Im}\,t_1$, the left cut would receive exact treatment. This is obviously not the case, as 
$G$ is unknown \textit{a priori}. Therefore, we examine the contributions of both $G$ and $t_1$ there.

\subsubsection{Inspection of the left-cut integral for $t_1(s)$}

First, let us address $t_1$, {\it i.e.} the NLO term of the partial wave amplitude. To assess which part of the left cut integral is numerically most relevant  we use, as a typical case, the $\mathcal{O}(p^4)$ ChPT partial wave amplitude for the $IJ=11$ channel in the chiral limit from \cite{Gasser:1983yg}.
The low-$|z|$ part of the $LC(t_1)$ integral is suppressed by the derivative coupling in the effective theory, but the high-$|z|$ one is suppressed by the $z^3(z-s)$ factor from the dispersion relation. We should like to see which one contributes the most, so we split the integration interval as
\begin{align}
\frac{s^3}{\pi}LC(t_1)=&\frac{s^3}{\pi}LC_{\text{far}}(t_1)+\frac{s^3}{\pi}LC_{\text{near}}(t_1)  \nonumber \\   
=&\frac{s^3}{\pi}\int_{-\infty}^{-\lambda^2}d s' \dfrac{\text{Im } t_1(z) }{z^3 (z-s)}     +\frac{s^3}{\pi}\int_{-\lambda^2}^{0} d s'\dfrac{\text{Im } t_1(z) }{z^3 (z-s)} \;, \label{eq:LC_split}
\end{align}
with a contribution coming from a region near to the origin $s=0$, $[-\lambda^2,0]$, and a second from higher energies, $(-\infty,-\lambda^2]$. We choose $\lambda$ to be $470\;\text{MeV}$, i.e. the scale where the $\mathcal{O}(p^4)$ ChPT amplitude separates from the IAM amplitude and the scale where ChPT is supposed to give a good prediction of the full amplitude ($200$ MeV above the two pion threshold). This is presented in Fig. where the moduli of $t_{IAM}(s) $ and $t_{\text{ChPT}}(s)$ are plotted along the RC.

\begin{figure}[ht!]
    \centering
    \includegraphics[width=85mm]{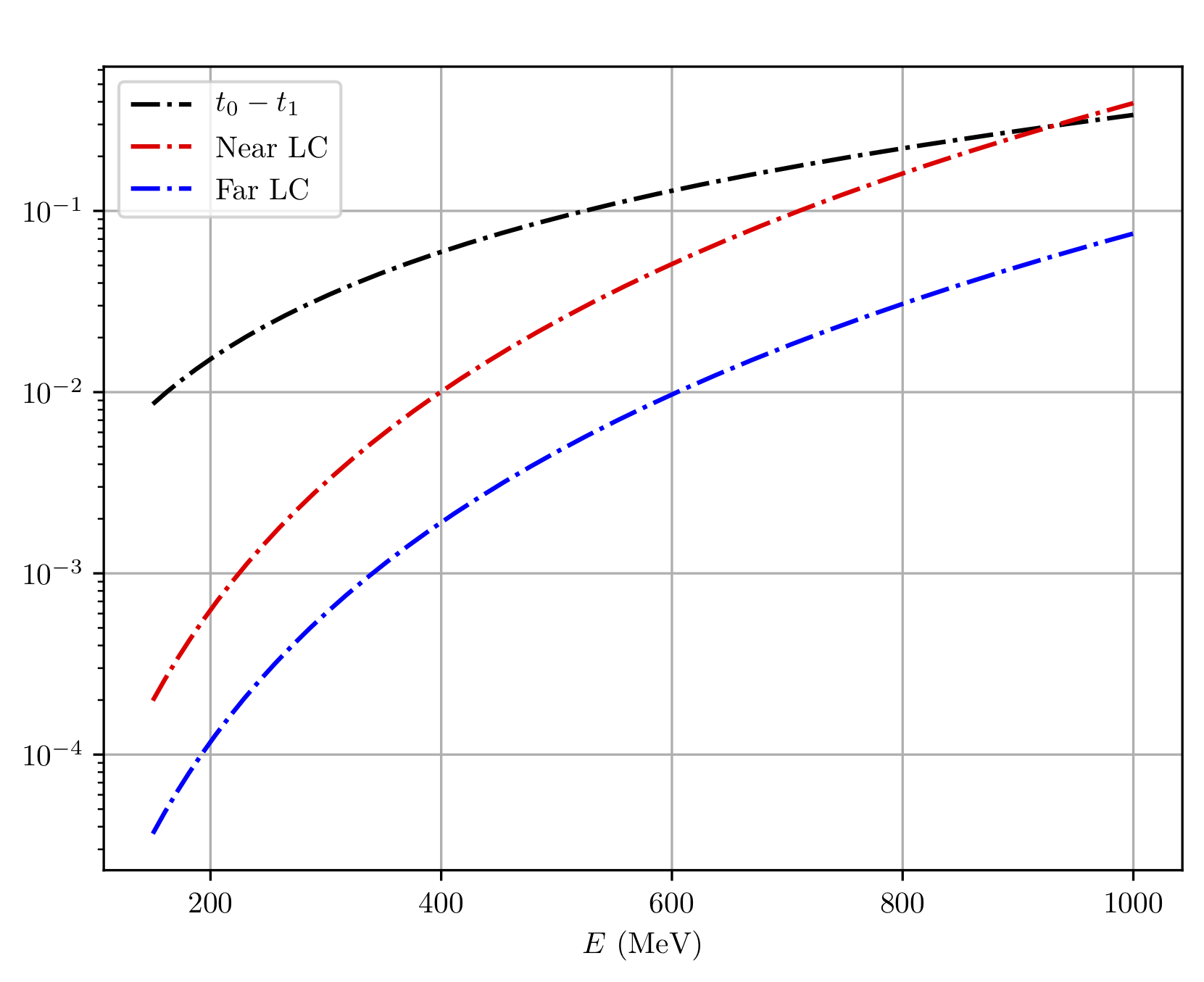}
    \caption[Comparison between the moduli of the near and far left cut]{Comparison between the moduli of the near and far left cut pieces of eq.~(\ref{eq:LC_split}) and $t_0-t_1$ for the $IJ=11$ channel.}
    \label{fig:LCs}
\end{figure}

Once $\lambda$ has been chosen, we compare the near and far (respect to the origin $s=0$) left cut contributions to eq.~(\ref{eq:LC_split}) (with $t_1$ in the integrand) in Fig.~\ref{fig:LCs}. Because of the structure of eq.~(\ref{intermediateG}), we also include a line to compare $t_0-t_1$. Note that around the resonance region the near left cut is about one order of magnitude larger than the far left cut.

The contribution from the near left cut of $t_1$ will be treated together with the integral of $\text{Im }G$ over the same region of integration. This is convenient since in the region $[-\lambda^2,0]$ the function $G$ is well approximated by the NLO ChPT expansion and the combination $\text{Im }G+\text{Im }t_1$ is a NNLO remnant contribution. This is discussed in the next section.

Part of the contribution from the far left cut of $t_1$ to eq.~(\ref{eq:LC_split}) will be treated together with a contribution coming from the $LC$ of $G$ in eq. (\ref{IGT11}). The rest will be accounted for in eq.~(\ref{201024.3}).

\subsubsection{Left Cut integral for $G+t_1$}

We now turn to the part of the integrand in the RHS of eq.~(\ref{IAMerr}).

For low-$|s|$, the  quantity ${\text {Im }}(G+t_1)$ in  eq.~(\ref{IAMerr}) is controlled \textit{a priori}, as perturbation theory is a good approximation to the scattering amplitude, and thus to $G$. This uncertainty is broadly included in the uncertainty counting of subsection  ~\ref{subsec:pert}.

At high-$|s|$ we can exploit the asymptotic behavior of the amplitude brought about by the generic relation between the left and the right cuts, described shortly, 
and in the case of an underlying non-Abelian gauge theory such as QCD, that is asymptotically free, the Brodsky-Farrar counting rules \cite{Farrar:1979aw}.

This suggests that we can divide the integration 
domain and thus the integral for $G+t_1$ in the RHS of eq.~(\ref{IAMerr}) as
\begin{align}
\frac{s^3}{\pi}\int_{LC}dz&\frac{\text{Im}\,(G+t_1)}{z^3(z-s)}\equiv I_1[G+t_1]+I_2[G+t_1]+I_3[G+t_1]\;. \label{threepieces}
\end{align}
into three regions and discuss them separately.
These will be an ultraviolet region extending through $(-\infty,-\Lambda^2)$, yielding $I_1$, an intermediate region $(-\Lambda^2,-\lambda^2)$ contributing to $I_2$, and the infrared region $(-\lambda^2,0)$ that returns $I_3$. These $I_i$ are functionals of an amplitude-like function and functions of $s$ and their respective interval cutoffs. \\

\paragraph{\textbf{Low-$|z|$ region ($|I_3|$).}}
\label{par:lowz}

The lowest dividing scale, $\lambda$, is chosen to allow the use of ChPT  for $|s|\!<\!\lambda^2$ since, as $|s|\!\to\! 0$,
 \begin{align}
\label{201019.4}
 \text{Im}\,G(s)=-\text{Im}\,t_1(s)+\mathcal{O}(s^3)&\nonumber\\
 \Rightarrow \;\text{Im}\,G(s)+\text{Im}\,t_1(s)=\mathcal{O}(s^3)&\;.
\end{align}
We naturally choose $\lambda$ to coincide with the value where we split the $LC(t_1)$ contribution in eq.~(\ref{eq:LC_split}) ($\lambda=470$~MeV). Automatically we can estimate $I_3$ 
from the near-to-the-origin contribution to $LC(t_1)$. That is, since the chiral counting is certainly valid up to this scale $\lambda$, our estimate to the error introduced by $G+t_1$ can be based on exposing the order of that counting,
\begin{align}
\label{jeronero}
|I_3[G+t_1](s,\lambda)|
&\simeq
\left|\frac{s^3}{\pi}\int^{0}_{-\lambda^2}dz\frac{k|z^3|}{z^3(z-s)}\right|= 
 \frac{ks^3}{\pi}\log \Big(1+
 \frac{\lambda^2}{s}
 \Big)\;,
\end{align}
where $k $ is a real positive constant encoding the third-order $\mathcal{O}\big(({s}/{4\pi f_\pi^2})^3\big)$ remnant dynamical contributions. For example, the uncertainty introduced by this part of the left cut far in the right $s$-cut on the resonance region,  $\lambda^2\ll s\simeq M_R^2$, can be given as
(note that $I_3[t_1]=\frac{s^3}{\pi}LC_{\text{near}}(t_1)(s)$)
\begin{equation}
    |I_3[G+t_1]|=|I_3[G]+I_3[t_1] |\propto \frac{s^2 \lambda^2}{\pi(4\pi f_\pi)^6}
\end{equation}
which is indeed of the same order as would be suggested by ChPT itself, if unitarity played no role in the discussion. \\

With the counting $k\sim 1/(4\pi f_\pi)^6$,
in hadron physics $\sqrt{s}\sim 0.77$ GeV, $f_\pi\sim 0.092$ GeV and $\lambda\sim 0.47 $ GeV, this treatment leads to a very acceptable uncertainty $I_3\sim 1\%$. In HEFT this is presumably similar upon rescaling $\lambda$ and $s$ to account for $v=246$ GeV instead of $f_\pi$.

In spite of $LC_{\text{near}}> LC_{\text{far}}$ when $t_1$ alone is considered, more generally, because of the near cancellation of ${\text{Im }G=-\text{Im }t_1}$, 
we find the low-$|z|$ contribution to be rather small.
\paragraph{\textbf{High-$|z|$ region ($|I_1|$).}}

The larger dividing scale $\Lambda$, defining region 1, is chosen to be able to invoke the Sugawara-Kanazawa (SK) theorem \cite{Sugawara:1961zz}. 
This theorem may be useful because it relates the asymptotic behavior of the right cut (dominated by unitarity, or further out by Regge theory, and in any case constrained by experimental data) to that of the left cut (about which much less is known and that we wish to constrain). 

To be precise, the theorem states that if the function $f(z)$ has the same analytic structure as the partial waves in Fig.~\ref{contour}; if it diverges at most as a finite power of $z$ as $|z|\to\infty$; if it has finite limits in the RC as $z\to\infty\pm i\epsilon$, $f(\infty\pm i\epsilon)$, and the limits at infinity in the LC exist, then

 \begin{align}
& \lim_{|z|\to\infty} f(z)=f(\infty+i\epsilon)\;\;\;\;\text{Im}\,z>0\nonumber\\
  &\lim_{|z|\to\infty} f(z)=f(\infty-i\epsilon)\;\;\;\;\text{Im}\,z<0\;.
  \end{align}
For example, the direct application of the SK theorem to the scattering amplitude, satisfying eq.~(\ref{unitarity}) (unitarity bound $|t|\leq 1/\sigma$) would entail that $\lim_{|z|\to\infty}t(z)/z=0$. However, it is not clear that the second of the hypotheses of the theorem, polynomial boundedness, is satisfied by partial waves in the scattering on composite objects~\cite{Llanes-Estrada:2019ktp}, since an exponential growth is expected for large imaginary $z$.
Luckily, for  
the inverse amplitude the exponential factor is actually damping. So that, for composite objects, $G$ vanishes as ${\rm Im}\;z\to  \infty$. 

Indeed, our interest is to apply the SK theorem  directly to $G$. 
In the entire construction of the IAM, from eq.~(\ref{disprel}) on, we have assumed that $t(s)$ does not decrease along the right cut faster than or proportionally to $1/s$, since  we subtracted $G$ three times (there is no obstacle in increasing the number of subtractions if $t$ decreases at large energy, but Regge theory suggests that this is not the case).
Actually, it is known that Regge theory gives a scaling of  $t(s)\propto \text{constant}\neq 0$
in the limit of large $s$ over the right cut for the $IJ=11$ channel \cite{Yndurain:2002ud}.

Standard deployment of the SK theorem also requires that any poles of $G(s)$
should lie between the cuts: however, the function $G(s)$  could present CDD poles along the cuts.
If their number  $N_C$ is {\it finite}, after identifying them  we can still use the theorem for $G$ by constructing an auxiliary function  of complex $s$ where the poles have been shifted to an acceptable position,
\begin{align}
\label{200910.1}
\widetilde{G}(z)=& G(z)\prod_{j=1}^{N_C}\frac{z-s_j}{z-\epsilon_j}.
\end{align}
There, $s_j$ are the original positions of the CDD poles of $G$ along the cuts and $\epsilon_j$ are the positions of an equal number of added poles that lie between the cuts, so that they  are accounted for by the SK theorem in standard form. 
In this
way, the auxiliary function $\widetilde{G}(z)$ has no CDD poles along the cuts and,
because of the SK theorem 
we then conclude that for $z\to -\infty\pm i\epsilon$ it has the same limit as in the RHC for $z\to +\infty\pm i\epsilon$, respectively.

Then, the function $G$ diverges  as $s^2$ and for fixed $s$ is bounded in magnitude by  $ k's^2\propto  s^2/(4\pi f_\pi)^4$.
Similarly,  we also know that $G$ is bounded by a constant times $|z|^2$ for $z\to \infty$ in any direction.

As a result of this analysis on the asymptotic behavior of $G(z)$, we conclude that
in the high energy regime we have the following bound on $|I_1[G]|$, 
\begin{eqnarray}
|I_1[G](s,\Lambda)|&\equiv&\frac{s^3}{\pi}\Big|\int_{-\infty}^{-\Lambda^2}dz\frac{\text{Im}\,G(z)}{z^3(z-s)}\Big|\leq \frac{k' s^2}{\pi}\log\left( 1+ \frac{s}{\Lambda^2}\right)\ .
\label{IG1}
\end{eqnarray}

To estimate the logarithm we choose $\Lambda=1.4 \text{ GeV}{\simeq \sqrt{2~{\rm GeV}^2}}$. 
This reasonable threshold for the division of intervals was used in earlier literature~\cite{Yndurain:2003vk,Oller:2007xd} to match the phase of the $\pi\pi$ scalar form factor with its smooth asymptotic expression given by the Brodsky-Farrar quark-counting rules \cite{Farrar:1979aw} for $s> \Lambda^2$, with good phenomenological success. 
Although these counting rules do not strictly apply to partial-wave amplitudes, that pick up Regge contributions due to the angular integration, 
(semi-local) duality also suggests that intermediate-energy resonances' contributions would average out already for $s\lesssim\Lambda^2$ to the smooth (Regge) asymptotic expression. Indeed Ref.~\cite{Yndurain:2002ud} also argues that above energies $|s|^{1/2}$ between  1.3 to 2~GeV, either perturbative QCD -in fixed angle problems- or
Regge theory -for partial wave amplitudes- are applicable.

An estimate of $k'$ in eq.~\eqref{IG1} can be obtained by matching the inferred asymptotic behavior of $\text{Im} G$ along the LC with an estimate of its value at  $s=-\Lambda^2$ ($|\text{Im}G(-\Lambda^2)|<k'\Lambda^4$). We use the following approximation,
\begin{align}
\label{201019.2}
k'\sim \frac{|\text{Im}G(-\Lambda^2)|}{\Lambda^4}\to\frac{|\text{Im} t_1(-\Lambda^2)|}{\Lambda^4}~,
\end{align} 
where we have taken into account that both $t_1$ and $G$ have the same $\propto s^2$ asymptotic behavior for $s\lesssim -\Lambda^2$, and we employ the former to ascertain the order of magnitude for $k'$.

This estimate of $k'$, once more in our reference scenario of ChPT and at the $\rho$ meson scale becomes 
$k'\simeq 0.15$ at $\Lambda$.
This figure can of course be changed by shifting $\Lambda$, passing to or bringing contributions from the intermediate energy region in the next paragraph.
This yields a variation of $0.05$ above or below,
taking $s$ in the range   $(-2~$GeV$^2$,$-1$~GeV$^2$).  
 eq.~( \ref{IG1}) then yields
$|I_1[G]|\sim 0.01 $
and we conclude that $|I_1[G]|$ is roughly $\mathcal{O}(1\%)$

We now combine eq.~(\ref{IG1}) with an equivalent piece of the left cut carrying $t_1$,
\begin{equation}
  (s^3/\pi)LC_{\text{far}}(t_1)\equiv 
I_1[t_1](s,\Lambda)  \ .
\end{equation}

Since for $I_1[G]$ we only estimated a quote for its modulus we add it in quadrature  with $I_1[t_1]$,
so that 
\begin{align}
&|I_1[G+t_1](s,\Lambda)| \lesssim
\left(\left[\frac{k' s^2}{\pi}\log\left( 1+ \frac{s}{\Lambda^2}\right)\right]^2+\frac{s^6}{\pi^2}\Big|\int_{-\infty}^{-\Lambda^2}dz\frac{\text{Im }t_1(z)}{z^3(z-s)}\Big|^2 \right)^{1/2}~.\label{IGT11}
\end{align}
$|I_1[t_1]|$ is evaluated, in ChPT and for the $IJ=11$ channel ($s=(0.77{\text{ GeV}})^2$), to be  $4.6\cdot10^{-3}$.  
Altogether, we have from eq.~\eqref{IGT11}  $|I_1[G+t_1]|\sim 0.01$.

\paragraph{\textbf{Intermediate-$|z|$ region.}($|I_2|$)}

This is the energy range that we have found most difficult to discuss, as the amplitude has no simple power-law behavior. 
We have attempted to lean on Mandelstam and Chew's work to express, through crossing, $\text{Im }t(s)$ over the left cut as a combination of partial waves over right cuts in different $IJ$ channels (see Eq (IV.6) in \cite{Chew:1960iv}). However, this characterization is not easily controllable for numerical computations. Therefore, we show three  possible simpler estimates that are broadly consistent.

For the first estimate we change in the quantity to be reduced,
\begin{align}
I_2[G+t_1](s,\Lambda,\lambda)&\equiv \frac{s^3}{\pi}\int_{-\Lambda^2}^{-\lambda^2}dz\frac{\text{Im}\,(G+t_1)}{z^3(z-s)} \ , 
\label{lodificil}
\end{align}
 the integral over $G+t_1$ by the integral over $t_1$ as a presumed upper bound, observing that they partially cancel each other (since in the range of lowest energies involved $\text{Im}G=-\text{Im}t_1+{\cal O}(s^3)$) and that $G$ is expected to be of the same order than $t_1$. We would have the bound
\begin{align}
I_2[G+t_1](s,\Lambda,\lambda)
&< |I_2[t_1](s,\Lambda,\lambda)|~.
\label{I2}
\end{align}

 This is the weakest point of the reasoning and could fail should $G$ turn out much stronger than $t_1$, which can happen for example  if a zero of $t$ resides in the intermediate left cut.  In such a case, 
 one should isolate it by considering  $\Im(G+t_1)(s-s_{\rm zero})/(s-s_1)$, with $0<s_1<4m_\pi^2$. We ignore how likely this is, so that in a practical uncertainty analysis it needs to be checked case by case employing eq.~\eqref{200725.9}. With the caveat, this should be sufficient for an uncertainty estimate with eq.~\eqref{I2}  ascertaining its typical order of magnitude.

Evaluating again eq.~\eqref{I2} for $s=(0.77{\text{ GeV}})^2$ in ChPT for the $IJ=11$, we see easily that
$|I_2[G+t_1]|\lesssim6\cdot 10^{-2} \simeq 6\%$. This would make it the biggest contribution to the uncertainty of all those examined, deserving further scrutiny.

We now follow a second, different path to $I_2[G+t_1](s,\Lambda,\lambda)$ so as to estimate the typical size expected for this contribution $I_2$, which avoids the discomfort of assuming that $\text{Im}G$ and $\text{Im}t_1$ are of similar magnitude. 
 Our point now is that higher derivatives respect to $s$ of the integral in eq.~\eqref{lodificil} are dominated by the integrand near the upper integration limit $z\lesssim -\lambda^2$ for positive values of $s\ll \Lambda^2$. 

To arrive to this result 
analytically, and for this purpose only, 
we take the low-energy expression $\text{Im}(G+t_1)= ks^3={\cal O}(s^3)$ with $k\simeq 1/(4\pi f_\pi)^6$, which is valid for $z\lesssim -\lambda^2$ in the upper part of the integrand. Thus,  
\begin{align}
\label{201024.1}
\int_{-\Lambda^2}^{-\lambda^2}dz\frac{\text{Im}\,(G+t_1)}{z^3(z-s)}
\to k\log\frac{s+\Lambda^2}{s+\lambda^2}~.
\end{align}
The $n_{\rm th}$ order derivative of the RHS is
\begin{align}
\label{201024.2}
&(-1)^{n-1} (n-1)!\left(\frac{1}{(s+\lambda^2)^n}-\frac{1}{(s+\Lambda^2)^n} \right)\approx \frac{(-1)^{n-1} (n-1)!}{(s+\lambda^2)^n}~, 
\end{align}
where the last step is valid for $s,\;\lambda^2 \ll \Lambda^2$, being quantitatively sensible  already for $n=1$. However, its primitive given by the RHS of eq.~\eqref{201024.1} reflects the assumption done for the higher-energy region of $\text{Im}(G+t_1)$. The general form for the primitive of the last term in eq.~\eqref{201024.2}  for $n=1$ is $\log(s+\lambda^2)+C$, where $C$ is a constant that, because of dimensional analysis, should be written as $-\log(L^2)$. Here $L$ is a hard scale reflecting the deeper energy tail of the integration interval. By estimating it we compute $C$, which can then be considered as a counterterm that we calculate in terms of a natural-size scale $L$. For numerical computations we allow  $L^2$ to take values between $\Lambda^2=2$~GeV$^2$ up to $2\Lambda^2=4$~GeV$^2$. The latter is  motivated because in many QCD sum rules  the onset of perturbative QCD input for the spectral functions is precisely taken at 4~GeV$^2$ \cite{Jamin:1994vr}. Then we have the following expression for $I_2[G+t_1]$ in eq.~(\ref{lodificil})  for positive $s,\,\lambda^2\ll \Lambda^2$,
\begin{align}
\label{201024.3}
I_2[G+t_1](s,\Lambda,\lambda)=&\frac{s^3 k}{\pi}\log\frac{s+\lambda^2}{L^2}~,~L^2\in[2,4]~\text{GeV}^2~.
\end{align}

Evaluating eq.~\eqref{201024.3} at  $s=(0.77{\text{ GeV}})^2$, we have 
$|I_2[G+t_1]|$ ranging between 1.7-3.5\%. 
As a check of consistency, we would like to mention that our estimate for $|I_2[G+t_1]|$ is consistent with the more reliable one for $|I_1[G+t_1]|$. Taking into account the scales driving both contributions, one would expect that $|I_2|\sim |I_1|\Lambda^4/(4\pi f_\pi)^4\simeq 2.4\%$, which is well inside our estimate for $|I_2|$.

Those two estimates can be deployed for HEFT as soon as data beyond the Standard Model is available. But in hadron physics we can perform a third, independent check with
an explicit calculation of  this intermediate energy region in eq.~(\ref{lodificil}). 

We have used,
for $I=J=1$, the amplitude $t(s)$ from Ref.~\cite{Pelaez:2019eqa},\footnote{We thank Jacobo Ruiz de Elvira for providing the data.} where the $\pi\pi$ $S$ and $P$ waves are parameterized analytically {\it in the physical region} based on the GKPY equation \cite{Garcia-Martin:2011iqs} and reproduce  $\pi\pi$ scattering data up to $\sqrt{s}=2$~GeV. 
The integral $|I_2[G+t_1]|$ evaluates to 0.08, of the same size but on the larger side of the first estimate. 
The reason is that this channel is close to the worst scenario because the $\pi\pi$ $P$ wave, as calculated from  Ref.~\cite{Pelaez:2019eqa}, becomes very small around $-0.36$~GeV$^2$, so that $|\text{Im } G|$ is much larger in this region than the typical values for any of the $S$ waves from the same reference.  This is due to the presence of zeroes in both the real and imaginary parts of the $I=J=1$ partial-wave amplitude that are almost coincident in energy. The exact relative difference between these zeroes strongly affects the result of $I_2[G]$. Therefore, an error estimate would be desirable (if baroque: an uncertainty on the uncertainty!) to provide a numerically accurate calculation of $I_2[G+t_1]$ for this partial-wave amplitude, beyond estimating its  order of magnitude as done here. Unfortunately, the uncertainty in the parameterization of~\cite{Pelaez:2019eqa} is presently not known for the left cut. Nonetheless, this estimate is conservative enough to fairly account for the magnitude.

\paragraph{\textbf{Total Left Cut uncertainty.}}\label{montarLC}

We finally put together the three  pieces $I_i[G+t_1]$ from eq.~(\ref{jeronero}), (\ref{IGT11}) and \eqref{201024.3} 
yielding the uncertainty to eq.~\eqref{IAMerr},
\begin{equation}
\Delta(s) G(s) = \sum_i I_i[G+t_1] = \sum_i I_i[G] + \sum_i I_i[t_1]    
\end{equation}


 Proceeding to the quantity controlling the displacement of the pole in eq.~\eqref{corrpoleposition}, we add in quadrature the different sources of uncertainties which moduli have been bounded previously.  Attending to the estimated typical size for them we have
 \begin{align}
\label{201019.5}
&\left|\Delta(s)G(s)\right|\leq 
\sqrt{0.01^2+0.03^2+0.01^2} \simeq 0.04~. 
\end{align}
This is then propagated to eq.~(\ref{corrpoleposition}) for estimating the displacement of the pole (see also eq.~(\ref{LCuncertain}) in subsection~\ref{scaling} below). The resulting typical size expected for the uncertainty in the pole position of the $\rho(770)$ is around a 
17\%, which is given also in the last line of Table~\ref{tab:wrapup}. 
 
In the extreme case that we use the parametrization for the $I=J=1$ $\pi\pi$ partial-wave amplitiude provided in Ref.~\cite{Pelaez:2019eqa} we have the larger uncertainty
\begin{align}
\label{201019.10}
&\left|\Delta(s)G(s)\right|\leq 
\sqrt{0.01^2+0.08^2+0.01^2} \simeq 0.08~, 
\end{align}
which would double the value of the relative error in the pole displacement up to $34\%$. However, this calculation is very sensitive to the {\it exact} position of the zeroes in the real and imaginary parts of the partial-wave amplitude, whose uncertainty in position is not provided. 

{\it A posteriori} in hadron physics we do of course know that the IAM \emph{predicts} the $\rho(770)$ at $710$ MeV with the  central values from threshold determination of the ChPT low energy constants\cite{Dobado:1997jx}, so that the error is of order 10\% and not 30\%.

This left-cut uncertainty is by far the largest entry in the table that can only be improved partially by including higher-orders in the chiral series.  This is because the largest uncertainty stems from the intermediate-energy region along the LHC, $I_2[G+t_1]$, which could be further suppressed, albeit only slowly, by including more subtractions since $m_\rho \lesssim \Lambda\sim 4\pi f_\pi$, being the latter the natural scale in ChPT suppressing loop contributions (like those generating $I_2$).  Nonetheless, a  better knowledge of this part of the LC could  sharpen our estimate of the uncertainty. 
This improvement would be likely associated with a better unitarization method compared to the NLO IAM.

\subsection{A comment on crossing symmetry violation}

While the Effective Theory may be crossing symmetric, as it is a Lorentz-invariant local field theory, the unitarization of its amplitudes is a procedure that, as we have seen, treats the left and right cuts in a different manner. This leads to questions about to what extent can crossing symmetry be respected~\cite{Cavalcante:2001yw,Cavalcante:2002bit}.

Crossing symmetry is manifest when comparing the three isospin amplitudes in $SU(2)$ Goldstone boson scattering,
\begin{align} \label{isospindec}
&T_0(s,t) = 3A(s,t,u) + A(t,s,u) + A(u,t,s) \nonumber \\
&T_1(s,t) = A(t,s,u) - A(u,t,s) \nonumber \\
&T_2(s,t) = A (t,s,u) + A(u,t,s)
\end{align}
all three of which are written in terms of only one function $A(s,t,u)$ analytically extended to different Mandelstam kinematic regions and with the arguments swapped.

If $A(s,t,u)$, $A(t,s,u)$ and $A(u,t,s)$ were three independent complex functions $A$, $B$ and $C$, then they could be reconstructed by separately applying the IAM to each partial wave of fixed isospin, and employing the partial-wave projection in each channel to reconstruct the three $T_0$, $T_1$, $T_2$, then solving the linear system in eq.~(\ref{isospindec}) to obtain all three. But the fact that $A$, $B$, $C$ are analytic extensions of each other with the arguments swapped leads to subtle relations between them. 

A practical way to expose them is to encode the information in integral relations between the partial wave amplitudes, the Roy equations~\cite{Roy:1971tc}. However, these equations relate partial waves with different angular momenta and the number of computations accessible to the NLO IAM is limited to $J<2$. Therefore, neither a reconstruction of $A$ nor a test of the Roy equations is really sensible: the IAM at NLO is not predictive enough to be tested by crossing symmetry. What it does, parametrize a few low-lying partial waves into the resonance region, it does reliably enough as we have exposed through the work, but its reach is not sufficiently extended  to make a statement about crossing.

Nevertheless, a direct attempt to test crossing in the physical region from eq.~(\ref{isospindec}) found sizeable violations~\cite{Cavalcante:2001yw};  But any reasonable parametrization of the data, not only the IAM, will find very similar results. Either crossing cannot be precisely tested without higher angular momentum waves, or the data points themselves are violating crossing symmetry (!).

Such attempt at testing crossing with the IAM at very low $s$ where only the first partial waves contribute to the amplitude~\cite{Cavalcante:2001yw} employs the Roskies relations~\cite{Roskies:1970uj}. These are integral relations for the amplitudes between $s=0$ and threshold. The first few relations are very well satisfied~\cite{Nieves:2001de} to $\mathcal{O}(1\%)$, and the agreement only deteriorates upon increasing their order. But this is a region where the IAM (once the Adler zero is taken care of: the mIAM was not known in 2001, but already then an approximate subtraction of the sub-threshold zeroes was carried out~\cite{Cavalcante:2001yw}) essentially coincides with chiral perturbation theory, so that the test should be ascribed to the uncertainty of the EFT itself, and not to that of the unitarization method.

\subsection{Behavior and scaling of pole position uncertainty}\label{scaling}

Complementing subsection~\ref{subsec:KK} we here present a simplified analytical derivation of the behavior of the pole position and its relative displacement due to the two kaon inelastic channel. These simple results are consistent with a numerical check on the displacement of the $\rho$ (this derivation is extensible to all the other sources of uncertainty). \par
 In the chiral limit, the real part of the standard IAM equation for the pole position, $s_R$, eq.~(\ref{poleposition}) can be parametrized as
 \begin{equation}
     as_R-bs_R^2=0\;.\label{pato0}
 \end{equation}
 (The logarithm multiplying the NLO $s^2$ piece, subleading to the power, need not be kept for the purpose of estimating the uncertainty.)
 This is solved by $s_R=a/b$. When the two body coupled channel correction in (\ref{2body}) is included we have to modify  eq.~(\ref{pato0}), using  eq.~(\ref{corrpoleposition}), as
  \begin{equation}
     as_R-bs_R^2=cs_R^2\;.\label{pato}
 \end{equation}
 Assuming the correction is sufficiently small, the pole displacement is
 \begin{equation}
     s_R\rightarrow\frac{a}{b}\Big(\frac{1}{1+\frac{c}{b}}\Big)\simeq s_R\Big(1-\frac{c}{b}\Big)=s_R\Big(1-s_R\frac{c}{a}\Big)\;.\label{alo}
     \end{equation}
The constant $a$ can be taken from LO ChPT, for example, in the vector channel, $t_0=as\Rightarrow a=-1/96\pi f_\pi^2\simeq0.4\text{ GeV}^{-2}$, and to estimate the correction term $c$ (sloppily, since this was an upper bound and we ignore the expression's sign) $cs_R^2 \sim \Delta(s_R)G(s_R)=0.08\frac{s_R^3}{\pi}RC(t_1)(s_R)\Rightarrow c\simeq 2\cdot10^{-4} \text{GeV}^{-4}$. This means that we can compute the displacement of the pole in eq. (\ref{alo})
 \begin{equation}
     s_R\frac{c}{a}\simeq0.001\;.
 \end{equation}
 So that  the $\rho$ pole is uncertain at order $10^{-3}$ by a possible kaon coupled-channel induced displacement. {The behavior with energy of this uncertainty is readily obtained
 \begin{equation}
  |\Delta(s)G(s)|=   |cs^2|\simeq 5\cdot10^{-4}\Big(\frac{\sqrt{s}}{4\pi f_\pi}\Big)^4\;.
 \end{equation}}
 
 This treatment can be immediately extended to the uncertainties coming from four particle coupled channels and from the next order in perturbation theory in the EFT, as both sources scale as $s^2$. For the first we obtain 
 a displacement of the pole of order $10^{-4}$. 
The second source of uncertainty affects the pole's position at order $10^{-2}$. \par

The largest error in the budget comes from the left cut. We have that
$c s_R^3\sim 0.04$, hence the pole gets displaced
\begin{equation}
     s_R^2\frac{c}{a}\simeq0.17\;.\label{LCuncertain}
 \end{equation}

 So that the pole may get displaced at order $10^{0}$ upon approximating the left cut. The energy behavior of this uncertainty is $(m_\rho/f_\pi)^6$ due to the scaling of the intermediate part of the left cut uncertainty (eq. (\ref{201024.3})).

\section{Outlook} \label{sec:outlook}

\subsection{Summary of systematic uncertainty sources}

  \begin{table*}[]
   \makebox[\linewidth]{ \begin{tabular}{lc|cc|l} \hline
        Source of uncertainty & Equation & Behavior & Pole displacement at $\sqrt{s}=m_\rho$  &  Can it be improved?\\ \hline\hline
        Adler zeroes of $t$  & (\ref{mIAM}) & $(m_\pi/m_\rho)^4$  & $10^{-3}$ - $10^{-4}$ &  Yes: mIAM \\
        CDD poles at $M_0$&
        (\ref{200725.6}) &
        $M_R^2/M_0^2$ &
        $0\,$-$\,\mathcal{O}(1)$ &
        Yes: extract zero \\
        Inelastic 2-body & (\ref{2body})  & $(m_\rho/ f_\pi)^4$  &  $10^{-3}$ & Yes: matrix form \\
        Inelastic 4...-body & (\ref{4body}) &$(m_\rho/ f_\pi)^8$ &$10^{-4}$ &  Partially \\
        $O(p^4)$ truncation  &(\ref{3PT})  & $(m_\pi^2 m_\rho^4)/ f_\pi^6$ & $10^{-2}$ & Yes: $O(p^6)$ IAM 
        \\
        Approximate Left Cut  & (\ref{201019.5})  & $(m_\rho/ f_\pi)^6$ & $0.17$ 
        & Partially  \\
    \hline
    \end{tabular}}
     \caption[Different sources of uncertainty for the Inverse Amplitude Method]{Different sources of uncertainty for the Inverse Amplitude Method in Hadron Physics, their scaling if relevant, the order of magnitude of the error they introduce in the resonance region and whether the basic method can be improved to remove that source of uncertainty if need be.    
}  \label{tab:wrapup}    
\end{table*}

It is looking increasingly likely that the LHC will not find new resonances, but it may still have a chance of revealing non-resonant~\cite{Folgado:2020utn} separations from Standard Model cross-sections, particularly in the EW SBS discussed in the previous chapter. 

If that is the case, extrapolation of the data to higher energies will be needed to determine the new physics scale (see the author's handbook on how to do so \cite{Salas-Bernardez:2022xk} with the IAM). One tool to do this is unitarization of the low-energy EFT, and controlling the uncertainties in the method is therefore necessary. Among the many versions of unitarization, those that have an underlying derivation in terms of a dispersion relation are more amenable to controlled systematics on the theory side. We have examined the Inverse Amplitude Method, salient among unitarization procedures and well understood in theoretical hadron physics, but whose uncertainties have not yet been systematically listed. 

Table~\ref{tab:wrapup} summarizes the sources of uncertainty that we have identified and the status of each (whether it is or not amenable to systematic improvement).

In the table we have explicitly spelled out our knowledge of the scaling with energy of the various contributions to the uncertainty (see subsection \ref{scaling}). If instead of the $\rho(770\ {\rm MeV})$ in QCD, or an analogous particle with proportional mass ({\it e.g.} $Z'$) in EW model extensions, we had taken a resonance with a larger mass 
(above that unitarity scale $\Lambda_{\rm U}$ of $4\pi f_\pi$ in QCD, or the generic $4\pi v$ in a BSM theory)
the entries in table~\ref{tab:wrapup}  labelled ``Partially'' would become a definite ``No''. For the remaining entries one could set in the estimates the mass of the resonance around 1.5~GeV (around twice the $\rho$ mass) and calculate the numbers.

For resonances originating below the fundamental cutoff,  such as the renowned $\sigma/f_0(500)$ meson of QCD,  the estimated uncertainty would be even smaller than found here (for the $\rho(770)$), even though it is a wide resonance. 
Specifically, the estimate of the largest uncertainty source along the 
LC coming from the intermediate energy region in eq.~\eqref{201024.3} scales as $s^3$.
Because $|s_\sigma/M^2_\rho|^3\approx 0.11\ll 1$, with $s_\sigma$ the pole position of the $\sigma/f_0(500)$ resonance \cite{Garcia-Martin:2011nna}, the uncertainty is indeed much smaller.

It is true that dealing with such a broad resonance could modify the straightforward applicability of eq.(\ref{poleposition}) to propagate the error bar to the pole position. But even then, we think that the calculated uncertainties would be sensible at the semiquantitatively level at least. In any case, one could exquisitely perform the error propagation by attending directly to the resonance pole position in the second Riemann sheet by extending eq.~(\ref{corrpoleposition}).

Thus, as long as we are below the unitarity scale $4\pi f_\pi$ (or $4\pi v$), the scaling of the uncertainty can be estimated directly from table~\ref{tab:wrapup}, although the numerical coefficient of that scaling does depend on the $IJ$ channel.

The dominant $\left( m_{\rm resonance}/f_\pi \right)^6 \sim \left( m_{\rm V}/v \right)^6$ dependence (coming from the left cut uncertainty) eventually becomes of order 1 and the method is overpowered by the error. It then loses its applicability in its known form as presented here.
In QCD this would happen when $1\sim 0.17 \times (m_{\rm max}/m_V)^6$, {\it i.e.}, for $m_V\simeq 770$ MeV, 
$m_{\rm max}\simeq 1050$ MeV (that is in the ballpark of $4\pi f_\pi$ where the entire EFT setup becomes dubious anyway).\par 

Also interesting is the scaling of the low-energy constants~\cite{Guo:2009hi} in table~\ref{tab:constants} that we have used to estimate the uncertainty in the truncation at a given perturbative order (NLO). The NNLO constants governing the uncertainty behave as 
$f_\pi^4/m_V^4$, 
which is a very steep dependence: the NLO $l_i\propto f^2/m_V^2$ see a slower fall off and therefore are more important at low energy. 
The NNLO constants tend to become less visible as the mass of the resonance increases.
\par
And last, the inelastic four-Goldstone boson channel becomes of the same order as the two-Goldstone boson channel at about the same energy $\Lambda_{\rm U}$ (1.2 GeV for hadron physics) as inferred from our computation of the relevant phase space in subsection~\ref{subsec:fourpi}. 
An extension of the IAM to a more complicated method is then mandatory.

\subsection{Comparison to statistical uncertainties from eventual measurements}

We have dealt with theoretical systematic uncertainties. But eventual experimental measurements of the low-energy constants will have an uncertainty, presumably very dominated by statistical fluctuations at initial stages.

\begin{figure}[ht!]
\centerline{\includegraphics[width=0.65\columnwidth]{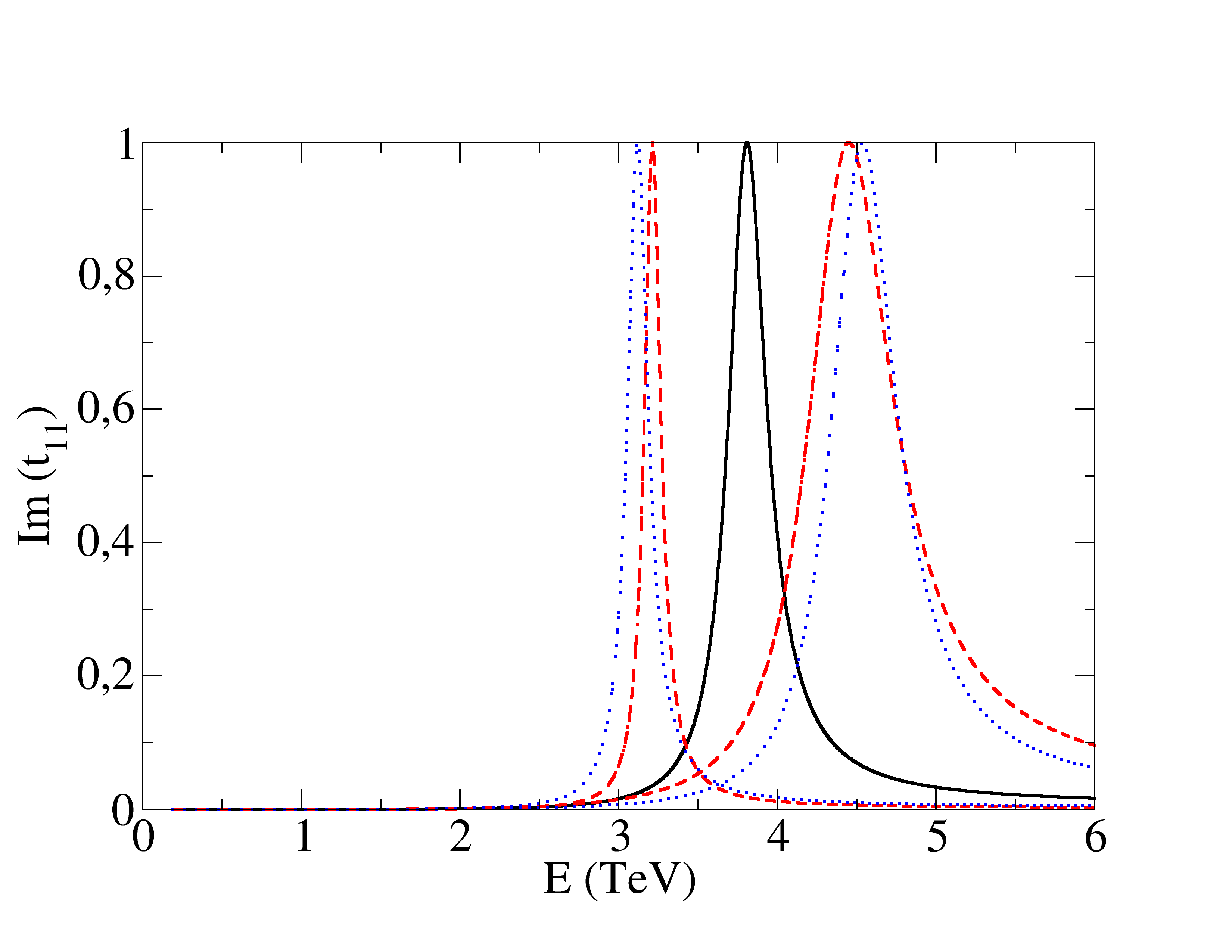}}
\caption[Vector-isovector resonance generated with the HEFT Lagrangian]{\label{fig:lecuncertainty}
Solid line in the middle: vector-isovector resonance generated with the HEFT Lagrangian with parameters $a=0.95$ and $(a_4-2a_5)=10^{-4}$ at $\mu=3$ TeV.
The mass of the resonance turns out to be about 3.8 GeV. Dashed lines to either side (red): recalculated resonance with $a=0.965$ and $a=0.93$. Dotted lines (blue): alternatively, $a$ is fixed at 0.95 and $a_4-2a_5$ increased by 50\% or decreased by 30\%, respectively, with similar result. (The larger the separation from the Standard Model values $a=1$, $a_4-2a_5=0$, the further to the left).
}
\end{figure}

The question of interest here is to what level do these  uncertainties on the parameters of the low energy theory have to be pushed down so that the theory uncertainties in this chapter become the pressing issue. This is exemplified in figure~\ref{fig:lecuncertainty}.

The plot shows a computed elastic $\omega_L\omega_L$ vector resonance chosen to show up at 3.8 TeV, approximately. A 17\%-sized uncertainty band, as suggested  by the largest entry in table~\ref{tab:wrapup}, would make the resonance mass uncertain between about 3.2 and 4.4 TeV.
This may seem like a large uncertainty, but note that in the graph there are additional plots obtained by varying the low-energy constants that govern this $J=1=I$ channel. The variation levels that are roughly equivalent to the IAM systematics are
\begin{equation}
\frac{\Delta (1-a)}{(1-a)} \simeq 
\left\{ \begin{tabular}{c} $+30\%$ \\ $-40\%$ \end{tabular} \right\}\ ; \ \ \ \ \
\frac{\Delta (a_4-2a_5) }{(a_4-2a_5)} \simeq \left\{ \begin{tabular}{c} $+50\%$ \\ $-30\%$ \end{tabular} \right\}\ .
\end{equation}

This means that, before the IAM systematic uncertainty became dominant, the measurement of the LO BSM coefficient $(1-a)$ as well as the NLO coefficient $(a_4-2a_5)$ would need to be accurate at a level of 30\%. Presently, $a$ is compatible with 1 at the 10\% level and no deviation from the SM is seen at LO;
but how soon that 30\% might be achieved for the NLO combination cannot be easily answered. 

This comparison would have to be carried out in each particular application of the analysis for a given experimental channel.

\subsection{Conclusion}

We hope to have mapped the grounds that systematize the study of the IAM's uncertainties and that it may one day be useful at the LHC and inspire further studies for this or other unitarization alleys.  
Our numerical estimates have most often  relied on known hadron-physics parameters and phenomena, but they are expected to be similar in Higgs Effective Field Theory beyond the Standard Model (because the interplay between derivative couplings and unitarity is universal once the correct scale has been fixed) so they can be used as a rough guide in BSM searches if any HEFT parameter separates from the Standard Model.

{\it A posteriori}, we know that the IAM in hadron physics yields, in the channel where we have concentrated most, the vector-isovector $\rho$ resonance, a calculated mass $m_\rho\simeq 710$ MeV instead of $770$ MeV~\cite{Dobado:1997jx}, when using the central values of the chiral counterterms fitted to low-energy data with ChPT.  This is an error of order 8\%, that is way better than the $ 20\%$  uncertainty that we have estimated {\it a priori} for the largest source thereof,
and have given in table~\ref{tab:wrapup}.  Thus, the IAM seems to be faring even better than we have a right to state with the uncertainty analysis that we have carried out, and since this is dominated by the left-cut, it is likely that our bounds can be improved in the future.

Very clearly, what we estimate to be the major source of uncertainty, which should not surprise practitioners in this field, is the approximated left cut, which we have the most trouble constraining and easily brings a 10\% loss of precision in extrapolation to high energy. But this should be sufficient, if separations from the SM are identified, to ascertain what energy scale should a future collider reach to be able to study the underlying new physics.

As for the common usage of the Inverse Amplitude Method, we should also emphasize that, prior to a search for resonance poles, one needs to examine the amplitude (through its chiral expansion) for the possible presence of zeroes (see subsection~\ref{subsec:mIAM}). If one is present in the resonance region, typically driving the partial-wave amplitude to having a narrow resonance with a zero close to its mass, the expansion and construction of the IAM have to be modified to account for it. 
We have introduced here the necessary methodological modification for such circumstances.

\setcounter{chapter}{5}
\setcounter{section}{0}
\chapter{QCD and QFTs in coordinate space}\label{ch:CS}
\setcounter{equation}{0}
\setcounter{table}{0}
\setcounter{figure}{0}
In Quantum Field Theory (QFT), Green's functions are the mathematical entities used to relate the theoretical framework to physical processes (as introduced in chapter 2). In principle, if all the Green's functions of a particular theory are known, then the theory is solved completely (since these functions are introduced to solve the quantum equations of motion for the fields). Usually, the computation of Green's functions is performed in momentum space due to mathematical simplicity. For example, the momentum space DSE that these functions satisfy were deployed in chapter 2. Furthermore, scattering and collider experiments, which are among the main methods and tools to unravel the structure of elementary particles, are naturally formulated and studied in momentum space. Here, the requirement of unitarity (forcing conservation of probabilities) manifests itself as the optical theorem and helps identifying resonances or bound states that the theory possesses (see chapter 4). Moreover, huge efforts were made at the end of last century to justify the consistent use of perturbation theory in Quantum Chromodynamics (QCD). These resulted in the factorization theorems for amplitudes in momentum space, which, by means of Resummation, ensured the convergence of the perturbation series in the weak coupling regime (see \cite{Collins:1989bt} or \cite{Sterman:1995fz}). On the other hand, analogue results in coordinate space only appeared in recent years \cite{Erdogan:2013bga}, signaling the little attention that the coordinate space description has received in high-energy physics research. Seeking new perspectives and interpretations, we will review several aspects of QCD amplitudes from the coordinate-space viewpoint.

Based in the author's publication \cite{Salas-Bernardez:2022cuw}, in section \ref{sec:explicitcomputationjetCS} we will study the topics of factorization of QCD amplitudes in both momentum and coordinate spaces, aiming at explicitly computing some contributions to the jet function in coordinate space and gaining some intuition on the relevant physical configurations involved. In section \ref{sec:fopt}, also based in the author's original article \cite{Borinsky:2022msp}, we introduce a novel representation of massless scalar field theories in coordinate space. This representation equips us with a nice picture in terms of energy flows through Feynman graphs as well as having promising features in terms of factorization of IR singularities on a per-diagram level, among other curious features.

\section{Explicit computation of jet-functions} \label{sec:explicitcomputationjetCS}
The factorization theorems mentioned above are produced in the context of another EFT appropriate for the study of hadronic jets (usually in $e^-e^+\to 2$ jets processes). This EFT is the so called Soft Collinear EFT (SCET), the small invariant mass of a jet, $m_\text{jet}$, compared to the centre of mass squared energy of the process, $Q^2$, provides a natural expansion parameter $\lambda \sim m_\text{jet}/Q$. In the centre of mass, the two jets produced travel back-to-back with directions $n^\mu=(1,\hat{n})$ and $\bar{n}^\mu=(1,-\hat{n})$, we can hence decompose any momentum $p^\mu$ as $p^\mu=(n\cdot p,\bar{n}\cdot p,p_\perp)=(p^+,p^-,p_\perp)$. If we consider now an energetic quark moving along the $n$ direction, this particle can split into several particles moving approximately in the same direction. These particles will be called collinear and have a momentum scaling of
\begin{equation}
    p_\text{col}\sim Q(\lambda^2,1,\lambda)\;.\label{eq:colscaling}
\end{equation}
Also, it is possible that particles emit low energy radiation with scaling
\begin{equation}
    p_\text{soft}\sim Q(\lambda^2,\lambda^2,\lambda^2)\;.\label{eq:softscaling}
\end{equation}
Usually, poles will arise in loop integrals of a corresponding Feynman diagram when the loop momenta lie in hypersurfaces defined by (\ref{eq:colscaling}), (\ref{eq:softscaling}) and $\lambda \to 0$. These will be identified as ``pinch surfaces'' below and they produce leading contributions to the cross section of the process. The SCET aims hence at capturing the behavior of QCD's high energy dynamics, enabling us to compute IR singularities of QCD from the SCET. Physically, collinear modes encode the interactions of particles inside a jet. While the low energy interaction between jets is encoded in the soft modes. We will explain these concepts in detail below.

This section is organized as follows: subsection \ref{momfact} is devoted to offer a concise introduction to the topics regarding Leading Power (LP) factorization of QCD amplitudes in the perturbative regime. Here Landau's equations are introduced as a condition for having an unavoidable singularity in a momentum space Feynman amplitude. These equations for the one-loop quark electromagnetic form factor, together with the Coleman-Norton picture, already elucidate the all-order structure of divergences in the $q\bar{q}\gamma$ vertex, \textit{i.e.} a hard function, two jets and a soft function which will be explained. After power counting the possible singularities, factorization of the vertex is possible. The momentum space factorization results were translated to coordinate space amplitudes in O. Erdogan's work \cite{Erdogan:2013bga} and are summarized in subsection \ref{coordfact} . Using these results we compute in subsection \ref{computations} the one-loop jet function in coordinate space. \\

In recent years, within the community of QCD phenomenologists, interest has grown in the all-order structure of the so called subleading regions. These kinematical regions can enhance non-analyticites in certain observables such as cross sections (see f.e. the first equation in \cite{Bonocore:2015esa}), usually related to the emission of soft and collinear gluons. Explicitly, at Next-to-Leading Power (NLP) in the soft and collinear expansion, one deals with the so-called ``radiative jets'' (introduced for the first time in \cite{DelDuca:1990gz}). 
Given that the study of radiative jet functions in coordinate space could give interesting insights on the still incomplete NLP all-order factorization of QCD (many recent developments show how rich and complicated is the structure of these subleading regions \cite{Gervais:2017yxv,Beneke:2017ztn,Moult:2019mog,Laenen:2020nrt,Liu:2021mac}), in section \ref{radiativesection} we will reduce to Feynman parameter integrals two contributions to the radiative one-loop jet function in coordinate space. We also analyze Landau's equations for all contributions to the Abelian one-loop radiative jet function in coordinate space.

\subsection{Landau Equations and Leading Power Factorization}\label{momfact}
Amplitudes in QFT encode the probabilities of certain processes contributing to a particular outcome of a collider experiment with known initial conditions. Usually, amplitudes present singularities, and their characterization is important to identify their associated kinematical configurations. Each contributing process to an amplitude is represented by a Feynman diagram or graph. A general Feynman diagram in $d\equiv4-2\epsilon$ spacetime dimensions $G(p_1,...,p_n)$ is given by (leaving out coupling constants for now)
\begin{equation}
G(p_1,...,p_n)=\prod_{i=1}^L\int \frac{d^dk_i}{(2\pi)^d} \prod_{j=1}^I\frac{N(\{k_i\},\{p_r\})}{l^2_j+m_j^2-i\eta}\;,\label{graph}
\end{equation}
where the $\{p_r\}$ are the external momenta, $\{k_i\}$ the loop momenta, $\{l_j\}$ the line momenta (which are linear combinations of the loop and external momenta) and $m_j$ is the current mass for the particle propagating in line $j$. $L$ is the number of loops,  $N(\{k_i\},\{p_r\})$ is an arbitrary polynomial in the momenta and $I$ the number of internal lines. Introducing Feynman's parametrization in (\ref{graph}) we obtain
\begin{align}
G=&(I-1)!\prod_{i=1}^L\int d^dk_i\prod_{j=1}^I\int_0^1d\alpha_j\delta(1-\sum_{j=1}^I\alpha_j)\frac{N( \{k\},\{p\})}{(D(\{\alpha\}, \{k\},\{p\}))^I}.\label{eq:land1}
\end{align}
The graph $G$ will be infrared (IR) divergent if its denominator unavoidably vanishes, $D\equiv\sum_{j=1}^I\alpha_j(l_j^2+m_j^2)=0$. To see what ``unavoidably'' means notice that $D$ is a quadratic function in the momenta and therefore it has no more than two zeroes in any momentum component $l_i^\mu$. If these zeroes are at real values of $l_i^\mu$ we will encounter singularities of the integrand in (\ref{eq:land1}). However, if the contour of integration can be deformed in each of the integration variables' complex plane, then by virtue of Cauchy's theorem the integral will be well defined and will not present singularities. \par 
There can be two cases where deforming the integration contour will not be possible:
\begin{itemize}
\item If the two zeroes in $k^\mu_i$ merge at the same point and ``pinch'' the contour of integration. This condition amounts to, 
\begin{equation}
\frac{\partial D(p_r,k_i,\alpha_j)}{\partial k_i^\mu}\Big|_{D=0}=0\;,\label{EQ:111}
\end{equation}
which by the definition of $D$ means
\begin{equation}
\sum_{j=1}\alpha_j\frac{\partial l^2_j(p_r,k_i)}{\partial k_i^\mu}=\sum_{i\in loop \;j}\alpha_jl_j^\mu\epsilon_{j,i}=0\;.
\end{equation}
Where the incidence matrix element $\epsilon_{j,i}$ is $+1$ if the line momentum $l_j$ in the loop $i$ flows in the same direction as the loop momentum $k_i$, $-1$ if in the opposite direction and zero otherwise. The sum runs over all the edges in loop $j$.
\item If the singularity is at the endpoints of the contour of integration then we will not be able to apply Cauchy's theorem: we cannot modify the endpoints of integration without affecting the result of the integral. Since $k^\mu_i\in \mathbb{R}$ this type of singularities corresponds to Ultraviolet (UV) divergences and are taken care by renormalization. For $\alpha_j$ integrations these singularities are important when $\alpha_j=0$ or, if $D$ does not depend on $\alpha_j$ (meaning that $l_j^2=-m^2_j$). Either one of these two conditions on all of the $\alpha$s has to apply in order to have an unavoidable singularity.
\end{itemize}
Hence, we can condense the necessary conditions for unavoidable divergences in the Landau Equations, that must be simultaneously satisfied,
\begin{empheq}[left=\empheqlbrace]{align}
\sum_{j\in loop \;i}\alpha_jl_j^\mu\epsilon_{j,i}=0 &\;\;\;\; \forall \mu,i \nonumber\\
  \alpha_j(l_j^2+m_j^2)=0 & \;\;\;\;\forall j\;\;\;.   \label{eq:Landau}
\end{empheq}
 To verify that the solutions are indeed unavoidable singularities we have to resort to the method of power counting,  i.e. count the divergence of the integrand and the volume of integration, to see if they give indeed a power or logarithmic divergence when approaching certain divergent kinematical configurations. Furthermore, in the next section we will see that these configurations must be allowed by classical free-particle propagation of the internal lines.

\subsubsection{Coleman-Norton picture in momentum space}
Finding solutions to Landau's equations (\ref{eq:Landau}) by hand is not an easy task, especially for higher-order diagrams. However, owing to Coleman and Norton (CN) \cite{Coleman:1965xm}, there is a much easier and intuitive procedure to solve the equations.\\

Recall that, in a solution to Landau's equations, for off-shell lines we have $\alpha_j=0$ while for an on-shell internal line we will have that $\alpha_j\neq0$ and ${\partial D}/{\partial k_i^\mu}=0$. Now, if we identify the products $\alpha_j l_j$ for each on-shell line with a spacetime vector (introducing a parameter $\kappa$)
\begin{equation}
 \Delta x_j^\mu\equiv\kappa\alpha_jl_j^\mu\;,
\end{equation}
and $\lambda\alpha_j=\Delta x_j^0/l_j^0$ as the Lorentz invariant ratio of the time component of $\Delta x_j^0$ to the energy $l_j^0$, then we have that
\begin{equation}
 \Delta x_j^\mu=\Delta x_j^0v_j^\mu\;,
\end{equation}
with the four-velocity $v_i^\mu=(1,\vec{l}_j/l_j^0)$.  Notice that the parameter $\kappa$ has dimensions of length squared to keep the Feynman parameter dimensionless and its introduction has a subtle meaning. For collinear divergences it can be set to unity but it is necessary in the soft case to keep the displacement $\Delta x^\mu$ finite even when all components of $l^\mu$ go to zero. Since soft gluons have almost infinite wavelength, it is natural to think that they will have a finite displacement as classical particles.

 Summarizing, $\Delta x_j^\mu$ may be thought of as a four-vector describing the free propagation of a classical on-shell particle with momentum $l_j$. In this way, Landau's Equations become

\begin{empheq}[left=\empheqlbrace]{align}
\sum_{j\in loop \;i}\Delta x_j^\mu\epsilon_{j,i}=0 &\;\;\;\; \text{if}\;\;l^2_j=-m_j^2 \nonumber\\
  \Delta x_j^\mu=0\;\;\;& \;\;\;\;\text{if}\;\;l^2_j\neq-m_j^2\;.   \label{eq:Landau2}
\end{empheq}

This means that the ``pinch'' condition for on-shell lines amounts to the requirement that every loop made out of these lines is a closed classical path and that off-shell lines are shrunk to a point (i.e. they do not propagate a finite distance). We hence construct all the possible displacement configurations inside the loops (shrinking or taking the lines as soft) and then decide if these are allowed by the C-N picture. We will regard these as \textbf{reduced diagrams}. To illustrate this method we now turn to an example.
\paragraph{One-loop Quark Electromagnetic Form Factor reduced diagrams}

Let's now work on an illustrative example of the application of the Coleman-Norton trick to find possible ``pinched'' singularities. In Feynman gauge, the only relevant diagram contributing to the QCD one loop correction to the quark-quark-photon ($q\bar{q}\gamma$) vertex is
\begin{equation}
\centering
\begin{tikzpicture}
\begin{feynman}
\vertex (a);
\vertex [below=1.3 cm of a] (b);
\vertex [below=0.1cm of b] (H){};
\vertex [right=2.5cm of b] (Hh){$\equiv-ie\,\Gamma_{(1)}^\mu(p_+,p_-)$};
\vertex [below right=1.4cm of b] (k);
\vertex [below right=1cm of k] (t);
\vertex [below left=1.4cm of b] (f1);
\vertex [below left=1cm of f1] (c);

\diagram* {
(a) -- [photon] (b);  (k)-- [gluon] (f1);
(b)-- [] (k);
(k)-- [fermion] (b);
(k)-- [momentum=$p_-$] (t);
(t)-- [fermion] (k);
(f1)-- [fermion] (c);
(c)-- [momentum=$-p_+$] (f1);
(f1) -- [] (b);
(b) -- [fermion] (f1)};
\end{feynman}
\end{tikzpicture}\nonumber\;.
\end{equation}
In physical gauges, there are extra contributions to the form factor coming from pinch surfaces in self energy diagrams.
There are several phenomenological uses of this vertex, some examples can be found in \cite{Bicudo:1998qb,Brodsky:1983vf}.\\

Which reduced diagrams (i.e. diagrams with shrinked lines or with soft lines) of the diagram above are allowed by classical free propagation between vertices?
\begin{figure}[ht!]
\centering
\includegraphics[width=0.7\columnwidth]{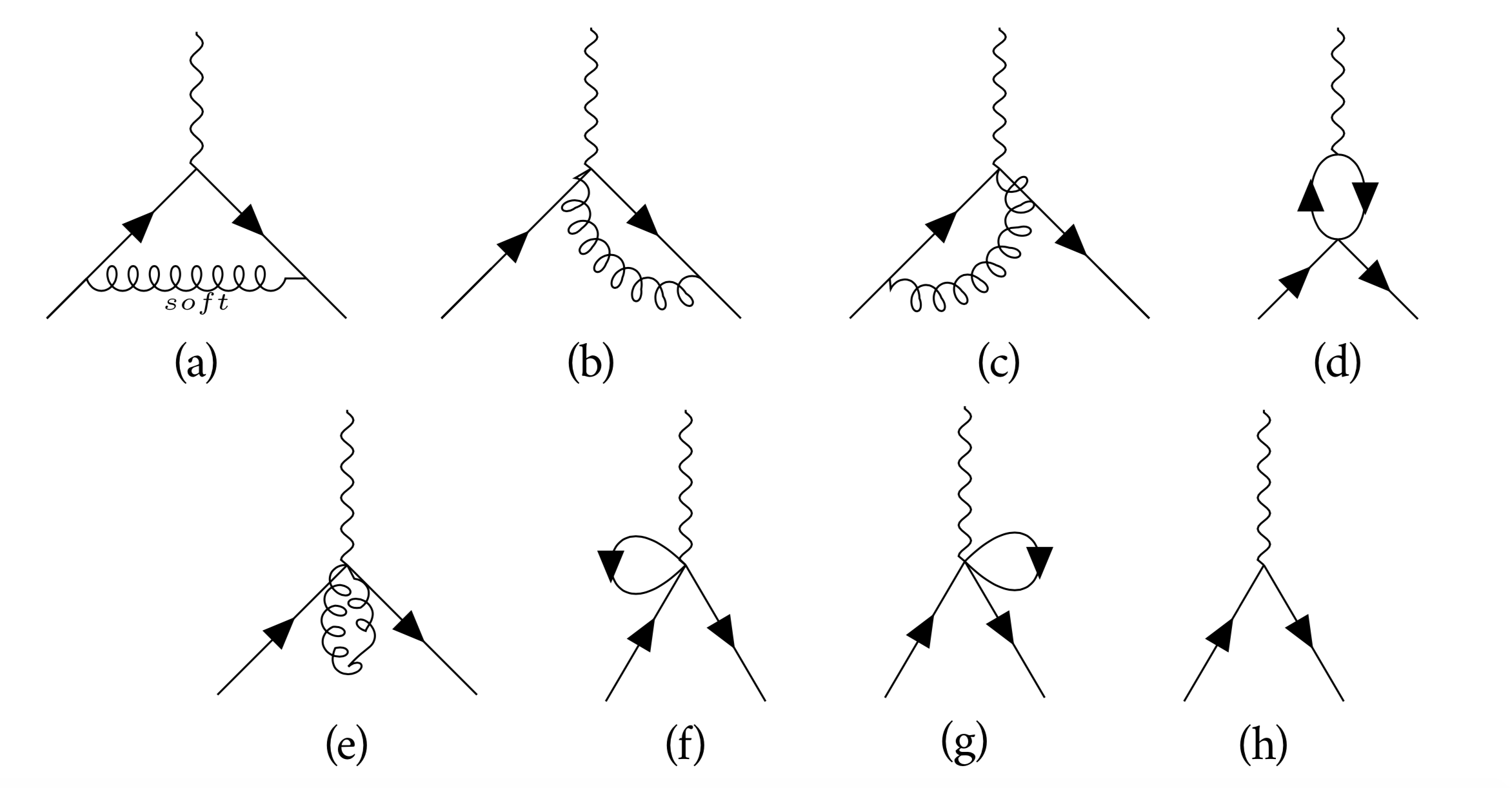}
\caption{Reduced diagrams for the one-loop QCD correction to the $qq\gamma$ vertex.}
\label{fig:reddiag}
\end{figure}

 We can see all the reduced diagrams in Fig. \ref{fig:reddiag}, each corresponding to an \textit{a priori} solution to Landau's equations. Thanks to the Coleman-Norton Trick we can immediately rule out the diagrams (e)-(g) since a particle that leaves a vertex cannot come back to the same vertex in classical free flight. Diagram (d) is also ruled out since the photon is taken to be off-shell and hence the particles leaving the vertex must have different directions and can never meet again in free-flight (here we assume that the quarks are massless). Diagram (h) corresponds to having all the internal particles off-shell, meaning that this solution to Landau's equations represents an UV divergence. In this way we are left only with diagrams (a)-(c) as possible candidates to IR soft and collinear divergences.

Already with this example it is possible to start picturing the all-order structure of the IR and UV pinched singularities for the electromagnetic quark form factor. In Fig. 1, (a) represents the first order contribution to what will be called the Soft function,  (b) and (c) are each the first order contribution to the two jet functions, and (h) corresponds to the Hard function encoding the UV singularities.

\subsubsection{Power-counting and Factorization}
We will deal now with the all order $q\bar{q}\gamma$ vertex defined as in eq. (\ref{eq:irrvertex}) and extract its Lorentz index as $\gamma^\mu\Gamma\equiv\Gamma^\mu$.

\begin{figure}[ht!]
\centering
\begin{tikzpicture}
\shadedraw[inner color=white,outer color=gray,draw=black,rotate=30.7]  (3,0.005) ellipse (40pt and 10pt);
\shadedraw[inner color=white,outer color=gray,draw=black,rotate=-30.7]  (3,-0.005) ellipse (40pt and 10pt);
\shadedraw[inner color=white,outer color=gray,draw=black,rotate=-30]  (0,0) ellipse (10pt and 10pt);
\shadedraw[inner color=white,outer color=gray,draw=black]  (3,0) ellipse (15pt and 15pt);
\draw [thick](0.3,-0.2) -- (4.5,-2.7);
\draw [thick](0.3,0.2) -- (4.5,+2.7);
\draw (4.5,+2.9) node {$p_+$};
\draw (4.5,-2.9) node {$p_-$};
\draw (2.6,-2.4) node {$J_{-}$};
\draw (2.6,2.4) node {$J_{+}$};
\draw (0,0) node {$H$};
\draw (3,0) node {$S$};
\draw [decoration={aspect=0, segment length=1.6mm, amplitude=0.7mm,coil},decorate] (-0.35,0)   -- (-1.5,0)node {\tiny$|$};

 \draw [thick, dotted] (0,0.35) .. controls (0,1.2) and (1.2,1.5) .. (1.7,1.35);
  \draw [thick, dotted] (0,-0.35) .. controls (0,-1.2) and (1.2,-1.5) .. (1.7,-1.35);
 \draw [thick, dotted] (0.35,0) -- (2.47,0);
 \draw [thick, dotted] (2.7,-1.2) -- (3,-0.53);
  \draw [thick, dotted] (2.7,1.2) -- (3,0.53);

\end{tikzpicture}
\caption[Pinch surfaces for the quark electromagnetic vertex before power counting]{Representation of the hard $(H)$, the soft $(S)$, and jet $J_{(\pm)}$ pinch surfaces for the quark electromagnetic vertex at all orders before power counting (see below). The dotted lines represent all the possible lines (fermionic or gluonic) connecting different pinch surfaces (in the sense of the reduced diagrams giving rise to divergences). }\label{pinch}
\end{figure}
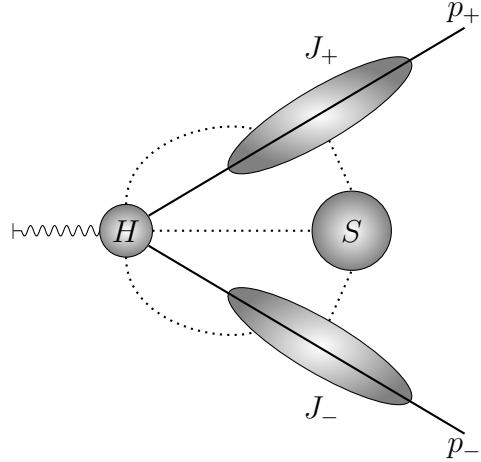

Thanks to the picture of reduced diagrams, we can study the IR and UV divergences in the elastic electromagnetic form factor ($q\bar{q}\gamma$ vertex) to \textbf{all orders in perturbation theory} without explicitly considering Landau's equations. It turns out that the structure of singularities we have already studied at one-loop is also present at higher orders. The possible reduced diagrams associated with pinch surfaces are all of the form shown in Figure \ref{pinch}.

The reduced diagram in Fig. \ref{pinch} corresponds to physical processes in which the photon decays into two jets $J_+$ and $J_-$ each with the total momenta of the two final state particles, $p_1$ and $p_2$. Between these two jets the only interaction is via zero-momentum soft particles, labeled $S$. This is due to the fact that once the jets are formed, they travel in different directions at the speed of light and hence no finite momentum transfer can occur between the two. These interactions result on possible phase shifts on the final states due to the inter-quark potential (see \cite{Laenen:2014jga}). Higher order off-shell, short-distance contributions coming from shrunk lines are encoded in the subdiagram $H$. The full derivation of this characterization of general pinch surfaces to all orders in perturbation theory is presented in \cite{Sterman:1978bi}. The term pinch surface is introduced to illustrate the fact that the singularity configurations constrain loop momenta, defining a hypersurface in the $(\{\alpha\},\{k\})$ space where the singularity pinches the contour of integration and makes the singularity unavoidable. 
\par

\begin{figure}[ht!]
\centering
\begin{tikzpicture}
\shadedraw[inner color=white,outer color=gray,draw=black,rotate=30.7]  (3,0.005) ellipse (40pt and 10pt);
\shadedraw[inner color=white,outer color=gray,draw=black,rotate=-30.7]  (3,-0.005) ellipse (40pt and 10pt);
\shadedraw[inner color=white,outer color=gray,draw=black,rotate=-30]  (0,0) ellipse (10pt and 10pt);
\shadedraw[inner color=white,outer color=gray,draw=black]  (3,0) ellipse (15pt and 15pt);
\draw [thick](0.3,-0.2) -- (4.5,-2.7);
\draw [thick](0.3,0.2) -- (4.5,+2.7);
\draw (4.5,+2.9) node {$p_+$};
\draw (4.5,-2.9) node {$p_-$};
\draw (2.6,-2.4) node {$J_{-}$};
\draw (2.6,2.4) node {$J_{+}$};
\draw (0,0) node {$H$};
\draw (3,0) node {$S$};
\draw [decoration={aspect=0, segment length=1.6mm, amplitude=0.7mm,coil},decorate] (-0.35,0)   -- (-1.5,0)node {\tiny$|$};
 \draw [decoration={aspect=0.4, segment length=1mm, amplitude=1mm,coil},decorate] (0,0.35) .. controls (0,1.2) and (1.2,1.5) .. (1.7,1.35);
  \draw [decoration={aspect=0.4, segment length=1mm, amplitude=1mm,coil},decorate] (0,-0.35) .. controls (0,-1.2) and (1.2,-1.5) .. (1.7,-1.35);
 \draw [decoration={aspect=0.4, segment length=1mm, amplitude=1mm,coil},decorate] (2.7,-1.2) -- (3,-0.53);
  \draw [decoration={aspect=0.4, segment length=1mm, amplitude=1mm,coil},decorate] (2.7,1.2) -- (3,0.53);
\end{tikzpicture}
\caption[Pinch surfaces for the quark electromagnetic vertex after power counting]{General pinch surface corresponding to logarithmic divergences \cite{Sterman:1995fz}. Only gluons connecting the hard and soft to the jet surfaces can be present to give a logarithmic divergence.}\label{pinch2}
\end{figure}
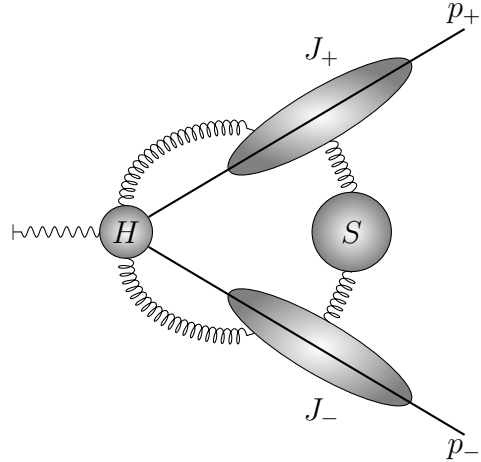

The next step towards the factorization of the $q\bar{q}\gamma$ amplitude is to power-count (this amounts to just counting the power of the so called normal variables, i.e. the ones whose vanishing defines the pinch surface, in the numerator and in the denominator of the integrand) and find the most divergent solutions to Landau's equations. In \cite{Sterman:1978bi} a proof can be found that in any covariant gauge, the divergences are logarithmic at worst and the general pinch surface is of the form presented in Fig. \ref{pinch2} in $d=4$ dimensions. Where the jet functions $J_\pm$ are only connected to the hard function $H$ through one fermion line and longitudinally polarized gluons, the jets are connected to each other through soft gluons attached to the soft function $S$ \cite{Collins:1989bt}. In all physical gauges the divergences are also logarithmic at worst and furthermore no lines except for the fermionic lines connect $H$ to the jets $J_\pm$ \cite{Sterman:1995fz}.

For the sake of resummation of the logarithmic divergences (a topic we do not delve into in this dissertation; for details, see \cite{Sterman:1995fz}),our goal now is to factorize the general pinch surface in Fig. \ref{pinch2} into contributions where in each of the regions ($H$, $J_\pm$ and $S$) the loop momenta are not restricted anymore in a gauge independent way (recall that in the soft and jet subdiagrams all the lines are soft and collinear respectively). This is performed through the introduction of the so called Wilson lines on a spacetime curve, $\gamma(t)$, with $t\in[t_1,t_2]$ between the two points $y=\gamma(t_1)$ and $z=\gamma(t_2)$,
\begin{equation}
\Phi(t_2,t_1)\equiv \mathcal{P} \Big\{\exp\Big(-ig\mu^{\epsilon}\int_{t_1}^{t_2}dt\frac{d\gamma^\mu}{dt} A_\mu(\gamma(t))\Big)\Big\}\;,\label{Wilsonline}
\end{equation}
where the symbol $\mathcal{P}$ is the path ordering operator which orders the $A_\mu(\gamma(t))$ so that the ones with higher $t$ stand to the left (remember that the $A$'s are matrices). Basically, the Wilson lines help in building gauge-invariant objects between two different points in space-time. For example, spacelike Wilson loops (which are closed Wilson lines) are used to obtain the static interquark QCD potential on the lattice \cite{Greensite:2011zz}.\par

The Wilson lines are introduced because they reproduce order by order the so called eikonal Feynman rule of soft gluon emissions. This is so because soft gluons only couple to the color and the direction of the jet they attach to \cite{Sterman:1995fz}. For this reason we will regard their interaction as eikonal.\par
 How to relate this fact to the Wilson line? Recall that, due to the Coleman-Norton picture, internal lines in reduced diagrams represent particles in free-flight so that their corresponding four-velocity $v^\mu=d\gamma^\mu(t)/dt$ is constant along their trajectory and we can hence write (\ref{Wilsonline}) for these particles as
\begin{equation}
\Phi_v(t_2,t_1)=\mathcal{P} \Big\{\exp\Big(-ig\mu^{\epsilon}\int_{t_1}^{t_2}dtv^\mu A_\mu(tv)\Big)\Big\}\;.
\end{equation}
The usual assumption in QFT is that the interaction (the $q\bar{q}\gamma$ vertex in our case) occurs inside a compact collision volume (set at the origin and $t_1=0$) and the final state particles travel out to infinity where they are detected. Hence we will deal with the special Wilson line $\Phi(\infty,0)$ and, expressing $A^\mu(x)$ as the sum of its Fourier coefficients, we see that it equals
\begin{equation}
\Phi_v(\infty,0)=\mathcal{P} \Big\{\exp\Big(-ig\mu^{\epsilon}\int_{0}^{\infty}dtv^\mu\int\frac{d^dk}{(2\pi)^d}\tilde{A}_\mu(k)e^{i t k\cdot v}\Big)\Big\} \;.
\end{equation}
To carry out the integration in $t$ we will Wick rotate $k^0\to ik^0$ so that the contribution from $t=\infty$ vanishes. In this way, counterrotating to real energy, we obtain
\begin{align}
\Phi_v(\infty,0)=\exp\Big(g\mu^{\epsilon}\int\frac{d^dk}{(2\pi)^d}\frac{v^\mu}{k\cdot v}\tilde{A}_\mu(k)\Big)\;,
\end{align}
which reproduces order by order the eikonal soft gluon emission Feynman rule from a fermion line (which is ${v^\mu}/{k\cdot v}$).\par
\begin{figure}[ht!]
\centering
\begin{tikzpicture}

\shadedraw[inner color=white,outer color=gray,draw=black,rotate=30]  (3.5,-0.) ellipse (20pt and 10pt);

\shadedraw[inner color=white,outer color=gray,draw=black,rotate=-30]  (3.5,-0.) ellipse (20pt and 10pt);
\shadedraw[inner color=white,outer color=gray,draw=black,rotate=-30]  (0,0) ellipse (10pt and 10pt);
\shadedraw[inner color=white,outer color=gray,draw=black,rotate=0]  (4.8,0) ellipse (10pt and 10pt);
\draw [thick](0.3,-0.2) -- (1.125,-0.675);
\draw [thick](0.3,0.2) -- (1.125,+0.675);

\draw [thick](1.6,-0.92) -- (4.05,-2.358);
\draw [thick](1.6,0.92) -- (4.05,2.358);
\draw [double](1.6,-0.88) -- (2.9,-0.2);
\draw [double](1.6,0.88) -- (2.9,0.2);
\draw (0,0) node {$H$};
\draw (4.8,0) node {$S$};
 \draw (1.5,-0.893) arc [start angle=180, end angle=-180, radius=0.05cm];
  \draw (1.5,0.893) arc [start angle=180, end angle=-180, radius=0.05cm];
\draw (2,-2) node {$J_-/\mathcal{J}_-$};
\draw (2,2) node {$J_+/\mathcal{J}_+$};
\draw [decoration={aspect=0, segment length=1.6mm, amplitude=0.7mm,coil},decorate] (-0.35,0)   -- (-1.5,0)node {\tiny$|$};

\draw [decoration={aspect=0.4, segment length=1mm, amplitude=1mm,coil},decorate] (4.1,0.3)   -- (4.47,0.1);
\draw [decoration={aspect=0.4, segment length=1mm, amplitude=1mm,coil},decorate] (4.1,-0.3)   -- (4.47,-0.1);
\draw [decoration={aspect=0.4, segment length=1mm, amplitude=1mm,coil},decorate] (4.75,-0.685)   -- (4.67,-0.325);
\draw [decoration={aspect=0.4, segment length=1mm, amplitude=1mm,coil},decorate] (4.75,0.685)   -- (4.67,0.325);

 \draw (3.6,0) arc [start angle=180, end angle=-180, radius=0.05cm];
 \draw [double](3.69,0.032) -- (5.05,0.9);
  \draw [double](3.69,-0.032) -- (5.05,-0.9);

\draw [decoration={aspect=0.4, segment length=1mm, amplitude=1mm,coil},decorate]  (3.2,-1.43) .. controls (3.2,-0.5) and (2.7,-0.6) .. (2.35,-0.5);
\draw [decoration={aspect=0.4, segment length=1mm, amplitude=1mm,coil},decorate]  (3.2,1.43) .. controls (3.2,0.5) and (2.7,0.6) .. (2.35,0.5);
\end{tikzpicture}
\caption[Factorized quark EM form factor]{Factorized quark EM form factor corresponding to equation (\ref{factorization}). The double lines represent Wilson lines.}\label{factorized}
\end{figure}
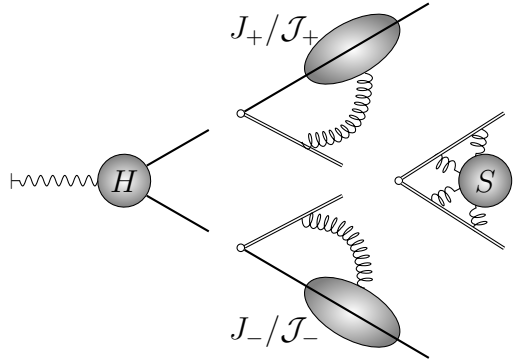

Thanks to the introduction of the Wilson lines, the so called eikonal identity (which represents the fact that at leading order in softness soft emissions factorize and are expressed in terms of independent emissions with eikonal vertices), and the use of Ward identities, it is possible to factorize the all order quark EM form factor \cite{Collins:1989bt}. This will lead us to express the general pinch surface in Fig. \ref{pinch2} in the factorized form in Fig. \ref{factorized}.\par
To get an idea how factorization of the $q\bar{q    }\gamma$ vertex will come about, notice that the interaction between two jets through soft gluons can be completely encoded in the soft function (when jets are produce and travel in different directions at the speed of light they can only exchange soft gluons), defined as
\begin{equation}
\mathcal{S}(\beta_+\cdot \beta_-,\alpha_s(\mu^2),\epsilon)\equiv\bra{0} \Phi_{\beta_+}(\infty,0) \Phi_{\beta_-}(\infty,0)\ket{0}\;,\label{softf}
\end{equation}
where $\beta_+$ and $\beta_-$ are velocities proportional to the jet momenta $p_+$ and $p_-$ respectively, $\mu^2$ is a renormalization scale, $\epsilon$ the dimensional regulator, and $\alpha_s(\mu^2)$ is the renormalization-scale-dependent strong-force coupling-constant. \par
Now we want to describe how the fermion 
 produced at the vertex becomes a final state fermion while interacting through soft gluons with the other fermion and hence define each jet leg as
\begin{equation}
J_\pm\Big(\frac{(p_\pm\cdot n_\pm)^2}{n_\pm^2\mu^2},\alpha_s(\mu^2),\epsilon\Big)u(p_\pm)={\bra{0} \Phi_{n_\pm}(\infty,0) \psi(0)\ket{p_\pm}}\,,\label{JET}
\end{equation}
where $\psi$ is the fermion field operator and $n_i$ is the direction of the Wilson line. To avoid spurious collinear singularities it is customary to choose $n_i^2\neq 0$ (although there are considerable computational advantages in setting $n^2=0$ with an easy fixing of the spurious singularities \cite{Bonocore:2015esa}).\par

Now we need to take into account the overlap of the soft and collinear regions to avoid double counting of divergences and also cancel graphs with eikonal self interactions \cite{Sterman:1995fz} The overlapping can be seen as a jet function whose collinear gluons become soft or a soft function whose gluons become collinear. In either case, it can be described by the \textit{eikonal jet function} defined as
\begin{equation}
\mathcal{J}_i\Big(\frac{(\beta_i\cdot n_i)^2}{n_i^2\mu^2},\alpha_s(\mu^2),\epsilon\Big)\equiv{\bra{0} \Phi_{n_i}(\infty,0) \Phi_{\beta_i}(\infty,0)\ket{0}}\,.\label{eikonaljet}
\end{equation}
 Since we will divide the jet function (\ref{JET}) by (\ref{eikonaljet}), the eikonal self interactions of the Wilson lines $\Phi_{n_i}$ and $\Phi_{\beta_i}$ in (\ref{JET}) and (\ref{softf}) are canceled out. 
Defining the hard function $\mathcal{H}$ as the result of dividing the form factor $\Gamma$ by $\mathcal{S}\prod_i(J_i/\mathcal{J}_i)$, we can finally write the formula for the factorized form factor
\begin{align}
\Gamma\Big(\frac{\mu^2}{Q^2},\alpha_s(\mu^2),\epsilon\Big)=&\mathcal{H}	\Big(\frac{\mu^2}{Q^2},\alpha_s(\mu^2),\epsilon\Big)\mathcal{S}(\beta_+\cdot \beta_-,\alpha_s(\mu^2),\epsilon) \prod_{i=\pm}\frac{J_i\Big(\frac{(p_i\cdot n_i)^2}{n_i^2\mu^2},\alpha_s(\mu^2),\epsilon\Big)}{\mathcal{J}_i\Big(\frac{(\beta_i\cdot n_i)^2}{n_i^2\mu^2},\alpha_s(\mu^2),\epsilon\Big)}\;,\label{factorization}
\end{align}
represented in Fig. \ref{factorized}.

In eq. (\ref{factorization}) it is implied that, since the soft and jet functions present in (\ref{factorization}) can generate spurious UV divergences, UV counterterms are introduced to cancel them.  Note that the functions defined to obtain (\ref{factorization}) depend only on general properties of the external particles like spin, charge or color, and collect all soft and collinear divergences. This dependence and some issues concerning the so called \textit{cusp anomaly} are studied in detail in \cite{Dixon:2009ur}. 
The factorization formula (\ref{factorization}) can be extended to more generic amplitudes. In cases with more legs, the color dependence of the amplitude is non-trivial but remains tractable. The presence of the so called Glauber gluons (see \cite{Collins:1989bt} in pg. 14 for details) might spoil factorization. This is still an open topic of research (see for example \cite{Catani:2011st}-\cite{Forshaw:2012bi} for more recent and refined results). Al things said, for fixed-angle scattering and the form-factor discussed here it is always possible to deform contours away from the Glauber region \cite{Collins:1989bt}.

\subsection{Coordinate space factorization}\label{coordfact}
The momentum space results presented above were extensively studied from the early 1980's to the present day, however, the coordinate space analogue results appeared only recently thanks to O. Erdogan's work \cite{Erdogan:2013bga}. There, Erdogan presents a precise derivation of the factorization formulas for the quark EM form factor at LP using the same steps as presented above. We will work with propagators and loop integrals in coordinate space: now, the variables used will be Minkowski space coordinates.

 First, the unavoidable singularities of coordinate space Feynman graphs (see sec. \ref{sec:fopt}),
\begin{equation}
I(\{x_i^\mu\})=\prod_{\substack{\text{vertices}\\k}}\int d^dy_k\prod_{\substack{\text{lines}\\j}}\frac{F(\{x_i\},\{y_k\})}{[z_j^2+i\eta]^{p_j}}\;,\label{fefo}
\end{equation}
 are identified leaning again on the Coleman-Norton interpretation (here the $\{y_k^\mu\}$ are the position of internal vertices and $\{x_i^\mu\}$ the positions of external points, and $z_j^\mu$ denotes the argument of the denominator in propagator of line $j$. $F(\{x_i\},\{y_k\})$ is a numerator factor containing all color factors,  constant and numerator factors (such as vertex derivative terms), and $p_j$ depends on whether the line is fermionic or bosonic, taking the values $p_j=1-\epsilon$ for bosonic lines and $p_j=2-\epsilon$ for fermionic lines).\par
 
For a general massless diagram, the pinched vanishing of the denominator of eq. (\ref{fefo}) defines Landau's equations in coordinate space 
\begin{empheq}[left=\empheqlbrace]{align}
\sum_{\substack{\text{lines } j\\ \text{at vertex } k}}\alpha_jz_j^\mu\epsilon_{kj}=0 &\;\;\;\;\text{and} \nonumber\\
  \alpha_j\,z^2_j=0 & \;\;,
\end{empheq} 
where $\epsilon_{ij}$ are incidence matrices in coordinate space.
It is of course intended that these equations must be satisfied together with the vanishing of the overall denominator obtained after Feynman parametrization. To connect with the Coleman-Norton picture, identify the product $\alpha_j z_j^\mu$ with a momentum vector $l^\mu\equiv \kappa\alpha_j z^\mu_j$ and $\kappa\alpha_j\equiv l_j^0/z^0_j$. Note that in the coordinate space picture the $\kappa$ parameter is interpreted in the inverse way as in the momentum space picture. This means that the soft singularities will have $\alpha_j=0$ or $\kappa$ going to zero. For a hard singularity, i.e. $z^0_j=|\vec{z}_j|\to0$, the Landau's Equations are satisfied automatically and $\kappa$ helps the $\alpha$ parameters remain finite. 
In this way we can see that each pinch singularity corresponds to massless particles propagating a finite distance on the lightcone between vertices with their momenta satisfying momentum conservation at each vertex. On the other hand off-lightcone particles have zero displacement when in divergent configurations \cite{Erdogan:2013bga}. \par
If one studies the pinch surfaces for the $q \bar{q}\gamma$ vertex one can easily recognize the same divergent surfaces in the integration domain as in the momentum picture \cite{Erdogan:2013bga}: two jets, one soft and one hard surface. Using power-counting techniques it is possible to see that the singularities of the vertex are,  at worst, logarithmic in $d=4$ dimensions and they correspond to the coordinate space analogue of Fig. \ref{pinch2}, where only gluons connect the hard and soft with jet surfaces respectively \cite{Erdogan:2013bga}. \par
Finally, after using the so called hard-collinear and soft-collinear approximations it is also possible to see that the jets are connected to the soft and hard surfaces only by longitudinally polarized gluons (also known as scalar gluons) in Landau Gauge. Due to this fact, Ward identities are used to factorize the coordinate space vertex function as in Fig. \ref{factorized} \cite{Erdogan:2013bga}.\par
In the same fashion, these results help to factorize cross sections for partonic processes such as Drell-Yan (see \cite{Contopanagos:1996nh}). Given Hermiticity of the interaction and using the Largest Time Equation \cite{Veltman:1963th}, the cancellation of IR divergences in inclusive cross sections can be shown in coordinate space \cite{Sterman:2016zby}. 
\subsection{One-loop jet function}\label{computations}
Next we will turn to computing some contributions to the jet function in coordinate space. To be able to compare our results we present here the results from momentum space calculations. There, the jet function defined in eq. (\ref{JET}) is organized as a perturbative expansion in the number of loops $n$,
\begin{equation}
    J= \sum_{n} J^{(n)}\;.\label{eq:perturbativejet}
\end{equation}

\subsubsection{Momentum space results for the one-loop jet function}
\begin{figure}[ht!]
\centering
\includegraphics[width=140mm]{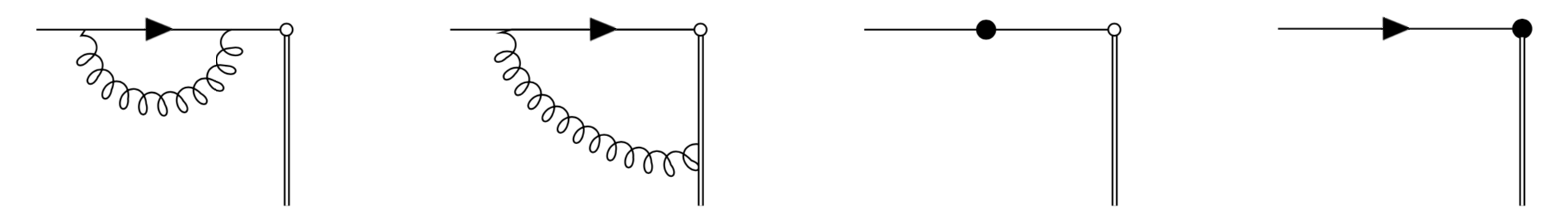}
\caption[Diagrams contributing to the first-order momentum space jet function]{Diagrams contributing to the first-order momentum space jet function, $J^(1)$ in eq. (\ref{eq:perturbativejet}). The solid dots represent UV counterterms and the double line is a Wilson line.}\label{jetff}
\end{figure}

For the jet function at one loop we will have the contributions from a self energy correction $J_p^{(1)}$ of the quark line and a gluon exchange vertex correction $J_V^{(1)}$ between the Wilson and the fermion line. All these contributions including the UV counterterms are presented in Fig. \ref{jetff}. The eikonal self interaction graphs do not appear due to the fact that we divided the jet function by the eikonal jet function in (\ref{eikonaljet}).

The first diagram in Fig. \ref{jetff} vanishes in dimensional regularization since it is scaleless  (remember the external momentum is lightlike, $p^2=0$) and there is no available quantity with non-zero mass dimension. This comes from a cancellation of UV and IR poles. Therefore, after introducing the UV counterterm in the third diagram of Fig. \ref{jetff}, we will obtain that $J_p^{(1)}=1/\epsilon_{\rm UV}$ (here we make explicit the UV nature of the pole in $\epsilon$). The vertex correction in the second diagram in Fig. \ref{jetff} amounts to
\begin{align}
J_V^{(1)}=&2 i\mu^{2\epsilon}g^2\int\frac{d^dk}{(2\pi)^d}\frac{(\slashed{k}-\slashed{p})\slashed{n}}{(k^2-i\eta)(k^2-2p\cdot k-i\eta)(2n\cdot k-i\eta)}\,,\label{eq:jetfdiagram}
\end{align}
where $n$ is the direction of the Wilson line and $n^2=0$. In the $\overline{\text{MS}}$ scheme, by using Dirac's equation on the spinor in the LHS of eq. (\ref{JET}),  adding the appropriate $\overline{\text{MS}}$ counterterm in the fourth diagram in Fig. \ref{jetff}  to eq. (\ref{eq:jetfdiagram}) 
 results in (see eq. (3.2) in \cite{Bonocore:2015esa}),
\begin{equation}
J_{V\rm r}^{(1)}\equiv J_V^{(1)}+J_{V,\rm CT}^{(1)}=2\left(\frac{\alpha_s}{4\pi }\right)\Big[\frac{1}{\epsilon^2}+\frac{1}{\epsilon}\Big(1-\gamma_E+\log\frac{4\pi\mu^2}{(-2p\cdot n)}\Big)+\mathcal{O}(\epsilon^0)\Big]\;.\label{eq:22}
\end{equation}

Now we turn to reproducing this result in coordinate space.

 \subsubsection{One-loop jet function in coordinate space}
 Once we have seen that factorization of the vertex function in coordinate space comes along pretty much as in the momentum space picture (see \cite{Erdogan:2013bga}), let us now compute the one-loop jet function in coordinate space and see whether, using the LSZ reduction formula, we can recover eq. (\ref{eq:22}) in momentum space. \par
 
 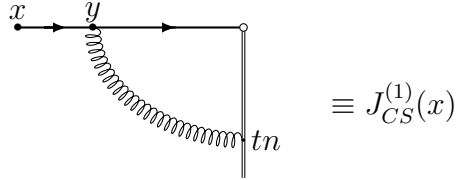
\begin{figure}[ht!]
 \centering
 \begin{tikzpicture}[>=stealth]
 \draw [decoration={aspect=0.4, segment length=1mm, amplitude=1mm,coil},decorate] (1,0) .. controls (1,-0.9) and (2.1,-1.5) .. (3,-1.5);
 \draw[-latex,thick] (2,0)--(2.1,0);
  \draw[thick] (2,0)--(2.95,0);
  \draw[-latex,thick] (0.35,0)--(.65,0);
   \draw[thick] (0.35,0)--(1,0);
    \draw [thick](0,0)--(0.55,0);
  \draw [thick](1,0)--(2.1,0);
   \draw (1,-0.00) node {\tiny$\bullet$};
      \draw (1,0.2) node {$y$};
          \draw (0,0.2) node {$x$};
          \draw (3.3,-1.5) node {$tn$};
          \draw (5,-1) node {$\equiv J^{(1)}_{CS}(x)$};
          
      \draw (0,-0.00) node {\tiny$\bullet$};
   \draw[double] (3,-0.05)--(3,-2);
   \draw (2.95,0) arc [start angle=180, end angle=-180, radius=0.05cm];
   \draw (3,-1.51) node {$\cdot$};
\end{tikzpicture}
\caption[Diagram representing the one-loop jet function in coordinate space]{Diagram representing the one-loop jet function (without UV conterterms) in coordinate space.}\label{repera}
\end{figure}

Reading from Fig. \ref{repera} (with the help of the coordinate space Feynman rules introduced in chapter 2 and explicitated below in section \ref{sec:fopt}) we have that, taking the Wilson line to be lightlike ($n^2=0$),
\begin{align}
J^{(1)}_{CS}(x)=&(-ig T^a\gamma_\mu)\frac{\Gamma(1-\epsilon)\Gamma^2(2-\epsilon)}{(2\pi^{2-\epsilon})^24\pi^{2-\epsilon}}\int d^d y \frac{-(\slashed{x}-\slashed{y})}{((x-y)^2+i\eta)^{2-\epsilon}}\frac{-\slashed{y}}{(y^2+i\eta)^{2-\epsilon}}\nonumber\times\\
&\times\int_0^{+\infty}dt \frac{(-igT^a)n^\mu}{((y-tn)^2+i\eta)^{1-\epsilon}}\,.\label{tonaser}
\end{align}

To carry out the whole integral we will introduce Feynman parameters in two steps, first combining the second and third denominators and then the resulting denominator with the first one in (\ref{tonaser}). In this way
\begin{align}
J^{(1)}_{CS}(x)=&-g^2 C_F\frac{\Gamma(1-\epsilon)\Gamma^2(2-\epsilon)}{16\pi^{6-3\epsilon}}\frac{\Gamma(5-3\epsilon)}{\Gamma^2(2-\epsilon)\Gamma(1-\epsilon)}\int d^d y\int_0^{+\infty}dt\int_0^1 d\alpha_1d\alpha_2\times\nonumber\\
&\times\frac{(\slashed{x}\slashed{n}\slashed{y}-\slashed{y}\slashed{n}\slashed{y})\alpha_1^{1-\epsilon}(1-\alpha_1)^{-\epsilon}\alpha_2^{1-\epsilon}(1-\alpha_2)^{2-2\epsilon}}{(y^2-2y\cdot(\alpha_2 x+(1-\alpha_1)(1-\alpha_2)tn)+\alpha_2 x^2+i\eta)^{5-3\epsilon}}\;.\label{eq:CSjet1}
\end{align}

 With this representation, Landau's equations read 
\begin{align}
&\alpha_2(x-y)^2+(1-\alpha_2)\alpha_1 y^2 +(1-\alpha_2)(1-\alpha_1)(y^2-2tn\cdot y)=0\;,\\
&\alpha_2(x-y)^\mu+(1-\alpha_2)\alpha_1 y^\mu+(1-\alpha_2)(1-\alpha_1)(y^\mu-tn^\mu)=0\,.
\end{align}
To study the region of Minkowski space where the gluon line becomes soft we take $\alpha_1=1$. In this case we do not expect the gluon to change the direction of the external line since it carries no momentum. This is indeed the case because the solution to Landau's equations tells us that 
\begin{equation}
y^\mu=\frac{\alpha_2}{2\alpha_2-1} x^\mu\,,\label{EQ:222}
\end{equation}
and that $x$ and $y$ must lie in the lightcone for $\alpha_2\neq 0,1$.
For the end-point singularities in the $\alpha$ parameters we see that for $\alpha_2=1$ the vertex $y$ migrates to the external point $x$, $y=x$. This makes sense since the fermion line from $0$ to $y$ will be carrying all the momentum (due to the CN picture) while the fermion line from $y$ to $x$ shrinks. Taking $\alpha_2=0$ means that the $y$ vertex migrates to the origin, $y^\mu=0$, signaling that we are dealing with an UV divergence. However in both cases there is not restriction to $y$ or $x$ to lie on the lightcone.
\par
For the case when $\alpha_1=0$, so that the exchanged gluon is not soft, Landau's equations tell us that
\begin{equation}
y^\mu=\frac{(1-\alpha_2)tn^\mu-\alpha_2 x^\mu}{(1-2\alpha_2)}\;.
\end{equation}
As we will see below, the only contribution to this amplitude will come when $t=0$ so that we will have a collinear pinch surface with the vertex $y$ obeying (\ref{EQ:222}) but this time the gluon emerges from the Wilson line cusp (the origin), making the gluon collinear. Both the fermion external line and gluon line are lightlike for $\alpha_2\neq0,1$. Again if $\alpha_2=1$ the internal vertex $y$ coincides with the external point $x$ and if $\alpha_2=0$ we will have that $y^\mu=tn^\mu$ signaling again a UV divergence for $t=0$.
\par

Now we shift the integration variable in eq. (\ref{eq:CSjet1}) as $y\to y-(\alpha_2 x+(1-\alpha_1)(1-\alpha_2)tn)$ and, by parity, drop odd powers of the new $y$ in the numerator. This yields
\begin{align}
J^{(1)}_{CS}(x)=&-g^2 C_F\frac{\Gamma(5-3\epsilon)}{16\pi^{6-3\epsilon}}\int d^d y\int_0^{+\infty}dt\int_0^1 d\alpha_1d\alpha_2\times\nonumber\\
&\times\frac{(\alpha_2(1-\alpha_2)\slashed{x}\slashed{n}\slashed{x}-\slashed{y}\slashed{n}\slashed{y})\alpha_1^{1-\epsilon}(1-\alpha_1)^{-\epsilon}\alpha_2^{1-\epsilon}(1-\alpha_2)^{2-2\epsilon}}{(y^2+\alpha_2(1-\alpha_2) x^2-2(1-\alpha_1)(\alpha_2-\alpha_2^2)tn\cdot x+i\eta)^{5-3\epsilon}}\;.
\end{align}
At this point, it is possible (taking $n\cdot x\neq0$) to carry out the integration in $t$ (contributing only at $t=0$). This gives
\begin{align}
J^{(1)}_{CS}(x)=&g^2 C_F\frac{\Gamma(5-3\epsilon)}{16\pi^{6-3\epsilon}}\int d^d y\int_0^1 d\alpha_1d\alpha_2\nonumber\times\\
&\times\frac{(\alpha_2(1-\alpha_2)\slashed{x}\slashed{n}\slashed{x}-\slashed{y}\slashed{n}\slashed{y})\alpha_1^{1-\epsilon}(1-\alpha_1)^{-1-\epsilon}\alpha_2^{-\epsilon}(1-\alpha_2)^{1-2\epsilon}}{(4-3\epsilon) 2n\cdot x(y^2+\alpha_2(1-\alpha_2) x^2+i\eta)^{4-3\epsilon}}\;.
\end{align}
Using standard Dimensional Regularization formulas we carry out the integral in $y$ and identify the Euler Beta functions in the Feynman parameters integrations to get
\begin{align}
J^{(1)}_{CS}&(x)=\frac{-i\pi^{\frac{d}{2}}g^2 C_F(x^2)^{2\epsilon-2}}{(2x\cdot n) 16\pi^{6-3\epsilon}}\frac{\Gamma(1-2\epsilon)\Gamma(2-\epsilon)\Gamma(-\epsilon)}{\Gamma(2-2\epsilon)}\Big(\frac{\Gamma(\epsilon)(1+\epsilon)}{\Gamma(2+\epsilon)}\Big)\Big(2x\cdot n \slashed{x}(1-2\epsilon)+\epsilon\slashed n x^2\Big)\;.\label{JCs}
\end{align}

The result of expanding in powers of $\epsilon$ eq. (\ref{JCs}) is,
\begin{align}
J^{(1)}_{CS}(x)=&\frac{i \alpha_s C_F}{4\pi}\frac{-\slashed x}{2\pi^2(x^2+i\eta)^2}\Big(-\frac{2}{\epsilon^2}+\frac{2(1-2\gamma_E)}{\epsilon}-(\slashed x)^{-1}\frac{\slashed n x^2}{n\cdot x}\frac{1}{\epsilon}\Big)+\mathcal{O}(\epsilon^0)\;.\label{eq:expandedresult}
\end{align}

This is a very nice result since the most divergent part is proportional to a Fermion propagator $S(x)$ (one can think of this result as follows: the gluon merges collinearly with the fermion producing the divergence times the fermion propagator) and this will allow us to use the LSZ formula to get the momentum space expression in a very simple manner. The LSZ formula relates the jet function in momentum space, $J_{MS}$, with the one in coordinate space as
\begin{align}
J_{MS}(p)=-i\int {d^dx} e^{-ip\cdot x} (i\slashed\partial)J^{(1)}_{CS}(x)\;.
\end{align}
By using the Dirac equation $\slashed\partial S(x)=i\delta^{(d)}(x)$ we find that the most divergent part of the jet function is
\begin{align}
\lim_{\epsilon\to0}J_{MS}(p)=\lim_{\epsilon\to0}\frac{\alpha_s C_F}{4\pi} \Big(\frac{2}{\epsilon^2}\Big)\;.
\end{align}
So that we recover the same leading pole structure as the one-loop gluon exchange vertex correction $J_{V\rm r}^{(1)}$ of eq. (\ref{eq:22}). The last subleading term inside the brackets (\ref{eq:expandedresult}) must contain the dependence on the collinear scale $p\cdot n$ after LSZ reduction (which is not straightforward to perform). \par

\subsection{Abelian radiative one-loop jet function}\label{radiativesection}

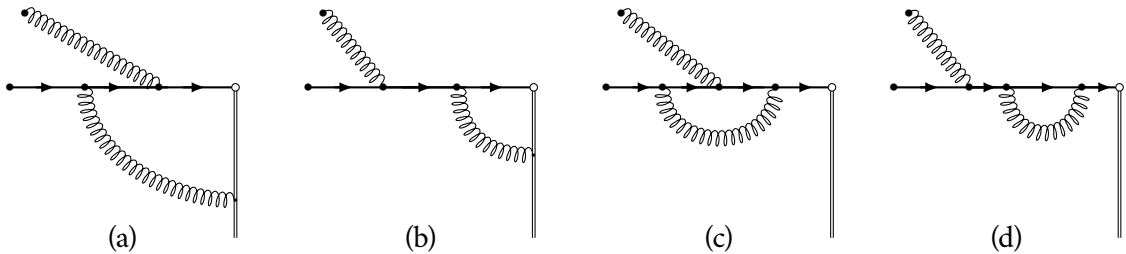
\begin{figure}[ht!]
 \centering
 \begin{tikzpicture}[>=stealth]
 \draw [decoration={aspect=0.4, segment length=1mm, amplitude=1mm,coil},decorate] (1,0) .. controls (1,-0.9) and (2.1,-1.5) .. (3,-1.5);
  \draw [decoration={aspect=0.4, segment length=1mm, amplitude=1mm,coil},decorate] (.2,1) -- (2,0);
 \draw[-latex,thick] (2.3,0)--(2.6,0);
 \draw[-latex,thick] (1.25,0)--(1.5,0);
  \draw[-latex,thick] (0.35,0)--(.6,0);
    \draw [thick](0,0)--(2.95,0);
  \draw [thick](1,0)--(2.1,0);
   \draw (1,-0.00) node {\tiny$\bullet$};
   \draw (1.98,-0.00) node {\tiny$\bullet$};
      \draw (.2,0.99) node {\tiny$\bullet$};
      \draw (0,-0.00) node {\tiny$\bullet$};
      \draw (1.5,-2.00) node {(a)};
   \draw[double] (3,-0.05)--(3,-2);
   \draw (2.95,0) arc [start angle=180, end angle=-180, radius=0.05cm];
   \draw (3,-1.51) node {$\cdot$};
\end{tikzpicture}\;\;\;\;
 \begin{tikzpicture}[>=stealth]
 \draw [decoration={aspect=0.4, segment length=1mm, amplitude=1mm,coil},decorate] (2,0) .. controls (2,-0.9) and (2.8,-.9) .. (3,-.9);
 \draw (1.5,-2.00) node {(b)};
  \draw [decoration={aspect=0.4, segment length=1mm, amplitude=1mm,coil},decorate] (.2,1) -- (1,0);
 \draw[-latex,thick] (2.3,0)--(2.6,0);
 \draw[-latex,thick] (1.25,0)--(1.65,0);
  \draw[-latex,thick] (0.35,0)--(.6,0);
    \draw [thick](0,0)--(2.95,0);
  \draw [thick](1,0)--(2.1,0);
   \draw (1,-0.00) node {\tiny$\bullet$};
   \draw (1.98,-0.00) node {\tiny$\bullet$};
      \draw (.2,0.99) node {\tiny$\bullet$};
      \draw (0,-0.00) node {\tiny$\bullet$};
   \draw[double] (3,-0.05)--(3,-2);
   \draw (2.95,0) arc [start angle=180, end angle=-180, radius=0.05cm];
   \draw (3,-.91) node {$\cdot$};
\end{tikzpicture}\;\;\;\;
\begin{tikzpicture}[>=stealth]
\draw (1.5,-2.00) node {(c)};
 \draw [decoration={aspect=0.4, segment length=1mm, amplitude=1mm,coil},decorate] (0.75,0) .. controls (0.75,-0.9) and (2.25,-.9) .. (2.25,0);
  \draw [decoration={aspect=0.4, segment length=1mm, amplitude=1mm,coil},decorate] (.2,1) -- (1.5,0);
  \draw[-latex,thick] (2.6,0)--(2.75,0);
  \draw[-latex,thick] (1,0)--(1.25,0);
    \draw[-latex,thick] (1.75,0)--(2,0);
  \draw[-latex,thick] (0.35,0)--(.6,0);
    \draw [thick](0,0)--(2.95,0);
  \draw [thick](1,0)--(2.1,0);
   \draw (0.75,-0.00) node {\tiny$\bullet$};
   \draw (2.25,-0.00) node {\tiny$\bullet$};
   \draw (1.5,-0.00) node {\tiny$\bullet$};
      \draw (.2,0.99) node {\tiny$\bullet$};
      \draw (0,-0.00) node {\tiny$\bullet$};
   \draw[double] (3,-0.05)--(3,-2);
   \draw (2.95,0) arc [start angle=180, end angle=-180, radius=0.05cm];
\end{tikzpicture}\;\;\;\;
 \begin{tikzpicture}[>=stealth]
 \draw (1.5,-2.00) node {(d)};
 \draw [decoration={aspect=0.4, segment length=1mm, amplitude=1mm,coil},decorate] (1.5,0) .. controls (1.5,-0.8) and (2.5,-.8) .. (2.5,0);
  \draw [decoration={aspect=0.4, segment length=1mm, amplitude=1mm,coil},decorate] (.2,1) -- (1,0);
  \draw[-latex,thick] (2.,0)--(2.15,0);
    \draw[-latex,thick] (2.5,0)--(2.9,0);
 \draw[-latex,thick] (1.,0)--(1.375,0);
  \draw[-latex,thick] (0.35,0)--(.6,0);
    \draw [thick](0,0)--(2.95,0);
  \draw [thick](1,0)--(2.1,0);
   \draw (1,-0.00) node {\tiny$\bullet$};
   \draw (1.5,-0.00) node {\tiny$\bullet$};
    \draw (2.5,-0.00) node {\tiny$\bullet$};
      \draw (.2,0.99) node {\tiny$\bullet$};
      \draw (0,-0.00) node {\tiny$\bullet$};
   \draw[double] (3,-0.05)--(3,-2);
   \draw (2.95,0) arc [start angle=180, end angle=-180, radius=0.05cm];
\end{tikzpicture}
\caption[Diagrams contributing to the one-loop radiative jet function in coordinate space]{Diagrams contributing to the one-loop radiative jet function in coordinate space. No external leg corrections are included.}\label{radiatedjetAbelian}
\end{figure}

As stated in the introduction, the study of subleading regions (also called NLP regions, i.e. kinematical configurations that give rise to divergent subleading terms in observables, usually coming from emission of soft gluons) has gained attention in recent years \cite{Gervais:2017yxv,Beneke:2017ztn,Moult:2019mog,Laenen:2020nrt,Liu:2021mac}. For this reason, we will now turn on to analyze some features of the radiative one-loop jet function in coordinate space and try to reduce in quadrature the contributing diagrams. 
The contributions to the one-loop radiative jet function in the Abelian case are listed in Figure \ref{radiatedjetAbelian}. Here we do not include external leg corrections. 

We will also analyze Landau's equations for each diagram contributing to the radiative jet function. This will help us again to identify the configurations giving rise to collinear divergences and their overlapping with soft divergences of the emitted gluon (highlighted in the text below). To see this, as an oversimplified example, imagine that, after reducing to Feynman parameters a contribution to the radiative jet, one identifies through Landau's equations that $\alpha_1=0$ entails a collinear divergence and $\alpha_2=1$ a soft one. By separating the divergent part from the finite one on each of the two Feynman parameters (through the usual multiplication by $1=\alpha_i+(1-\alpha_i)$), one would get four terms
$\alpha_1\alpha_2+\alpha_1(1-\alpha _2) +\alpha_2 (1-\alpha_1) +(1-\alpha_1)(1-\alpha_2)$ encoding a collinear-finite, a collinear-soft overlapping, a finite-finite, and a finite-soft region respectively. This example shows the potential of the present treatment concerning the factorization and overlapping of divergent kinematical regions.

One should however be careful when analyzing divergences on a diagram by diagram basis, since results may suffer from artifacts due to the specific integral representation under treatment. All things said, the direct identification of individual Feynman diagrams with certain physical processes is a well established practice in theoretical particle physics. On the other hand, if one has access to the fully integrated final result in coordinate space, the soft and collinear limits can be studied by taking these in the values of the external momenta for the outgoing fermion and gluon (after performing LSZ reduction of the coordinate space, UV finite,  result).

\subsubsection{One-loop jet function with internal gluon emission in coordinate space}
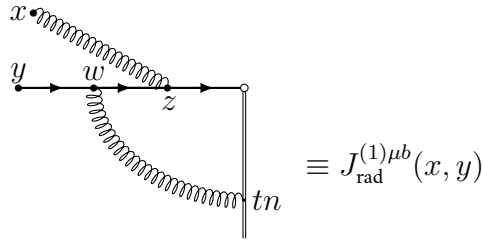
\begin{figure}[ht!]
 \centering
 \begin{tikzpicture}[>=stealth]
 \draw [decoration={aspect=0.4, segment length=1mm, amplitude=1mm,coil},decorate] (1,0) .. controls (1,-0.9) and (2.1,-1.5) .. (3,-1.5);
  \draw [decoration={aspect=0.4, segment length=1mm, amplitude=1mm,coil},decorate] (.2,1) -- (2,0);
 \draw[-latex,thick] (2.3,0)--(2.6,0);
 \draw[-latex,thick] (1.25,0)--(1.5,0);
  \draw[-latex,thick] (0.35,0)--(.6,0);
    \draw [thick](0,0)--(2.95,0);
  \draw [thick](1,0)--(2.1,0);
   \draw (1,-0.00) node {\tiny$\bullet$};
   \draw (1.98,-0.00) node {\tiny$\bullet$};
      \draw (.2,0.99) node {\tiny$\bullet$};
      \draw (1,0.2) node {$w$};
      \draw (2,-0.2) node {$z$};
      \draw (0,1) node {$x$};
          \draw (0,0.2) node {$y$};
          \draw (3.3,-1.5) node {$tn$};
          \draw (5,-1) node {$\equiv J^{(1)\mu b}_{\text{rad}}(x,y)$};
          
      \draw (0,-0.00) node {\tiny$\bullet$};
   \draw[double] (3,-0.05)--(3,-2);
   \draw (2.95,0) arc [start angle=180, end angle=-180, radius=0.05cm];
   \draw (3,-1.51) node {$\cdot$};
\end{tikzpicture}
\caption{Diagram representing the one-loop internal gluon emission from a jet.}\label{repera2}
\end{figure}

The first contribution to the Abelian one-loop radiative jet function in Fig. \ref{radiatedjetAbelian} is depicted in detail in Fig. \ref{repera2}. We consider a lightlike Wilson line, i.e. $n^2=0$. We will keep the non-Abelian numerical factors (such as the fundamental Casimir of the gauge group, $C_F$).

This diagram amounts to
\begin{align}
J^{(1)\mu b}_{\text{rad}}(x,y)=&(ig^3T^aT^bT^a)\Big(\frac{\Gamma(2-\epsilon)}{2\pi^{2-\epsilon}}\Big)^3\Big(\frac{\Gamma(1-\epsilon)}{4\pi^{2-\epsilon}}\Big)^2
\int d^dzd^dw \int_0^\infty d t \frac{-(\slashed y-\slashed w)}{((w-y)^2+i\eta)^{2-\epsilon}} \times\nonumber\\
&\times\frac{-\slashed n(\slashed w-\slashed z)}{((w-z)^2+i\eta)^{2-\epsilon}}\gamma^\mu\frac{-\slashed z}{(z^2+i\eta)^{2-\epsilon}}\frac{1}{((w-tn)^2+i\eta)^{1-\epsilon}}\frac{1}{((x-z)^2+i\eta)^{1-\epsilon}}\;.\label{radjet1}
\end{align}
The color structure above is easily computable, and gives $T^aT^bT^a=-T^b/(2N)$ for $SU(N)$. This factor is also easily  computed for $SO(N)$, where is just $T^b/4$ (giving no suppression of the diagram for large number of colors), and for $Sp(N)$ where it amounts to $-T^b/4$ (see the Appendix in \cite{Llanes-Estrada:2018azk} for a detailed derivation).\par

\paragraph{Analysis of Landau's equations}
We will analyze Landau's equations for this diagram by anticipating the way we will use Feynman parameters to solve eq. (\ref{radjet1}) (The explicit expression is in eqns. (\ref{radiativejet1a}) and (\ref{intermediatestep}) below). For the integration in $z$ we will first combine the second and third denominators and subsequently the resulting denominator with the last denominator in eq. (\ref{radjet1}). Landau's equations for the resulting denominator read
\begin{align}
&z^2-2z\cdot (\alpha_1\alpha_2 w+(1-\alpha_2)x)+\alpha_1\alpha_2 w^2+(1-\alpha_2)x^2=0\label{land1}\;,\\
&z^\mu=\alpha_1\alpha_2 w^\mu+(1-\alpha_2)x^\mu\label{land2}\;.
\end{align}
Let us analyze the case when $\boxed{\alpha_2=1}$, which corresponds to the case of a \textbf{soft emitted gluon} (since the Feynman parameter of the external gluon line is $1-\alpha_2$). In this case eq. (\ref{land2}) amounts to $z^\mu=\alpha_1 w^\mu$ so that the emitted gluon indeed does not change the direction of the fermion. Using this result in eq. (\ref{land1}) for a lightlike external point, $x^2=0$, one obtains $\alpha_1(1-\alpha_1)w^2=0$ which sets $w$ on the lightcone except for the endpoints of the integration in $\alpha_1$, these endpoints correspond to UV divergences (in two distinct fermion lines) $z^\mu=0$ for $\alpha_1=0$ and $z^\mu=w^\mu$ for $\alpha_1=1$. Next, let us take $\boxed{\alpha_2=0}$, this gives the UV condition that $x^\mu=z^\mu$ so that the emitted gluon travels no distance. 

Now, setting $\boxed{\alpha_1=1}$, eq. (\ref{land2}) tells us that $z$ must lie in the line connecting $x$ and $w$ since it equals $z^\mu=\alpha_2w^\mu+(1-\alpha_2)x^\mu$. Using this in eq. (\ref{land1}) one obtains the condition $ \alpha_2(1-\alpha_2)(w-x)^2=0$ which sets $w$ on the lightcone of $x$ except for the UV gluon with $\alpha_2=0$ and hence $z^\mu=x^\mu$ or the UV fermion with $\alpha_2=1$ and $z^\mu=w^\mu$. For the case $\boxed{\alpha_1=0}$ one has that $z^\mu=(1-\alpha_2)x^\mu$ which gives $\alpha_2(1-\alpha_2)x^2=0$, which is automatically fulfilled if the external point is taken on the lightcone.\\

Landau's equations for the other vertex $w$ in eq. (\ref{radjet1}), after the steps taken to reach eq. (\ref{intermediatestep}) below, read
\begin{align}
\alpha w^2+2w\cdot \theta+\kappa^2=&0\\
\alpha w^\mu+\theta^\mu=&0\;.
\end{align}
Here $\alpha=\alpha_3\alpha_4\alpha_5+\alpha_5(1-\alpha_3)\alpha_1\alpha_2(1-\alpha_1\alpha_2)+\alpha_5(1-\alpha_4)$, $ \theta^\mu=-(1-\alpha_5)n^\mu-\alpha_1\alpha_2(1-\alpha_2)x^\mu-\alpha_5(1-\alpha_4)y^\mu$, and $\kappa^2=\alpha_5(1-\alpha_3)\alpha_2(1-\alpha_2)x^2+\alpha_5(1-\alpha_4)y^2$. \\
We will set the external vertices to be on the lightcone $x^2=y^2=0$, which means $\kappa^2=0$. For the case $\boxed{\alpha\neq0}$ we will obtain that $\theta^2=0$ which gives the condition on the external vertices $(1-\alpha_5)n\cdot(-\alpha_1\alpha_2(1-\alpha_2)x-\alpha_5(1-\alpha_4)y)+\alpha_1\alpha_2(1-\alpha_2)\alpha_5(1-\alpha_4)x\cdot y=0$. Which can be solved for a collinear gluon emission with $x\propto y$ and $\alpha_5=1$. Another solution has the quarks and the gluon emitted from the vertex all traveling in the same direction, \textit{i.e.} $x,y\propto n$. The case $\boxed{\alpha=0}$ as seen from the result in eq. (\ref{eq:23}) entails several UV divergences arising from the fact that there is a Jacobian of $\alpha^{-2+\epsilon}$ (UV divergences are regulated for $\epsilon>0$). Some of them have $\alpha_4=\alpha_5=1$, $\alpha_3=0$ and $\alpha_1=\alpha_2=1$, $\alpha_1=\alpha_2=0$ or $\alpha_1=0$ and $\alpha_2=1$.

\paragraph{Computation} As in the previous subsection we introduce Feynman parameters in two steps for combining the second, the third, and last denominators above such that
\begin{align}
&J^{(1)\mu b}_{\text{rad}}(x,y)=F^b\int d^dw\int d^dz\int_0^\infty dt \int_0^1d\alpha_1d\alpha_2\frac{(\alpha_1^{1-\epsilon}(1-\alpha_1)^{1-\epsilon}\alpha_2^{3-2\epsilon}(1-\alpha_2)^{-\epsilon})}{((w-y)^2+i\eta)^{2-\epsilon}((w-tn)^2+i\eta)^{1-\epsilon}}\times\nonumber\\
&\times\frac{(\slashed y-\slashed w)\slashed n( \slashed z \gamma^\mu\slashed z-\slashed w\gamma^\mu \slashed z)}{(z^2-2z\cdot (\alpha_1\alpha_2 w+(1-\alpha_2)x)+\alpha_1\alpha_2 w^2+(1-\alpha_2)x^2)^{5-3\epsilon}}\;,\label{radiativejet1a}
\end{align}
where $F^b$ encodes the color structure and numerical factors. Now, to compute the integral in $z$, we shift it as $z\to z-(\alpha_1\alpha_2 w+(1-\alpha_2)x)$ and then drop odd powers of $z$ in the numerator due to parity. This integration results in
\begin{align}
J^{(1)\mu b}_{\text{rad}}(x,y)=&F'^b\int d^dw\int_0^\infty dt \int_0^1d\alpha_1d\alpha_2\frac{\alpha_1^{1-\epsilon}(1-\alpha_1)^{1-\epsilon}\alpha_2^{3-2\epsilon}(1-\alpha_2)^{-\epsilon}}{((w-y)^2+i\eta)^{2-\epsilon}((w-tn)^2+i\eta)^{1-\epsilon}} \nonumber \times\\
&\times\Big(K^\mu-(\slashed y-\slashed w)\slashed n \gamma^\mu (m^2)/2\Big)(m^2)^{2\epsilon-3}\;,
\end{align}
where $m^2=\alpha_1\alpha_2(1-\alpha_1\alpha_2)w^2+\alpha_2(1-\alpha_2)x^2-2\alpha_1\alpha_2(1-\alpha_2)  x\cdot w$ and $K^\mu=(\slashed y-\slashed w) \slashed n [((1+\alpha_1\alpha_2) \slashed w+(1-\alpha_2)\slashed x)\gamma^\mu(\alpha_1\alpha_2 \slashed w+(1-\alpha_2)\slashed x)$. Next we combine the second denominator above with the last one again (the one proportional to $m^2$) with Feynman parameters and then compute the integral in $t$ assuming we are not integrating $w$ in the time-like region where $w\cdot n\geq0$. From this integration we will obtain a denominator factor and a $-2n\cdot w$. We will combine these with the last propagator denominator left in two steps of Feynman parametrization, which gives
\begin{align}
J^{(1)\mu b}_{\text{rad}}(x,y)=&F''^b\int d^dw \int_0^1d\alpha_1...d\alpha_5{\alpha_1^{1-\epsilon}(1-\alpha_1)^{1-\epsilon}\alpha_2^{3-2\epsilon}(1-\alpha_2)^{-\epsilon}\alpha_3^{-1-\epsilon}(1-\alpha_3)^{2-2\epsilon}}\times\nonumber\\
&\times \frac{\alpha_4^{2-3\epsilon}(1-\alpha_4)^{1-\epsilon}\alpha_5^{4-4\epsilon}}{(\alpha w^2+2w\cdot  \theta+\kappa^2)^{6-4\epsilon}}\Big(K^\mu-(\slashed y-\slashed w)\slashed n \gamma^\mu (m^2)/2\Big)\;.\label{intermediatestep}
\end{align}
 Finally we shift the integration variable as $w\to w\alpha^{\frac{1}{2}}+ \theta\alpha^{-\frac{1}{2}}$ and carry out the $w$ integration finding
\begin{align}
J^{(1)\mu b}_{\text{rad}}&(x,y)=\frac{ig^3\pi^{3\epsilon-6}T^b}{2^8N(3-3\epsilon)}\times\int_0^1d\alpha_1...d\alpha_5\alpha_1^{1-\epsilon}(1-\alpha_1)^{1-\epsilon}\alpha_2^{3-2\epsilon}(1-\alpha_2)^{-\epsilon}\alpha_3^{-1-\epsilon}(1-\alpha_3)^{2-2\epsilon} \times\nonumber\\
&\times\alpha_4^{2-3\epsilon}(1-\alpha_4)^{1-\epsilon}\alpha_5^{4-4\epsilon}\alpha^{-2+\epsilon}\Big(\alpha g^{\rho\sigma}A^\mu_{\rho\sigma}\Gamma(3-3\epsilon)(M^2)/2+B^\mu\Gamma(4-3\epsilon)\Big)(M^2)^{3\epsilon-4}\;,\label{eq:23}
\end{align}
for $SU(N)$.
The expressions for the $A_{\rho\sigma}^\mu$, $B^\mu$, and $M^2$ are shown next (here $\beta^\mu\equiv\alpha^{-\frac{1}{2}}\theta^\mu$):
\begin{align}
M^2=\kappa^2-\beta^2\;,\label{M2}
\end{align}
\begin{align}
B^\mu=&(\slashed y-\slashed\beta)\slashed n\Big[ \big( (1+\alpha_1\alpha_2)\slashed\beta+(1-\alpha_2)\slashed x \big)\gamma^\mu \big( \alpha_1\alpha_2\slashed\beta +(1-\alpha_2)\slashed x\big)-\nonumber\\
&-\gamma^\mu \big( \alpha_1\alpha_2(1-\alpha_1\alpha_2)\beta^2+\alpha_2(1-\alpha_2)x^2-2\alpha_1\alpha_2(1-\alpha_2)\beta\cdot x\big)/2\Big]\,,
\end{align}
\begin{align}
A_{\rho\sigma}^\mu=&(\slashed y-\slashed\beta)\big((1+\alpha_1\alpha_2)\alpha_1\alpha_2\gamma_\rho\gamma^\mu\gamma_\sigma-\alpha_1\alpha_2(1-\alpha_1\alpha_2)\gamma^\mu g_{\rho\sigma}/2\big)-\nonumber\\
&-\gamma_\rho \slashed n\Big[ (1+\alpha_1\alpha_2)\gamma_\sigma\gamma^\mu\big((1-\alpha_2)\slashed x +\alpha_1\alpha_2 \slashed \beta \big)+\alpha_1\alpha_2\big((1+\alpha_1\alpha_2)\slashed \beta+(1-\alpha_2)\slashed x\big)\gamma^\mu\gamma_\sigma-\nonumber\\
&-\alpha_1\alpha_2\gamma^\mu\big((1-\alpha_1\alpha_2)\beta_\sigma-(1-\alpha_2) x_\sigma\big)\Big]\;.
\end{align}

This result is rather cumbersome and will be left for a future work to  fully solve the integrations over Feynman parameters in eq. (\ref{eq:23}). Nonetheless, by inspection, one straightforward possible divergence is identified by noticing that in eq. (\ref{eq:23}) there is a $\alpha_3^{-1-\epsilon}$ factor that will produce a divergence whenever $\alpha_3=0$. Since $\alpha_3$ multiplies the denominator $((w-tn)^2+i\eta)^{1-\epsilon}$, this limit can be identified with the configuration where the internal gluon line becomes soft (indeed it is regulated for $\epsilon<0$). This is always the case for gluon lines, due to the power of their propagator. This happens too for the external gluon, which produces a soft divergence whenever $\alpha_2=1$ and overlaps with a possible collinear singularity whenever $\alpha_5=1$ at the same time.  \\
As already stated above, the Jacobian $\alpha^{-2+\epsilon}$ in eq. (\ref{eq:23}) encodes several UV divergences (regulated for $\epsilon>0$). Also, the appearance of the $M^2$ term in eq. (\ref{eq:23}) also helps identifying further divergent configurations: for example, $M^2$ vanishes for lightlike and collinear external points, $x^2=y^2=x\cdot y=0$, (in this case one needs to set also $\alpha_5=1$). 

\subsubsection{One-loop jet function with external gluon emission in coordinate space}
We will continue by reducing in quadrature the second contribution to the Abelian one-loop radiative jet function in Fig. \ref{radiatedjetAbelian}.
\begin{figure}[ht!]
 \centering
 \begin{tikzpicture}[>=stealth]
 \draw [decoration={aspect=0.4, segment length=1mm, amplitude=1mm,coil},decorate] (2,0) .. controls (2,-0.9) and (2.8,-.9) .. (3,-.9);
  \draw [decoration={aspect=0.4, segment length=1mm, amplitude=1mm,coil},decorate] (.2,1) -- (1,0);
 \draw[-latex,thick] (2.3,0)--(2.6,0);
 \draw[-latex,thick] (1.25,0)--(1.65,0);
  \draw[-latex,thick] (0.35,0)--(.6,0);
    \draw [thick](0,0)--(2.95,0);
  \draw [thick](1,0)--(2.1,0);
   \draw (1,-0.00) node {\tiny$\bullet$};
   \draw (1.98,-0.00) node {\tiny$\bullet$};
      \draw (.2,0.99) node {\tiny$\bullet$};
      \draw (0,-0.00) node {\tiny$\bullet$};
   \draw[double] (3,-0.05)--(3,-2);
   \draw (2.95,0) arc [start angle=180, end angle=-180, radius=0.05cm];
   \draw (3,-.91) node {$\cdot$}; 
   \draw (5,-1) node {$\equiv J^{(1)\mu b}_{\text{ext, rad}}(x,y)$};
         \draw (1,-0.2) node {$z$};
      \draw (2,0.2) node {$w$};
      \draw (0,1) node {$x$};
          \draw (0,0.2) node {$y$};
          \draw (3.3,-0.9) node {$tn$};
\end{tikzpicture}
\caption{Diagram representing the one-loop external gluon emission from a jet.}\label{repera33}
\end{figure}
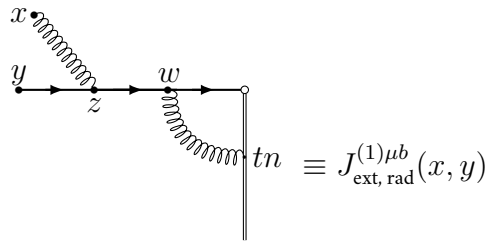

The diagram in Fig. \ref{repera33} equals
\begin{align}
J^{(1)\mu b}_{\text{ext, rad}}(x,y)=&(ig^3T^bC_F)\Big(\frac{\Gamma(2-\epsilon)}{2\pi^{2-\epsilon}}\Big)^3\Big(\frac{\Gamma(1-\epsilon)}{4\pi^{2-\epsilon}}\Big)^2
\int d^dzd^dw \int_0^\infty d t \frac{-(\slashed y-\slashed z)}{((z-y)^2+i\eta)^{2-\epsilon}} \times\nonumber\\
&\times \frac{-\gamma^\mu(\slashed z-\slashed w)}{((z-w)^2+i\eta)^{2-\epsilon}} \slashed n \frac{-\slashed w}{(w^2+i\eta)^{2-\epsilon}}\frac{1}{((w-tn)^2+i\eta)^{1-\epsilon}}\frac{1}{((x-z)^2+i\eta)^{1-\epsilon}}\;.\label{radjet2}
\end{align}
\paragraph{Analysis of Landau's equations} We will analyze Landau's equations for this diagram by anticipating the way we will use Feynman parameters to solve eq. (\ref{radjet2}). For the integration in $z$ we will first combine  the first and second denominator and subsequently the resulting denominator with the last denominator in eq. (\ref{radjet2}). Landau's equations for the resulting denominator for the $z$ integration read
\begin{align}
&z^2-2z\cdot (\alpha_2(\alpha_1y+(1-\alpha_1)w)+(1-\alpha_2)x)+\alpha_2(\alpha_1y^2+(1-\alpha_1)w^2)+(1-\alpha_2)x^2=0\label{land12}\\
&z^\mu=\alpha_2(\alpha_1y^\mu+(1-\alpha_1)w^\mu)+(1-\alpha_2)x^\mu\label{land22}
\end{align}
Let us analyze the case when $\boxed{\alpha_2=1}$, which corresponds to the case of a \textbf{soft emitted gluon} (since the Feynman parameter of the external gluon line is $1-\alpha_2$). In this case eq. (\ref{land22}) amounts to $z^\mu=\alpha_1y^\mu+(1-\alpha_1)w^\mu$ so that the emitted gluon does not change the direction of the fermion traveling from $w$ to $y$ as we expect from the momentum space picture. Using this result in eq. (\ref{land12}) for lightlike external points, $x^2=y^2=0$, one obtains  $\alpha_1(1-\alpha_1)(w-y)^2=0$, which sets $w-y$ on the lightcone except for the endpoints of the integration in $\alpha_1$, these endpoints correspond to UV divergences (in two distinct fermion lines) $z^\mu=w^\mu$ for $\alpha_1=0$ and $z^\mu=y^\mu$ for $\alpha_1=1$. Next, let us take $\boxed{\alpha_2=0}$, this gives the UV condition that $x^\mu=z^\mu$ so that the emitted gluon travels no distance. For general $\alpha_2$ and $\alpha_1=0,1$ we will have that $z$ will lie in the line connecting $x$ to $w$ and $y$ respectively.

Landau's equations for the other vertex, after the steps taken to reach eq. (\ref{intermediatestep2}) below, read
\begin{align}
\alpha w^2+2w\cdot \theta+\kappa^2=&0\;,\\
\alpha w^\mu+\theta^\mu=&0\;.
\end{align}
Here $\alpha=\alpha_4\alpha_5+(1-\alpha_1)\alpha_2(1-(1-\alpha_1)\alpha_2)$, $ \theta^\mu=-(1-\alpha_4)\alpha_5n^\mu-\alpha_1\alpha_2(1-\alpha_1)y^\mu-(1-\alpha_1)(1-\alpha_2)y^\mu$, and $\kappa^2=\alpha_1\alpha_2(1-\alpha_1\alpha_2)y^2+\alpha_2(1-\alpha_2)x^2-2\alpha_1\alpha_2(1-\alpha_2)x\cdot y$. Taking lightlike and collinear external points, one finds similar conclusions as in the previous example by following the same logic. 

\paragraph{Computation} As in the previous example we will use sequential Feynman parameters to perform the integration in $z$. We will first combine the first and second denominator and the resulting denominator with the last one in eq. (\ref{radjet2}), after shifting $z\to z-(\alpha_2(\alpha_1 y+(1-\alpha_1)w)+(1-\alpha_2)x)$, dropping odd powers and integrating in $z$ yields
\begin{align}
J^{(1)\mu b}_{\text{ext, rad}}(x,y)=&C^b
\int d^dw \int_0^\infty d t   \int_0^1d\alpha_1d\alpha_2\frac{\alpha_1^{1-\epsilon}(1-\alpha_1)^{1-\epsilon}\alpha_2^{3-2\epsilon}(1-\alpha_2)^{-\epsilon}}{(w^2+i\eta)^{2-\epsilon}((w-tn)^2+i\eta)^{1-\epsilon}} \times\nonumber\\
&\times\Big(K^\mu-\gamma^\mu\slashed n  \slashed w(m^2)/2\Big)(m^2)^{2\epsilon-3} \;.\label{radjet22}
\end{align}
where $C^b$ is a color constant, $m^2=\alpha_1\alpha_2(1-\alpha_1\alpha_2)y^2+(1-\alpha_1)\alpha_2(1-(1-\alpha_1)\alpha_2)w^2+\alpha_2(1-\alpha_2)x^2-2(\alpha_1\alpha_2 y\cdot((1-\alpha_1)w+(1-\alpha_2)x)+(1-\alpha_1)(1-\alpha_2)w\cdot x)$, and $K^\mu=(\alpha_2(\alpha_1  \slashed y+(1-\alpha_1) \slashed w)+ \slashed y)\gamma^\mu (\alpha_2(\alpha_1  \slashed y+(1-\alpha_1) \slashed w)+ \slashed w) \slashed n\slashed w$. Next we combine the first and second denominators in eq. (\ref{radjet22}) and compute the integral in $t$ in the spacelike region $w\cdot n<0$. We combine the resulting two denominators with the last denominator left in two steps to get 
\begin{align}
J^{(1)\mu b}_{\text{ext, rad}}(x,y)=&C''^b\int d^dw \int_0^1d\alpha_1...d\alpha_5{\alpha_1^{1-\epsilon}(1-\alpha_1)^{1-\epsilon}\alpha_2^{3-2\epsilon}(1-\alpha_2)^{-\epsilon}\alpha_3^{1-\epsilon}(1-\alpha_3)^{-1-\epsilon}}\times\nonumber\\
&\times \frac{\alpha_4^{1-2\epsilon}\alpha_5^{2-2\epsilon}(1-\alpha_5)^{2-2\epsilon}}{(\alpha w^2+2w\cdot  \theta+\kappa^2)^{6-4\epsilon}}\Big(K^\mu-\gamma^\mu\slashed n  \slashed w(m^2)/2\Big)\;.\label{intermediatestep2}
\end{align}
We are now ready to compute the integral in $w$ by shifting $w\to w\alpha^{\frac{1}{2}}+ \theta\alpha^{-\frac{1}{2}}$ to obtain
\begin{align}
J^{(1)\mu b}_{\text{rad}}&(x,y)=\frac{ig^3T^b C_F}{2^{15-4\epsilon}\pi^{14-7\epsilon}}\int_0^1d\alpha_1...d\alpha_5\alpha_1^{1-\epsilon}(1-\alpha_1)^{1-\epsilon}\alpha_2^{3-2\epsilon}(1-\alpha_2)^{-\epsilon}\alpha_3^{-1-\epsilon}(1-\alpha_3)^{2-2\epsilon} \times\nonumber\\
&\times\alpha_4^{2-3\epsilon}(1-\alpha_4)^{1-\epsilon}\alpha_5^{4-4\epsilon}\alpha^{-2+\epsilon}\Big(\alpha g^{\rho\sigma}A^\mu_{\rho\sigma}\Gamma(3-3\epsilon)(M^2)/2+B^\mu\Gamma(4-3\epsilon)\Big)(M^2)^{3\epsilon-4}\;.\label{eq:23222}
\end{align}
The expressions for the $A_{\rho\sigma}^\mu$, $B^\mu$, and $M^2$ are shown next (here $\beta^\mu\equiv\alpha^{-\frac{1}{2}}\theta^\mu$):
\begin{align}
M^2=\kappa^2-\beta^2\;,
\end{align}
\begin{align}
B^\mu=&\Big[-(\alpha_2(\alpha_1  \slashed y+(\alpha_1-1) \slashed \beta)+ \slashed y)\gamma^\mu (\alpha_2(\alpha_1  \slashed y+(\alpha_1-1) \slashed \beta)- \slashed \beta)+\gamma^\mu \big(\alpha_1\alpha_2(1-\alpha_1\alpha_2)y^2\nonumber+\\
&+(1-\alpha_1)\alpha_2(1-(1-\alpha_1)\alpha_2)\beta^2+\alpha_2(1-\alpha_2)x^2-2[\alpha_1\alpha_2y\cdot((1-\alpha_2)x-\beta)-\nonumber\\
&-(1-\alpha_1)(1-\alpha_2)x\cdot \beta]\big)/2\Big]\slashed n\slashed \beta\;,
\end{align}
\begin{align}
A_{\rho\sigma}^\mu=&(\alpha_2(\alpha_1  \slashed y+(\alpha_1-1) \slashed \beta)+ \slashed y)\gamma^\mu(1+\alpha_2(1-\alpha_1))\gamma_\rho\slashed n \gamma_\sigma+(1-\alpha_1)\alpha_2\gamma_\rho \gamma^\mu\times\nonumber\\
&\times \big[(\alpha_2(\alpha_1  \slashed y+(\alpha_1-1) \slashed \beta)- \slashed \beta)\slashed n \gamma_\sigma-(1+\alpha_2(1-\alpha_1))\gamma_\sigma \slashed n\slashed \beta \big]+\nonumber\\
&+\gamma^\mu\slashed n \big[(1-\alpha_1)\alpha_2(1-(1-\alpha_1)\alpha_2)\slashed \beta g_{\rho\sigma}/2+\gamma_\rho \big((1-\alpha_1)\alpha_2(1-(1-\alpha_1)\alpha_2)\beta_\sigma+\nonumber\\
&+\alpha_1\alpha_2(1-\alpha_1)y_\sigma +(1-\alpha_1)(1-\alpha_2)x_\sigma \big)\big]\;.
\end{align}
The result in eq. (\ref{eq:23222}) has the same form as eq. (\ref{eq:23}) and could be combined to produce a more tractable result, although the inclusion of the other two diagrams in the Abelian case (of Fig. \ref{radiatedjetAbelian}) is needed for gauge invariance. We will leave this for a future work.
Instead, to complete the discussion of the structures that may appear, we now turn to analyze Landau's equations for the self energy corrections to the one-loop radiative jet function.

\subsubsection{Landau's equations for the internal-emission self-energy correction of the one-loop radiative jet function}

\begin{figure}[ht!]
\begin{tikzpicture}[>=stealth]
\draw (1.5,-0.2) node {$v$};
\draw (2.25,0.2) node {$z$};
\draw (.75,0.2) node {$w$};
 \draw [decoration={aspect=0.4, segment length=1mm, amplitude=1mm,coil},decorate] (0.75,0) .. controls (0.75,-0.9) and (2.25,-.9) .. (2.25,0);
  \draw [decoration={aspect=0.4, segment length=1mm, amplitude=1mm,coil},decorate] (.2,1) -- (1.5,0);
  \draw[-latex,thick] (2.6,0)--(2.75,0);
  \draw[-latex,thick] (1,0)--(1.25,0);
    \draw[-latex,thick] (1.75,0)--(2,0);
  \draw[-latex,thick] (0.35,0)--(.6,0);
    \draw [thick](0,0)--(2.95,0);
  \draw [thick](1,0)--(2.1,0);
   \draw (0.75,-0.00) node {\tiny$\bullet$};
   \draw (2.25,-0.00) node {\tiny$\bullet$};
   \draw (1.5,-0.00) node {\tiny$\bullet$};
      \draw (.2,0.99) node {\tiny$\bullet$};
      \draw (0,-0.00) node {\tiny$\bullet$};
   \draw[double] (3,-0.05)--(3,-2);
   \draw (2.95,0) arc [start angle=180, end angle=-180, radius=0.05cm];
      \draw (0,1) node {$x$};
          \draw (0,0.2) node {$y$};
\end{tikzpicture}
\centering
\caption{Diagram representing the one-loop self energy with internal gluon emission from a jet.}\label{repera3}
\end{figure}
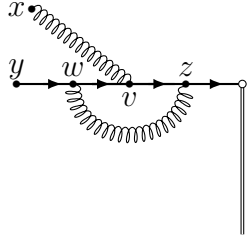
Next, we will analyze Landau's equations for the self-energy contributions to gain some intuition on some of their divergent configurations. The first one is depicted in Fig. \ref{repera3} and, after Feynman parametrization, equals 

\begin{align}
J^{(1)\mu b}_{\text{SE, int, rad}}&(x,y)=(ig^3T^aT^bT^a)\frac{\Gamma(10-6\epsilon)}{(2\pi^{2-\epsilon})^4(4\pi^{2-\epsilon})^2}
\int d^dzd^dw\int_0^1d\alpha'_1... d\alpha'_6\nonumber\times\\\times&{\alpha'_1}^{1-\epsilon}{\alpha'_2}^{1-\epsilon}{\alpha'_3}^{1-\epsilon}{\alpha'_4}^{-\epsilon}{\alpha'_5}^{1-\epsilon}{\alpha'_6}^{-\epsilon} \delta\left(1-\sum_{n=1}^6 \alpha'_n\right)\times\nonumber\\
\times &\frac{(\slashed y-\slashed w)\gamma^\nu(\slashed w-\slashed v)\gamma^\mu(\slashed v-\slashed z)\gamma_\nu \slashed z}{\big(\alpha'_1(y-w)^2+\alpha'_2(w-v)^2+\alpha'_3(v-z)^2+\alpha'_4(w-z)^2+\alpha'_5z^2+\alpha'_6(x-v)^2\big)^{10-6\epsilon}}\;.\label{radjet3}
\end{align}

The first of Landau's equations for the diagram in Fig. \ref{repera3} reads
\begin{align}
&&\alpha'_1(y-w)^2+\alpha'_2(w-v)^2+\alpha'_3(v-z)^2+\alpha'_4(w-z)^2+\alpha'_5z^2+\alpha'_6(x-v)^2=0\label{landauselfenergy1}\;,
\end{align}
which as usual sets all lines on the lightcone except if the lines are hard ($z_j\to0$) or soft ($\alpha_j=0$).  

Now, we will separately treat the integration in each vertex, effectively imposing conservation of momentum by employing Feynman parameters in separate steps (just as in the previous examples). The primed Feynman parameters in eq. (\ref{landauselfenergy1}) will be replaced later with the corresponding parameters after successive Feynman parametrizations. Furthermore, we will set all lines not connected to the vertex under analysis to be lightlike. 

For the integration in $z$, the second of Landau's equations gives (taking $\alpha_3'=\alpha_3(1-\alpha_4)$, $\alpha_5'=(1-\alpha_3)(1-\alpha_4)$ and $\alpha_4'=\alpha_4$)
\begin{align}
&z^\mu={\alpha_3(1-\alpha_4) v^\mu+\alpha_4 w^\mu}\;.
\end{align}
If either the emitted gluon or the emitted fermion at $z$ are soft ($\alpha'_3=0$ and $\alpha'_4=0$ respectively) then $z$ must lie in the direction of $w$ or $v$ respectively. In this way, the emitted particle does not change the direction of the particle emitting it. \\
If $\alpha'_5=1$ and hence $\alpha_3=\alpha_4=0$ (soft emitted lines from $z$) then there is a UV solution to Landau's equations with $z^\mu=0$. \\
For $\alpha'_5=0$ with $\alpha_3=1$ (so that the outgoing fermion from the Wilson line cusp is soft), $z^\mu$ will be the convex combination of the other two internal vertices (\textit{i.e.} on the line connecting them) and yields the condition that $(1-\alpha_4)\alpha_4(v-w)^2=0$. Therefore, if $\alpha_4\neq 0,1$, then $v$ and $w$ must be lightlike separated in order to obtain a divergence in the amplitude. 
For the case when the emitted gluon is soft, $\alpha_4=0$, then the gluon does not change the direction of the fermion, $z^\mu=\alpha_3 v^\mu$, and $v^2=0$ except for the UV cases where $\alpha_3=0$ ($z=0$) and $\alpha_3=1$ ($z=v$). If $\alpha_3=0$ then it is the outgoing fermion from $z$ that is soft and hence $z^\mu=\alpha_4 w^\mu$ and hence $w^2=0$ except for the UV cases.\\

For the integration in $v$, the second Landau equation gives (taking $\alpha_2'=\alpha_2(1-\alpha_6)$, $\alpha_3'=(1-\alpha_2)(1-\alpha_6)$ and $\alpha_6'=\alpha_6$)
\begin{align}
&v^\mu={\alpha_2(1-\alpha_6) w^\mu+(1-\alpha_2)(1-\alpha_6) z^\mu+\alpha_6 x^\mu}\label{eq:vint}\;.
\end{align}
Let us analyze the case when the emitted gluon traveling to $x$ becomes soft, \textit{i.e.} ${\alpha_6=0}$. In this case $v^\mu$ will lie on the line connecting the other two internal vertices not changing the direction of the fermion. Introducing eq. (\ref{eq:vint}) into eq. (\ref{landauselfenergy1}) we find $\alpha_2(1-\alpha_2)(w-z)^2=0$. This means that, if no fermion line connected to $v$ is soft, then the fermion lines connecting $z$ to $w$ must lie on the lightcone. If $\alpha_2$ or $(1-\alpha_2)$ are equal to zero then the fermion lines can be off the lightcone and there is a UV singularity since we will have that  $v=z$ or $v=w$ respectively. \\
The integration in $w$ has very similar Landau's equations as the one in $z$. Hence, we will move on to analyze these equations for the next and last contribution to the Abelian one-loop radiative jet function.

\subsubsection{Landau's equations for the external-emission self-energy correction of the one-loop radiative jet function}

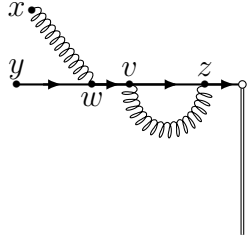
\begin{figure}[ht!]
\begin{tikzpicture}[>=stealth]
      \draw (0,1) node {$x$};
          \draw (0,0.2) node {$y$};
 \draw (1.5,.2) node {$v$};
 \draw (2.5,.2) node {$z$};
  \draw (1,-.2) node {$w$};
 \draw [decoration={aspect=0.4, segment length=1mm, amplitude=1mm,coil},decorate] (1.5,0) .. controls (1.5,-0.8) and (2.5,-.8) .. (2.5,0);
  \draw [decoration={aspect=0.4, segment length=1mm, amplitude=1mm,coil},decorate] (.2,1) -- (1,0);
  \draw[-latex,thick] (2.,0)--(2.15,0);
    \draw[-latex,thick] (2.5,0)--(2.9,0);
 \draw[-latex,thick] (1.,0)--(1.375,0);
  \draw[-latex,thick] (0.35,0)--(.6,0);
    \draw [thick](0,0)--(2.95,0);
  \draw [thick](1,0)--(2.1,0);
   \draw (1,-0.00) node {\tiny$\bullet$};
   \draw (1.5,-0.00) node {\tiny$\bullet$};
    \draw (2.5,-0.00) node {\tiny$\bullet$};
      \draw (.2,0.99) node {\tiny$\bullet$};
      \draw (0,-0.00) node {\tiny$\bullet$};
   \draw[double] (3,-0.05)--(3,-2);
   \draw (2.95,0) arc [start angle=180, end angle=-180, radius=0.05cm];
\end{tikzpicture}
\centering
\caption{Diagram representing the one-loop self energy with external gluon emission from a jet.}\label{repera4}
\end{figure}

The last contribution to the Abelian one-loop radiative jet function is shown in Fig. \ref{repera4} and, after Feynman parametrization, amounts to

\begin{align}
J^{(1)\mu b}_{\text{SE, ext, rad}}&(x,y)=(ig^3T^bC_F)\frac{\Gamma(10-6\epsilon)}{(2\pi^{2-\epsilon})^4(4\pi^{2-\epsilon})^2}
\int d^dzd^dw\int_0^1d\alpha'_1... d\alpha'_6 \times\nonumber\\
&\times{\alpha'_1}^{1-\epsilon}{\alpha'_2}^{1-\epsilon}{\alpha'_3}^{1-\epsilon}{\alpha'_4}^{-\epsilon}{\alpha'_5}^{1-\epsilon}{\alpha'_6}^{-\epsilon}\delta\left(1-\sum_{n=1}^6 \alpha'_n\right) \times\nonumber\\
&\times \frac{(\slashed w-\slashed y) \gamma^\mu(\slashed w-\slashed v)(d-2)(\slashed v-\slashed{z})\slashed{z}}{\big(\alpha'_1(y-w)^2+\alpha'_2(w-v)^2+(\alpha'_3+\alpha'_4)(v-z)^2+\alpha'_5z^2+\alpha'_6(x-w)^2\big)^{10-6\epsilon}}\;.\label{radjet4}
\end{align}

Hence, the first of Landau's equations for the diagram in Fig. \ref{repera4} reads
\begin{align}
&&\alpha'_1(y-w)^2+\alpha'_2(w-v)^2+(\alpha'_3+\alpha'_4)(v-z)^2+\alpha'_5z^2+\alpha'_6(x-w)^2=0\;,\label{landauselfenergy2}
\end{align}
which again sets all lines on the lightcone except those that are either hard ($z_j\to0$) or soft ($\alpha_j=0$). Notice that for the self energy piece $\alpha_3+\alpha_4$ combine into only one Feynman parameter.

For the integration in $w$, the second of Landau's equation gives (taking $\alpha_1'=\alpha_1(1-\alpha_6)$, $\alpha_2'=(1-\alpha_1)(1-\alpha_6)$ and $\alpha_6'=\alpha_6$)
\begin{align}
&w^\mu={\alpha_1(1-\alpha_6) y^\mu+(1-\alpha_1)(1-\alpha_6) v^\mu+\alpha_6 x^\mu}\label{eq:wint}\;.
\end{align}
This gives the usual solution for the soft-emitted gluon case ${\alpha_6=0}$, $w$ lying in the line connecting $v$ to $y$ and $\alpha_1(1-\alpha_1) (y-v)^2=0$. So that the fermion lines connecting $y$ and $v$ must lie on the lightcone and the soft gluon does not change the direction of the fermion. Again, UV solutions arise whenever $\alpha_1=0,1$ with $w=v$ and $w=y$ respectively. An analogue situation happens for a soft emitted fermion from $w$ to $y$ ($v$) corresponding to $\alpha_1=0$ ($\alpha_1=1$) for general $\alpha_6$. 
To finish, let us analyze Landau's equations for one vertex in the self energy part, for example $v$ (the one with $z$ is very similar). The second Landau's equation gives (taking $\alpha_3'=\alpha_3(1-\alpha_2)$, $\alpha_4'=(1-\alpha_3)(1-\alpha_6)$ and $\alpha_2'=\alpha_2$)
\begin{equation}
v^\mu=\alpha_2 w^\mu+(1-\alpha_2)z^\mu\;,
\end{equation}
so that $v$ must lie on the straight line connecting $w$ and $z$ with the usual condition $\alpha_2(1-\alpha_2)(w-z)^2=0$.

 \subsection{Summary}
In this section we have discussed the results leading to factorization of QCD amplitudes in both momentum and coordinate spaces and fully solved the one-loop contribution to the jet function in coordinate space, finding that the most divergent part of it is proportional to a fermion propagator (eq. (\ref{JCs})). This fact was identified as a configuration where the gluon emerging from the cusp of the Wilson line travelled collinearly with the fermion to the external point, hence producing the divergence. An LSZ reduction of this result shows that we recover the correct leading divergence expected from the previously known momentum space results.

This means that the theory of jets can, with confidence, be cast in the coordinate space formalism.

\newpage
\section{Flow Oriented Perturbation Theory}\label{sec:fopt}
In this section we will introduce a novel representation of Feynman graphs in coordinate space, which we call Flow Oriented Perturbation Theory (FOPT) based in the author's original paper \cite{Borinsky:2022msp}. We start by discussing the one-loop triangle diagram in some detail. A number of useful concepts for FOPT graphs, such as their completion, cycles and routes are introduced and explained with this example. Then we move on to introduce the general derivation of FOPT and discuss some of its interesting features. 

\subsection{Scalar QFT in coordinate space}
\label{sec:coordinate_space_feynman_rules}
The free massless coordinate-space Feynman propagator for a scalar field in $D=4$ dimensional space time reads 
\begin{align}
\Delta_F(z) = \int \frac{\dd^4 p}{(2\pi)^4} e^{-i p \cdot z}\frac{i}{p^2+ i \ep}
= \frac{1}{(2 \pi)^2} \frac{1}{-z^2 +  i \ep}\,.
\end{align}
In analogy to the momentum space formulation, the Feynman rules provide a recipe to translate a graph $G$ with sets of edges $\gE$, internal vertices $\gVI$ and external vertices $\gVE$ into an integral. Recall that external vertices are defined by the requirement that there is only one adjacent edge to each of them, therefore, they are guaranteed to make part of no loop. To complete the ones in introduced in subsection \ref{sec:PTphi4}, the usual coordinate-space Feynman rules (see for instance \cite[Ch.~6.1]{Itzykson:1980rh}) read
\begin{enumerate}
\item
Associate a coordinate vector to each internal or external vertex. We label the location of external vertices with $x_a$, $a\in V^{\text{ext}}$, and that of internal vertices with $y_v$, $v\in V^{\text{int}}$. 
\item
For each \emph{internal} edge $e=\{v,v'\}$ multiply by a Feynman propagator $\Delta_F(z_e) = \Delta_F(-z_e)$, where $z_e$ is the difference of the coordinates of the vertices to which the edge $e$ is incident. For example, if $e$ is an internal edge (none of the two vertices defining it is external), then $z_e=y_{v}-y_{v'}$. Alternatively, if $v$, for example, is instead an external vertex, then $z_e=x_v-y_{v'}$.
\item For each interaction vertex multiply by a factor  $-ig$.
\item For each internal vertex $v\in V^{\text{int}}$ integrate over all values of the components of $y_v$, i.e. over all possible locations of the internal vertex in $4$-dimensional Minkowski space. 
\end{enumerate}
The resulting expression is a function of the external coordinates $\{x_a\}_{a\in V^{\text{ext}}}$. To be explicit, the application of the coordinate-space Feynman rules to a generic graph $G$ contributing to a Green's function of a massless scalar theory gives
\begin{align}
A_G(x_1, \ldots, x_{|\gVE|})=
\frac{(-ig)^{|\gVI|}}{(2\pi)^{2|\gE|}}
\left[ \prod_{v \in \gVI} \int \dd^4 y_v \right]
\prod_{e \in \gE}\frac{1}{-z_e^2+i\ep}\;.\label{eq:FI}
\end{align}
One integrates over the four dimensional Minkowski space location of each internal vertex. Accounting for symmetry factors results in an expression for the scalar $n$-point correlation function,
\begin{align}\label{scalarcorrelator}
\Gamma(x_1,...,x_{|\gVE|})=\left\langle 0 | T ( \varphi(x_1) \cdots \varphi(x_{|\gVE|}) ) | 0 \right \rangle =
\sum_{G} \frac{1}{\Sym G} A_G(x_1, \ldots, x_{|\gVE|}),
\end{align}
where we sum over all graphs $G$ from a given scalar QFT with the given, fixed external vertices $\gVE$.

\subsection{Illustrative example: The triangle diagram in FOPT}

In this section we treat the FOPT representation of the triangle diagram. It will serve as a prototype for later derivations in this section, because it is simple enough to show all details, while also exhibiting most of the subtleties associated with the general arguments. In coordinate space the triangle diagram can be drawn as
\begin{center}
    \resizebox{5cm}{!}{%
\begin{tikzpicture}

    \begin{feynman}
        \vertex(1);
        \vertex[above right = 1.5cm and 2.5cm of 1](2);
        \vertex[below right = 1.5cm and 2.5cm of 1](3);
        
        \vertex[left = 1.5cm of 1](E1);
        \vertex[right = 1.5cm of 2](E2);
        \vertex[right = 1.5cm of 3](E3);
        
        \vertex[left = 0.2cm of E1](L1) {\scalebox{1.5}{$x_1$}};
        \vertex[right = 0.2cm of E2](L2) {\scalebox{1.5}{$x_2$}};
        \vertex[right = 0.2cm of E3](L3) {\scalebox{1.5}{$x_3$}};
        
        \vertex[below = 0.2cm of 1](L1_I) {\scalebox{1.5}{$y_1$}};
        \vertex[above = 0.2cm of 2](L2_I) {\scalebox{1.5}{$y_2$}};
        \vertex[below = 0.2cm of 3](L3_i) {\scalebox{1.5}{$y_3$}};
        
        \vertex[above left = 0.2cm and 0.5cm of 1](edge1) {\scalebox{1.5}{$e_1$}};
        \vertex[above right = 0.2cm and 0.5cm of 2](edge2) {\scalebox{1.5}{$e_2$}};
        \vertex[below right = 0.2cm and 0.5cm of 3](edge3) {\scalebox{1.5}{$e_3$}};
        
        \vertex[above right = 1cm and 1cm of 1](edge4) {\scalebox{1.5}{$e_4$}};
        \vertex[below right = 1cm and 1cm of 1](edge5) {\scalebox{1.5}{$e_5$}};
        \vertex[below right = 1.25cm and 0.2cm of 2](edge6) {\scalebox{1.5}{$e_6$}};

         \diagram*[large]{	
            (1) -- [-,line width=0.6mm] (2) -- [-,line width=0.6mm] (3) -- [-,line width=0.6mm] (1),

            (1) -- [-,line width=0.6mm] (E1),
            (2) -- [-,line width=0.6mm] (E2),
            (3) -- [-,line width=0.6mm] (E3),

        }; 
       \path[draw=black, fill=black] (E1) circle[radius=0.1];
       \path[draw=black, fill=black] (E2) circle[radius=0.1];
       \path[draw=black, fill=black] (E3) circle[radius=0.1];
       \path[draw=black, fill=black] (1) circle[radius=0.1];
       \path[draw=black, fill=black] (2) circle[radius=0.1];
       \path[draw=black, fill=black] (3) circle[radius=0.1];
       
    \end{feynman}

\end{tikzpicture}
}
\end{center}
We labelled the external vertex locations with the variables $x_1$, $x_2$ and $x_3$ and their adjacent internal vertices' locations with $y_1$, $y_2$ and $y_3$ respectively. To each edge we associate a label $e_i$, $i=1,...,6$.  
Given this labelling, the coordinate-space triangle diagram, according to the Feynman rules presented in sec.~\ref{sec:coordinate_space_feynman_rules}, reads
\begin{align}
\label{eq:triangle_position_space}
    A_G&(x_1,x_2,x_3)=\nonumber\frac{(-ig)^{3}}{(2\pi)^{12}}
\int \left[\prod_{v\in \gVI}\dd^4 y_v\right]\times
\\
 &\times\frac{1}{(x_1-y_1)^2(x_2-y_2)^2(x_3-y_3)^2 (y_1-y_2)^2 (y_2-y_3)^2 (y_1-y_3)^2 }\,,
\end{align}
where a positive $i\varepsilon$ prescription is assumed for all propagators. The expression $A_G$ is a function of the three external coordinates $x_1,x_2,x_3$.
Our first aim is to perform the time integrations $\left[ \int \mathrm{d} y_v^0 \right]$ in eq.~\eqref{eq:triangle_position_space} explicitly. We will do so by employing the \emph{residue theorem} together with a series of distributional identities, which will  allow us to cast the result in an especially elegant form. 

    We aim for an expression for 
the following partially integrated version of $A_G$, which we denote by $\AA_G$:
\begin{align}
        \AA_G(x_1,x_2,x_3,\vec{y}_1, \vec{y}_2, &\vec{y}_3)= \frac{(-ig)^{3}}{(2\pi)^{12}}
\int \left[\prod_{v\in \gVI}\mathrm{d} y_v^0\right]\times\nonumber\\
 &\times\frac{1}{(x_1-y_1)^2(x_2-y_2)^2(x_3-y_3)^2 (y_1-y_2)^2 (y_2-y_3)^2 (y_1-y_3)^2 }\,.\label{eq:mathcal_A_triangle}
\end{align}
Henceforth, we omit the variable dependence of $\AA_G$, but emphasize that it depends on the 4-vector coordinates of the external vertices and the 3-vector coordinates of the internal vertices.

To evaluate eq.~\eqref{eq:mathcal_A_triangle} using the residue theorem it is convenient to introduce auxiliary variables $z_e^0$, for each $e\in E$,  set equal to the time-difference between the vertices connected through $e$ via a delta function. These variables will be integrated from $-\infty$ to $\infty$. Thus we have the representation
\begin{align}
\label{eq:triangle_dirac_deltas}
    \AA_G=&
\frac{(-ig)^{3}}{(2\pi)^{12}}
\int \left[\prod_{v \in \gVI}\mathrm{d} y_v^0\right]\left[\prod_{e\in\gE}\frac{\mathrm{d} z_e^0}{-z_e^2 + i\varepsilon}\right] \times \nonumber \\ &
\times
\left[\prod_{i=1}^3\delta(z_i^0-x_i^0+y_i^0)\right] \delta(z_4^0-y_{12}^0)\delta(z_5^0-y_{13}^0)\delta(z_6^0-y_{23}^0),
\end{align}
where we reintroduced the $i\varepsilon$ prescription and uses the shorthand notation $y_{ab}^0 = y_{a}^0-y_b^0$. 

For each edge, we can use the integral representation of the delta function, 
$\delta(z) = \int_{-\infty}^\infty
\frac{\dd E}{2\pi} e^{iEz}$,
to write $\AA_G$ as integrals of oscillating exponentials
\begin{gather}
    \AA_G=
\frac{(-ig)^{3}}{(2\pi)^{12}}
\int \left[\prod_{e \in \gE}\frac{\dd z_e^0 \dd E_e/(2\pi)}{-z_e^2 + i\varepsilon}\right]
\left[\prod_{i=1}^3\dd y_i^0 e^{i E_i(z_i^0-x_i^0+y_i^0)} \right]  
e^{iE_4(z_4^0-y_{12}^0)+iE_5(z_5^0-y_{13}^0)+iE_6(z_6^0-y_{23}^0)}.\label{eq:intermediatesteptriangle}
\end{gather}
Eventually, we will interpret the auxiliary variables $E_e$ as the amount of energy that flows through the edge $e$ in the chosen orientation (notice the mass dimension of the $E_e$).
We are now ready to carry out the integration in the auxiliary variables $z_e^0$ of eq.~(\ref{eq:intermediatesteptriangle}). Reordering eq.~(\ref{eq:intermediatesteptriangle}) gives 
\begin{equation}
    \AA_G=
\frac{(-ig)^{3}}{(2\pi)^{12}}
\int\left[\prod_{e\in \gE} \frac{\dd E_e}{2\pi}\right] \left[\prod_{i=1}^3\dd y_i^0e^{iE_i(-x_i^0+y_i^0)}\right]  e^{-iE_4 y_{12}^0-iE_5y_{13}^0-iE_6y_{23}^0}\prod_{e\in \gE}\int\dd z_e^0\frac{ e^{i z_e^0 E_e}}{-z_e^2+i\varepsilon}.
\end{equation}
The integrals over $z_e^0$ can be performed using the residue theorem. As the denominators read $-z_e^2 + i\varepsilon = -(z_e^0)^2 + |\vec{z_e}|^2 +i\varepsilon$, the integrand has two poles, located at $z_e^0=\pm \sqrt{|\vec{z}_e|^2+i\varepsilon}$. For $E_e>0$, we close the contour of integration in the upper-half of complex plane, while for $E_e<0$, we close it in the lower half. In the former case, the pole contained in the integration contour is $z_e^0= \sqrt{|\vec{z}_e|^2+i\varepsilon}$, while in the latter case it is $z_e^0=- \sqrt{|\vec{z}_e|^2+i\varepsilon}$. Therefore, we have
\begin{equation}
\label{eq:contour}
    \int\dd z_e^0\frac{ e^{i z_e^0 E_e}}{-z_e^2+i\varepsilon} =
\frac{-2\pi i}{2|\vec{z}_e|}\left(e^{i(|\vec{z}_e|+i\ep)E_e}\Theta(E_e)+e^{-i(|\vec{z}_e|+i\ep)E_e}\Theta(-E_e)\right),
\end{equation}
where we respected the $i\ep$-prescription by adding a small imaginary part to $|\vec{z}_e|$.

The above expression means that we should differently treat negative and positive-energy flows through an edge. This is the first key step in the flow-oriented perturbation theory formalism.
We will make this aspect explicit by writing the product of the sums of two terms as a sum over $2^{|\gE|}$ terms. Each resulting term can be interpreted as an assignment of flow directions to each edge of the graph. 
We will denote such an assignment as $\bb\sigma$ that assigns  $\sigma_e =\pm 1$ to an edge $e$ to indicate a positive or negative energy flow. 
Hence, we write the product as,
\begin{equation}
    \prod_{e \in \gE}\int\dd z_e^0\frac{ e^{i z_e^0 E_e}}{-z_e^2+i\varepsilon}=\sum_{\bb\sigma\in\{\pm 1\}^6}\prod_{e \in \gE}\frac{-2\pi i}{2|\vec{z}_e|}e^{i \sigma_e( |\vec{z}_e| + i \ep)E_e}\Theta(\sigma_e E_e),
\end{equation}
where $\bb \sigma$ runs over all vectors of length $6$ with $\pm 1$ entries. Inserting this in the expression for $\AA_G$ gives
\begin{gather}
     \AA_G= \frac{(-ig)^{3}}{(2\pi)^{12}}
\sum_{\bb\sigma\in\{\pm 1\}^6}\int
    \left[\prod_{e\in \gE} \frac{\dd E_e}{2i|\vec{z}_e|}e^{i\sigma_e( |\vec{z}_e| + i \ep)E_e}\Theta(\sigma_e E_e)\right] \times \nonumber \\
 \times    \left[\prod_{i=1}^3\dd y_i^0e^{iE_i(y_i^0-x_i^0)}\right]   e^{-iE_4y_{12}^0-iE_5y_{13}^0-iE_6y_{23}^0}.
\end{gather}
Note that the $i\ep$ ensures convergence of the $E_e$ integrals. We can rearrange the exponentials and resolve the integration over the $y_i^0$ variables by using the integral representation of the delta function, but in reverse:
\begin{align}
    \AA_G=&
\frac{(-ig)^{3}}{i^6 (2\pi)^{9}}\sum_{\bb\sigma\in\{\pm 1\}^6}
\int \left[\prod_{e\in\gE} \frac{\dd E_e}{2|\vec{z}_e|}e^{i\sigma_e(|\vec{z}_e|+ i\ep)E_e}\Theta(\sigma_eE_e)\right]\left[\prod_{i=1}^3 e^{-iE_ix_i^0}\right]\times
  \nonumber\\ &\times
    \delta(E_1-E_4-E_5)\delta(E_2+E_4-E_6)\delta(E_3+E_5+E_6)\,.
\end{align}
We observe that each of the delta functions is associated to an internal vertex $y_v^0$ and that they enforce \emph{energy-conservation} at each internal vertex.
Performing the change of variables $E_e\rightarrow\sigma_e E_e$, and resolving the theta functions gives
\begin{align}
\label{eq:integralA123}
    \AA_G=&
\frac{(-ig)^{3}}{i^6 (2\pi)^{9}}
\sum_{\bb\sigma\in\{\pm 1\}^6}\left[ \prod_{e\in \gE} \int_{0}^\infty\frac{\dd E_e}{2|\vec{z}_e|}e^{i (|\vec{z}_e|+i\ep)E_e}\right]\left[\prod_{i=1}^3 e^{-i \sigma_i E_ix_i^0}\right]  \nonumber\times\\
    &\times
    \delta(\sigma_1E_1-\sigma_4E_4-\sigma_5E_5)\delta(\sigma_2E_2+\sigma_4E_4-\sigma_6E_6)\delta(\sigma_3E_3+\sigma_5E_5+\sigma_6E_6)  . 
\end{align}
Whenever the sign vector $\bb\sigma$ is such that the argument of a delta function is either a strictly negative or positive sum of energies, then the integral is zero, as such sums cannot vanish under the constraint $E_e > 0$. 
This implies that only a subset of the vectors $\bb\sigma$ contribute. We will make heavy use of a diagrammatic interpretation of this integral in order to correctly understand which such vectors lead to a non-zero contribution. When  we introduced in eq.~\eqref{eq:triangle_dirac_deltas} the auxiliary variables that correspond to time differences, we implicitly chose an orientation for the graph. If $z_j^0=x_j^0-y_j^0$, then we choose the orientation of the $j$-th edge such that energy flows from the vertex $y_j$ to the vertex $x_j$. Analogously, if $z_4^0=y_1^0-y_2^0$, then the $4$-th edge orientation flows from $y_2$ to $y_1$. We can depict this chosen orientation as a directed graph, or \emph{digraph},
\begin{equation}
\resizebox{5cm}{!}{%
\begin{tikzpicture}

    \begin{feynman}
        \vertex(1);
        \vertex[above right = 1.5cm and 2.5cm of 1](2);
        \vertex[below right = 1.5cm and 2.5cm of 1](3);
        
        \vertex[left = 1.5cm of 1](E1);
        \vertex[right = 1.5cm of 2](E2);
        \vertex[right = 1.5cm of 3](E3);
        
        \vertex[left = 0.2cm of E1](L1) {\scalebox{1.5}{$x_1$}};
        \vertex[right = 0.2cm of E2](L2) {\scalebox{1.5}{$x_2$}};
        \vertex[right = 0.2cm of E3](L3) {\scalebox{1.5}{$x_3$}};
        
        \vertex[below = 0.2cm of 1](L1_I) {\scalebox{1.5}{$y_1$}};
        \vertex[above = 0.2cm of 2](L2_I) {\scalebox{1.5}{$y_2$}};
        \vertex[below = 0.2cm of 3](L3_i) {\scalebox{1.5}{$y_3$}};
        
        \vertex[above left = 0.2cm and 0.5cm of 1](edge1) {\scalebox{1.5}{$e_1$}};
        \vertex[above right = 0.2cm and 0.5cm of 2](edge2) {\scalebox{1.5}{$e_2$}};
        \vertex[below right = 0.2cm and 0.5cm of 3](edge3) {\scalebox{1.5}{$e_3$}};
        
        \vertex[above right = 1cm and 1cm of 1](edge4) {\scalebox{1.5}{$e_4$}};
        \vertex[below right = 1cm and 1cm of 1](edge5) {\scalebox{1.5}{$e_5$}};
        \vertex[below right = 1.25cm and 0.2cm of 2](edge6) {\scalebox{1.5}{$e_6$}};

         \diagram*[large]{	
            (1) -- [->-,line width=0.6mm] (2) -- [->-,line width=0.6mm] (3) -- [-<-,line width=0.6mm] (1),

            (1) -- [-<-,line width=0.6mm] (E1),
            (2) -- [-<-,line width=0.6mm] (E2),
            (3) -- [-<-,line width=0.6mm] (E3),

        }; 
       \path[draw=black, fill=black] (E1) circle[radius=0.1];
       \path[draw=black, fill=black] (E2) circle[radius=0.1];
       \path[draw=black, fill=black] (E3) circle[radius=0.1];
       \path[draw=black, fill=black] (1) circle[radius=0.1];
       \path[draw=black, fill=black] (2) circle[radius=0.1];
       \path[draw=black, fill=black] (3) circle[radius=0.1];
       
    \end{feynman}
\end{tikzpicture}
}
\label{eq:triangle_original_orientation}
\end{equation}

The sign vector $\bb \sigma$ can now be interpreted as inverting the orientations of edges in this digraph. For example, if $\bb\sigma=(-1,...,-1)$, then the corresponding digraph would have all edge orientations inverted. If instead $\bb\sigma=(1,-1,-1,1,1,-1)$, then the corresponding digraph is
\begin{equation}
\resizebox{5cm}{!}{%
\begin{tikzpicture}
 
    \begin{feynman}
        \vertex(1);
        \vertex[above right = 1.5cm and 2.5cm of 1](2);
        \vertex[below right = 1.5cm and 2.5cm of 1](3);
        
        \vertex[left = 1.5cm of 1](E1);
        \vertex[right = 1.5cm of 2](E2);
        \vertex[right = 1.5cm of 3](E3);
        
        \vertex[left = 0.2cm of E1](L1) {\scalebox{1.5}{$x_1$}};
        \vertex[right = 0.2cm of E2](L2) {\scalebox{1.5}{$x_2$}};
        \vertex[right = 0.2cm of E3](L3) {\scalebox{1.5}{$x_3$}};
        
        \vertex[below = 0.2cm of 1](L1_I) {\scalebox{1.5}{$y_1$}};
        \vertex[above = 0.2cm of 2](L2_I) {\scalebox{1.5}{$y_2$}};
        \vertex[below = 0.2cm of 3](L3_i) {\scalebox{1.5}{$y_3$}};
        
        \vertex[above left = 0.2cm and 0.5cm of 1](edge1) {\scalebox{1.5}{$e_1$}};
        \vertex[above right = 0.2cm and 0.5cm of 2](edge2) {\scalebox{1.5}{$e_2$}};
        \vertex[below right = 0.2cm and 0.5cm of 3](edge3) {\scalebox{1.5}{$e_3$}};
        
        \vertex[above right = 1cm and 1cm of 1](edge4) {\scalebox{1.5}{$e_4$}};
        \vertex[below right = 1cm and 1cm of 1](edge5) {\scalebox{1.5}{$e_5$}};
        \vertex[below right = 1.25cm and 0.2cm of 2](edge6) {\scalebox{1.5}{$e_6$}};

         \diagram*[large]{	
            (1) -- [->-,line width=0.6mm] (2) -- [-<-,line width=0.6mm] (3) -- [-<-,line width=0.6mm] (1),

            (1) -- [-<-,line width=0.6mm] (E1),
            (2) -- [->-,line width=0.6mm] (E2),
            (3) -- [->-,line width=0.6mm] (E3),

        }; 
       \path[draw=black, fill=black] (E1) circle[radius=0.1];
       \path[draw=black, fill=black] (E2) circle[radius=0.1];
       \path[draw=black, fill=black] (E3) circle[radius=0.1];
       \path[draw=black, fill=black] (1) circle[radius=0.1];
       \path[draw=black, fill=black] (2) circle[radius=0.1];
       \path[draw=black, fill=black] (3) circle[radius=0.1];
       
    \end{feynman}

\end{tikzpicture}
}\label{eq:triangle_right_orientation}
\end{equation}

As we can see from this example, the sum over all vectors $\bb\sigma$ is actually equal to the sum over all possible orientations of the triangle graph. The energy-conservation conditions imposed by the delta functions in eq.~\eqref{eq:integralA123} can then be interpreted as enforcing the conservation of energies at any internal vertex. As an example, let us look at the contribution of the sign vector $\bb\sigma=(1,\ldots,1)$ orientation to eq.~\eqref{eq:triangle_original_orientation},
\begin{equation}
\resizebox{10cm}{!}{%
\begin{tikzpicture}
    \begin{feynman}
        \vertex(1);
        \vertex[above right = 1.5cm and 2.5cm of 1](2);
        \vertex[below right = 1.5cm and 2.5cm of 1](3);
        
        \vertex[left = 1.5cm of 1](E1);
        \vertex[right = 1.5cm of 2](E2);
        \vertex[right = 1.5cm of 3](E3);

        \vertex[above left = 0.5cm and 0.cm of E1](mom1) {\scalebox{1.5}{$\color{col1}E_1-E_4-E_5=0$}};
        \vertex[above right = 0.5cm and 0.cm of E2](mom2) {\scalebox{1.5}{$\color{col2} -E_6+E_2+E_4=0$}};
        \vertex[below right = 0.5cm and 0.cm of E3](mom3) {\scalebox{1.5}{$\color{col3} E_3+E_5+E_6=0$}};
        
        \vertex[left = 0.2cm of E1](L1) {\scalebox{1.5}{$x_1$}};
        \vertex[right = 0.2cm of E2](L2) {\scalebox{1.5}{$x_2$}};
        \vertex[right = 0.2cm of E3](L3) {\scalebox{1.5}{$x_3$}};
        
        \vertex[below = 0.2cm of 1](L1_I) {\scalebox{1.5}{$y_1$}};
        \vertex[above = 0.2cm of 2](L2_I) {\scalebox{1.5}{$y_2$}};
        \vertex[below = 0.2cm of 3](L3_i) {\scalebox{1.5}{$y_3$}};
        
        \vertex[above left = 0.2cm and 0.5cm of 1](edge1) {\scalebox{1.5}{$e_1$}};
        \vertex[above right = 0.2cm and 0.5cm of 2](edge2) {\scalebox{1.5}{$e_2$}};
        \vertex[below right = 0.2cm and 0.5cm of 3](edge3) {\scalebox{1.5}{$e_3$}};
        
        \vertex[above right = 1cm and 1cm of 1](edge4) {\scalebox{1.5}{$e_4$}};
        \vertex[below right = 1cm and 1cm of 1](edge5) {\scalebox{1.5}{$e_5$}};
        \vertex[below right = 1.25cm and 0.2cm of 2](edge6) {\scalebox{1.5}{$e_6$}};

         \diagram*[large]{	
            (1) -- [->-,line width=0.6mm] (2) -- [->-,line width=0.6mm] (3) -- [-<-,line width=0.6mm] (1),

            (1) -- [-<-,line width=0.6mm] (E1),
            (2) -- [-<-,line width=0.6mm] (E2),
            (3) -- [-<-,line width=0.6mm] (E3),

        }; 
       \path[draw=black, fill=black] (E1) circle[radius=0.1];
       \path[draw=black, fill=black] (E2) circle[radius=0.1];
       \path[draw=black, fill=black] (E3) circle[radius=0.1];
       \path[draw=black, fill=black] (1) circle[radius=0.1];
       \path[draw=black, fill=black] (2) circle[radius=0.1];
       \path[draw=black, fill=black] (3) circle[radius=0.1];
       
       \path[draw=col1] (1) circle[radius=1.];
       \path[draw=col2] (2) circle[radius=1.];
       \path[draw=col3] (3) circle[radius=1.];
    \end{feynman}
\end{tikzpicture}
}\label{eq:triangle_zero_vertex}
\end{equation}
where we included the conditions imposed by the delta functions into the graphical representation.
The energy conservation condition in orange of the bottom right vertex is a sum of positive energies. Such an energy conservation condition can never be satisfied. It follows that this orientation gives no contribution to the FOPT representation of the triangle.

It is actually quite easy to see that the solution to the delta function constraint coincides with the physically intuitive picture 
of a realizable energy flow through the diagram along the indicated directions.
A digraph gives a non-zero contribution if
edges are followed in the positive orientation, and two connectivity conditions are fulfilled: (i) we must be able to reach each vertex by starting from some external vertex, and (ii) we must be able to reach some external vertex if we start form any vertex. Even though the $y_3$ vertex in the digraph in eq.~\eqref{eq:triangle_zero_vertex} can be reached from many external vertices by following a positive route, we cannot 
reach any of the other external vertices if we start from it. Hence, there is no proper energy-flow possible with the assigned orientation.
The orientation of eq.~\eqref{eq:triangle_right_orientation} can be depicted with its associated delta function arguments as
\begin{equation}
\resizebox{10cm}{!}{%
\begin{tikzpicture}
    \begin{feynman}
        \vertex(1);
        \vertex[above right = 1.5cm and 2.5cm of 1](2);
        \vertex[below right = 1.5cm and 2.5cm of 1](3);
        
        \vertex[left = 1.5cm of 1](E1);
        \vertex[right = 1.5cm of 2](E2);
        \vertex[right = 1.5cm of 3](E3);

        \vertex[above left = 0.5cm and 0.cm of E1](mom1) {\scalebox{1.5}{$\color{col1}E_1-E_4-E_5=0$}};
        \vertex[above right = 0.5cm and 0.cm of E2](mom2) {\scalebox{1.5}{$\color{col2} E_6-E_2+E_4=0$}};
        \vertex[below right = 0.5cm and 0.cm of E3](mom3) {\scalebox{1.5}{$\color{col3} -E_3+E_5-E_6=0$}};
        
        \vertex[left = 0.2cm of E1](L1) {\scalebox{1.5}{$x_1$}};
        \vertex[right = 0.2cm of E2](L2) {\scalebox{1.5}{$x_2$}};
        \vertex[right = 0.2cm of E3](L3) {\scalebox{1.5}{$x_3$}};
        
        \vertex[below = 0.2cm of 1](L1_I) {\scalebox{1.5}{$y_1$}};
        \vertex[above = 0.2cm of 2](L2_I) {\scalebox{1.5}{$y_2$}};
        \vertex[below = 0.2cm of 3](L3_i) {\scalebox{1.5}{$y_3$}};
        
        \vertex[above left = 0.2cm and 0.5cm of 1](edge1) {\scalebox{1.5}{$e_1$}};
        \vertex[above right = 0.2cm and 0.5cm of 2](edge2) {\scalebox{1.5}{$e_2$}};
        \vertex[below right = 0.2cm and 0.5cm of 3](edge3) {\scalebox{1.5}{$e_3$}};
        
        \vertex[above right = 1cm and 1cm of 1](edge4) {\scalebox{1.5}{$e_4$}};
        \vertex[below right = 1cm and 1cm of 1](edge5) {\scalebox{1.5}{$e_5$}};
        \vertex[below right = 1.25cm and 0.2cm of 2](edge6) {\scalebox{1.5}{$e_6$}};

         \diagram*[large]{	
            (1) -- [->-,line width=0.6mm] (2) -- [-<-,line width=0.6mm] (3) -- [-<-,line width=0.6mm] (1),

            (1) -- [-<-,line width=0.6mm] (E1),
            (2) -- [->-,line width=0.6mm] (E2),
            (3) -- [->-,line width=0.6mm] (E3),

        }; 
       \path[draw=black, fill=black] (E1) circle[radius=0.1];
       \path[draw=black, fill=black] (E2) circle[radius=0.1];
       \path[draw=black, fill=black] (E3) circle[radius=0.1];
       \path[draw=black, fill=black] (1) circle[radius=0.1];
       \path[draw=black, fill=black] (2) circle[radius=0.1];
       \path[draw=black, fill=black] (3) circle[radius=0.1];
       
       \path[draw=col1] (1) circle[radius=1.];
       \path[draw=col2] (2) circle[radius=1.];
       \path[draw=col3] (3) circle[radius=1.];
       
    \end{feynman}
\end{tikzpicture}
}\label{eq:triangle_flipped_deltas}
\end{equation}

In this case, the connectivity condition is fulfilled and we can 
comply with the previously problematic condition imposed by the delta function from the bottom right vertex. 
Hence, this orientation provides a non-zero contribution. 

This utility of the graphical representation suggests a definition of the Feynman integral $\AA_{G,\bb \sigma}$ associated to a single digraph (i.e.~a graph $G$ with an orientation $\bb \sigma$). We define
\begin{gather}
\begin{gathered}
    \AA_{G,\boldsymbol{\sigma}}(x_1,x_2,x_3,\vec{y}_1,\vec{y}_2,\vec{y}_3)=
\frac{(-ig)^{3}}{i^6 (2\pi)^{9}}
\left[ \prod_{e\in \gE} \int_{0}^\infty\frac{\dd E_e}{2|\vec{z}_e|}e^{i (|\vec{z}_e|+ i\ep)E_e}\right]\left[\prod_{i=1}^3 e^{-i \sigma_i E_ix_i^0}\right] \times \\    \times\,
    \delta(\sigma_1E_1-\sigma_4E_4-\sigma_5E_5)\delta(\sigma_2E_2+\sigma_4E_4-\sigma_6E_6)\delta(\sigma_3E_3+\sigma_5E_5+\sigma_6E_6)\,,
\end{gathered}
\end{gather}
such that 
\begin{align}
\AA_{G}(x_1,x_2,x_3,\vec{y}_1,\vec{y}_2,\vec{y}_3) = 
\sum_{\bb\sigma} 
\AA_{G,\bb \sigma} (x_1,x_2,x_3,\vec{y}_1,\vec{y}_2,\vec{y}_3)\,,
\end{align}
where  the sum runs over all possible orientations of the graph $G$. Each such orientation gives rise to a digraph $(G,\bb \sigma)$.

We will  derive a compact representation for the functions $\AA_{G,\bb \sigma} (x_1,x_2,x_3,\vec{y}_1,\vec{y}_2,\vec{y}_3)$,
and illustrate the derivation for this compact representation with an example. Take the orientation from eq.~\eqref{eq:triangle_right_orientation} and set $\bb\sigma=(1,-1,-1,1,1,-1)$.

In this case we have
\begin{align}
\AA_{G,\bb \sigma} = &
\frac{(-ig)^{3}}{i^6 (2\pi)^{9}}
 \left[\prod_{e\in\gE} \int_0^\infty \frac{\dd E_e}{2|\vec{z}_e|}e^{i (|\vec{z}_e|+i\ep)E_e}\right]\left[\prod_{i=1}^3 e^{-i \sigma_i E_i x_i^0}\right]  \times\nonumber\\
    &\times\delta(E_1-E_4-E_5)\delta(-E_2+E_4+E_6)\delta(-E_3+E_5-E_6). \label{eq:trangle_ori123}
\end{align}
To resolve the delta functions we need to choose a set of linearly independent energies. We choose $E_3, \, E_4, \, E_6$, which gives as linearly dependent energies
\begin{align}
    E_1=E_3+E_4+E_6, \quad
    E_2=E_4+E_6, \quad
    E_5=E_3+E_6.
    \label{eq:basis_choice_triangle}
\end{align}
We see that our choice has the property of expressing the dependent energies $E_1,E_2,E_5$ as strictly \emph{positive sums} of independent energies $E_3,E_4,E_6$. This is an important property that shall feature prominently in our derivation. 
In order to achieve a diagrammatic understanding of this property, let us look at the \emph{closed graph} for this orientation. 
The closed graph is obtained from the original one by gathering all external vertices into one special vertex $\circ$. For a graph $G$ with an orientation $\bb \sigma$, i.e. $(G,\bb \sigma)$, we denote the associated closed digraph as $(G,\bb \sigma)^\circ$.
For our present triangle example from eq.~(\ref{eq:triangle_flipped_deltas}), the closed graph looks as follows: 
\begin{equation}
\resizebox{2.5cm}{!}{%
\begin{tikzpicture} 
    \begin{feynman}
        \vertex(1);
        \vertex[above right = 1.5cm and 2.5cm of 1](2);
        \vertex[below right = 1.5cm and 2.5cm of 1](3);
        
        \vertex[left = 1.5cm of 1](E1);
        \vertex[right = 1.5cm of 2](E2);
        \vertex[right = 1.5cm of 3](E3);
        
         \diagram*[large]{	
            (1) -- [->-,line width=0.6mm] (2) -- [-<-,line width=0.6mm] (3) -- [-<-,line width=0.6mm] (1),

            (1) -- [-<-,line width=0.6mm] (E1),
            (2) -- [->-,line width=0.6mm] (E1),
            (3) -- [->-,line width=0.6mm] (E1),

        }; 
       \path[draw=black, fill=white,thick] (E1) circle[radius=0.1];
       \path[draw=black, fill=black] (1) circle[radius=0.1];
       \path[draw=black, fill=black] (2) circle[radius=0.1];
       \path[draw=black, fill=black] (3) circle[radius=0.1];
       
    \end{feynman}
\end{tikzpicture}}\;.
\end{equation}

The condition that a proper energy-conserving flow exists on the original graph translates to a graph-theoretical property of the 
closed graph, viz.~the requirement that the closed graph is \emph{strongly connected}. 
A digraph is strongly connected if we can reach each vertex from any other vertex by taking some positively oriented route. 
The contribution of a digraph in FOPT will only be non-zero if the associated closed graph is strongly connected. Furthermore, a strongly connected digraph has a unique set of \emph{cycles}. A cycle is defined as a subset of edges of an oriented graph that compose a positive-energy oriented path starting at a vertex and coming back to that same vertex.

In the running example of the triangle, this graph has exactly three oriented cycles $\{\pp_1,\pp_2,\pp_3\}$ (depicted with coloured edges): 
\begin{equation}
 \pp_1: \  
    \raisebox{-0.8cm}{\resizebox{2.5cm}{!}{%
\begin{tikzpicture}
    

    \begin{feynman}
        \vertex(1);
        \vertex[above right = 1.5cm and 2.5cm of 1](2);
        \vertex[below right = 1.5cm and 2.5cm of 1](3);
        
        \vertex[left = 1.5cm of 1](E1);
        \vertex[right = 1.5cm of 2](E2);
        \vertex[right = 1.5cm of 3](E3);
        
         \diagram*[large]{	
            (1) -- [->-,line width=0.6mm, col1] (2) -- [-<-,line width=0.6mm] (3) -- [-<-,line width=0.6mm] (1),

            (1) -- [-<-,line width=0.6mm, col1] (E1),
            (2) -- [->-,line width=0.6mm, col1] (E1),
            (3) -- [->-,line width=0.6mm] (E1),

        }; 
       \path[draw=black, fill=white,thick] (E1) circle[radius=0.15];
       \path[draw=black, fill=black] (1) circle[radius=0.1];
       \path[draw=black, fill=black] (2) circle[radius=0.1];
       \path[draw=black, fill=black] (3) circle[radius=0.1];
       
    \end{feynman}

\end{tikzpicture}
}}\;\;\;\;  \pp_2: \  
    \raisebox{-0.8cm}{\resizebox{2.5cm}{!}{%
\begin{tikzpicture}
    

    \begin{feynman}
        \vertex(1);
        \vertex[above right = 1.5cm and 2.5cm of 1](2);
        \vertex[below right = 1.5cm and 2.5cm of 1](3);
        
        \vertex[left = 1.5cm of 1](E1);
        \vertex[right = 1.5cm of 2](E2);
        \vertex[right = 1.5cm of 3](E3);
        
         \diagram*[large]{	
            (1) -- [->-,line width=0.6mm] (2) -- [-<-,line width=0.6mm] (3) -- [-<-,line width=0.6mm, col2] (1),

            (1) -- [-<-,line width=0.6mm, col2] (E1),
            (2) -- [->-,line width=0.6mm] (E1),
            (3) -- [->-,line width=0.6mm, col2] (E1),

        }; 
       \path[draw=black, fill=white,thick] (E1) circle[radius=0.15];
       \path[draw=black, fill=black] (1) circle[radius=0.1];
       \path[draw=black, fill=black] (2) circle[radius=0.1];
       \path[draw=black, fill=black] (3) circle[radius=0.1];
       
    \end{feynman}

\end{tikzpicture}
}}\;\;\;\; \pp_3: \  
    \raisebox{-0.8cm}{\resizebox{2.5cm}{!}{%
\begin{tikzpicture}
    

    \begin{feynman}
        \vertex(1);
        \vertex[above right = 1.5cm and 2.5cm of 1](2);
        \vertex[below right = 1.5cm and 2.5cm of 1](3);
        
        \vertex[left = 1.5cm of 1](E1);
        \vertex[right = 1.5cm of 2](E2);
        \vertex[right = 1.5cm of 3](E3);
        
         \diagram*[large]{	
            (1) -- [->-,line width=0.6mm] (2) -- [-<-,line width=0.6mm, col3] (3) -- [-<-,line width=0.6mm, col3] (1),

            (1) -- [-<-,line width=0.6mm, col3] (E1),
            (2) -- [->-,line width=0.6mm, col3] (E1),
            (3) -- [->-,line width=0.6mm] (E1),

        }; 
       \path[draw=black, fill=white,thick] (E1) circle[radius=0.15];
       \path[draw=black, fill=black] (1) circle[radius=0.1];
       \path[draw=black, fill=black] (2) circle[radius=0.1];
       \path[draw=black, fill=black] (3) circle[radius=0.1];
       
    \end{feynman}

\end{tikzpicture}
}}\;.
\end{equation}
Each of these three cycles has exactly one edge that is not contained in any other cycle. For $\pp_1$, using the original labels, it is $e_4$. For $\pp_2$, it is $e_3$ and for $\pp_3$, it is $e_6$. 
This choice of edges gives exactly the basis of energies that we used to write eq.~\eqref{eq:basis_choice_triangle}, namely $E_3, \, E_4, \, E_6$. It turns out that for any strongly connected orientation such a choice can be made.
Furthermore, the three cycles above are \emph{canonical}. We can only find exactly these three cycles of the graph if we insist on the property of positive-energy flow. This is in contrast to the usual covariant momentum representation, where we have many choices to route the momenta through the diagram. In more mathematical terms, there is (up to permutation) a unique basis of the first homology\footnote{The first homology of a graph is the vector space spanned by all its loops.} of $G^\circ$ in which each basis vector is a simple, positively oriented cycle.
Opening up the $\circ$-vertex leads to an interpretation of the cycles $\pp_1,\pp_2,\pp_3$ as three distinct paths through the diagram that connect different external vertices,
\begin{align}
\label{eq:triangle_paths}
\centering
\begin{array}{cccccccccccc}
&\def\scale{.8}
\begin{tikzpicture}[baseline={([yshift=-0.7ex]0,0)}] 
\coordinate (i00) at (-2*\scale,0);
    \coordinate (i1) at (-\scale,0);
\draw[->-,line width=0.3mm] (i00) -- (i1) node[midway,right] {};
    \filldraw (i00) circle (1.3pt);
    \coordinate[] (i2) at (0,.5*\scale);
    \coordinate[] (i21) at (0,-.5*\scale);
    \coordinate[] (v1) at (\scale,\scale);
    \coordinate[] (v2) at (\scale,-\scale);
    \draw[-<-,line width=0.3mm] (i2) -- (i1) node[midway,above left] {};
    \draw[-<-,line width=0.3mm] (i21) -- (i1) node[midway,below left] {};
    \draw[->-,line width=0.3mm] (i21) -- (v2) node[midway,below left] {};
    \draw[->-,line width=0.3mm] (i21) -- (i2) node[midway,right] {};
    \draw[->-,line width=0.3mm] (i2) -- (v1) node[midway,above left] {};
        \filldraw (v1) circle (1.3pt);
        \filldraw (i21) circle (1.3pt);
    \filldraw (v2) circle (1.3pt);
    \filldraw (i1) circle (1.3pt);
    \filldraw (i2) circle (1.3pt);
\end{tikzpicture}&\longrightarrow&\def\scale{.8}
\begin{tikzpicture}[baseline={([yshift=-0.7ex]0,0)}] 
\coordinate (i00) at (-2*\scale,0);
    \coordinate (i1) at (-\scale,0);
\draw[col1,->-,line width=0.3mm] (i00) -- (i1) node[midway,right] {};
    \filldraw (i00) circle (1.3pt);
    \coordinate[] (i2) at (0,.5*\scale);
    \coordinate[] (i21) at (0,-.5*\scale);
    \coordinate[] (v1) at (\scale,\scale);
    \coordinate[] (v2) at (\scale,-\scale);
    \draw[col1,-<-,line width=0.3mm] (i2) -- (i1) node[midway,above left] {};
    \draw[-<-,line width=0.3mm] (i21) -- (i1) node[midway,below left] {};
    \draw[->-,line width=0.3mm] (i21) -- (v2) node[midway,below left] {};
    \draw[->-,line width=0.3mm] (i21) -- (i2) node[midway,right] {};
    \draw[col1,->-,line width=0.3mm] (i2) -- (v1) node[midway,above left] {};
        \filldraw (v1) circle (1.3pt);
        \filldraw (i21) circle (1.3pt);
    \filldraw (v2) circle (1.3pt);
    \filldraw (i1) circle (1.3pt);
    \filldraw (i2) circle (1.3pt);
\end{tikzpicture}&\def\scale{.8}
\begin{tikzpicture}[baseline={([yshift=-0.7ex]0,0)}] 
\coordinate (i00) at (-2*\scale,0);
    \coordinate (i1) at (-\scale,0);
\draw[col2,->-,line width=0.3mm] (i00) -- (i1) node[midway,right] {};
    \filldraw (i00) circle (1.3pt);
    \coordinate[] (i2) at (0,.5*\scale);
    \coordinate[] (i21) at (0,-.5*\scale);
    \coordinate[] (v1) at (\scale,\scale);
    \coordinate[] (v2) at (\scale,-\scale);
    \draw[-<-,line width=0.3mm] (i2) -- (i1) node[midway,above left] {};
    \draw[col2,-<-,line width=0.3mm] (i21) -- (i1) node[midway,below left] {};
    \draw[col2,->-,line width=0.3mm] (i21) -- (v2) node[midway,below left] {};
    \draw[->-,line width=0.3mm] (i21) -- (i2) node[midway,right] {};
    \draw[->-,line width=0.3mm] (i2) -- (v1) node[midway,above left] {};
        \filldraw (v1) circle (1.3pt);
        \filldraw (i21) circle (1.3pt);
   \filldraw (v2) circle (1.3pt);
    \filldraw (i1) circle (1.3pt);
    \filldraw (i2) circle (1.3pt);
\end{tikzpicture}&\def\scale{.8}
\begin{tikzpicture}[baseline={([yshift=-0.7ex]0,0)}] 
\coordinate (i00) at (-2*\scale,0);
    \coordinate (i1) at (-\scale,0);
\draw[col3,->-,line width=0.3mm] (i00) -- (i1) node[midway,right] {};
    \filldraw (i00) circle (1.3pt);
    \coordinate[] (i2) at (0,.5*\scale);
    \coordinate[] (i21) at (0,-.5*\scale);
    \coordinate[] (v1) at (\scale,\scale);
    \coordinate[] (v2) at (\scale,-\scale);
    \draw[-<-,line width=0.3mm] (i2) -- (i1) node[midway,above left] {};
    \draw[col3,-<-,line width=0.3mm] (i21) -- (i1) node[midway,below left] {};
    \draw[->-,line width=0.3mm] (i21) -- (v2) node[midway,below left] {};
    \draw[col3,->-,line width=0.3mm] (i21) -- (i2) node[midway,right] {};
    \draw[col3,->-,line width=0.3mm] (i2) -- (v1) node[midway,above left] {};
        \filldraw (v1) circle (1.3pt);
        \filldraw (i21) circle (1.3pt);
    \filldraw (v2) circle (1.3pt);
    \filldraw (i1) circle (1.3pt);
    \filldraw (i2) circle (1.3pt);
\end{tikzpicture}\\
&&&\pp_1&\pp_2&\pp_3
\end{array}
\end{align}

With our chosen basis, we are now ready to solve the delta functions in eq.~\eqref{eq:trangle_ori123}. Carrying out the 
$E_1,E_2,E_5$ integrals we obtain
\begin{align}
\AA_{G,\bb \sigma} =& 
\frac{(-ig)^{3}}{i^6 (2\pi)^{9}}
\left[ \int_{0}^\infty \frac{\mathrm{d}E_3\mathrm{d}E_4\mathrm{d}E_6}{\prod_{e\in\gE}2|\vec{z}_e|}\right]e^{iE_4(|\vec{z}_1|+|\vec{z}_2|+|\vec{z}_4|+x_2^0-x_1^0+i\ep)}
\times\nonumber\\ &\times 
e^{iE_3(|\vec{z}_1|+|\vec{z}_3|+|\vec{z}_5|+x_3^0-x_1^0+i\ep)}
e^{iE_6(|\vec{z}_1|+|\vec{z}_2|+|\vec{z}_5|+|\vec{z}_6|+x_2^0-x_1^0+i\ep)}\;.\label{eq:triangle_ori123_fin}
\end{align}
The remaining integrations are readily performed, which gives 
\begin{gather}
\begin{gathered}
\label{eq:triangle_fopt}
\AA_{G,\bb \sigma} = 
\frac{(-ig)^{3}}{i^6 (2\pi)^{9}}
\frac{i^3}{\prod_{e\in \gE}2|\vec{z}_e|} \frac{1}{(\gamma_{\pp_1}+x_{12}^0+i\ep)(\gamma_{\pp_2}+x_{13}^0+i\ep)(\gamma_{\pp_3}+x_{12}^0+i\ep)}
\end{gathered}
\intertext{where $\gamma_{\pp_1},\gamma_{\pp_2},\gamma_{\pp_3}$ are the \emph{path lengths} associated to the cycles $\pp_1, \pp_2, \pp_3$,}
\begin{aligned}
\gamma_{\pp_1} &= |\vec{z}_1|+|\vec{z}_4|+|\vec{z}_2|,\\
\gamma_{\pp_2} &= |\vec{z}_1|+|\vec{z}_5|+|\vec{z}_3|,\\
\gamma_{\pp_3} &= |\vec{z}_1|+|\vec{z}_5|+|\vec{z}_6|+|\vec{z}_2|.
\end{aligned}
\end{gather}
Eq.~\eqref{eq:triangle_fopt} is the full FOPT expression associated 
to the digraph~\eqref{eq:triangle_right_orientation}.

\subsection{Derivation of the general FOPT Feynman rules}
\label{sec:derivation}
Having discussed the triangle diagram in detail, we will proceed to the general derivation of the 
FOPT Feynman rules \cite{Borinsky:2022msp}. 

\subsubsection{Cauchy integrations}
We will perform the integrals over the time components $y^0_v$ of eq.~\eqref{eq:FI} analytically via the residue theorem. In fact, we are only interested in the partially integrated version $\AA_G$ of $A_G$ which we already discussed in the triangle example:
\begin{gather}
\begin{gathered}
    \label{eq:mathcal_A}
        \AA_G(x_1,\ldots,x_{|\gVE|},\vec{y}_1, \ldots, \vec{y}_{|\gVI|})= 
\frac{(-ig)^{|\gVI|}}{(2\pi)^{2|\gE|}}
\left[ \prod_{v \in \gVI} \int \dd y_v^0 \right]
\prod_{e \in \gE}\frac{1}{-z_e^2+i\ep}\;,
\end{gathered}
\end{gather}
with the relation 
$A_G(x_1,\ldots,x_{|\gVE|}) = 
\left[ \prod_{v \in \gVI} \int \dd^3 \vec y_v \right]
        \AA_G(x_1,\ldots,x_{|\gVE|},\vec{y}_1, \ldots, \vec{y}_{|\gVI|})$.

To perform the integration in eq.~\eqref{eq:mathcal_A} in full generality, it is convenient to introduce some additional notation.
The edge displacement four-vectors $z_e^\mu$ can be written  as
\begin{equation}
z_e^\mu= \sum_{v \in \gVI} \EI_{e,v} \, y_v^\mu + \sum_{a \in \gVE} \EE_{e,a} \, x_a^\mu\;,
\end{equation}
where $\EI_{e,v}$ and $\EE_{e,a}$ are \emph{incidence matrices} of the graph $G$. To calculate these matrices, we have to pick some arbitrary orientation of the edges of the underlying graph and set $\EI_{e,v} = +1$ $(-1)$ if the edge $e$ is directed away from (towards) the internal vertex $v$. A matrix entry $\EI_{e,v}$ is $0$ if the edge $e$ is not incident to the vertex $v$. The matrix $\EE_{e,a}$ is defined analogously, but only for external vertices labeled by the $a$ index. The initial choice of an orientation of the edges, which is necessary to define these matrices, is arbitrary and the value of the integral in eq. (\ref{eq:mathcal_A}) does not depend on this choice.

Slightly abusing the previous notation, we can introduce one auxiliary integration variable $z_e^0$ for each edge, fixed to be the time difference between its incident vertices,
\begin{equation}
    \AA_G=
\frac{(-ig)^{|\gVI|}}{(2\pi)^{2|\gE|}}
\left( \prod_{v \in \gVI} \int \dd y_v^0 \right)
\left( \prod_{e \in \gE} 
\int_{-\infty}^{\infty}
\dd z^0_e 
\frac{
\delta
\left( z^0_e - 
 \EI_{e,v} \, y_v^0 -  \EE_{e,a} \, x_a^0 \right)
}{-{z_e^0}^2 + {\vec z_e}^{\,2} +i\ep}
\right)\,,
\end{equation}
where we implicitly sum over the indices $v$ and $a$ in the argument of the delta function.
We then again use the integral representation of the $\delta$ function, $\delta(z) = \int_{-\infty}^\infty \frac{\dd E}{2 \pi} e^{i E z}$ (where $E$ is an auxiliary variable with dimensions of energy),  to arrive at
\begin{equation}
    \AA_G=
\frac{(-ig)^{|\gVI|}}{(2\pi)^{2|\gE|}}
\left( \prod_{v \in \gVI} \int \dd y_v^0 \right)
\left( \prod_{e \in \gE} 
\int_{-\infty}^{\infty}
\dd z^0_e 
\int_{-\infty}^{\infty}
\frac{\dd E_e}{2\pi}
\frac{
e^{
 iE_e
\left( z^0_e - 
\EI_{e,v} \, y_v^0 - \EE_{e,a} \, x_a^0 \right)
}
}{-{z_e^0}^2 + {\vec z_e}^{\,2} +i\ep}
\right)\;.
\end{equation}
Carrying out the $z_e^0$ integrations using Cauchy's theorem is now straightforward (see eq.~\eqref{eq:contour}). It gives rise to the sum of two terms that we will interpret as a positive (closing the contour in the upper half-plane) and a negative energy contribution (closing the contour in the lower half-plane):
\begin{gather}
\AA_G=
\frac{(-ig)^{|\gVI|} (-i)^{|\gE|}}{(2\pi)^{|\gE|}}
\left( \prod_{v \in \gVI} \int \dd y_v^0 \right)
 \nonumber\times \\
\times
\left[ \prod_{e \in \gE} 
\int_{-\infty}^{\infty}
\frac{\dd E_e}{2 |\vec z_e|}
e^{ 
-iE_e
\left(  
\EI_{e,v} \, y_v^0 + \EE_{e,a} \, x_a^0 
\right)
}
\left(
\Theta(E_e)
e^{
iE_e (|\vec z_e|+ i\ep)
}
+
\Theta(-E_e)
e^{
-iE_e (|\vec z_e| + i\ep)
}
\right)
\right],\label{eq:flowseparation}
\end{gather}
where the $i \ep$ can be dropped in the 
denominators as it only matters if $\vec z_e = 0$, which is an end-point singularity. 
\subsubsection{Energy flows and digraphs}
We now interpret the different terms of the integral in eq.~(\ref{eq:flowseparation}) (coming from expanding the product over edges $e$) as corresponding to different energy flows and split the integral into $2^{|E|}$ terms:
\begin{equation}
\label{eq:flowsum}
    \AA_G(x_1,\ldots,x_{|\gVE|},\vec{y}_1, \ldots, \vec{y}_{|\gVI|})
=\sum_{\bb{\sigma}\in\{\pm 1\}^{|E|}}\AA_{G,{\bb\sigma}}(x_1,\ldots,x_{|\gVE|},\vec{y}_1, \ldots, \vec{y}_{|\gVI|})\,,
\end{equation}
where (with suppressed dependence on the arguments of $\AA_{G,\bb\sigma}$),
\begin{equation}
    \AA_{G,{\bb\sigma}}= 
\frac{(-ig)^{|\gVI|}}{(2\pi)^{|\gE|}}
\int\left[\prod_{v \in \gVI}  \dd y_v^0 \right]
\left[ \prod_{e \in \gE} 
\int_{-\infty}^{\infty}
\frac{\dd E_e }{2 i |\vec z_e|}
e^{ 
iE_e
\left(  
-\EI_{e,v} \, y_v^0 - \EE_{e,a} \, x_a^0 
+\sigma_e (|\vec z_e| + i \ep) \right)
}
\Theta(\sigma_e E_e)
\right].
\end{equation}
We can identify the $y^0_v$ integrations with Fourier representations of the $\delta$ function. These $\delta$ functions give rise to energy conservation constraints at each internal vertex and cast $\AA_{G,{\bb\sigma}}$ into the form
\begin{equation}
\AA_{G,{\bb\sigma}}=
\frac{(-2\pi ig)^{|\gVI|}}{(2\pi)^{|\gE|}}
\int
\prod_{e \in \gE}\left[\frac{\dd E_e}{2 i|\vec z_e|}
e^{ 
iE_e(
-\EE_{e,a} \, x_a^0 
+\sigma_e (|\vec z_e|  + i\ep)
)
}
\Theta(\sigma_e E_e)
\right]
\prod_{v \in \gVI}
\delta\left( \sum_{e \in \gE} E_e \, \EI_{e,v} \right).
\end{equation}
The change of variables $\sigma_e E_e \rightarrow E_e$ resolves the $\Theta$ function, and because $\sigma_e^2 = 1$ we get
\begin{equation}
\label{eq:AGnoori}
\AA_{G,{\bb\sigma}}=
\frac{(-2\pi ig)^{|\gVI|}}{(2\pi)^{|\gE|}}
\int_{\mathbb{R}_+^{|E|}}
\prod_{e \in \gE}\left[\frac{\dd E_e}{2i |\vec z_e|}
e^{ 
iE_e
(
-\sigma_e \EE_{e,a} \, x_a^0 
+ |\vec z_e|  + i \ep
)
}
\right]
\prod_{v \in \gVI}
\delta\left( \sum_{e \in \gE} \sigma_eE_e \, \EI_{e,v} \right),
\end{equation}
where we integrate over all positive energies $E_e$.
Note that the $\sigma_e$ only appears in front of incidence matrices $\EI_{e,v}$ or $\EE_{e,a}$. Changing the sign of some $e$-indexed row in these incidence matrices just changes the previously chosen arbitrary orientation by inverting the direction of the $e$-th edge.
It is clear that the starting orientation does not matter, as we eventually sum over all orientations by reversing each edge in all possible ways. 
Therefore, we can forget about the $\sigma_e$-sums and sum over all different overall orientations of the graph instead by always changing the $\mathcal E$ and $\mathcal R$-matrices accordingly. Such an orientation of the graph shall be denoted by $\bb \sigma$, in the obvious way. For each orientation $\bb \sigma$, we have different $\mathcal E$ and $\mathcal R$-matrices. The data of the integrand is therefore combinatorially encoded in the graph $G$ with an assigned orientation $\bb \sigma$. As before we will denote the resulting digraph as $(G,\bb \sigma)$. 
We can thus rewrite the integral in eq.~\eqref{eq:AGnoori} as
\begin{equation}
\label{eq:orientation_with_deltas}
   \AA_{G,\bb\sigma}=
\frac{(-2\pi ig)^{|\gVI|}}{(2\pi)^{|\gE|}}
\int_{\mathbb{R}_+^{|E|}}\left[ \prod_{e \in \gE} 
\frac{\dd E_e}{2 i |\vec z_e|}
e^{ 
iE_e \left(
- \EE_{e,a}^{\bb \sigma} \, x_a^0 
+
|\vec z_e| + i \ep
\right)
}
\right]
\prod_{v \in \gVI}
\delta\left( \sum_{e \in \gE} E_e \, \EI_{e,v}^{\bb \sigma} \right),
\end{equation}
where we absorbed the entire dependence on $\bb \sigma$ into the incidence matrices. That means $\EI_{e,v}^{\bb \sigma}$ is $+1$ if under the orientation $\bb \sigma$ the edge $e$ is pointing away from  the vertex $v$,  $-1$ if it points towards $v$ and $0$ if $e$ is not incident to $v$; $\EE_{e,a}^{\bb \sigma}$ is defined analogously.

Standard symmetry factor arguments also allow us to rewrite eq.~\eqref{eq:flowsum} as
\begin{equation}
\label{eq:AAsym}
\frac{\AA_G(x_1, \ldots, x_{|\gVE|}, \vec y_{1},\ldots, \vec y_{|\gVI|})}{\Sym G}
=
\sum_{ \langle \bb \sigma\rangle} \frac{\AA_{G,\bb \sigma}(x_1, \ldots, x_{|\gVE|},\vec y_{1},\ldots, \vec y_{|\gVI|})}{\Sym(G,\bb \sigma)} ,
\end{equation}
where we sum over all \emph{nonequivalent} orientations $\bb \sigma$ of the graph $G$.
$\Sym(G,\bb \sigma)$ is the symmetry factor of the digraph $(G,\bb \sigma)$. The calculation 
of a digraph symmetry factor, $\Sym (G,\bb \sigma) $, is the same as for covariant diagrams if all edges were associated to charged particles.

There is a positive-energy flow on each edge of the graph in the dictated direction, which is conserved at each vertex by the $\delta$-functions in eq.~\eqref{eq:orientation_with_deltas}. 
Due to this conservation law, not all possible orientations of a graph give a non-trivial contribution as we illustrated in detail in the last section and will expand upon next. 

\subsubsection{Canonical cycle basis and admissible paths}
\begin{figure}[ht!]
\begin{align*}
   \begin{tikzpicture}[baseline={([yshift=10ex]current bounding box.south)}]
    \coordinate (v) at (0,0);
    \pgfmathsetmacro{\rad}{.6}
    \pgfmathsetmacro{\rud}{1.6}
   \draw[pattern=north west lines]  (v) circle (\rad);
    \draw ([shift=(150:{\rad})]v) -- ([shift=(150:{\rud})]v) node[left] {$x_1$};
    \draw ([shift=(180:{\rad})]v) -- ([shift=(180:{\rud})]v) node[left] {$x_2$};
    \draw ([shift=(210:{\rad})]v) -- ([shift=(210:{\rud})]v) node[left] {$x_3$};
    \draw ([shift=(30:{\rad})]v) -- ([shift=(30:{\rud})]v) node[right] {$x_{|\gVE|}$};
    \draw ([shift=(0:{\rad})]v) -- ([shift=(0:{\rud})]v) node[right] {$x_{|\gVE|-1}$};
    \draw ([shift=(-30:{\rad})]v) -- ([shift=(-30:{\rud})]v) node[right] {$x_{|\gVE|-2}$};
    \node (w) at ([shift=(-90:{1.3})]v) {$\bullet \bullet \bullet$};
    \filldraw ([shift=(30:{\rud})]v) circle (1.3pt);
    \filldraw ([shift=(0:{\rud})]v) circle (1.3pt);
    \filldraw ([shift=(-30:{\rud})]v) circle (1.3pt);
    \filldraw ([shift=(150:{\rud})]v) circle (1.3pt);
    \filldraw ([shift=(180:{\rud})]v) circle (1.3pt);
    \filldraw ([shift=(210:{\rud})]v) circle (1.3pt);
    \node (u) at ([shift=(225:{2})]v) {$G$};
   \end{tikzpicture}&
&\rightarrow&&
   \begin{tikzpicture}[baseline={([yshift=10ex]current bounding box.south)}]
    \coordinate (v) at (0,0);
    \coordinate (w) at (0,1.3);
    \pgfmathsetmacro{\rad}{.6}
    \pgfmathsetmacro{\rud}{1.6}
   \draw[pattern=north west lines]  (v) circle (\rad);
    \draw ([shift=(150:{\rad})]v) to[looseness=1,out=150,in=210] (w);
    \draw ([shift=(180:{\rad})]v) to[looseness=2,out=180,in=180] (w);
    \draw ([shift=(210:{\rad})]v) to[looseness=3,out=210,in=150] (w);
    \draw ([shift=(30:{\rad})]v) to[looseness=1,out=30,in=-30] (w);
    \draw ([shift=(0:{\rad})]v) to[looseness=2,out=0,in=0] (w);
    \draw ([shift=(-30:{\rad})]v) to[looseness=3,out=-30,in=30] (w);
    \filldraw[fill=white] (w) circle (3pt);
    \node (w) at ([shift=(-90:{1.3})]v) {$\bullet \bullet \bullet$};
    \node (u) at ([shift=(225:{2})]v) {$G^\circ$};
   \end{tikzpicture}
\end{align*}
\caption[Illustration of the \emph{closed} graph $G^\circ$]{Illustration of the \emph{closed} graph $G^\circ$ that is obtained after joining all external vertices at the artificial vertex $\circ$.}
\label{fig:completion}
\end{figure}
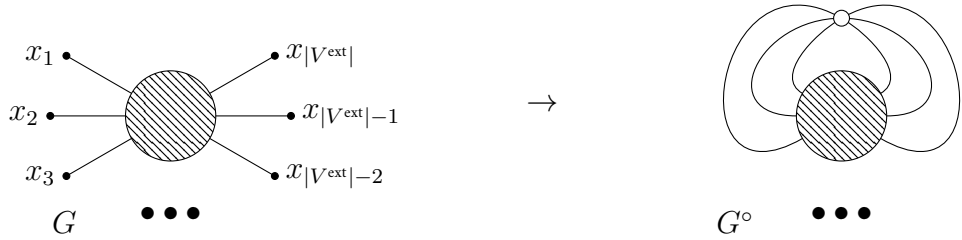
Recall that for each graph $G$ we can form the associated closed graph $G^\circ$ by joining all the external vertices into a new vertex $\circ$. This process is illustrated in Figure~\ref{fig:completion} and works analogously for digraphs.
A digraph is \emph{strongly connected} if there is a positively oriented path between any ordered pair of vertices
\cite[Ch.~10]{bondy1976graph}. Such a strongly connected digraph comes with a \emph{canonical cycle basis}, which we can find as follows: start with some directed edge $e_1^{(1)}$ pointing from some vertex $v_a$ to another vertex $v_b$. If $v_a=v_b$, we have a tadpole cycle, which we take as one of our canonical base cycles. If we assume that $v_a \neq v_b$, then, by strong connectivity, there must be some oriented path in $(G,\boldsymbol{\sigma})^\circ$ that we can follow to go back from $v_b$ to $v_a$ in a full oriented cycle. Moreover, we can require that such a cycle consisting of a set of edges $\pp = \{e_1^{(\pp)}, e_2^{(\pp)},\ldots \}$ does not visit any vertex twice. We can pick such a path and declare it to be our first independent oriented cycle $\pp_1$. Next, we pick some edge that was not in this first cycle and construct another closed oriented cycle that contains this edge. For this second cycle $\pp_2$ we are allowed to revisit edges that have been in $\pp_1$, but clearly $\pp_1 \neq \pp_2$ as $\pp_2$ contains at least one edge that has not been in $\pp_1$. Continuing this, we start again with another edge that has neither been visited by the oriented cycle $\pp_1$, nor $\pp_2$, to find a third cycle and so on until no edges are left. In each step, we are guaranteed to find a closed oriented cycle by the strong connectivity requirement. The remarkable observation is that the resulting set of cycles $\pp_1,\ldots,\pp_L$ is unique up to renumbering of the path labels. In the sense that, no matter in what order the edge numbering is chosen, the same set of cycles appears.

The contribution of the directed graph $(G,\bb\sigma)$ to eq. (\ref{eq:flowsum}) will only be non-zero if the associated closed digraph $(G,\bb \sigma)^\circ$ is strongly connected. 
We will denote the canonical cycle basis of the closed digraph $(G,\bb \sigma)^\circ$ as $\Gamma = \{\pp_1,\ldots,\pp_{|\Gamma|}\}$ where we omit the explicit reference to the digraph $(G,\bb \sigma)$ if it is clear from the context.
Opening up the closed graph again, gives rise to a set of paths in the (open) digraph $(G,\bb \sigma)$. In this context we will also call the elements of $\Gamma$ the set of admissible paths of $(G,\bb \sigma)$. Note that every admissible path is either a completely internal and closed \emph{cycle} in $G$, which does not pass any external vertex, or it is a path that starts and ends at  external vertices without passing another external vertex in-between. We will call such open paths \emph{routes} through the graph.

Since the graph $(G,\bb \sigma)^\circ$ has $|\gVI|+1$ vertices including the special vertex $\circ$, and $|\gE|$ edges, one can conclude, using the graph's Euler characteristic, that the digraph has at most  $|\gE|-|\gVI|-1+1=|\gE|-|\gVI|$ independent cycles. That means there are $|\gE|-|\gVI|$ admissible paths of the digraph $(G,\bb \sigma)$. The energy integral in eq.~\eqref{eq:orientation_with_deltas} is effectively $|\gE|-|\gVI|$ dimensional, due to the $|\gVI|$ delta functions.  This calculation suggests therefore that we can associate each admissible path of $(G,\bb \sigma)$ with an independent energy integration.

Just as in the momentum space loop integral case, we can resolve the $\delta$ functions in eq.~\eqref{eq:orientation_with_deltas} by introducing one integration variable $E_{\pp}$ for each admissible path $\pp \in \Gamma$. 

The delta functions are resolved with the choice of coordinates
\begin{align}
E_e = \sum_{\substack{\pp\in \Gamma \\ \text{s.t. } e \in \pp}} E_\pp\,,
\end{align}
where we sum over all admissible paths of $(G,\bb \sigma)$, that contain the edge $e$. This is still analogous to the loop momentum integral case, except for the fact that we have a dictated orientation that we need to follow in our cycles.
Using these energy variables in eq.~\eqref{eq:orientation_with_deltas} resolves the delta functions and gives
\begin{gather}
\begin{gathered}
   \AA_{G,\bb\sigma}=
\frac{(-2\pi i g)^{|\gVI|}}{(2\pi i)^{|\gE|}}
\int_{\mathbb{R}_+^{|\Gamma|}}
\left[\prod_{\pp \in \Gamma} \dd E_\pp \right]\left[\prod_{e\in \gE}
\frac{1}{2 |\vec z_e|}\right]
\exp\left(
i
\sum_{e\in \gE}
\sum_{\substack{\pp\in \Gamma \\ \text{s.t. } e \in \pp}} E_\pp
\left(
-\EE_{e,a}^{\bb \sigma} \, x_a^0 
+
|\vec z_e| 
+i\ep
\right)
\right)\,.
\end{gathered}
\end{gather}
Recall that there is an implicit summation over $a$. Changing the order of summation in the exponential renders integration in the remaining energy variables straightforward, and gives a product of trivial oscillatory integrals
\begin{equation}
   \AA_{G,\bb\sigma}=
\frac{(-2\pi i g)^{|\gVI|}}{(2\pi i)^{|\gE|}}
    \left[\prod_{e\in \gE}
\frac{1}{2 |\vec z_e|}\right]
 \prod_{\pp \in \Gamma} 
\int_0^\infty
\dd E_\pp 
\exp\left(
i
E_\pp
\sum_{e\in \pp}
\left(
-\EE_{e,a}^{\bb \sigma} \, x_a^0 
+
|\vec z_e| 
+i\ep
\right)
\right)\,.
\end{equation}
We can now perform the $E_\pp$ integrations. The boundary term at infinity is going to vanish by the $\ep > 0$ assumption. The final expression for the FOPT representation of a digraph $(G,\bb \sigma)$ is remarkably simple and amounts to
\begin{equation}
   \AA_{G,\bb\sigma}=
\frac{(2\pi g)^{|\gVI|}}{(-2\pi)^{|\gE|}}
\left[
\prod_{e\in \gE}
\frac{1}{2 |\vec z_e|}
\right]
 \prod_{\pp \in \Gamma} 
\frac{1}
{
\sum_{e\in \pp}
\left(
- \EE_{e,a}^{\bb \sigma} \, x_a^0 
+
|\vec z_e|
+ i \ep
\right)
}.\label{eq:FOPTresult}
\end{equation}
The term $\EE_{e,a}^{\bb \sigma} \, x_a^0$ is only non-zero if the admissible path $\pp$ goes from external to external vertex. It can be identified with the \emph{time difference} of the two corresponding vertices. The other term in the denominator is the total Euclidean path length of the admissible path.
We will discuss this entirely combinatorial formula in the next section.

\subsection{FOPT Feynman rules}
\label{sec:flowfeynmanrules}
The procedure illustrated in the last subsection
generalizes to all Feynman diagrams. It provides an 
alternative perturbative decomposition of correlation functions as
\begin{align}
\label{eq:FOPT_greens_functions}
\Gamma(x_1,...,x_{|\gVE|})=\left\langle 0 | T ( \varphi(x_1) \cdots \varphi(x_{|\gVE|}) ) | 0 \right \rangle =
\sum_{(G,\bb \sigma)} \frac{1}{\Sym (G,\bb \sigma)} A_{G,\bb \sigma}(x_1, \ldots, x_{|\gVE|}),
\end{align}
where we sum over all topologically different digraphs $(G,\bb \sigma)$, i.e.~graphs $G$ from the given scalar QFT with a specified energy flow orientation on the propagators.
Note that in contrast to `old-fashioned', or time-ordered perturbation theory, where each covariant integral is replaced by $|\gVI|!$ time-ordered integrals, we get at most $2^{|\gE|}$ \emph{energy-flow-oriented}  integrals in our coordinate-space approach. 

By eq.~\eqref{eq:AAsym}, the FOPT Feynman rules provide a way of decomposing an individual covariant Feynman integral into its different flow-oriented components:
\begin{align}
\frac{1}{\Sym G} A_{G}(x_1, \ldots, x_{|\gVE|})
=
\sum_{\langle \bb \sigma \rangle}
\frac{1}{\Sym (G,\bb \sigma)} A_{G,\bb \sigma}(x_1, \ldots, x_{|\gVE|}),
\end{align}
where we sum over all nonequivalent ways to orient the graph $G$ via $\bb \sigma$.

An integral expression for $A_{G,\bb \sigma}(x_1, \ldots, x_{|\gVE|})$ can be found using the following, entirely combinatorial recipe:

\begin{enumerate}
        \item $A_{G,\bb \sigma} =0$ if the closed directed graph $(G,\bb \sigma)^\circ$ is not strongly connected.
        \item Multiply by a factor of $-i g$ for each interaction vertex. 
        \item For each edge $e$ of $G$ multiply by a factor $\frac{-i}{(8\pi^2)|\vec{z}_e|}$ where $\vec{z}_e = \vec{y}_{v} - \vec{y}_{u}$ and $\vec{y}_v,\vec{y}_u$ are the coordinates of the internal or external vertices to which the edge $e$ is incident. 
        \item For each admissible path $\pp$ of $(G,\bb \sigma)$ (i.e.~for each cycle in the 
        canonical cycle basis of $(G,\bb \sigma)^\circ$) multiply by a factor of $
i/\left(
{
\gamma_\pp
+
\tau_{\pp} + i\ep
}
\right),
$ where
\begin{equation}
    \gamma_\mathrm{p} =\sum_{e\in \pp}
|\vec{z}_e|
\end{equation} is the sum over all edge lengths that are in the cycle and $\tau_{\pp}$ is either the time passed between the starting and ending external vertices of the path, or zero if the cycle does not go through the $\circ$ vertex.
        \item For each internal vertex $v$ of the graph $G$ integrate over three-dimensional space $\int \dd^3 \vec{y}_v$ and multiply by $2 \pi$.
\end{enumerate}
Note that the $i\ep$ can be dropped for cycles that do not go through the special vertex $\circ$ as the denominator is only positive and the corresponding pole is an end-point singularity.

We can summarize these Feynman rules as follows. For a given 
digraph $(G,\bb \sigma)$ with cycle basis $\Gamma$, where all interaction vertices in $G$ are internal vertices and vice-versa, we have
\begin{equation}
    A_{G,\bb \sigma}(x_1, \ldots, x_{|\gVE|})=
\frac{(2\pi g)^{|\gVI|}}{(-4\pi^2)^{|\gE|}}
\left( \prod_{v \in \gVI} \int \dd^3 \vec y_v 
\right)
\left(
\prod_{e\in \gE}
\frac{1}{2 |\vec z_e |}
\right)
 \prod_{\pp \in \Gamma} 
\frac{1}
{
\gamma_\pp
+\tau_{\pp} 
+ i \ep
}.
\label{eq:delta_free_rep}
\end{equation}

\subsection{A simple example: The self-energy graph}
We now treat the simple example of the bubble (self-energy) graph to show the use of the FOPT Feynman rules. It will
also show how UV singularities are conveniently isolated in FOPT. The graph is
\begin{gather}
\label{eq:bubble_graph}\includegraphics[width=0.35\columnwidth]{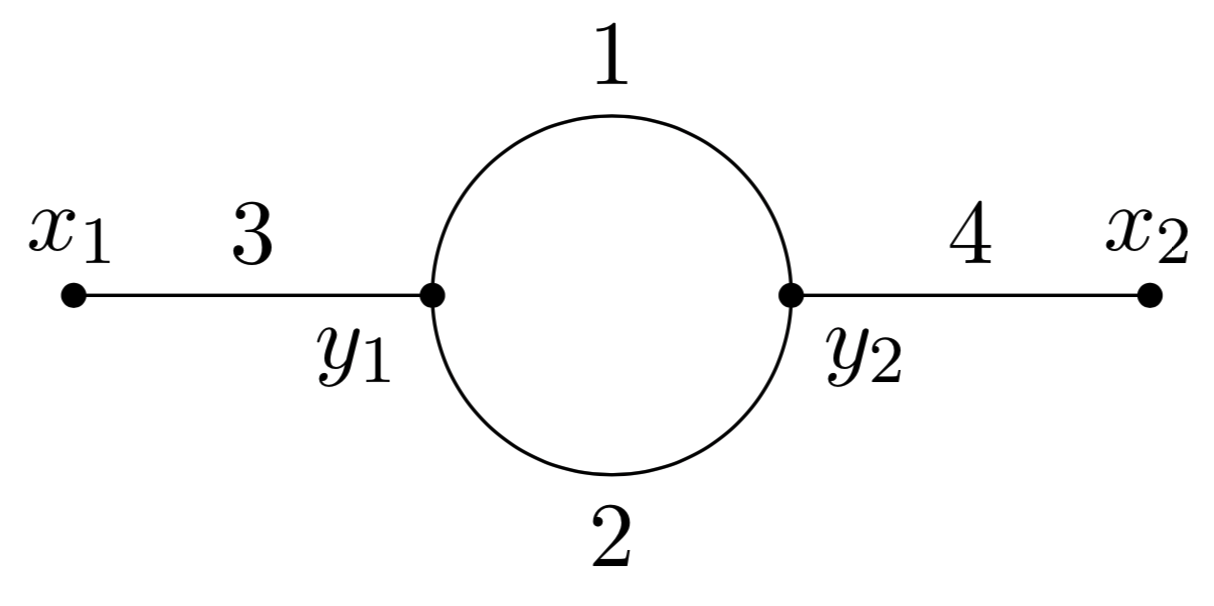}
\end{gather}
The traditional coordinate-space Feynman integral associated to this diagram reads,
\begin{align}\label{eq:bubble1}
A_G(x_1,x_2) = \frac{(-i g)^2}{(4 \pi^2)^4}
\int \dd^4 y_1 \dd^4 y_2
\,\frac{1}{-z_1^2+i \ep}
\,\frac{1}{-z_2^2+i \ep}
\,\frac{1}{-z_3^2+i \ep}
\,\frac{1}{-z_4^2+i \ep}\,.
\end{align} 
The symmetry factor of the graph is $2$, because we can exchange both edges of the bubble without altering eq. (\ref{eq:bubble1}). 
Adding the $\circ$ vertex that joins the external vertices results in the closed graph
\begin{align}\label{eq:bubbleclosed}
\includegraphics[width=.17\columnwidth]{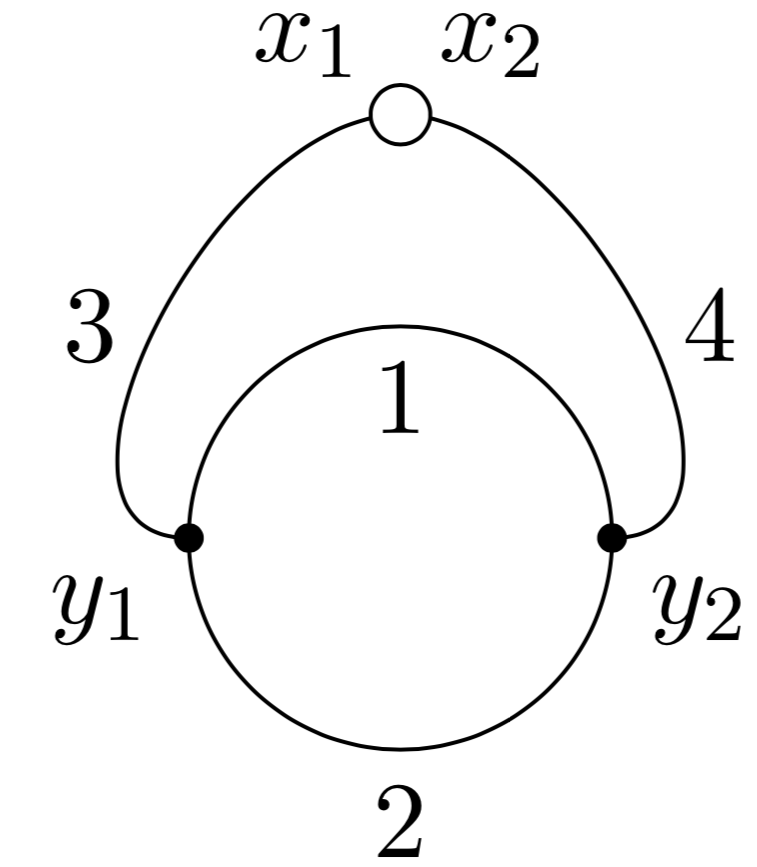}
\end{align}

There are 12 topologically distinguishable ways to give an orientation to the edges of this graph:
\begin{align}
    \includegraphics[width=0.6\columnwidth]{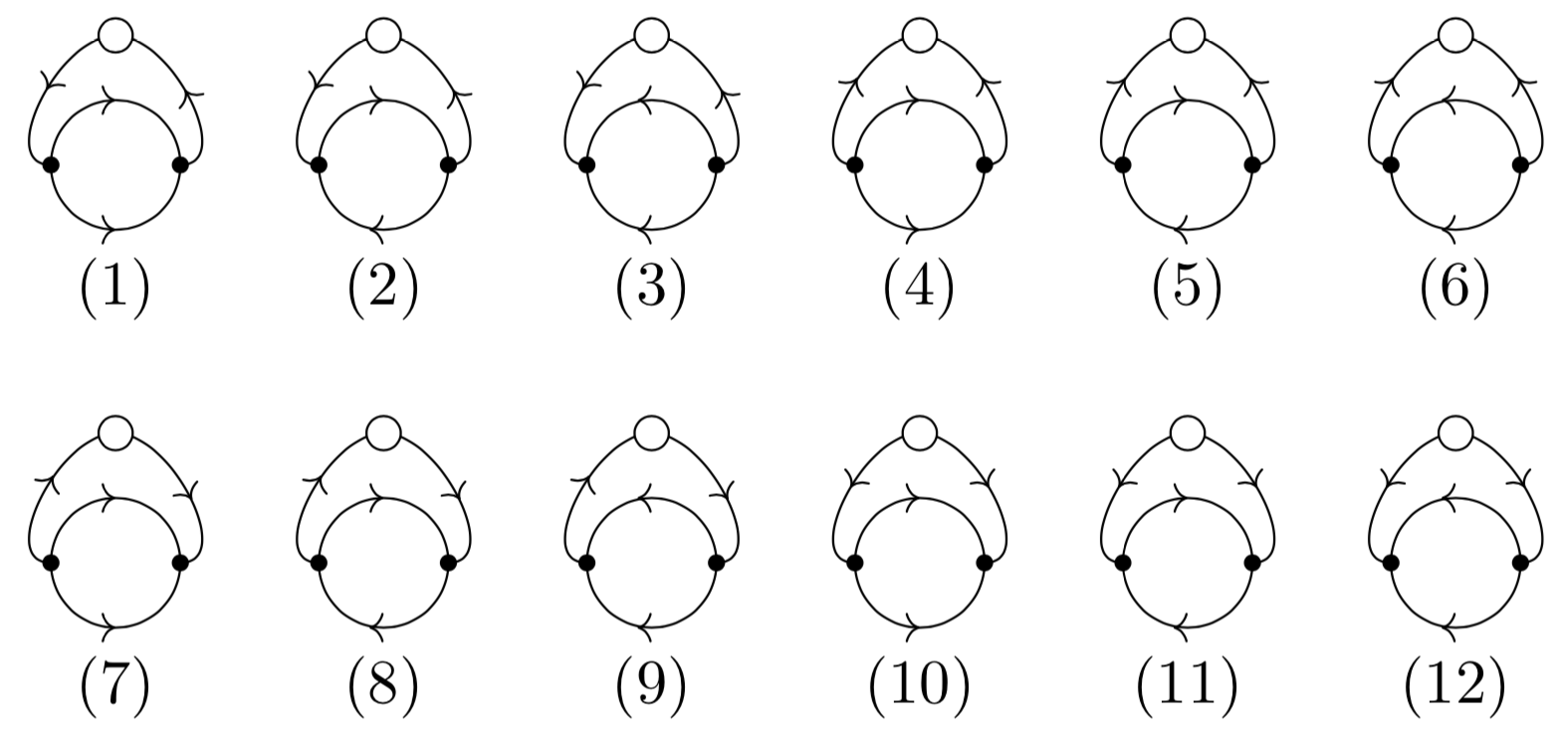}
\end{align}

Note that we are not allowed to permute the edges that are incident to the $\circ$ vertex as this would correspond to a permutation of the external vertices of the original graph. We can think of the $\circ$ vertex and the edges that are incident to it as fixed while computing the symmetry factor. The edge-oriented graphs $(1),(3),(4),(6),(7),(9),(10),(12)$ have therefore a symmetry factor of $2$ and the other graphs have a trivial symmetry factor of $1$. 
Of these directed graphs only $(1),(2),(8)$ and $(9)$ are strongly connected as one can easily check. In the remaining directed graphs, we can always find a vertex (including $\circ$) that has only in-going or out-going incident edges. These graphs are forbidden as a consequence of energy conservation and positivity.
At this point it is good to remark that there are also non-strongly connected graphs that do not have a vertex with only in- or out-going incident edges. Examples are graphs that consist of two cycles, each with positive cyclic energy flow, which are connected  by edges only pointing from cycle 1 to cycle 2. 

The closed directed graph $(1)$ has two independent cycles, $314$ and $324$, with the edge numbering as indicated in~\eqref{eq:bubbleclosed},
\begin{align}
\includegraphics[width=0.35\columnwidth]{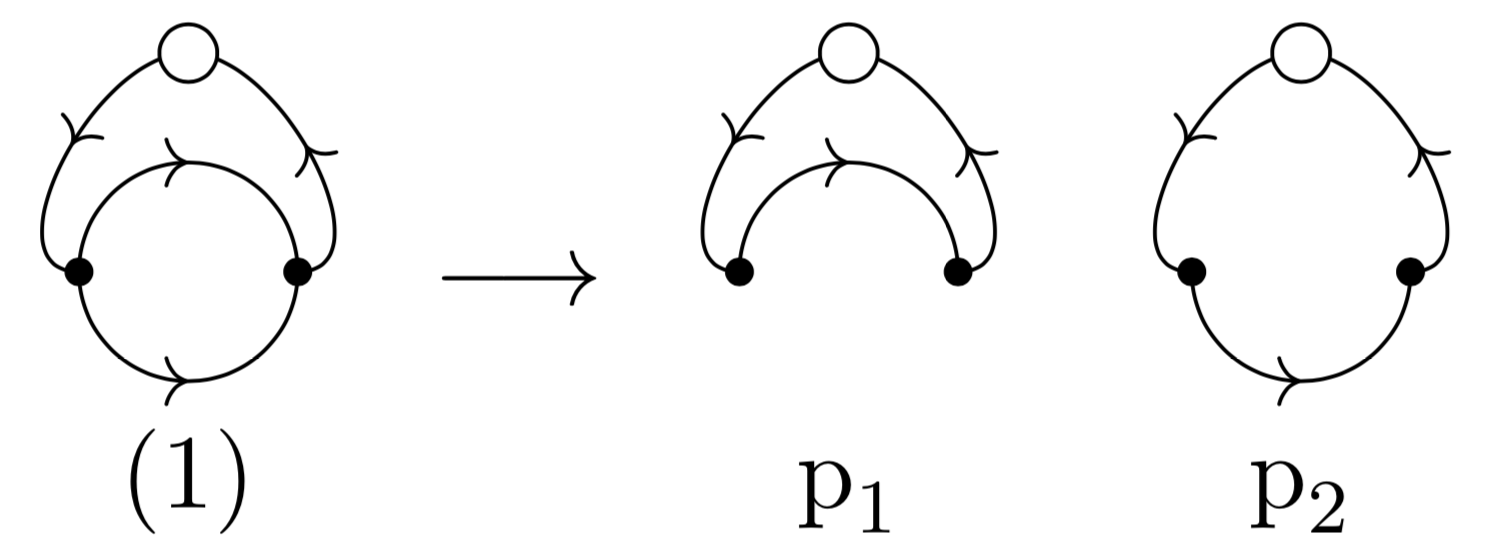}
\end{align}

Both cycles pass the $\circ$ vertex.
We find by the FOPT Feynman rules from sect.~\ref{sec:flowfeynmanrules} that
\begin{gather}
\begin{gathered}
A_{G,\bb \sigma_{(1)}}(x_1,x_2) = 
\\
\frac{(2\pi g)^{2}}{(8\pi^2)^{4}}
\int  \frac{\dd^3 \vec y_1 \dd^3 \vec y_2}{|\vec z_1| | \vec z_2 |  | \vec z_3 | | \vec z_4 |}
\frac{1}{| \vec z_3| + | \vec z_1 | + | \vec z_4 | + \tau + i \ep}
\frac{1}{| \vec z_3| + | \vec z_2 | + | \vec z_4 | + \tau + i \ep},
\end{gathered}
\end{gather}
where we defined $\tau = x_{2}^0 - x_{1}^0$.
The graph $(2)$ has the independent cycles $314$ and $12$. The latter cycle does not pass the $\circ$ vertex. Therefore,
\begin{gather}
\begin{gathered}
A_{G,\bb \sigma_{(2)}} (x_1,x_2)= \frac{(2\pi g)^{2}}{(8\pi^2)^{4}}
\int  \frac{\dd^3 \vec y_1 \dd^3 \vec y_2}{|\vec z_1| | \vec z_2 |  | \vec z_3 | | \vec z_4 |}
\frac{1}{| \vec z_3| + | \vec z_1 | + | \vec z_4 | + \tau + i \ep}
\frac{1}{| \vec z_1| + | \vec z_2 |}
\end{gathered}\,.
\end{gather}
For the graph $G$ we have $\vec z_1 = \vec z_2$. 
By changing integration variables from $\vec y_1,\vec y_2$ to $\vec y_1, \vec z_1 = \vec y_2 - \vec y_1$, we find that
\begin{gather}
\begin{gathered}
A_{G,\bb \sigma_{(2)}} (x_1,x_2)= \frac{(2\pi g)^{2}}{(8\pi^2)^{4}}
\int  \frac{\dd^3 \vec y_1 \dd^3 \vec z_1}{ | \vec z_3 | | \vec z_4 |}
\frac{1}{| \vec z_3| + | \vec z_1 | + | \vec z_4 | + \tau + i \ep}
\frac{1}{2| \vec z_1|^3}\,.
\end{gathered}
\end{gather}
Basic power counting reveals that the integrand features 
a logarithmic UV singularity for $\vec z_1 \rightarrow 0$, just like in momentum space.
The cycle that loops between the vertices $\vec y_1$ and $\vec y_2$ 
is associated to such a singularity. The intuitive explanation of this in the FOPT formalism is that the energy that flows through this cycle is unbounded and can lead to a UV divergence\footnote{We do not address the issue of regularization of divergent FOPT graphs in this thesis. We comment on possible approaches to UV renormalization in the summary.}. 

The directed graph $(8)$ has exactly the inverted orientation of $(2)$. Therefore
\begin{gather}
\begin{gathered}
A_{G,\bb \sigma_{(8)}}(x_1,x_2) = \frac{(2\pi g)^{2}}{(8\pi^2)^{4}}
\int  \frac{\dd^3 \vec y_1 \dd^3 \vec y_2}{|\vec z_1| | \vec z_2 |  | \vec z_3 | | \vec z_4 |}
\frac{1}{| \vec z_3| + | \vec z_1 | + | \vec z_4 | - \tau + i \ep}
\frac{1}{| \vec z_1| + | \vec z_2 |}\,.
\end{gathered}
\end{gather}
Analogously the graph $(9)$ has the inverted orientation of $(1)$
\begin{gather}
\begin{gathered}
A_{G,\bb \sigma_{(9)}}(x_1,x_2) = 
\\
=
\frac{(2\pi g)^{2}}{(8\pi^2)^{4}}
\int  \frac{\dd^3 \vec y_1 \dd^3 \vec y_2}{|\vec z_1| | \vec z_2 |  | \vec z_3 | | \vec z_4 |}
\frac{1}{| \vec z_3| + | \vec z_1 | + | \vec z_4 | - \tau + i \ep}
\frac{1}{| \vec z_3| + | \vec z_2 | + | \vec z_4 | - \tau + i \ep}.
\end{gathered}
\end{gather}
Observe that inverting the complete orientation only results in a sign change of all external time differences.
By collecting the overall and individual symmetry factors, we have
\begin{equation}
\frac{1}{2}A(x_1,x_2)=\frac{1}{2} A_{G,\bb \sigma_{(1)}}+ A_{G,\bb \sigma_{(2)}}+A_{G,\bb \sigma_{(8)}}+\frac{1}{2} A_{G, \bb \sigma_{(9)}}\;.
\end{equation}
The digraphs $(G,\bb \sigma_{(2)})$ and $(G,\bb \sigma_{(8)})$ feature UV singularities, as expected as $G$ is a UV singular graph in $D=4$; the other contributions are finite.

\subsection{Singularities in FOPT and a novel representation of the \texorpdfstring{$S$}{S}-matrix}
In our publication \cite{Borinsky:2022msp} the UV and IR singularities of FOPT diagrams are thoroughly analyzed. The UV divergences were there found to match the ones expected in the covariant formulation if the subgraphs generating divergences are strongly connected. On the other hand, the analysis of IR and finite distance singularities pointed towards the need of focusing on the $S$-matrix in the FOPT representation, since no previously known results were reproduced (the ones in \cite{Erdogan:2013bga} and momentum space results). The details of this representation of the $S$-matrix are beyond the scope of this thesis and we quote here the main benefits of it. For this representation one Fourier transforms the FOPT contributing amplitude with respect to the external points of the diagrams to express them in momentum space (this is needed for computing cross sections from the $S$-matrix). The resulting $S$-matrix is expressed in a hybrid representation, it is hybrid in the sense that external kinematics are given in momentum space while internal integrations are performed in three-dimensional coordinate space. It happens that the hybrid $S$-matrix depends on the Fourier transform of the well studied \textit{flow polytope}. The appearance of this polytope's Fourier transform nicely explains the cancellation of spurious singularities of this representation. Furthermore, IR singularities in this representation have a factorization property on a diagram per diagram basis.

To finish this section we present a peculiar feature of the FOPT in the context of unitarity-cut integrals, Cutkowsky's theorem \cite{Cutkosky:1960sp} and the Largest Time equation \cite{Veltman:1963th}. It happens that both virtual and real corrections to a certain process can be put all under the same integral sign, which is the whole effort of Loop Tree Duality (LTD) in momentum space \cite{Catani:2008xa,Bierenbaum:2010cy,Runkel:2019zbm,Capatti:2019ypt}, in order to show the manifest cancellation of IR divergences. 

\subsection{Unitarity, cut integrals and Cutkosky's Theorem}
\label{app:cutkosky}

The FOPT representation that we presented above in sect.~\ref{sec:fopt} takes an interesting form for unitarity-cut based phase space integrals.  In this subsection, we  discuss this viewpoint along the lines of a classic exposition of phase space integrals focused on the largest time equation \cite{Veltman:1963th}. The result is also related to Cutkosky's theorem \cite{Cutkosky:1960sp}.  
This FOPT-cut representation has some remarkable properties. For instance the integrals associated to virtual and real corrections turn out to have the same integral measure. Hence, the following considerations might be useful to pursue explicit computations of phase space integrals with manifest cancellation of real and virtual singularities (which is the aim of LTD in momentum space). Here, we briefly present this FOPT-based representation of cut integrals.

Given a subset of vertices $V \subset V_G$ of a graph $G$, the \emph{subgraph induced} from $V$ is the subgraph of $G$ that contains all edges whose both end-vertices lie in $V$. A \emph{cut} $\mathfrak C$ of a Feynman graph $\Gamma$ partitions the vertices $V_G$ into two parts $V_\sunny, V_\shady$ such that the respective induced subgraphs with edges $E_{\sunny},E_{\shady}$ from both parts are connected and each part contains at least one external vertex. The edges that have one end in $V_\sunny$ and one in $V_\shady$ are \emph{cut edges}, $E_{\mathfrak{C}}$. Intuitively, we think of energy flowing from the $\sunny$-side to the $\shady$-side (just as in the largest time equation).

A \emph{cut propagator} is given by
\begin{gather*}
\Delta^\pm(z) = 
\int  
\frac{\dd^{4} p}{(2\pi)^4}
2 \pi 
\delta(p^2) \theta(\pm p_0)
e^{-i p \cdot z }
=
-\frac{1}{(2\pi)^{2}}
 \left( 
\frac{1}{(z^0\mp i\ep)^2 -  |\vec z\,|^2 }
 \right).
\end{gather*}
It follows, from the largest time equation of Veltman \cite{Veltman:1963th}, that the real part\footnote{Due to an extra conventional factor of $i$, this real part of the integral contributes to the \emph{imaginary} part of the Feynman amplitude.} of a Feynman integral can be expressed in terms of a sum over \emph{cut integrals}:
\begin{align}
A_{G}(x_1, \ldots, x_{|\gVE|})
+
A_{G}^*(x_1, \ldots, x_{|\gVE|})
=
-
\sum_{\text{cuts } \mathfrak C \text{ of } G}
A_{G,\mathfrak C}(x_1, \ldots, x_{|\gVE|})\,,
\end{align}
where the cut integrals $A_{G,\mathfrak C}(x_1, \ldots, x_{|\gVE|})$ are given by the expression
\begin{gather}
\begin{gathered}
\label{eq:cutintegral}
A_{G,\mathfrak C}(x_1, \ldots, x_{|\gVE|})=\\=
(-ig)^{|\gVI_\sunny|}
(ig)^{|\gVI_\shady|}
\left( \prod_{v\in \gVI} \int \dd^4 y_v \right)
\left(
\prod_{e \in E_{\sunny}} \Delta_F(z_e)
\right)
\left(
\prod_{e \in E_{\mathfrak C}} 
\Delta^+(z_e)
\right)
\left(
\prod_{e \in E_{\shady}} \Delta^*_F(z_e)
\right).
\end{gathered}
\end{gather}
In contrast to the original Feynman integral, the propagators on the cut are replaced with the positive frequency $\Delta^+$ cut propagator and the Feynman propagators on the $\shady$-side of the cut are replaced with the complex conjugate Feynman propagator.

We have the following expressions for the Fourier transform of the coordinate space propagator
\begin{align}
\int_{-\infty}^\infty \dd z^0_e e^{i E_e z^0_e}
\Delta^{+}(z_e) 
&= - \frac{1}{(2 \pi)^2} 
\int_{-\infty}^\infty \dd z^0_e 
\frac{e^{i E_e z^0_e}}{(z^0_e - i \ep)^2 - \vec z_e^2}\nonumber=
\\
&= - \frac{1}{(2 \pi)^2} 
\int_{-\infty}^\infty \dd z^0_e 
\frac{e^{i E_e z^0_e}}{2|\vec z_e|}
\left(
\frac{1}{z^0_e - |\vec z_e| - i \ep}
-
\frac{1}{z^0_e + |\vec z_e| - i \ep}
\right)\nonumber=
\\
&= - \frac{1}{(2 \pi)^2} 
\frac{2 \pi i}{2|\vec z_e|}
\theta(E_e)
\left(
e^{i E_e (|\vec z_e| + i \ep)}
-
e^{i E_e (-|\vec z_e| + i \ep)}
\right)\,,
\end{align}
with the analogous Cauchy integrals
\begin{align}
&\begin{aligned}
\int_{-\infty}^\infty \dd z^0_e e^{i E_e z^0_e}
\Delta_{F}(z_e) 
&= - \frac{1}{(2 \pi)^2} 
\frac{2 \pi i}{2|\vec z_e|}
\left(
\theta(E_e)
e^{i E_e (|\vec z_e| + i \ep)}
+
\theta(-E_e)
e^{-i E_e (|\vec z_e| + i \ep)}
\right)\,,\\
\int_{-\infty}^\infty \dd z^0_e e^{i E_e z^0_e}
\Delta_{F}^*(z_e) 
&= - \frac{1}{(2 \pi)^2} 
\frac{-2 \pi i}{2|\vec z_e|}
\left(
\theta(E_e)
e^{i E_e (-|\vec z_e| + i \ep)}
+
\theta(-E_e)
e^{-i E_e (-|\vec z_e| + i \ep)}
\right)\,.
\end{aligned}
\end{align}
Repeating the derivation in sect.~\ref{sec:derivation}, now for the cut integral in eq.~\eqref{eq:cutintegral}, while using the Fourier transforms of the respective propagators, results in a representation of a cut integral as a sum over FOPT-cut integrals.
\subsubsection{FOPT Feynman rules for cut integrals}
\label{sec:cutrules}
The result is the following set of FOPT-cut integral Feynman rules for a digraph $(G,\boldsymbol{\sigma})$ with a cut $\mathfrak C$.
\begin{enumerate}
\item The integral vanishes if the closed directed graph $(G,\boldsymbol{\sigma})^\circ$ is not strongly connected or if the admissible paths on the cut do not go from the $\sunny$-side to the $\shady$-side of the graph.
\item Multiply a factor of $-ig$ ($ig$) for each $\sunny$-side ($\shady$-side) interaction vertex. 
\item For each internal vertex $v\in V^{\text{int}}$ of the digraph $(G,\boldsymbol{\sigma})$ integrate over $3$-dimensional space with the measure $2\pi \int \dd^3 \vec{y}_v$.
\item For each edge $e$ of the graph multiply a factor of $\frac{\mp i}{8\pi^2|\vec{z}_e|}$ with a $-$ sign for a $\sunny$-side or a cut edge, and a $+$ sign for a $\shady$-side edge. 
\item For each entirely uncut directed path, $ \pp_\ell$, of $(G,\boldsymbol{\sigma})^\circ$ multiply a factor of 
\begin{align*}
\frac{i}
{
\sum_{e\in \pp_\ell}
|\vec z_e| 
+
\tau_{\mathcal \pp_\ell} 
+i\ep
}
&&\text{ if $\pp_\ell$ consists entirely of $\sunny$-side edges}
\\
\frac{i}
{
-\sum_{e\in \pp_\ell}
|\vec z_e| 
+
\tau_{\pp_\ell} 
+i\ep
}
&&\text{ if $\pp_\ell$ consists entirely of $\shady$-side edges}
\end{align*}
where the sum in the denominator goes over all edges that are in the admissible path $\pp_\ell$ and $\tau_{\mathcal \pp_\ell}$ is the time difference that has passed between extremes of the path if it passes through the $\circ$ vertex, or zero if the admissible path does not go through the $\circ$ vertex, i.e.~is a cycle.
\item For each directed admissible path $\pp_{\ell}$ of $(G,\boldsymbol{\sigma})^\circ$ that passes the cut $\mathfrak C$, multiply a factor of 
\begin{align*}
\frac{-2 i |\vec z_{e_\mathfrak C}|}
{
\left(
\sum_{e\in \pp_\ell^\sunny}
|\vec z_e| 
-\sum_{e\in \pp_\ell^\shady}
|\vec z_e| 
+
\tau_{\pp_\ell} 
+i\ep
\right)^2
-
{\vec z_{e_\mathfrak C}}^{\,2}
}\,,
\end{align*}
where we sum over the uncut $\sunny$-side and $\shady$-side edges in $\pp_\ell$, $\pp_\ell^\sunny$ and $\pp_\ell^\shady$, and where $e_\mathfrak C$ denotes the unique edge of the admissible path that is on the cut. 
The edge is unique because, once the path passes over the cut edge, the energy cannot flow back through the cut.
\end{enumerate}

\subsubsection{Example}

We consider the cut integrals associated to the following graph.
\begin{align}
    \label{eq:theta_graph}
\includegraphics[width=0.2\columnwidth]{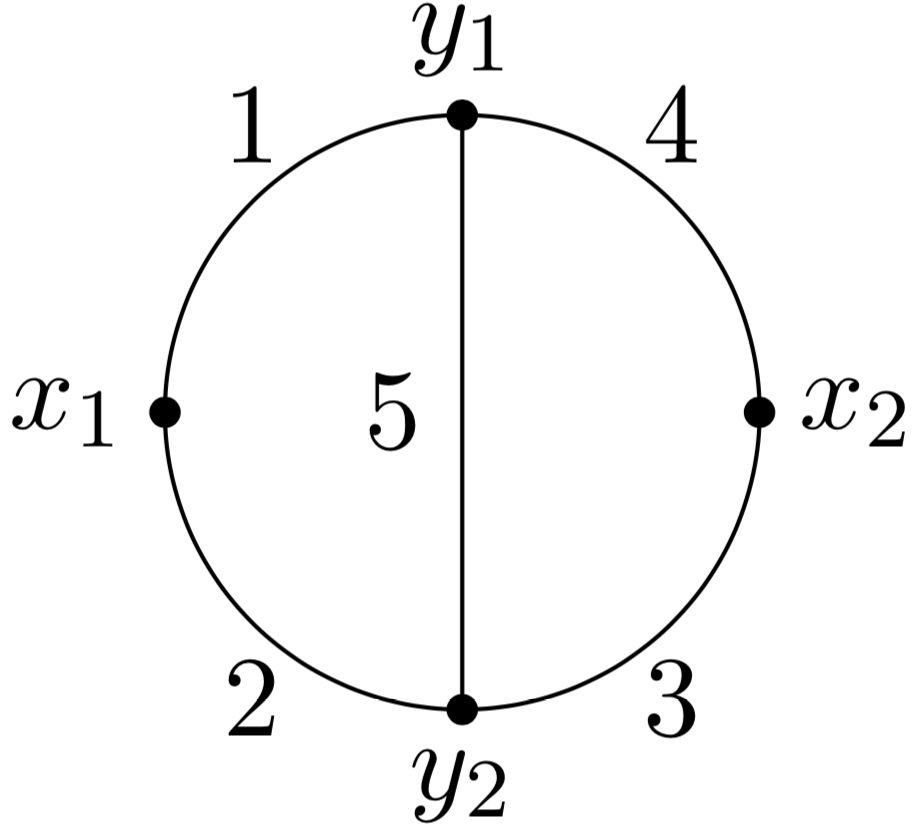}
\end{align}
Note that $x_1$ and $x_2$ are both external and interaction vertices. 
The associated closed graph is
\begin{align}  \label{eq:theta_graph_compl}
\includegraphics[width=0.2\columnwidth]{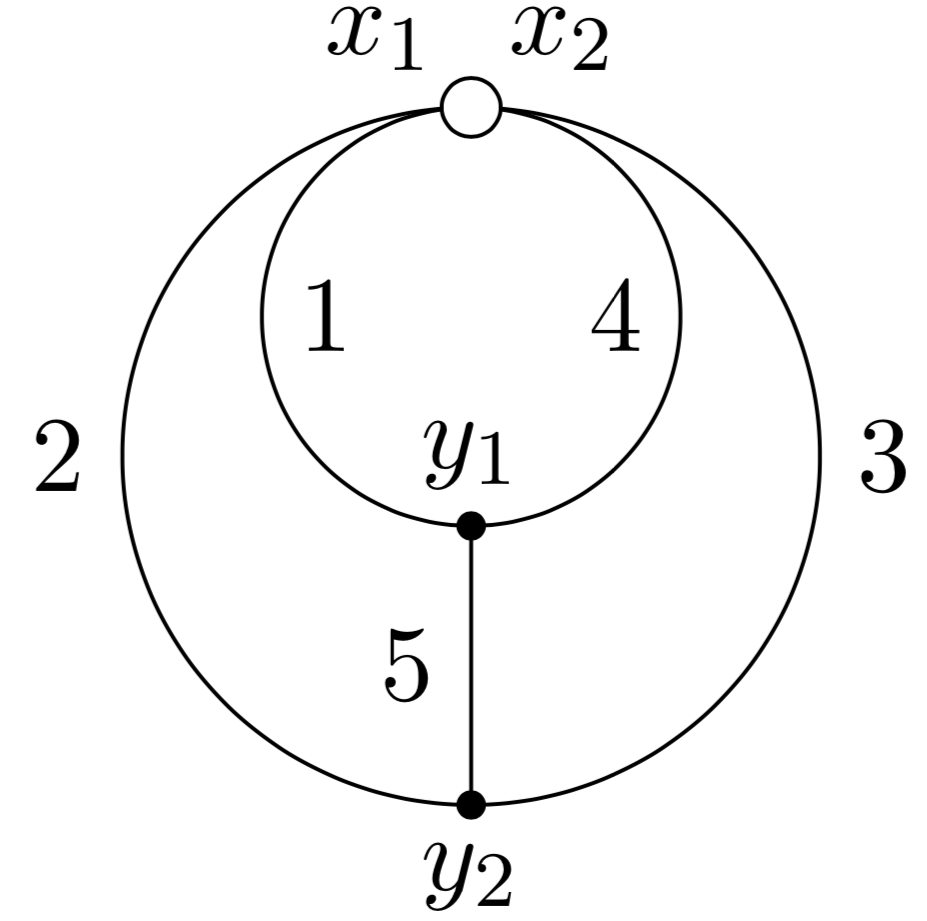}
\end{align}
and we have the following three different admissible cuts as permutations of the internal vertices result in topologically indistinguishable graphs (remember that each cut subdiagram must contain on of the external vertices),
\begin{align}
&    \def\scale{1}
\begin{tikzpicture}[baseline={([yshift=-0.7ex]0,0)}] 
    \coordinate (i1) at (-\scale,0);
    \coordinate (i2) at (+\scale,0);
    \coordinate (v1) at (0,\scale);
    \coordinate (v2) at (0,-\scale);
    \draw (i1) arc (180:90:\scale);
    \draw (i1) arc (180:270:\scale);
    \draw (i2) arc (0:-90:\scale);
    \draw (i2) arc (0:90:\scale);
    \filldraw (v1) circle (1.3pt);
    \filldraw (v2) circle (1.3pt);
    \filldraw (i1) circle (1.3pt);
    \filldraw (i2) circle (1.3pt);
    \draw (v1) -- (v2);
    \draw[dashed] (-\scale,-\scale) -- (\scale,\scale);
    \node at (-\scale/2,\scale/2) {$\sunny$};
    \node at (\scale/2,-\scale/2) {$\shady$};
\end{tikzpicture}%
&
&    \def\scale{1}
\begin{tikzpicture}[baseline={([yshift=-0.7ex]0,0)}] 
    \coordinate (i1) at (-\scale,0);
    \coordinate (i2) at (+\scale,0);
    \coordinate (v1) at (0,\scale);
    \coordinate (v2) at (0,-\scale);
    \draw (i1) arc (180:90:\scale);
    \draw (i1) arc (180:270:\scale);
    \draw (i2) arc (0:-90:\scale);
    \draw (i2) arc (0:90:\scale);
    \filldraw (v1) circle (1.3pt);
    \filldraw (v2) circle (1.3pt);
    \filldraw (i1) circle (1.3pt);
    \filldraw (i2) circle (1.3pt);
    \draw (v1) -- (v2);
    \draw[dashed] (-\scale/2,-\scale) -- (-\scale/2,\scale);
    \node at (-\scale/4*3,0) {$\sunny$};
    \node at (\scale/2,0) {$\shady$};
\end{tikzpicture}%
&
&    \def\scale{1}
\begin{tikzpicture}[baseline={([yshift=-0.7ex]0,0)}] 
    \coordinate (i1) at (-\scale,0);
    \coordinate (i2) at (+\scale,0);
    \coordinate (v1) at (0,\scale);
    \coordinate (v2) at (0,-\scale);
    \draw (i1) arc (180:90:\scale);
    \draw (i1) arc (180:270:\scale);
    \draw (i2) arc (0:-90:\scale);
    \draw (i2) arc (0:90:\scale);
    \filldraw (v1) circle (1.3pt);
    \filldraw (v2) circle (1.3pt);
    \filldraw (i1) circle (1.3pt);
    \filldraw (i2) circle (1.3pt);
    \draw (v1) -- (v2);
    \draw[dashed] (\scale/2,-\scale) -- (\scale/2,\scale);
    \node at (-\scale/2,0) {$\sunny$};
    \node at (\scale/4*3,0) {$\shady$};
\end{tikzpicture}%
.
\intertext{In addition to the positivity requirements, only energy flows from $\sunny$ to $\shady$ are allowed on cut edges. Therefore only the following energy flows are compatible with the cuts and the positive energy requirement:}
\label{eq:theta_cuts1}
&    \def\scale{1}
\begin{tikzpicture}[baseline={([yshift=-0.7ex]0,0)}] 
    \coordinate (i1) at (-\scale,0);
    \coordinate (i2) at (+\scale,0);
    \coordinate (v1) at (0,\scale);
    \coordinate (v2) at (0,-\scale);
    \draw[->-] (i1) arc (180:90:\scale);
    \draw[->] (i1) arc (180:270:\scale);
    \draw[->] (v1) arc (90:0:\scale);
    \draw[->-] (v2) arc (-90:0:\scale);
    \draw[->] (v1) -- (v2);
    \filldraw (v1) circle (1.3pt);
    \filldraw (v2) circle (1.3pt);
    \filldraw (i1) circle (1.3pt);
    \filldraw (i2) circle (1.3pt);
    \draw[dashed] (-\scale,-\scale) -- (\scale,\scale);
\end{tikzpicture}%
&
&    \def\scale{1}
\begin{tikzpicture}[baseline={([yshift=-0.7ex]0,0)}] 
    \coordinate (i1) at (-\scale,0);
    \coordinate (i2) at (+\scale,0);
    \coordinate (v1) at (0,\scale);
    \coordinate (v2) at (0,-\scale);
    \draw[->-] (i1) arc (180:90:\scale);
    \draw[->-] (i1) arc (180:270:\scale);
    \draw[->-] (v1) arc (90:0:\scale);
    \draw[->-] (v2) arc (-90:0:\scale);
    \draw[->-] (v1) -- (v2);
    \filldraw (v1) circle (1.3pt);
    \filldraw (v2) circle (1.3pt);
    \filldraw (i1) circle (1.3pt);
    \filldraw (i2) circle (1.3pt);
    \draw[dashed] (-\scale/2,-\scale) -- (-\scale/2,\scale);
\end{tikzpicture}%
&
&    \def\scale{1}
\begin{tikzpicture}[baseline={([yshift=-0.7ex]0,0)}] 
    \coordinate (i1) at (-\scale,0);
    \coordinate (i2) at (+\scale,0);
    \coordinate (v1) at (0,\scale);
    \coordinate (v2) at (0,-\scale);
    \draw[->-] (i1) arc (180:90:\scale);
    \draw[->-] (i1) arc (180:270:\scale);
    \draw[->-] (v1) arc (90:0:\scale);
    \draw[->-] (v2) arc (-90:0:\scale);
    \draw[->-] (v1) -- (v2);
    \filldraw (v1) circle (1.3pt);
    \filldraw (v2) circle (1.3pt);
    \filldraw (i1) circle (1.3pt);
    \filldraw (i2) circle (1.3pt);
    \draw[dashed] (\scale/2,-\scale) -- (\scale/2,\scale);
\end{tikzpicture}%
\\
(1a) && (1b) && (1c)
\notag
\\
\label{eq:theta_cuts2}
&&
&    \def\scale{1}
\begin{tikzpicture}[baseline={([yshift=-0.7ex]0,0)}] 
    \coordinate (i1) at (-\scale,0);
    \coordinate (i2) at (+\scale,0);
    \coordinate (v1) at (0,\scale);
    \coordinate (v2) at (0,-\scale);
    \draw[->-] (i1) arc (180:90:\scale);
    \draw[->-] (i1) arc (180:270:\scale);
    \draw[-<-] (v1) arc (90:0:\scale);
    \draw[->-] (v2) arc (-90:0:\scale);
    \draw[->-] (v1) -- (v2);
    \filldraw (v1) circle (1.3pt);
    \filldraw (v2) circle (1.3pt);
    \filldraw (i1) circle (1.3pt);
    \filldraw (i2) circle (1.3pt);
    \draw[dashed] (-\scale/2,-\scale) -- (-\scale/2,\scale);
\end{tikzpicture}%
&
&    \def\scale{1}
\begin{tikzpicture}[baseline={([yshift=-0.7ex]0,0)}] 
    \coordinate (i1) at (-\scale,0);
    \coordinate (i2) at (+\scale,0);
    \coordinate (v1) at (0,\scale);
    \coordinate (v2) at (0,-\scale);
    \draw[->-] (i1) arc (180:90:\scale);
    \draw[-<-] (i1) arc (180:270:\scale);
    \draw[->-] (v1) arc (90:0:\scale);
    \draw[->-] (v2) arc (-90:0:\scale);
    \draw[->-] (v1) -- (v2);
    \filldraw (v1) circle (1.3pt);
    \filldraw (v2) circle (1.3pt);
    \filldraw (i1) circle (1.3pt);
    \filldraw (i2) circle (1.3pt);
    \draw[dashed] (\scale/2,-\scale) -- (\scale/2,\scale);
\end{tikzpicture}%
\\
&& (2b) && (2c)
\notag
\end{align}
In the depiction above, each row features only one orientation of the graph and each column a possible cut. In this example, there are only two admissible paths that are compatible with a cut. 
The cut diagram $(1a)$ has the following three routes
{
\def\scale{.8}
\begin{align*}
\begin{array}{ccccc}
\begin{tikzpicture}[baseline={([yshift=-0.7ex]0,0)}] 
    \coordinate (i1) at (-\scale,0);
    \coordinate (i2) at (+\scale,0);
    \coordinate (v1) at (0,\scale);
    \coordinate (v2) at (0,-\scale);
    \draw[->-] (i1) arc (180:90:\scale);
    \draw[->] (i1) arc (180:270:\scale);
    \draw[->] (v1) arc (90:0:\scale);
    \draw[->-] (v2) arc (-90:0:\scale);
    \draw[->] (v1) -- (v2);
    \filldraw (v1) circle (1.3pt);
    \filldraw (v2) circle (1.3pt);
    \filldraw (i1) circle (1.3pt);
    \filldraw (i2) circle (1.3pt);
    \draw[dashed] (-\scale,-\scale) -- (\scale,\scale);
\end{tikzpicture}%
& \longrightarrow
&\;\;\;\;  \begin{tikzpicture}[baseline={([yshift=-0.7ex]0,0)}] 
    \coordinate (i1) at (-\scale,0);
    \coordinate (i2) at (+\scale,0);
    \coordinate (v1) at (0,\scale);
    \coordinate (v2) at (0,-\scale);
    \draw[->] (i1) arc (180:270:\scale);
    \draw[->-] (v2) arc (-90:0:\scale);
    \filldraw (v1) circle (1.3pt);
    \filldraw (v2) circle (1.3pt);
    \filldraw (i1) circle (1.3pt);
    \filldraw (i2) circle (1.3pt);
    \draw[dashed] (-\scale,-\scale) -- (\scale,\scale);
\end{tikzpicture}%
&\;\;\;\;  \begin{tikzpicture}[baseline={([yshift=-0.7ex]0,0)}] 
    \coordinate (i1) at (-\scale,0);
    \coordinate (i2) at (+\scale,0);
    \coordinate (v1) at (0,\scale);
    \coordinate (v2) at (0,-\scale);
    \draw[->-] (i1) arc (180:90:\scale);
  \draw[->-] (v2) arc (-90:0:\scale);
    \draw[->] (v1) -- (v2);
    \filldraw (v1) circle (1.3pt);
    \filldraw (v2) circle (1.3pt);
    \filldraw (i1) circle (1.3pt);
    \filldraw (i2) circle (1.3pt);
    \draw[dashed] (-\scale,-\scale) -- (\scale,\scale);
\end{tikzpicture}%
&\;\;\;\; \begin{tikzpicture}[baseline={([yshift=-0.7ex]0,0)}] 
    \coordinate (i1) at (-\scale,0);
    \coordinate (i2) at (+\scale,0);
    \coordinate (v1) at (0,\scale);
    \coordinate (v2) at (0,-\scale);
    \draw[->-] (i1) arc (180:90:\scale);
    \draw[->] (v1) arc (90:0:\scale);
    \filldraw (v1) circle (1.3pt);
    \filldraw (v2) circle (1.3pt);
    \filldraw (i1) circle (1.3pt);
    \filldraw (i2) circle (1.3pt);
    \draw[dashed] (-\scale,-\scale) -- (\scale,\scale);
\end{tikzpicture}%
\\
&&&&
\\
(1a) & &\;\;\;\;\pp_1 &\;\; \;\;\pp_2 &\;\;\;\; \pp_3
\end{array}
\end{align*}
}
Hence, applying the FOPT-cut Feynman rules from above to the cut diagram $(1a)$ 
results in the following expression 
\begin{gather}
\begin{gathered}
A_{(\boldsymbol{\sigma},\mathfrak{C})_{(1a)}} 
=
-8
\frac{(2\pi)^2 g^{4}}{(8\pi^2)^{5}}
\int \frac{\dd^3 \vec y_1 \dd^3 \vec y_2}{|\vec z_1| |\vec z_2| |\vec z_3| |\vec z_4| |\vec z_5| }
\\
\times
\frac{|\vec z_2|}{(-|\vec z_3|+\tau+i\ep)^2-\vec z_2^{\,2}}
\frac{|\vec z_5|}{(|\vec z_1|-|\vec z_3|+\tau+i\ep)^2-\vec z_5^{\,2}}
\frac{|\vec z_4|}{(|\vec z_1|+\tau+i\ep)^2-\vec z_4^{\,2}}.
\end{gathered}
\intertext{
where we accounted for the admissible paths through the cut, $23$, $153$ and $14$ via the appropriate denominators, and $\tau = x_2^0 - x_1^0$. 
Analogously, applying the Feynman rules to the FOPT-cut graphs $(1b)$ and $(1c)$ results in
}
\begin{gathered}
A_{(\boldsymbol{\sigma},\mathfrak{C})_{(1b)}} 
=
8
\frac{(2\pi)^2 g^{4}}{(8\pi^2)^{5}}
\int \frac{\dd^3 \vec y_1 \dd^3 \vec y_2}{|\vec z_1| |\vec z_2| |\vec z_3| |\vec z_4| |\vec z_5| }\times
\\
\times
\frac{|\vec z_1|}{(-|\vec z_4|+\tau+i\ep)^2-\vec z_1^{\,2}}
\,\frac{|\vec z_1|}{(-|\vec z_5|-|\vec z_3|+\tau+i\ep)^2-\vec z_1^{\,2}}
\,\frac{|\vec z_2|}{(-|\vec z_3|+\tau+i\ep)^2-\vec z_2^{\,2}}\,,
\end{gathered}
\\
\begin{gathered}
A_{(\boldsymbol{\sigma},\mathfrak{C})_{(1c)}} (x_1,x_2)
=
-8
\frac{(2\pi)^2 g^{4}}{(8\pi^2)^{5}}
\int \frac{\dd^3 \vec y_1 \dd^3 \vec y_2}{|\vec z_1| |\vec z_2| |\vec z_3| |\vec z_4| |\vec z_5| }
\times \\
\times
\frac{|\vec z_4|}{(|\vec z_1|+\tau+i\ep)^2-\vec z_4^{\,2}}
\,\frac{|\vec z_3|}{(|\vec z_1| +|\vec z_5|+\tau+i\ep)^2-\vec z_3^{\,2}}
\,\frac{|\vec z_3|}{(|\vec z_2|+\tau+i\ep)^2-\vec z_3^{\,2}}\,.
\end{gathered}
\end{gather}
These three FOPT-cut integrals all correspond to the same directed graph with different cuts on it.  Each integrand features three factors in the denominator, which each corresponding to a unique route from the left-most to the right-most vertex.

For the other possible flow orientations of the graph, we only have two routes through the diagram and one closed cyclic energy flow:
\begin{gather}
\begin{gathered}
A_{(\boldsymbol{\sigma},\mathfrak{C})_{(2b)}} (x_1,x_2)
=
-4
\frac{(2\pi)^2 g^{4}}{(8\pi^2)^{5}}
\int \frac{\dd^3 \vec y_1 \dd^3 \vec y_2}{|\vec z_1| |\vec z_2| |\vec z_3| |\vec z_4| |\vec z_5| }
\times \\
\times
\frac{|\vec z_1|}{(-|\vec z_5|-|\vec z_3|+\tau+i\ep)^2-\vec z_1^{\,2}}
\frac{|\vec z_2|}{(-|\vec z_3|+\tau+i\ep)^2-\vec z_2^{\,2}}
\frac{1}{-|\vec z_5| - |\vec z_3| -|\vec z_4|}\,,
\end{gathered}
\\
\begin{gathered}
A_{(\boldsymbol{\sigma},\mathfrak{C})_{(2c)}} (x_1,x_2)
=
4
\frac{(2\pi)^2 g^{4}}{(8\pi^2)^{5}}
\int \frac{\dd^3 \vec y_1 \dd^3 \vec y_2}{|\vec z_1| |\vec z_2| |\vec z_3| |\vec z_4| |\vec z_5| }\times
\\
\times
\frac{|\vec z_4|}{(|\vec z_1|+\tau+i\ep)^2-\vec z_4^{\,2}}
\frac{|\vec z_3|}{(|\vec z_1| +|\vec z_5|+\tau+i\ep)^2-\vec z_3^{\,2}}
\frac{1}{|\vec z_2|+|\vec z_5|+|\vec z_1|}\,.
\end{gathered}
\end{gather}
The energy-flow-oriented cut graph $(2b)$ has the cycle $534$ and the graph $(2c)$ the cycle $152$.

Even though, e.g., $(1a)$ and both $(1b)$ and $(1c)$ have differently sized cuts, the corresponding integrands are of the same dimension. This is a convenient situation from the perspective of the numerical evaluation of these integrals, as we can put all $(1a)$, $(1b)$ and $(1c)$ under the same integral sign. We expect IR singularities to cancel locally in our proposed representation, but we postpone the detailed analysis of this conjecture to a future work.

\subsection{Summary}
In this section we have presented the general derivation of a novel representation of Feynman diagrams in terms of energy flows, which we called Flow Oriented Perturbation Theory (Which is related to Lightcone Ordered Perturbation theory \cite{Erdogan:2017gyf}). This representation is interesting by itself and has several promising features such as per diagram IR factorization at the level of the $S$-matrix among others (see \cite{Borinsky:2022msp} for details). It will be the topic of future work to extend this treatment to $D$ dimensions in order to be able to regularize UV divergences as well as exploring the cancellation of IR divergences when summing real and virtual corrections in FOPT.
\chapter{General Summary}
\setcounter{chapter}{5}
In this thesis we have covered various relevant topics in Quantum Field Theories and their application to the Phenomenology of Particle Physics. We hope to have illustrated the wide range of applicability of QFTs for unraveling the structure of matter and Nature's forces, as well as systematizing the quest for new physics.
\subsubsection*{Objectives}
The goal of this dissertation was to work out different aspects of quantum field theory, particularly in non-perturbative but also in perturbative regimes, applied to the intellectual construction that is the Standard Model, but also its extension via effective field theories.

The following are practical contributions that we wanted to develop for different subfields:
\begin{itemize}
    \item Qualitatively assess why the SM might have its specific symmetries.
    \item Propose observables to experimentally distinguish Electroweak Effective Field Theories at accelerators.
    \item Extrapolate eventual beyond the SM LHC data (low energies) to new physics resonant regions (high energies) with controlled uncertainties.
    \item Study QCD precision (high energy) calculations in coordinate space.
\end{itemize}

\subsubsection*{Methodology}
The methodology used throughout this thesis is based on the application of mathematical analysis to the Quantum Field Theory framework. More specifically, this thesis combines both analytical and numerical methods to address the various objectives.
\subsubsection*{Results}

\begin{itemize}
    \item Combining perturbative and non-perturbative QCD, we have studied the dynamical mass generation of fermions charged under large Lie groups and shown they acquire large masses (under mild hypotheses) that might push them above the observed spectrum.
    \item We have shown how chiral symmetry breaking structures of the non-perturbative $q\bar{q} g$ vertex seem to be dominant at the production threshold of $q\bar{q}$ pairs that trigger meson decays.
    \item We have elucidated, from a simple analytical criterion, the correlations that Standard Model EFT (SMEFT) induces into Higgs EFT (HEFT) parameters. We have assessed how to experimentally test these correlations, obtaining the amplitudes and cross sections for the relevant processes.
    \item To extend this EFTs beyond the lowest energy range (few hundreds of GeV), we have systematized the theory uncertainties of unitarization with the Inverse Amplitude Method (IAM).
    \item We have explicitly computed jet functions in coordinate space. We have also explored novel ways of working with perturbation theory. In particular we have derived a new approach to Feynman amplitudes called Flow-Oriented Perturbation Theory (FOPT).
\end{itemize}

\subsubsection*{Conclusions}

In chapter 1 we have introduced the status of Particle Physics Phenomenology and the main features of the Standard Model (SM): its forces and matter content. There, we have discussed the experiments that are at odds with it and also several other reasons that point to the possibility that the SM is not a complete theory for Nature's particle interactions. 

Chapter 2, introduced the theoretical framework of this thesis. We reviewed the basic and fundamental definitions of Quantum Field Theories (QFTs): the gauge principle, quantization, the $S$-matrix and cross sections. Then, we focused our attention on both the perturbative and non-perturbative regimes of QFTs and more specifically QCD. In subsection \ref{subsc:CTP} we used both regimes (perturbation theory at two loops and Dyson-Schwinger equations (DSEs)) in order to partially answer the question: \textit{``Why does the SM have such specific symmetries?''}, finding that big Lie groups generate to much mass (from a Grand Unification scale) to be observed at current accelerators. In this chapter we also studied in section \ref{sc:Confinement} the topic of Confinement and briefly introduced some of its partial explanations. This DSEs can also be used to describe the main features of QCD's confining properties. Finally, in this chapter we introduced the main features of Effective Field Theories (EFTs) and more specifically Chiral Perturbation Theory (ChPT), which is very useful for both hadron and beyond the SM electroweak physics.

We discuss physics beyond the SM, but instead of looking for specific models to complete it, in chapter 3 we studied its EFT extensions of the Electroweak Symmetry Breaking Sector. This is interesting for the quest of new physics since it belongs to the most uncharted ones in the SM, with many parameters still unconstrained. To extend this sector, two effective field theories have been recently put forward: the SMEFT and the HEFT. It happens that SMEFT is a special case of HEFT, as can be shown by powerful geometric methods. However, in order to stay close to accelerator physics, we have derived simpler and analytical conditions on a HEFT Lagrangian in order to be expressed in the SMEFT language. These conditions produce testable correlations on parameters that are accessible at experiments such as LHC's ATLAS and CMS or future accelerators, these correlations can be assessed in processes involving several Higgs bosons. For these we have obtained amplitudes and cross sections whose measurement will help distinguishing between SMEFT and HEFT frameworks.

To extrapolate eventual low energy data to higher energies and predict resonant new physics, in chapter 4 we have analyzed the most promising unitarization extension of the EFTs and its uncertainties.  These uncertainties were not analyzed before. It is crucial to control them since, once LHC data suggests a new physics scale, we need to assess the uncertainty on the prediction to see if the new physics may be within the reach of a future accelerator. In this chapter we have consistently systematized the analysis of the method's uncertainties, using hadron physics as an experimental ``handle'' (since data is known here), based on the universality of ChPT. Our hope is that this type of analyses start to be generalized when analyzing experimental data.

Finally, in chapter 5 we have laid out our contribution to perturbation theory with a study of QFTs from the coordinate space viewpoint. In section \ref{sec:explicitcomputationjetCS} we have, after introducing the topic of factorization of infrared (IR) singularities in both momentum and coordinate spaces, explicitly obtained expressions for the jet functions in coordinate space, comparing the results to known momentum space results. In section \ref{sec:fopt} we have introduced a novel approach to coordinate space QFT amplitudes. This approach, which we called Flow Oriented Perturbation Theory (FOPT) produces a nice picture of Feynman graphs in terms of energy flows on the diagram and has very promising features regarding the cancellation and factorization of IR singularities in Feynman amplitudes.

\setcounter{chapter}{1}
\renewcommand{\thechapter}{\Alph{chapter}}%
\chapter{Appendices}
\setcounter{section}{0}
\setcounter{equation}{0}
\setcounter{table}{0}
\setcounter{figure}{0}
\section{Schwarz's lemma in complex analysis and its corollaries}
\label{app:Lemma}

This brief appendix provides a short overview of Schwarz's lemma, which we quickly use to demonstrate the corollary of interest for subsection~\ref{subsec:Schwarz}. 
Again, the idea is whether the image set under $\mathcal{F}$ includes or not a disk around $\mathcal{F}=1$ large enough
to encompass the origin $\mathcal{F}=0$. The second corollary below gives a sufficient condition for this to be true. 
The point $h^*$ which is the preimage of $\mathcal{F}=0$ is the symmetric point around which the SMEFT expansion can be constructed. If the conditions of the second corollary are met, we know that $\mathcal{F}$ will be analytic in a region broad enough to guarantee the power-expansion.

To start, take a disk $D_z(0,1)$  around the origin in the preimage complex space (in our application, the extension $h\to z$ of the singlet Higgs field to be a complex variable). 
Second, $D_{f}(0,1)$ is a disk in the image complex space (also extending $\mathcal{F}\to f \in \mathbb{C}$), both disks having radius 1 and being centered around 0 as the notation indicates. We can then state the lemma {\cite{analisis}} as follows.

\subsection{Schwarz's Lemma} 

Let $f\;:\;D_z(0,1)\;\to \; D_f(0,1)$ be holomorphic with $f(0)=0$. Then $|f(z)|\leq |z| $ and  $|f'(0)|\leq 1$. 
Furthermore, if $|f(z_0)|= |z_0| $ for some $z_0\in D_z(0,1)$, then $|f(z)|=1 $   $\forall z\in D(0,1).$\\

\paragraph{Proof} 

Given those $f(0)$ and $f'(0)$, write $f(z)=zg(z)$:  $g$ is also holomorphic. 
Take $r<1$, if $|z|=r$ we have that $|g(z)|=\frac{|f(z)|}{r}$ and hence $|g(z)|= \frac{|f(z)|}{r}\leq \frac{1}{r}$ (since the image of $f$ is $D_f(0,1)$). The inequality $|g(z)|\leq \frac{1}{r}$  is satisfied for all $z\in  \bar{D}(0,r)$ thanks to the \textit{Maximum modulus principle} (if $f$ is a holomorphic function, then the modulus $|f|$ cannot exhibit a strict local maximum that is in the interior of the domain of $f$). This means that if $g(z_0)=1/r$, that is, it reaches its maximum for some $z_0$ satisfying $|z_0|< r$, then the function $g$ must be a constant (and the maximum is reached at the boundary anyway).  Now, taking the limit $r\to 1$ from the left we obtain $|g(z)|\leq 1$ and consequently $|f(z)|\leq |z|$ for all $z\in D_z(0,1)$. 

Noticing that $f'(z)=g(z)+zg'(z)$ it is immediate to prove that $|f'(0)|\leq 1$.\\

\subsection{Corollaries}
\paragraph{First Corollary} 

Let $f\;:\;D_z(0,1)\;\to \; D_f(0,M)$ analytic such that $f(0)=0$ and {$|f'(0)|=1$}. Then we will have that $M\geq 1$ and $D_f(0,\frac{\sqrt{2}}{\left(\sqrt{2}+1\right) \left(\sqrt{2}+2\right)M})\subset f(D_z(0,1))$ ( the open disc of radius $\frac{\sqrt{2}}{\left(\sqrt{2}+1\right) \left(\sqrt{2}+2\right)M}$ is contained in the image through $f$ of the open unit disc).\\
\paragraph{Proof} Thanks to Schwarz's lemma we know that $M\geq 1$ since otherwise $|f'(0)|<1$. We can then write $f$ as
\begin{equation}f(z):=z+\sum_{n=2}^\infty a_n z^n\end{equation}

 The triangular inequality of the complex norm yields
\begin{align}
|z|=|f(z)-\sum_{n=2}^\infty a_n z^n|\leq |f(z)|+|\sum_{n=2}^\infty a_n z^n|\ .
\end{align}

Choosing to evaluate with $|z_M|:=\frac{1}{\alpha M}$ with $\alpha> 1$, we find 
\begin{align}
|f(z_M)|&\geq |z_M|-|\sum_{n=2}^\infty a_n z_M^n|\geq |z_M|-\sum_{n=2}^\infty |a_n| |z_M|^n=\frac{1}{\alpha M}-\sum_{n=2}^\infty \frac{|a_n| }{(\alpha M)^n}\ .
\end{align}

Thanks to Cauchy's estimates (cancelling factorials) we know that for all $r<1$, $|a_n|\leq \frac{M}{r^n}$ and hence $|a_n|\leq M$; we may take the worst bound with $r\to 1$. Then,
\begin{align}
|f(z_M)| &\geq\frac{1}{\alpha M}-M\sum_{n=2}^\infty \frac{1}{(\alpha M)^n} =\frac{1}{\alpha M}-\frac{M}{(\alpha M)^2}\frac{1}{1-\frac{1}{\alpha M}}=\frac{1}{\alpha M}\Big(1-\frac{M}{\alpha M-1}\Big)\;,
\end{align}
(having reconstructed the geometric series).
Now, since $M\geq 1$ and we have taken $\alpha>1$ we have that
\begin{align}
  \frac{1}{\alpha M}\Big(1-\frac{1}{\alpha -1/M}\Big)\geq   \frac{1}{\alpha M}\Big(1-\frac{1}{\alpha-1}\Big)=\frac{1}{M}\Big(\frac{\alpha-2}{\alpha^2-\alpha}\Big),
\end{align}
giving the highest lower bound for $\alpha = 2+\sqrt{2}$.
Hence, taking  $|z_M|=\frac{1}{(2+\sqrt{2})M}$, we have
\begin{align}
|f(z_M)|&\geq \frac{\sqrt{2}}{\left(\sqrt{2}+1\right) \left(\sqrt{2}+2\right)M}\;.
\end{align}

We now proceed to prove that the image of the disk in the Higgs field
$f(D_z(0,1))$ contains the disk of the $\mathcal{F}$ function, namely $D_f(0,\frac{\sqrt{2}}{\left(\sqrt{2}+1\right) \left(\sqrt{2}+2\right)M})$.

Use for this an auxiliary 
$w_f\in D_f(0,\frac{\sqrt{2}}{\left(\sqrt{2}+1\right) \left(\sqrt{2}+2\right)M})$,
that is, $|w_f|\leq |f(z_0)|$;
the function $g(z)=w_f-f(z)$ verifies then
\begin{equation}|f(z_M)+g(z_M)|= |w|< \frac{\sqrt{2}}{\left(\sqrt{2}+1\right) \left(\sqrt{2}+2\right)M}\leq |f(z_M)|\;\;\; \text{for }\;\;|z_M|=\frac{1}{(2+\sqrt{2})M}\;.
\end{equation}
Now we make use of Rouché's theorem to state that $f$ and $g$ have the same number of zeroes in $D(0,\frac{1}{(2+\sqrt{2})M})$, i.e. at least one by definition (because by hypothesis $f(0)=0$). As a consequence there exists $z_0\in D(0,1)$ such that $f(z_0)=w_f$, in other words $f(D_z(0,1))$ contains $D_f(0,\frac{\sqrt{2}}{\left(\sqrt{2}+1\right) \left(\sqrt{2}+2\right)M})$.

\paragraph{Second Corollary} 
Let $g \;:\; D(0,R) \;\to\;D_g(0,M)$ analytic, such that $g(0)=0$ and $|g'(0)|=\mu>0$. Then $g(D(0,R))$ contains
another disk where $g$ is analytic,
\begin{equation}  \label{2nddisk}
g(D(0,R)) \supset
D_g(0,\frac{\sqrt{2}R^2\mu^2}{\left(\sqrt{2}+1\right) \left(\sqrt{2}+2\right)M})
\end{equation}
The $R$ appearing there is what can be tested to guarantee that the image 
includes a disk that in turn includes 0 (and therefore, $\exists h_* | \mathcal{F}(h_*)=0$ ), and the function is analytic between the vacuum and that symmetric point.

\paragraph{Proof} It follows in a relatively straightforward manner by applying the first corollary to the auxiliary function  
\begin{equation}  
f(z):=\frac{1}{Rg'(0)}g(R z)\;\;\text{ for }\;\; |z|<1\;.
\end{equation}

\section{Illustration of the IAM's inefficiency
if a CDD pole is present} 
\label{CDDfailure}
\subsection{General discussion}

To illustrate how to proceed to overcome the CDD difficulty in the IAM, 
it is enough to consider the following simple  toy model of an unspecified partial wave~\footnote{The model is obviously inspired by $K$-matrix considerations, so it has theoretical deficiencies that are of no concern for this discussion, such as inadequate analytical behavior on the first Riemann sheet -poor implementation of causality- with a second pole, that fortunately is far from the physical region of interest.}
that features both a resonance and a CDD pole,
\begin{align}
\label{200725.1}
t(s)=&\frac{1}{\frac{{\mathfrak{f}}^4}{s(s-M_0^2)}-i\sigma}~,
\end{align}
with $\mathfrak{f}$ being an energy scale.
We notice two zeroes of $t(s)$, the Adler zero at $s=0$ and the CDD pole at $s=M_0^2$. 
In the second Riemann sheet the equation for the resonance pole is
\begin{align}
\label{200725.2}
\mathfrak{f}^4-i s(s-M_0^2)\sigma(s)=0~,
\end{align}
with $\text{Im}[\sigma(s-i\ep)]<0$. 
To keep the discussion simple, let us take the chiral limit  $m_\pi\to 0$ to calculate the pole positions. For $m_\pi\to 0$, the analytical extrapolation of $\sigma(s)$ to  $\text{Im}(s)<0$ is $+1$ in the second and $-1$ in the first Riemann sheet, respectively. 
As a result, the secular equation to be solved for $\text{Im}( s_R)<0$ is $\mathfrak{f}^4/s_R(s_R-M_0^2)-i=0$, with the solution
\begin{align}
\label{200725.3}
s_R=&\frac{M_0^2}{2}\left(
1+\sqrt{1-i\frac{4\mathfrak{f}^4}{M_0^4}}\right)~.
\end{align}

Let us examine what would the IAM predict for such amplitude.
First, let us construct the  chiral expansion (in powers of $s$) of $t(s)$ in eq.~\eqref{200725.1}, which is all that would be accessible from a low-energy measurement, 
\begin{align}
\label{200725.5}    
t(s)=-\frac{s M_0^2}{\ff^4}
+\frac{s^2}{\ff^4} \left( 1+i\frac{M_0^4}{\ff^4}\sigma(s)\right)
+{\cal O}(s^3)~.
\end{align}
To more clearly express this expansion in notation familiar in ChPT and HEFT, we rewrite
\begin{align}
\label{210126.1}
\frac{M_0^2}{{\ff}^4}&\equiv \frac{\lambda }{v^2}\ \implies \  v^2=\frac{\lambda \ff^4}{M_0^2}~,
\end{align}
where $\lambda$ is a positive numerical coefficient,\footnote{The factor $M_0^2$ relating $\ff^4$ to $v^2$ appears because the $t(s)$ in eq.~\eqref{200725.1} can be considered a $K$-matrix unitarization of the chiral $s$-channel exchange of a resonance with constant propagator  \cite{Oller:1998zr,Guo:2011pa}.} e.g. for the $P$-wave $\pi\pi$ scattering $\lambda=1/6$, while $v$ is equivalent to $f_\pi$ \cite{Oller:1998zr}. 
Then, the expansion in eq.~\eqref{200725.5} becomes
\begin{align}
\label{210126.2}
t(s)=&
-\frac{\lambda s}{v^2}+\frac{\lambda s^2}{M_0^2v^2}+i\sigma(s)\left(\frac{\lambda s}{v^2}\right)^2+{\cal O}(s^3)~.
\end{align}

It is clear that the scale driving the chiral expansion is $M_0^2$.  For instance, the NLO counterterm that can be read from eq.~(\ref{210126.2}) scales as $v^2/M_0^2$, of typical size in any chiral expansion.
In terms of these parameters the  pole position $s_R$ in eq.~\eqref{200725.3} reads
\begin{align}
\label{210127.1}
s_R=&\frac{M_0^2}{2}\left(1+\sqrt{1-i\frac{4v^2}{\lambda M_0^2}}\right)~.
\end{align}

The IAM construction would then extrapolate this expansion to higher physical $s$  taking the form
\begin{align}
\label{200725.6}
t_{IAM}(s)=&
-\left( \frac{\mathfrak{f}^4}{sM_0^2}\left(1+\frac{s}{M_0^2}\right)+i\sigma\right)^{-1}
=-\left( \frac{v^2}{\lambda s}\left(1+\frac{s}{M_0^2}\right)+i\sigma\right)^{-1}~.
\end{align}
The Adler zero at $s=0$ is recognizable, but the second zero, the CDD pole at $s=M_0^2$, has not been recovered by the IAM based on the expansion of eq.~(\ref{200725.5}): 

Compare this to the analogous form that eq.~(\ref{200725.1}) takes in terms of $v$,
\begin{equation}
    t(s)=
-\left( \frac{v^2}{\lambda s}\left(\frac{M_0^2}{M_0^2-s}\right)+i\sigma\right)^{-1}~.
\end{equation}

Failure to reproduce the Adler zero by the standard IAM comes along with the lack of the original pole in the second Riemann sheet, suffering instead from one in the first Riemann sheet at

\begin{equation} s_R=-\frac{M_0^2}{1+iM_0^4/\mathfrak{f}^4}=-\frac{M_0^2}{1+i\lambda M_0^2/ v^2}~.
\end{equation}

The criterion of eq.~\eqref{polepositionb} to identify a resonance near the real axis,  
\begin{equation}t_0(s_R)-\text{Re}t_1(s_R) =
-\frac{M_0^2 s_R}{\mathfrak{f}^4}
-\frac{s_R^2}{\mathfrak{f}^4}=0\ ,
\end{equation} whose solutions are $0$ and $-M_0^2$,  does not yield any sensible result with ${\rm Re}(s_R)>0$ either. It appears clear to us~\cite{Oller:1999me} that this difference between the actual wanted amplitude in eq.~(\ref{200725.1}) and the IAM is due to the near proximity of the  CDD zero of the amplitude to the resonance pole, which is then masked to the IAM.

\subsection{Numerical implementation of the method in subsection.~\ref{subsec:modifyIAMCDD}}

\begin{figure}[ht!]
    \centering
    \includegraphics[width=0.65\columnwidth]{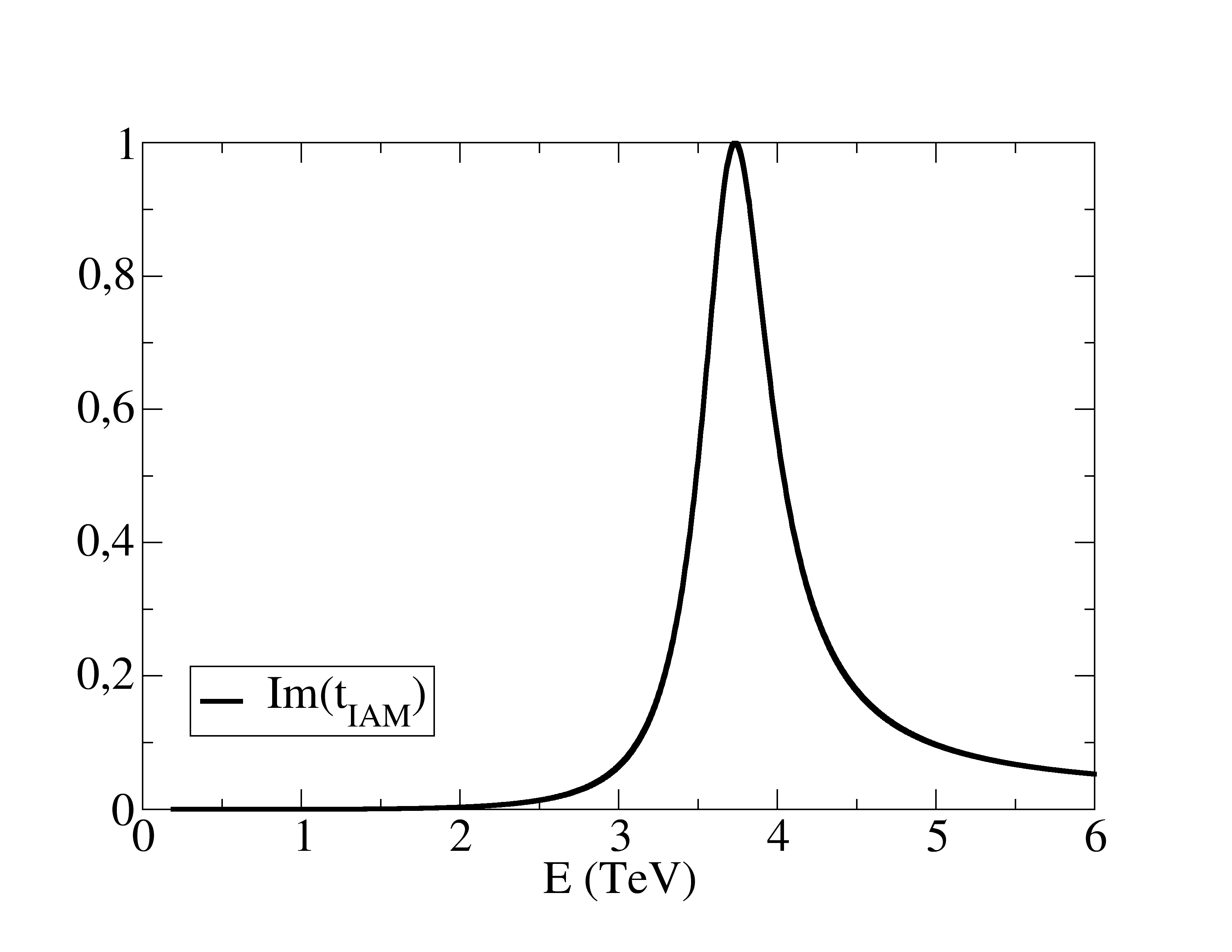}
    \caption{A simple $\rho$-like resonance in $\omega\omega\simeq W_LW_L$ scattering in the HEFT formulation.}
    \label{fig:demoCDD1}
\end{figure}

To illustrate the procedure taking care of the CDD pole, let us look at a computation for the $I,J=1,1$ vector isovector channel with HEFT as deployed in~\cite{Delgado:2015kxa}.
The LO constants are chosen as $a=0.9$, $b=a^2$, and the NLO ones as $a_5(\mu=3{\rm TeV})=-1.75\times 10^{-4}$, $a_4(\mu=3{\rm TeV})=-1.5\times 10^{-4}$ (with all others set to zero). This set yields a $\rho$-like resonance around 3-4 TeV as shown in figure~\ref{fig:demoCDD1}.

\begin{figure}
    \centering
    \includegraphics[width=0.6\columnwidth]{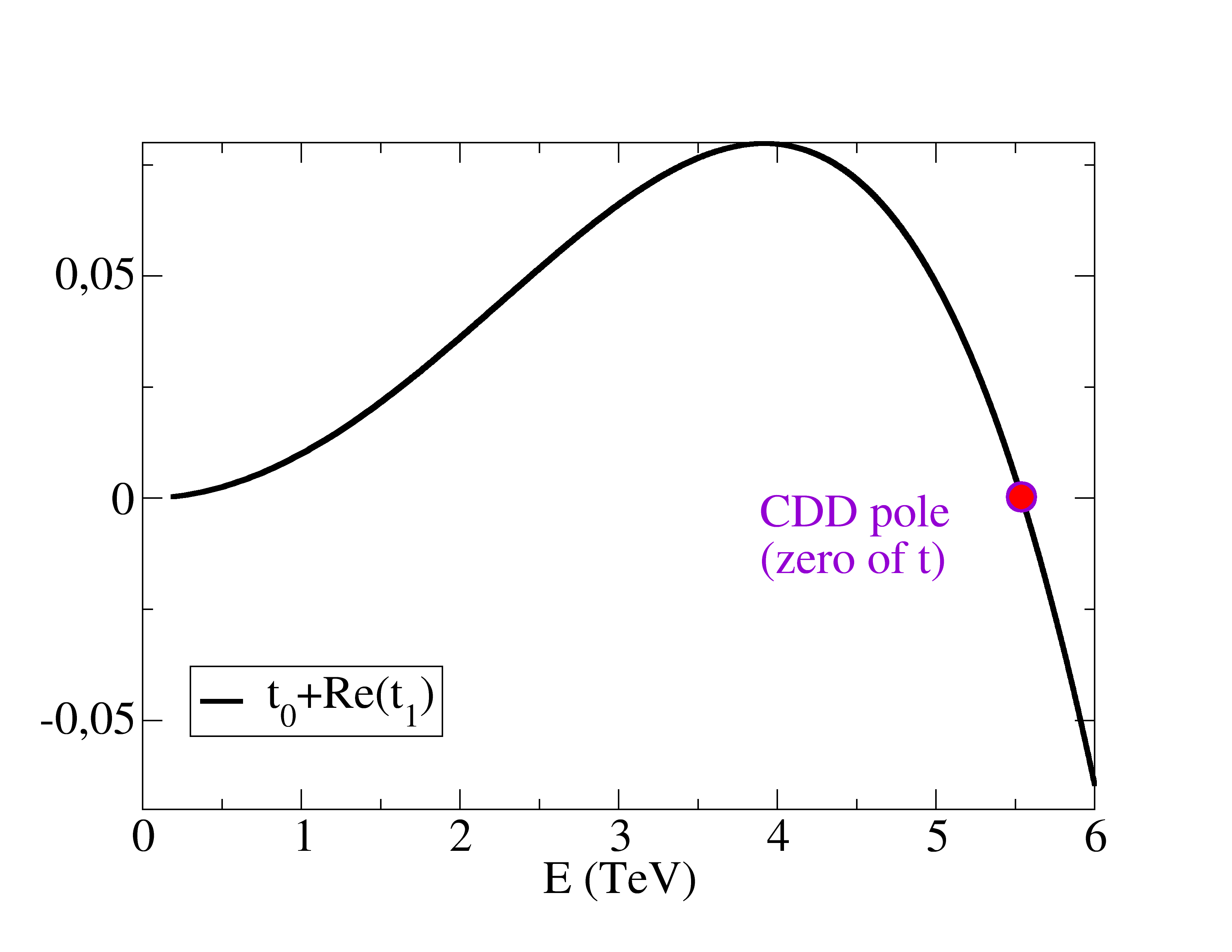}
    \caption[CDD pole appearance]{If the value of $a_4$ is made  more negative than in figure~\ref{fig:demoCDD1}, a CDD pole (zero of the partial wave amplitude) appears, as shown by $t_0+{\rm Re}(t_1)=0$. }
    \label{fig:demoCDD2} 
\end{figure}

\begin{figure}[ht]
    \centering
    \includegraphics[width=0.6\columnwidth]{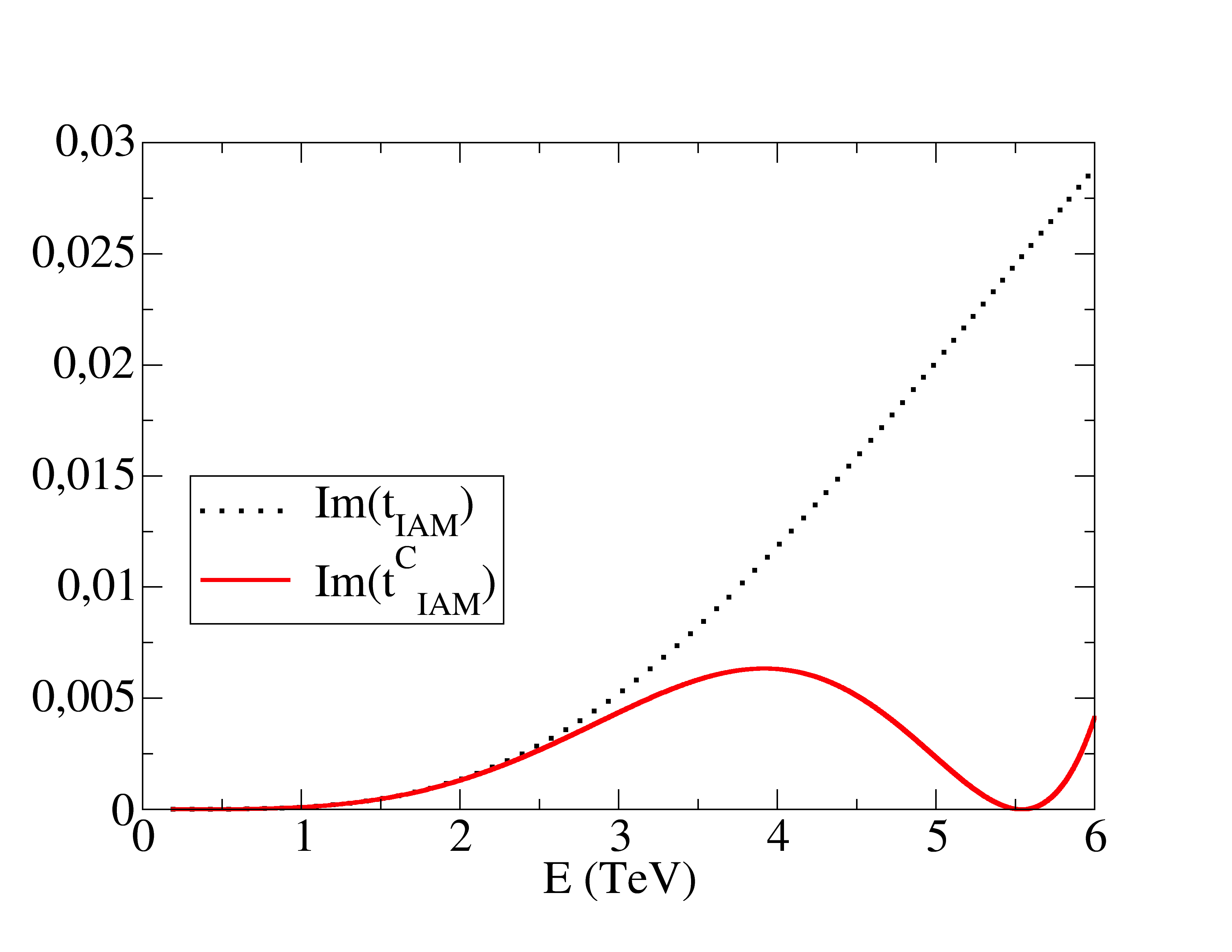}
    \caption[Failure of the IAM when a CDD pole is present]{The conventional IAM (dotted) applied to the amplitude $t$ fails to reproduce the zero associated to a CDD pole. However, applying the method to the amplitude $t/(s-s_C)$, leading to the small modification of eq.~(\ref{modIAM2}), correctly reproduces the CDD zero in the amplitude (solid line).
    \label{fig:demoCDD3}}
\end{figure}

For those values of the low-energy constants, there is no CDD pole so the standard IAM suffices~\footnote{ In figure~\ref{fig:movpolo} below we show a calculation of the pole position of the resonance in the second Riemann sheet. For the same parameter set, the real part of the pole position $s_P$ differs from the nominal mass on the real axis by 5\%. Therefore, at this width, the uncertainty of computing on the real $s$ axis and not in the complex plane is subleading to the uncertainty introduced by the left cut (at 17\%). In any case, a more detailed analysis can beat that 5\% by propagating errors taking into account the complex-number nature of the resonance position.}.
If we increase the absolute value of $a_4(\mu=3{\rm TeV})$ to $-4.5\times 10^{-4}$,
a CDD pole appears satisfying eq.~(\ref{200725.9}), as shown in figure~\ref{fig:demoCDD2}.
We then calculate the conventional IAM of eq.~(\ref{usualIAM}) and the CDD-modified IAM of eq.~(\ref{modIAM2}), and plot it in Figure~\ref{fig:demoCDD3}.

 \begin{figure}[ht!]
  \centerline{\includegraphics[width=0.45\columnwidth]{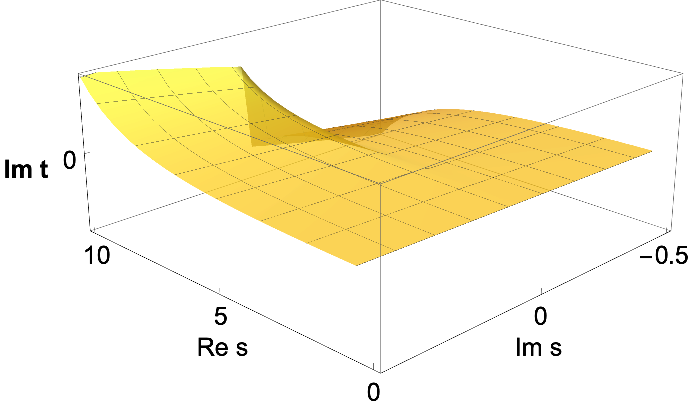}
  \includegraphics[width=0.45\columnwidth]{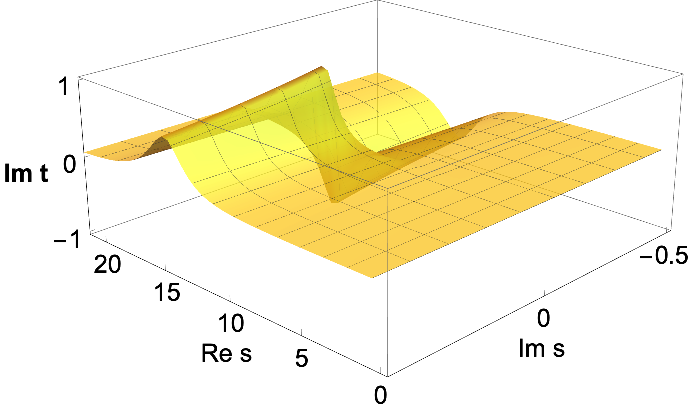}}
\centerline{
\includegraphics[width=0.45\columnwidth]{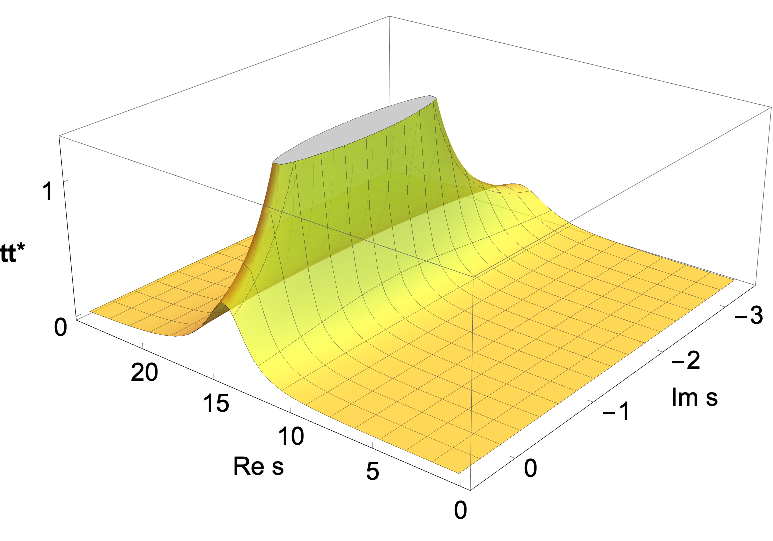}}
\caption[Resonances using the IAM extended to the complex $s$ plane]{\label{fig:poleposition2ndsheet}
The resonance of figure~\ref{fig:lecuncertainty}  using the IAM extended to the complex $s$ plane, with the  parameters $a=0.95$,
$a_4=-2.5\cdot 10^{-4}$ and $a_5=-1.75\cdot 10^{-4}$. 
The two top plots show the first Riemann sheet, cut along the real axis ${\rm Im}(s)=0$. The left one is limited to low and moderate $s$, and only the cut is visible; the right one extends to the resonance energy that, although it does not leave a pole in this first sheet, it is seen to saturate unitarity (${\rm Im}(s)\simeq 1$ for a certain real $s$ (the scale does not allow to visualize the cut). 
The plot in the bottom line is the extension to the second Riemann sheet and clearly features a pole for negative ${\rm Im} (s)<0$.
}
\end{figure}
The difference between the IAM and the CDD-IAM is clear: the second naturally reproduces the CDD pole (zero of the amplitude), while the first fails to do so and starts taking off towards a resonance such as that of figure~\ref{fig:demoCDD1}, just at higher mass (whether and when we would trust the method at that high energy is another discussion about predictivity reach not relevant for this point of the CDD zero; once this first structure is well behind, the CDD-IAM should not be used.).

To clarify what happened to the resonance of figure~\ref{fig:demoCDD1} upon changing the low-energy constant so a CDD pole of the inverse amplitude appears in this channel, we need to pursue the computation of the pole in the second Riemann sheet.

Figure~\ref{fig:poleposition2ndsheet} shows the amplitude in the first (top plots) and second (bottom plot) Riemann sheets for $a_4(\mu=3{\rm TeV})=-2.5\times 10^{-4}$. A branching point and cut along the real $s$ axis is clearly visible in the first one, whereas the second shows an amplitude growing towards a pole for negative imaginary $s$ as expected.

\begin{figure}[ht!]
\centering
\includegraphics[width=0.75\textwidth]{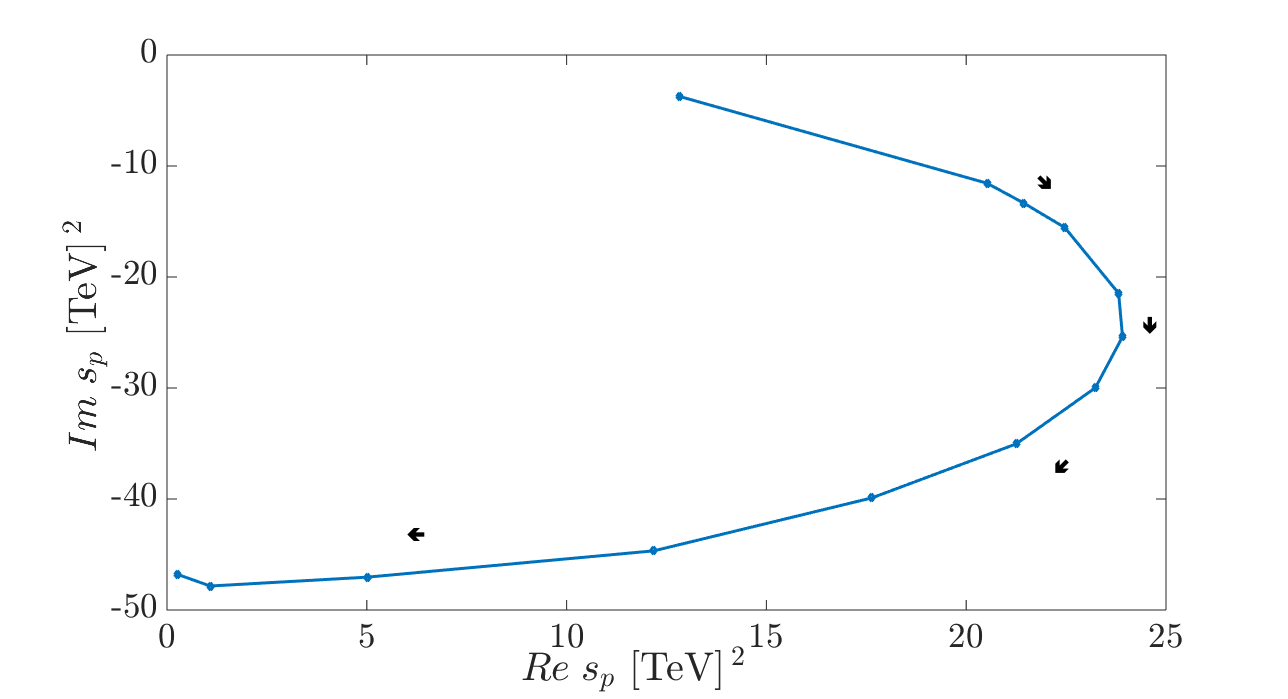}
\caption[Motion of a resonance pole in the complex plane]{ \label{fig:movpolo}
Motion of a resonance pole in the complex plane, for 
$a_4\in [-3.56,-1.5pi]\cdot 10^{-4}$, $a_5=-1.75\cdot 10^{-4}$ and $a=b^{1/2}=0.9$ for the basic IAM.  The arrows indicate the flow of the pole position with increasing absolute value for $a_4$. 
}
\end{figure}

Once this has been visually checked, we can follow the movement of the resonance pole in the complex plane in the second Riemann sheet, shown in figure~\ref{fig:movpolo}. The computed data has employed the bare IAM: because the resonance is broad and far from the axis for a large swath of the parameter space described in the figure, the method is able to capture it in spite of not reproducing the shape along the physical real axis. 

If we instead employ the modified IAM from eq.~(\ref{modIAM2}), with parameters
$a=0.9$, $a_4=-4.5\times 10^{-4}$, $a_5=-1.5\times 10^{-4}$, the pole sits, approximately, at $s=(15.7,-96.1){\rm TeV}^2$ on the second Riemann sheet, corresponding to $M\simeq 4$ TeV, $\Gamma\simeq 24$ GeV, that is, extremely deep in the complex $s$-plane, in agreement with the basic IAM. The difference is sufficiently substantial to much prefer the use of the modified method also for the broadest resonances, due to the effect of the missed zero in the basic IAM.

%
\section{Left-cut partial wave amplitude: a new characterization}

In this appendix we provide a (possibly new) characterization of the partial wave over the left cut, which might be useful for uncertainty estimates, but we have not yet fully exploited it and have decided to relegate it off the main text.

\begin{figure}[ht!]
 \centering
\includegraphics[width=0.6\columnwidth]{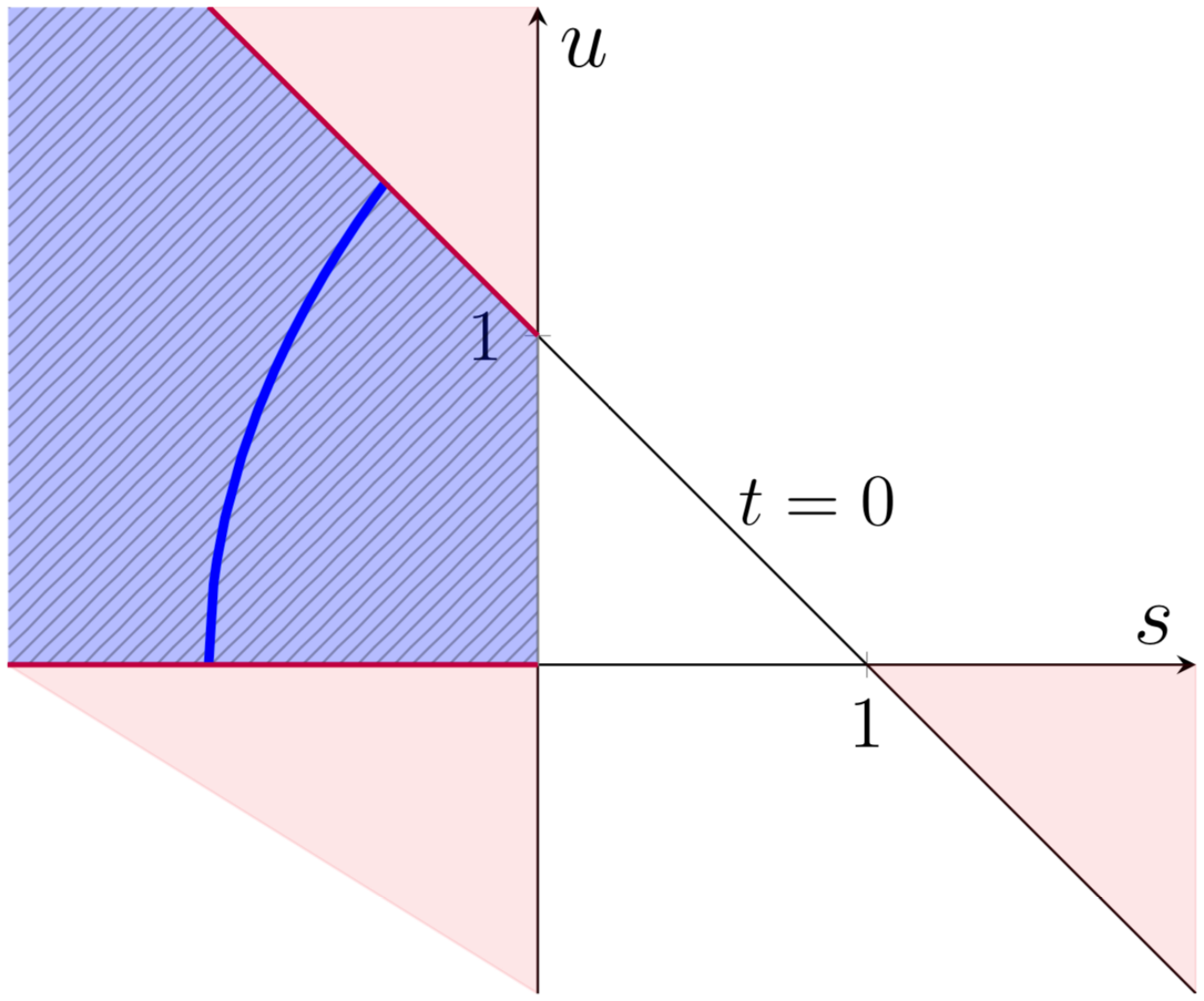}
 \caption[Mandelstam plane for two identical particles]{Mandelstam plane for two identical particles (in units of $4m^2$, $t=1-s-u$). The three $2\to 2$ physical regions (misty rose-coloured) extend outside the triangle vertices.
 In the center-left (patterned, purplish), we show the unphysical region where the integrand in eq.~(\ref{lcpartialwv}) is evaluated. The arch (thick, blue) represents the contour of integration for $x\in[-1,1]$; the argument of the partial wave on the left cut, $s$, is negative.  Note that along the slanted thick segments (velvet) the
 kinematics correspond to the physical $u$- and $t$-channels, so the amplitude can be evaluated from data at the end points of the arch.
 }\label{region}
 \end{figure}

The partial wave over the left cut, with $s\leq0$, is
\be
t_{IJ}(s)= \frac{1}{32\pi\eta}\int_{-1}^{+1} dx \,P_J(x)	\,T^I(s,t(x),u(x))\label{lcpartialwv}
 \ee\par
where $t(x)=(2m^2-s/2)(1-x)$ and $u(x)=(2m^2-s/2)(1+x)$. Following Mandelstam, we assume that there is a unique analytic amplitude $T(s,t,u)$ such that
   \[
T(s,t,u)=    \left\{
   \begin{array}{l}
T_s(s,t,u) \,\, \text{for}\,\,s\geq4m^2,\,t\leq0,\,u\leq0\ ,\nonumber\\
T_{t\,}(t,s,u) \,\, \text{for}\,\,t\geq4m^2,\,s\leq0,\,u\leq0\ ,\nonumber\\
 T_u(u,t,s) \, \text{for}\,\,u\geq4m^2,\,t\leq0,\,s\leq0\ ,
              \end{array}
    \right.
  \]
  with these three physical regions of different channels 
  shaded in Fig.~\ref{region} (misty rose color) .

 Notice that the partial-wave interval of integration in eq.~(\ref{lcpartialwv}) corresponds to a line crossing the patterned region (purplish) of the Mandelstam plane shown in Fig. \ref{region}, with endpoints fixed at the physical $u$- and $t$-channel borderlines (velvet).
 
Hence we see that, for the integral which defines the partial wave over the left cut, the endpoint values of the integrand are amenable to treatment from experimental data.  These values correspond to the physical amplitude below its right cut in the $u$- and $t$-channels respectively,
  \[
    \left\{
                \begin{array}{l}
                 T(s,x=+1)=T(s,0,4m^2-s)=T_u(4m^2-s,0,s)\\
                  T(s,x=-1)=T(s,4m^2-s,0)=T_{t\,}(4m^2-s,s,0)\;,
                \end{array}
    \right.
  \]
  since the left cut integrand of the $s$-channel (on the left-hand side) is evaluated at $s+i\epsilon$ with $s\leq0$ and therefore the combination $4m^2-s-i\epsilon$ lies below the right cut of the $u$ or $t$ channels on the RHS.
 \par
  
  The partial-wave of eq.~(\ref{lcpartialwv}) is a linear combination of the auxiliary functions (omitting isospin indices),
  \begin{align}
  \psi_J(s)\equiv\int_{-1}^{+1} dx \,x^J	\,T(s,t(x),u(x))\;,\label{si}
  \end{align}
  which we will integrate by parts \textit{ad infinitum}. 
  
  These infinitely many integrations by parts can be carried out due to Cauchy's inequalities, since they guarantee that 
 the $n$-th derivative  respect to $x$  of the amplitude is bounded in modulus by a quantity of order $n!\times |\text{sup}\; (T)|$ in the region of the complex $x$-plane where $T$ is analytic. Successively integrating ($n$ times) the monomial $x^J$ yields $x^{n+J}J!/(n+J)!$. Observe then that the denominator controls the $n$-th derivative of the amplitude and the integral of the monomial $x^{J+n}$ tends to zero when $n\to\infty$.  Hence,
    \begin{align}
  &\psi_J(s)=\sum_{n=0}^\infty \frac{(-1)^n J!}{(n+1+J)!}\Big(\frac{\partial^n}{\partial x^n}T(s,x)\Big|_{x=+1}+(-1)^{n+J}\frac{\partial^n}{\partial x^n}T(s,x)\Big|_{x=-1}\Big)\;.\label{series}
  \end{align}

Note that
\begin{equation}
\frac{\partial^n}{\partial x^n}T(s,x)\Big|_{x=+1}=(2m^2-s/2)^n\frac{\partial^n}{\partial u^n}T_u(u,0,s)\Big|_{u=4m^2-s}\;,
\end{equation}

with $u=4m^2-s$, and
\begin{equation}
\frac{\partial^n}{\partial x^n}T(s,x)\Big|_{x=-1}=(s/2-2m^2)^n\frac{\partial^n}{\partial t^n}T_t(t,s,0)\Big|_{t=4m^2-s}
\end{equation}
with $t=4m^2-s$.
So that these derivatives are to be taken from the amplitudes in both $u$- and $t$- physical channels. In the case of purely elastic scattering the only branch points are at each channel's two pion threshold. Hence, we exclude $s=0$ where the branch points of $u$- and $t$- channels sit for $x=\pm1$ respectively from this treatment. The derivatives can be taken since the functions are evaluated in the first Riemann sheet, where the amplitude is analytic. Here we are also assuming that the supremum of $T$, $\sup{T}$, exists in a neighbourhood of the cuts.
\par

  To pass from the monomials $x_J$ to the Legendre polynomials, we represent them as
 \be
 P_J(x)=2^J\sum_{k=0}^J \binom{J}{k}\binom{\frac{J+k-1}{2}}{J}_g x^k\,,
 \ee
 where the generalized binomial coefficient is
  \be
 \binom{\alpha}{k}_g=\frac{\alpha(\alpha-1)(\alpha-2)...(\alpha-k+1)}{k!}\,,
 \ee
 to express the partial wave over the left cut ($s<0$) as
 \begin{align}
 t_J(s)=&\frac{2^J}{32\pi\eta}\sum_{k=0}^J\binom{J}{k}\binom{\frac{J+k-1}{2}}{J}_g\sum_{n=0}^\infty \frac{(-1)^n k!}{(n+1+k)!}\times\nonumber\\
 &\times \Big(\frac{\partial^n}{\partial x^n}T(s,x)\Big|_{x=+1}+(-1)^{n+J}\frac{\partial^n}{\partial x^n}T(s,x)\Big|_{x=-1}\Big)\;.\label{leftcutt}
 \end{align}
 
 The advantage of this expression is that the partial wave over an unphysical line (the left cut) is expressed in terms of quantities evaluated a two physical points corresponding to the $u$- and $t$-channel right cuts respectively. Studies such as \cite{Oehme:1956zz} review dispersion relations for the derivatives of the forward amplitude appearing in eq. (\ref{leftcutt}) and could be useful for future studies.
Now, we could be tempted to use the unitarity relations for the physical amplitudes,
 \be
 -2\,\text{Im}\,T_{u}(4m^2-s,0,s)=T_{u}(4m^2-s,0,s)T_u^*(4m^2-s,0,s)
 \ee
 and
  \be
 -2\,\text{Im}\,T_{t}(4m^2-s,s,0)=T_{t}(4m^2-s,s,0)T_t^*(4m^2-s,s,0)\,,
 \ee
 
 to relate $\text{Im}\,t_{J}(s)$ to $|t_{J}(s)|^2$ over the left cut (remember that $\text{Im}\,G=-(t_0^2\,\text{Im}\, t)/|t|^2$), and use derivatives of these expressions to address eq.~(\ref{leftcutt}). We may pursue this line of thought in future work.

\printbibliography
\end{document}